\documentclass{emulateapj}

\shorttitle{HAe/Be Spectropolarimetry II. Survey Comparisons.}
\shortauthors{Harrington \& Kuhn}

\begin{document}

\title{Spectropolarimetric Observations of Herbig Ae/Be Stars. II. Comparison of Spectropolarimetric Surveys: HAeBe, Be and Other Emission-Line Stars.}
\author{D. M. Harrington and J.R. Kuhn}
\affil{Institute for Astronomy, University of Hawaii, Honolulu-HI-96822}
\email{dmh@ifa.hawaii.edu}

\begin{abstract}

	The polarization of light across individual spectral lines contains information about the circumstellar environment on very small spatial scales. We have obtained a large number of high precision, high resolution spectropolarimetric observations of Herbig Ae/Be, Classical Be and other emission-line stars collected on 117 nights of observations with the HiVIS spectropolarimeter at a resolution of R=13000 on the 3.67m AEOS telescope. We also have many observations from the ESPaDOnS spectropolarimeter at a resolution of R=68000 on the 3.6m CFH telescope. In roughly $\sim$2/3 of the so-called ``windy'' or ``disky'' Herbig Ae/Be stars, the detected H$_\alpha$ linear polarization varies from our typical detection threshold near 0.1\% to over 2\%. In all but one HAe/Be star the detected polarization effect is not coincident with the H$_\alpha$ emission peak but is detected in and around the obvious absorptive part of the line profile. The qu-loops are dominated by the polarization in this absorptive region. In several stars the polarization varies in time mostly in the absorptive component and is not necessarily tied to corresponding variations in intensity. This is a new result not seen at lower resolution. In the Be and emission-line stars, 10 out of a sample of 30 show a typical broad depolarization effect but 4 of these 10 show weaker effects only visible at high resolution. Another 5 of 30 show smaller amplitude, more complex signatures. Six stars of alternate classification showed large amplitude (1-3\%) absorptive polarization effects. These detections are largely inconsistent with the traditional disk-scattering and depolarization models.	
		
\end{abstract}


\section{Introduction}

	Understanding how stars interact with their environment is a general problem important for many astrophysical systems. Many types of stars are surrounded by shells, envelopes or disks of material. Such circumstellar material often participates in accretion, polar outflows, winds and disk-star interactions. In young stars, these processes directly effect star and planet formation and evolution. Even for the closest young stars we study here, the stellar radius is an important spatial scale that is smaller than 0.1 milliarcseconds and it will not be imaged directly even by the next generation of telescopes. Other methods like interferometry can yield useful constraints, but the indirect techniques of spectroscopy and spectropolarimetry have much to offer. In particular, spectropolarimetric measurements can put unique constraints on the geometry of the system as well as thermodynamic and radiative environments of circumstellar material. 

	High-resolution linear spectropolarimetry measures the change in linear polarization across a spectral line. Typical spectropolarimetric signals are very small, often a few tenths of a percent change in polarization across a spectral line. Measuring these signals requires very high signal to noise observations and careful control of systematics to measure signals at this level. In order to address these issues, we built a dedicated spectropolarimeter for the High-resolution Visible and Infrared Spectrograph (HiVIS) on the 3.7m AEOS telescope and performed a telescope polarization calibration and cross-instrument comparison (Harrington et al. 2006, Harrington \& Kuhn 2008). In addition, new and archival observations with the ESPaDOnS spectropolarimeter will be presented which will complement our HiVIS observations. 
	
	Many models show spectropolarimetry is a potentially useful probe of circumstellar environments at small spatial scales. Circumstellar disks, rotationally distorted winds, magnetic fields, asymmetric radiation fields (optical pumping), and in general, any scattering asymmetry can produce a change in linear polarization across a spectral line (cf. McLean 1979, Wood et al. 1993, Harries 2000, Ignace et al. 2004, Vink et al. 2005a, Kuhn et al. 2007). These spectropolarimeric signatures can directly constrain the physical properties of the circumstellar environment once sufficiently detailed models are developed. There have been several studies of young stars to date using medium resolution spectropolarimetry (Vink et al. 2002, 2005b, Mottram et al. 2007). These studies focused on young stars and are based on only a few nights of observations. Though spectropolarimetric signatures have been detected in many classes of star, a many-epoch, high-resolution comparison of different stellar classes has yet to be done.
			
	We collected a large number of stellar observations on over 100 nights from 2004 to 2008, monitoring 29 Herbig Ae/Be stars and 30 Be and emission-line stars. These stellar types have very different circumstellar environments with various combinations of winds, disks, accretion, and jets (cf. Porter \& Rivinius 2003 and Waters \& Waelkens 1998). However, there is still much debate over the details of these processes. Thomson scattering theory is the current framework used to describe linear spectropolarimetric signatures. A ``depolarization" scattering theory of spectropolarimetric line profiles was originally developed for Be stars (McLean 1979). Another scattering model we will call the ``disk-scattering" model has been advanced for other star types. However, neither model significantly matches most of our Herbig Ae/Be observations, and many observations of other star types. 

The observing campaign and basic data processing will be presented in section 2 along with the spectroscopy of the Herbig Ae/Be stars. A quick overview of current spectropolarimetric theories will be presented in section 3. The HAe/Be spectropolarimetry will be detailed in section 4. All the HiVIS and ESPaDOnS data will be presented with individual examples for each star. Section 5 will present the spectroscopy and spectropolarimetry of the Be and Emission-line stars for comparison. A comparison with other stellar types will also be made. These observations will then be compared and contrasted in the discussion of section 6.

\section{The Observing Campaign}

	Spectropolarimetry is a photon-hungry technique and the AEOS telescope suffers from significant pointing-dependent polarization properties. The robust results presented here have been obtained only after many epochs of observations on most targets (Harrington et al. 2006, Harrington \& Kuhn 2008). We collected observations of 29 Herbig Ae/Be stars from 2004 to 2008. The minimum acceptable signal-to-noise ratio to detect a 1\% signature is roughly 300. Our HiVIS observations have a raw signal-to-noise ratio of 100-1000 at a spectral resolution of 13000 with a 1.5'' slit. In using the 1.5'' slit, the detector significantly over-samples the spectral orders (see Harrington \& Kuhn 2008). Thorium-argon lines have a 16-pixel full-width-half-max near H$_\alpha$. This data is typically binned to lower sampling while preserving full spectral resolution in order to achieve high precision measurements at full spectral resolution. 
	
	In some targets there is obvious need for higher resolution spectropolarimetry. Our HiVIS data has been confirmed and extended using our own and archival ESPaDOnS observations. These have even higher signal-to-noise ratios of 700-3000 and will be presented unbinned at a full spectral resolution of 68000. This large dataset with high resolution is reliable for the detection of a diverse set of stellar spectropolarimetric signatures and robust detection statistics. In this section, the observing campaign, spectroscopy and information about individual Herbig Ae/Be stars is presented. The HiVIS instrument properties, telescope polarization calibration and reduction methods are outlined completely in Harrington \& Kuhn 2008. All the ESPaDOnS data presented was reduced with the dedicated Libre-Esprit reduction software at CFHT (Donati et al. 1999).
	
	The spectropolarimetric survey was conducted over three years while simultaneously doing instrument development and engineering. In early 2004 the spectropolarimeter was being assembled. Some test observations were done during the engineering of the HiVIS spectropolarimeter in late 2004. During this period, the original science camera failed and was pulled off for repairs. In the summer of 2005 the reassembled camera was mounted. There was a large polarization calibration effort in this season associated with the spectropolarimetry of comet 9P/Tempel 1 for the Deep Impact event (Meech et al. 2005, Harrington et al. 2007). In mid-2006 there were a number of instrument fixes to the spectrograph. There were then two major Herbig Ae/Be observing seasons once the instrument configuration stabilized: September 2006 to February 2007 and July 2007 to January 2008. Table \ref{obslog} lists the observing and engineering dates. Each line gives a description of what was done on each night. There were a number of nights spent on component testing or experimental programs such as the initial spectrograph, spectropolarimeter and IR spectrograph testing. There were also tests of the IR dichroic, new spectrograph components, and sensitivity experiments. There are also notes in the table showing when the various improvements, tests, and repairs took place.  
	
	In the first part of the survey, few bright targets were chosen for close monitoring.  AB Aurigae and MWC 480 were monitored almost every night over the 2006-2007 winter.  During the same period, MWC 120, MWC 158 and HD 58647 were less well covered, but still monitored heavily. AB Aurigae or MWC 480 were observed continuously for several hours on some nights, and all 5 targets intermittently on others.  There are a total of 148 polarization measurements for AB Aurigae, 58 for MWC 480, 24 for MWC120, 39 for MWC 158, 19 for HD 58647, plus another 33 unpolarized standard star observations taken over the 40 nights in winter 2006-2007.    
	
	The H$_\alpha$ line for these five stars showed significant variability in intensity, width, and profile shape that is entirely consistent with other spectroscopic variability studies (cf. Beskrovnaya et al. 1995, Beskrovnaya \& Pogodin 2004, Catala et al. 1999).  On some nights, AB Aurigae and MWC480 showed dramatic spectroscopic variation on a timescale of minutes to hours, mostly in the blueshifted absorption trough.  Some general line width and line strength variability was also seen on short timescales.  All nights showed much smaller but significant variation.  
	
	The campaign was broadened in July 2007 to include many other Herbig Ae/Be stars. During the summer, HD 163296, HD 179218, HD 150193 and V1295 Aql were chosen for closer study. Another 111 unpolarized standard star observations were also taken since the instrument had changed significantly. Table \ref{tab-obs} shows a list of the Herbig Ae/Be stars and their properties taken from the literature. During the fall of 2007 and winter of 2008, a much larger target list of Herbig Ae/Be stars was used as well as bright Be and emission-line stars for comparison. Throughout the survey MWC 361 was used as a detailed stable comparison between HiVIS and ESPaDOnS. All of the H$_\alpha$ line profiles for the Herbig Ae/Be stars are shown in figures \ref{fig:haebe-lprof1} and \ref{fig:haebe-lprof2}

\begin{table*}
\begin{center}
\vspace{-5mm}
\caption[Observing Log]{Observing Log \label{obslog}}
\begin{scriptsize}
\begin{tabular}{ll|ll}
\hline
{\bf Date} & {\bf Description} & {\bf Date}& {\bf Description}           \\
\hline
\hline
2004 02 27 & Testing Specpol				        & 2006 11 23 & Bad Weather \\									      	
2004 04 29 & Testing Specpol					& 2006 11 28 & Specpol	\\									      
2004 05 26 & Testing Specpol, Waveplate				& 2006 11 29 & Specpol	\\									      
2004 06 23 & Testing Specpol, Waveplate				& 2006 11 30 & Specpol	\\									      
2004 08 18 & Specpol						& 2006 12 09 & Specpol	\\									      
2004 08 19 & Specpol						& 2006 12 20 & Pixis Shutter Died \\							      
2004 08 22 & Specpol						& 2006 12 23 & IR 	\\							      
2004 10 04 & HiVIS Shakedown					& 2006 12 27 & Install Spare Pixis, Specpol	\\						      
2004 10 06 & Specpol						& 2006 12 28 & Specpol					\\					      
2004 10 07 & Specpol						& 2006 12 29 & Specpol                                    \\        				      
2004 10 13 & Bad Weather						      & 2007 01 02 & Specpol				         \\			      
2004 10 20 & Specpol							      & 2007 01 03 & Specpol, IR			      	 \\
2004 10 22 & Specpol						      & 2007 01 04 & Bad Weather				      	 \\			      
2004 10 28 & Bad Weather,  IterGain Testing			      & 2007 01 05 & Bad Weather				      	 \\
2004 10 29 & Bad Weather,  IterGain Testing				      & 2007 01 09 & Bad Weather				      	 \\
2004 11 26 & Specpol, Unpol, IterGain Test				      & 2007 01 10 & Bad Weather				      	 \\
2004 11 27 & Specpol, Unpol, IR Dichroic	              	      & 2007 01 17 & Specpol				      	 \\			      
2004 12 02 & Specpol, IR Testing					      & 2007 01 18 & Specpol, IR   	 \\	
2004 12 03 & Bad Weather						      & 2007 01 19 & Specpol				      	 \\			      
2004 12 09 & Specpol							      & 2007 01 29 & IR 			      	 \\			      
2004 12 10 & Specpol							      & 2007 01 30 & IR			      	 \\				      
2004 12 15 & Specpol							      & 2007 02 04 & IR			      	 \\				      
2004 12 16 & Bad Weather						      & 2007 02 05 & IR			      	 \\				      
&									      & 2007 02 06 & IR			      	 \\				      
&  									      & 2007 02 08 & Bad Weather				      	 \\		      
2005 06 27  & Reassemble VisCam, Shakedown.				      & 2007 06 18 & Specpol		      	 \\			      
2005 06 28  & Realign Polarimeter,  Efficiency		                   &2007 06 19 & Specpol				      	 \\			      
2005 06 30  & Unpol Stds, Gain Cal, Sky pol, Tempel			      & 2007 06 20 & Specpol				      	 \\			      
2005 07 01  & Sky pol, Unpol Stds, Tempel 				      & 2007 06 21 & Specpol				      	 \\			      
2005 07 02  & Sky pol, Unpol Stds					      & 2007 07 27 & Sky pol, Bad Weather			      	 \\		      
2005 07 03  & Sky pol, Unpol Stds, Tempel				      & 2007 07 28 & Sky pol, Specpol			      	 \\			      
2005 07 04  & Sky pol, Unpol Stds, Tempel	 	      & 2007 07 31 & Specpol 				      	 \\			      
2005 07 05  & Sky pol, Tempel						      & 2007 08 01 & Specpol				      	 \\			      
2005 07 06  & Tempel						 	        & 2007 08 28 & Specpol				      	 \\			      
2005 07 07  & Tempel, Specphot, Sky glow			                   & 2007 08 29 & Specpol			      	 \\		      
&									      & 2007 08 30 & Specpol				      	 \\			      
&  								                      & 2007 08 31 & Bad Weather				      	 \\			      
2006 05 08 & IR Shakedown - SlitViewer, VidFeed		       	      & 2007 09 01 & Bad Weather   \\		      
2006 05 09 & IR testing - Chip noise, 		        	      & 2007 08 17 & Bad Weather				      	 \\		      
2006 05 15 & IR testing - Chip noise, J Sensitiv        	    & 2007 09 17 & AEOS Down - Engineering	      	 \\			
2006 05 16 & IR testing - Bkgnd, H\&K sensitivity 	       	      & 2007 09 18 & Specpol		      	 \\		      
2006 08 09 & Specpol, Unpol 						      & 2007 09 19 & Bad Weather				      	 \\		      
2006 08 16 & Bad Chip RFI - No Observing				      & 2007 09 20 & Specpol			      	 \\			      
2006 08 21 & Specpol							      & 2007 09 21 & Specpol				      	 \\			      
2006 08 26 & Bad Weather						      & 2007 10 30 & Specpol				      	 \\			      
2006 09 06 & Bad Weather					              & 2007 10 31 & Specpol				      	 \\			      
2006 09 12 & Fix Vis Leach, Specpol, Unpol				      & 2007 11 01 & Bad Weather	    	 \\		      
2006 09 13 & Bad Chip RFI - No Observing			& 2007 11 02 & Bad Weather				       \\				      
2006 09 14 & Bad Chip RFI - No Observing			              & 2007 11 03 & Bad Weather      \\		      
2006 09 18 & Pixis Mounted, Specpol, Unpol			          & 2007 11 21 & Specpol				      	 \\			      
2006 09 19 & Specpol, Unpol                                     & 2007 11 22 & Bad Weather				      	 \\				      
2006 09 21 & Specpol, Unpol						      & 2007 11 23 & Specpol				      	 \\			      
2006 09 22 & Specpol, Unpol						      & 2007 11 24 & Specpol				      	 \\			      
2006 09 28 & Pixis Remounted, Specpol					      & 2007 11 27 & Bad Weather			      	 \\		      
2006 10 27 & Specpol, VisMotor, PolMount, Grating	        	      & 2007 11 28 & Bad Weather				      	 \\		      
2006 11 03 & Bad Weather						      & 2008 01 14 &	Specpol			      	 \\			      
2006 11 07 & Specpol, Unpol Stds				      & 2008 01 15 & Specpol				      	 \\				      
2006 11 17 & Specpol, CalStage Lenses, New Dekkar	    & 2008 01 16 & Specpol					      	 \\
2006 11 18 & Specpol							      & 2008 01 17 & Specpol			      	 \\			      
2006 11 21 & Specpol						      & 2008 01 19 &  Specpol   \\			      
2006 11 22 & Specpol						         & & 								 \\			      
\hline
\end{tabular}
\end{scriptsize}
\end{center}
\end{table*}

\begin{table*}[!h,!t,!b]
\begin{center}
\begin{footnotesize}
\caption{Herbig Ae/Be Stellar Properties \label{tab-obs}}
\begin{tabular}{lrccccrccccccc}
\hline
\hline
{\bf Name}  & {\bf HD} &  {\bf MWC} & {\bf V} & {\bf ST} & {\bf $T_{eff}$} & {\bf $M_\odot$} & {\bf D} & {\bf B(G)} & {\bf Age} & {\bf R} & {\bf  L} \\
\hline
\hline
MWC442    & 13867    &   442    &7.7      &B5Ve           &             &                  &                  &                   &                   &              &             \\
AB Aur        & 31293     &  93       & 7.1     & A0Vpe        &  9.6     & 2.4            & 144         & $<$100    & 2.0            &  2.7v     & 1.68           \\
MWC480    & 31648     &  480    & 7.7     & A3pshe+    & 8.7      &  2.2           & 131         & $<$100    & 2.5             &               & 1.51            \\
HD35187   &  35187    &              &7.8      &A2e             & 9.5       &                  &                  &                  &                    &             & \\
HD35929    & 35929    &               &8.1     &A5               &  7.2      &                   &$>$360   &                  &                    &             & \\
MWC758      & 36112    &   758    &8.3    & A3e            &  8.1      &  2.0            & 200         &                 & 3.1              &             & \\
KMS 27         & 37357    &              &8.9    &A0e             &             &                     &                &                 &                    &             & \\
MWC120     & 37806      &  120    & 7.9    & A0              & 8.9      & 3.0d           & $>$230  & $<$100  & 0.37e 2.6w& 2.4w   & $>$1.51     \\
HD38120     & 38120    &               &9.1     &A0              &             &                    &                 &                 &                    &             & \\
FS CMa       & 45677     &   142    & 8.1     & Bpshe       &  21.4   &                   &$>$300   &                 &                    &             & \\ 
MWC158     & 50138    &  158     & 6.6     & B9               &15.5     &  5.0         & 290        &                  & 0.10          &              &  2.85               \\
GU CMa       & 52721    & 164      &6.6      &B2Vne       & 21.9    &                     &               &                 &                    &            & \\
MWC 166     & 53367    &   166    &7.0     &B0IVe         &            & 31.6            &                &                 &                    &            & \\
27 CMa        & 56014     & 170      &  4.7   &  B3IIIe       &             &                     &                &                 &                    &            & \\
HD58647    & 58647       &            & 6.8     & B9IV          & 10.7   & 4.2              & 280         &                & 0.16           &              & 2.48                \\
HD141569  & 141569    &            & 7.0     & B9.5e         & 6.3      & 2.3            &  99           &                &  $>$10       &            & \\
HD142666  & 142666    &            &8.8      & A8Ve          &             &                  &                  &               &                    &             & \\ 
HD144432  & 144432    &            &8.2       & A9/F0V     &  8.1       &                 & $>$200   &               &                     &            & \\
V2307 Oph & 150193     & 863    &  8.9     & A1Ve         & 9.3       &  2.3          & 150         &               & $>$6.3       &             & 1.47            \\
51Oph          & 158643    &            &4.8       &A0V            &  10.0    &   4.0         & 131         &               &  0.31           &            & \\
MWC275    & 163296     &  275   & 6.9      & A1Ve          & 9.3      & 2.3            & 122        & -60h      & 3.00e          &             &   1.48                  \\
HD169142  & 169142    & 925    &8.2      &B9Ve           &               &                 &                &                &                     &            & \\
MWC614    & 179218     &  614   & 7.2     & B9e             &  10.5     & 4.3          & 240        &                & 1.06e          &             &   2.50                 \\
V1295 Aql   & 190073     & 325   & 7.8     & A2IVpe      & 8.9         & 2.85c       &$>$290  & $<$100 & 1.0e 1.3w & 3.3w    &                           \\
MWC361    & 200775     &  361   & 7.4     &  B2Ve        &  20.4      &  10           & 430        &                & 0.02           &              &    3.89              \\
Il Cep           & 216629    & 1075   &9.3     &B2IV-Vne   &18.6       &                  &$>$160   &              &                      &            & \\
V700 Mon    &259431   & 147      & 8.8     &  B6pe        &              &                   &                &               &                      &            & \\
XY Per         & 275877   &              &  9.4   & A2IIv          & 14.1     &                   &                &                &                      &            & \\
T Ori              &                 &  763     &9.5     & A3V           &               &                   &                &               &                      &            & \\
\hline
\hline
\end{tabular}
\end{footnotesize}
\end{center}
\tablecomments{The names, HD and MWC catalog numbers are listed for each star. The V magnitudes and spectral types are from the Simbad catalog.  The effective temperature in thousands of Kelvin ($T_{eff}$), mass in solar units ($M_\odot$), distances in parsecs (D), magnetic field strengths in Gauss, B(G), age in millions of years (Age), radius compared to one solar radii (R) and log luminosity in solar units (L), and are from van den Ancker et al. 1998 unless noted.  Magnetic field strengths in Gasus, B(G), are from Wade et al. 2007 unless noted.  c is Catala et al. 2007, d is Bohm \& Catala 1995, e is Manoj et al. 2006, h is Hubrig et al. 2006 and w is Wade et al. 2007}
\end{table*}

\subsection{Comments on Individual Stars}
	
	In this section, we will discuss the properties of many individual targets and present the spectroscopy of several variable stars. The long-term targets AB Aurigae, MWC 480, MWC 120, MWC 158 and HD 58647 are quite different from one another. Some present stable evidence for strong winds with P-Cygni profiles while others show strong central reversals or variable absorption properties. HD 58647 was essentially invariant. Other stars were so variable that the overall morphology of the line changed significantly over a few days. Many stars deserve an in-depth discussion and a description of other significant results from the literature. This is presented below.

\subsection{AB Aurigae - HD 31293 - MWC 93}

	The Herbig Ae star, AB Aurigae (HD31293, HIP22910) is the brightest of the Northern hemisphere Herbig Ae stars (V=7.1) and is one of the best studied intermediate-mass young stars. It has a near face-on circumstellar disk resolved in many wavelengths (eg: Grady et al. 2005, Fukagawa et al. 2004). It also has an active stellar wind with it's strong emission lines often showing strong P-Cygni profiles. Spectroscopic measurements put AB Aurigae somewhere between late B and early A spectral types (B9 in Th\'{e} et al. 1994, B9Ve in Beskrovnaya et al. 1995, A0 to A1 Fernandez et al. 1995) with an effective temperature of around 10000K. Bohm \& Catala (1993, 1995) present a complete spectral atlas and followup work on stellar activity. 
	
	The star has a wind that is not spherically symmetric with a mass loss rate of order $10^{-8} {M_\odot}$ per year, and an extended chromosphere reaching $T_{max}\sim$17000K  at 1.5$R_\ast$ (Catala \& Kunasz 1987, Catala et al. 1999). However, Bouret et al. 1997 also found N V in the wind which require clumps of T$\sim$140,000K material at a filling factor of $\sim10^{-3}$. A short-term variability study done by Catala et al. (1999) showed that an equatorial wind with a variable opening angle, or a disk-wind originating 1.6$R_\ast$ out with a similar opening angle could explain the variability. Bohm et al. 1996 describe strong changes in the P-Cygni absorption with a 32-hour period. There is evidence presented in Hubrig et al. 2006b showing AB Aurigae has significant inhomogeneities in the distribution of elements in its atmosphere. Baines et al. 2006 present spectroastrometry also showing evidence for binarity. Wade et al. 2007 did not detect a magnetic field and had an upper limit of roughly 50-250G depending on the field type. Rodgers 2001 derive an accretion rate of -6.85 Log(M$_\odot$/yr) using Bracket-$\gamma$ emission at 2.166$\mu$m.

	We have studied this star in detail and have observed P-Cygni profiles in H$_\alpha$. The spectroscopic variability was presented in Harrington \& Kuhn 2007. In complete agreement with many other spectroscopic studies, there was short time-scale variability (10-minute) in the blue-shifted absorptive component of the P-Cygni profile. Over time-scales of days to years the line varied in overall intensity and width as well.

\begin{figure*}
\begin{center}  
\includegraphics[width=0.24\linewidth, angle=90]{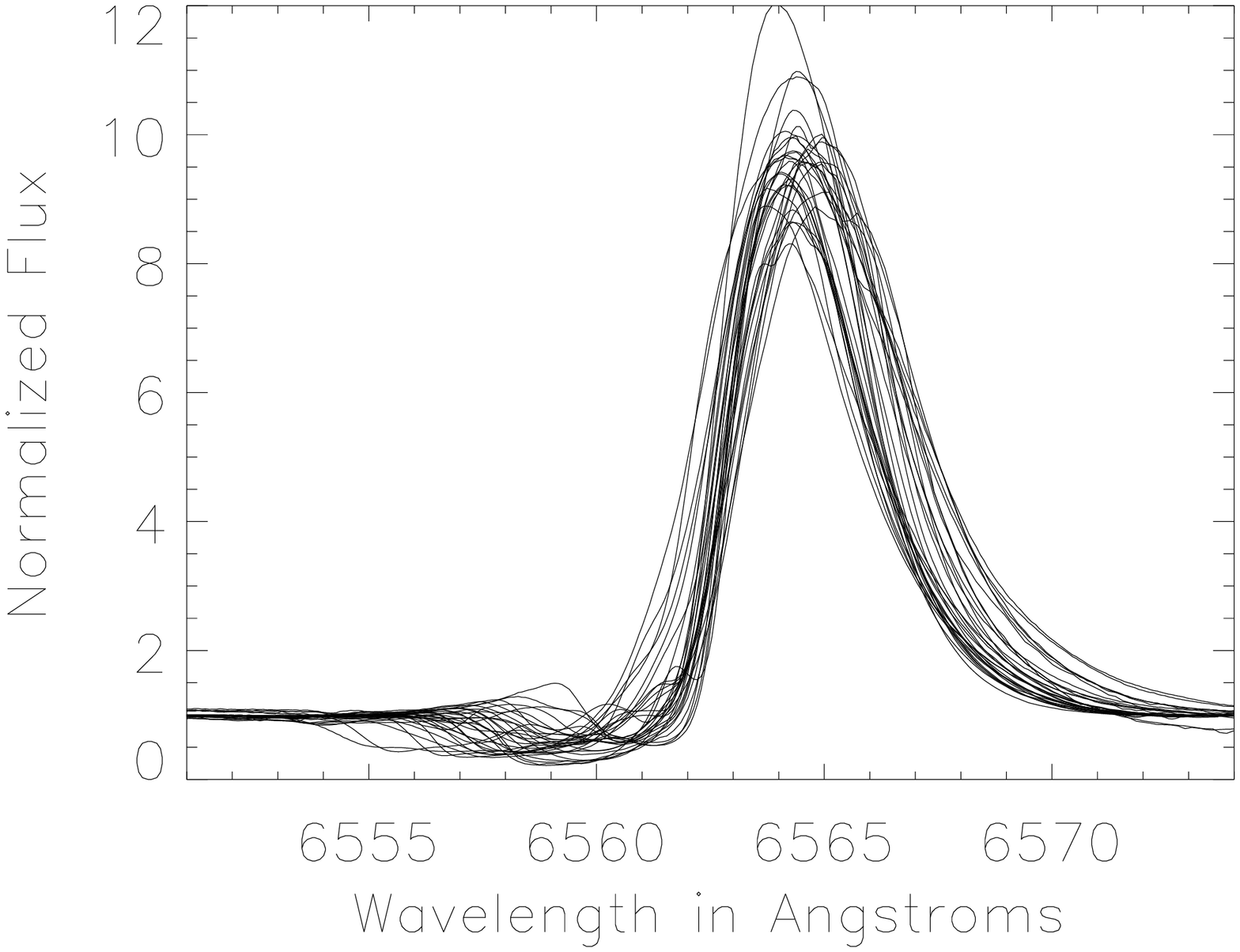}  
\includegraphics[width=0.24\linewidth, angle=90]{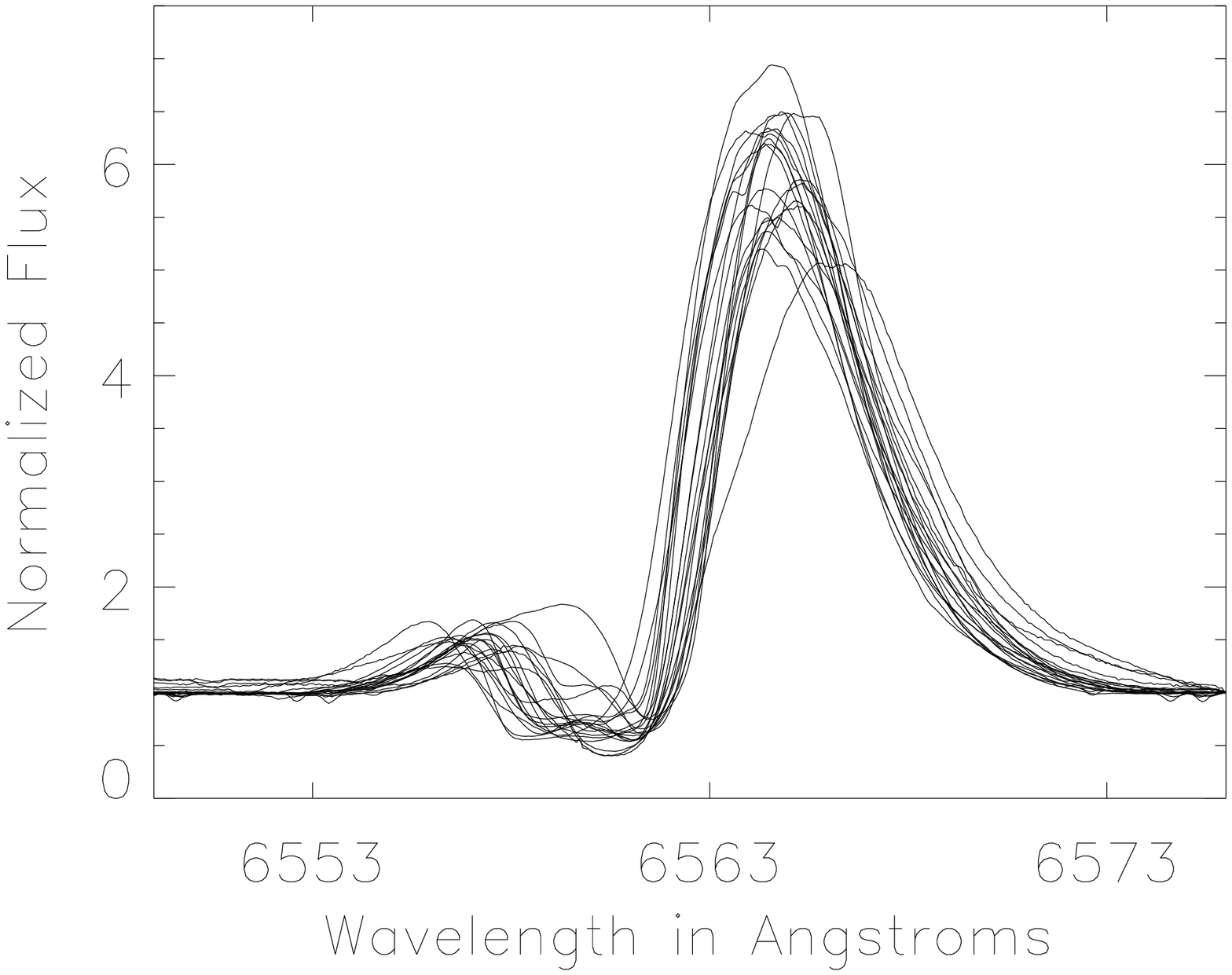}  
\includegraphics[width=0.24\linewidth, angle=90]{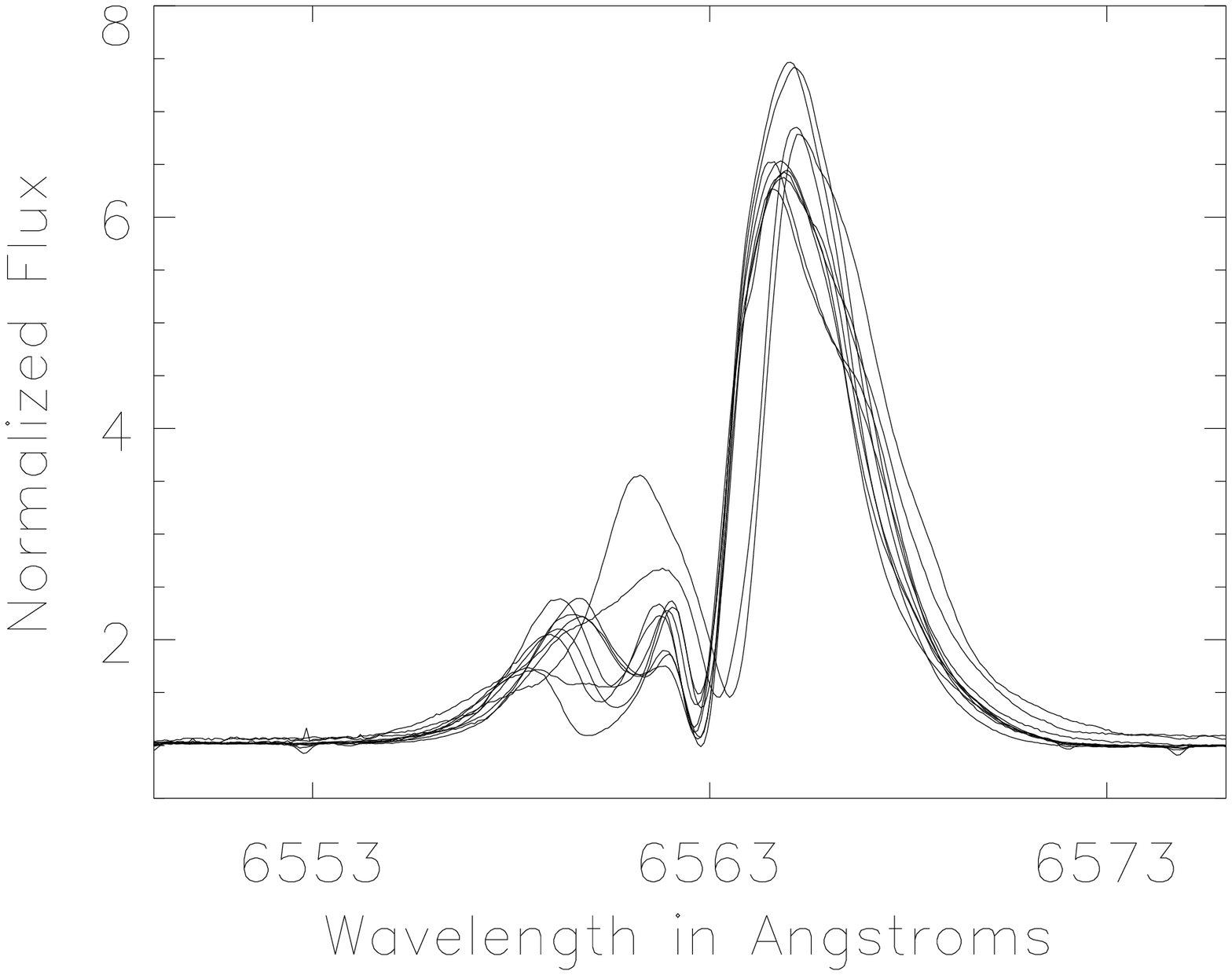}  \\
\includegraphics[width=0.24\linewidth, angle=90]{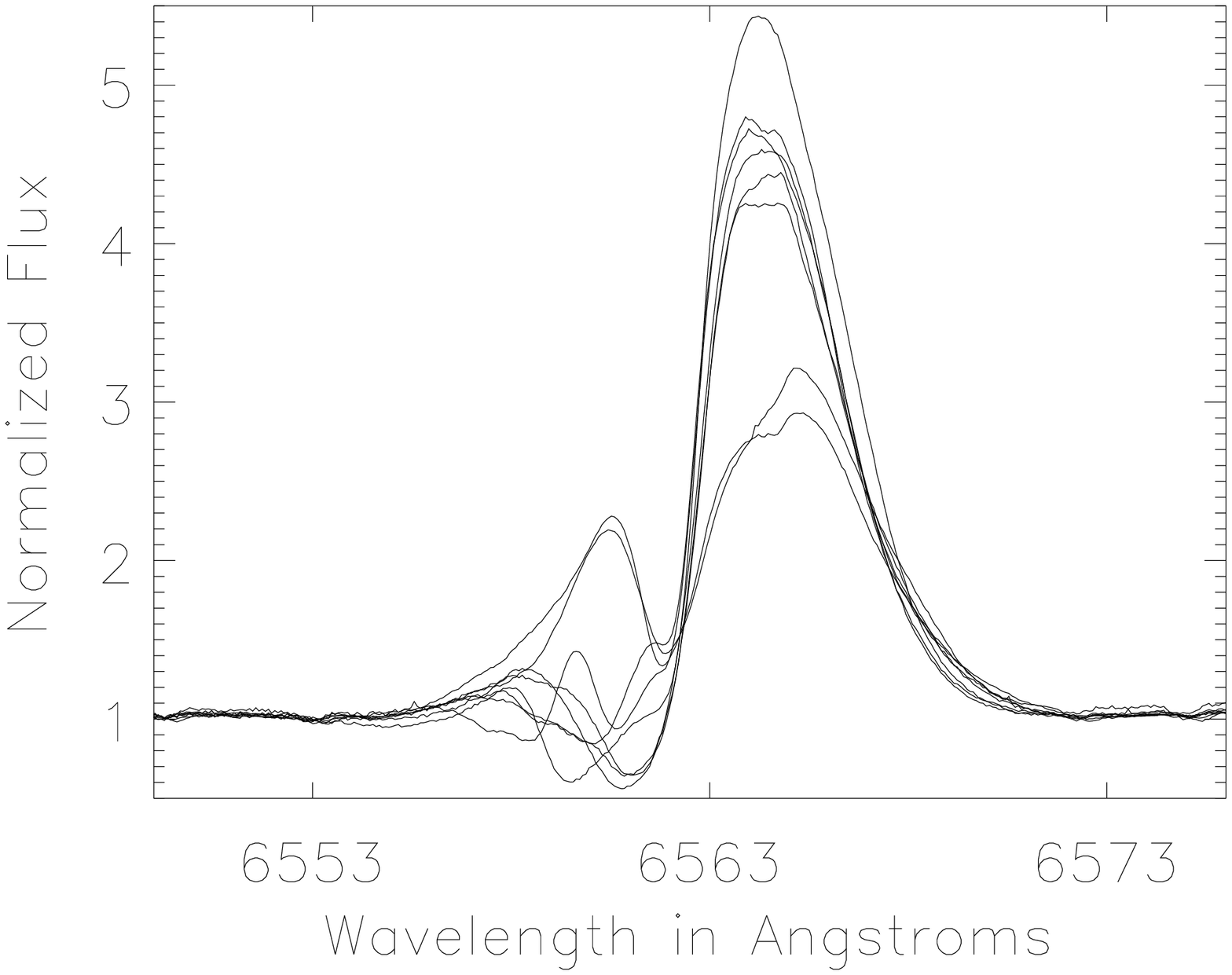}  
\includegraphics[width=0.24\linewidth, angle=90]{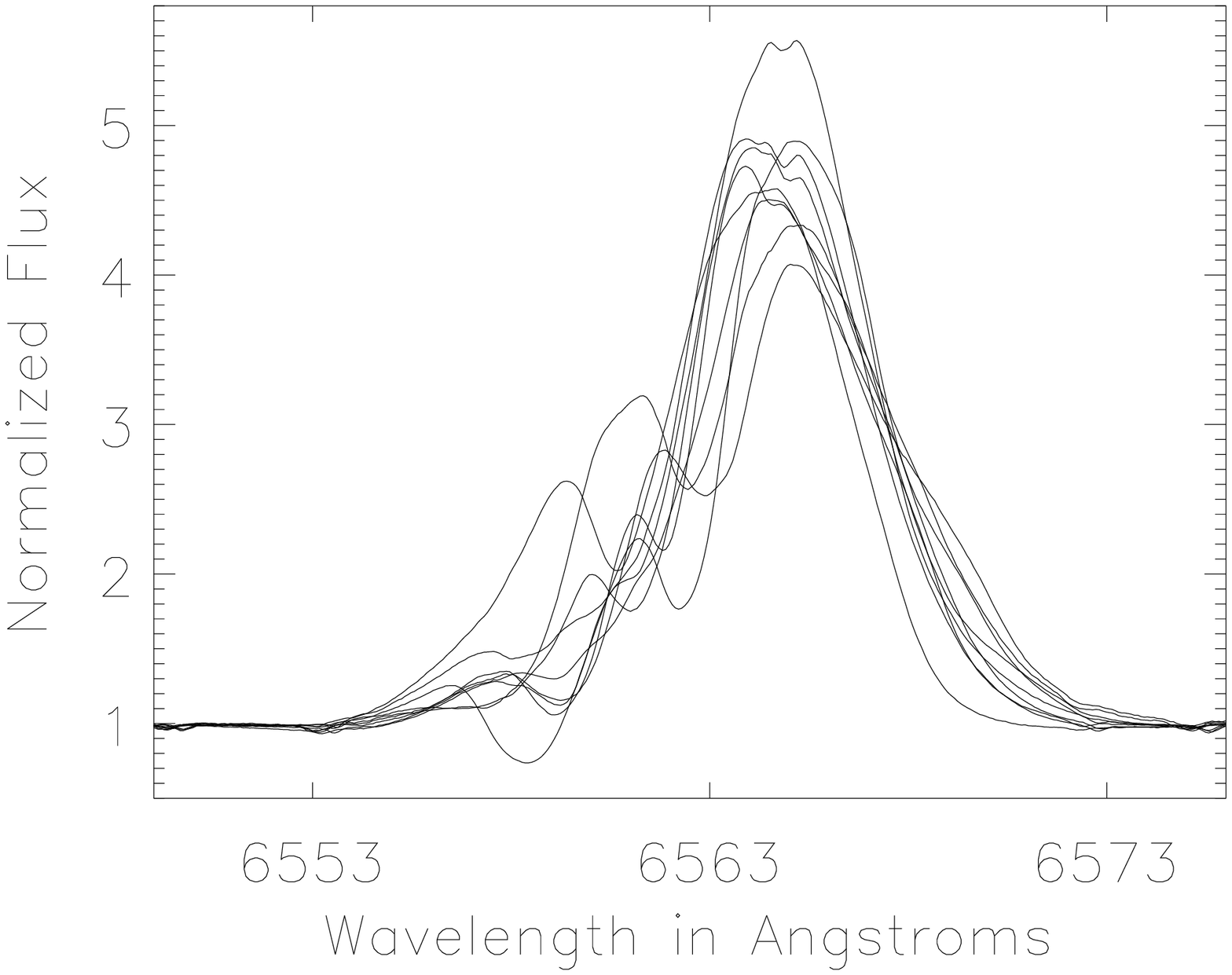}  
\includegraphics[width=0.24\linewidth, angle=90]{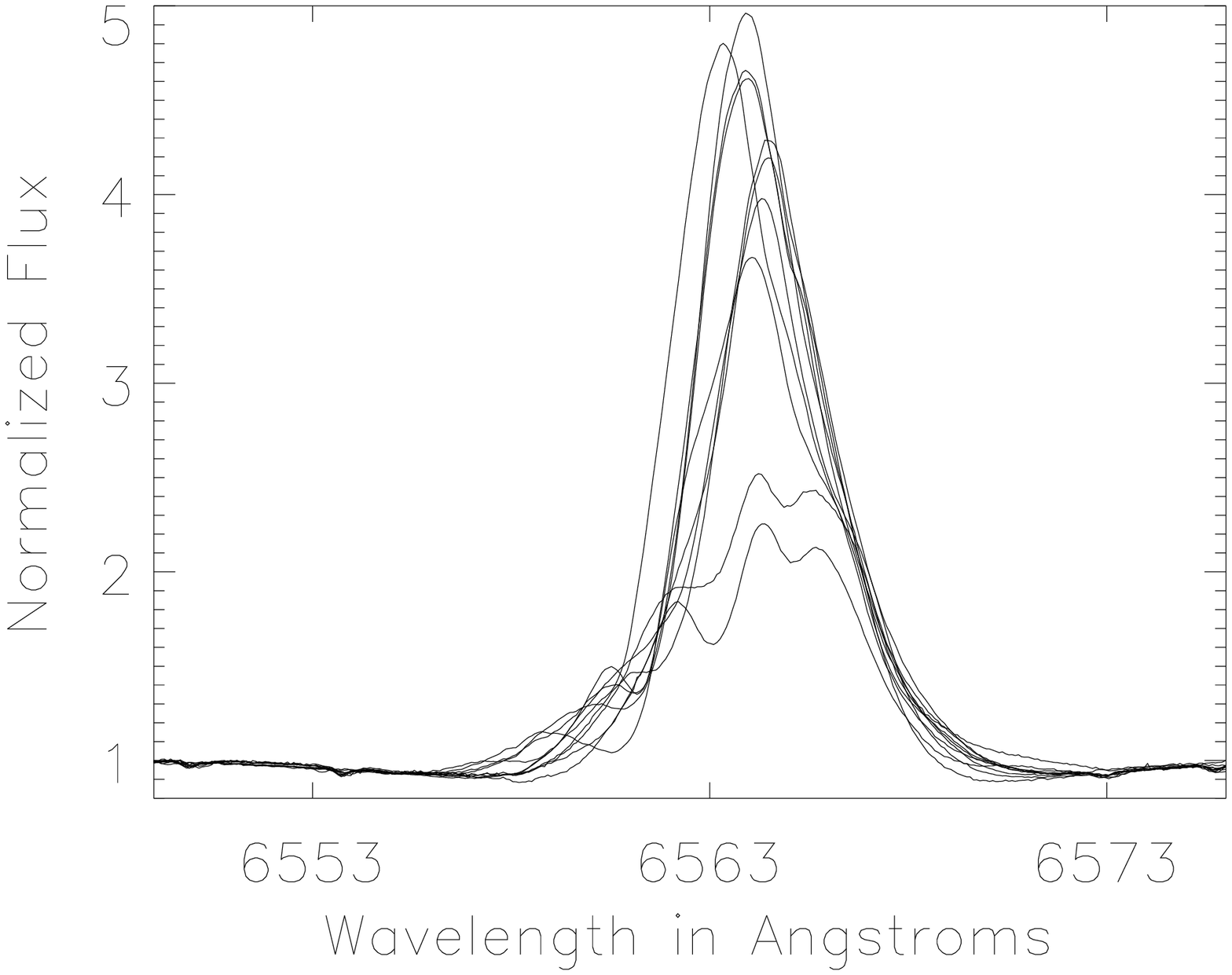}  \\
\includegraphics[width=0.24\linewidth, angle=90]{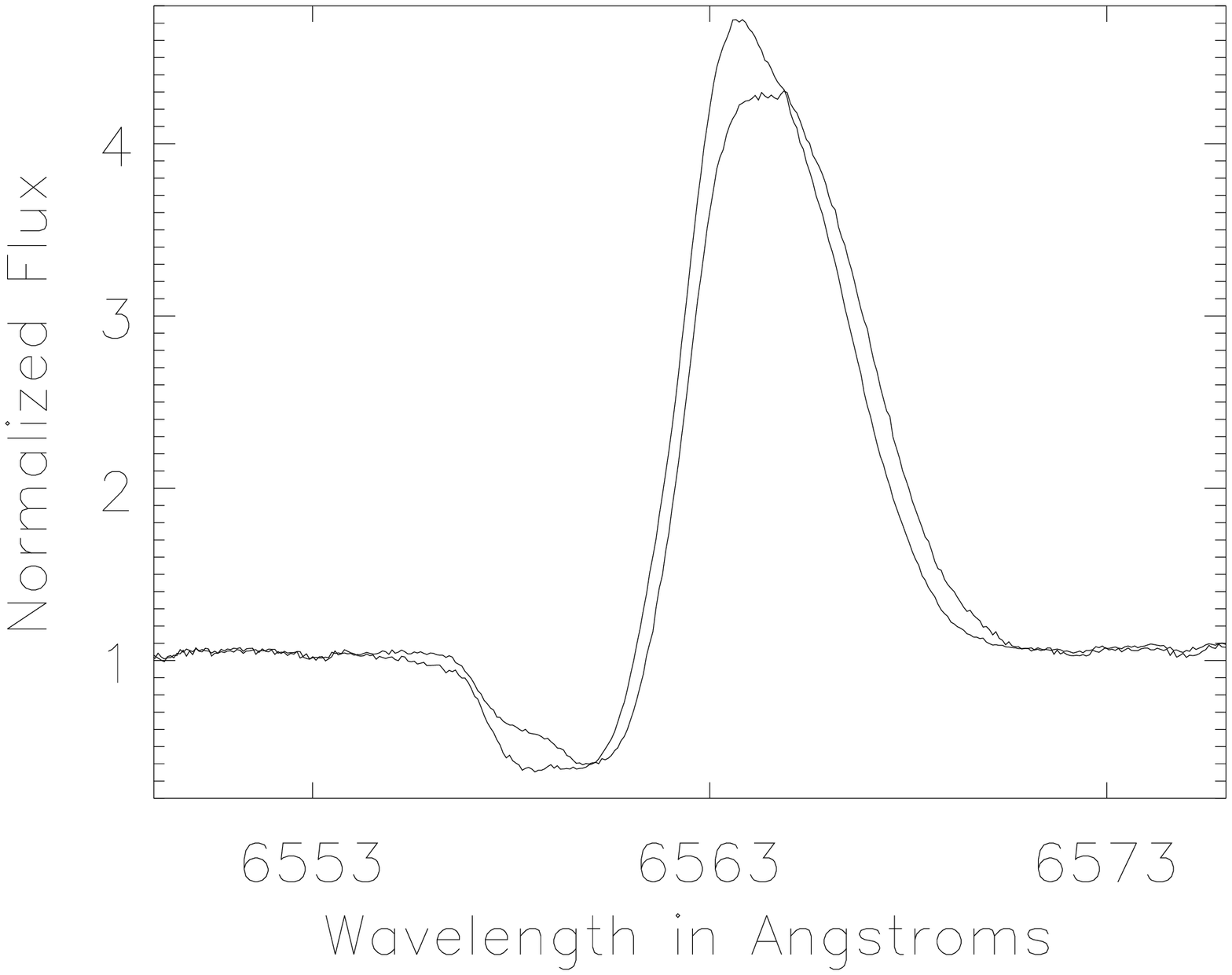} 
\includegraphics[width=0.24\linewidth, angle=90]{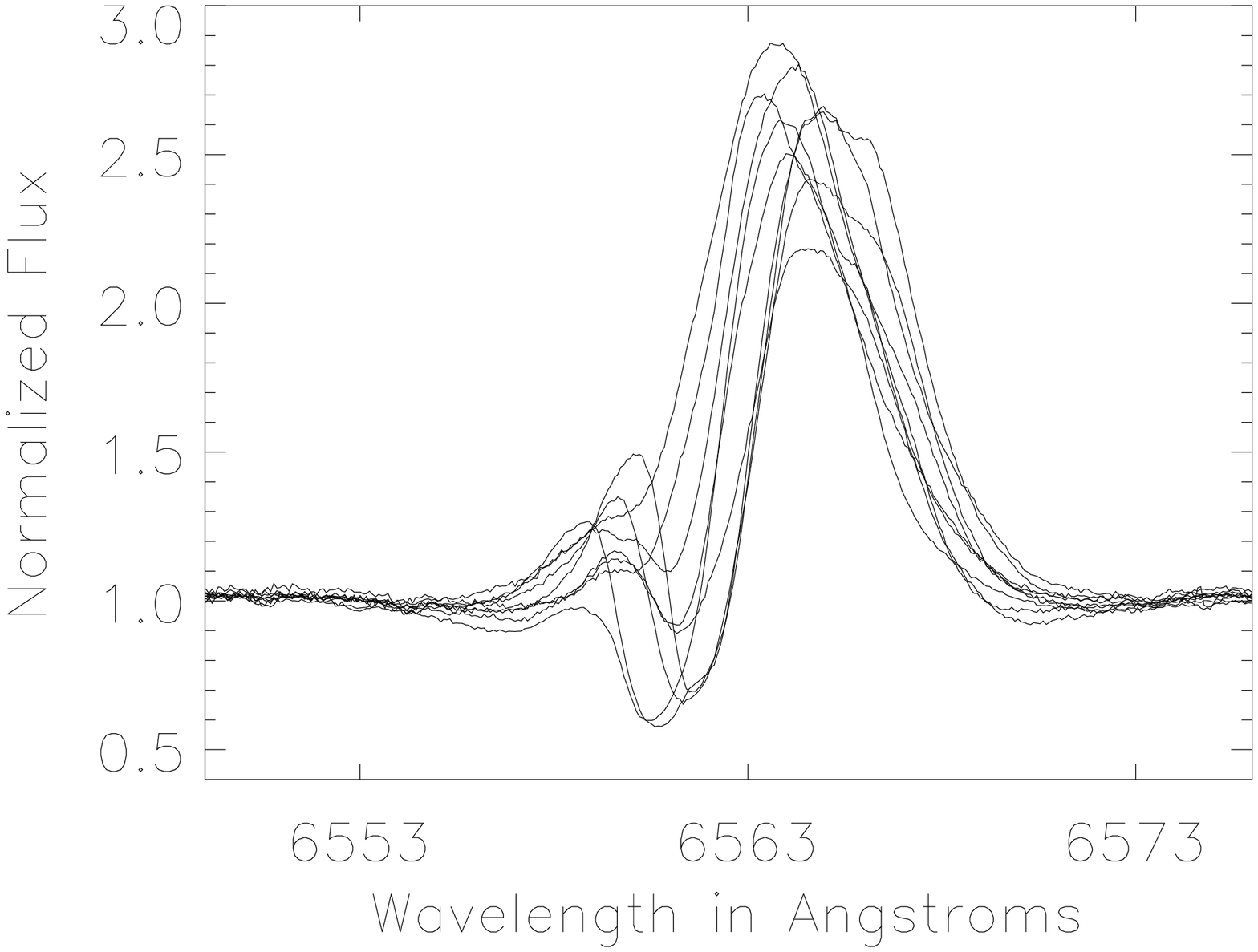}
\includegraphics[width=0.24\linewidth, angle=90]{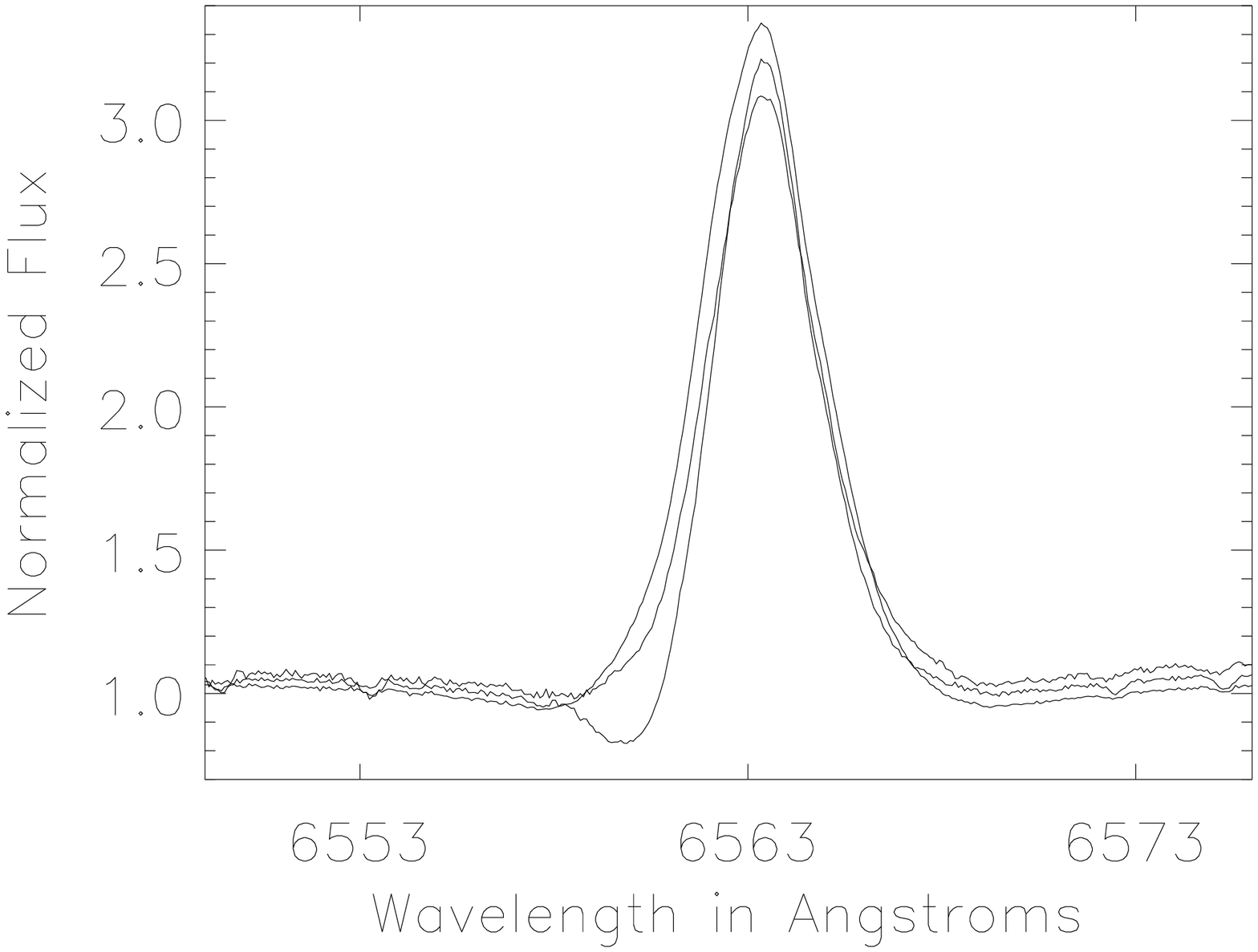}  \\
\includegraphics[width=0.24\linewidth, angle=90]{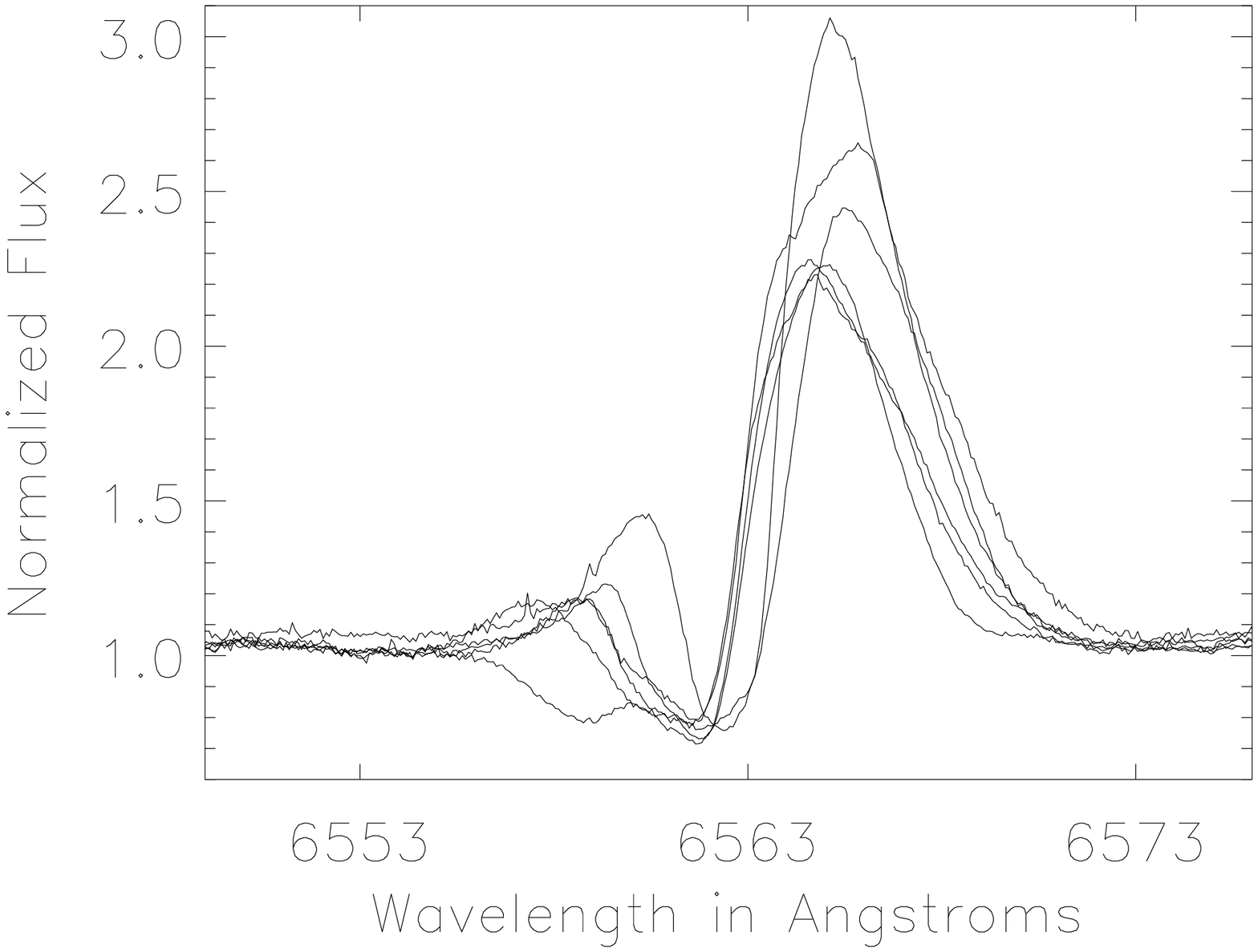}
\includegraphics[width=0.24\linewidth, angle=90]{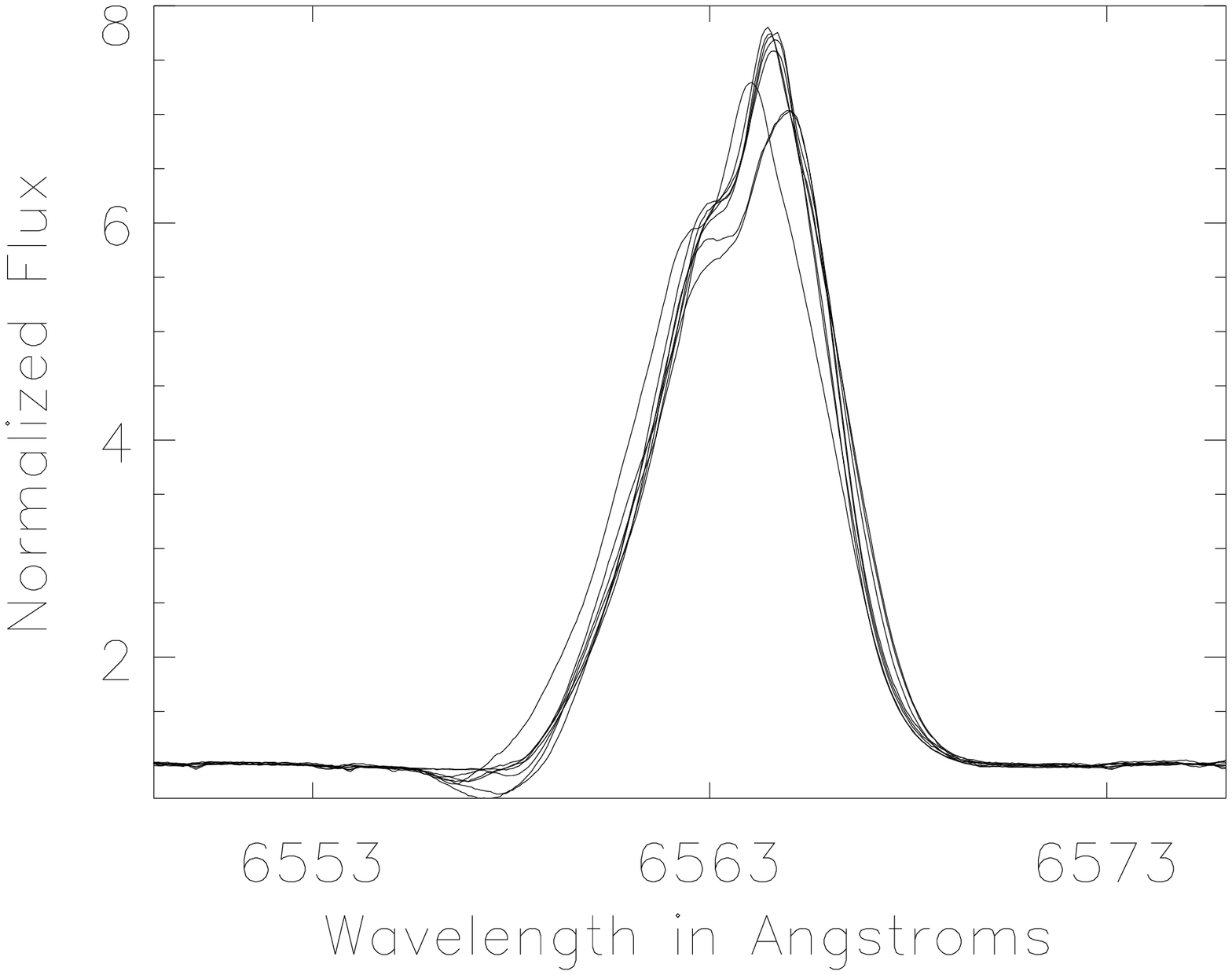}
\includegraphics[width=0.24\linewidth, angle=90]{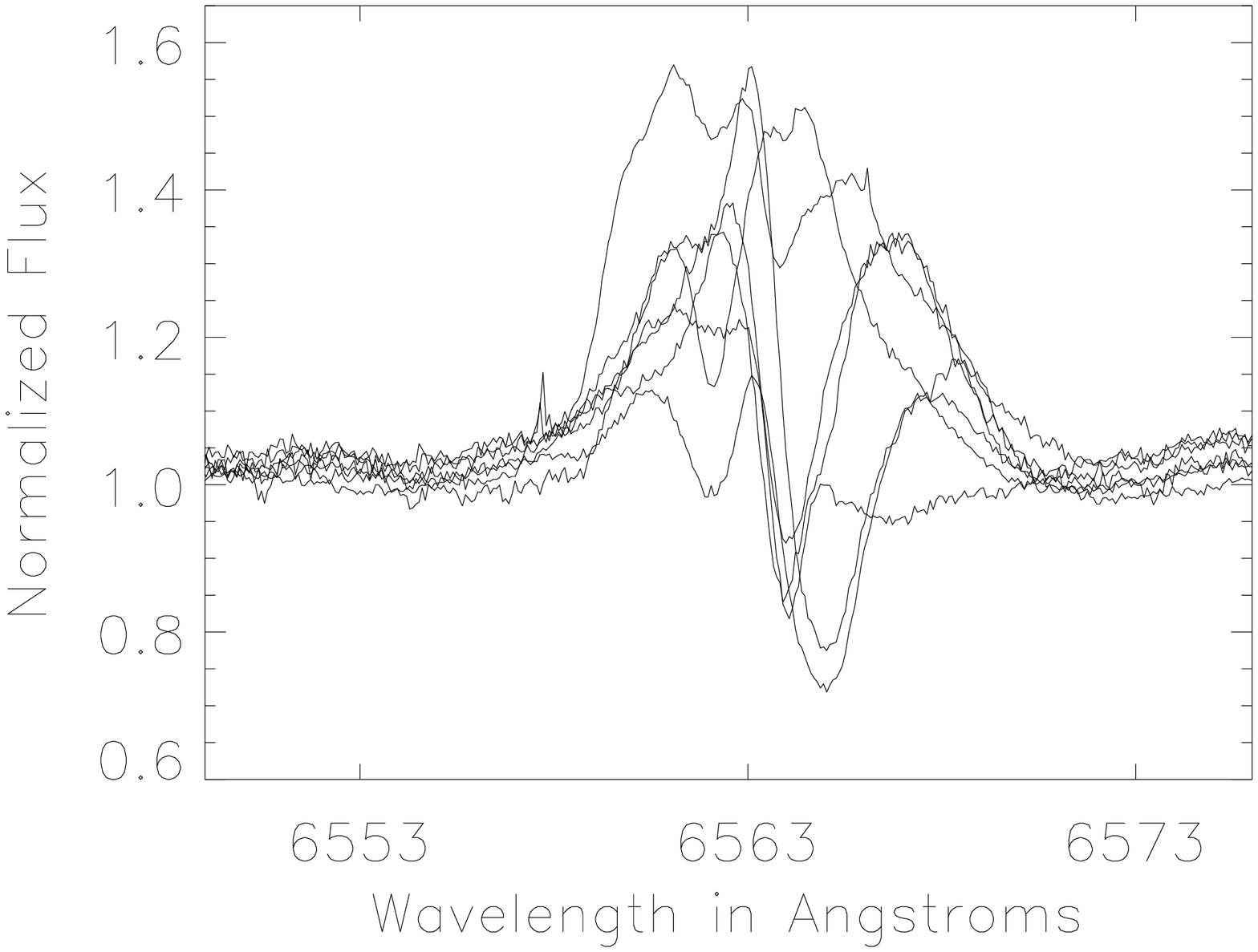}  \\
\includegraphics[width=0.24\linewidth, angle=90]{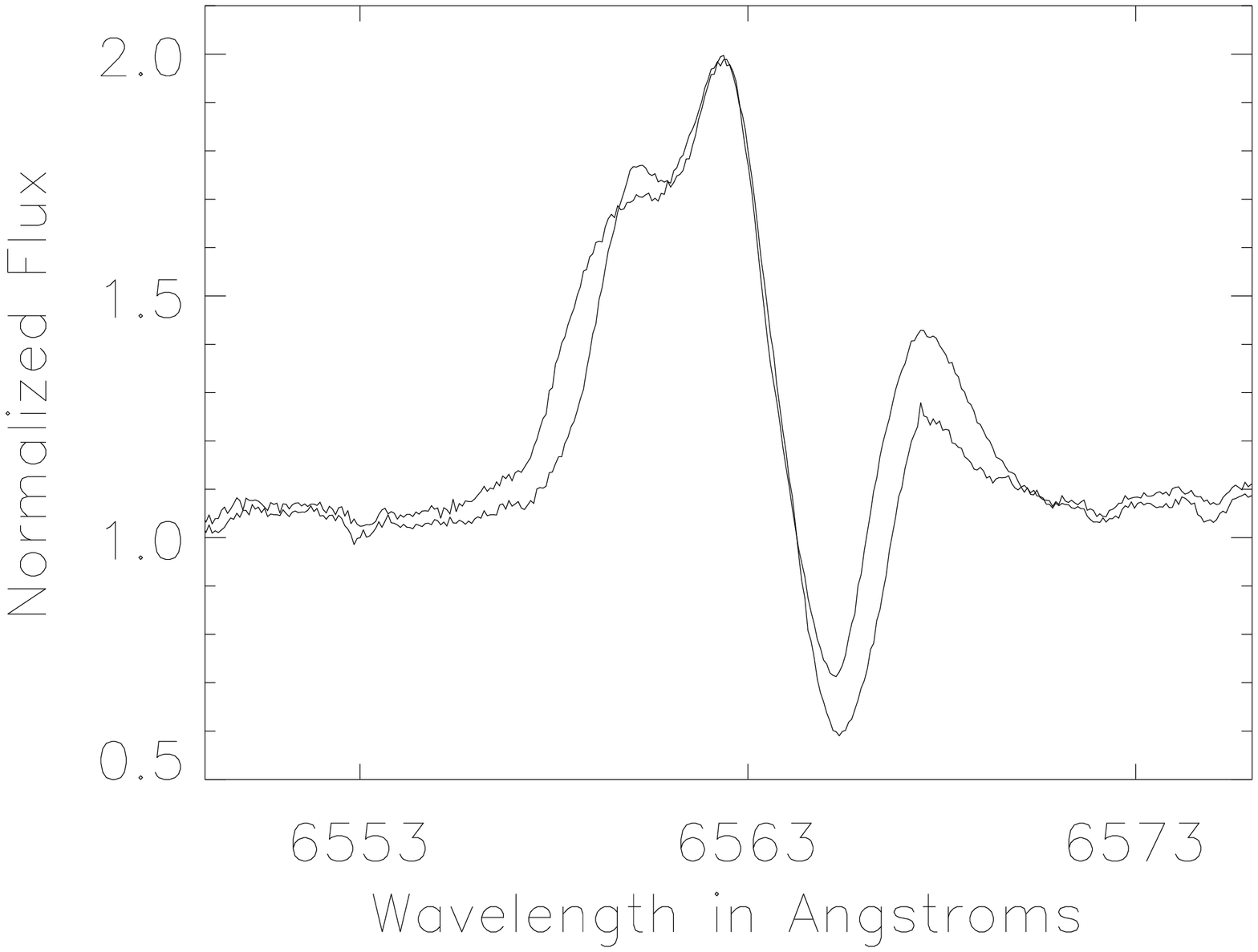}
\includegraphics[width=0.24\linewidth, angle=90]{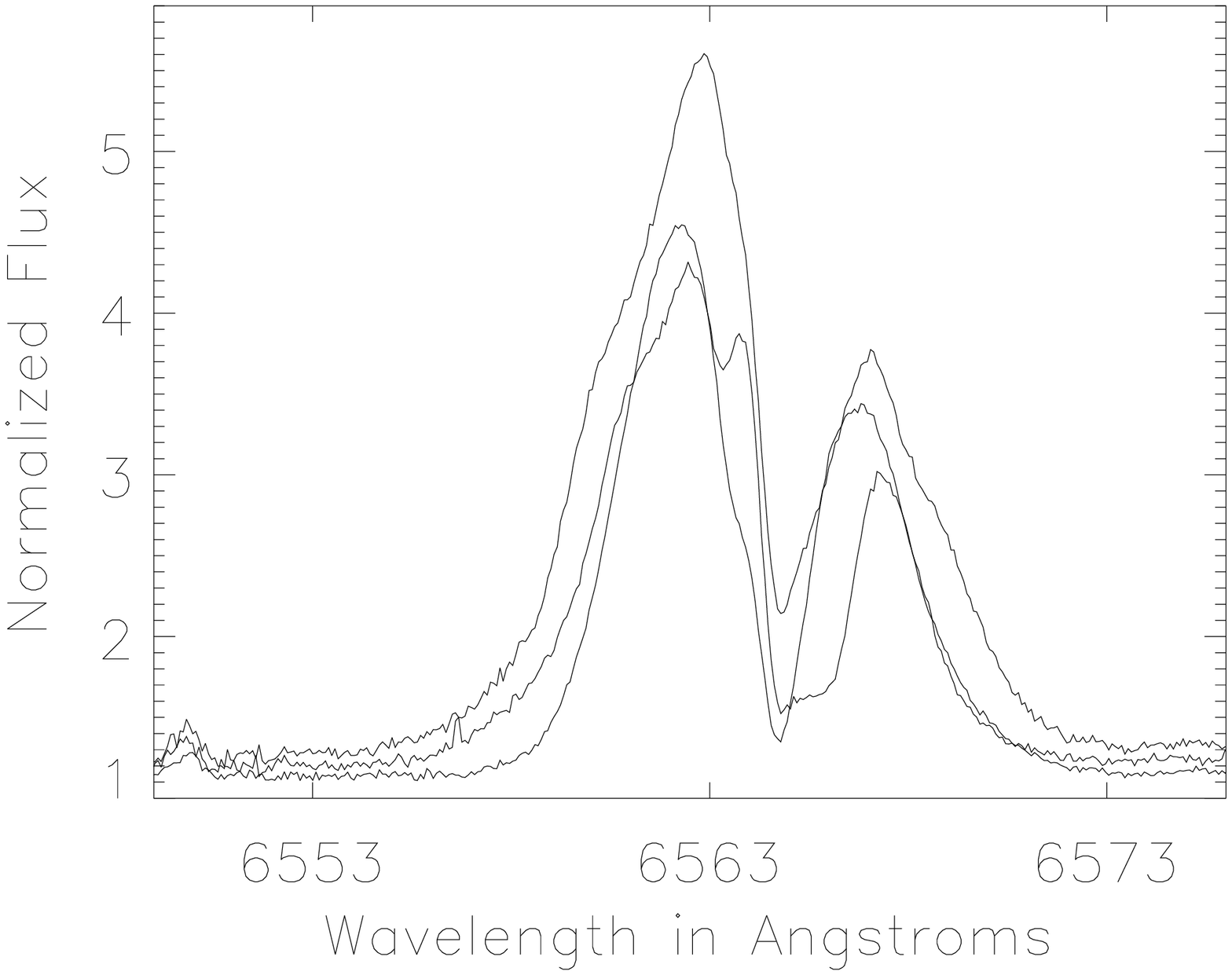}  
\caption{Herbig Ae/Be Line Profiles I. The stars, from left to right, are:  AB Aurigae, MWC 480, MWC 120, HD 150193, HD 163296, HD 179218, HD 144432, MWC 758, HD 169142, KMS 27, V 1295 Aql, HD 35187, HD 142666 and T Ori}
\label{fig:haebe-lprof1}
\end{center}
\end{figure*}

\begin{figure*}
\begin{center}
\includegraphics[width=0.24\linewidth, angle=90]{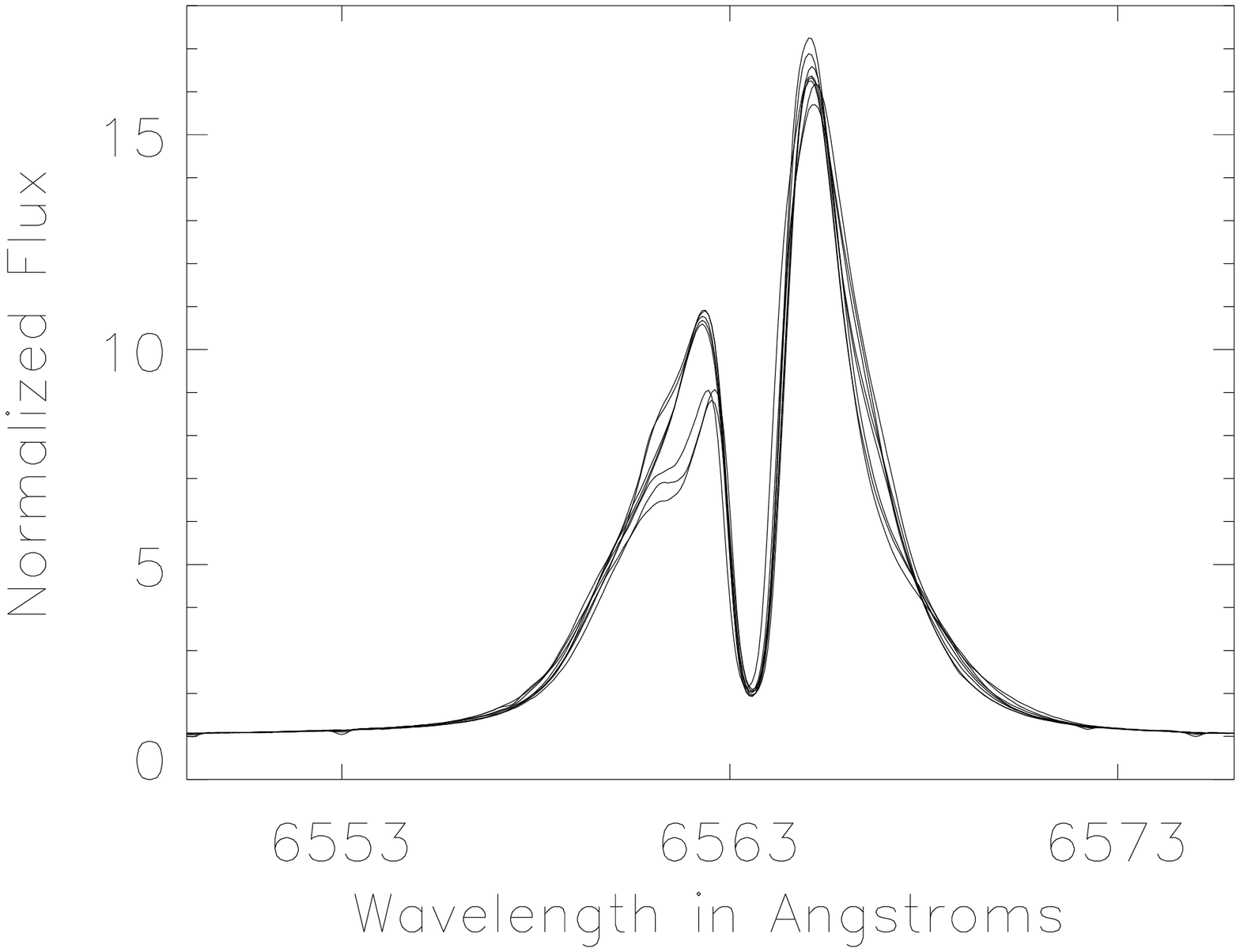}
\includegraphics[width=0.24\linewidth, angle=90]{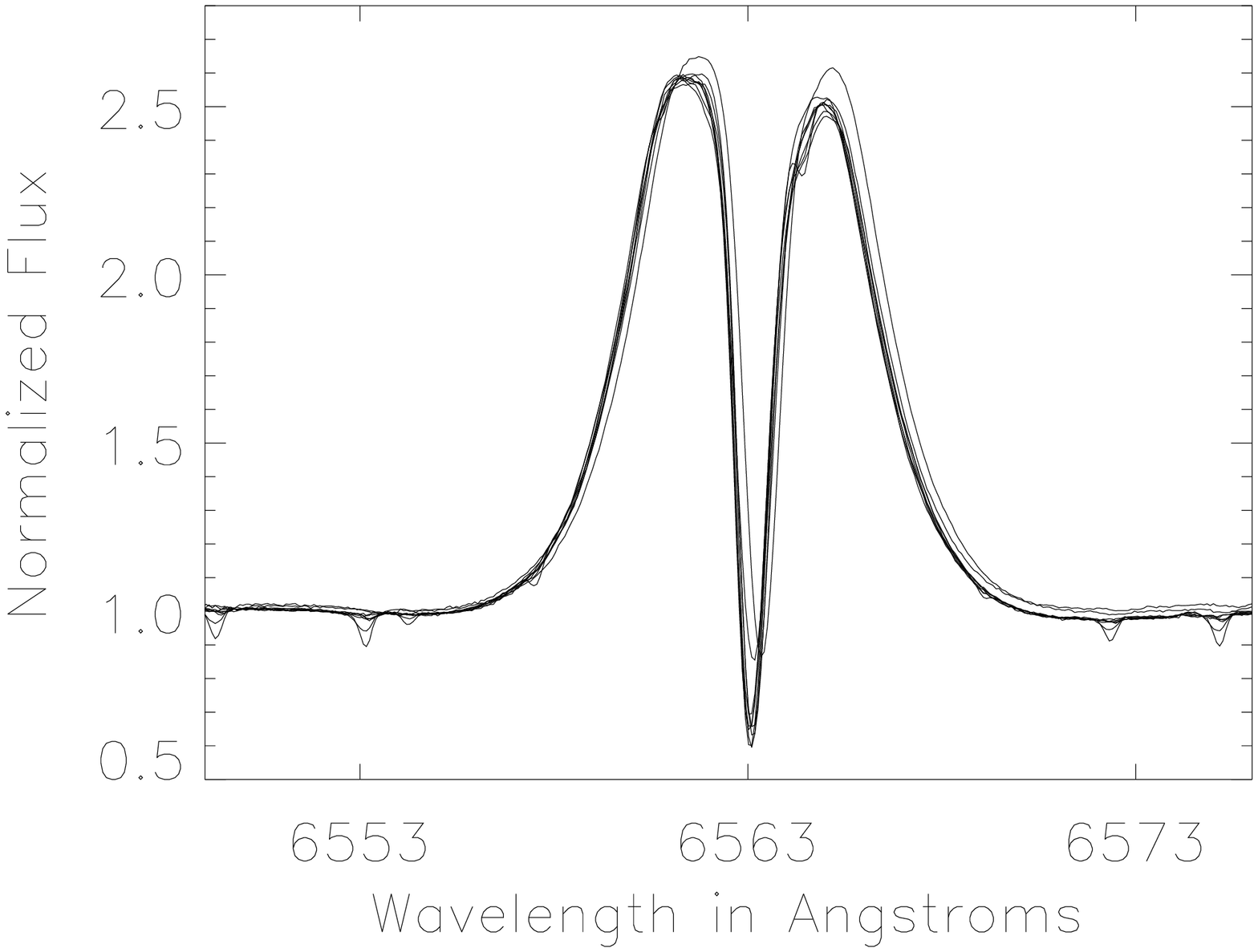}
\includegraphics[width=0.24\linewidth, angle=90]{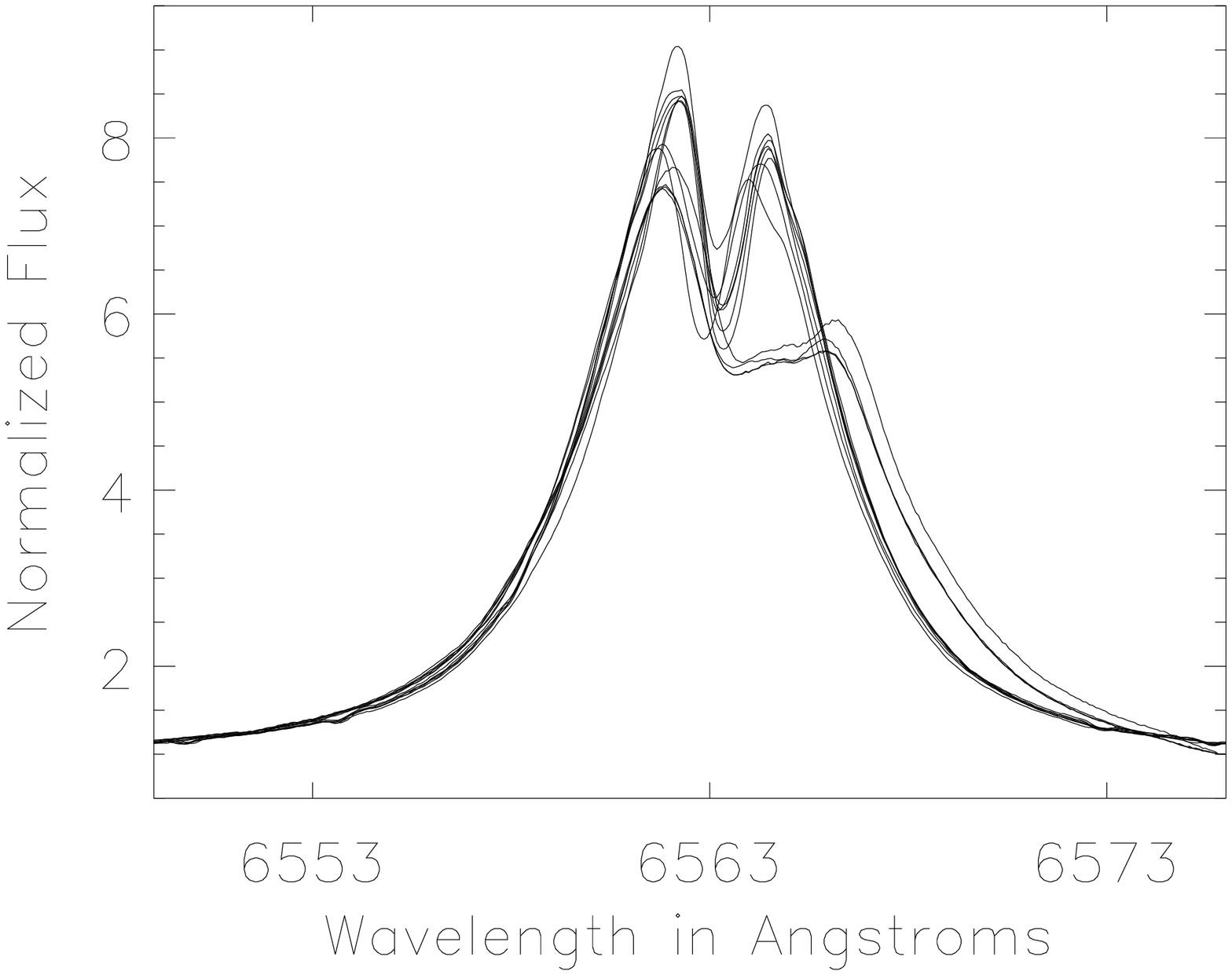} \\
\includegraphics[width=0.24\linewidth, angle=90]{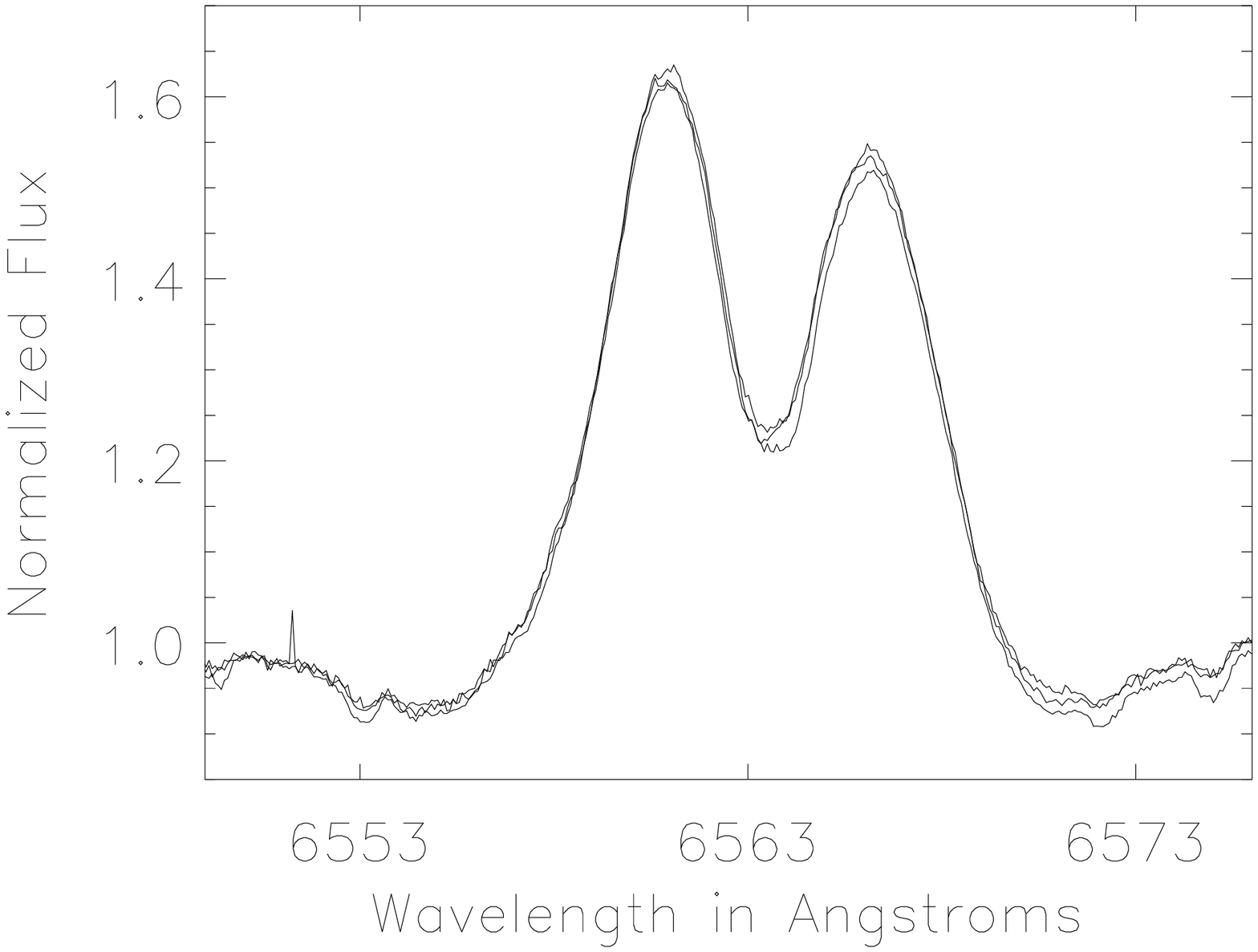}
\includegraphics[width=0.24\linewidth, angle=90]{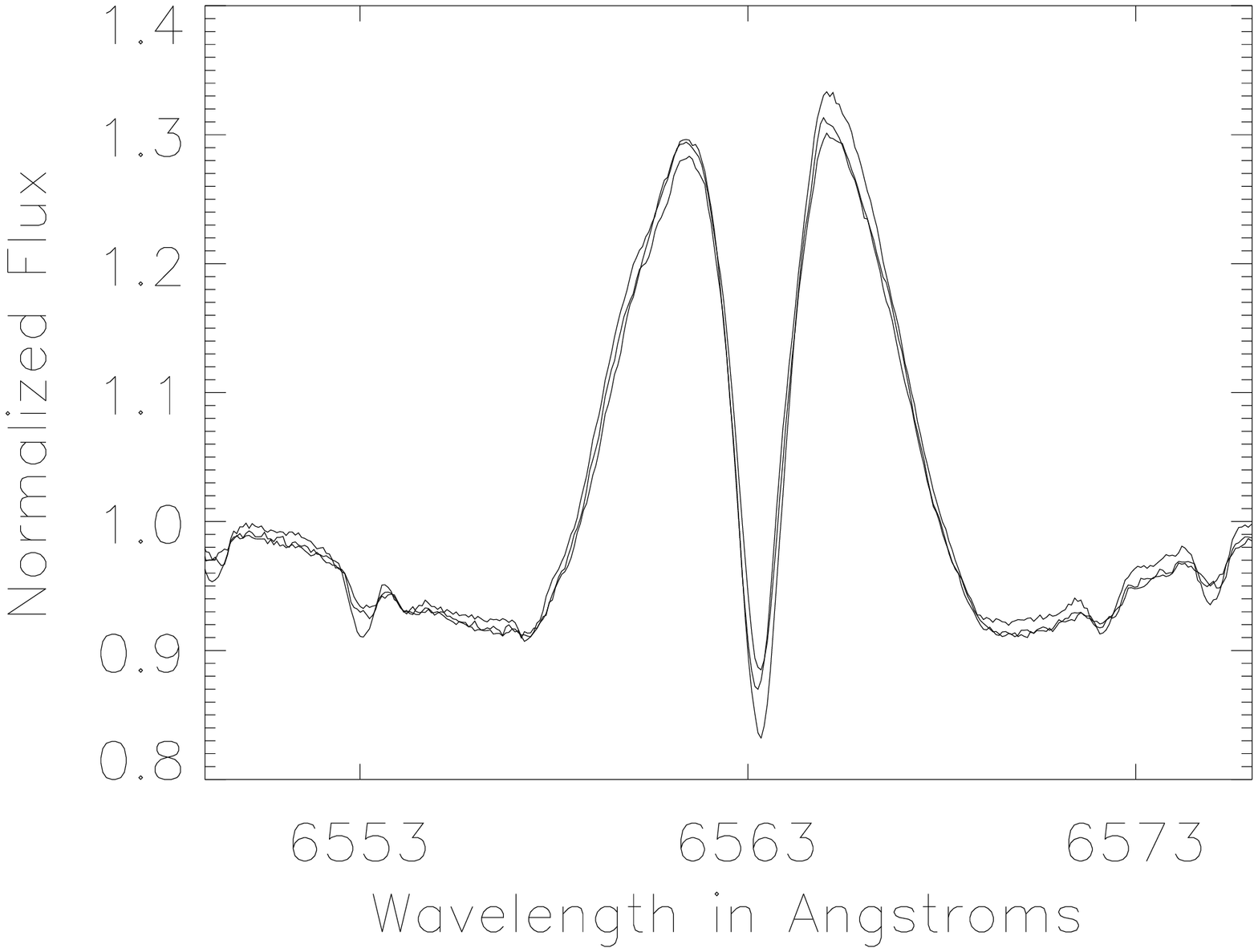}
\includegraphics[width=0.24\linewidth, angle=90]{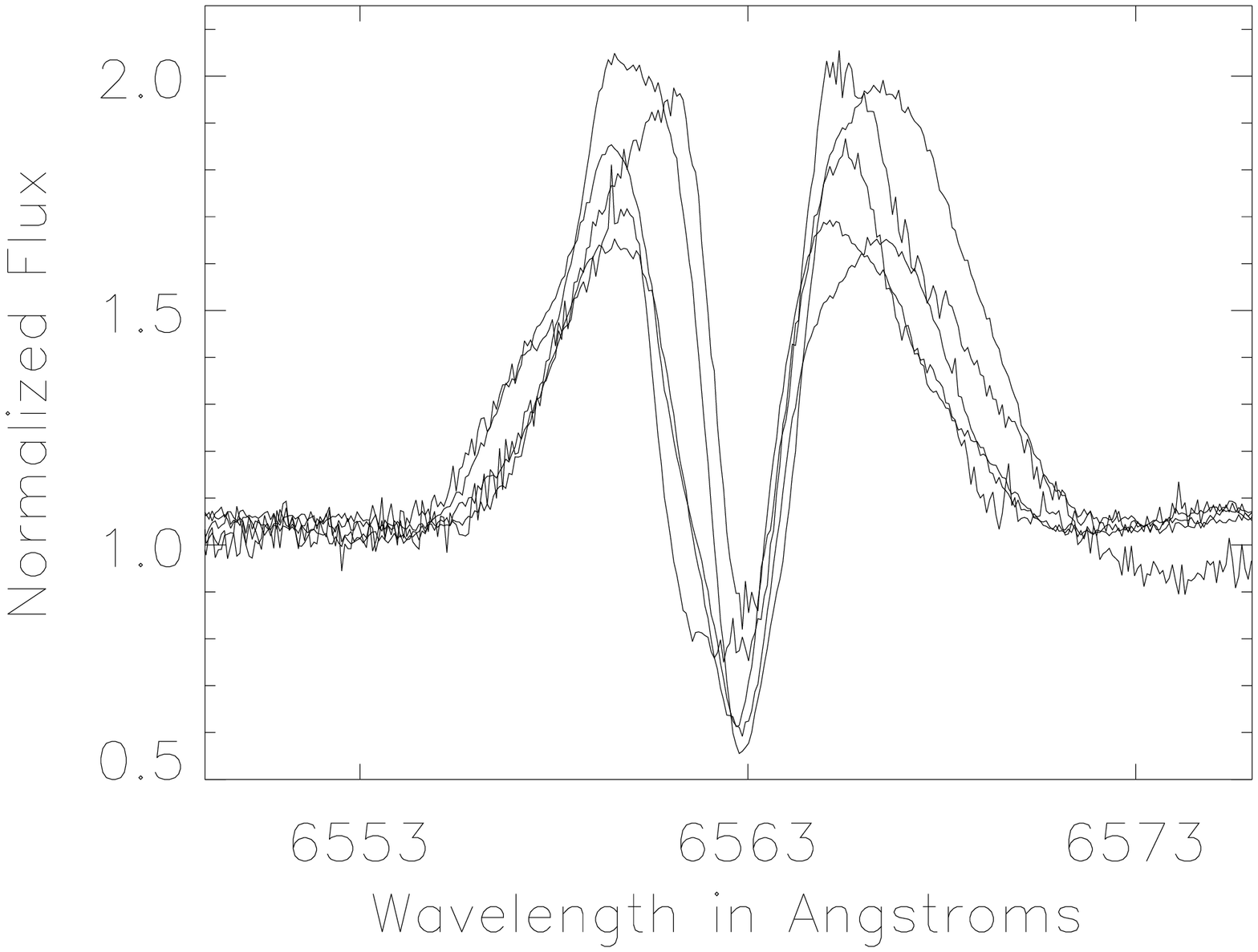} \\
\includegraphics[width=0.24\linewidth, angle=90]{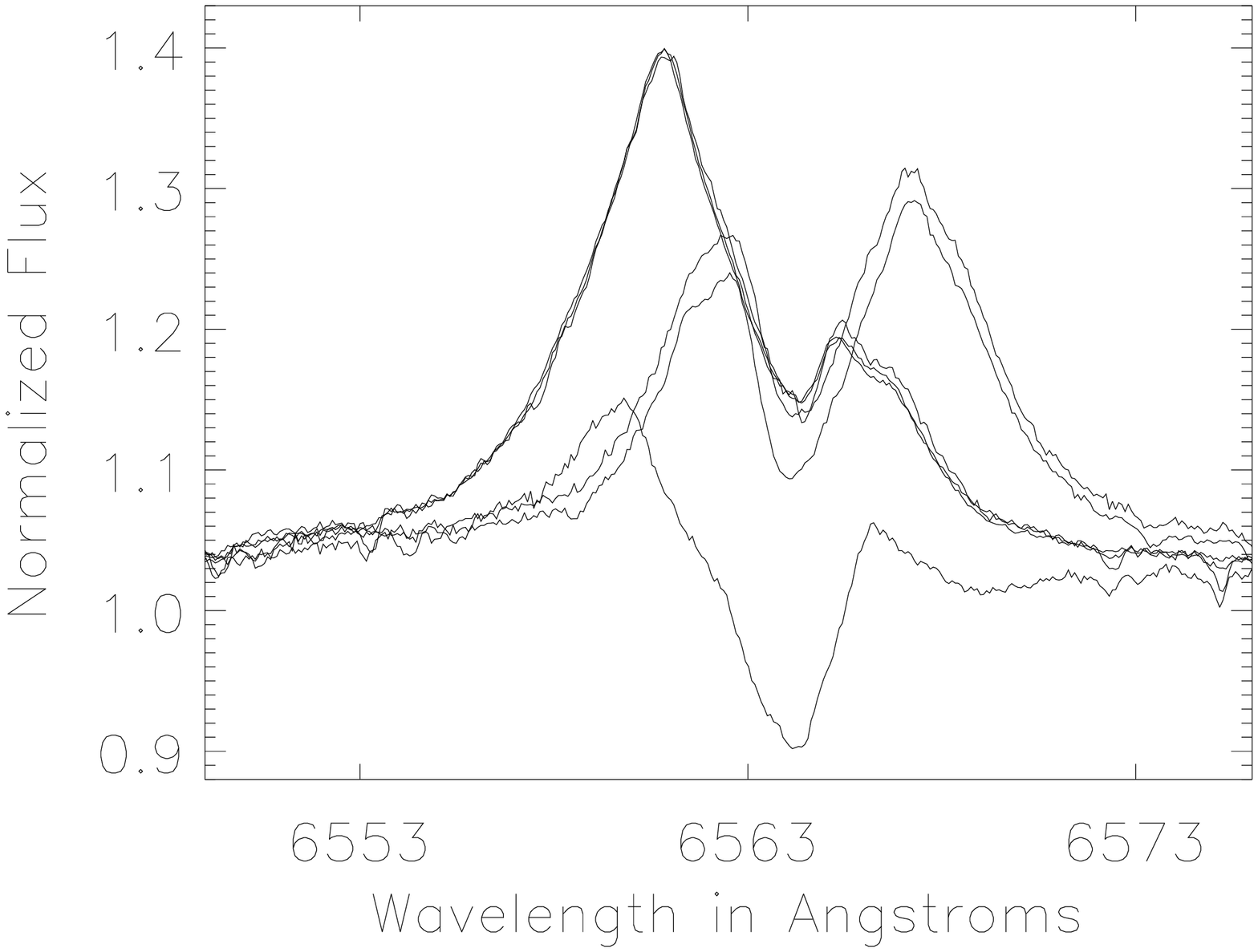}
\includegraphics[width=0.24\linewidth, angle=90]{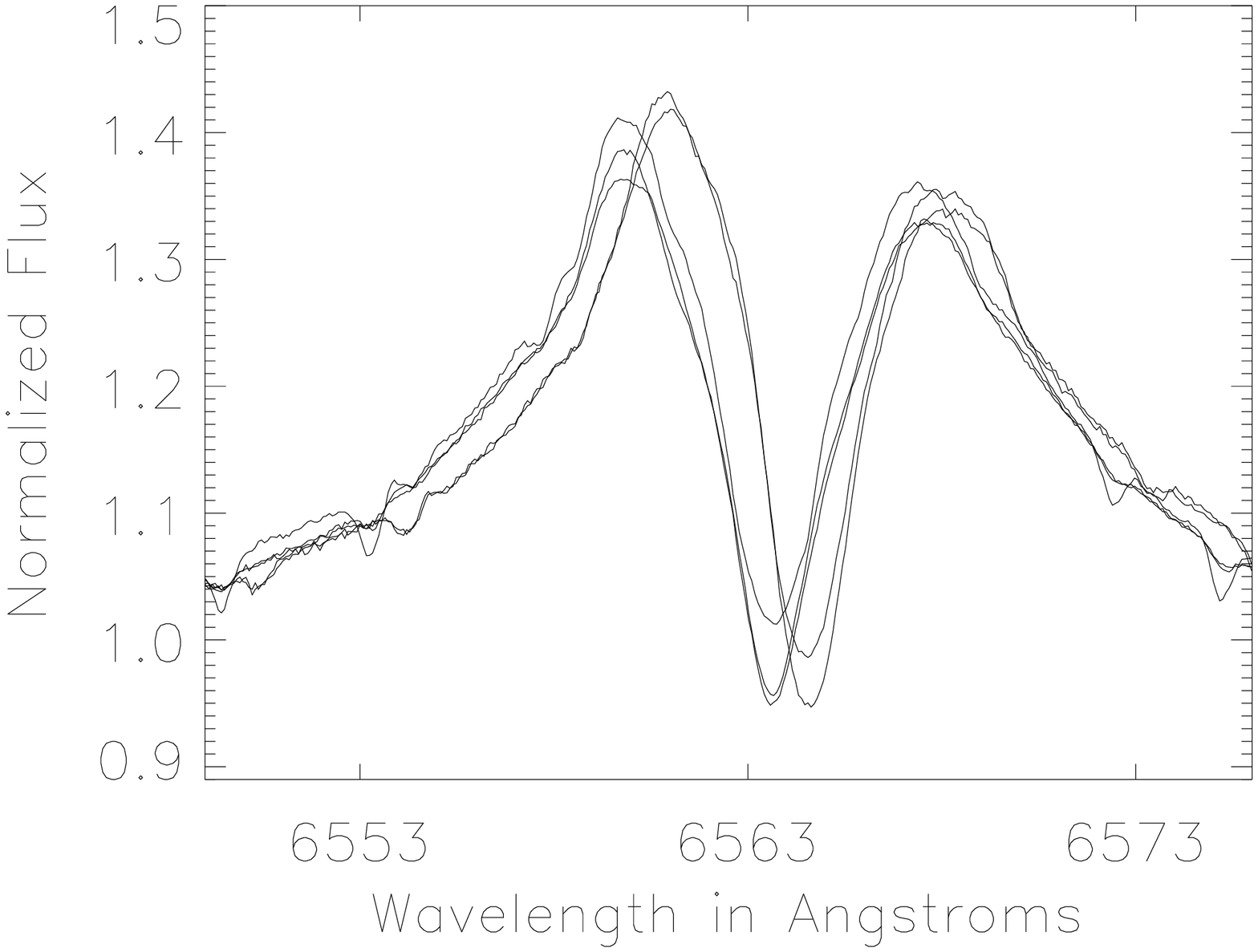}
\includegraphics[width=0.24\linewidth, angle=90]{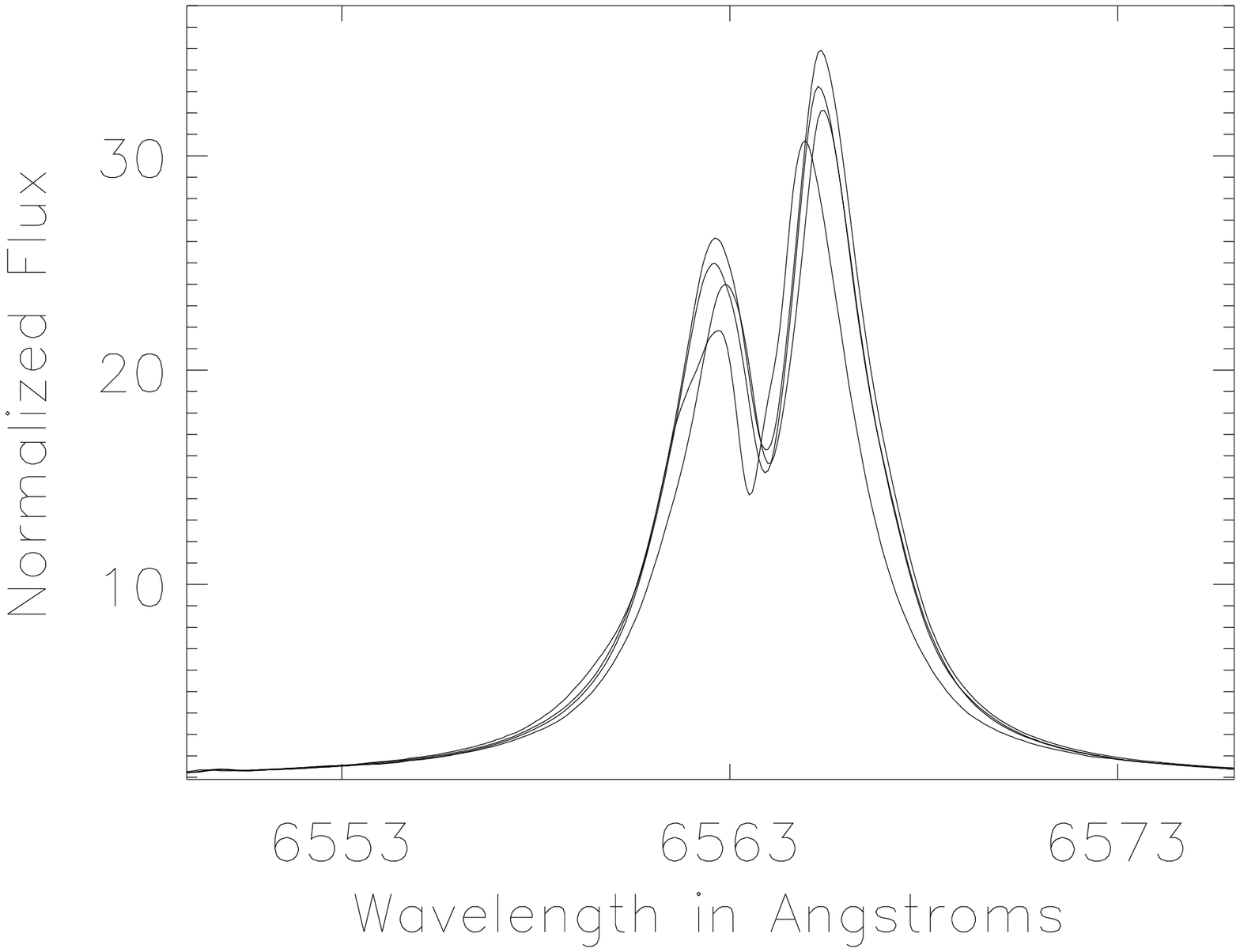} \\
\includegraphics[width=0.24\linewidth, angle=90]{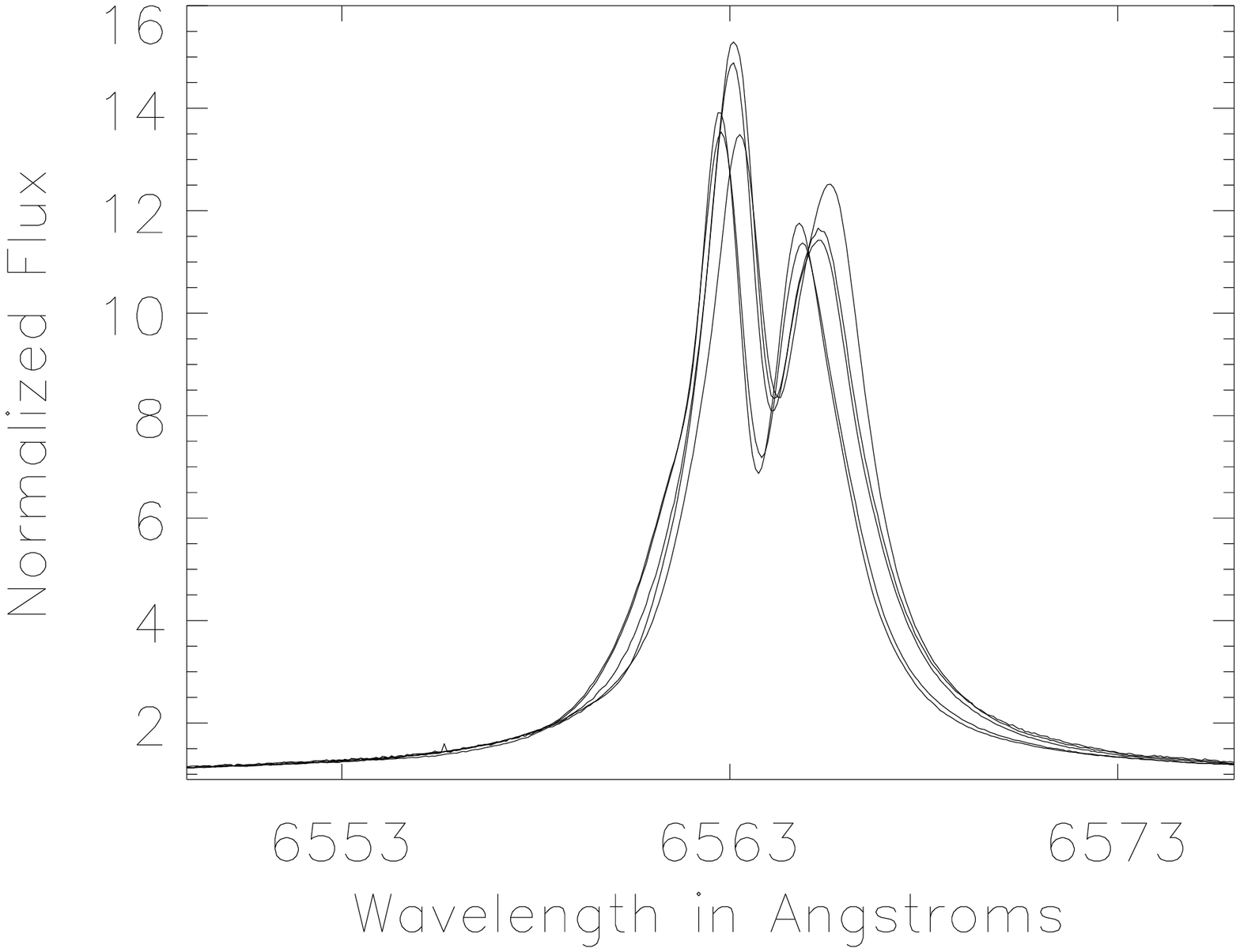}
\includegraphics[width=0.24\linewidth, angle=90]{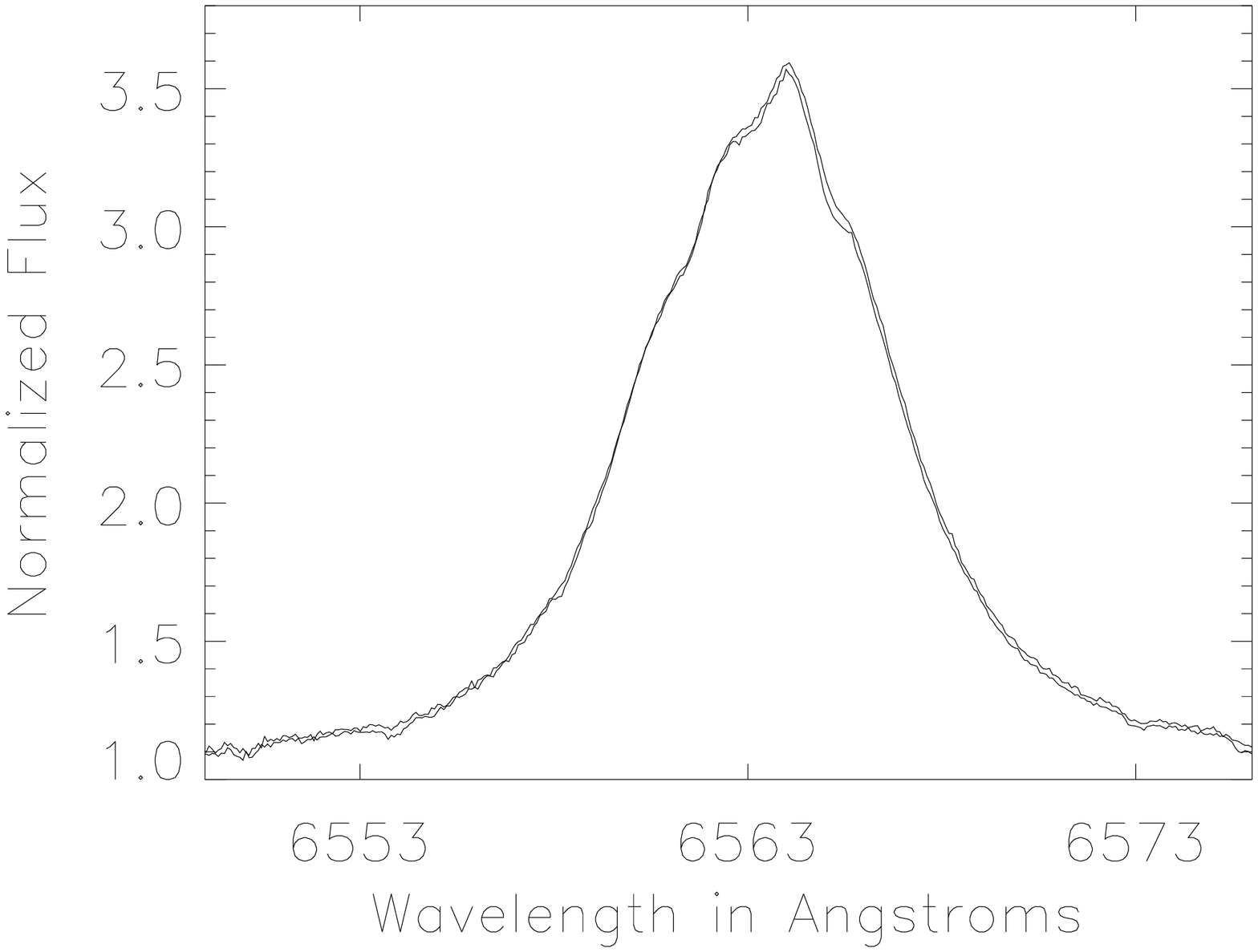}
\includegraphics[width=0.24\linewidth, angle=90]{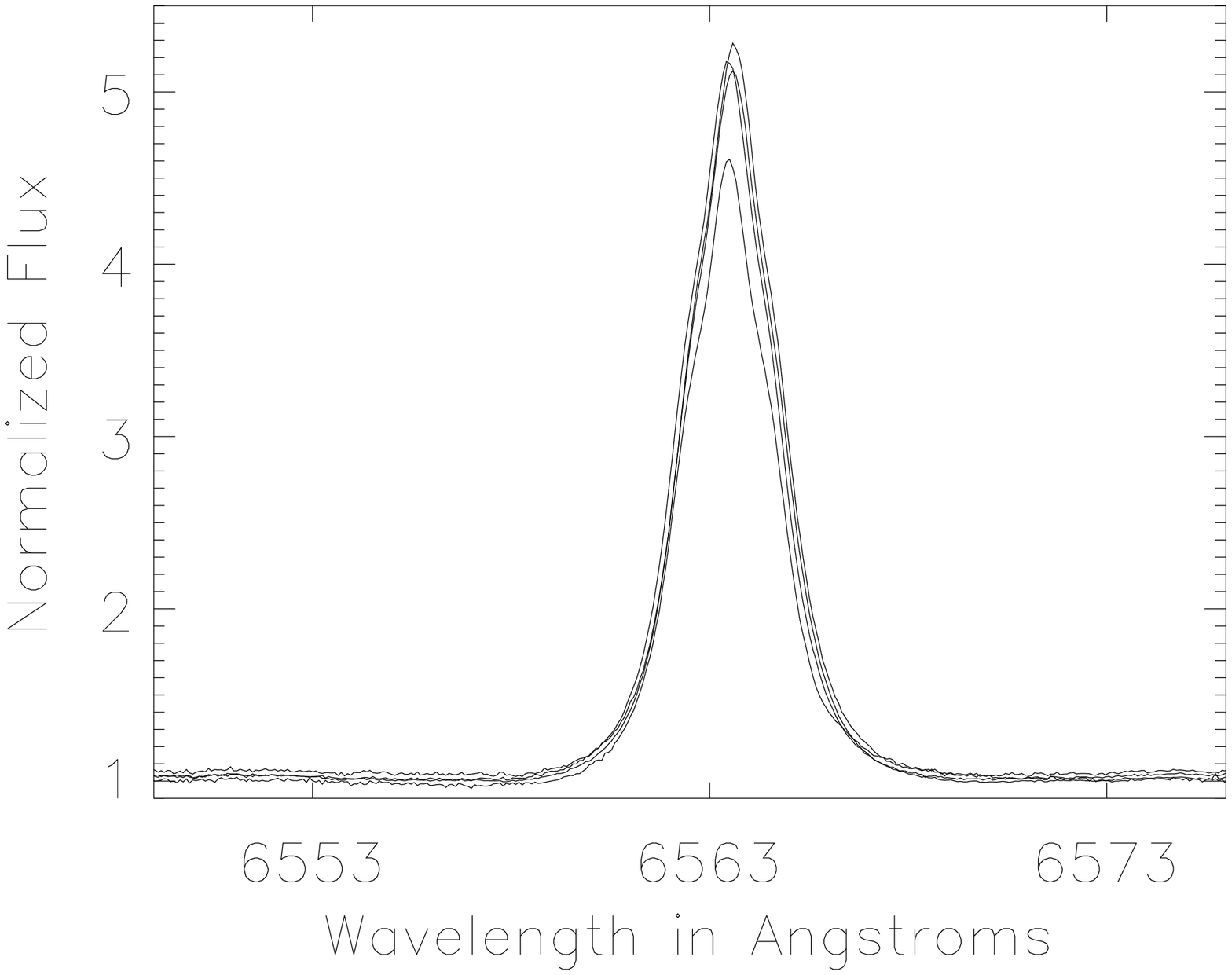} \\
\includegraphics[width=0.24\linewidth, angle=90]{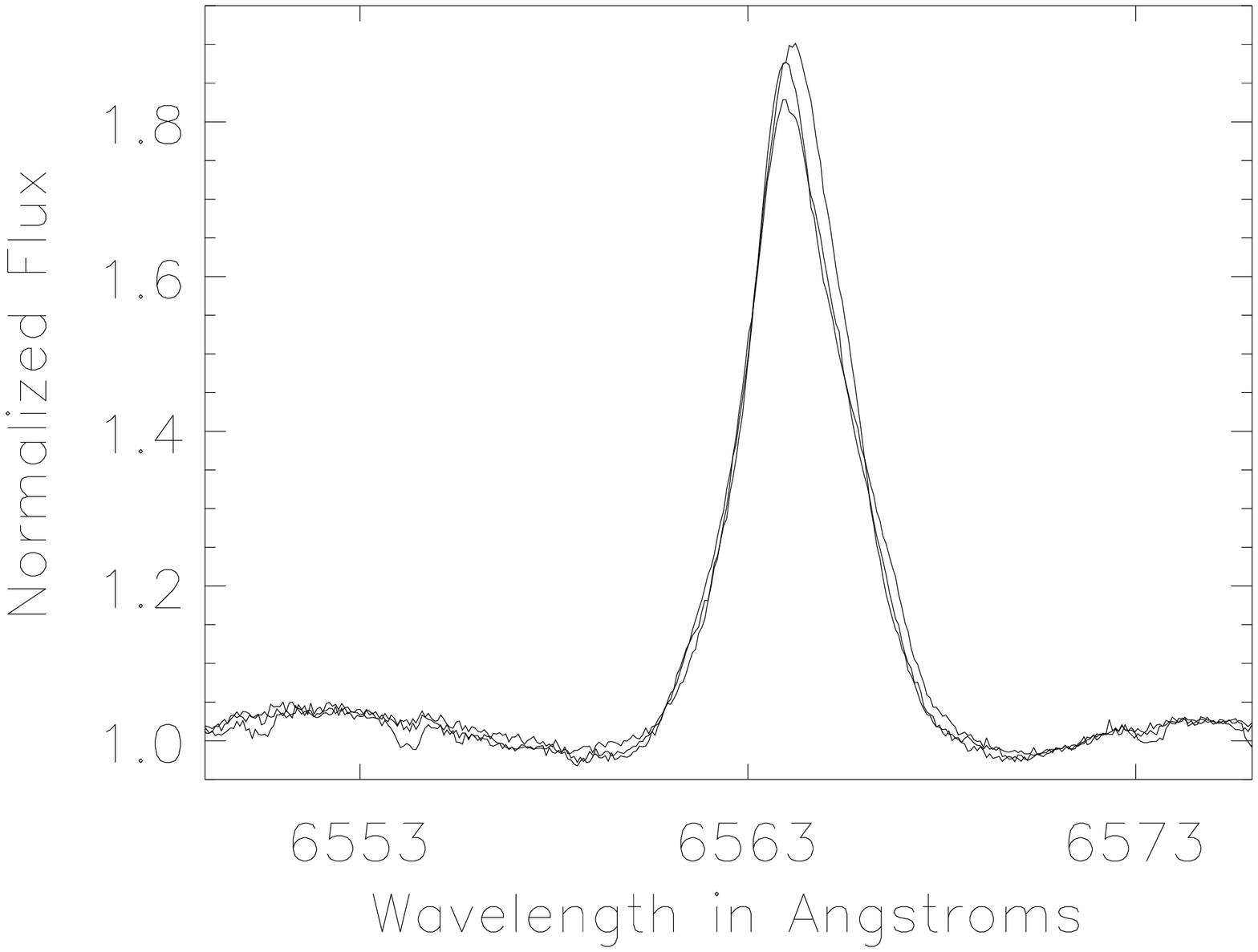}
\includegraphics[width=0.24\linewidth, angle=90]{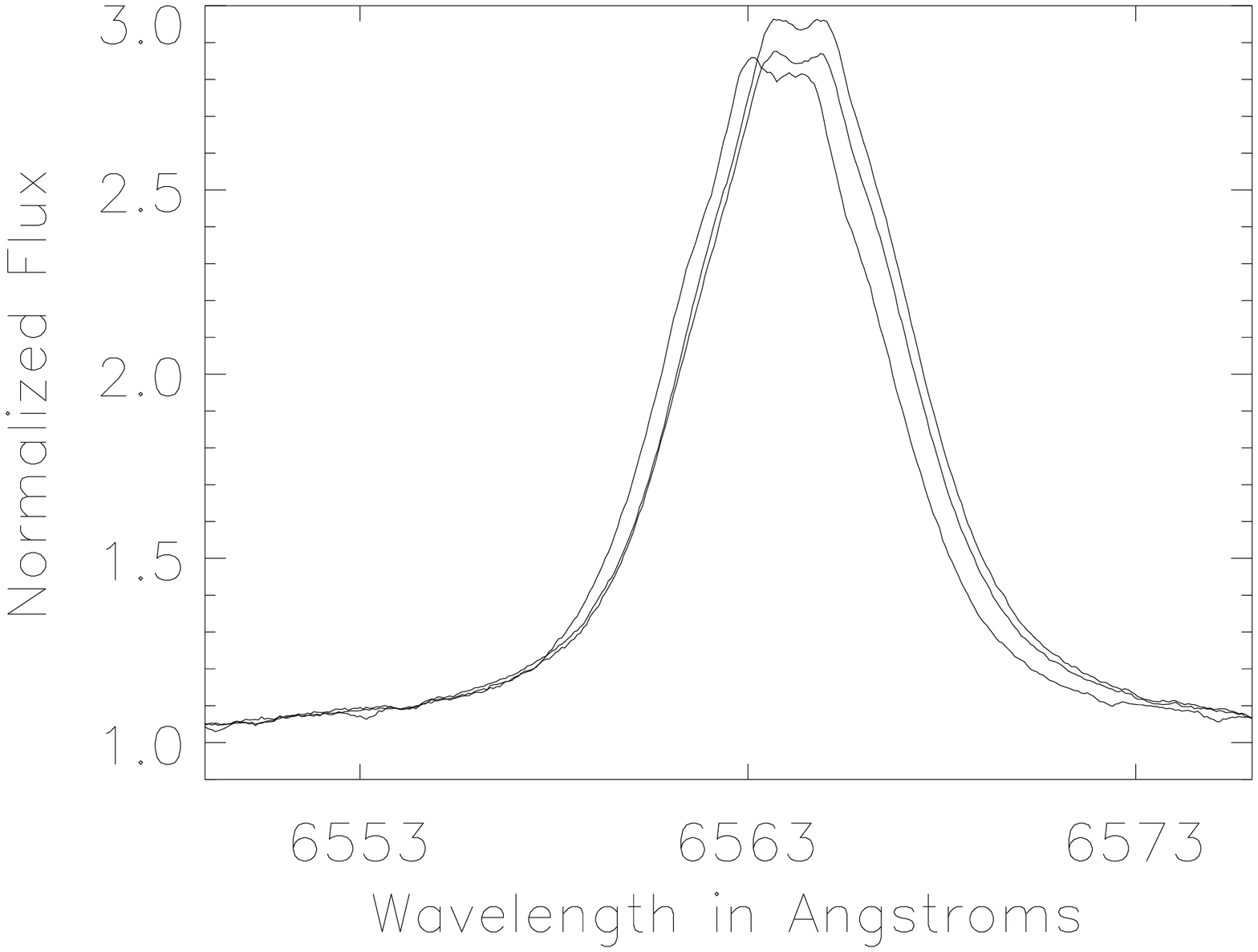}
\includegraphics[width=0.24\linewidth, angle=90]{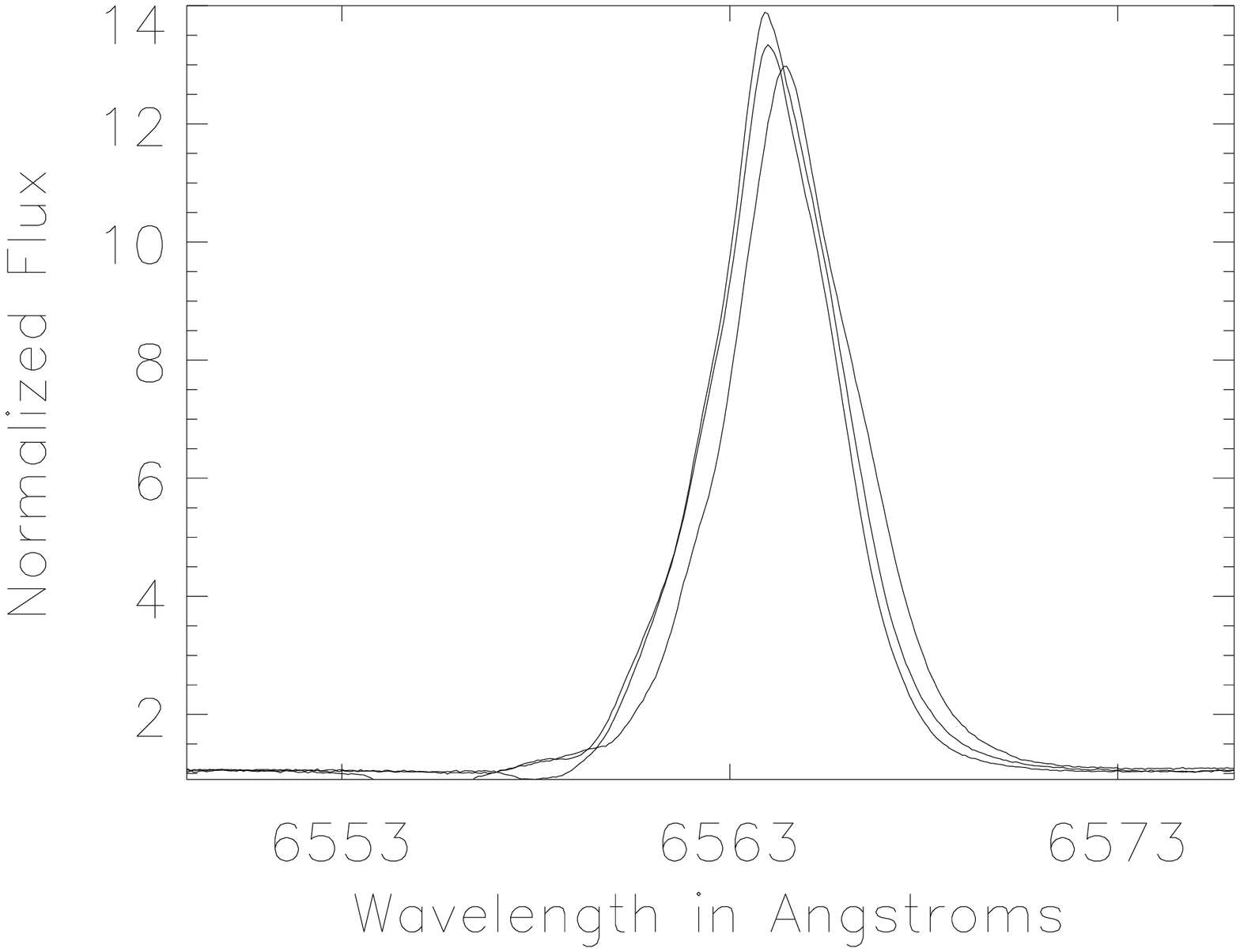} 
\caption{HAe/Be Line Profiles II: The stars, from left to right, are:  MWC 158, HD 58647, MWC 361, HD 141569, 51 Oph, XY Per, MWC 166, MWC 170, HD 45677, MWC 147, Il Cep, MWC 442, HD 35929, GU CMa and HD 38120}
\label{fig:haebe-lprof2}
\end{center}
\end{figure*}

\begin{figure*}
\begin{center}  
\includegraphics[width=0.23\linewidth, angle=90]{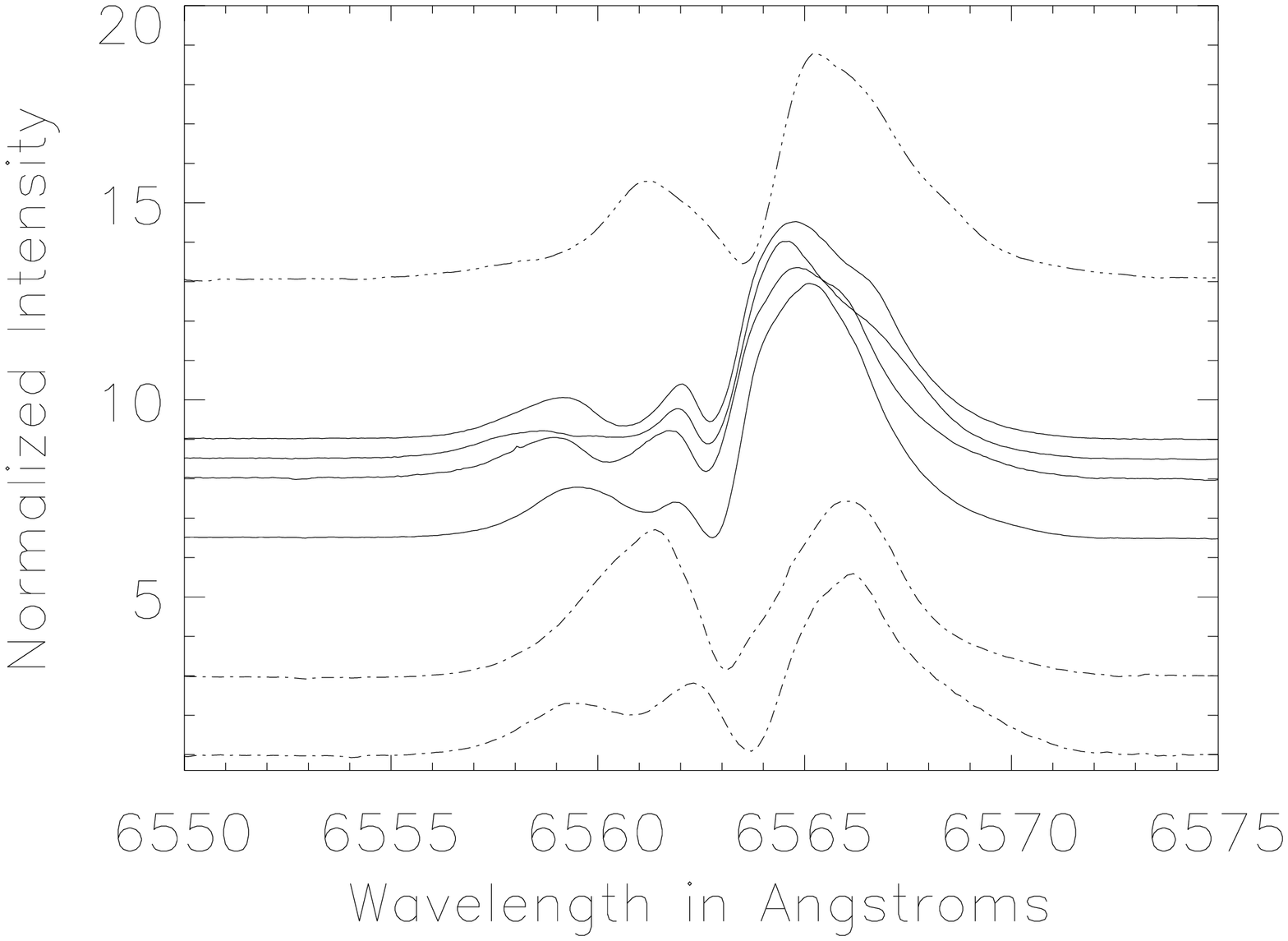} 
\includegraphics[width=0.23\linewidth, angle=90]{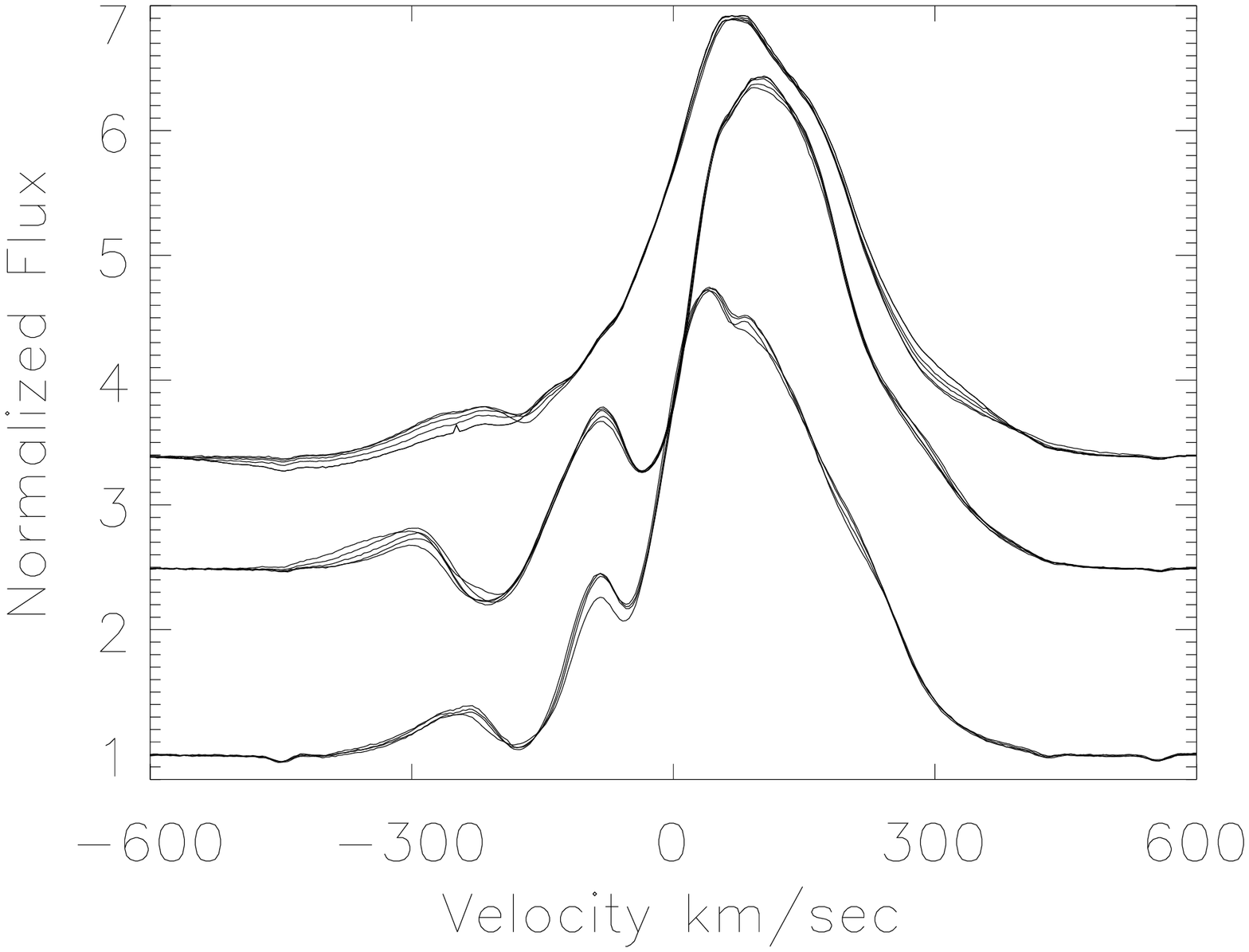} \\
\includegraphics[width=0.23\linewidth, angle=90]{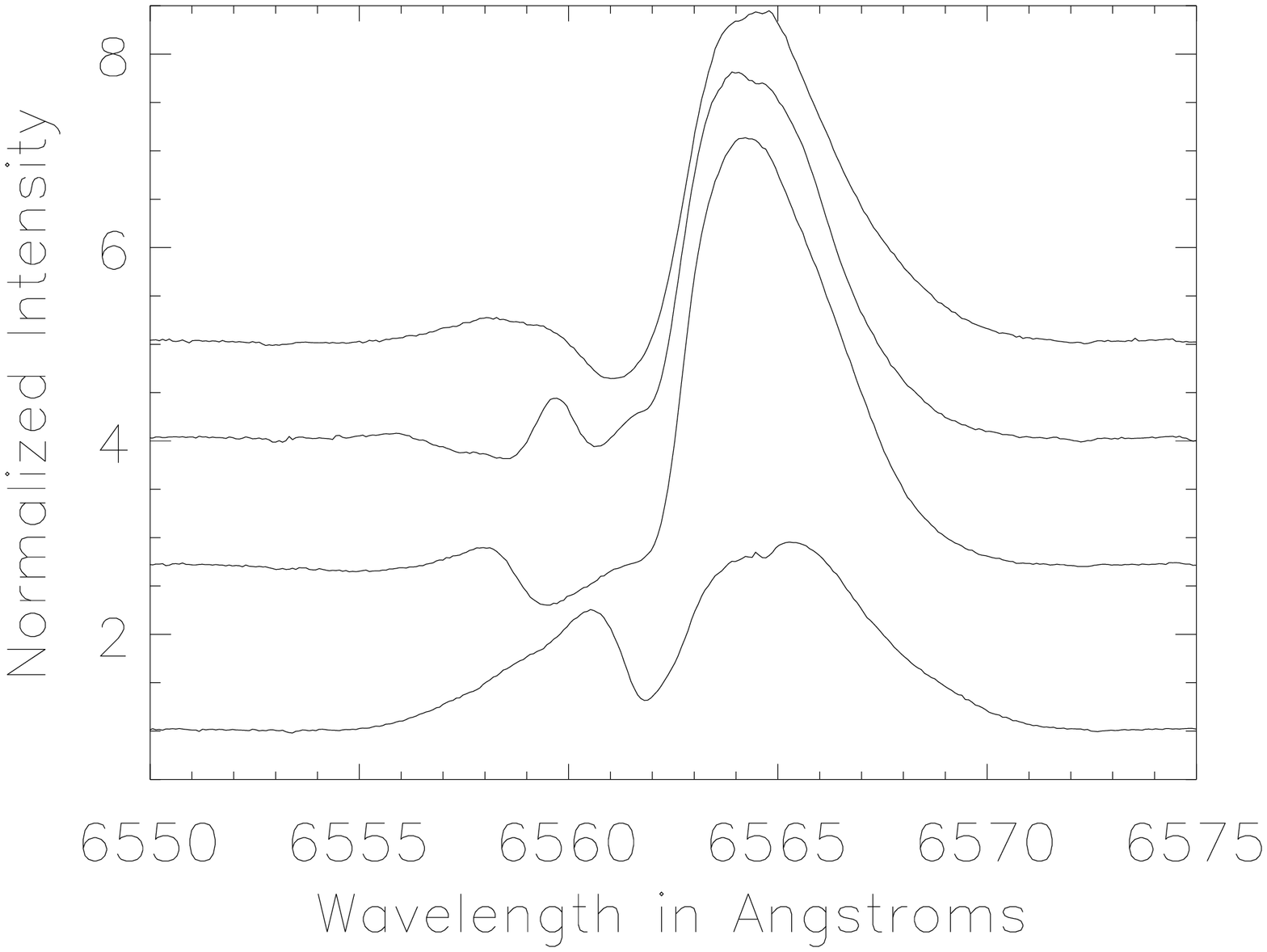} 
\includegraphics[width=0.23\linewidth, angle=90]{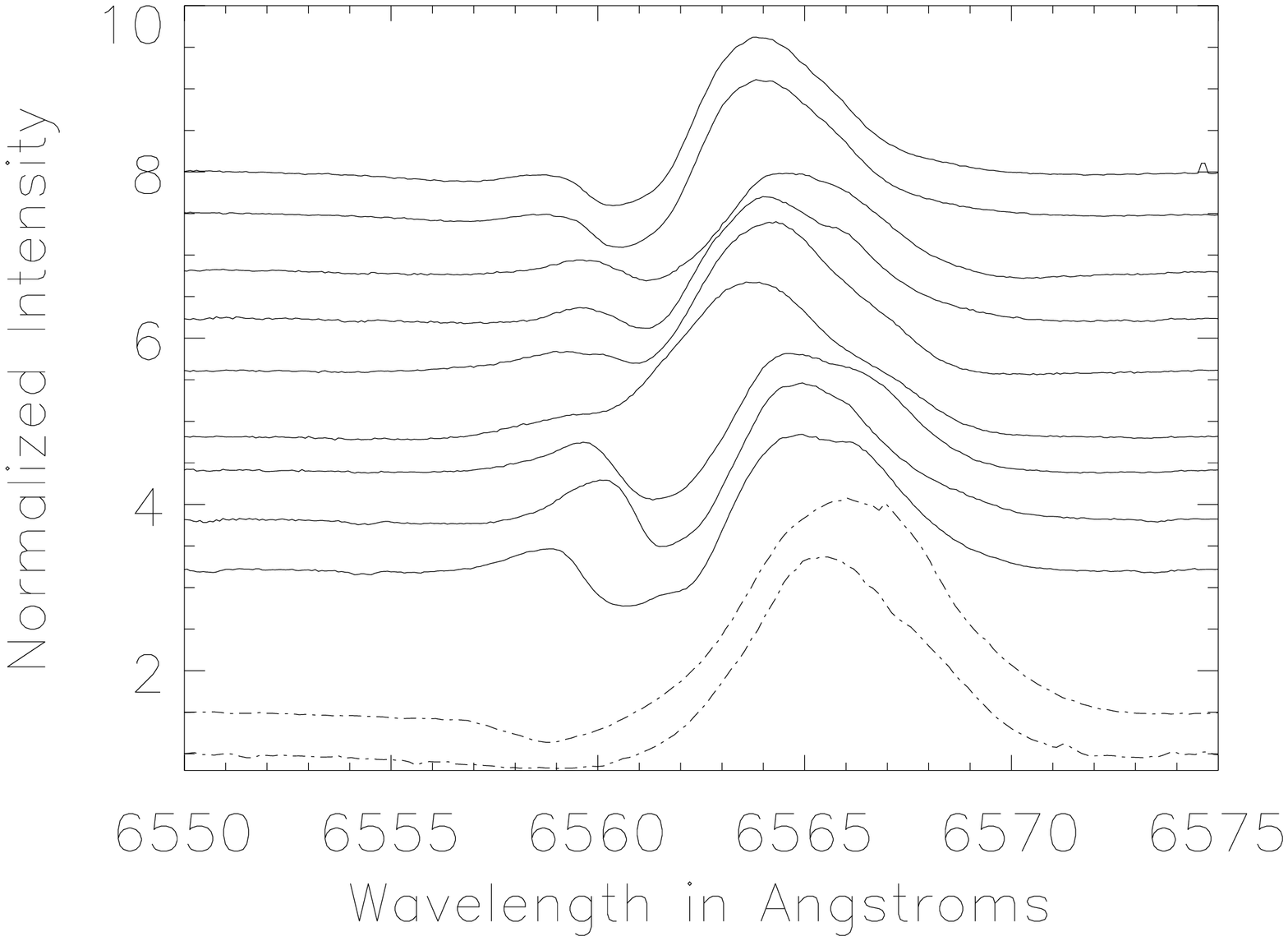} \\
\includegraphics[width=0.23\linewidth, angle=90]{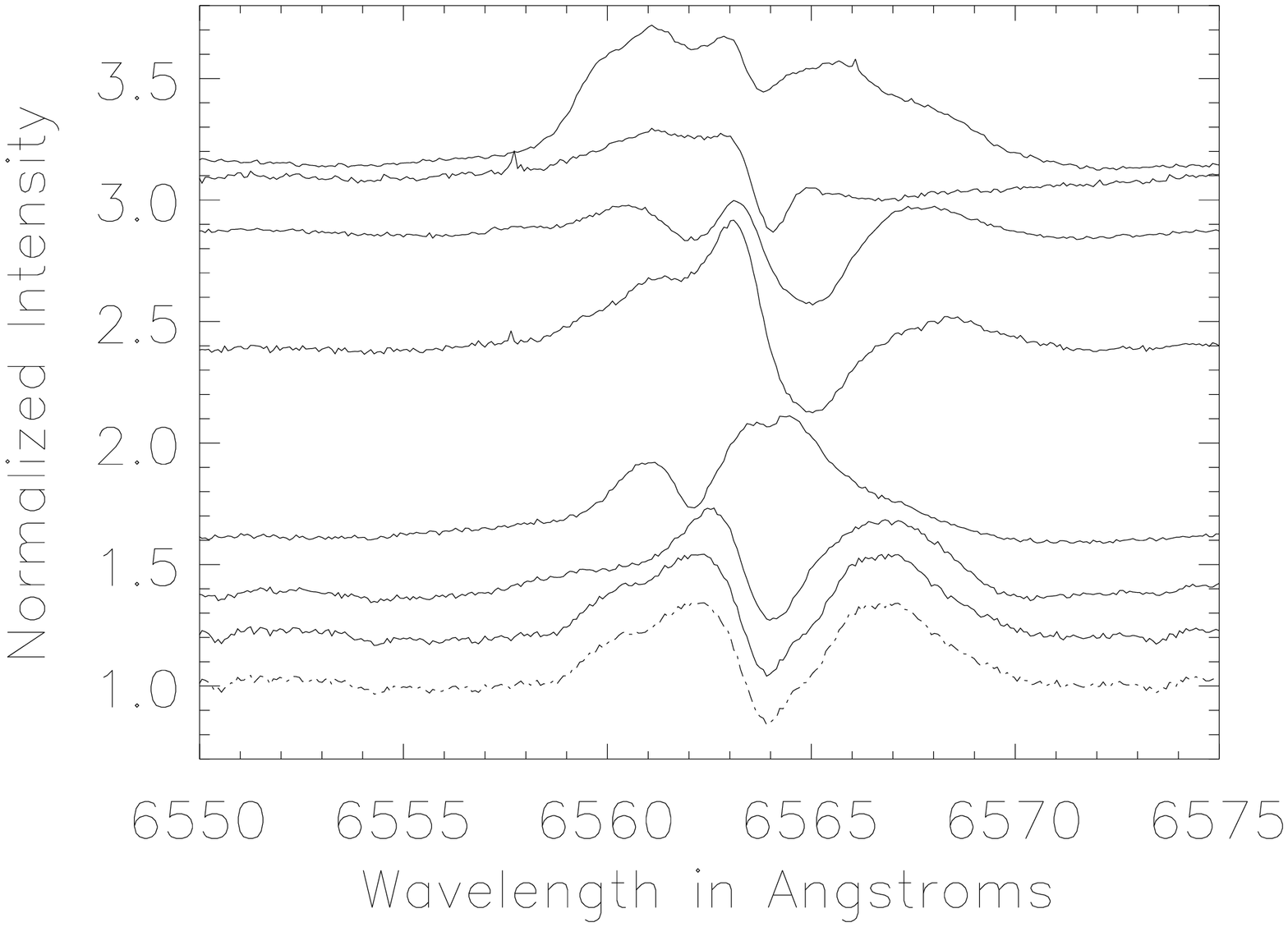} 
\includegraphics[width=0.23\linewidth, angle=90]{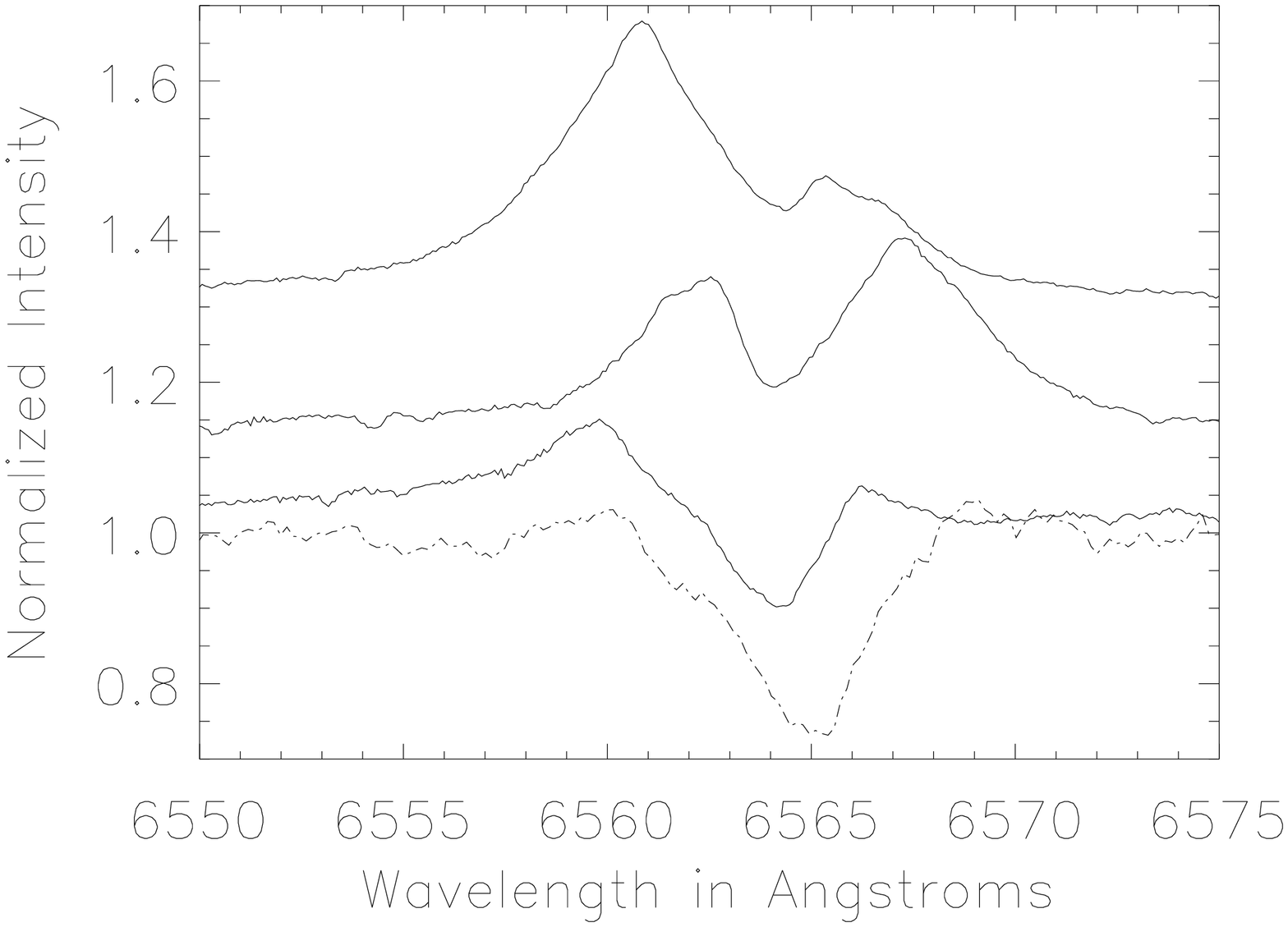}  
\caption{The spectroscopic variability of 6 stars. From left to right: {\bf a)} The variability of the H$_\alpha$ line for MWC 120 in 2004, 2006 and 2007 vertically offset for clarity. The dashed lines on the bottom show spectra from engineering runs on October 20th (bottom) and December 9th (top) 2004. The solid lines clustered in the middle of the plot from bottom to top are December 27th 2006, January 3rd, 18th, and 19th 2007. The final solid line on top is August 29th 2007. {\bf b)} The H$_\alpha$ for HD 163296 on three of the most variable nights where there was good coverage. {\bf c)} The variability of the H$_\alpha$ line for HD 150193 in 2007. The lines from bottom to top are June 20th, July 28th, August 1st and August 29th. The star changed from an emission line with a small, nearly central absorption to a mostly classical P-Cygni structure with a much greater intensity. {\bf d)} The variability of the H$_\alpha$ line for MWC 758 in 2004 and 2007. The dashed lines on the bottom show spectra from engineering runs on October 20th (bottom) and December 15th (top) 2004. The solid lines from bottom to top are August 28 and 29, September 20, October 30 \& 30, November 21 \& 24. {\bf e)} The variability of the H$_\alpha$ line for HD 35187 in 2007. From bottom to top the dates are August 28 and 29, September 20, October 30 \& 31, and finally November 21. {\bf f)} The variability of the H$_\alpha$ line for MWC 166. The dashed line on the bottom shows a single spectrum from an engineering run on October 20th 2004. The solid line in the middle is December 27th 2006 and the top line is September 20th 2007.}
\label{fig:var}
\end{center}
\end{figure*}

\subsection{HD 31648 - MWC 480}

	  MWC 480 showed strong blue-shifted absorption components in the H$_\alpha$ line. Mannings et al. 1997 showed the presence of a circumstellar disk inclined at 30$^\circ$. Kozlova et al. 2003 presented a spectroscopic study of many lines. They found a v-sini of 90km/s. Mannings et al. 1997 found an inclination of roughly 30$^\circ$. They concluded that the stellar wind is the inner layers of the accretion disk. The high velocity component of the wind were interpreted as jets ejected from this region seen in projection. The H$_\alpha$ spectrum and the continuum polarization had also been studied in detail by Beskrovnaya \& Pogodin (2004) who concluded that MWC 480 also had an inhomogenious, azimuthally-structured wind which was variable on short timescales. Kozlova 2006 presented a H$_\alpha$ monitoring program with observations from 1998 to 2005. The H$_\alpha$ line shows continuum-normalized intensities of roughly 4-6 with a complicated blue-shifted absorption profile and variability on the timescale of hours. A low-velocity ($\sim$-100km/s) absorption is almost always present while higher velocity (more blue-shifted) absorptions are more infrequent and more variable. Wade et al. 2007 did not detect a magnetic field and had an upper limit of roughly 50-150G depending on the field type but Hubrig et al. 2006 present a magnetic field measurement of +87G $\pm$22.

\subsection{HD 37806 - MWC 120}

 	MWC 120 showed strong blue-shifted absorption components. H$_\alpha$ line spectropolarimetry from 1995 and 1996 was presented in Oudmaijer \& Drew 1999. The emission line profiles they observed changed drastically between the years. In January 1995, the line was double-peaked but with a much weaker blue-shifted emission. In December 1996, the emission line was evenly double-peaked. Wade et al. 2007 did not detect a magnetic field and had an upper limit of roughly 50-100G depending on the field type.

	We observed strong morphological changes over the course of three years. Figure \ref{fig:var} shows that, while the line maintains a normalized intensity of roughly 5-7 times continuuum, the blue-shifted absorption varies quite strongly.

 \subsection{HD 50138 - MWC 158} 

	  MWC 158 is a mid-B type star which had a very strong emission line which showed a very strong central absorption. The peak intensities were roughly 15-17 times continuum while the central absorption was typically 2 times continuum. MWC 158 was previously studied for spectroscopic variability as well as low-resolution spectropolarimetry (Bjorkman et al. 1998, Pogodin 1997, Jaschek \& Andrillat 1998). Pogodin 1997 concluded that there is evidence for winds as well as in-falling matter from the envelope lines with the H$_\alpha$ absorption component existing from -200km/s to +70km/s. Bjorkman et al. 1998 found a nearly wavelength-independent continuum polarization and concluded that electron scattering (rather than dust scattering) had to be the polarizing mechanism. H$_\alpha$ line spectropolarimetry from 1995 and 1996 was presented by Oudmaijer et al. (1999).  Baines et al. 2006 showed spectroastrometry as evidence for binarity. They found a change in centroid and equivalent-width of the point-spread function across H$_\alpha$.  	  

 \subsection{HD58647} 

	  HD 58647 is a late B type star (B9 in Th\'{e} et al. 1994).  Baines et al. 2006 presented spectroastrometry claiming this star is a binary. They found a change in centroid and equivalent-width of the point-spread function across H$_\alpha$. Alhough this star was monitored for over 2 years, the H$_\alpha$ line was essentially invariant. 
	  
\subsection{HD 163296 - MWC 275}

	This star has shown very strong H$_\alpha$ variability across the entire line. Figure \ref{fig:var} shows three nights during June and July of 2007 with good coverage. On each occasion, there is significant variability in relatively narrow wavelength ranges on both blue-shifted and red-shifted sides of the line.
	
	Periodic variations variations in UV lines showed 35 and 50 hour periodicities (Catala et al. 1989). Pogodin 1994 studied the H$_\alpha$ and H$_\beta$ lines and found an active region of the stellar wind near the star with a stable shell surrounding it. Longer lived rotating jets as well as short-timescale clumps were concluded to cause the variability. The H$_\alpha$ line profiles had roughly the same shape and variability as were observed in figure \ref{fig:haebe-lprof1}. Garcia Lopez et al. 2006 derive an accretion rate of -7.12 Log(M$_\odot$/yr) using Bracket-$\gamma$ emission at 2.166$\mu$m. Grady et al. 2000 showed an azimuthally symmetric disk around this star inclined at 60$^\circ$ and chain of Herbig-Haro objects oriented perpendicular to the disk. Catala et al. 1989 reported UV spectroscopy showing a variable wind. Hillenbrand et al. 1992 derived an accretion rate from NIR emission of 1.3 10$^{-6}M_\odot$/yr but Skinner et al. 1993 get $<9.1\times 10^{-9}M_\odot$/yr using radio continuum emission. Devine et al. 2000 reported images of a bipolar outflow that was traced to within 60mas (7.3AU) of the star. The jets from this star are seen in Ly$_\alpha$ to extend from at least as close as  0.06$''$ (7.3AU) out to 6$''$ (725AU) forming a collimated, bipolar outflow designated HH409. Thus, this star is one of the more complex with claimed detections of a wind, disk, accretion, and a bipolar outflow.

\subsection{HD 179218 - MWC 614}

	This star is an isolated Herbig Ae star with a vsini 60km/s. Miroshnichenko et al. 1999 presented spectroscopy and low-resolution spectropolarimetry from 1995-1997. The H$_\alpha$ line was typically only 2 or 3 times continuum with a complicated absorption structure overlying the emission. They detected an R-band polarization of 0.45\% at an angle of 102$^\circ$ in August and September of 1997. They concluded that the star was 0.9 magnitudes above the zero-age main-sequence and had a luminosity greater than pre-main-sequence stars of similar temperature such as AB Aurigae. They argued for the presence of an inhomogeneous flattened circumstellar envelope. Kozlova 2004 presented spectroscopy of the H$_\alpha$ and Na I D lines from 1999 to 2003. They showed that the typical profile for this period was a 3 or 4 times continuum line with typically small but significant blue-shifted absorption. One extremely absorbed, nearly flat-topped emission line was reported in August 1999, similar to the two lower profiles in figure \ref{fig:haebe-lprof1} or those of Miroshnichenko et al. 1999. Our line profiles were typically 4 or 5 times continuum but a consistent small amount of blue-shifted absorption was also observed.
	
	Liu et al. 2007 resolved warm circumstellar material around this star on the 10's of AU scale using 10$\mu$m nulling interferrometry. Garcia Lopez et al. 2006 derive an accretion rate of -6.59 Log(M$_\odot$/yr) using Bracket-$\gamma$ emission at 2.166$\mu$m which is quite high.

\subsection{HD 150193 - MWC 863}

	This star underwent a quite strong change in line profile. Figure \ref{fig:var} shows the change from a low-intensity line with a relatively narrow, mildly blue-shifted absorption to a full P-Cygni profile. Carmona et al. 2007 reported FORS2 spectroscopy of the TTauri-component: HD 150193B at a distance of 1.1$''$ from the primary. The companion had a spectral type of F9Ve and had H$_\alpha$ emission of roughly 2.5 times continuum while the primary had a ration of 2.2:1. Garcia Lopez et al. 2006 derive an accretion rate of -7.29 Log(M$_\odot$/yr) using Bracket-$\gamma$ emission at 2.166$\mu$m.

\subsection{V1295 Aql - HD 190073 - MWC 325}

	This star is far away from the well known star formation regions, but shows a large far-IR excess from cool dust and is more luminous than other Herbig Ae stars (Sitko 1981). Pogodin et al. 2005 presented a spectroscopic study of this star from 1994-2002 showing a very pronounced unstable wind with an optically thick equatorial disk. The star displays a wealth of lines in emission. The Pogodin et al. 2005 H$_\alpha$ profiles looked fairly similar to our observations presented in figure \ref{fig:haebe-lprof1} though with slightly stronger and more complicated absorption as well as a slightly lower intensity. They did not find any short-term variability, and derived a vsini of 12km/s. Wade et al. 2007 did not detect a magnetic field with an upper limit of roughly 50G but Catala et al. 2007 did claim a detection of $\sim$70G. Catala et al. 2007 fit the H$_\alpha$ line with a 1.4 10$^{-8}M_\odot$/yr wind having a terminal velocity of 290km/s and a base temperature of roughly 18000K. Baines et al. 2006 suggested this is a possible binary star based on a change in width of the psf across the H$_\alpha$ line without a corresponding change in centroid location.

\subsection{HD 200775 - MWC 361}

	This Herbig Be star has a fairly stable, double-peaked emission line except when it undergoes a period of strong variability every 3.68-years (Beskrovnaya et al. 1994, Miroshnichenko et al. 1998, Pogodin et al. 2000, 2004) and is a known binary at 2.25$''$ and 164$^\circ$ (Pirzkal et al. 1997). Ismailov \& Aliyeva 2005 reported H$_\alpha$ lines of 8.5-10 times continuum with no significant short-timescale variability but a slow drift over 20 days of individual line components. Pogodin et al. 2004 used radial velocity measurements to derive an orbital solution for the binary as an eccentric binary at e$\sim$0.3 with the mean separation at 1000R$_\odot$ with the secondary most likely to be a $\sim$3.5M$_\odot$ pre-main-sequence star.

\subsection{HD 36112 - MWC 758}

	This star had a significantly variable H$_\alpha$ line. Figure \ref{fig:var} shows a progression from 2004 to 2008 as both the intensity and absorption change significantly. Beskrovnaya et al. 1999 presented spectroscopy as well as multi-color photometry and polarimetry. They conclude that the star is an unreddened A8V star of 5Myr age, close to the ZAMS. They present evidence for a gaseous dusty envelope and a variable stellar wind with an extended acceleration zone. The wind dominates the inner envelope and there is a high temperature zone, likely chromospheric in origin. They find small but irregular photometric variability (0.2m). A P-Cygni type H$_\alpha$ line was observed in the 1994-1996 period with rapid variability, thought to be caused by jet-like inhomogeneities. They also concluded that this envelope is the inner part of a disk that is close to edge on. A vsini was calculated as 60km/s $\pm$6 and the H$_\alpha$ line had an intensity of roughly 2-3. They report the SED is 'flat' in the IR region which is typical of younger HAe/Be stars like AB Aurigae with hot dust. This star also has barium and silicon abundance anomalies common in chemically-peculiar stars.

\subsection{HD 35187}

	This star is quite variable on monthly timescales. Figure \ref{fig:var} shows the line profiles in timeseries. The star is an obvious binary. Wade et al. 2007 did not detect a magnetic field in either binary component and had an upper limit of roughly 50-100G depending on the field type. Rodgers 2001 derive an accretion rate of -7.29 Log(M$_\odot$/yr) using Bracket-$\gamma$ emission at 2.166$\mu$m.

\subsection{HD 35929}

	HD 35929 is nominally classified as a pre-main-sequence Herbig Ae star candidate because it exhibits H$_\alpha$ emission and a weak IR excess. Miroshnichenko et al. 2004 presented spectroscopic analysis that concluded that this star is F2 III spectral type with a T$_{eff}$ of around 6900, vsini of 70km/s with a mass of 2.3M$_\odot$ at roughly 350pc. They supported an argument by Marconi et al. 2000 that this star is not a young object, but post main-sequence giant in the instability strip with significant mass loss.

\subsection{HD 144432}

  	HD 144432 is a binary at 1.4$''$ with a K2 to K7 type star (Carmona et al. 2007). They reported that the H$_\alpha$ emission of the primary had a peak of 3.4:1 and the secondary also had H$_\alpha$ emission at roughly 2 times continuum. 

\subsection{HD 45677}

	In the line profiles presented in Oudmaijer \& Drew 1999, the H$_\alpha$ line in this star is single peaked and asymmetric. This is quite different than the double-peaked line profile of figure \ref{fig:haebe-lprof2}. Grady et al. 1993 reported the presence of redshifted absorption lines caused by infalling bodies. Baines et al. 2006 claimed this star to be a binary based on their spectroastrometry. 

\subsection{ HD 53367 - MWC 166 }

	This star is a known binary at a separation of 0.65$''$ (Fu et al. 1997). Baines et al. 2006 used this star as a test of their spectroastrometry and showed a strong change in both point-spread function width and centroid across the H$_\alpha$ line. This star showed a strong change from 2004 to 2008 as shown in figure \ref{fig:var}. The bottom dashed line from 2004 is very similar to the spectrum observed in December 2003 by Vink et al. 2005. However, the emission increased in 2007. In January 2008, the emission on the blue-shifted side of the line increased dramatically, while the red-shifted component was significantly reduced compared to 2007.  
		  
\subsection{Campaign Summary}

	The collection of stars in this sample is quite diverse. There is a wide range of morphologies and H$_\alpha$ line types. Some showed remarkable stability in line-profile shape over more than a year. HD 58647 showed an essentially invariant line profile from December of 2006 to January of 2008. Others were quite variable and some even showed a wide range of morphologies. Many of the targets are known or suspected binaries. Some are significantly variable on very short time-scales. In general, the blue-shifted absorption is much more variable. This is entirely expected because the absorption comes from a comparatively small volume. However, some stars showed significant variability across the entire line, particularly HD 35187. This large collection allows a solid characterization of each star and will be useful in discussing the spectropolarimetric effects.

\section{Current Theory of Linearly Polarized Spectropolarimetric Line Profiles}

	Before discussing any specific spectropolarimetric observations, we review the leading theory for calculating spectropolarimetric line profiles. This gives us a framework for comparing and contrasting the observations which may fit this theories from those that don't. In this paper, we will denote a spectropolarimetric profile as $q(\lambda)$ where $q$ represents the percent polarization as a function of wavelength. The symbol $Q(\lambda)$ will represent a spectropolarimetric profile in percent that is normalized by the continuum flux, so that $q=\frac{Q(\lambda)}{I_c(\lambda)}$ where $I_c$ represents the continuum-normalized line profile. See Harrington \& Kuhn 2008 for detailed examples.

\begin{figure*}
\begin{center}
\includegraphics[width=0.23\linewidth, angle=90]{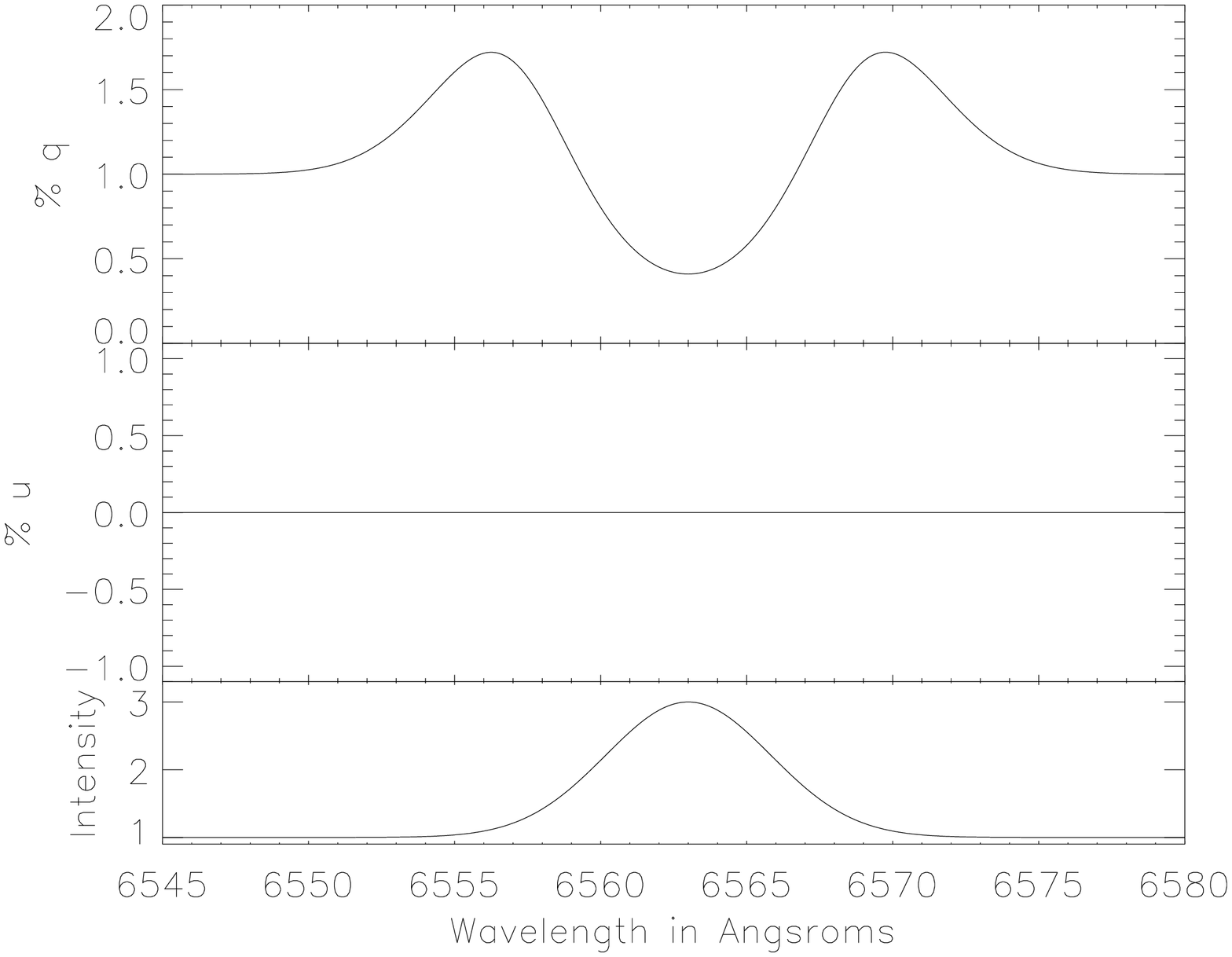}
\includegraphics[width=0.23\linewidth, angle=90]{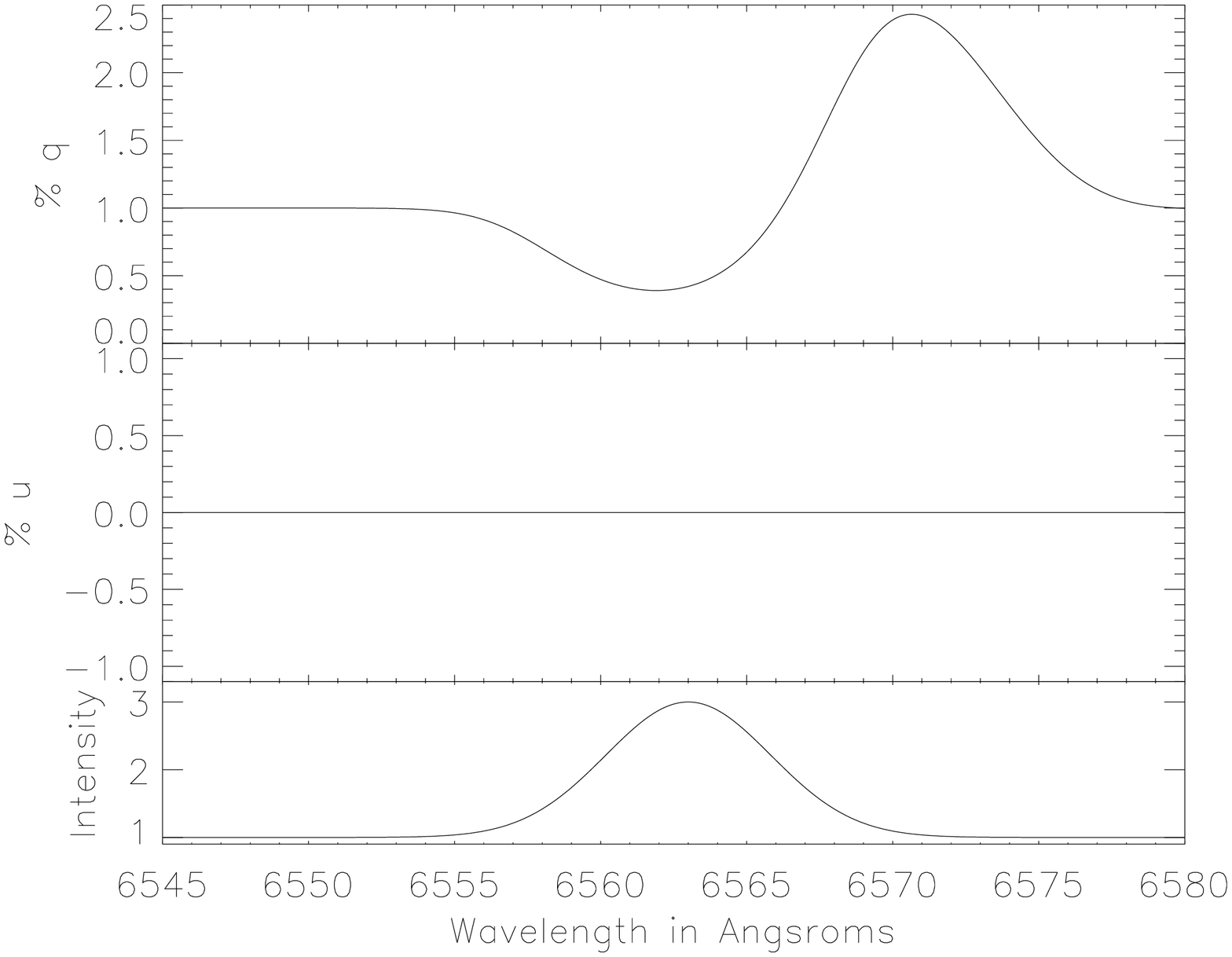} \\
\includegraphics[width=0.23\linewidth, angle=90]{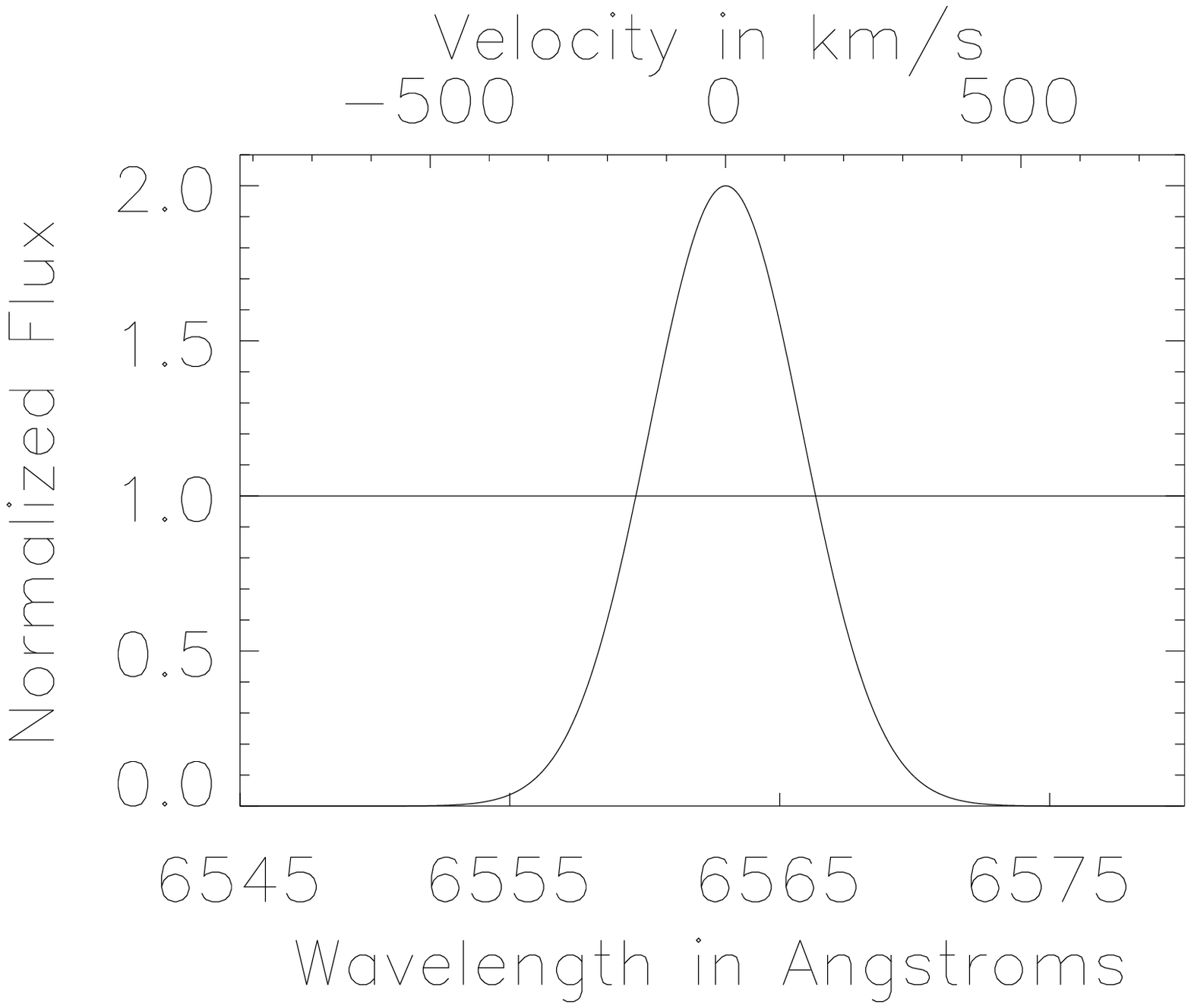} 
\includegraphics[width=0.23\linewidth, angle=90]{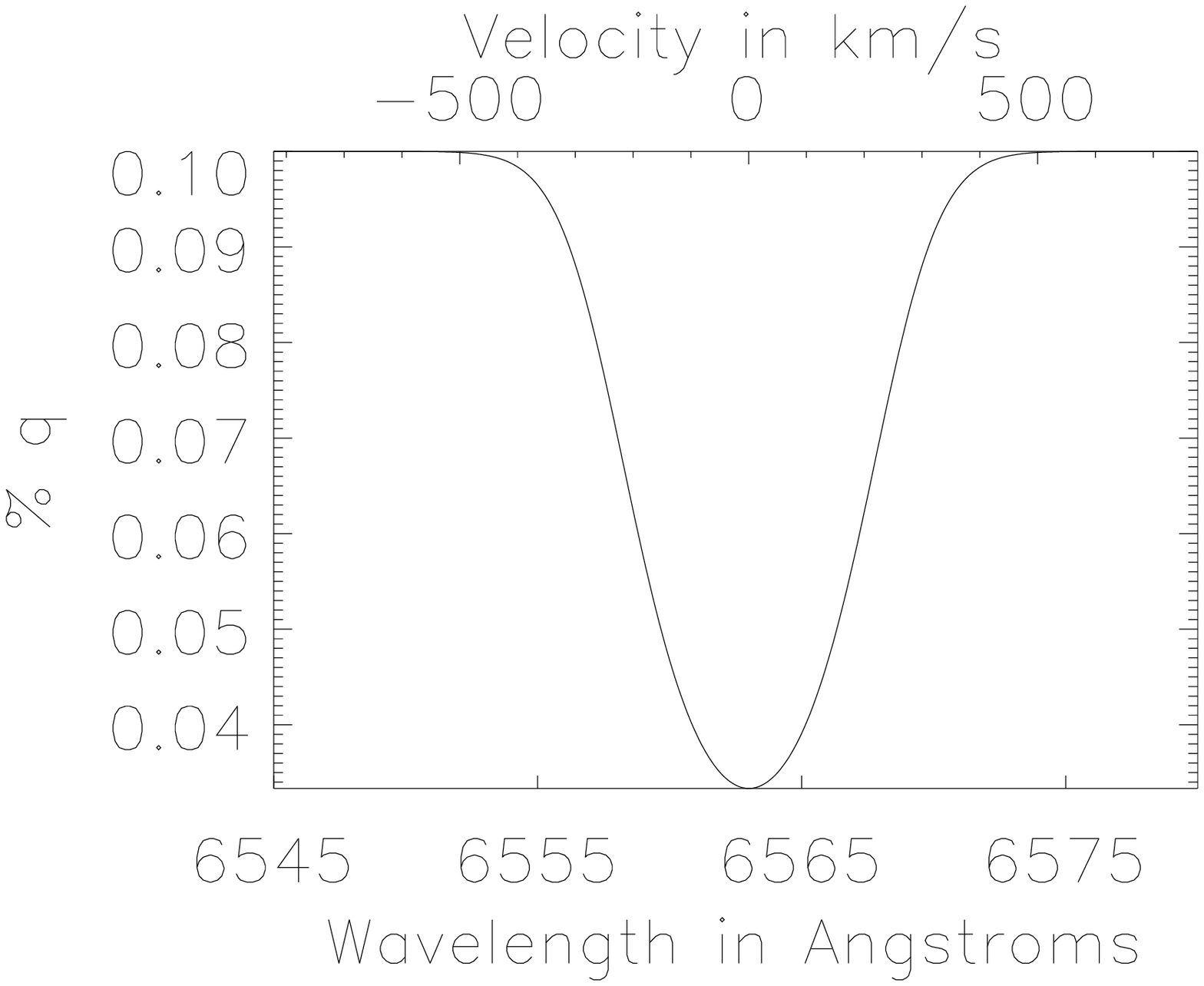} \\ 
\includegraphics[width=0.23\linewidth, angle=90]{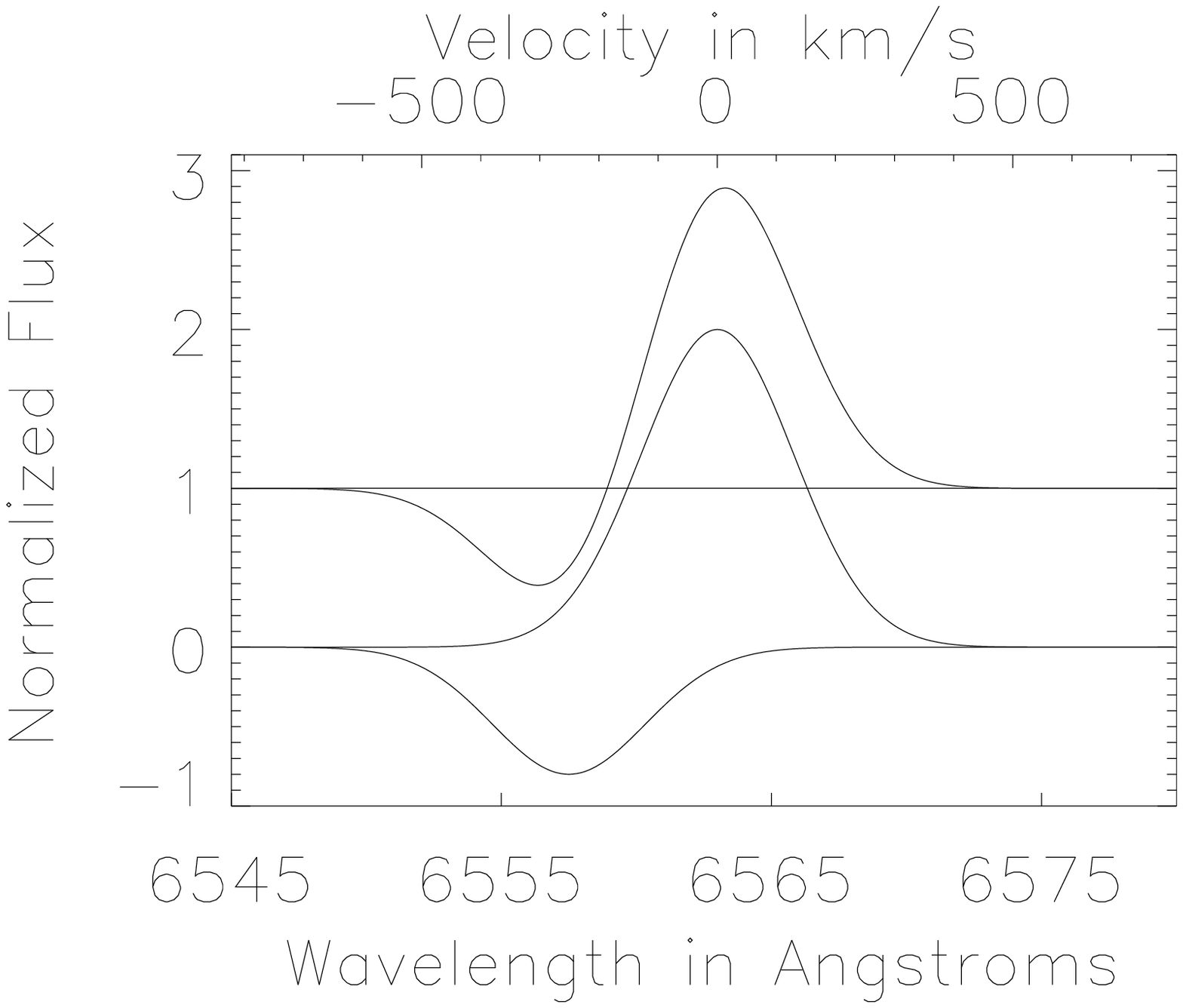}
\includegraphics[width=0.23\linewidth, angle=90]{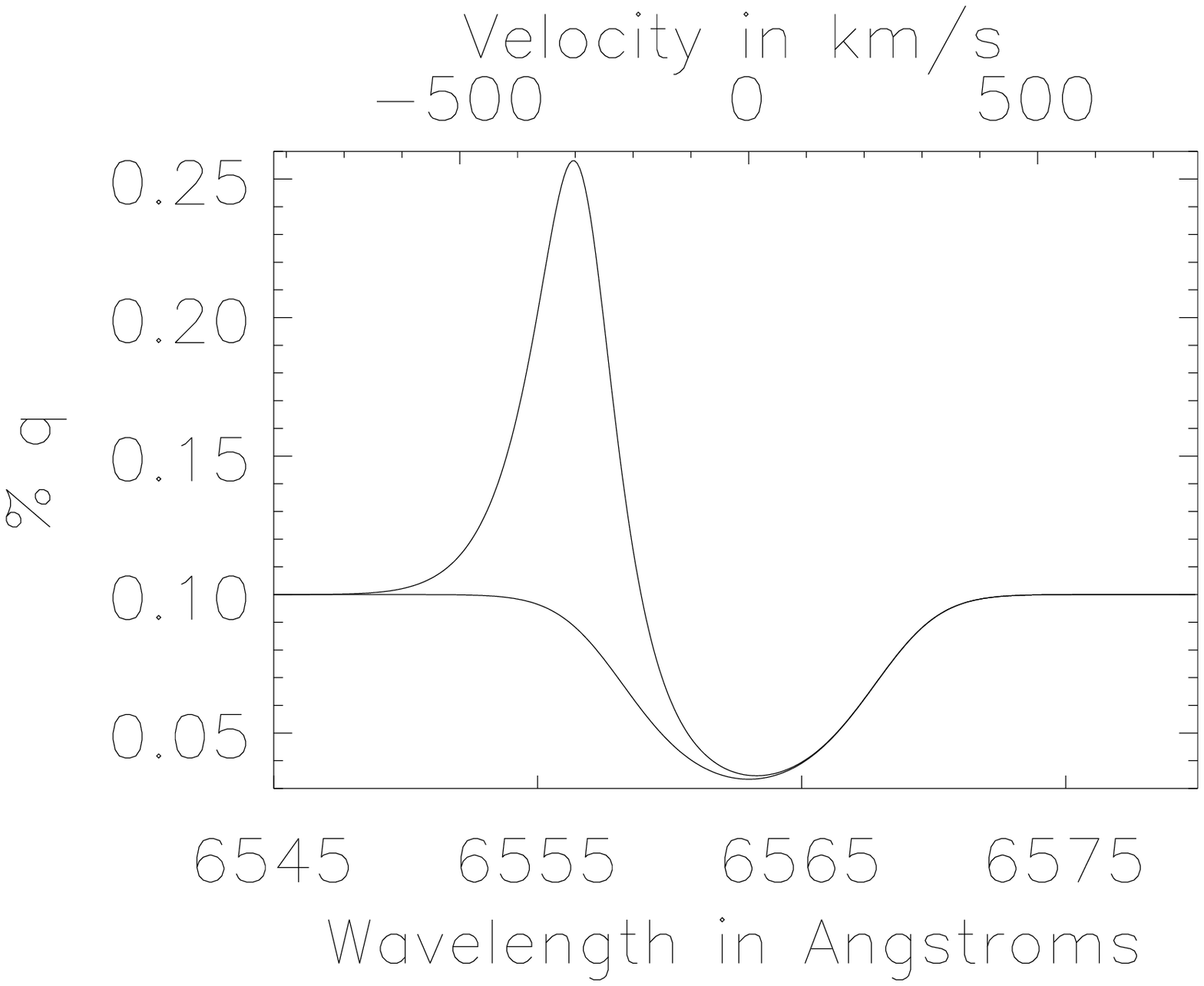}
\caption{The spectropolarimetric model parameters for a 60$^\circ$ inclined ring. From left to right: {\bf a)} The spectropolarimetric signature for a disk (orbital motion only) {\bf b)} The spectropolarimetric signature for a wind (radial outward motion only). {\bf c)}  The intensities from the stellar continuum and the broad line formation region. This simple example assumes a simple continuum, line-free spectrum that is 0.1\% polarized. An emission line with the line:continuum ratio of 2:1 forms outside the polarizing region and is entirely unpolarized.  {\bf d)}  The polarization spectrum assuming the star is 0.1\% polarized in +Q. The unpolarized emission dilutes the continuum polarization. {\bf e)}  A P-Cygni type profile as the sum of intensities from the stellar continuum and the broad line formation region as well as blue-shifted absorption. This simple example assumes a simple continuum, line-free spectrum that is 0.1\% polarized. An emission line with the line:continuum ratio of 2:1 forms outside the polarizing region and is entirely unpolarized. The blue-shifted absorption removes unpolarized light from the system.  {\bf f)} The polarization spectrum assuming the stellar continuum is 0.1\% polarized entirely in +Q. The unpolarized emission dilutes the continuum polarization but the absorptive component removes unpolarized light, increasing the polarization across the absorptive component. The original depolarization signature is overplotted to illustrate the effect of the absorption in the depolarized trough.}
\label{fig:scatpolsig}
\end{center}
\end{figure*}

	The existing framework (aside from Kuhn et al. 2007) for describing linear spectropolarimetric effects in hot stars has relied on the scattering of starlight off electrons in an asymmetric circumstellar environment. Rayleigh or Mie scattering is treated the same way. The important physics here is just that continuum scattered light is linearly polarized with a dependence on the scattering angle. Asymmetries in the circumstellar environment lead to a net scattered light polarization. A wavelength dependence over typical lineprofiles results from resonant absorption, unpolarized line emission, and systemic velocities within the scattering environment. 

	Some of these models were developed to explain Be-star observations (cf. McLean \& Brown 1978, McLean \& Clarke 1979, McLean 1979, Poeckert \& Marlborough 1976, 1977, 1978, 1979). They included an asymmetric-envelope theory for continuum polarization and the ``depolarization'' effect, especially the theories developed in McLean 1979. An analytical ``disk-scattering" model was presented in Wood et al. 1993 and Wood \& Brown 1994 for a thin equatorial disk of material moving orbitally or radially. Vink et al. 2005a extended this to more realistic geometries using numerical and monte-carlo calculations. Their approach was subsequently used to explain observations of Herbig Ae/Be stars in Vink et al. 2002, 2005b and Mottram et al. 2007 as well as hot-massive stars in Oudmaijer et al. 2005. 
	
	A conceptual understanding of the disk-scattering effect follows from Wood et al. 1993. A simple way to conceptualize this linear spectropolarimetric signature is to imagine a disk inclined to the line of sight. The scattered and doppler-shifted light from the disk causes the spectropolarimetric effect. Imagine that the stellar rotation axis is +q in the instrument reference frame. If simple Rayleigh scattering is the cause for polarization, the approaching and receding parts of the disk will be polarized vertically (+q) and will be blue and red shifted respectively. This results in a positive q polarization on the red and blue sides of the line. The line-center component of the scattered light will be along the line of sight, in front of and behind the star, and will be horizontally polarized leading to a negative q polarization. If the disk is symmetric, all Stokes u terms cancel. The resulting spectropolarimetric signature is a double-peaked, symmetric function like that in figure \ref{fig:scatpolsig}. 
	
	Now, imagine the same thin circumstellar disk with all motion being radially outwards (i.e. an equatorial wind).  The line-center component of the scattered light (from the circumstellar disk moving radially outward) is the component where the red-shift from the star-cloud motion and blue-shift from cloud-telescope motion will cancel most completely. This is the material closest to the line of sight, in front of the star. Scattered light from this region will still be polarized as negative q. However, both sides of the disk are moving tangentially on the sky (outward from the star) and see a red-shifted stellar spectrum with no shift with respect to the telescope. This leaves positive q only on the red side of the line. There is no scattered light on the blue side of the line. This is seen in figure \ref{fig:scatpolsig}. Thus any type of outward radial motion serves to shift the linear scattering signature towards the red. 
	 
	The main concept to take away from this is that the disk-scattering effect produces double-peaked, symmetric spectropolarimetric profiles. When there is radial motion, the amplitude of the polarization on the blue side of the line is decreased, while the polarization on the red side of the line remains present. The Vink et al. 2005 paper shows that the change in position angle of polarization also spans the entire line. There is a transition from a single qu-loop over the entire line to a double-qu-loop as the size of an inner hole increases. It is important to note that these qu-loops also come from the polarization change across the entire line. This will be critical for comparing the disk-scattering theory with our spectropolarimetric observations. It should also be noted that Vink et al. 2002 and 2005b briefly describe some the spectropolarimetric effects of non-uniform compact sources.

\subsection{Depolarization Effects}

	In the case of emission lines that form over broad regions, a depolarization effect is possible. If somehow the stellar continuum polarization forms interior to the line-formation region, the less-polarized emission will depolarize the stellar light within the emission line producing a broad decrease in polarization. This was described in Poeckert \& Marlborough 1977 and McLean 1979. Several authors use the depolarization effect in Be stars to separate intrinsic and interstellar polarization. Electron scattering models that assume asymmetric envelopes fit low-resolution spectropolarimetric observations quite well. The depolarization effect was conceived in this context - the H$_\alpha$ emission region is thought to be quite large in Be stars, and flattened, large emission regions have been resolved interferometrically (cf. Quirrenbach et al. 1994). The electron scattering optical depth estimates showed that the bulk of the intrinsic continuum polarization came from the inner region of the flattened envelope. Since the H$_\alpha$ emission formed outside this region, it would be unpolarized.
	
	In another simple example we assume a 0.1\% continuum polarization is formed near the star by scattering in an asymmetric envelope and that the emission line is formed entirely outside this polarizing region. The depolarization is calculated from the continuum polarized flux added to the unpolarized emission line flux. If we assume the asymmetric interior region produces only +Q polarization then the corresponding intensities and polarization spectra are shown in figure \ref{fig:scatpolsig}.

If the H$_\alpha$ photons come from outside the asymmetric (polarized continuum source) region and the emission is (assumed to be but isn't always) unpolarized and the absorption predominantly affects the polarized continuum then the fractional polarization will be proportional to $1/I$. In a ``disky" central absorber or a ``windy" P-Cygni type profile it follows that the largest polarization can be centered on the deepest absorptive feature of the intensity line profile. 

	An example of this effect can be constructed for a P-Cygni profile. We use a 0.1\% continuum polarization and geometrical assumptions following McLean 1979. We assume that the blue-shifted absorption only removes unpolarized light. In this model, the absorption occurs only along the line-of-sight to the photosphere and the continuum polarization must come from a broad asymmetric envelope inside the emission region. The observed line profile is built as a superposition of stellar continuum, emission and a smaller blue-shifted absorption, as in figure \ref{fig:scatpolsig}. Here the absorption is blue-shifted by 250km/s with the same width as the stellar spectrum. Since both the absorption and emission are only affecting unpolarized light, and the continuum is assumed to be polarized independent of wavelength, the observed spectropolarimetric effect ($q(\lambda )=Q/I$) is proportional to $1/I(\lambda )$. The resulting spectropolarimetric effect is shown in figure \ref{fig:scatpolsig} with the corresponding simple depolarization effect for reference. Essentially, the removal of unpolarized light increases the polarization from continuum values by a factor of 1/I. This model is quite simple, but a more advanced model incorporating more realistic scenario's has not been created. An important thing to note is the morphology of the depolarization effect for a P-Cygni profile. In general this effect strictly follows $1/I(\lambda )$ so there must be a spectropolarimetric effect across the entire line for this explanation to work.

\subsection{Intrinsic Polarized Absorption}

	There is  another important mechanism that leads to linearly polarized stellar line profiles that is unrelated to scattering. This effect (called optical pumping) causes an intrinsic absorptive linear polarization (see Happer 1972).  It is a non-local thermodynamic effect that has been calculated and applied to describe a system like AB Aurigae (Kuhn et al. 2007). Calculations yield linear polarization of order 1$\%$ that is proportional to the wavelength dependent intrinsic absorption near the star.  The polarization originates from radiation anisotropy that causes anisotropy in the groundstate magnetic sublevels of the resonant transition. Since the absorptive contribution to a complex $H_\alpha$ line profile is often unknown apriori, this mechanism may be a powerful tool for understanding the radiation environment and absorptive linear polarization of obscured stellar systems.

\begin{figure*}
\begin{center}
\includegraphics[width=0.23\linewidth, angle=90]{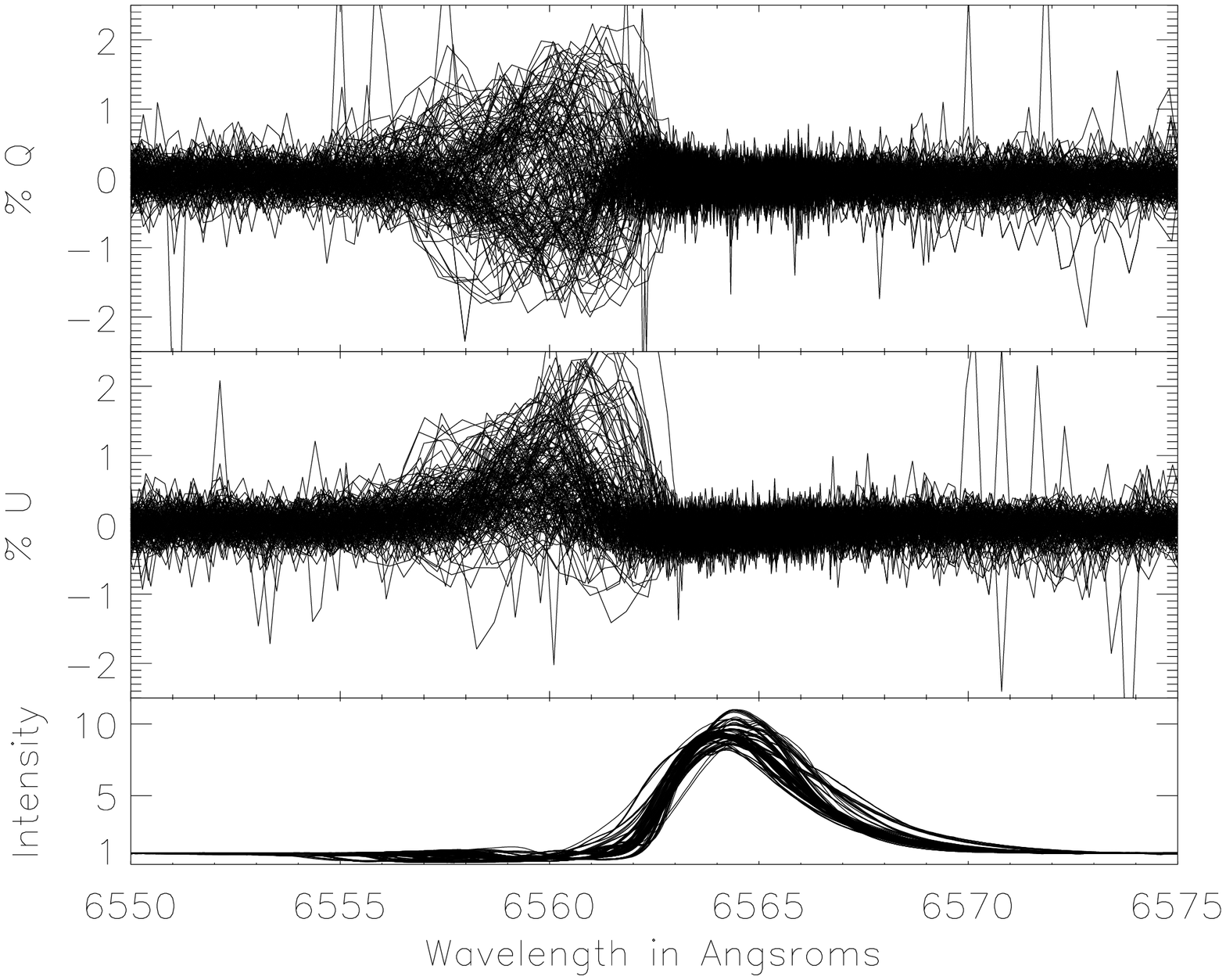}
\includegraphics[width=0.23\linewidth, angle=90]{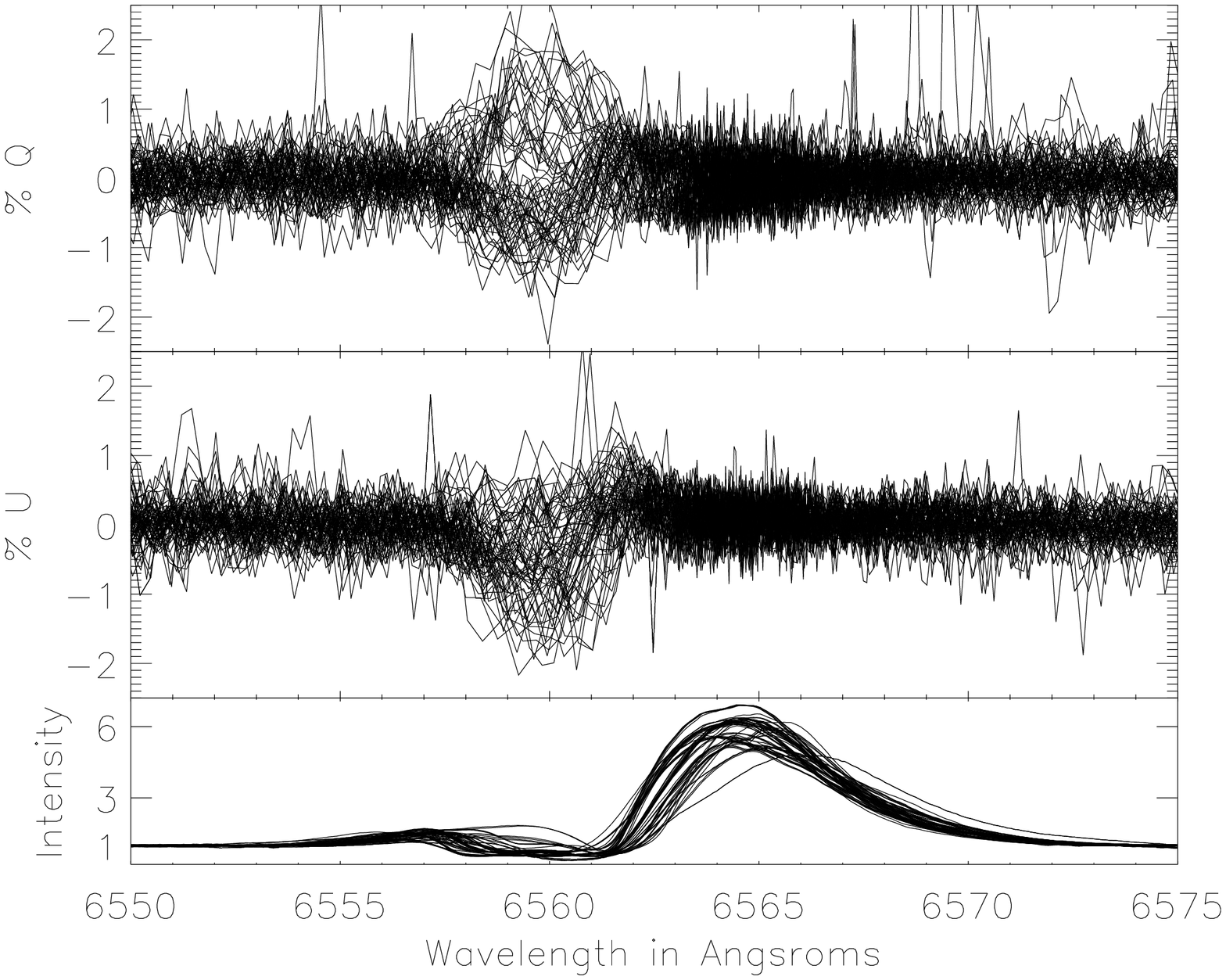}
\includegraphics[width=0.23\linewidth, angle=90]{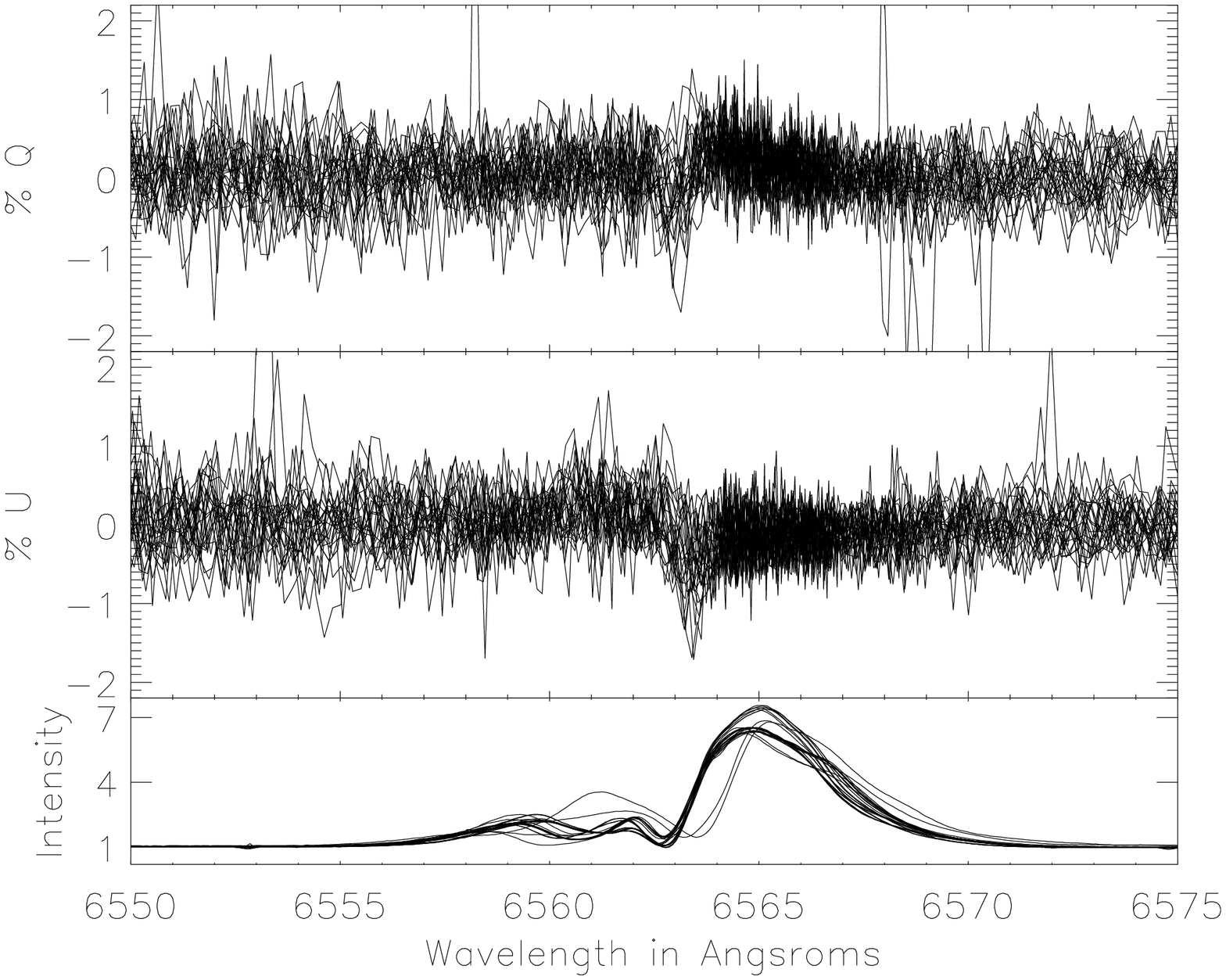} \\
\includegraphics[width=0.23\linewidth, angle=90]{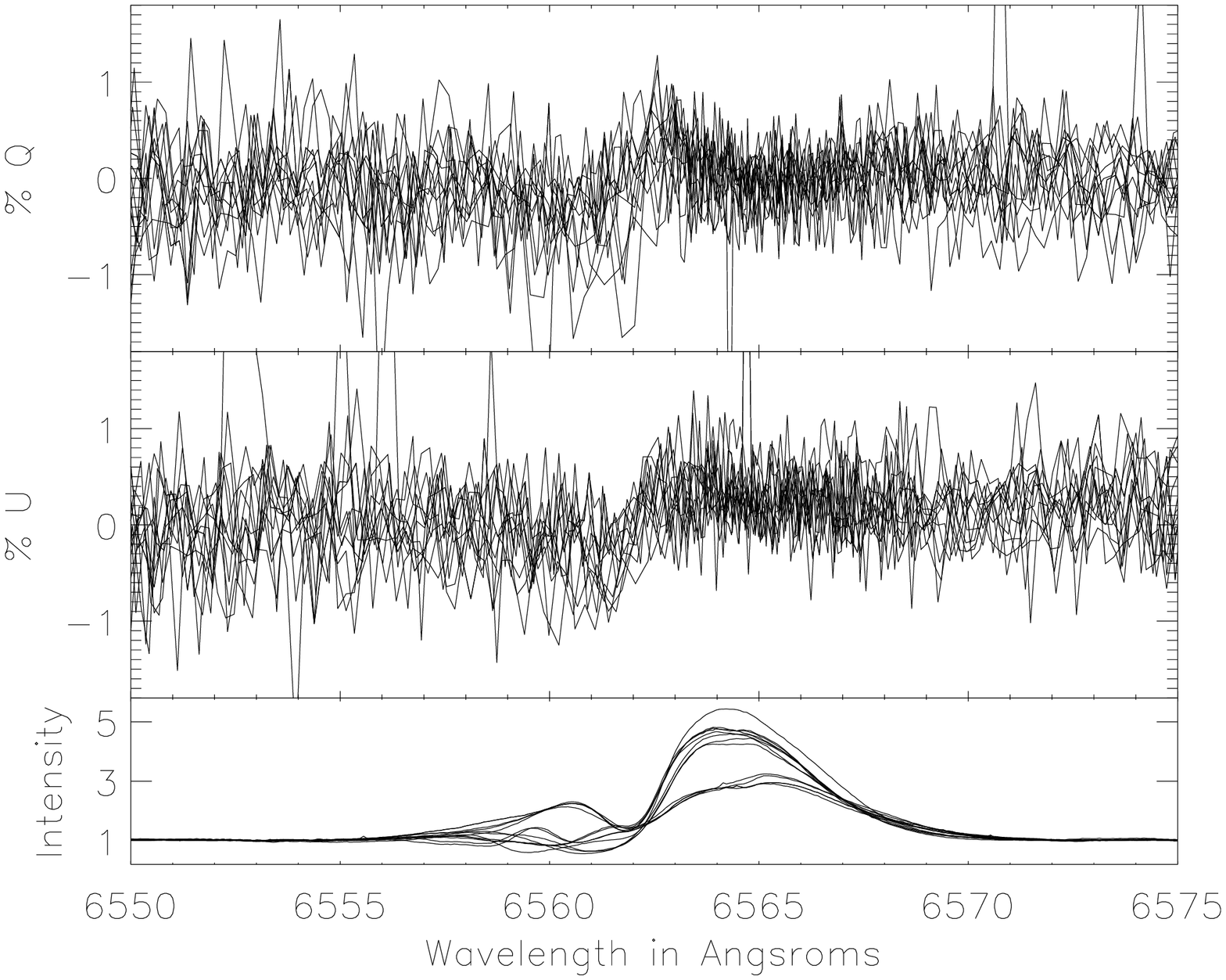}
\includegraphics[width=0.23\linewidth, angle=90]{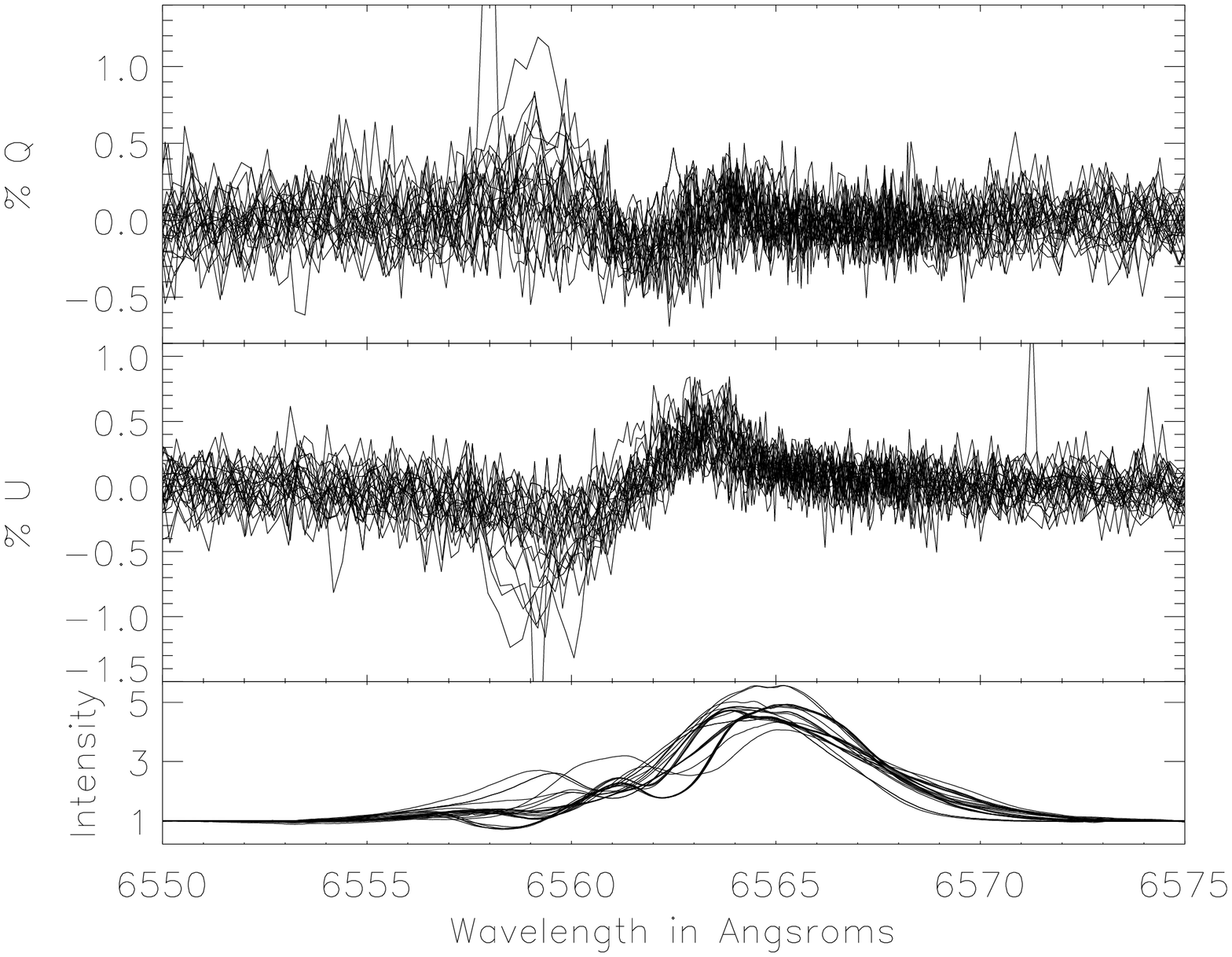}
\includegraphics[width=0.23\linewidth, angle=90]{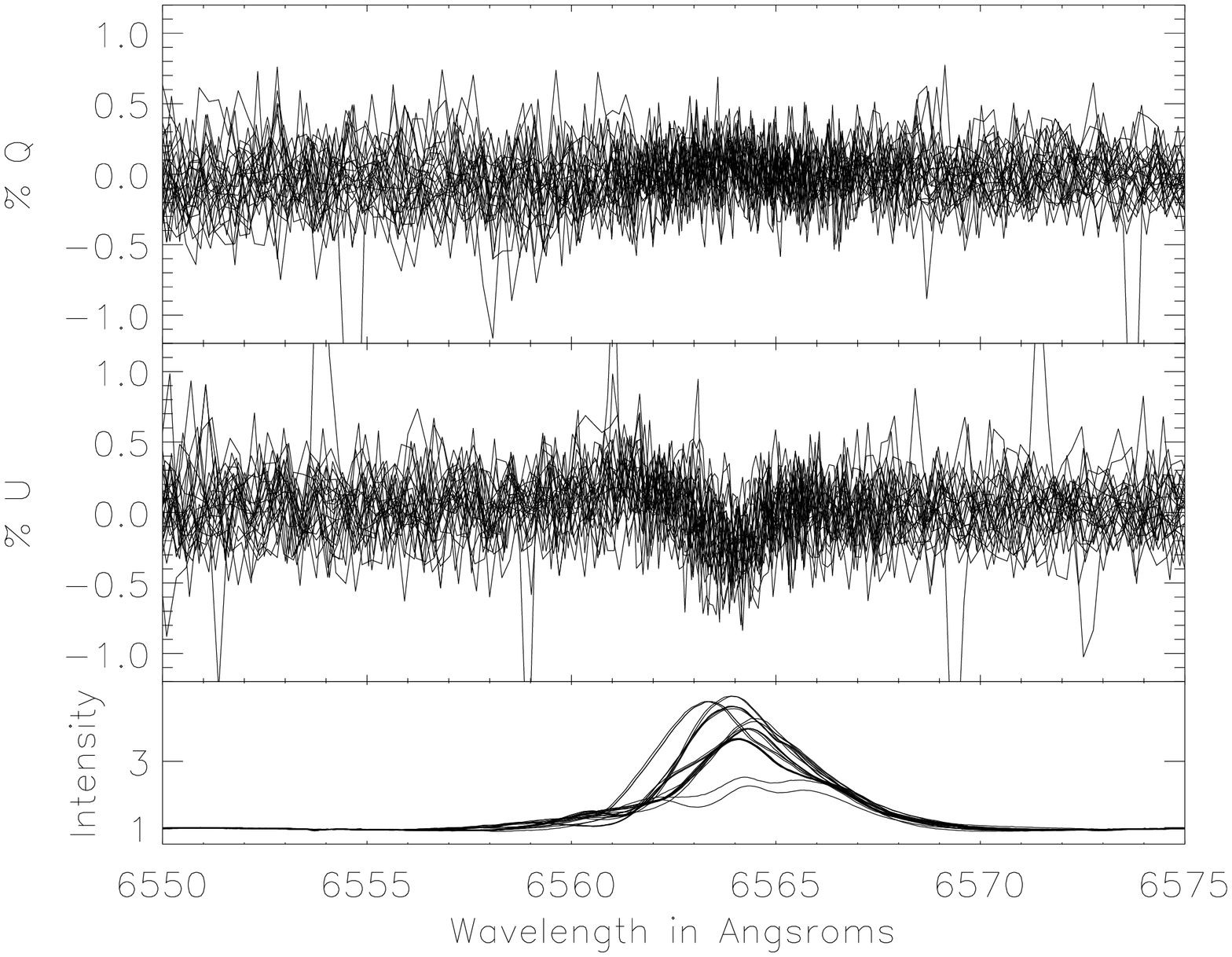} \\
\includegraphics[width=0.23\linewidth, angle=90]{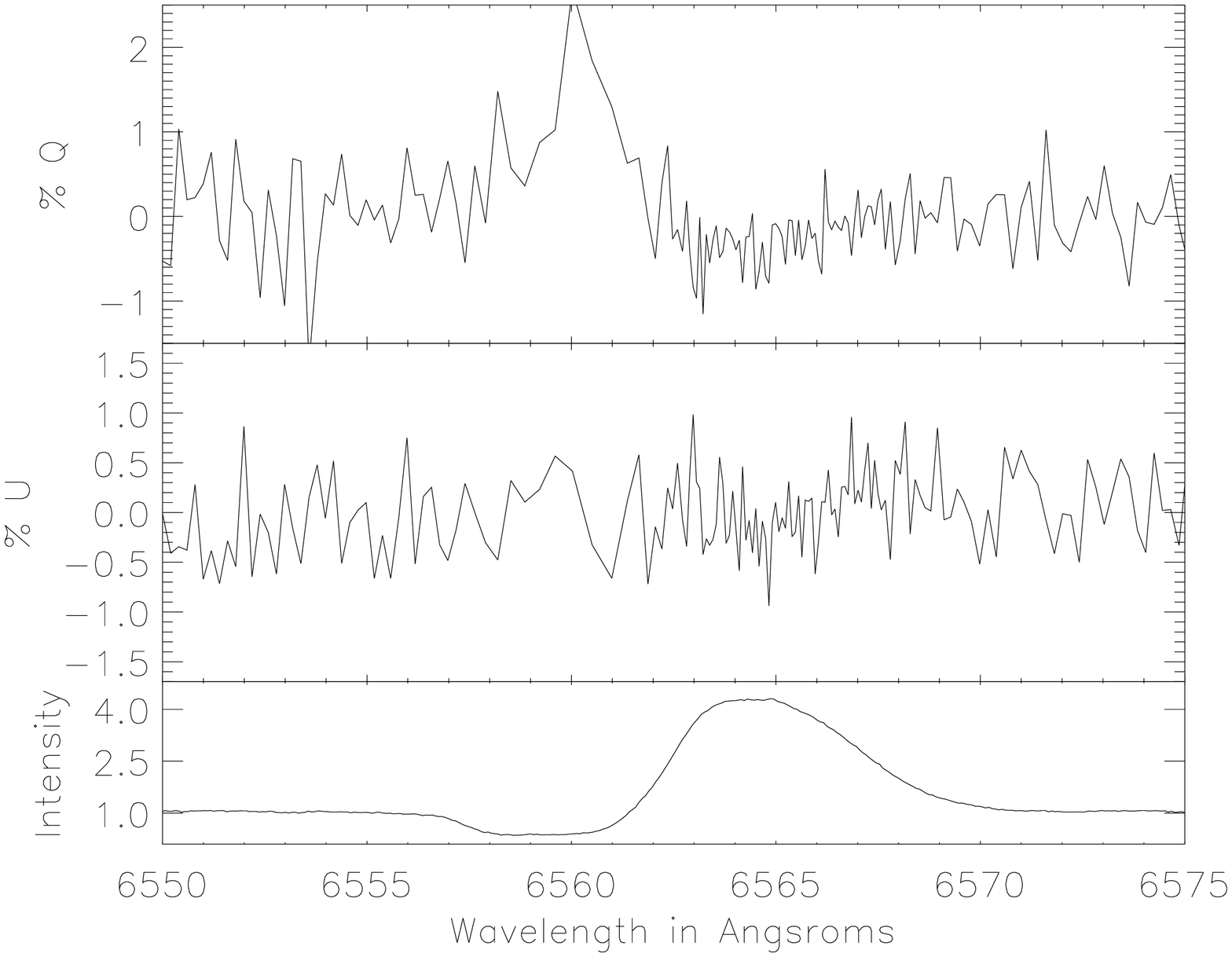}
\includegraphics[width=0.23\linewidth, angle=90]{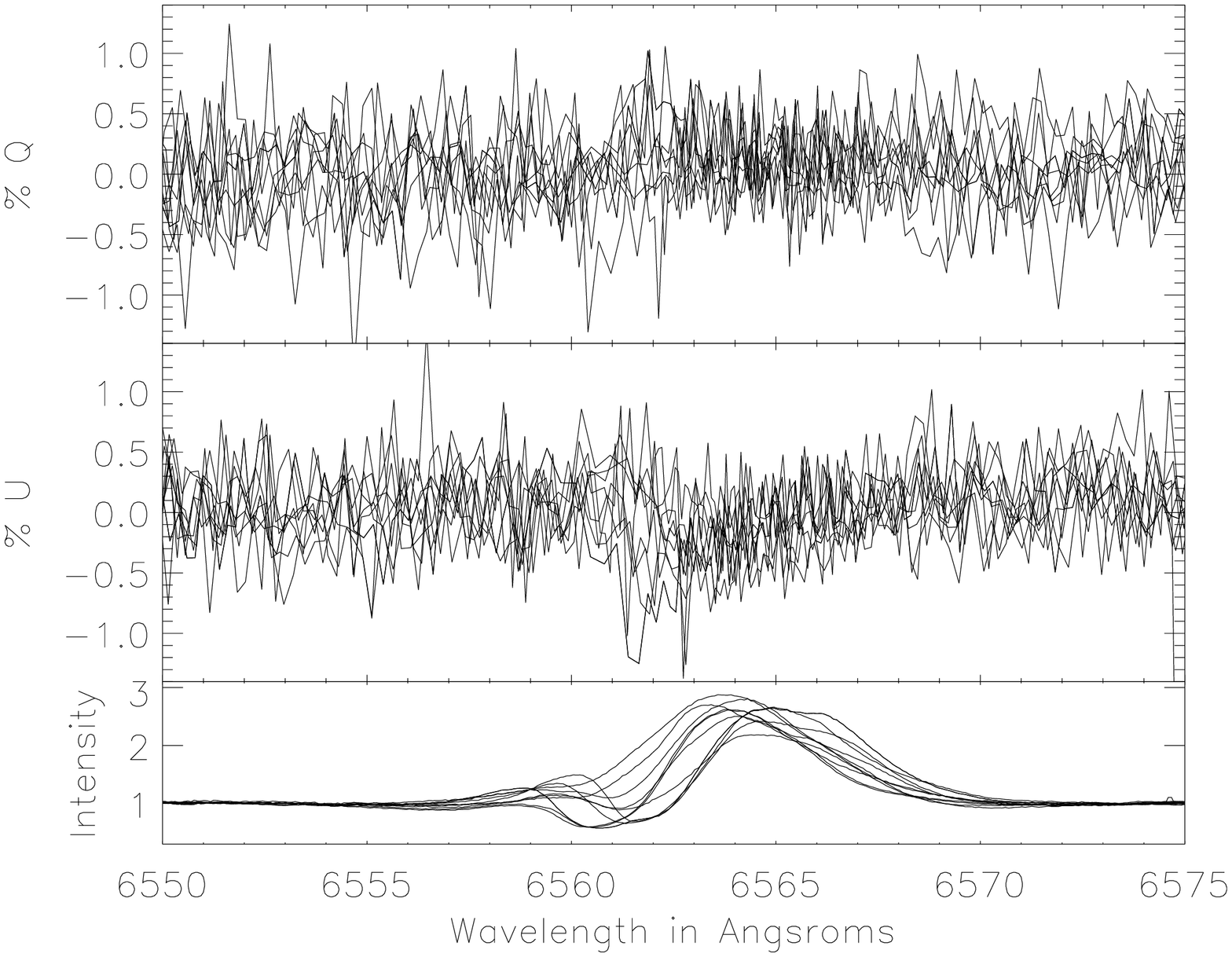}
\includegraphics[width=0.23\linewidth, angle=90]{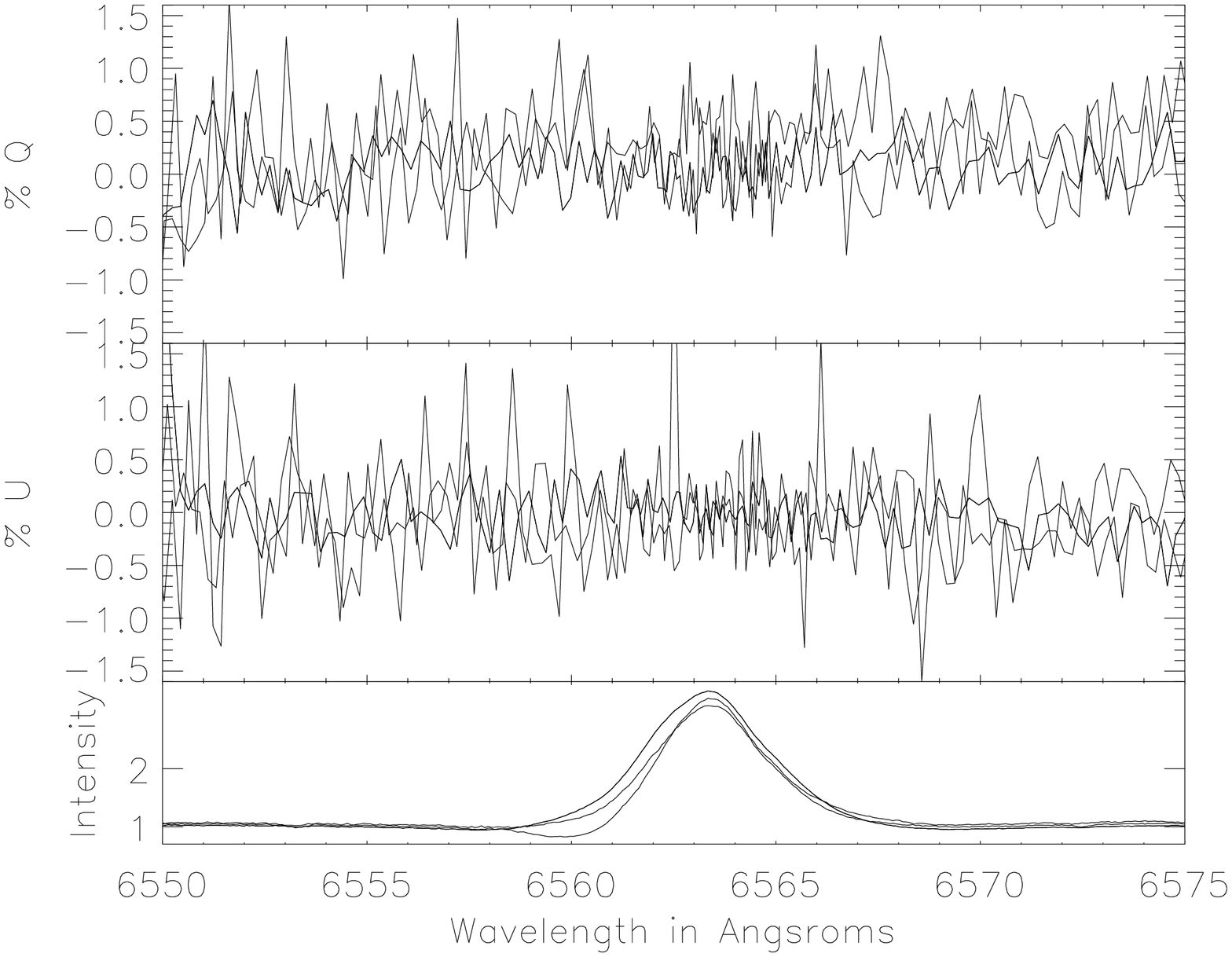} \\
\includegraphics[width=0.23\linewidth, angle=90]{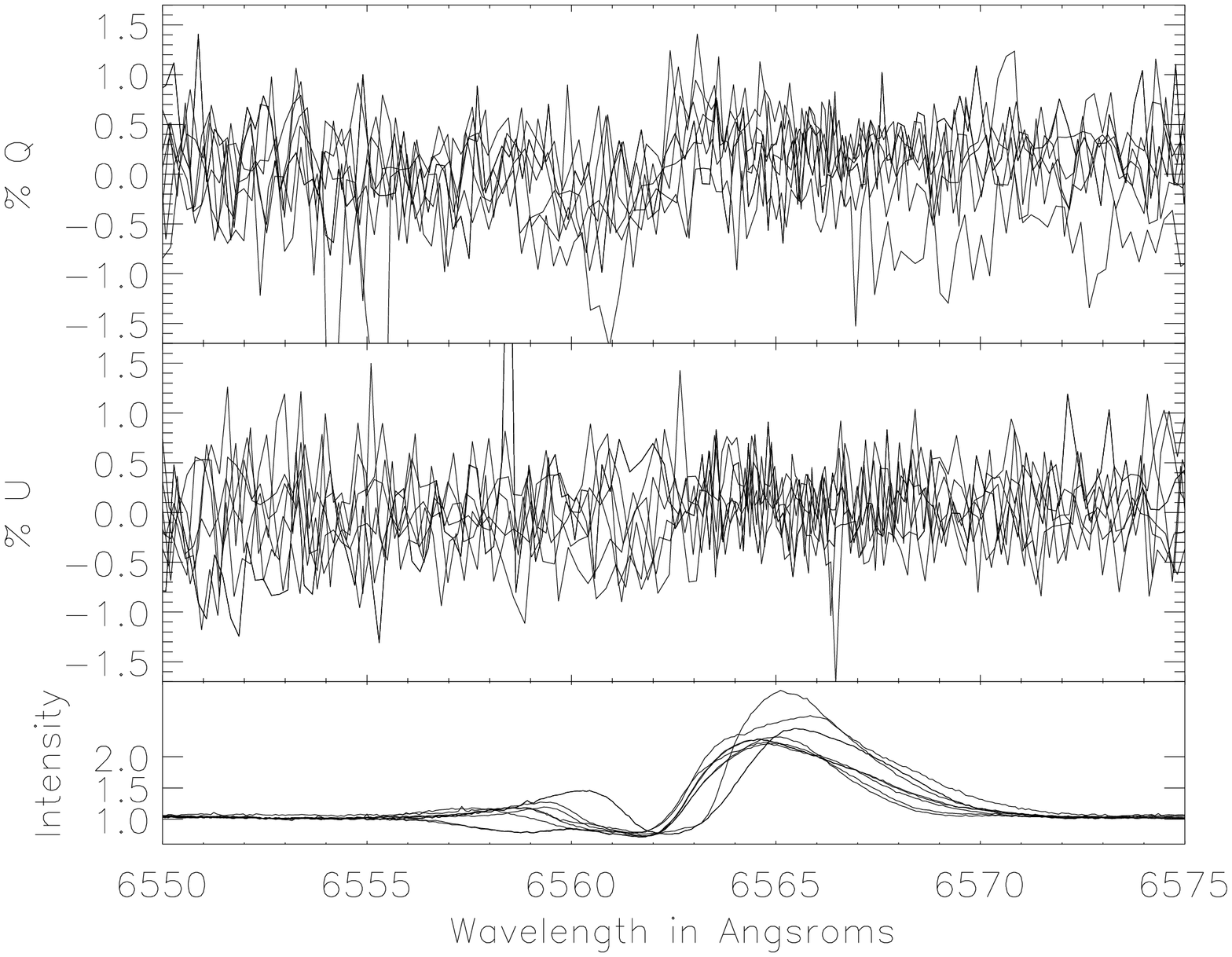}
\includegraphics[width=0.23\linewidth, angle=90]{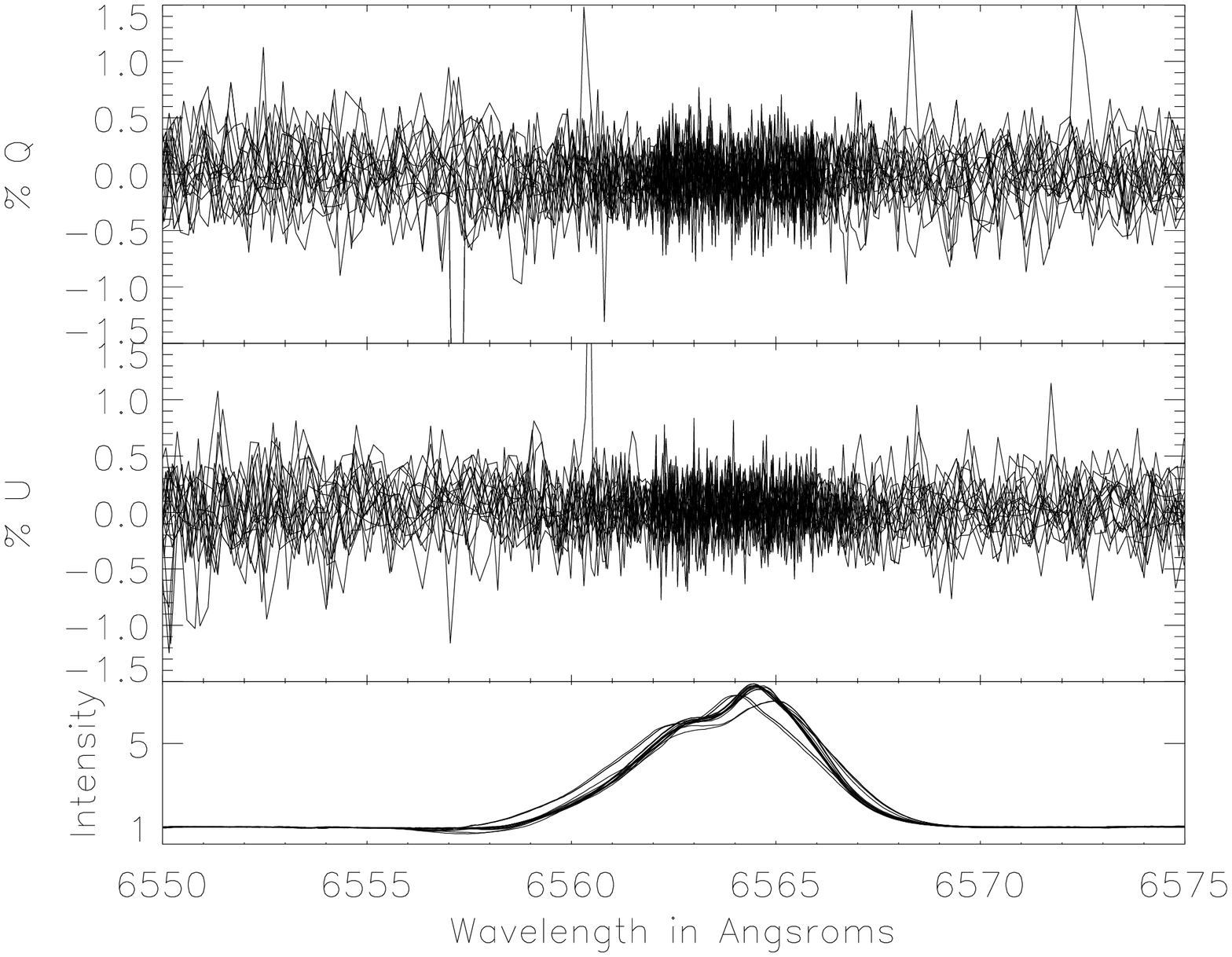}
\includegraphics[width=0.23\linewidth, angle=90]{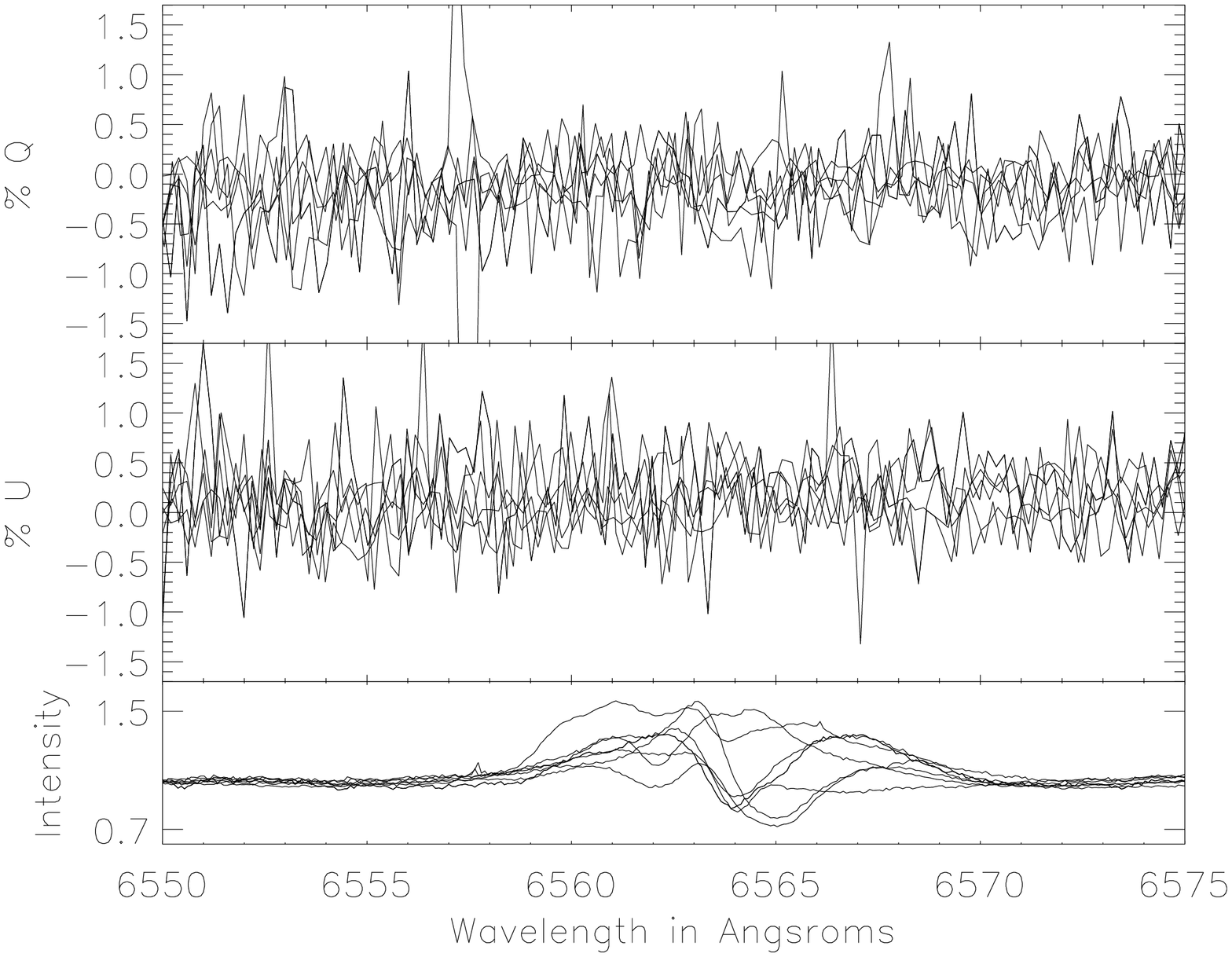} \\
\includegraphics[width=0.23\linewidth, angle=90]{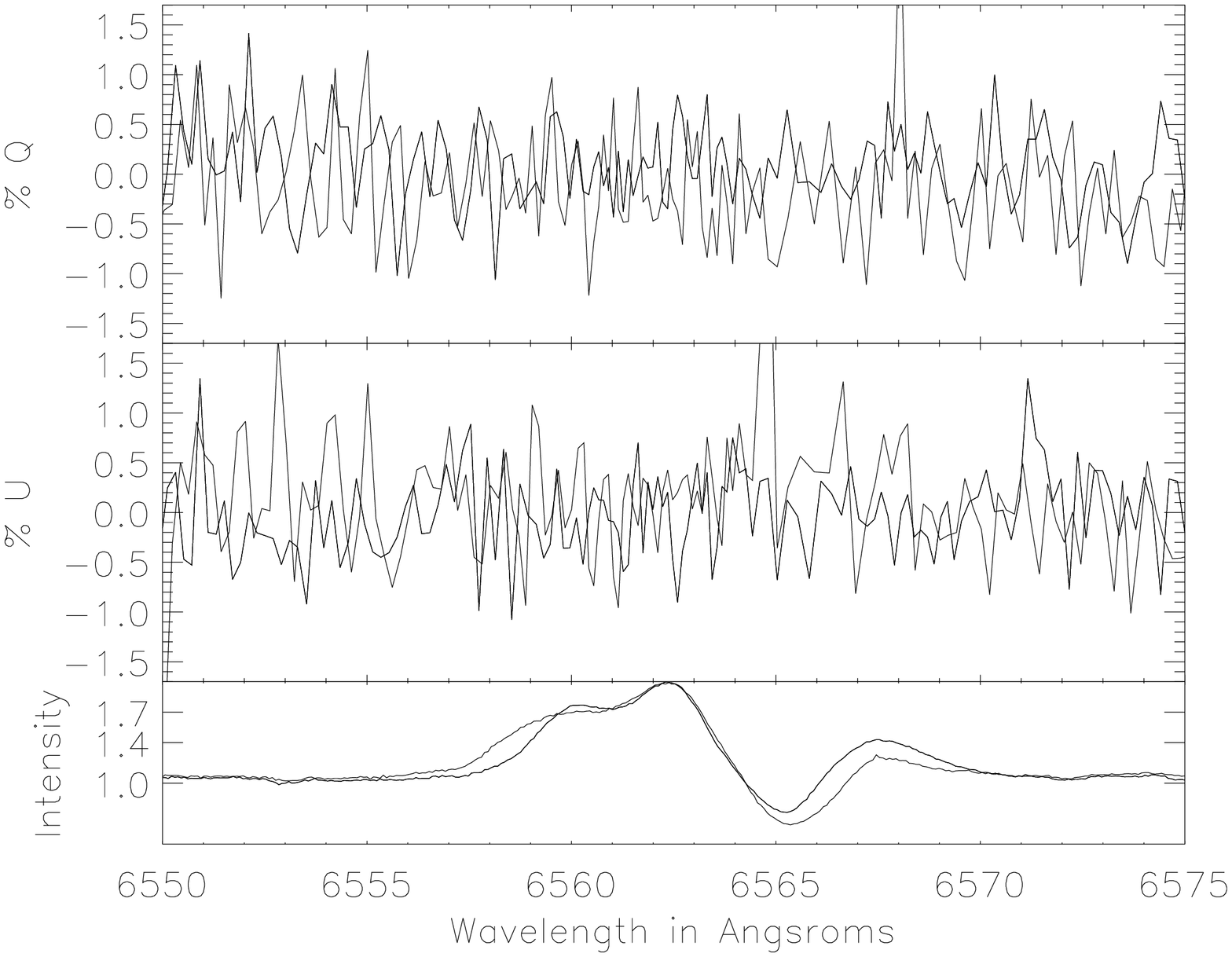}
\includegraphics[width=0.23\linewidth, angle=90]{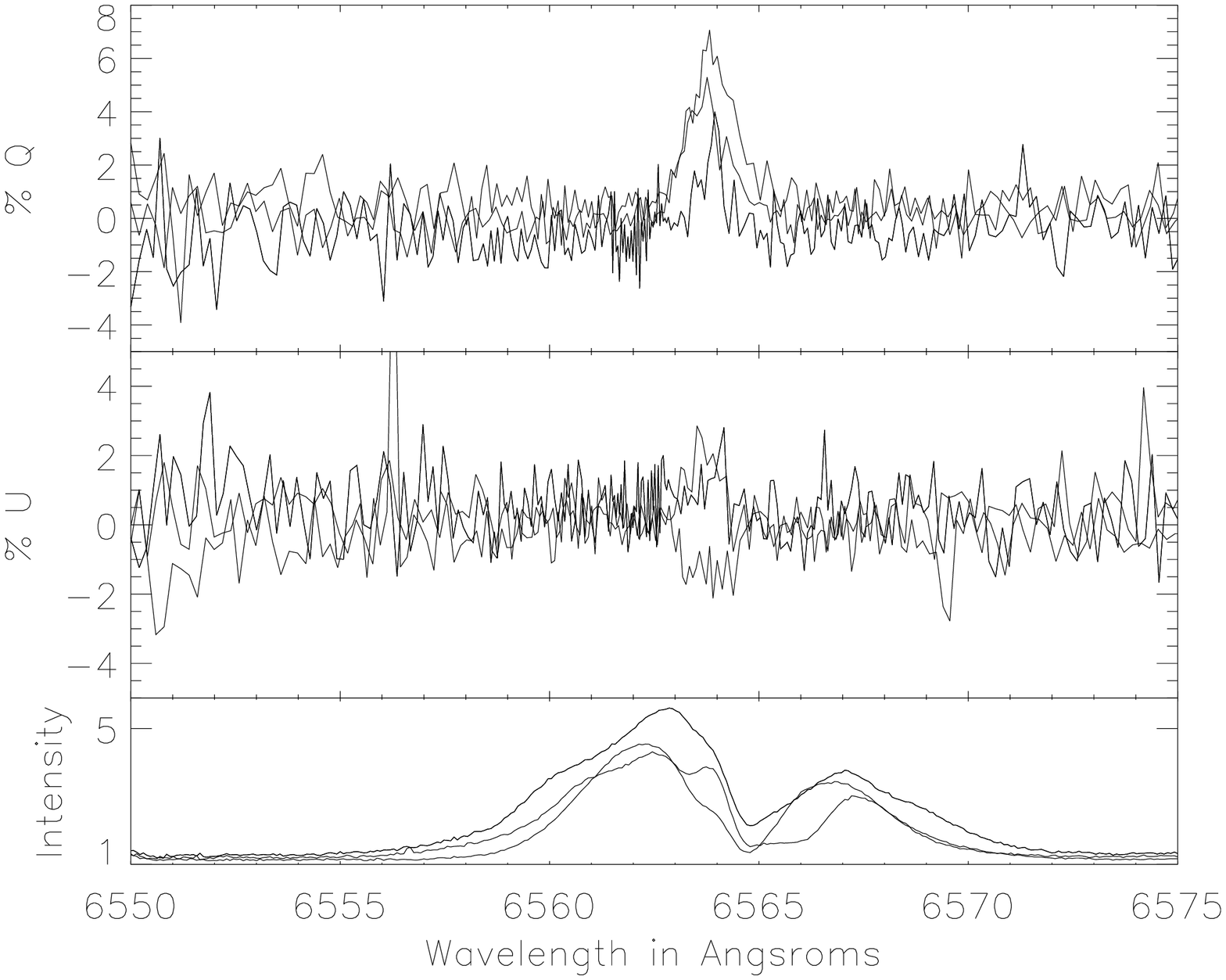}
\caption{Herbig Ae/Be Spectropolarimetric Profiles I. The stars, from left to right, are:  AB Aurigae, MWC 480, MWC 120, HD 150193, HD 163296, HD 179218, HD 144432, MWC 758, HD 169142, KMS 27, V 1295 Aql, HD 35187, HD 142666 and T Ori}
\label{fig:haebe-specpol1}
\end{center}
\end{figure*}

\begin{figure*}
\begin{center}
\includegraphics[width=0.23\linewidth, angle=90]{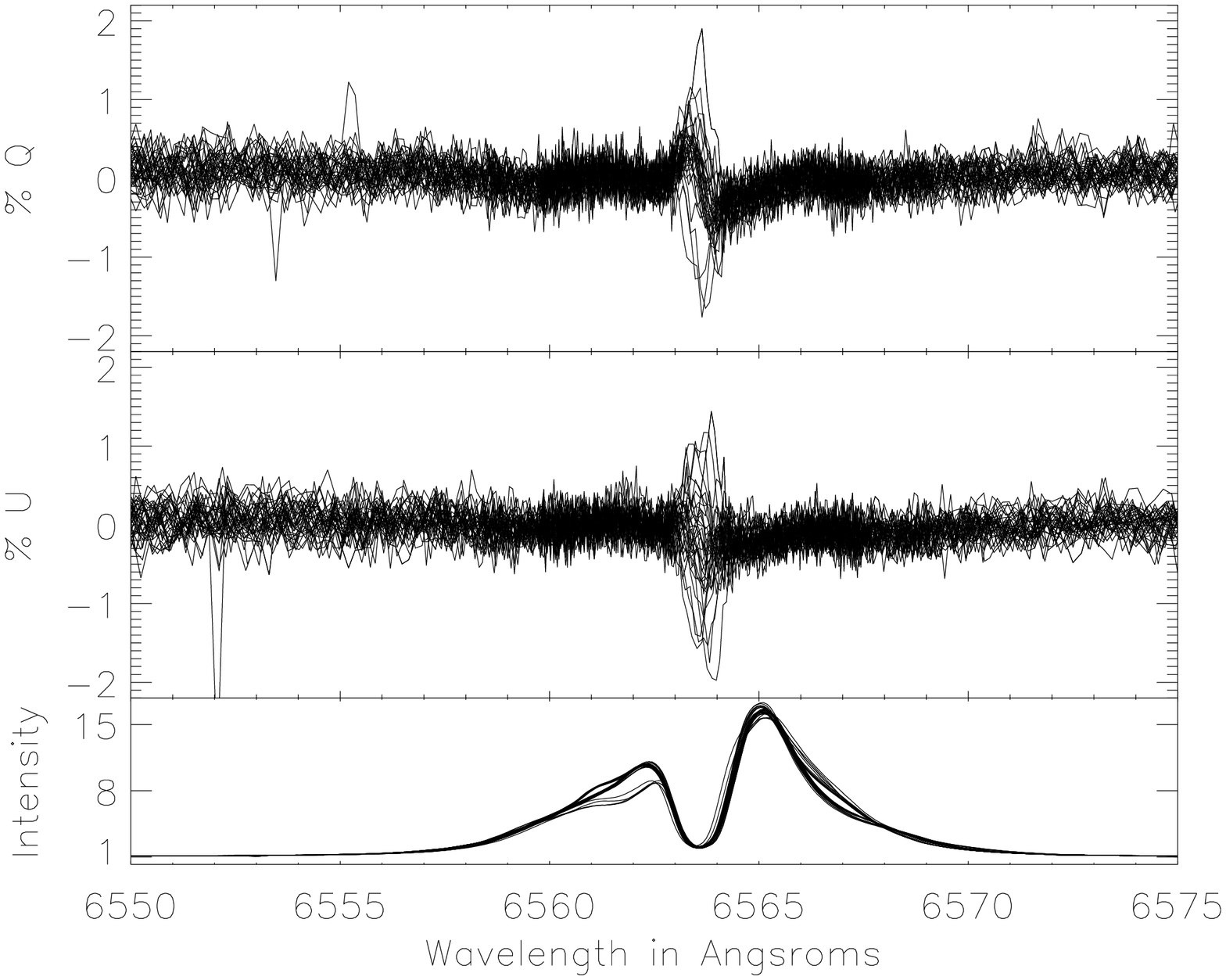}
\includegraphics[width=0.23\linewidth, angle=90]{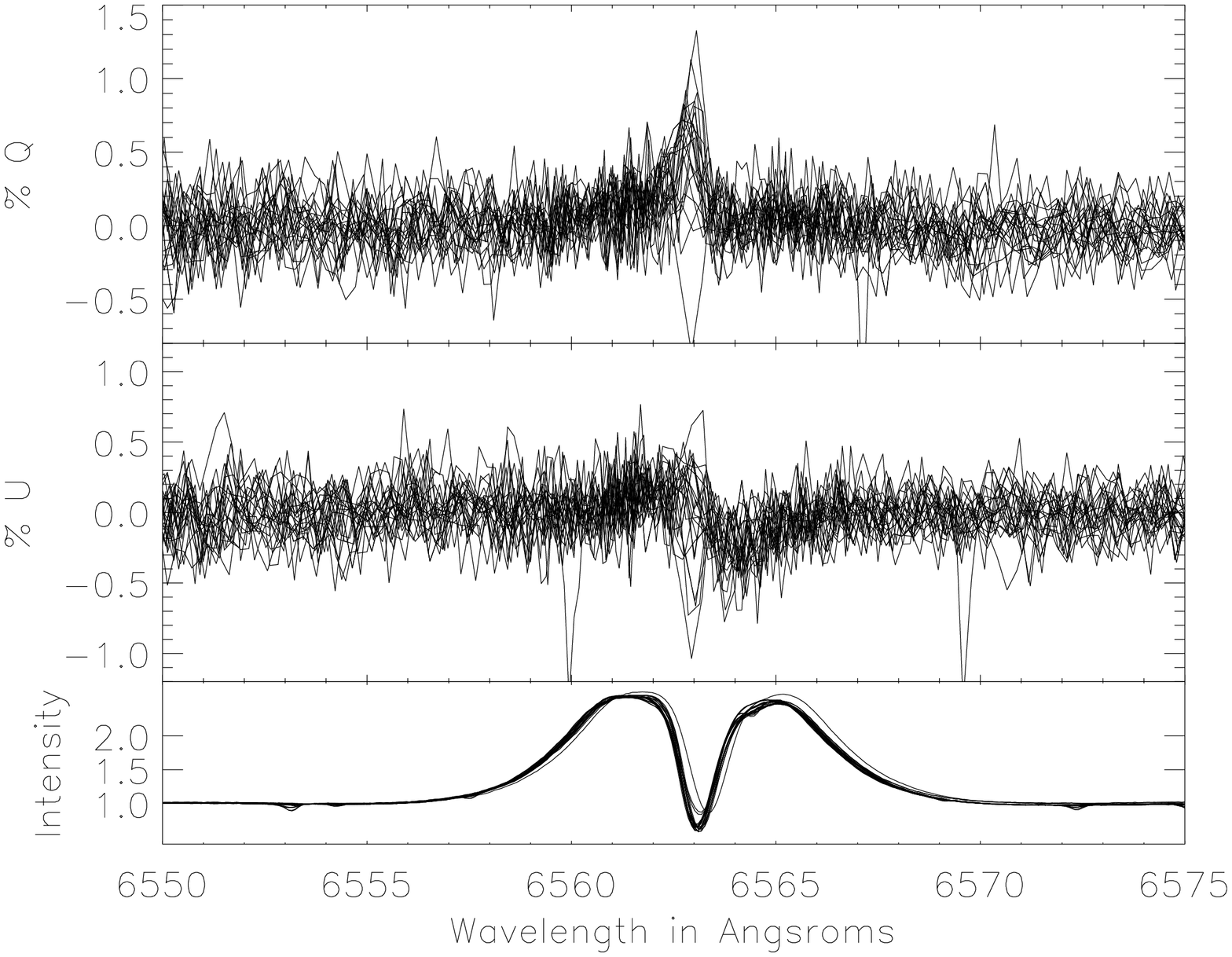}
\includegraphics[width=0.23\linewidth, angle=90]{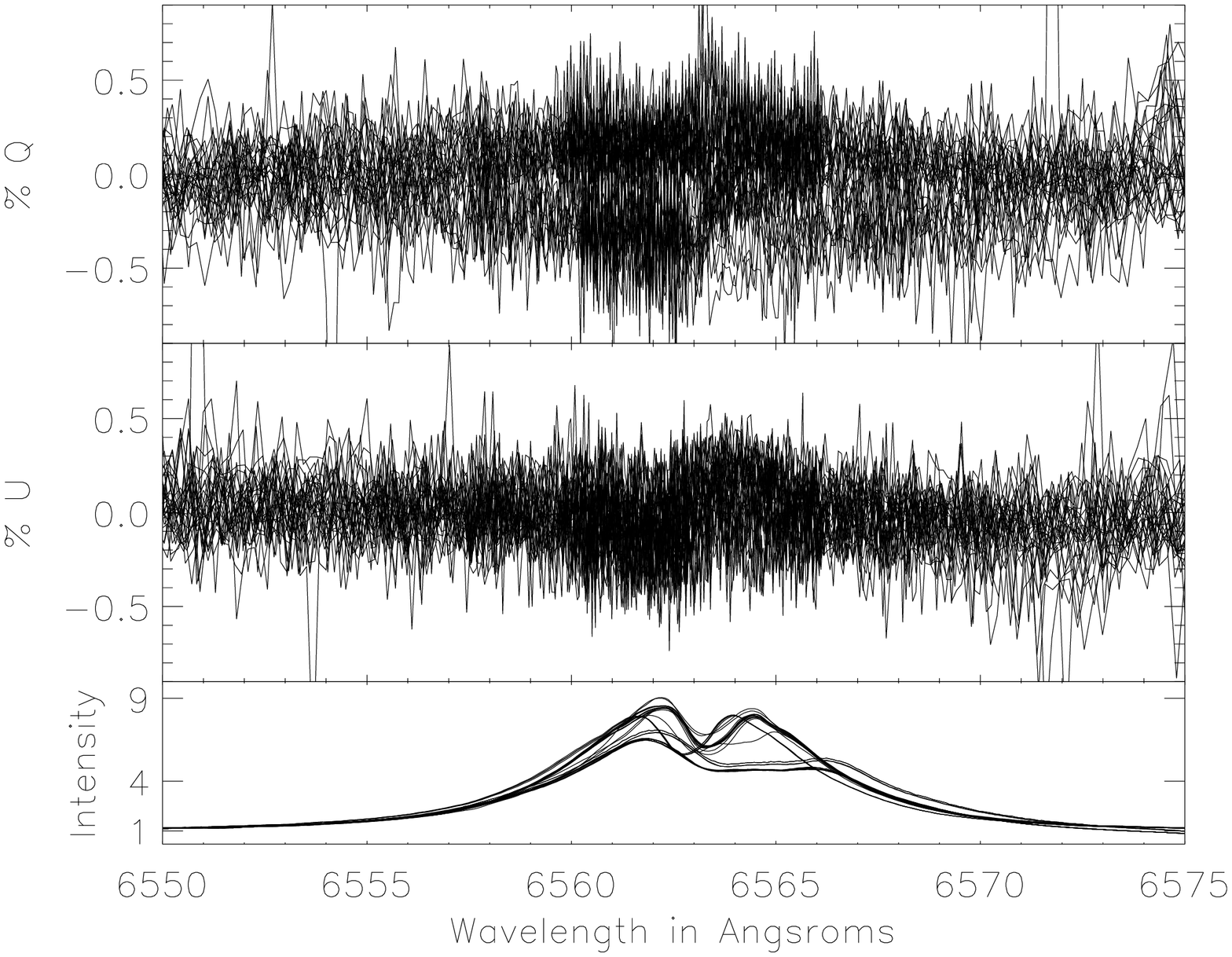}
\includegraphics[width=0.23\linewidth, angle=90]{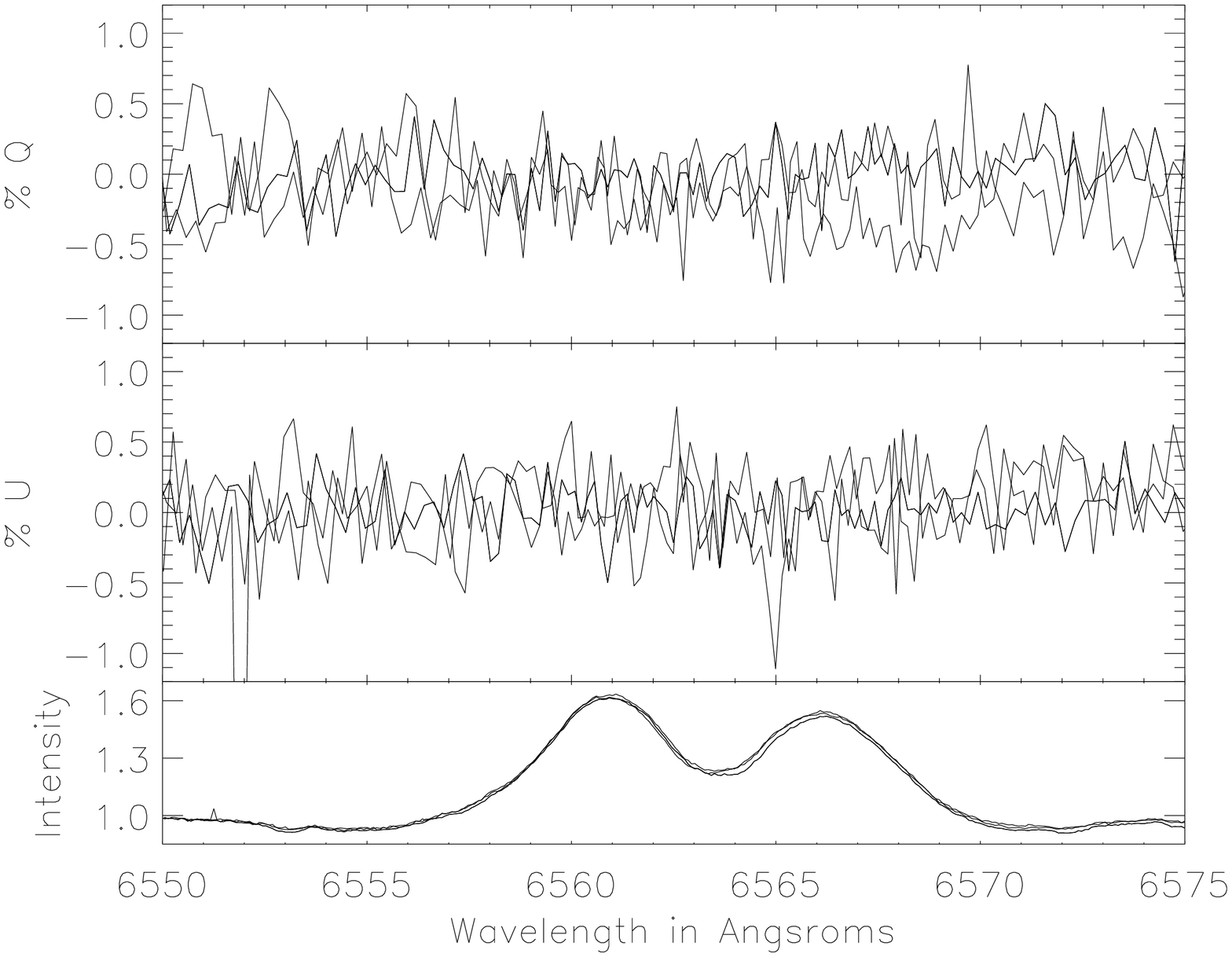}
\includegraphics[width=0.23\linewidth, angle=90]{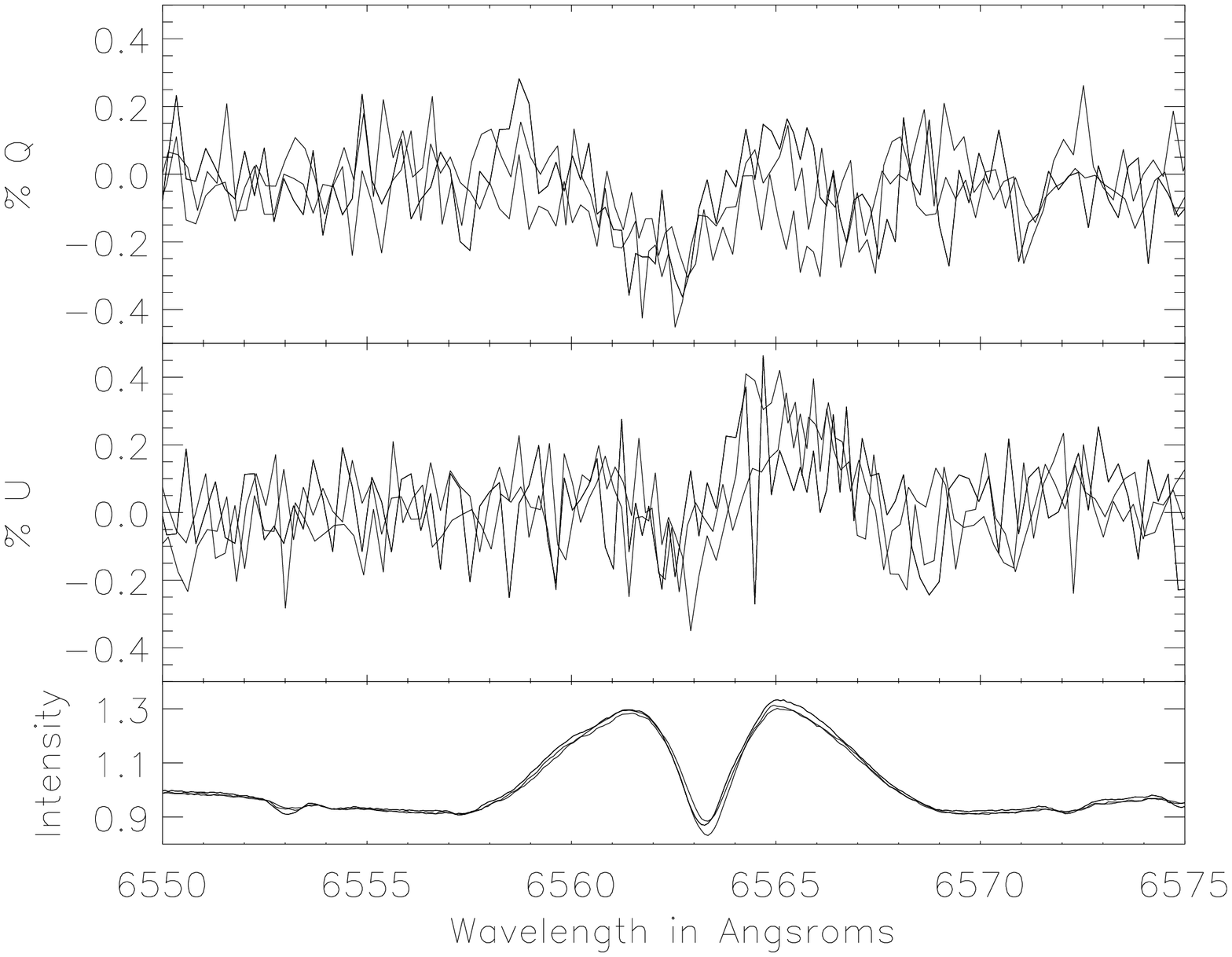}
\includegraphics[width=0.23\linewidth, angle=90]{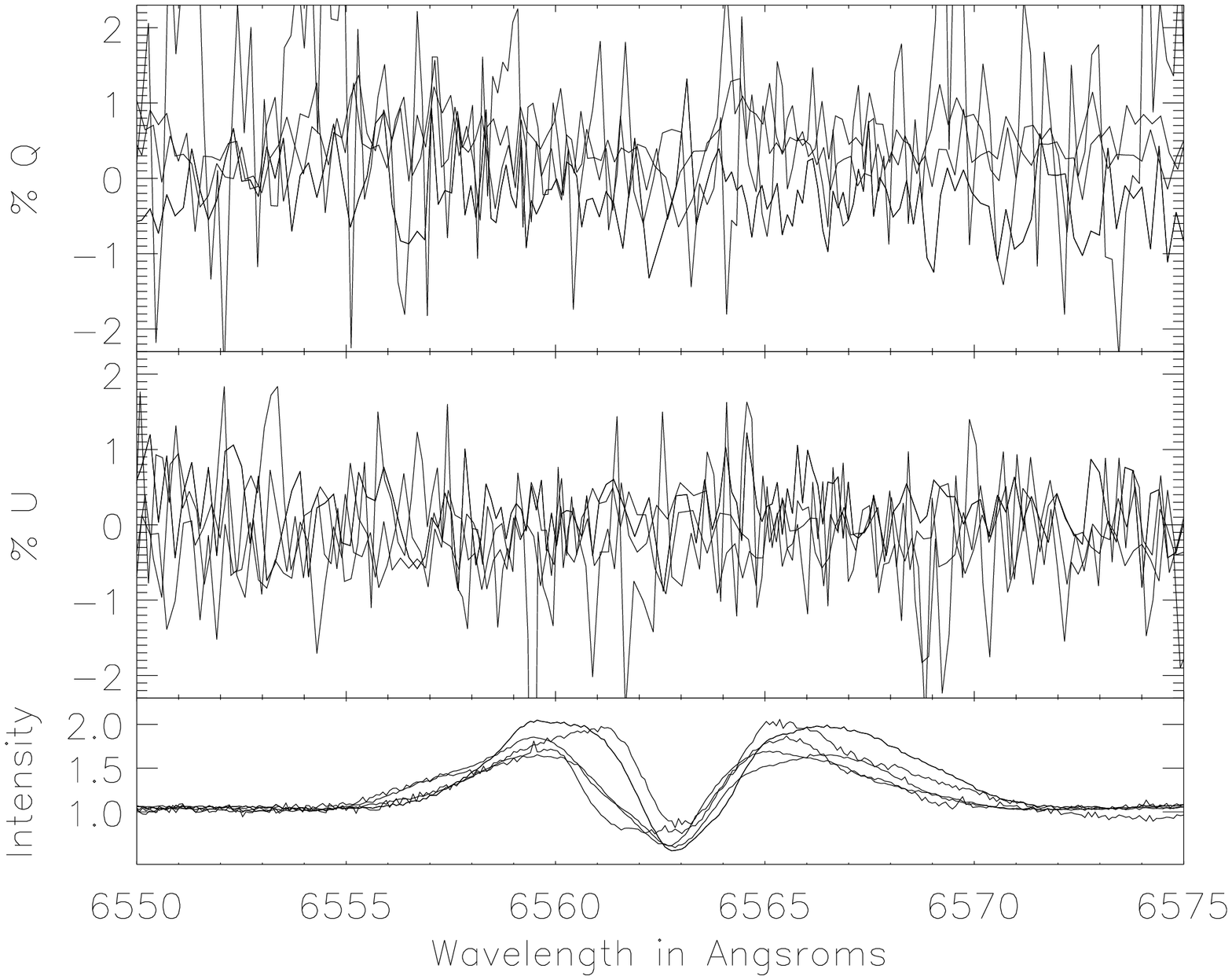}
\includegraphics[width=0.23\linewidth, angle=90]{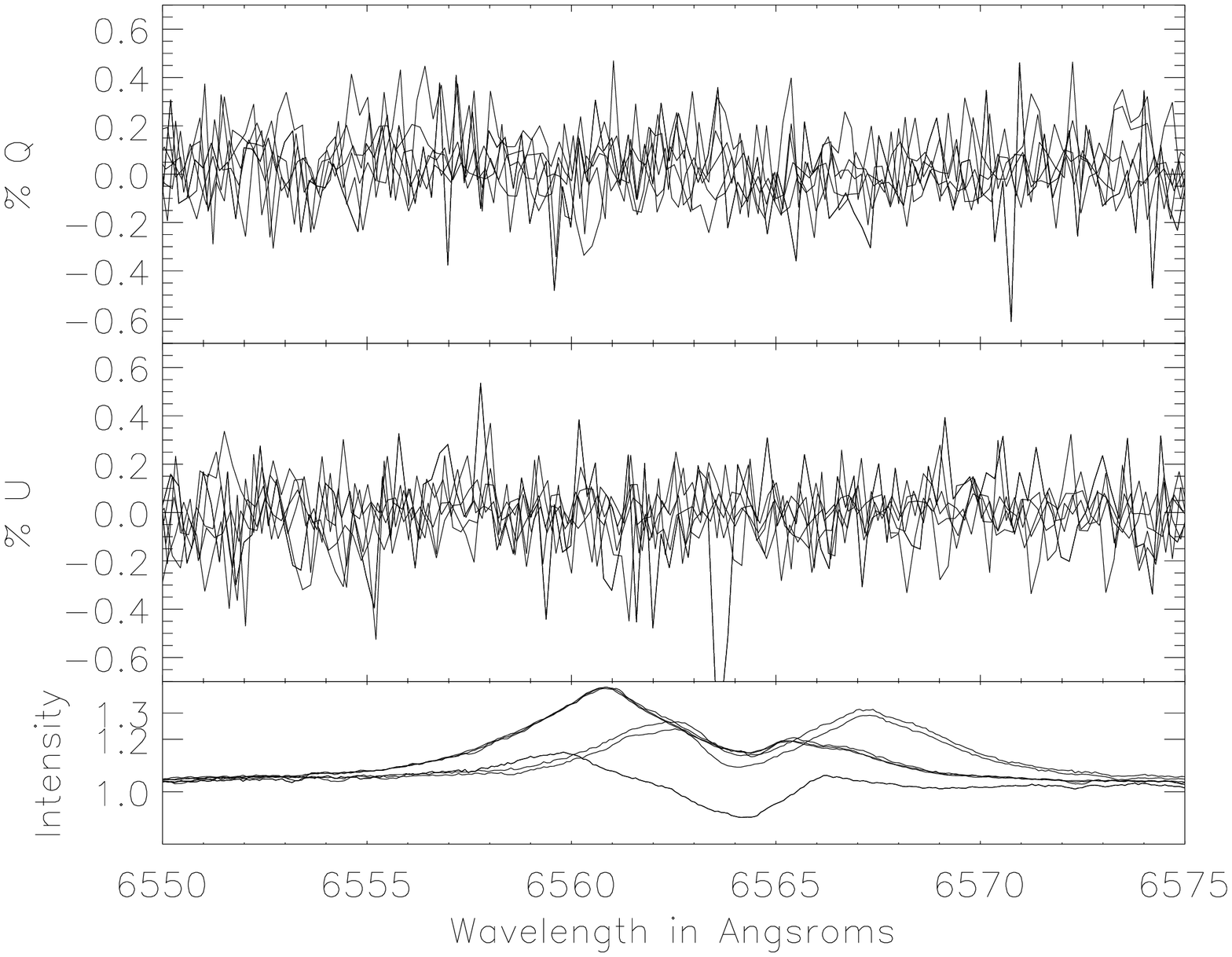}
\includegraphics[width=0.23\linewidth, angle=90]{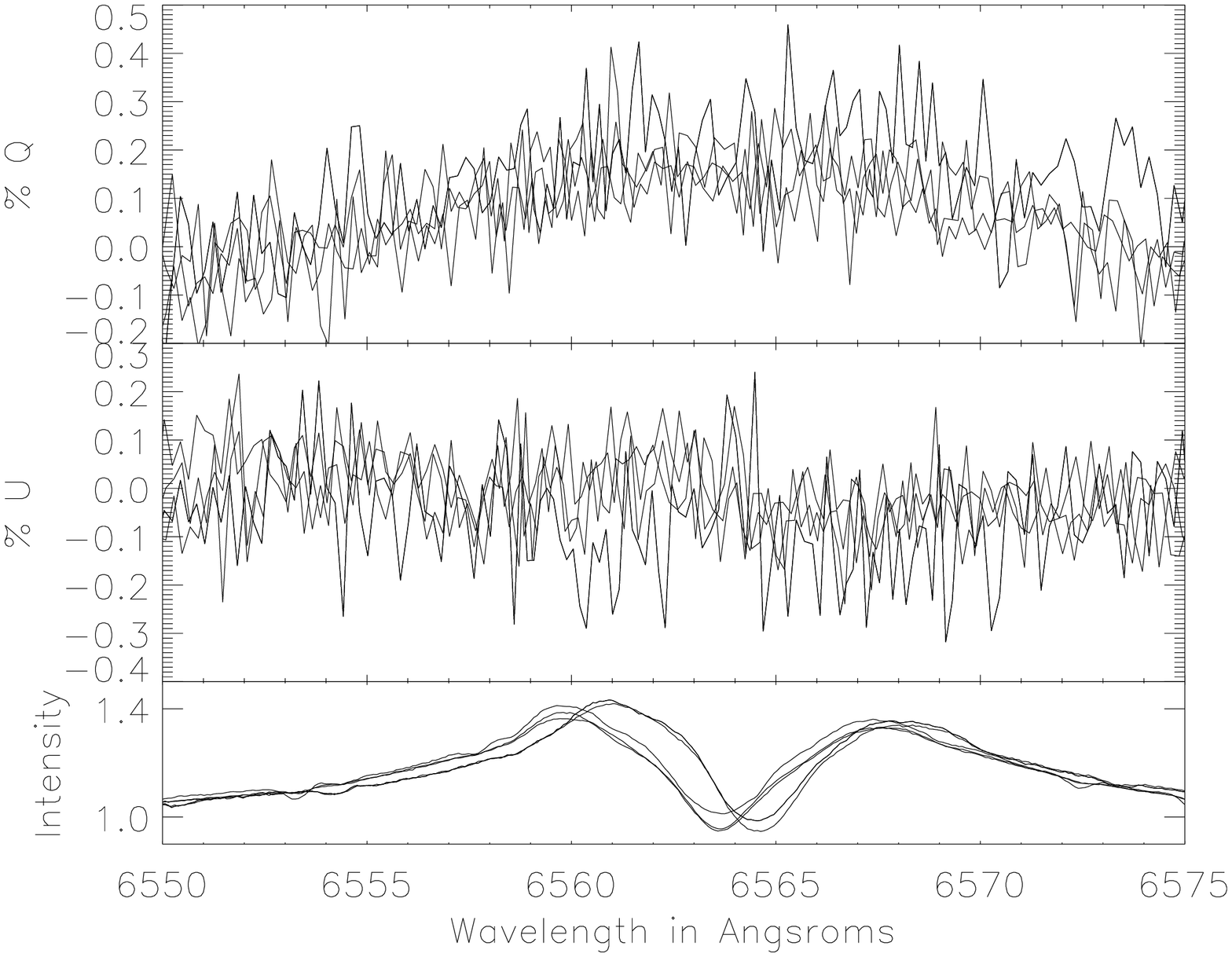}
\includegraphics[width=0.23\linewidth, angle=90]{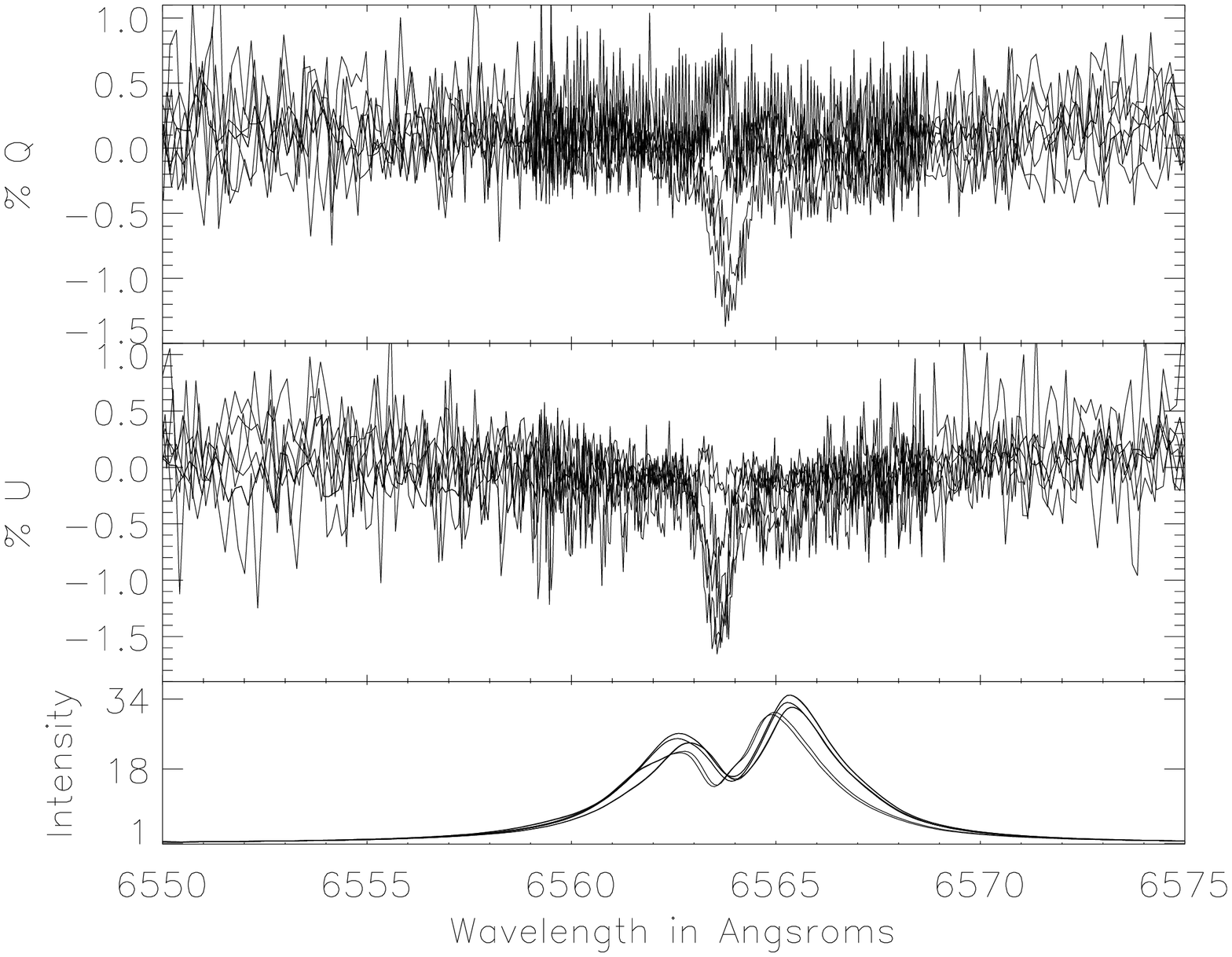}
\includegraphics[width=0.23\linewidth, angle=90]{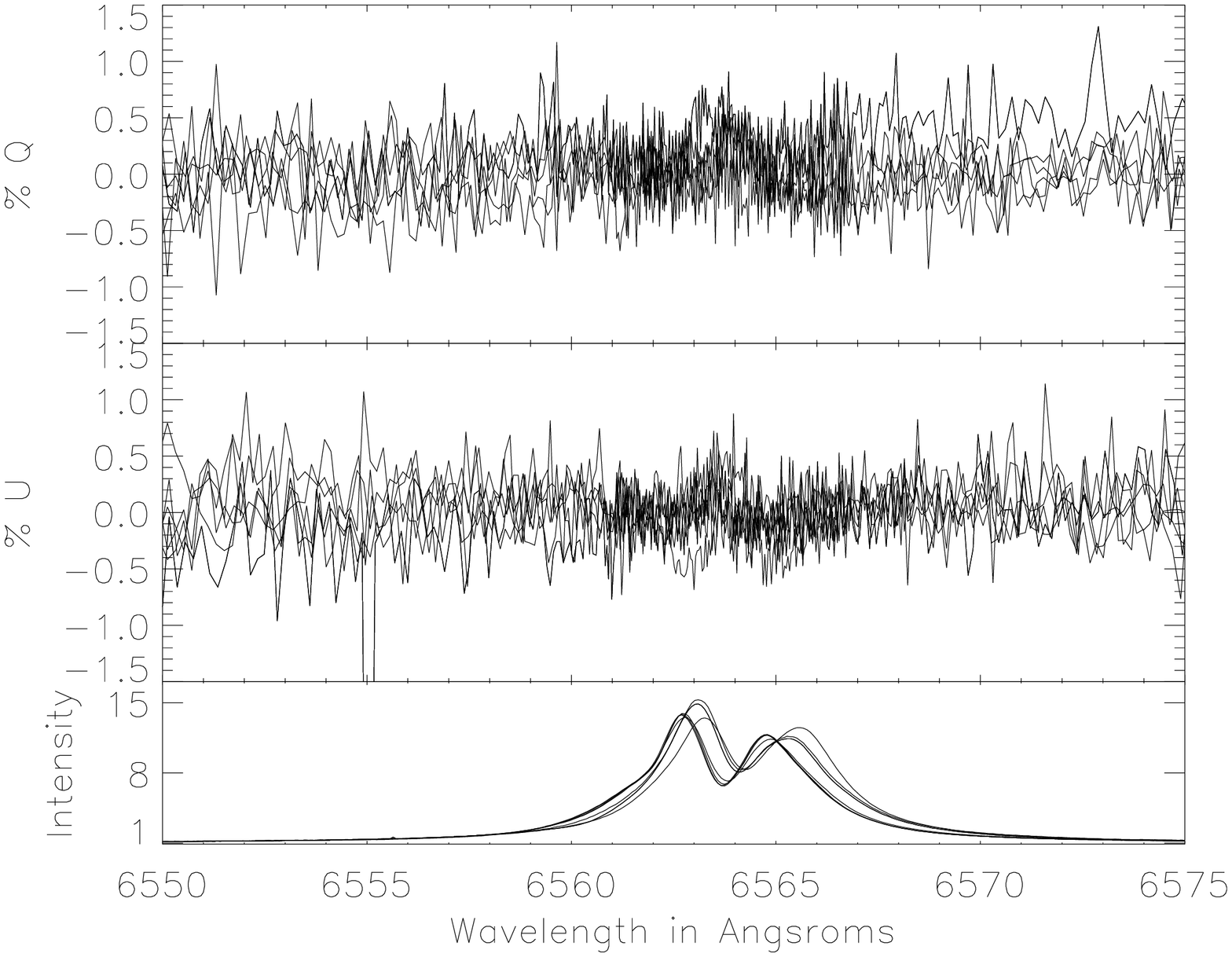}
\includegraphics[width=0.23\linewidth, angle=90]{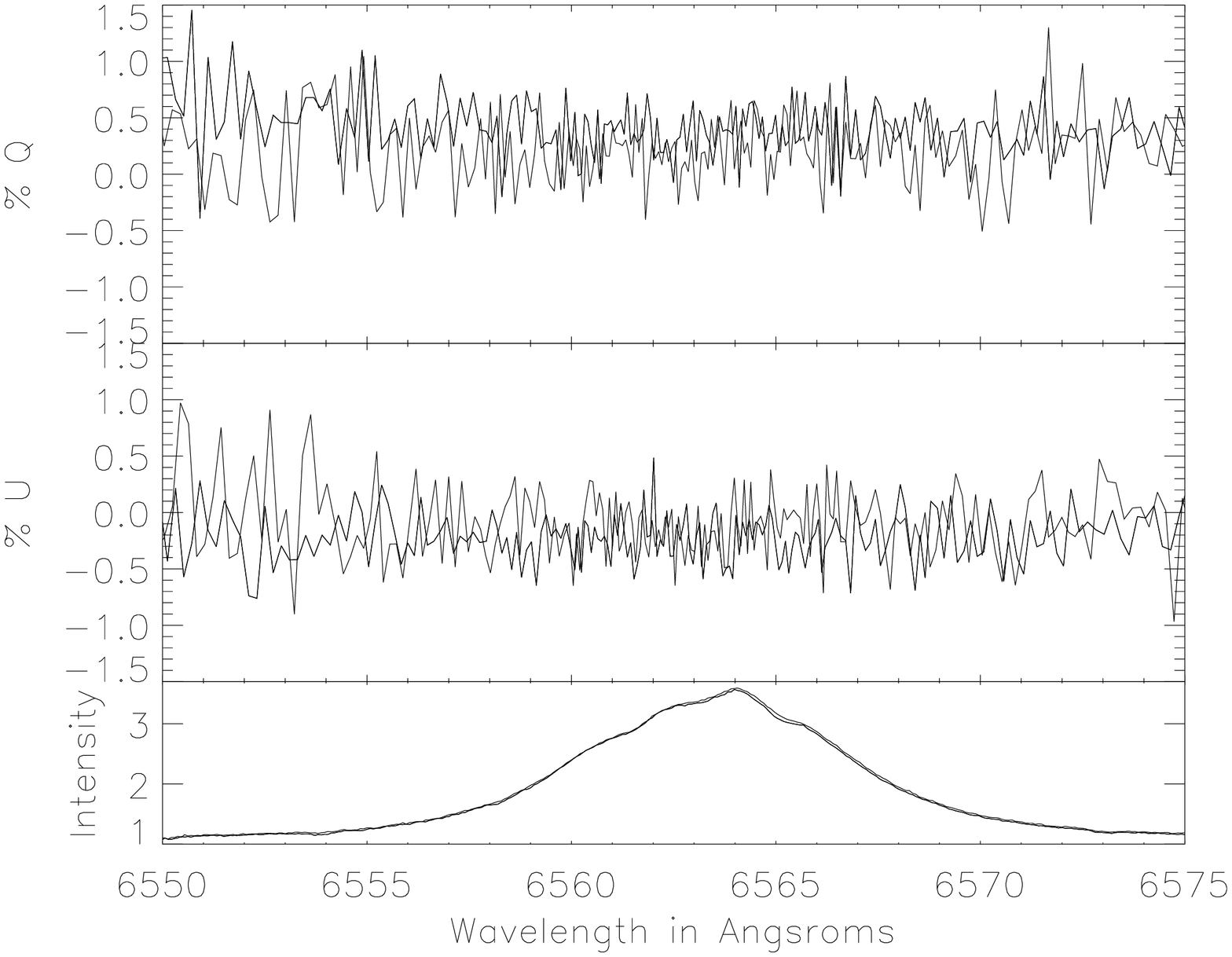}
\includegraphics[width=0.23\linewidth, angle=90]{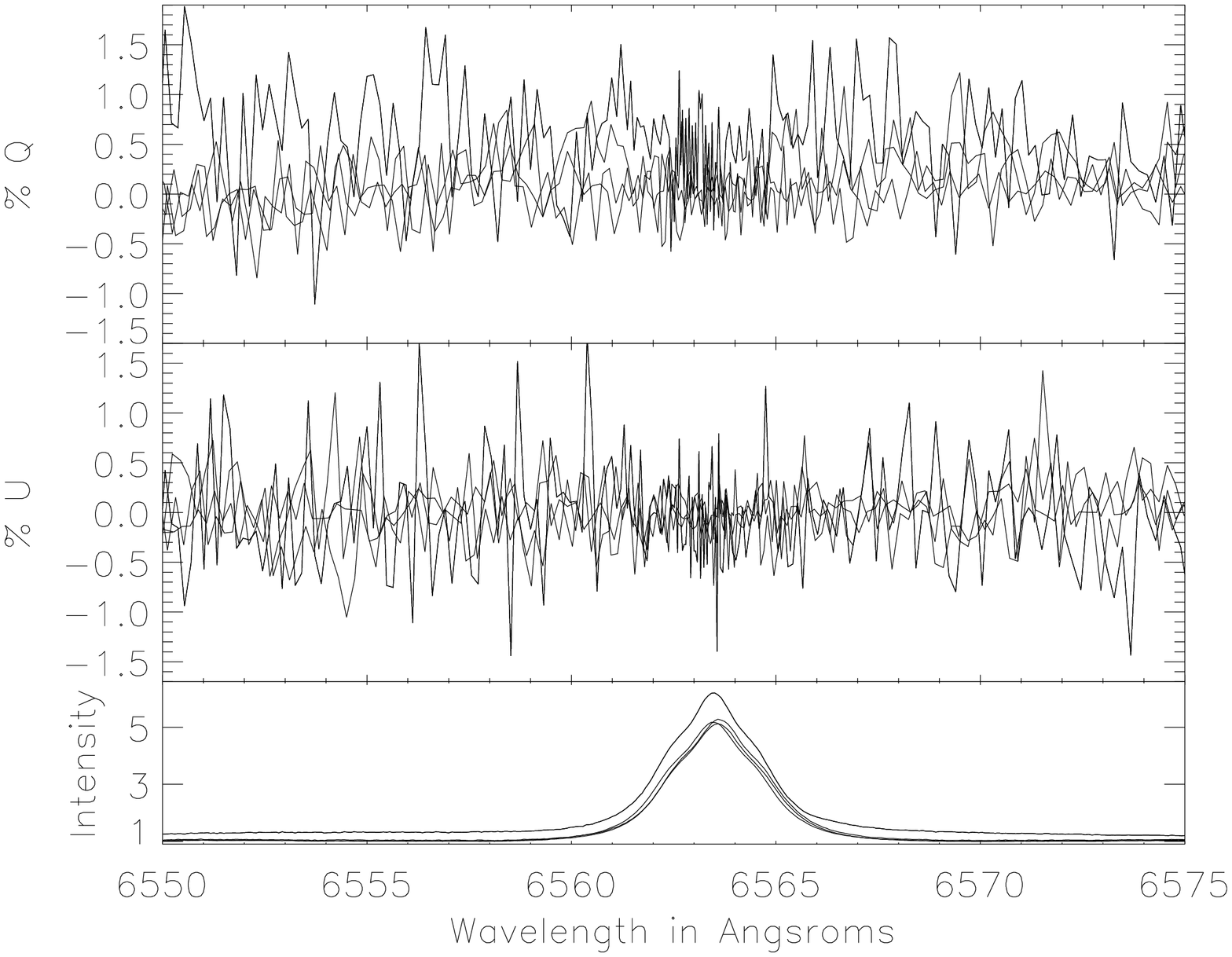}
\includegraphics[width=0.23\linewidth, angle=90]{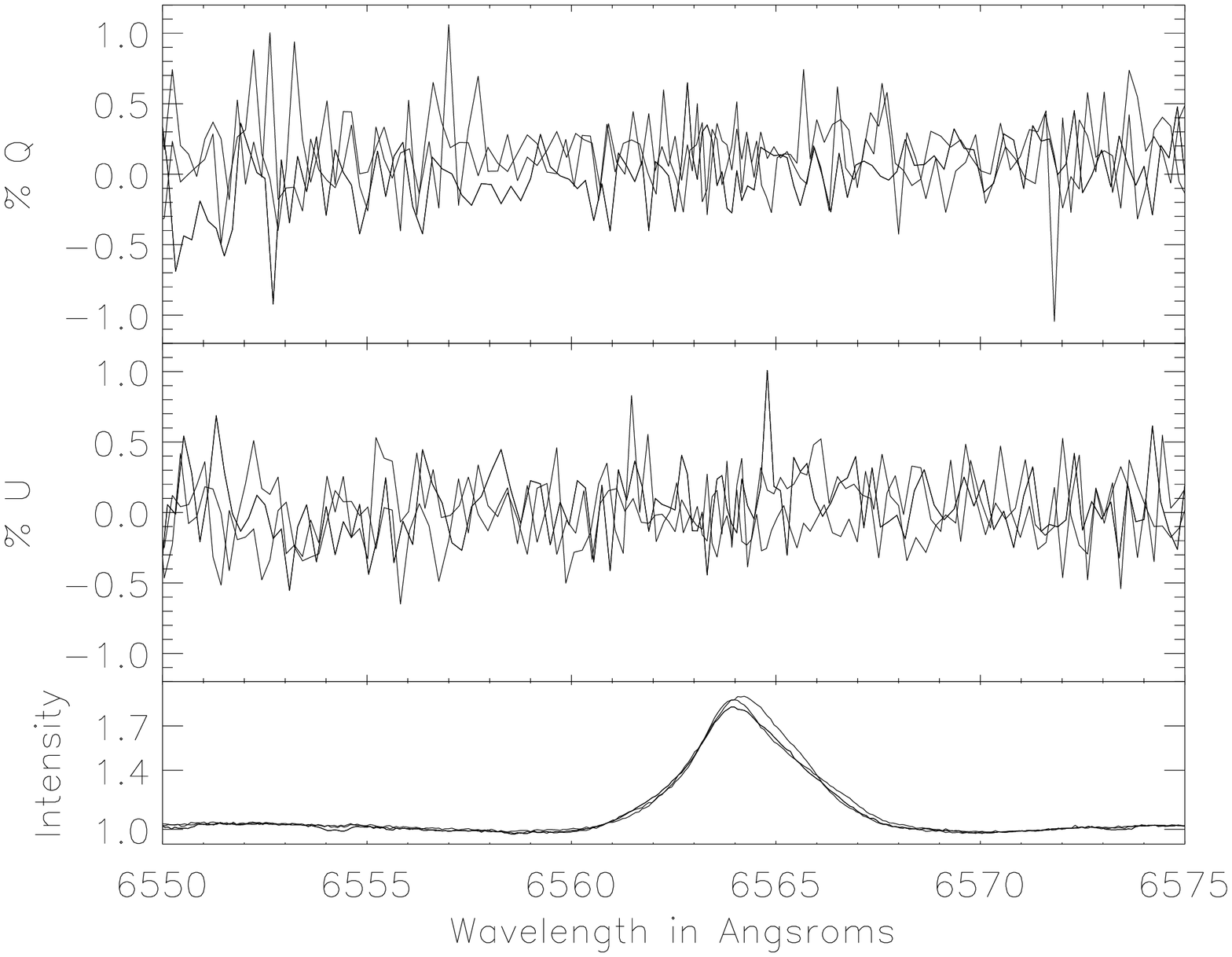}
\includegraphics[width=0.23\linewidth, angle=90]{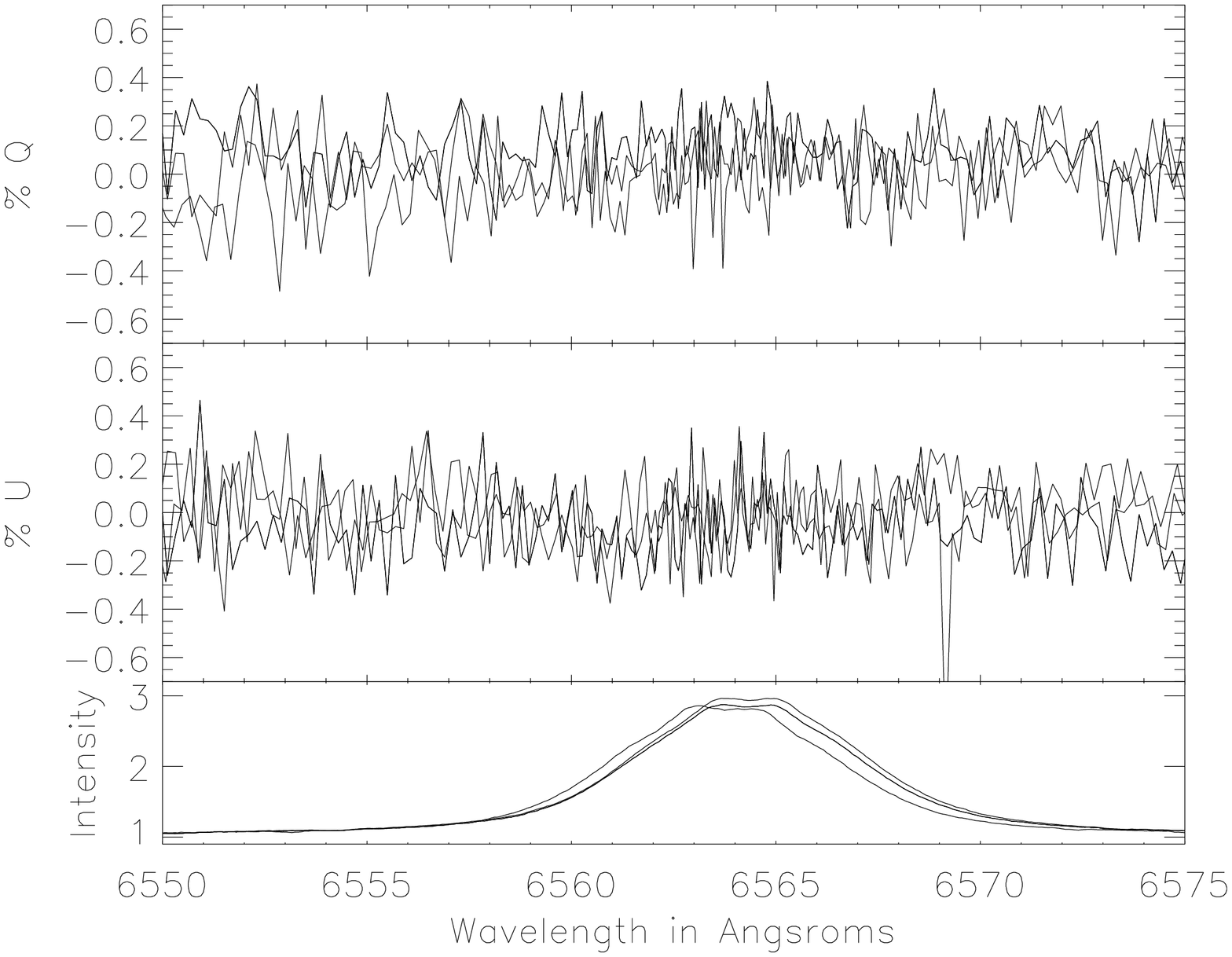}
\includegraphics[width=0.23\linewidth, angle=90]{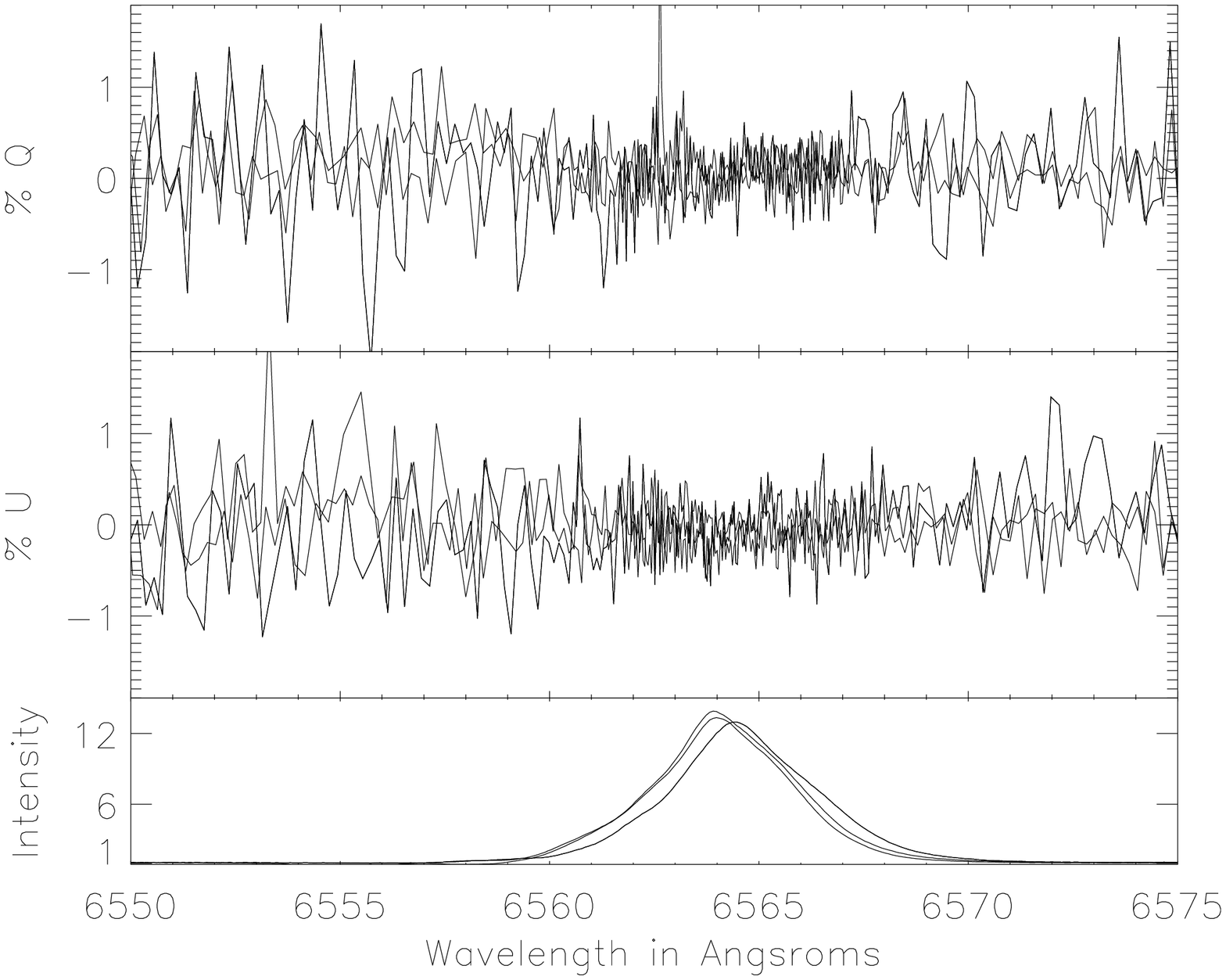}
\caption{HAe/Be Spectropolarimetric Profiles II: The stars, from left to right, are:  MWC 158, HD 58647, MWC 361, HD 141569, 51 Oph, XY Per, MWC 166, MWC 170, HD 45677, MWC 147, Il Cep, MWC 442, HD 35929, GU CMa and HD 38120}
\label{fig:haebe-specpol2}
\end{center}
\end{figure*}

\section{Herbig Ae/Be Spectropolarimetry}

	In this section, the spectropolarimetry of 29 Herbig Ae/Be stars will be presented. The compiled spectropolarimetry for each target will be presented as an overview followed by a discussion of individual targets and individual observations. A recurring theme in the spectropolarimetric morphology of Herbig Ae/Be stars is the strong presence of spectropolarimetric signatures in and around absorptive components of the emission line and the detected amplitudes are 0.2\% to 2\%. The presence of absorptive effects is not new. McLean 1979 proposed an absorptive modification to the depolarization theory. Vink et al. (2002, 2005b) use this and related theories stated in those references in interpreting many of their observations. However, scattering models make specific predictions about the polarization effects caused by absorption that are not seen in many of our observations.  Scattering theory shows no natural amplitude for a spectropolarimetric signature as the amplitude is simply set by the amount of scattered light. The qu-loops arise over a wavelength range wider than the entire line, not in absorptive effects.  Furthermore, the order-of-magnitude increase in spectral resolution (comparing ESPaDOnS at 68000 to ISIS after the bin-by-flux procedure of 2000-4000 in Vink et al. 2002 \& 2005b) allows temporal variability and previously un-detectable morphologies to be seen that may have confused earlier studies. The detailed morphology of individual lines presented below will highlight this ``polarization-in-absorption" theme.

	There is at least one good measurement for each star but typically we have 3-10. AB Aurigae has by far the largest number of observations at 162, with MWC 480 coming second at 67 observations. Since the polarization properties of the telescope conspire to rotate the plane of polarization in a complex manner, no attempt to de-rotate the results in any major systematic way will be made. There are a large number of observing conditions and exposure times. In figures \ref{fig:haebe-specpol1} and \ref{fig:haebe-specpol2} all good spectropolarimetric data sets are shown. Each polarized spectrum was binned-by-flux to a continuum-threshold of 5, as described in Harrington \& Kuhn 2008. Since the individual emission lines have continuum-normalized intensities of 0.2 to nearly 40, this binning will give a much more uniform signal-to-noise ratio for each measurement while preserving good wavelength coverage. We also note that since the slit-detector combination we used significantly over-samples the spectral orders, this binning does not effect the spectral resolution. After a 5:1 binning there are still typically 2.2 spectral resolution elements full-width-half-max for thorium-argon lines. Table \ref{aebe-res} shows the stars observed, the type and strength of the H$_\alpha$ line and the detected polarization effect. The actual signal-to-noise ratios in a polarization spectrum are dependent on four individual exposures, and measurements were performed through a wide range of observing conditions. The unbinned signal-to-noise ratio's range from 100 to over 1000 per pixel depending on the seeing, exposure time, and binning. Individual examples at the highest precisions and a discussion of each star with significant detections will illustrate these conditions.

\subsection{ESPaDOnS Archive Comparisons}

\begin{figure*}
\begin{center}
\includegraphics[width=0.35\linewidth, angle=90]{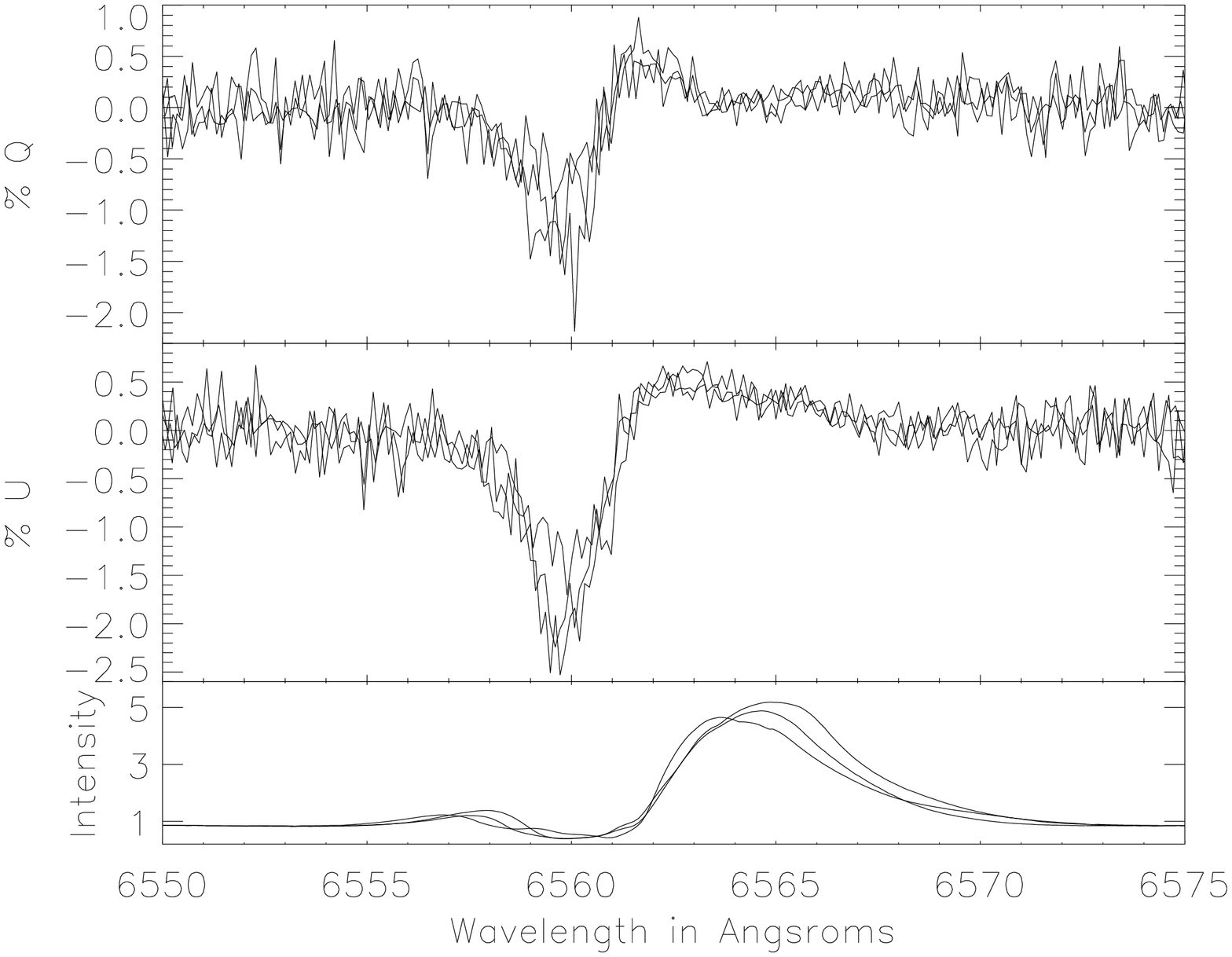}
\includegraphics[width=0.35\linewidth, angle=90]{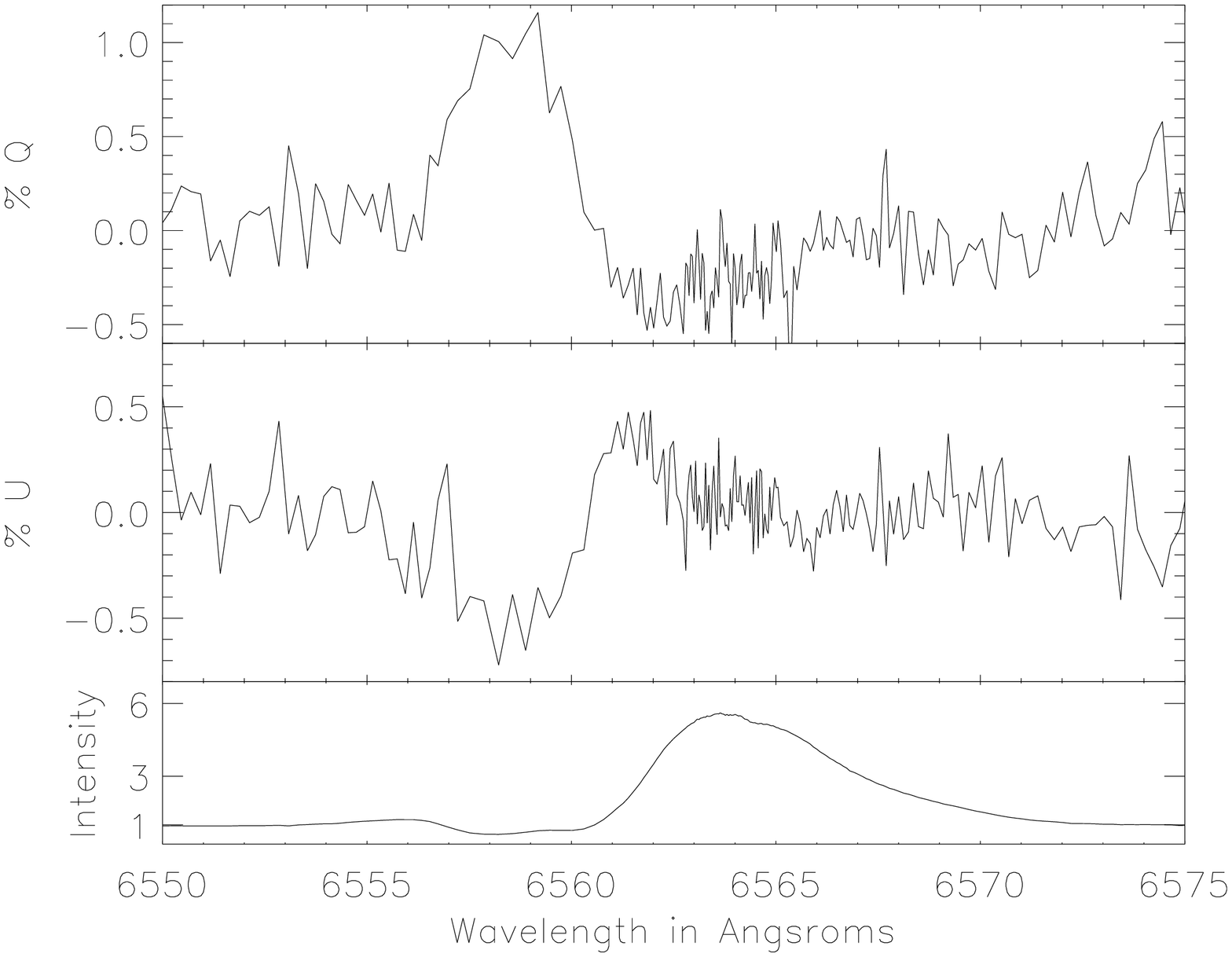}  \\
\includegraphics[width=0.35\linewidth, angle=90]{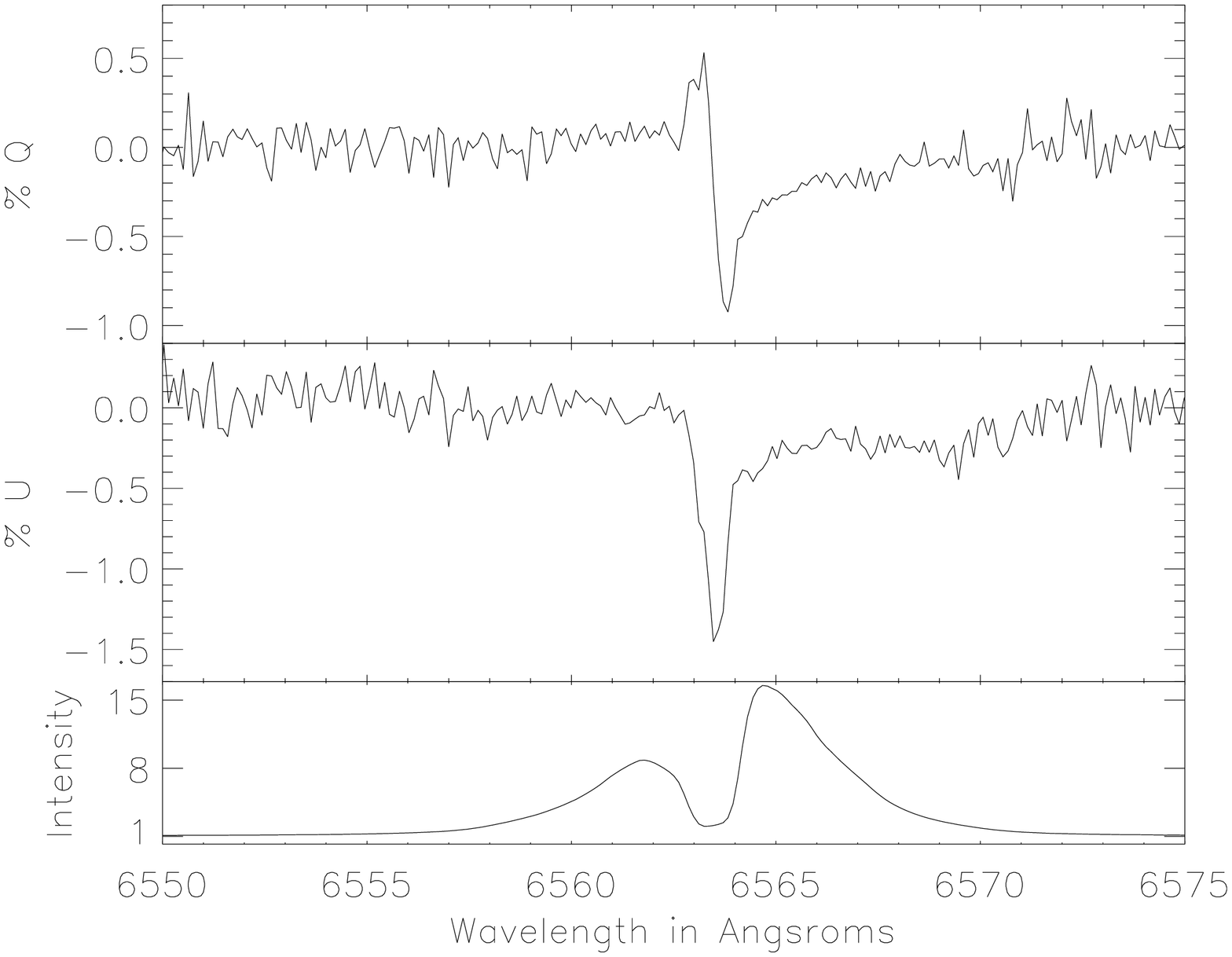}
\includegraphics[width=0.35\linewidth, angle=90]{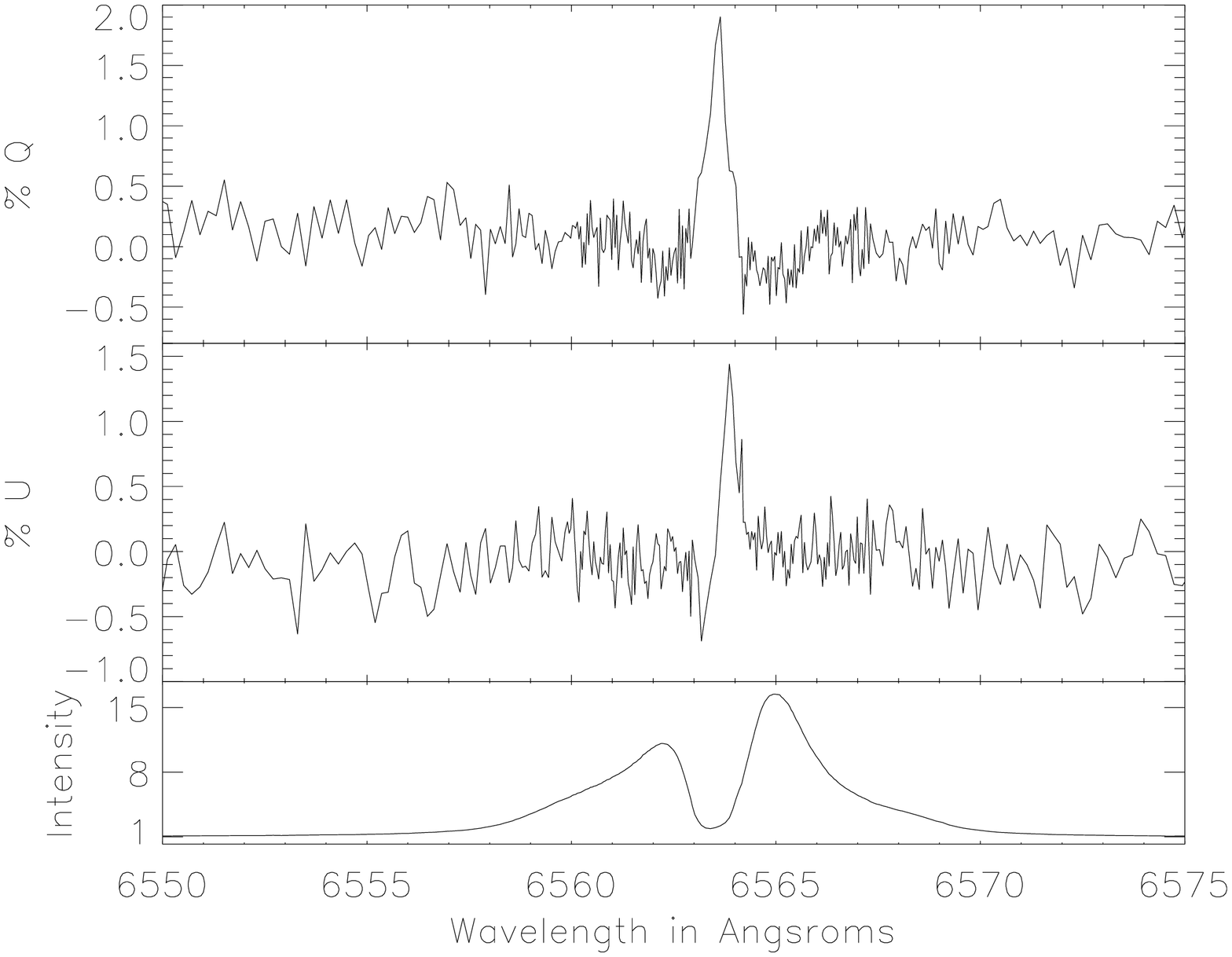}  \\
\includegraphics[width=0.35\linewidth, angle=90]{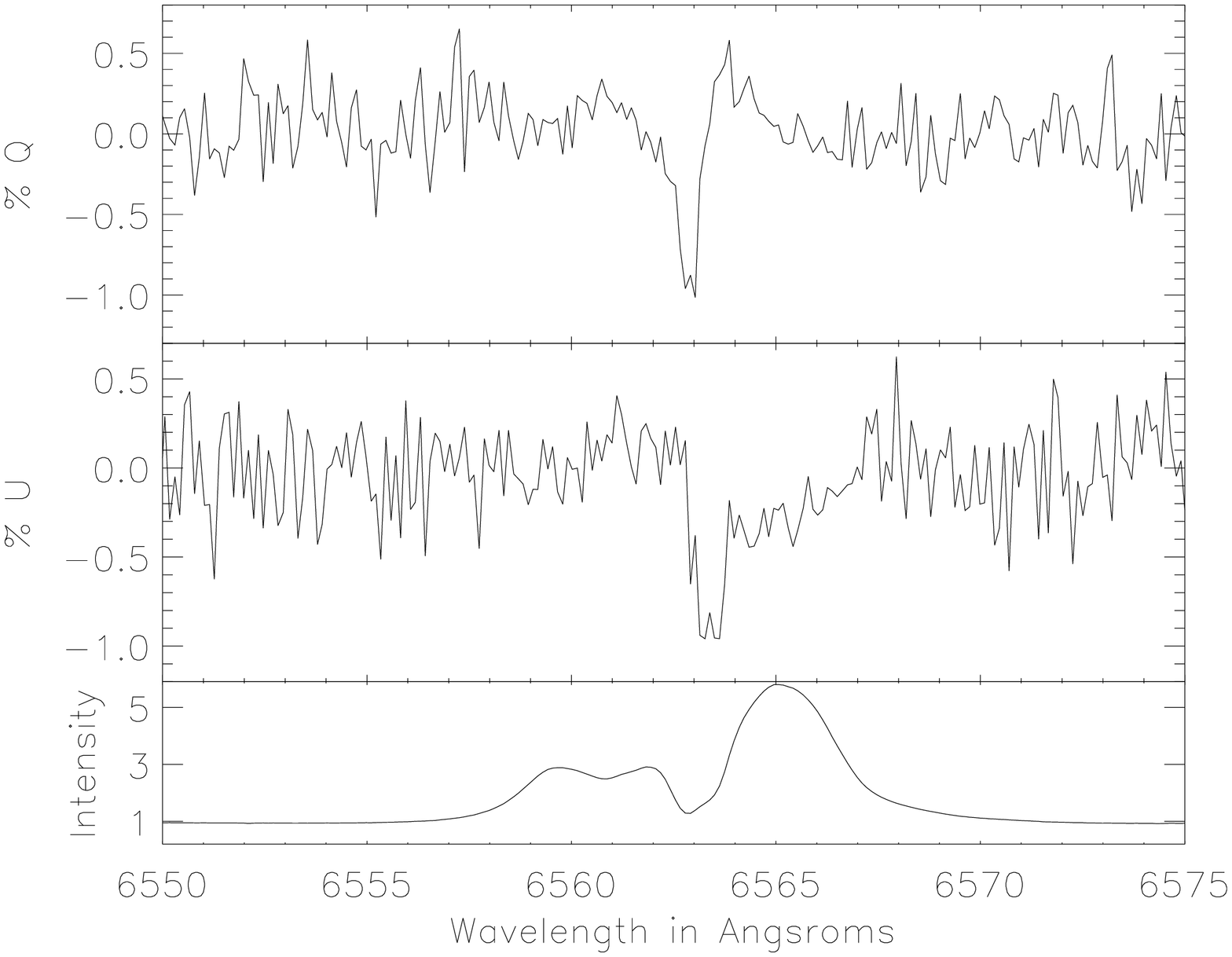}
\includegraphics[width=0.35\linewidth, angle=90]{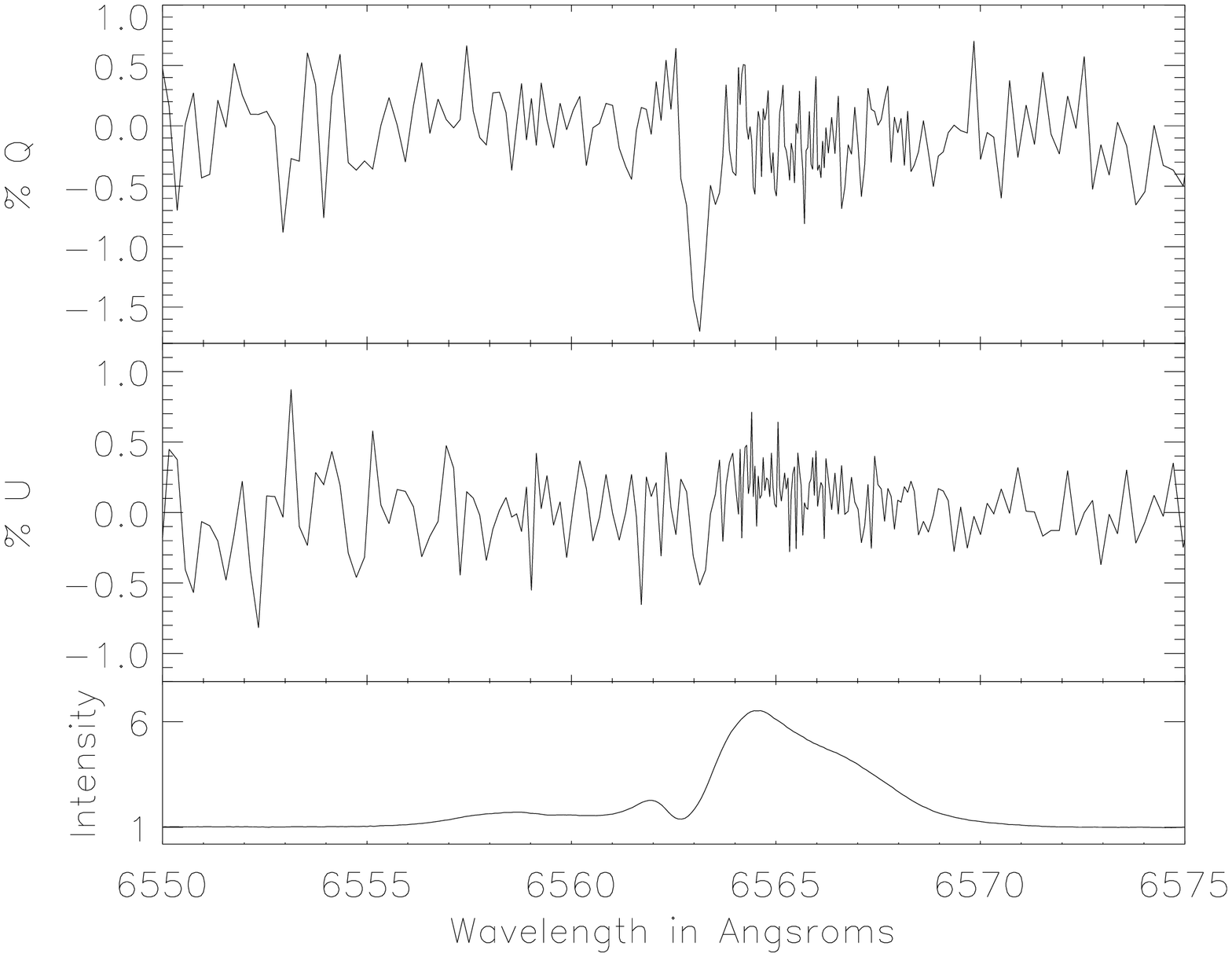}
\vspace{5mm}
\caption{A comparison of ESPaDOnS and HiVIS - MWC 480, MWC 158 \& MWC 120. From left to right: {\bf a)}  The spectropolarimetry of MWC 480 from the CFHT archive and {\bf b)}  HiVIS. {\bf c)}  The spectropolarimetry of MWC 158 from the CFHT archive and {\bf d)}  HiVIS.  {\bf e)}  The spectropolarimetry of MWC 120 from the CFHT archive and {\bf f)}  HiVIS. No rotation has been applied to the HiVIS data, so the exact form of the HiVIS qu spectra is subject to rotation, but the overall magnitude and location of the polarization effects will remain unchanged. The observations match in magnitude and wavelength quite well.}
\vspace{5mm}
\label{fig:esp-hiv-mwc}
\end{center}
\end{figure*}

	Before discussing individual examples, the cross-comparison of our HiVIS data with that of ESPaDOnS must be made. For many stars we have one or two ESPaDOnS measurements that will be presented side-by-side. The comparison between HiVIS and ESPaDOnS is outlined for three individual stars. Figure \ref{fig:esp-hiv-mwc} shows a side-by-side comparison of MWC 480, MWC 158 and MWC 120 spectropolarimetry with ESPaDOnS and HiVIS. We took archival ESPaDOnS data from February and August 2006 to compare with our October to December 2006 HiVIS observations. Many individual examples will be discussed thoroughly in later sections, but a few examples now can illustrate the rotation of the plane of polarization and the reduction in signature magnitude caused by the AEOS telescope effects. The observations for each star were taken at different times, 6-months to more than a year apart. Though the general H$_\alpha$ line types are the same, the line profiles have changed significantly between the observations making exact comparisons difficult. There has been no polarization-plane rotation applied to the HiVIS data. We did not attempt to align the polarization measurements so that the style of comparison used in later sections can be illustrated.
	
	MWC 480 shows a clear P-Cygni H$_\alpha$ line in both ESPaDOnS and HiVIS observations. The polarization change is over 1\% in each, with the change centered on the P-Cygni absorption. The ESPaDOnS observations show a large magnitude, 2\% in u and 1.5\% in q on two occasions and a smaller 0.5\% q, 1\% u on another. The polarization at the peak of the emission is at most 0.3\% away from continuum while the polarization on red side of the line, $>$6567$\AA$, is identical to continuum. The HiVIS observations show 1\% changes in q and 0.5\% in u with a very similar morphology. The polarization changes are largest in the absorptive component with a small sign change on the blue side of the emission, 6563$\AA$. The emission peak is roughly 0.2\% different from continuum, with the return to continuum polarization happening on the red side of the line, near 6566$\AA$. By observing that the operation q $\rightarrow$ u and u $\rightarrow$ -q matches the ESPaDOnS data with the HiVIS data, a crude estimate of 135$^\circ$ rotation between the two observations can be inferred, assuming the star is not variable. Though we can't disentangle telescope polarization effects from actual stellar variability, the spectropolarimetric form and magnitudes are similar between the instruments. 
	
	MWC 158 also shows a very strong polarization effect in the central absorption features of figure \ref{fig:esp-hiv-mwc}. The ESPaDOnS observations show an antisymmetric change of 0.5\% in q with a narrow 1.5\% spike in u. The HiVIS observation shows a 1.5\% spike in q with a 0.5\% drop then 1.5\% rise in u over the same narrow wavelength range. In this case the HiVIS amplitudes are greater than the ESPaDOnS data. Since the AEOS telescope can only reduce the magnitude of a spectropolarimetric effect, this star must have a variable spectropolarimetric signature with the HiVIS observation being greater in magnitude than ESPaDOnS. 
	
	The MWC 120 line profiles changed significantly as seen in figure \ref{fig:esp-hiv-mwc}. The ESPaDOnS observations show a much weaker blue-shifted absorption compared to the HiVIS observations. The ESPaDOnS observations, as in MWC 158, showed a strong change in the central absorptive component. Both q and u show decreases of roughly 1\%. HiVIS shows a 1.5\% drop at the same wavelengths in q only. There is also a smaller but significant broad change on the red side of the line profile in the ESPaDOnS observations. This can also be seen as a small 0.3\% rise in the HiVIS u spectrum. This comparison with the archival data shows that the spectropolarimetry from HiVIS, though subject to magnitude-reduction and rotation from the telescope polarization, can still provide a good measurement of the wavelength and relative magnitude of the polarization changes across a line. Both the narrow large-amplitude change in the absorption and the broad low-amplitude change across the line are reproduced.

\subsection{Comments on Individual Targets}

	In this section, detailed descriptions of the HiVIS, ESPaDOnS and literature spectropolarimetric results will be discussed for each individual star. Examples of spectropolarimetric effects will be shown for each and a description of the morphology will be given.

\begin{table}[!h,!t,!b]
\begin{center}
\begin{normalsize}
\caption{Herbig Ae/Be H$_\alpha$ Results \label{aebe-res}}
\begin{tabular}{lcccc}
\hline    
\hline    
{\bf Name}    &{\bf H$_\alpha$}    &  {\bf Effect?}          & {\bf Mag}            & ${\bf Type}$                  \\
\hline
\hline
AB Aur           & 8-11 &   Y                                           &    1.5\%               &   Wind                             \\
MWC480        & 4-7  &    Y                                        &    1.5\%                &   Wind                              \\
MWC120        & 5-8  &    Y                                        &   1.0\%                 &   Wind                            \\
HD163296    & 6-8  &  Y                                           &     1.0\%                &   Wind                          \\
HD179218    & 2-5   &   Y                                          &   0.5\%                 &   Wind*                         \\
HD150193     &3-5   & Y                                             &  0.5\%                 &  Wind*                            \\
MWC758        &2.5-3 & Y                                            &  0.5\%                   & Wind                             \\
HD144432     & 4       & Y                                            &  2.0\%                  & Wind                              \\
MWC158        & 15-17 &  Y                                          &    1.0\%                &   Disk                             \\
HD58647       & 2.6      &  Y                                           &    0.5\%                &   Disk                            \\
MWC361        & 8-10   &   Y                                          &   0.3\%                 &  Disk                            \\
51 Oph           & 1.3    & Y                                             &  0.3\%                  & Disk                             \\
HD45677      & 36    & Y                                              & 1.0\%                  &  Disk                            \\   
MWC147       & 15    & Y                                           &  0.3\%                   & Disk                            \\
MWC170       & 1.4  & Y                                              & 0.3\%                  & Disk                            \\
\hline
V1295Aql      &7-8   & N                                            &   $<$0.2\%                        &  Wind                           \\
KMS 27          &2.5   & N                                            &   $<$0.3\%                         & Wind                          \\
HD169142    &3.1-3.4 & N                                        &  $<$0.3\%                       &  Wind                            \\
HD35929      &1.9    & N                                            &  $<$0.2\%                     &  Wind                            \\
HD141569    &1.6   & N                                             &  $<$0.2\%                    &  Disk                            \\
XY Per           &2.0    & N                                            &   $<$0.5\%                    &  Disk                            \\
MWC166      &1.3     & N                                            &   $<$0.1\%                    &  Disk                            \\
Il Cep            &3.5     & N                                            &    $<$0.2\%                     &  Emis                            \\
MWC442      &6.2    & N                                             &  $<$0.2\%                   &  Emis                            \\
HD38120     &13    & N                                              &   $<$0.4\%                   &  Emis                            \\
GU CMa       &2.9    & N                                             &  $<$0.1\%                   &  Emis                            \\
HD35187      &1.1-1.6  & N                                       &  $<$0.5\%                   &  Var                            \\
HD142666   &2.0    & N                                            &   $<$0.5\%                  &  Acc                            \\
T Ori              &4-6    & N                                            &   $<$7\%                    & Disk**                           \\
\hline
\hline
\end{tabular}
\end{normalsize}
\end{center}
\tablecomments{The name and typical amplitude of the H$_\alpha$ line are presented along with detection statistics. The H$_\alpha$ column is the normalized line intensity, Effect? is the presence (Y) or absence (N) of a spectropolarimetric effect, Mag is the amplitude of the spectropolarimetric effect and Type is the H$_\alpha$ line type. }
\end{table}

\begin{figure}
\begin{center}
\includegraphics[ width=0.65\linewidth, angle=90]{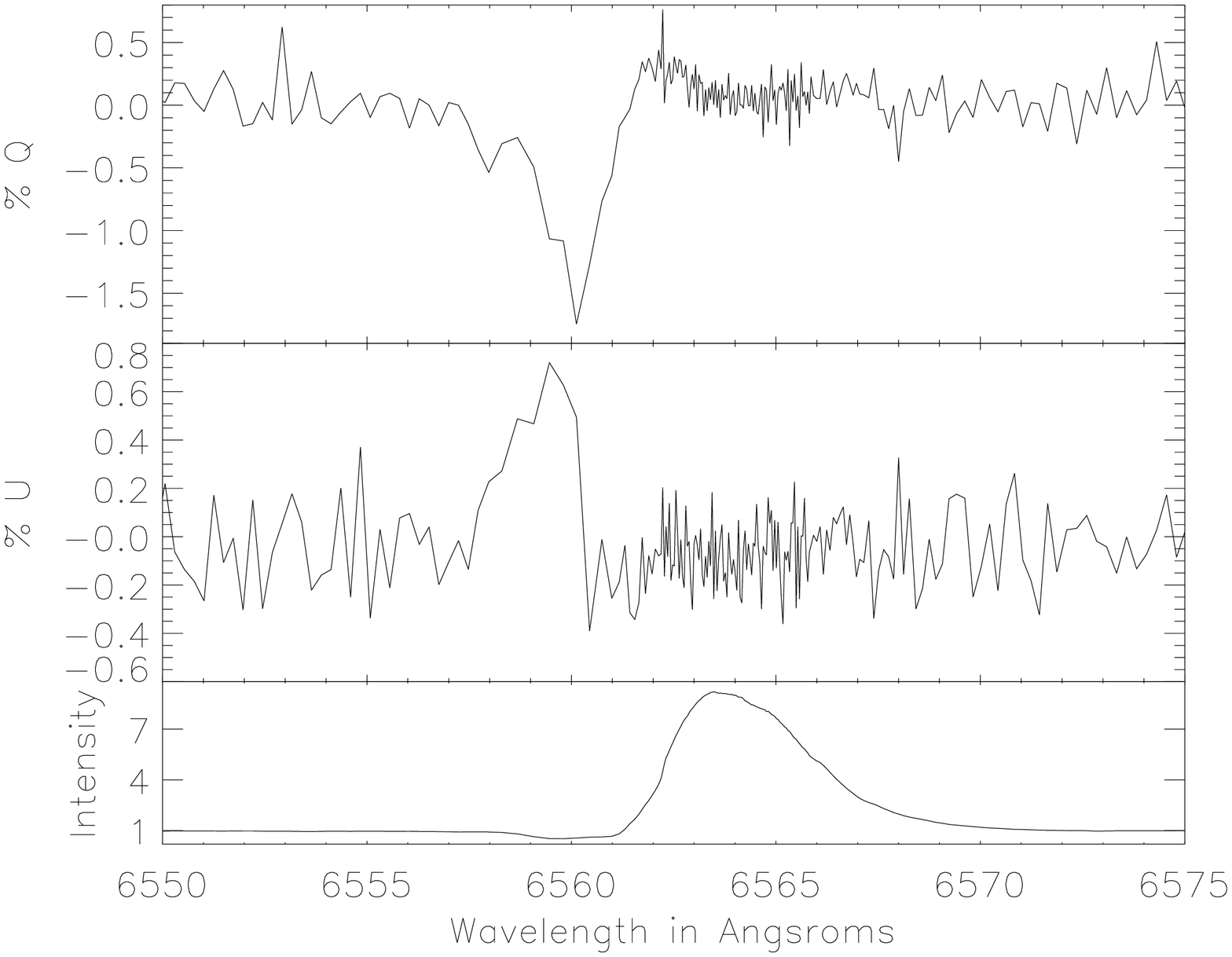} \\
\includegraphics[ width=0.65\linewidth, angle=90]{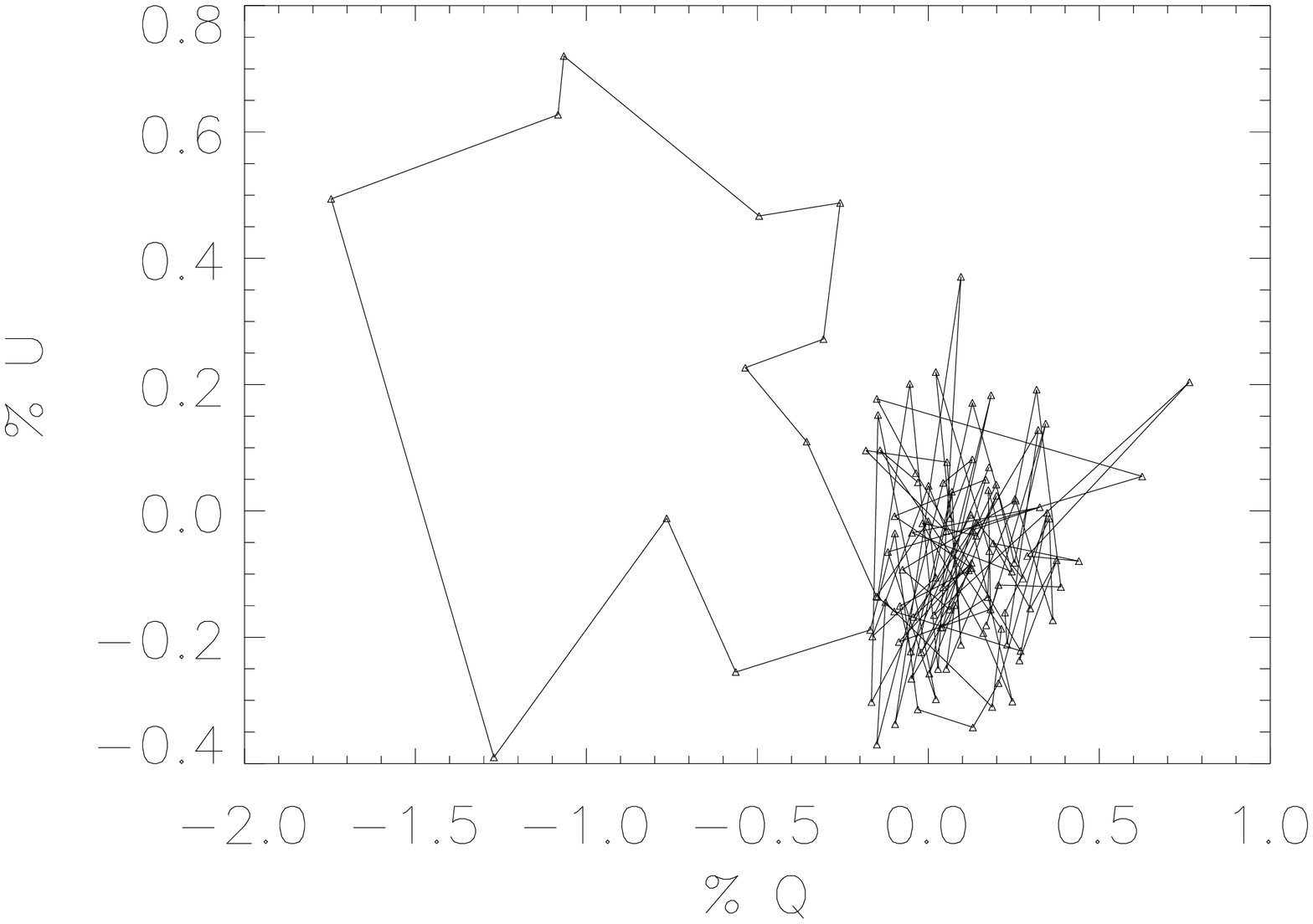}
\caption{An individual AB Aurigae polarization example. From left to right: {\bf a)} An example polarized spectrum for the AB Aurigae H$_\alpha$ line. The spectra have been binned to 5-times continuum. The top panel shows Stokes q, the middle panel shows Stokes u and the bottom panel shows the associated normalized H$_\alpha$ line. There is clearly a detection in the blue-shifted absorption of -1.5\% in q and 0.7\% in u. {\bf b)} This shows q vs u from 6547.9{\AA} to 6564.1{\AA}. The knot of points at (0.0,0.0) represents the continuum.}
\label{fig:swp-abaur}
\end{center}
\end{figure}

\begin{figure}
\begin{center}
\includegraphics[width=0.75\linewidth, angle=90]{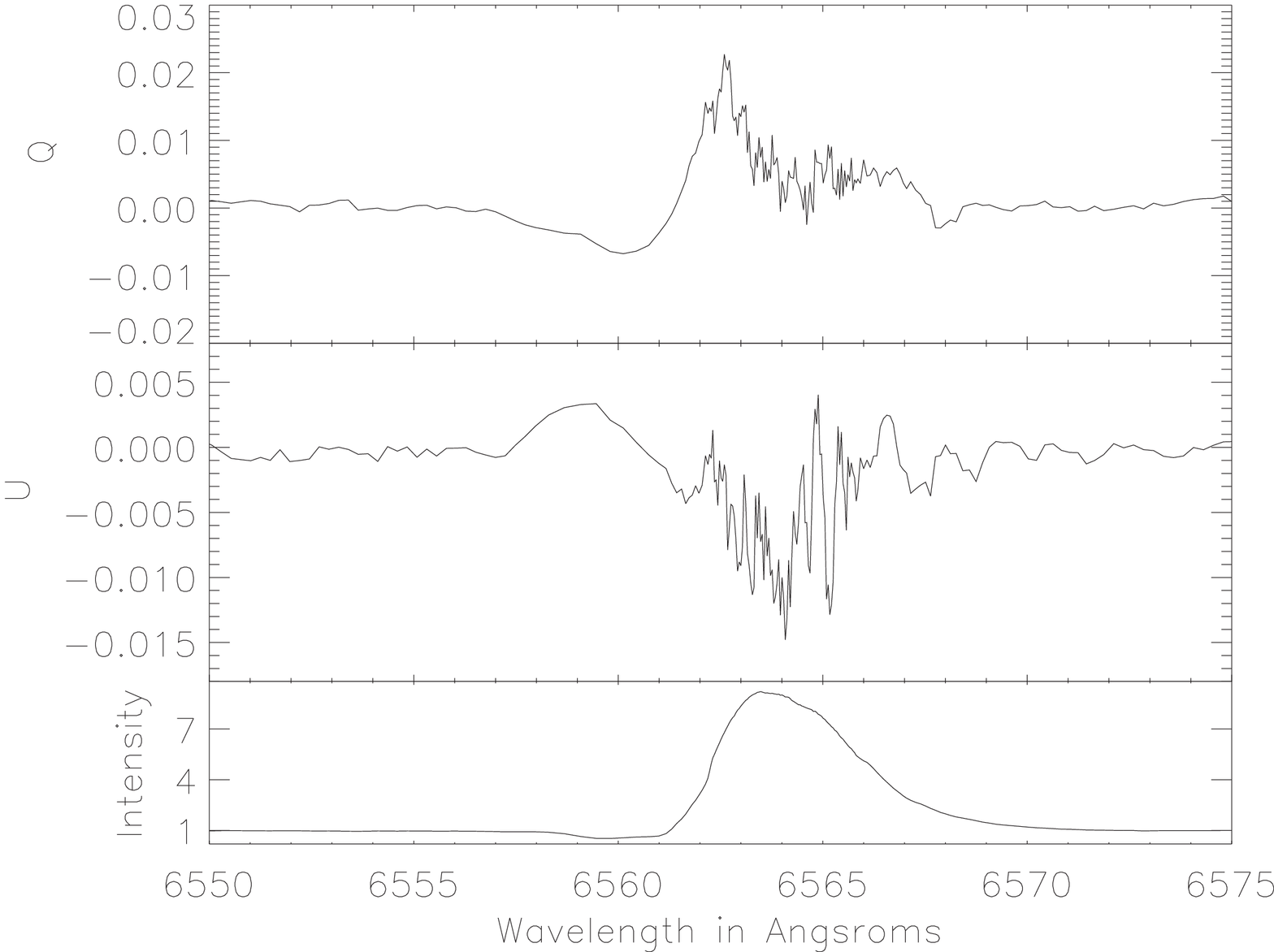}
\caption[AB Aurigae Polarized Flux]{The polarized flux (\% q*I) for the AB Aurigae spectropolarimetry in the previous plot. The polarized spectra are multiplied by the normalized intensity profile to produce a polarized spectrum normalized by the continuum flux. The vertical axis is fractional polarization.}
\label{fig:swp-abaur-pfx}
\end{center}
\end{figure}

\subsection{AB Aurigae - HD 31293 - MWC 93}

	There were three previous publications of spectropolarimetric observations showing 0.4\% to 0.7\% polarization across the H$_\alpha$ line. One observation performed in 1999 was presented in two different papers showing two somewhat different results (Pontefract et al. 2000 and Vink et al. 2002). It is unclear what caused this difference in results when using the same data, but both publications show a line effect. Another set of observations taken in 2001 and 2003 were described as showing the McLean effect (Vink et al. 2005b). In that publication, the phrase was used when a change in polarization was detected across the absorptive component of the line profile. The final observations, taken in 2004, were presented in Mottram et al. 2007 showing a change from the 1999 observations presented in both the Pontefract et al. 2000 and Vink et al. 2002 publications. The optical continuum polarization was 0.8\% $\pm$ 0.1 January 9th 1999 (Ashok et al. 1999).  Beskrovnaya et al. 1995 report R-band 0.3\% in 1993 and 0.15\% in 1988 with substantial variability on a daily basis. 
	
	There are 166 HiVIS polarization measurements presented in figure \ref{fig:haebe-specpol1}. Figure \ref{fig:swp-abaur} shows an individual polarized spectrum that shows a very clear detection of polarization at the 1\% level in both q and u. The polarization change is largest in the absorptive component of the P-Cygni profile and the polarization at maximum emission is indistinguishable from the continuum polarization. Figure \ref{fig:swp-abaur} also shows the corresponding qu-plot for the line, 6547.9 to 6564.1{\AA}. The polarization from 6558-6554 {\AA} dominates the loop. There is a knot of points near (0,0) that represents the continuum. As wavelength increases, q and u both increase in amplitude but not entirely at the same wavelengths.  This gives rise to the loop - if both q and u increased simultaneously, the qu-plot would show a line.

\begin{figure}
\begin{center}
\includegraphics[width=0.75\linewidth, angle=90]{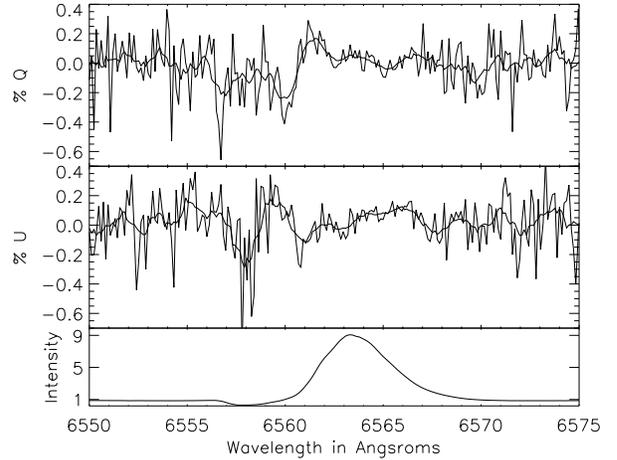}
\caption[AB Aurigae Archive ESPaDOnS Spectropolarimetry.]{The ESPaDOnS archive data for AB Aurigae on February 7th 2006. The solid lines in the spectropolarimetry show a lower resolution smoothed data set. There is a clear, low-amplitude detection across the blue-shifted absorption.}
\label{fig:swp-abaur-esp}
\end{center}
\end{figure}

	With a strong P-Cygni profile having absorption/emission ratio around 20, the possibility of a 1/I systematic error must be explored. Figure \ref{fig:swp-abaur-pfx} shows the normalized flux polarized change, basically the difference between the detected Stokes Q and Stokes Q for a line with a constant degree of polarization. The polarized flux for a purely 1/I error would be completely flat. That the polarized flux has both positive and negative components with different structure from Q to U shows that this is not a systematic effect. No shift in continuum polarization before computing q*I could create this Q spectrum.

	Polarization in absorption was the only type of signature detected for this system. A correlation between the width of the blue-shifted P-Cygni absorption and the width of the detected spectropolarimetric signature was discussed in Harrington \& Kuhn 2007. The detection was always around 1\% but the exact value of the position angle of the change is dependent on the rotation of the plane of polarization by the telescope as well as the source. Though the width of the polarization effect in all HiVIS observations is strongly correlated with the width of the absorptive component, there are a number of observations which show stronger polarization in the absorptive component closer to line center. 
	
	Curiously, the ESPaDOnS archive observations at a signal-to-noise of 900 shown in figure \ref{fig:swp-abaur-esp}, only a small polarization is detected at the 0.2\% level with a more complicated morphology. Given the overwhelming number of detections with HiVIS over a wide range of conditions, and the variability of other sources detected with ESPaDOnS, it must be concluded that this star is strongly variable in its spectropolarimetric signatures. Another thing that must be pointed out is the very unique morphology of the spectropolarimetric effects. Most of the HiVIS observations, when showing a clear spectropolarimetric signature, are of the same morphology as figure \ref{fig:swp-abaur}. The polarization in the center of the line and on the red-shifted side of the line is identical to continuum.

\subsection{HD 31648 - MWC 480}

	  This star had a large amplitude ($\sim$0.9\%) polarization increase in the blue-shifted absorption as well as a 0.3\% decrease across the emission line from a continuum of 0.4\% in Vink et al. 2002. In Vink et al. 2005b, the polarization increase in the absorption was 0.8\% with the continuum varying from 0.18\% to 0.30\%. The star showed a 0.4\% increase in the blue-shifted absorption with no signature in the emission line on a continuum of 0.2\% in Mottram et al. 2007, pointing to variability. Beskrovnaya \& Pogodin 2004 also found R-band polarization of roughly 0.3\% in most of their measurements.
	  
\begin{figure*}
\begin{center}
\includegraphics[width=0.35\linewidth, angle=90]{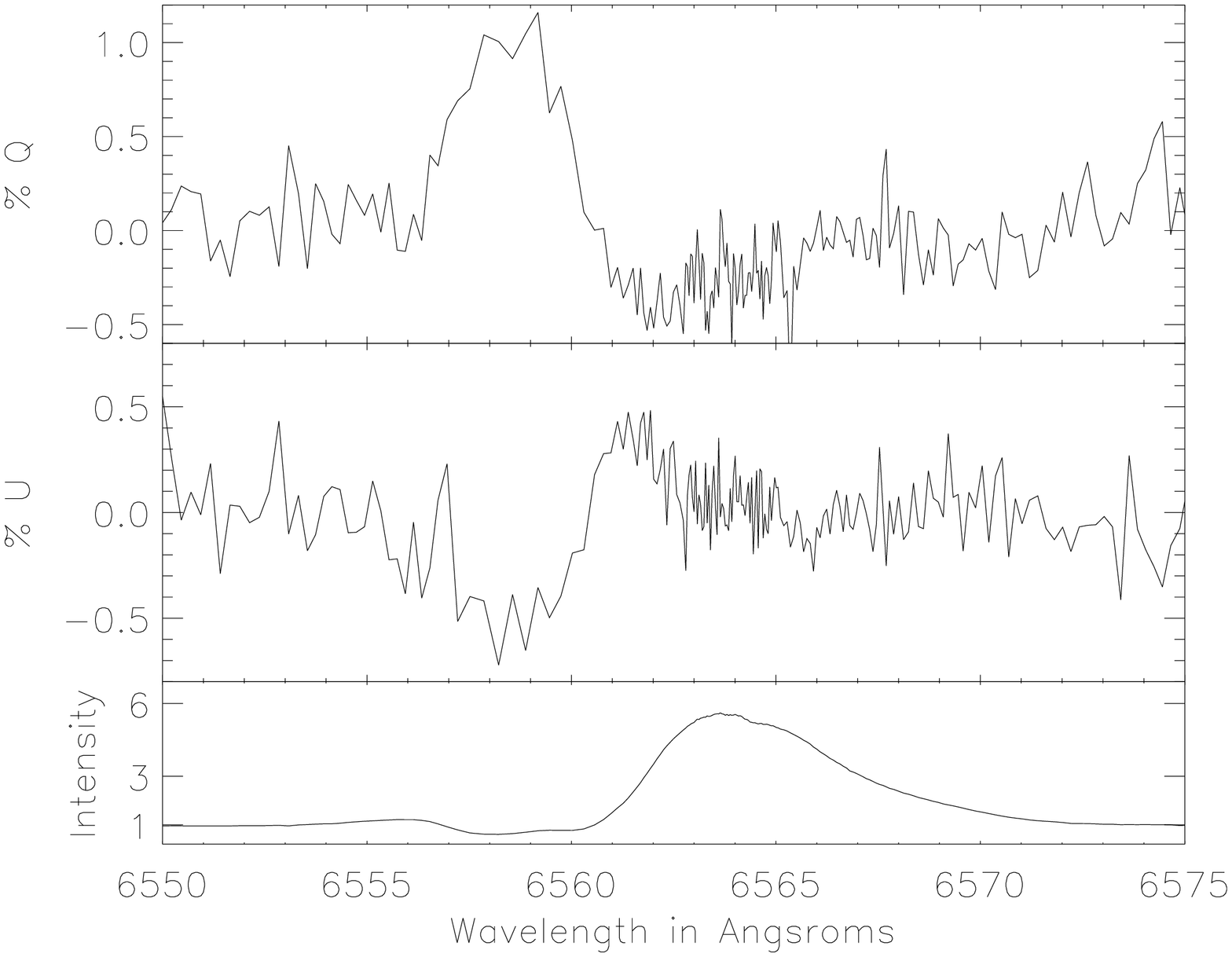}
\includegraphics[width=0.35\linewidth, angle=90]{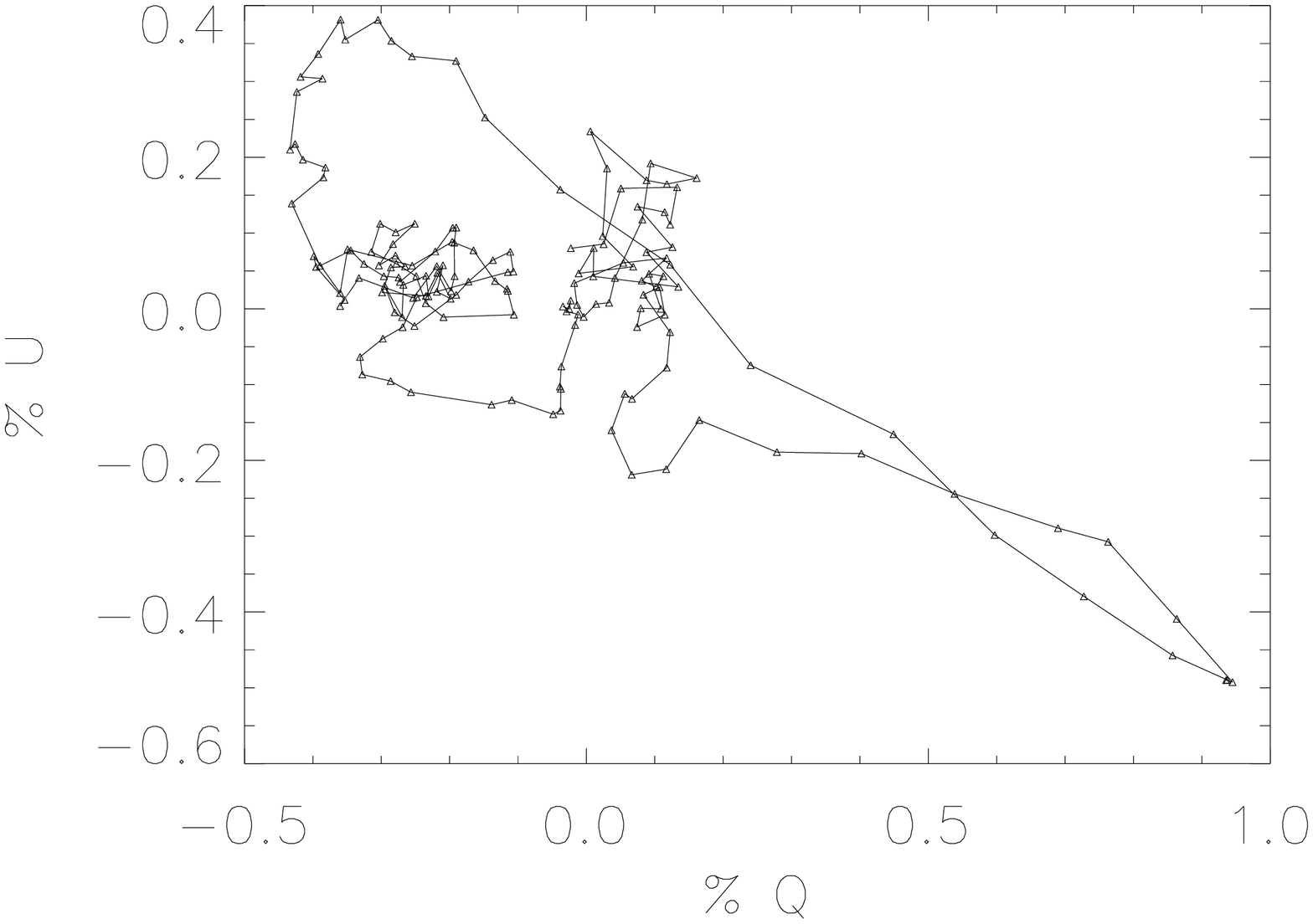} \\
\includegraphics[width=0.35\linewidth, angle=90]{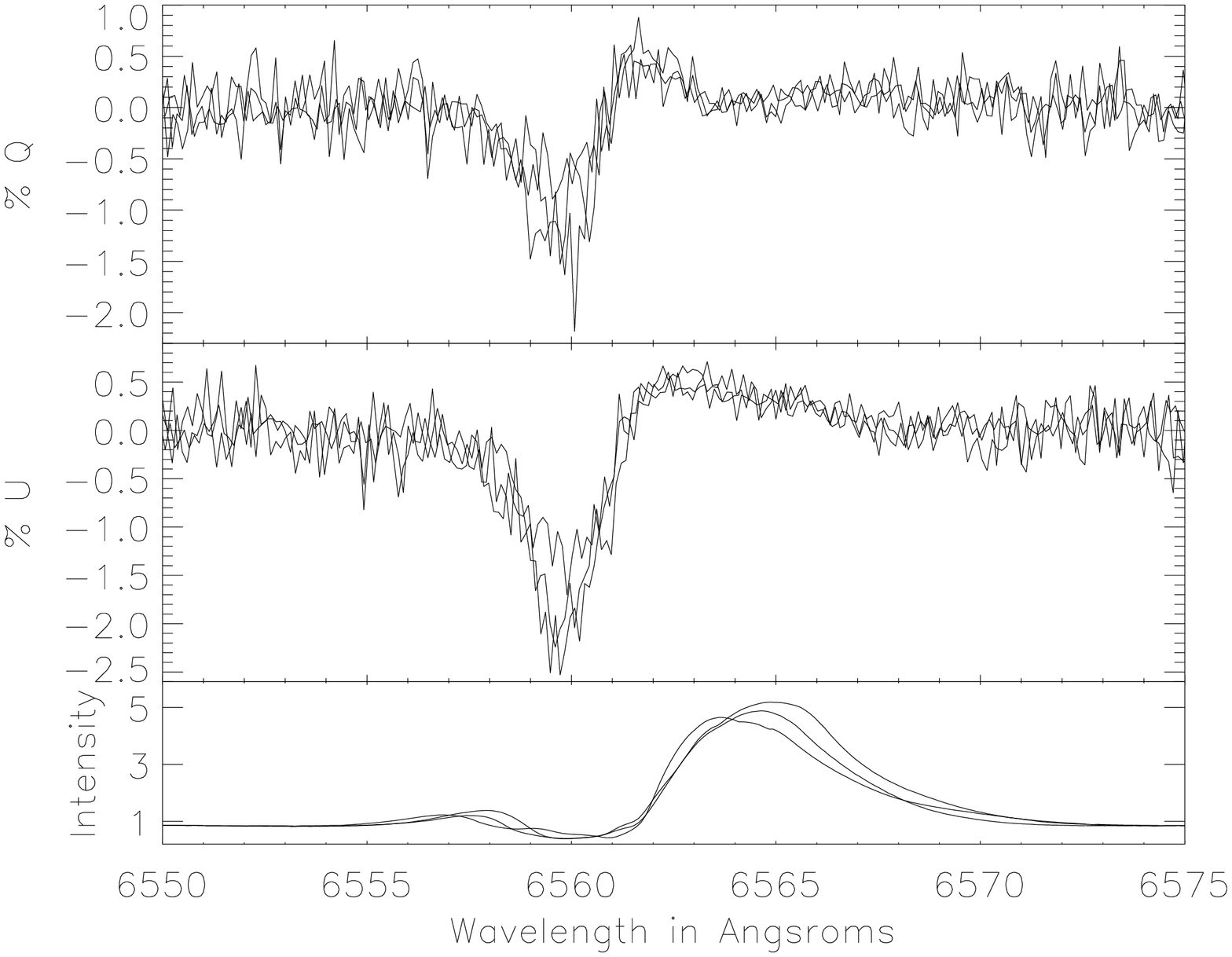} 
\includegraphics[width=0.35\linewidth, angle=90]{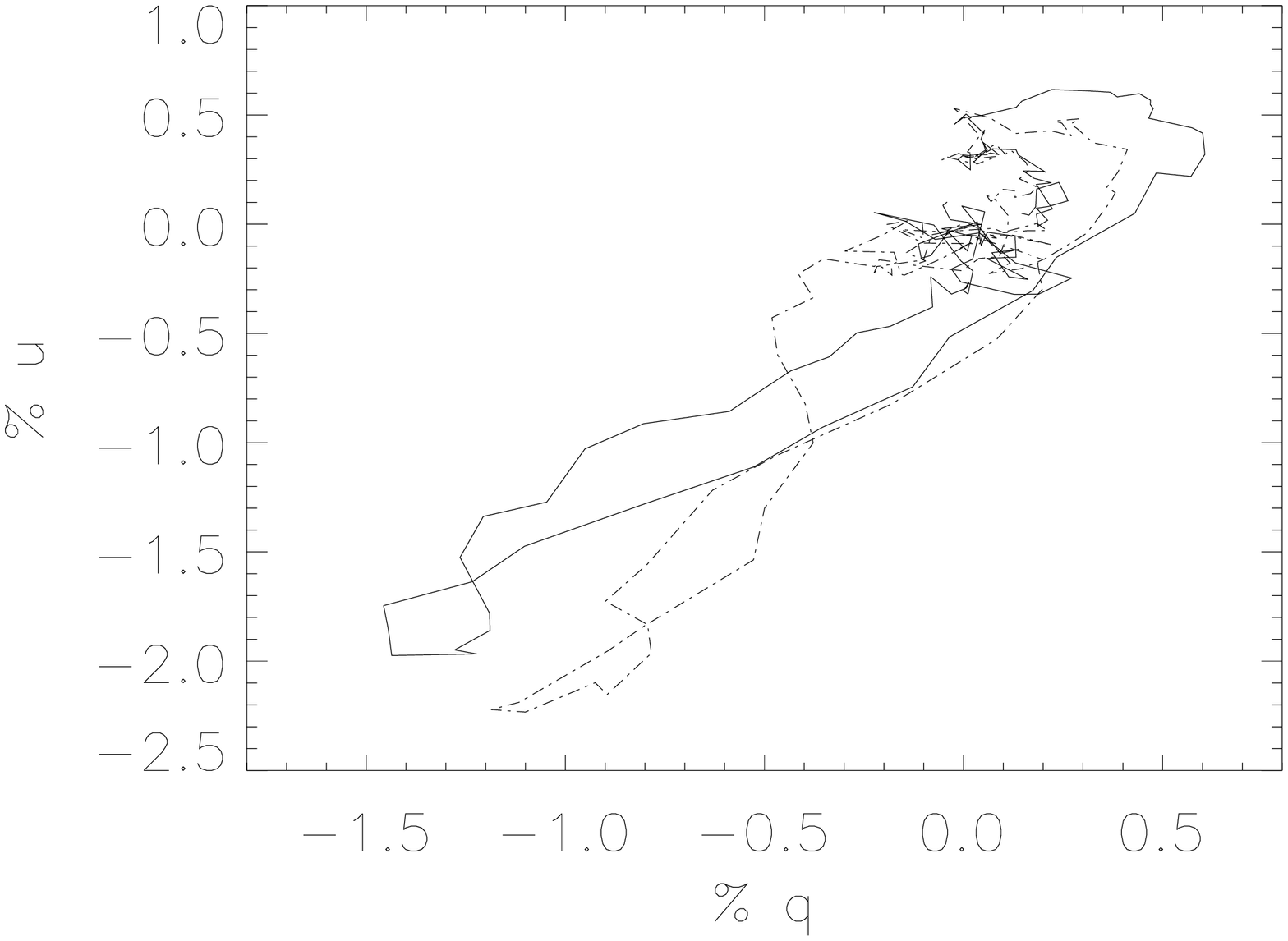} \\
\includegraphics[width=0.35\linewidth, angle=90]{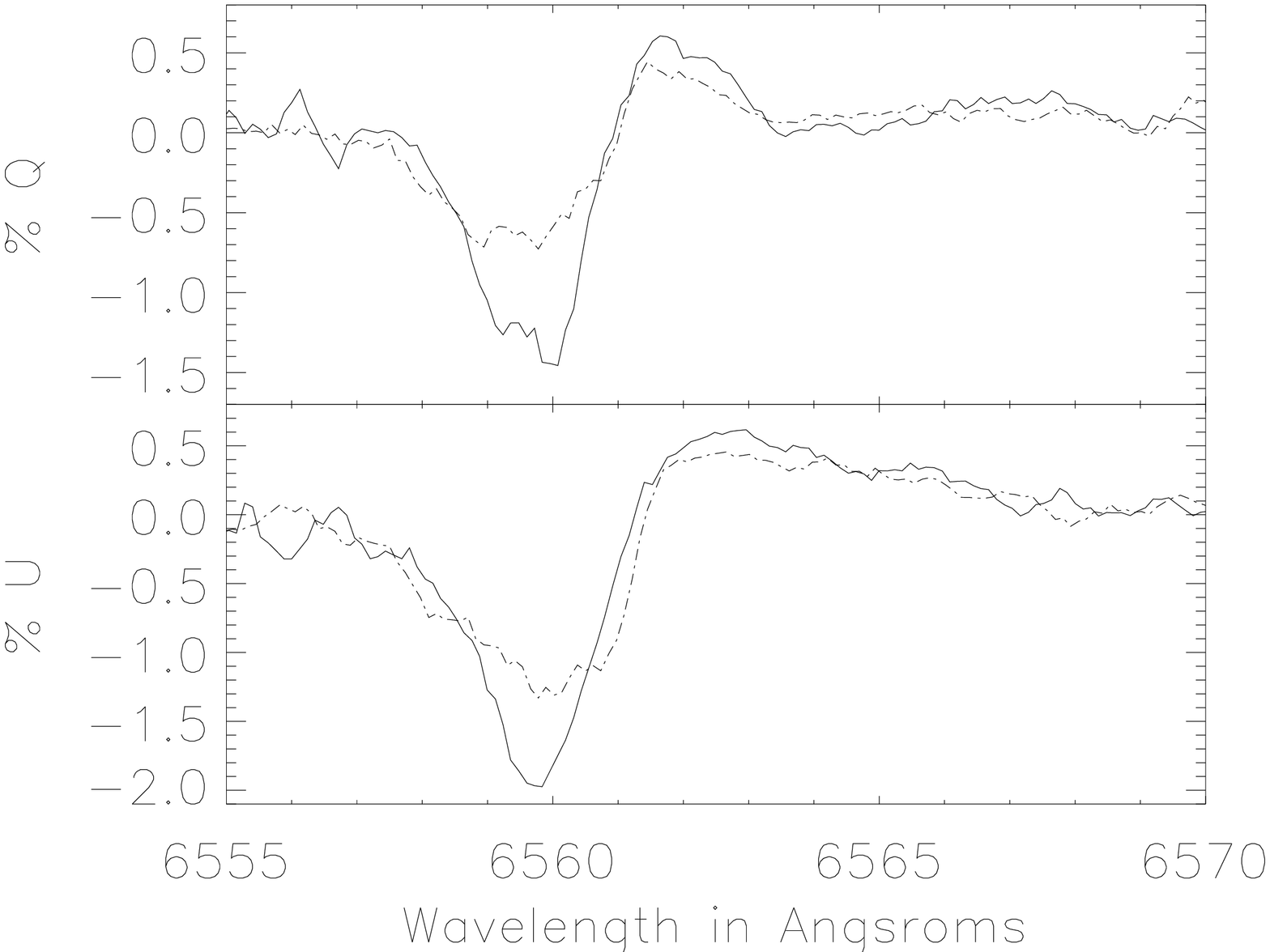}
\includegraphics[width=0.35\linewidth, angle=90]{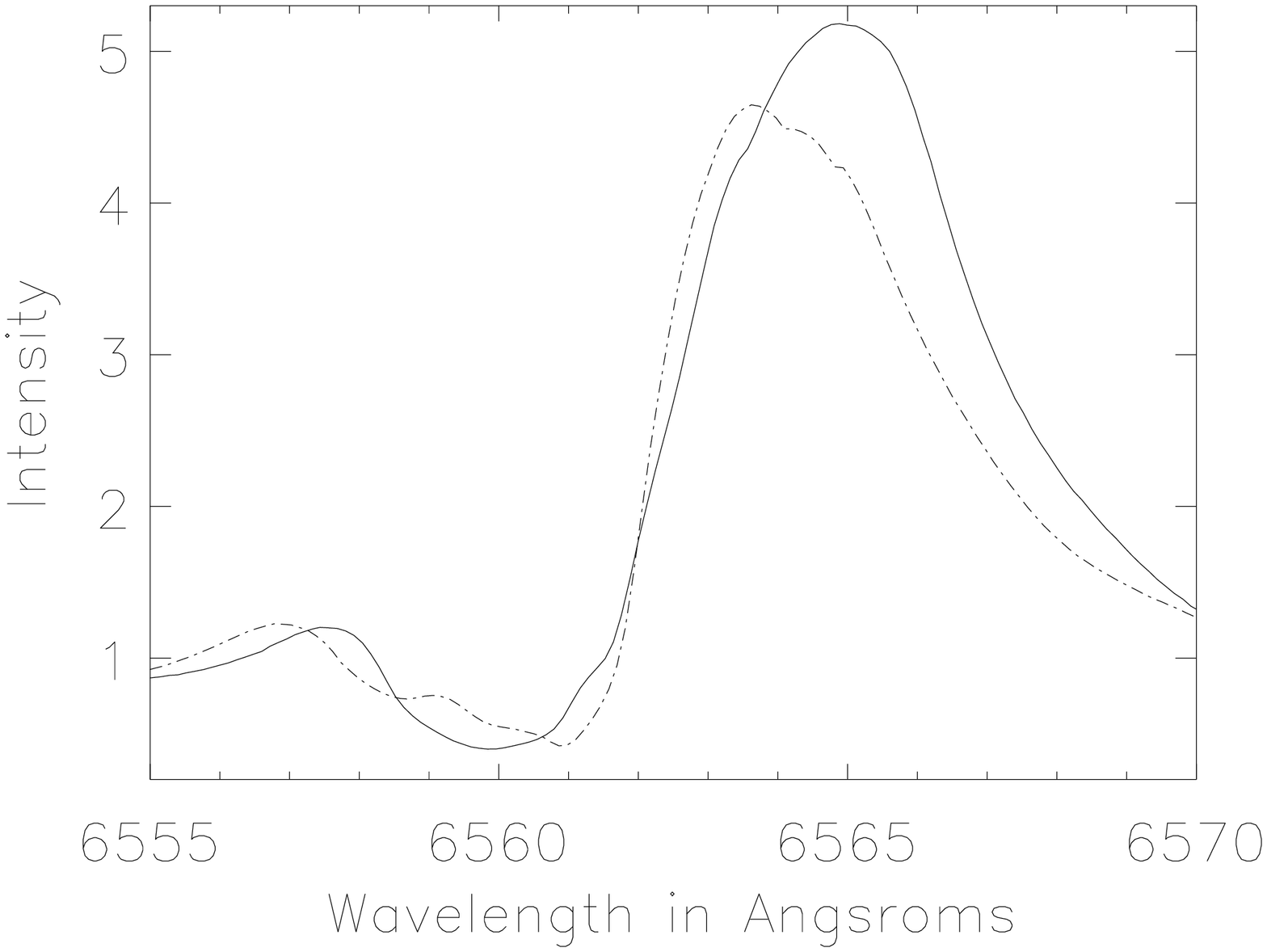}
\caption{An example of MWC 480 spectropolarimetry. From left to right: {\bf a)} An example polarized spectrum for the MWC 480 H$_\alpha$ line. The spectra have been binned to 5-times continuum. The top panel shows Stokes q, the middle panel shows Stokes u and the bottom panel shows the associated normalized H$_\alpha$ line. There is clearly a detection in the blue-shifted absorption of 1.0\% in q and -0.6\% in u. {\bf b)} This shows q vs u from 6545.7{\AA} to 6566.6{\AA}. The knot of points at (0.0,0.0) represents the continuum. There is another knot of points near (-0.2,0.0) which represents the non-zero q value at the emissive peak. The non-linear wavelength coverage created in the bin-by-flux procedure highlights this effect and shows distinctly that the emissive peak does have a small non-zero polarization. {\bf c)} The ESPaDOnS archive data for MWC 480 on February 7th, 8th, and August 13th 2006. The two polarized spectra from February are nearly identical and show a larger magnitude signature in the absorption trough. The August observations show a similar shape but the magnitude of the signature is smaller. {\bf d)} This shows the corresponding qu-loops. A significant deviation in shape and angle between the individual ESPaDOnS observations is more apparent in this plot.  {\bf e)} Shows the archival data changing over 7 months - from February 7th (solid) to August 13th 2006 (dashed). The polarization change across the line has been smoothed with a 5-pixel boxcar. The difference is strongest in the P-Cygni absorption where the magnitude of polarization change decreases by a factor of two, from (-1.5,-2.0) to (-0.7,-1.0). {\bf f)} Shows the corresponding line profiles. In the August observations, the absorption was not quite as deep but was more broad.}
\label{fig:swp-mwc480}
\end{center}
\end{figure*}

	  This star also showed polarization in absorption, as AB Aur, but the change was significantly wider and did extend toward the emissive peak. There are 67 measurements shown in figure \ref{fig:haebe-specpol1}, most of which show polarization in the absorption. Figure \ref{fig:swp-mwc480} shows a polarization spectrum where there is a large 1\% change in the absorptive component but with the change in Stokes u extending much wider to the blue and both q and u showng a change on the blue side of the emission peak. Figure \ref{fig:swp-mwc480} also shows the qu-plot for this polarization spectrum from 6545.7 to 6566.6{\AA}. There is a continuum knot near (0,0) but there is another knot near (-0.2,0) which corresponds to the change in q across the emission peak. The qu-plot is non-linear in wavelength because of the flux-dependent binning. The qu-plot shows a linear extension of (+q, -u) from the absorption trough but it also shows a smaller (-q, +u) loop from the blue side of the emission peak. The qu-plot does not return to continuum until the red side of the emission line near 6566.6{\AA}

\begin{figure*}
\begin{center}
\includegraphics[ width=0.35\linewidth, angle=90]{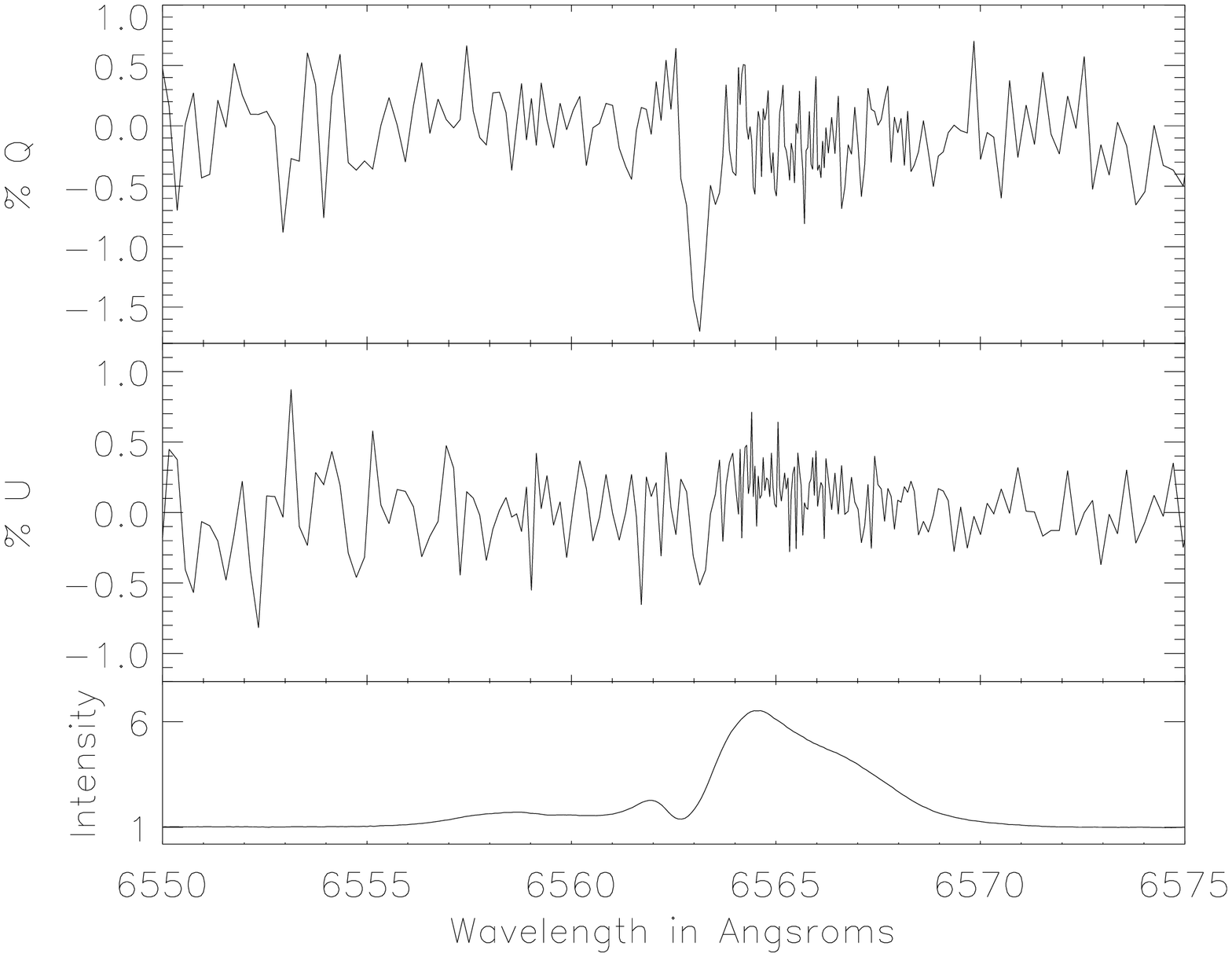}
\includegraphics[ width=0.35\linewidth, angle=90]{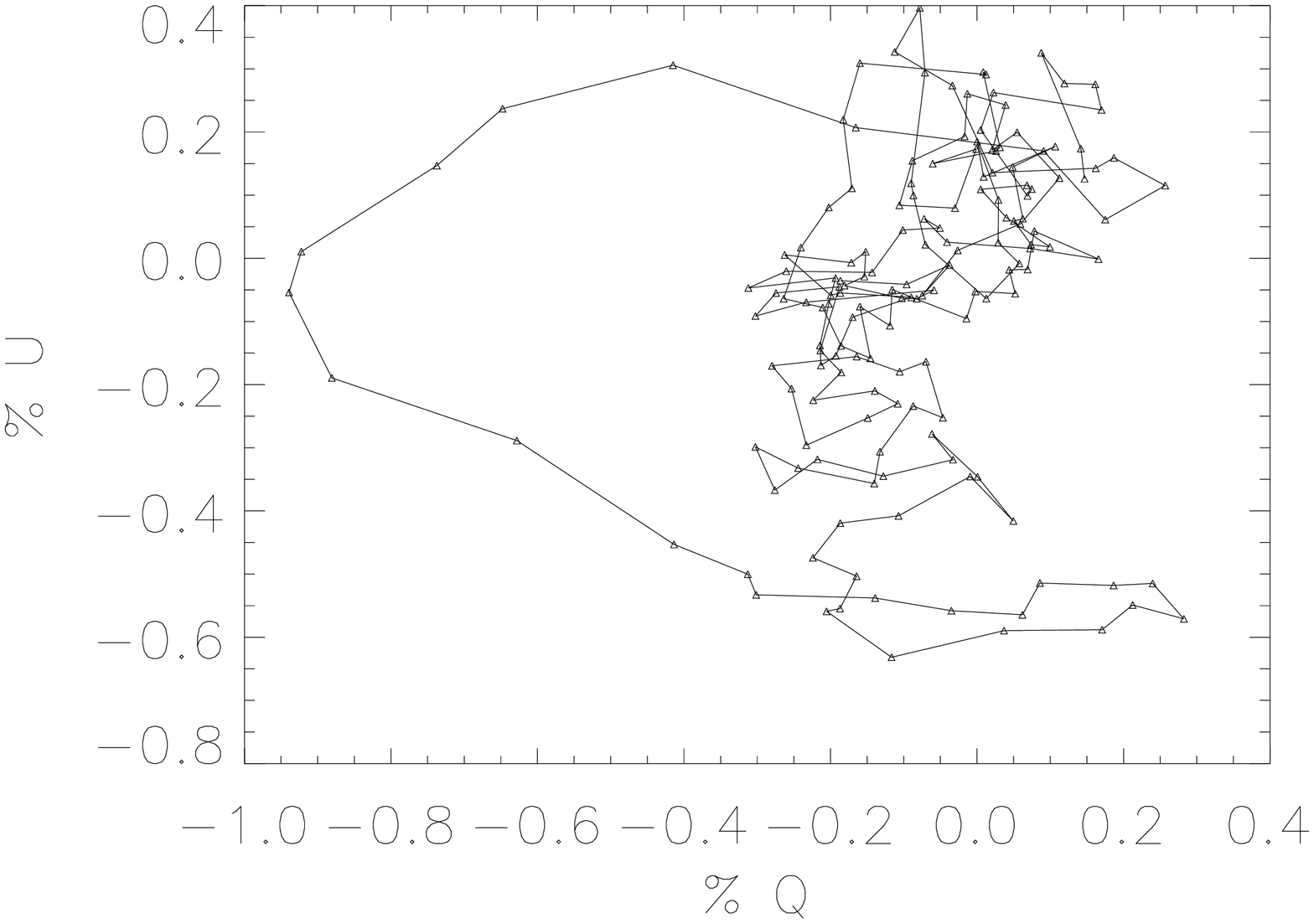} \\
\includegraphics[ width=0.35\linewidth, angle=90]{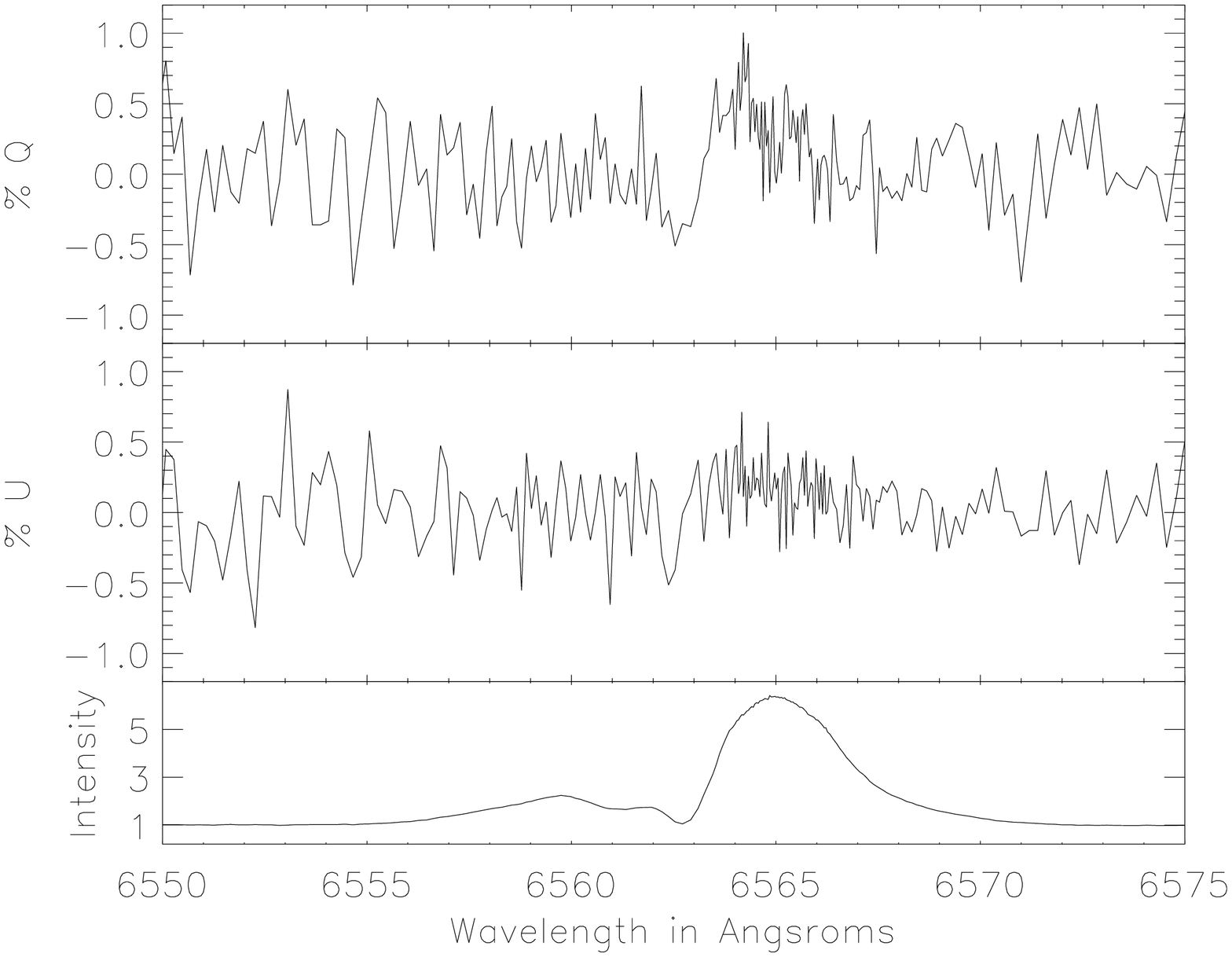}
\includegraphics[ width=0.35\linewidth, angle=90]{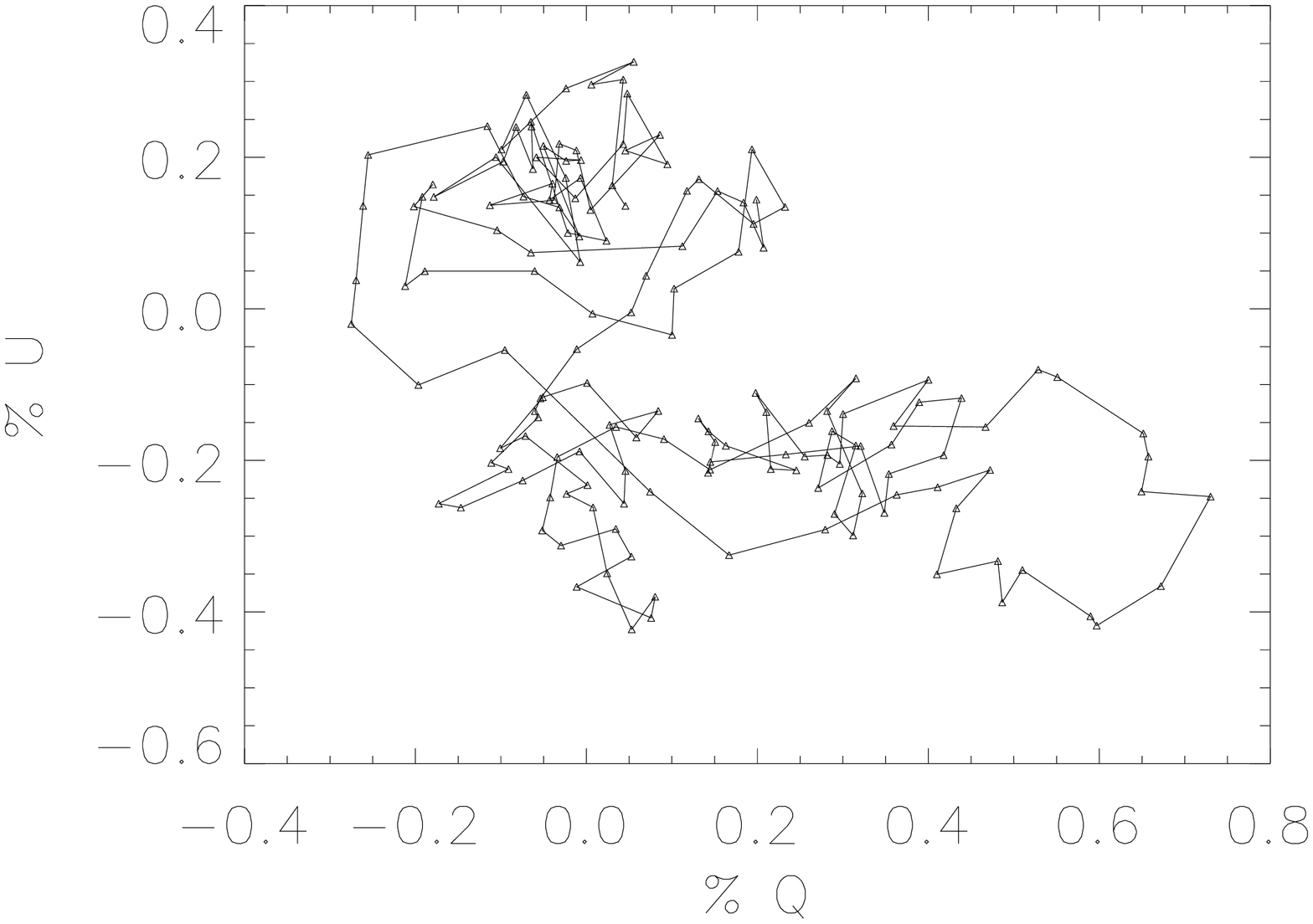} \\
\includegraphics[ width=0.35\linewidth, angle=90]{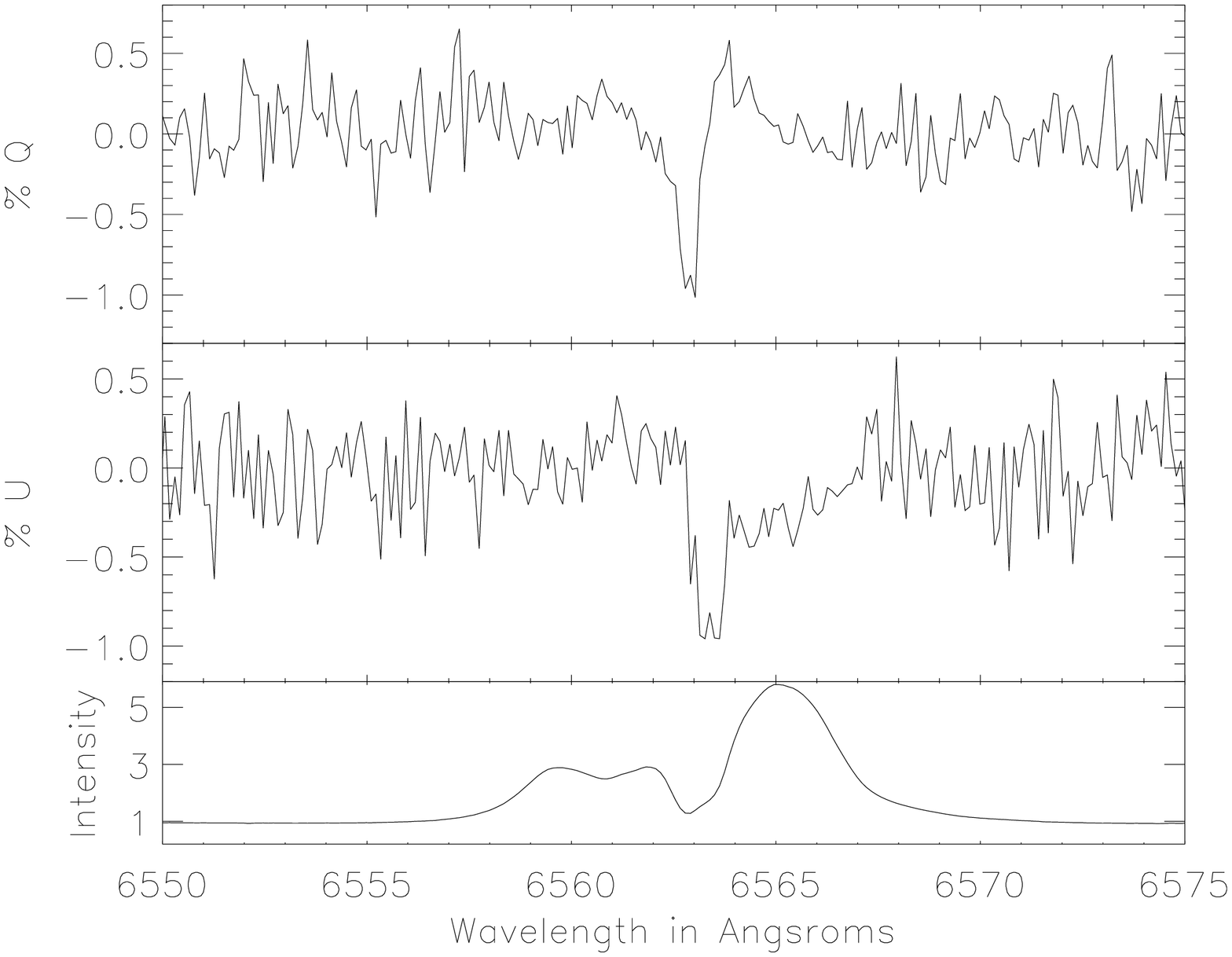}
\includegraphics[ width=0.35\linewidth, angle=90]{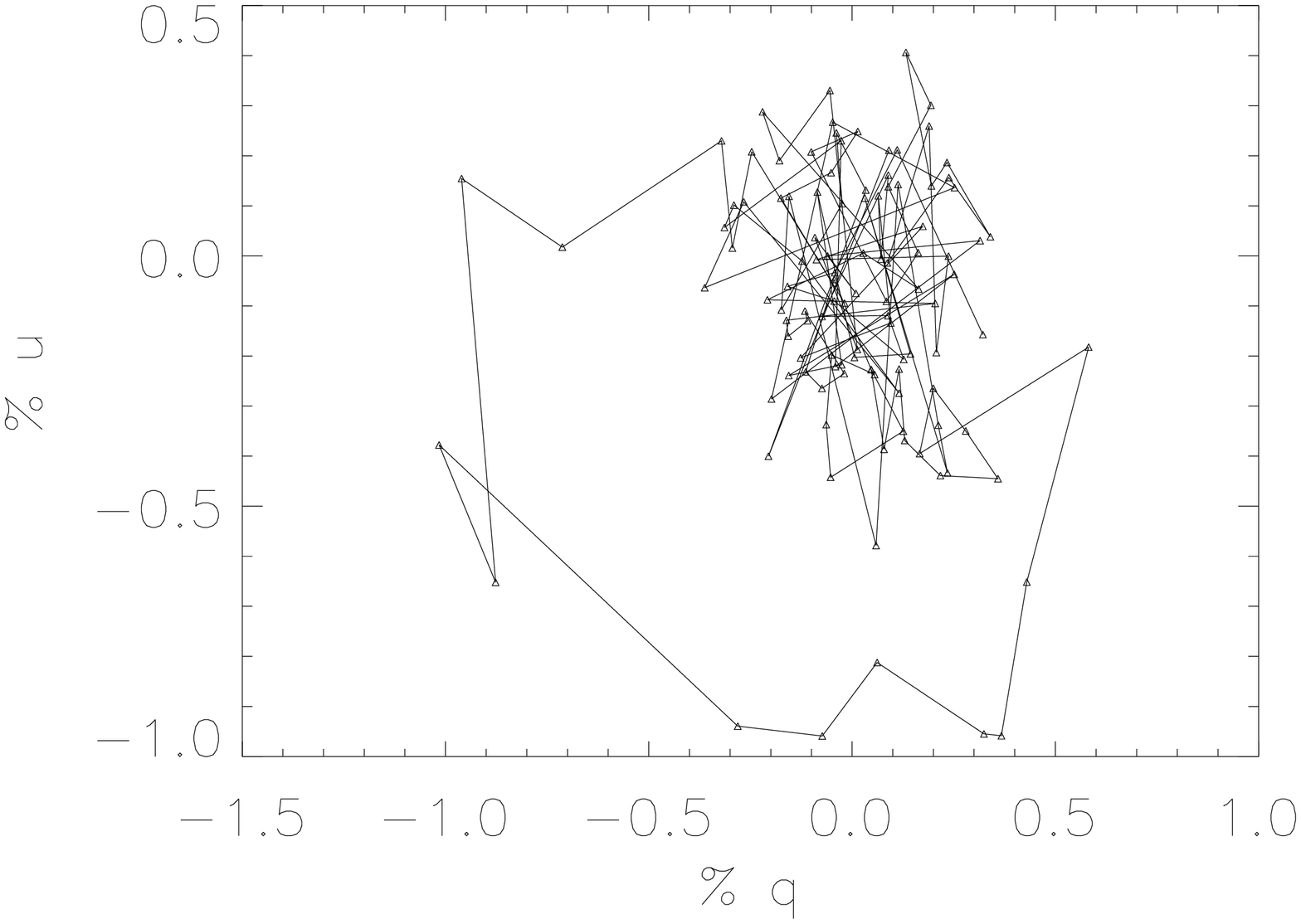}
\caption{An example of MWC 120 spectropolarimetry. From left to right: {\bf a)} An example polarized spectrum for the MWC 120 H$_\alpha$ line. The spectra have been binned to 5-times continuum. The top panel shows Stokes q, the middle panel shows Stokes u and the bottom panel shows the associated normalized H$_\alpha$ line. There is clearly a detection in the central absorption of -1.5\% in q. {\bf b)} This shows q vs u from 6557.2{\AA} to 6573.1{\AA}.  The knot of points at (0.0,0.0) represents the continuum. {\bf c)} This shows another epoch of HiVIS observations with a different character - a 0.5\% detection in q and {\bf d)} is the corresponding qu loop. {\bf e)} The ESPaDOnS archive data for MWC 120 on February 9th, 2006. There is a very strong, narrow polarization change near the line center with a more broad change on the red wing of the line. {\bf f)} Shows the corresponding qu-loop, demonstrating that the changes in q and u are out of phase.}
\label{fig:swp-mwc120}
\end{center}
\end{figure*}

\begin{figure*}
\begin{center}
\includegraphics[width=0.35\linewidth, angle=90]{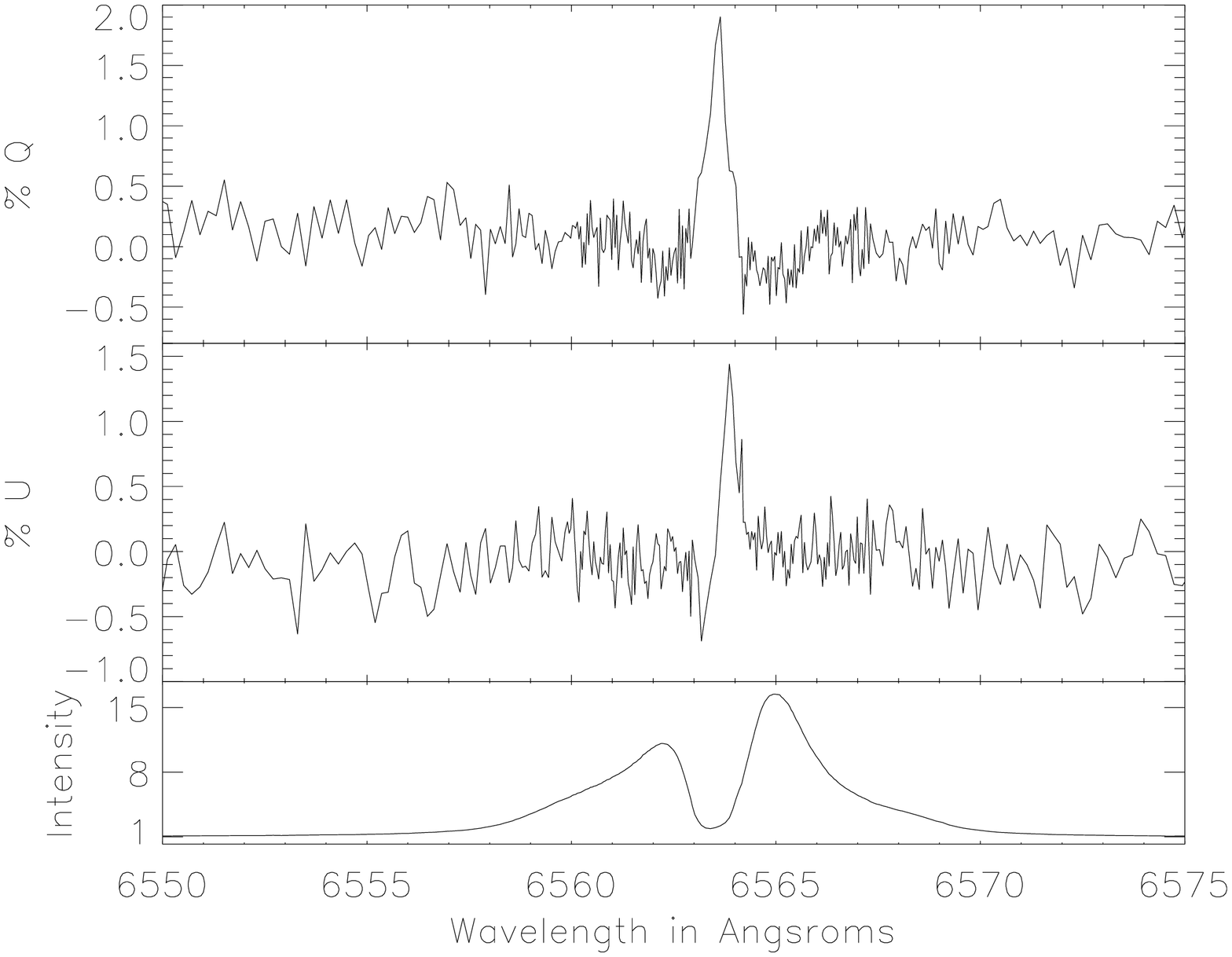}
\includegraphics[ width=0.35\linewidth, angle=90]{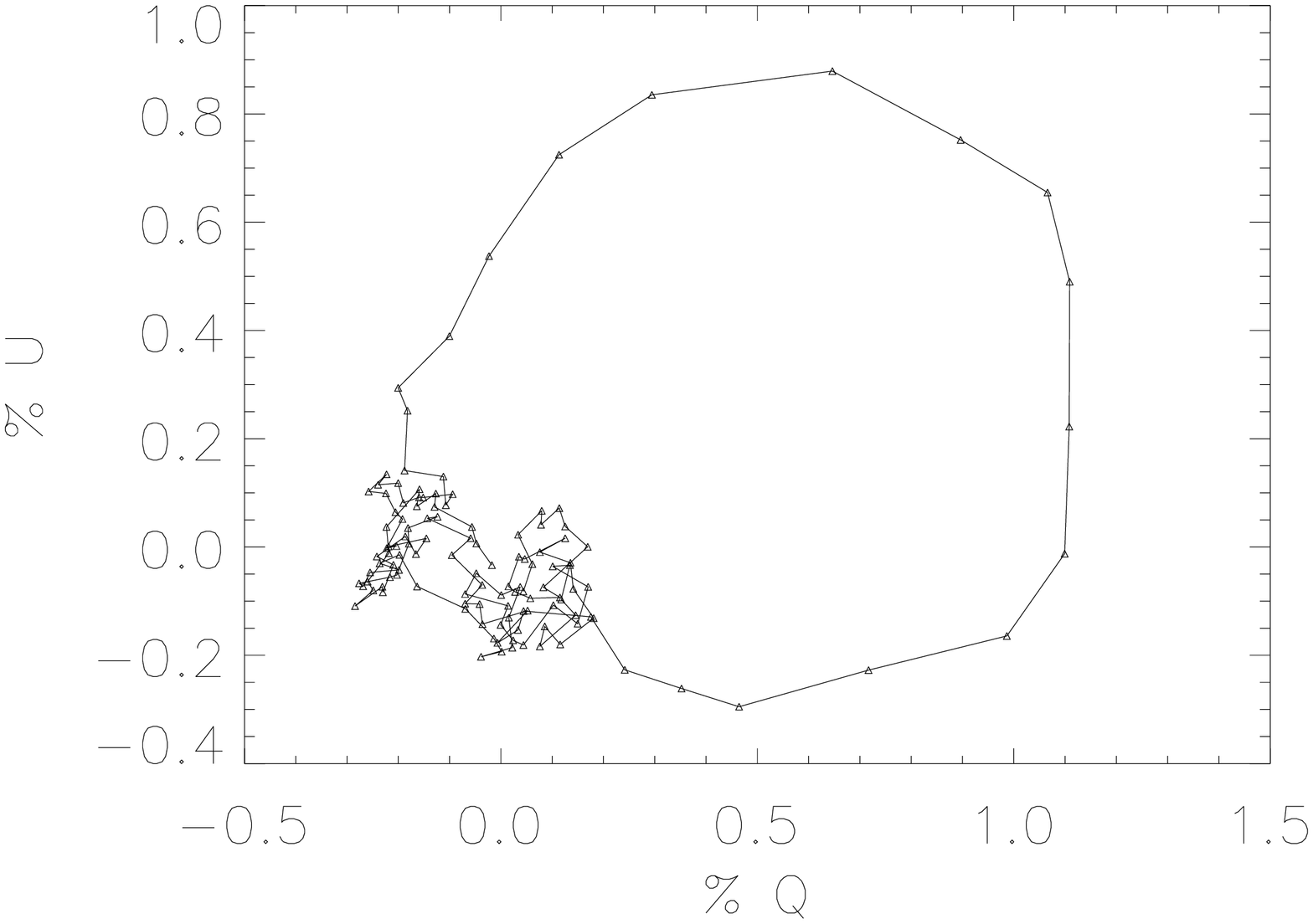} \\
\includegraphics[width=0.35\linewidth, angle=90]{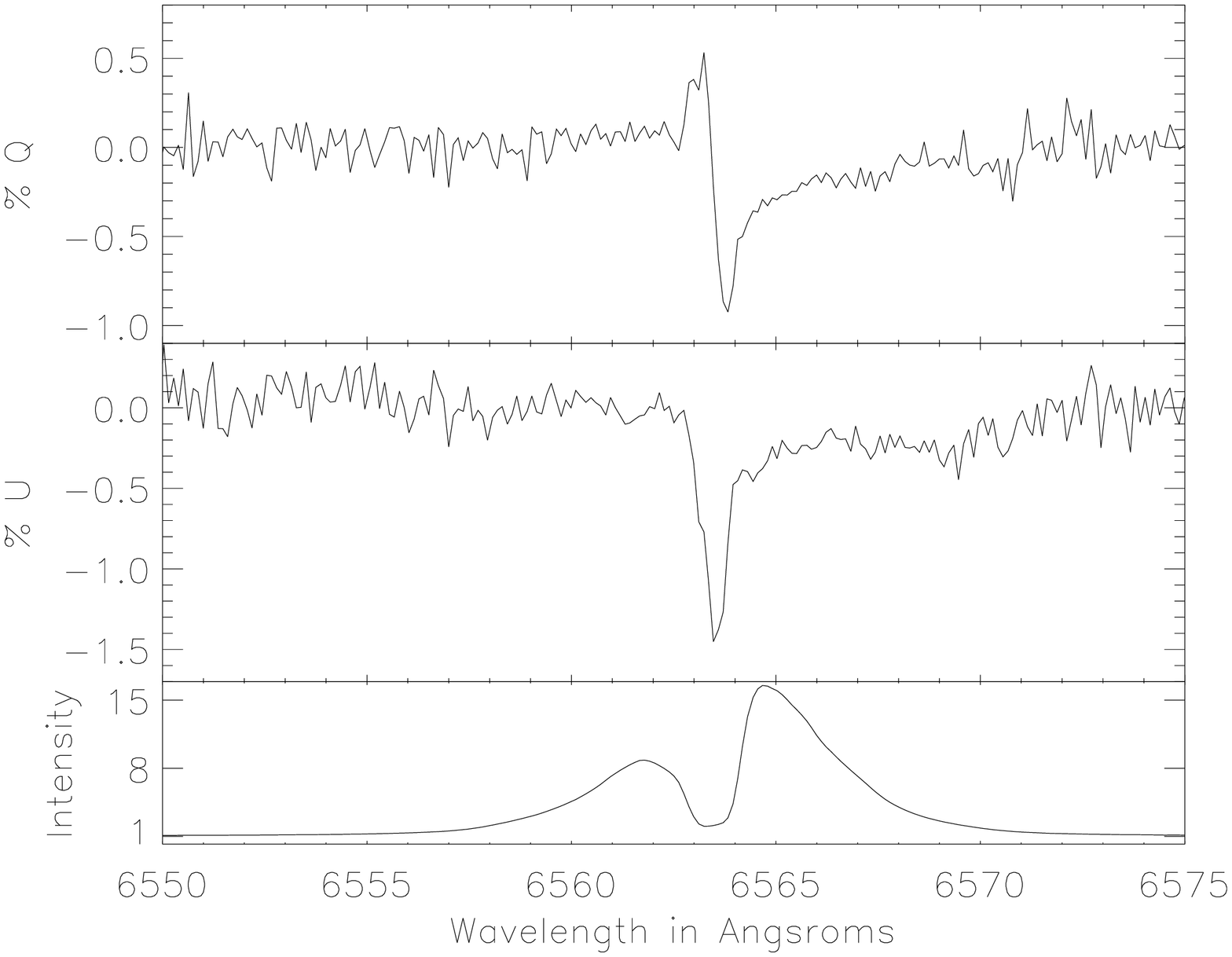}
\includegraphics[width=0.35\linewidth, angle=90]{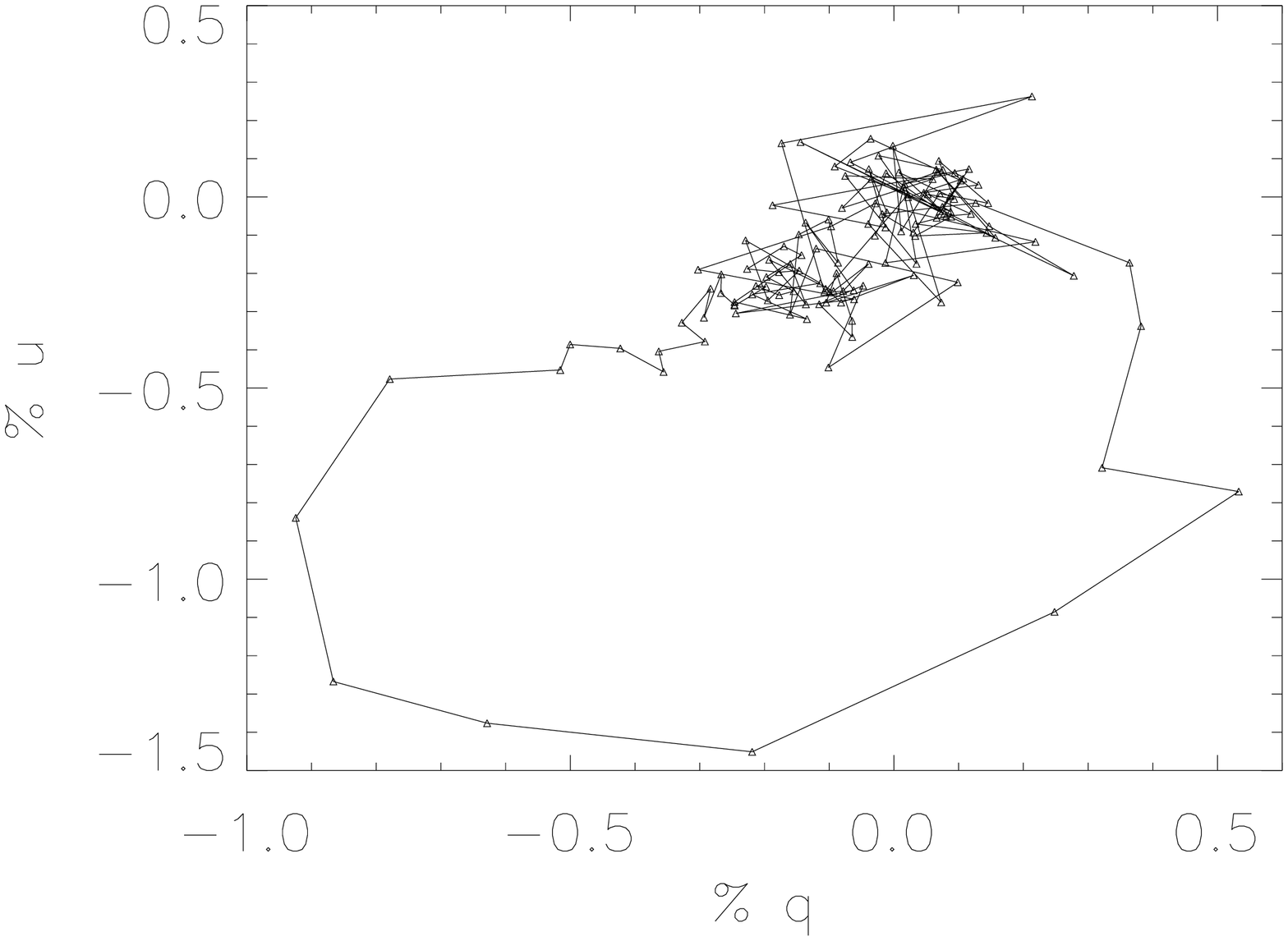}
\vspace{8mm}
\caption{An example of MWC 158 spectropolarimetry. From left to right: {\bf a)} An individual polarized spectrum for the MWC 158 H$_\alpha$ line. The spectra have been binned to 5-times continuum. The top panel shows Stokes q, the middle panel shows Stokes u and the bottom panel shows the associated normalized H$_\alpha$ line. There is clearly a detection in the central absorption of 2.0\% in q and an more complex 1.5\% detection in u. {\bf b)} This shows q vs u from 6560.1{\AA} to 6565.6{\AA}. The knot of points at (0.0,0.0) represents the continuum. {\bf c)} The ESPaDOnS archive data for MWC 158 on February 9th, 2006.  {\bf d)} This shows q vs u. The knot of points at (0.0,0.0) represents the continuum. Since both the magnitudes and widths of q and u are not similar, the qu-loop is wide and flat. The small, broad change seen on the red side of the emission in both q and u is seen as a separate cluster of points adjacent to the continuum knot in towards the lower left (-q, -u).}
\vspace{8mm}
\label{fig:swp-mwc158-esp}
\end{center}
\end{figure*}

	There is a very clear detection in ESPaDOnS archival observations as well. Three separate observations on February 7th, 8th, and August 13th 2006 all showed very strong spectropolarimetric signatures. The line profiles were similar in shape to those observed with HiVIS, but with a slightly lower amplitude. The spectropolarimetry, also shown in figure \ref{fig:swp-mwc480}, shows a very strong variability of the polarization even though the line profile itself is not. The two February observations show a roughly 1.5\% decrease in q and a 2\% decrease in u, both centered on the blue-shifted side of the absorption trough. There is a sign-change to a 0.5\% increase in both q and u, with the zero-point occurring on the red-shifted side of the absorption trough, though still on the blue side of the emission peak. By the center of the emission peak, both polarizations are quite close to continuum, although Stokes u does not reach continuum until 6567{\AA}, on the red-shifted side of the emission peak.

	These differences are highlighted in the bottom panels of figure \ref{fig:swp-mwc480}. The H$_\alpha$ line profiles show a much more significant change in the emissive component of the line whereas the blue-shifted absorption only mildly varies shape with the depth being roughly 0.5 in both. In contrast, the polarization values in the blue-shifted absorption vary by a factor of two over the 6-month baseline.  In the qu-loop of figure \ref{fig:swp-mwc480} a significant width on the emissive peak (the increase in q and u) is seen. The polarization in the absorption, though mostly linear in the wavelengths of greatest polarization, does show some significant width in one of the observations.

\subsection{HD 37806 - MWC 120}

 	H$_\alpha$ line spectropolarimetry from 1995 and 1996 was presented in Oudmaijer \& Drew 1999. The emission line profiles they observed changed drastically between the years. In January 1995, the line was double-peaked but with a much weaker blue-shifted emission. In December 1996, the emission line was evenly double-peaked. They observed a change in polarization angle only in both cases. In their December 1996 observations the change in PA is clearly seen in the central absorption. In the December 1995 observations the angle change is wider, but still centered on the absorption. The continuum polarization is 0.29\% and 0.36\% for the two data sets. Vink et al. 2002 present observations that show a double-peaked emission line with a 0.4\% increase in polarization on the red red-shifted emission peak off a continuum of 0.4\%. Mottram et al. 2007 report 2004 observations that show a 0.2\% decrease in polarization from a continuum of 0.4\% on the red-shifted emission peak. The line was much more symmetric, having a narrow mildly blue-shifted absorption, but nothing resembling the strong windy profiles observed in 2006. 

	This star had a more complex polarization spectrum, with strong polarization changes in the blue-shifted absorption, but also showed some significant change across the emission line. This can been seen in the overall shape of q and u in figure \ref{fig:haebe-specpol1} which shows 26 epochs. The figure shows one polarized spectrum where there is a strong change in q over a very narrow wavelength range. The change is on the blue side of the emission line but is not coincident with the strong notch near 6563{\AA}. There is a subtle change in u across the emission line that is better seen in the qu-plot. There is a strong change in q of -1\% that forms the wide loop but there is a significant elongation in u across a much wider range of wavelengths. The change in q starts near u of -0.5\% and ends near 0.2\%. The u values don't return to continuum until the 6573{\AA}. This morphology is also seen very clearly in the ESPaDOnS archive observations in the bottom row of figure \ref{fig:swp-mwc120} from February 9th 2006. The signal-to-noise ratio and spectral resolution are higher and show the complicated morphology of spectropolarimetric effect. The qu-loop for the archive data is dominated by the two large 1\% excursions that form the loop in the archive loop panel of figure \ref{fig:swp-mwc120}. There is significant polarization seen at low levels in a wider wavelength range that are lost among the continuum-knot in the qu-plot but are seen in the original polarized spectra.

	An entirely different example from HiViS can be seen in another example spectrum shown in figure \ref{fig:swp-mwc120}. In this measurement, the change in q is much more broad with a decrease on the blue side of the line and an increase at the emissive peak. There is a small but significant change in u across the emission peak as well. In the qu-plot of figure \ref{fig:swp-mwc120} there is a complicated structure. Simplistically, it shows that both q and u decrease then increase together, shown by the flattened elongation in qu-space, but there is no single dominating change at any one wavelength.
	
	Even though there is polarization across a broad range of wavelengths at a low amplitude, the red-shifted side of the line profile shows significant deviation from continuum while the blue does not. Also, the difference between the absorptive changes in q and u are significantly different. Stokes u decreases in emission and then decreases further in absorption. Also, the broad changes in q and u are also out of phase, u being more on the red side and q being more blue.

 \subsection{HD 50138 - MWC 158} 

\begin{figure}
\begin{center}
\includegraphics[width=0.75\linewidth, angle=90]{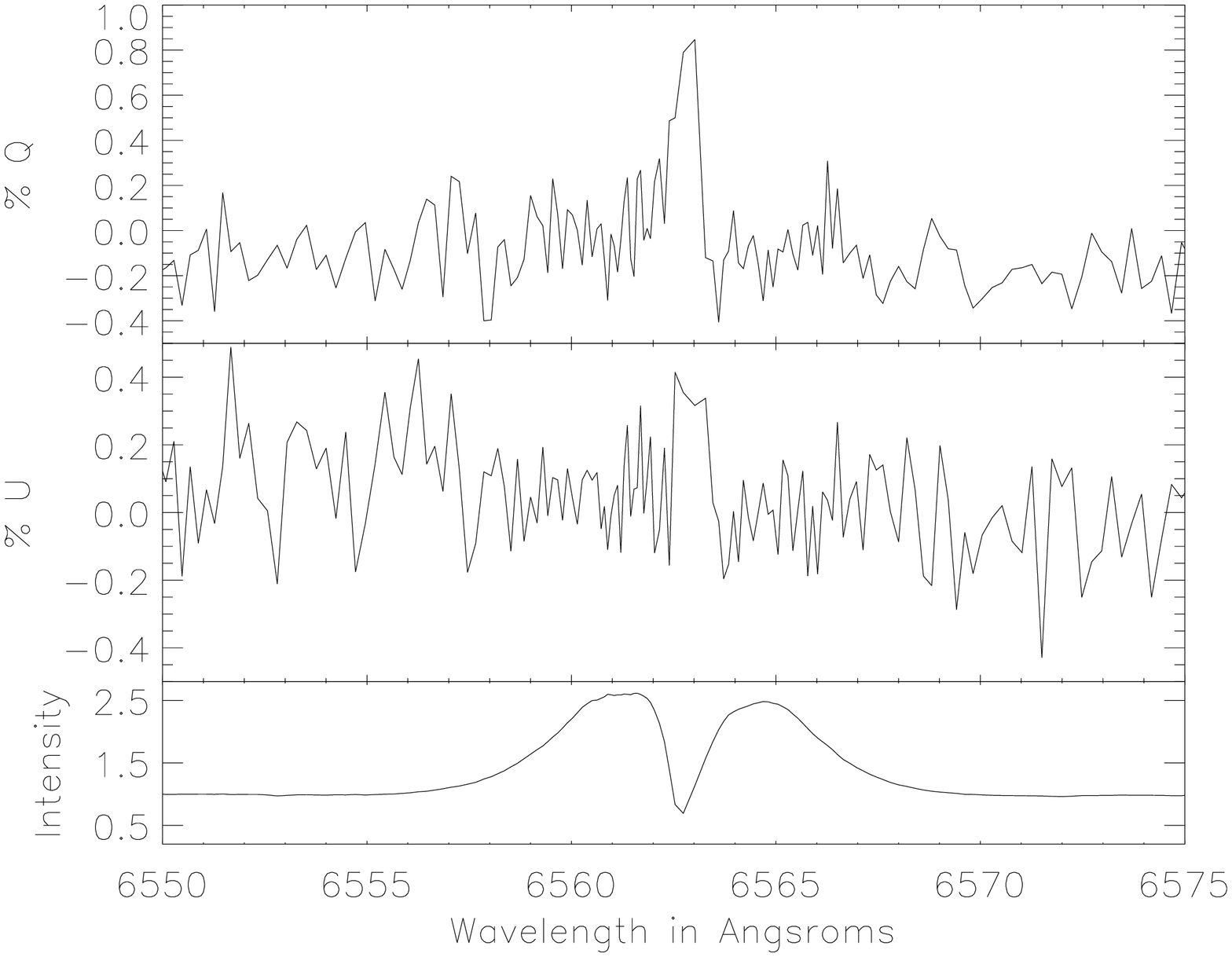}
\includegraphics[width=0.75\linewidth, angle=90]{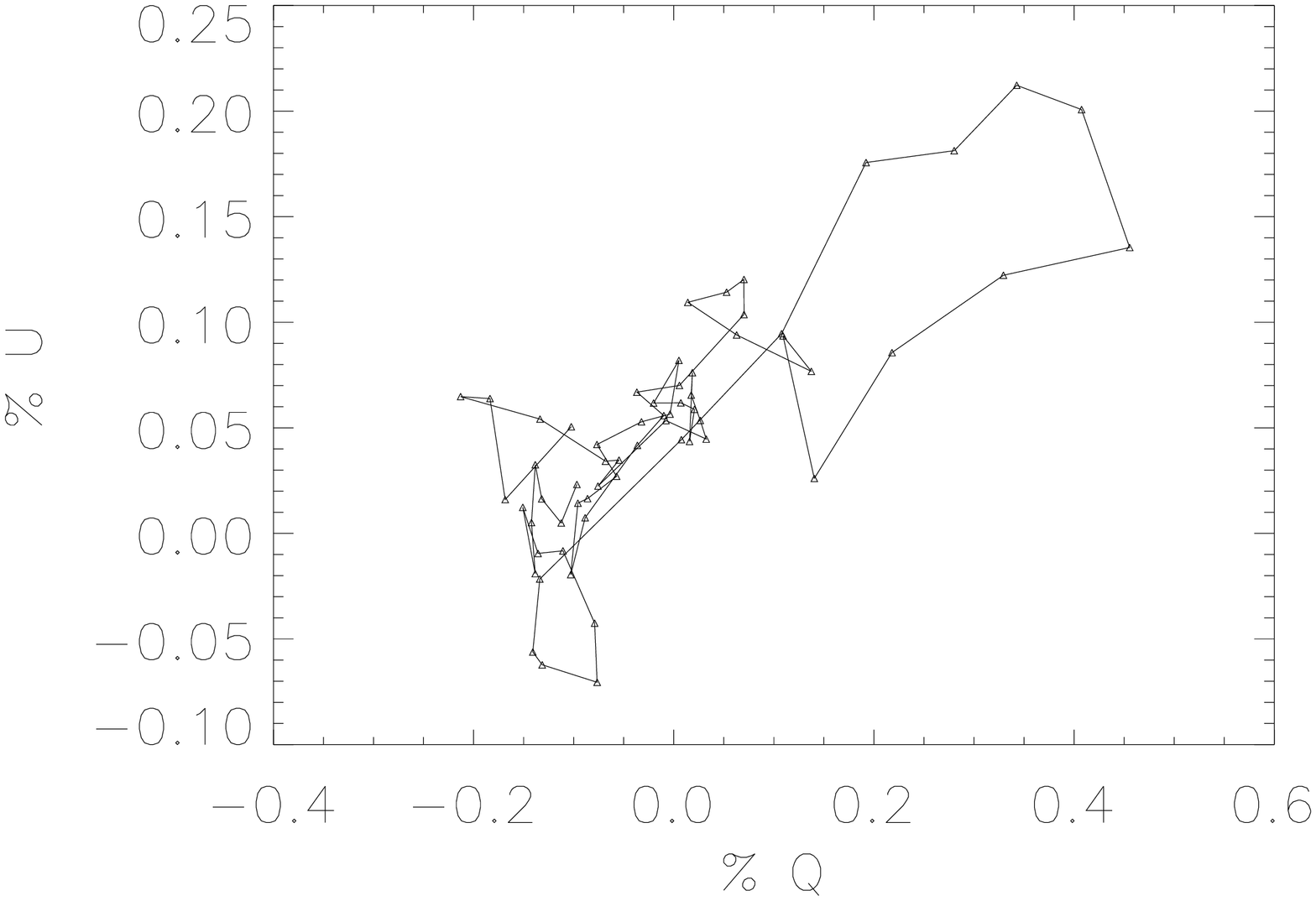}
\caption{An individual example of HD 58647 spectropolarimetry. From left to right {\bf a)} An example polarized spectrum for the HD 58647 H$_\alpha$ line. The spectra have been binned to 5-times continuum. The top panel shows Stokes q, the middle panel shows Stokes u and the bottom panel shows the associated normalized H$_\alpha$ line. There is clearly a detection in the central absorption of 0.8\% in q and 0.4\% in u. {\bf b)} This shows q vs u from 6557.9{\AA} to 6565.3{\AA}.  The knot of points at (0.0,0.0) represents the continuum.}
\label{fig:swp-hd}
\end{center}
\end{figure}

\begin{figure*}
\begin{center}
\includegraphics[width=0.35\linewidth, angle=90]{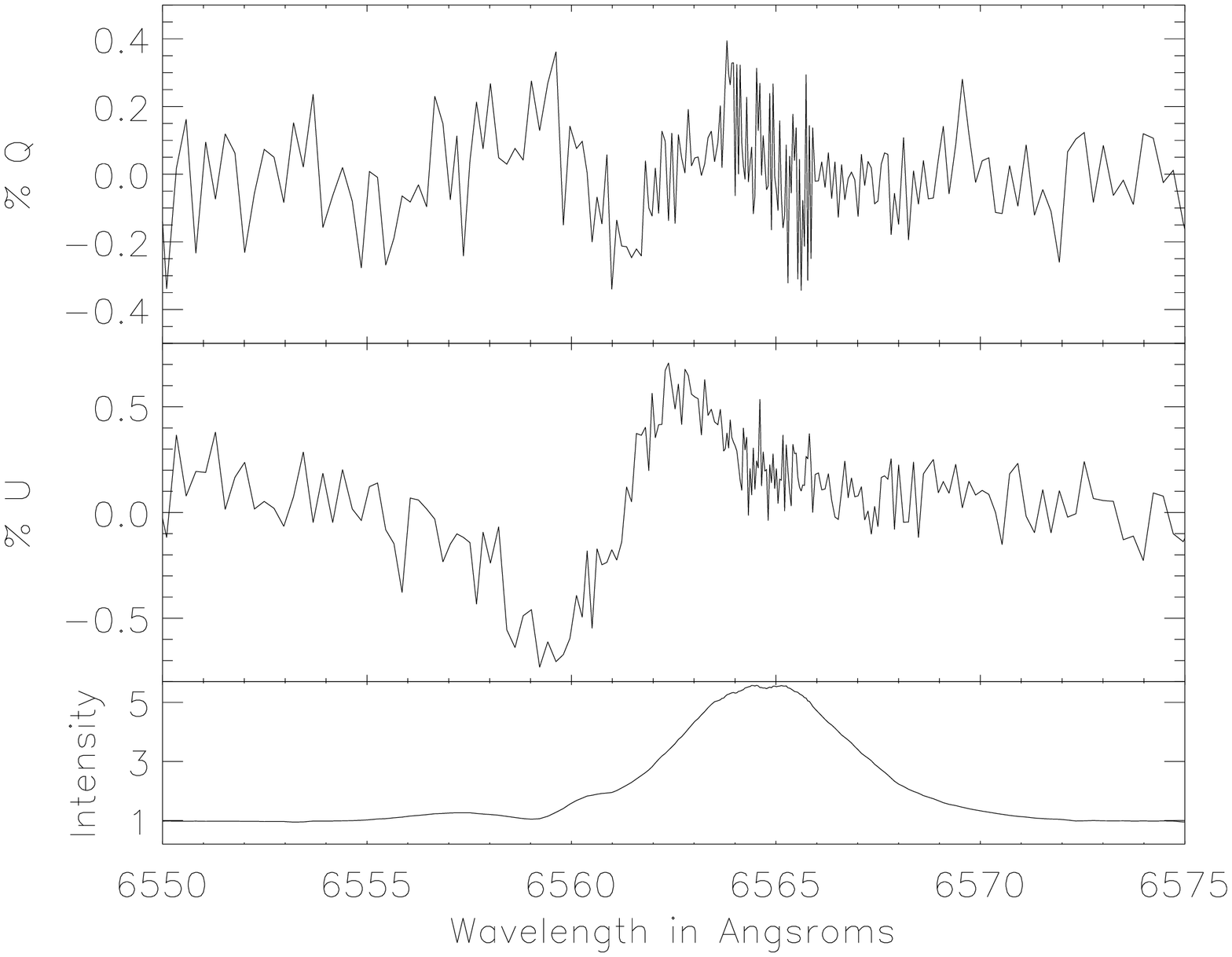}
\includegraphics[width=0.35\linewidth, angle=90]{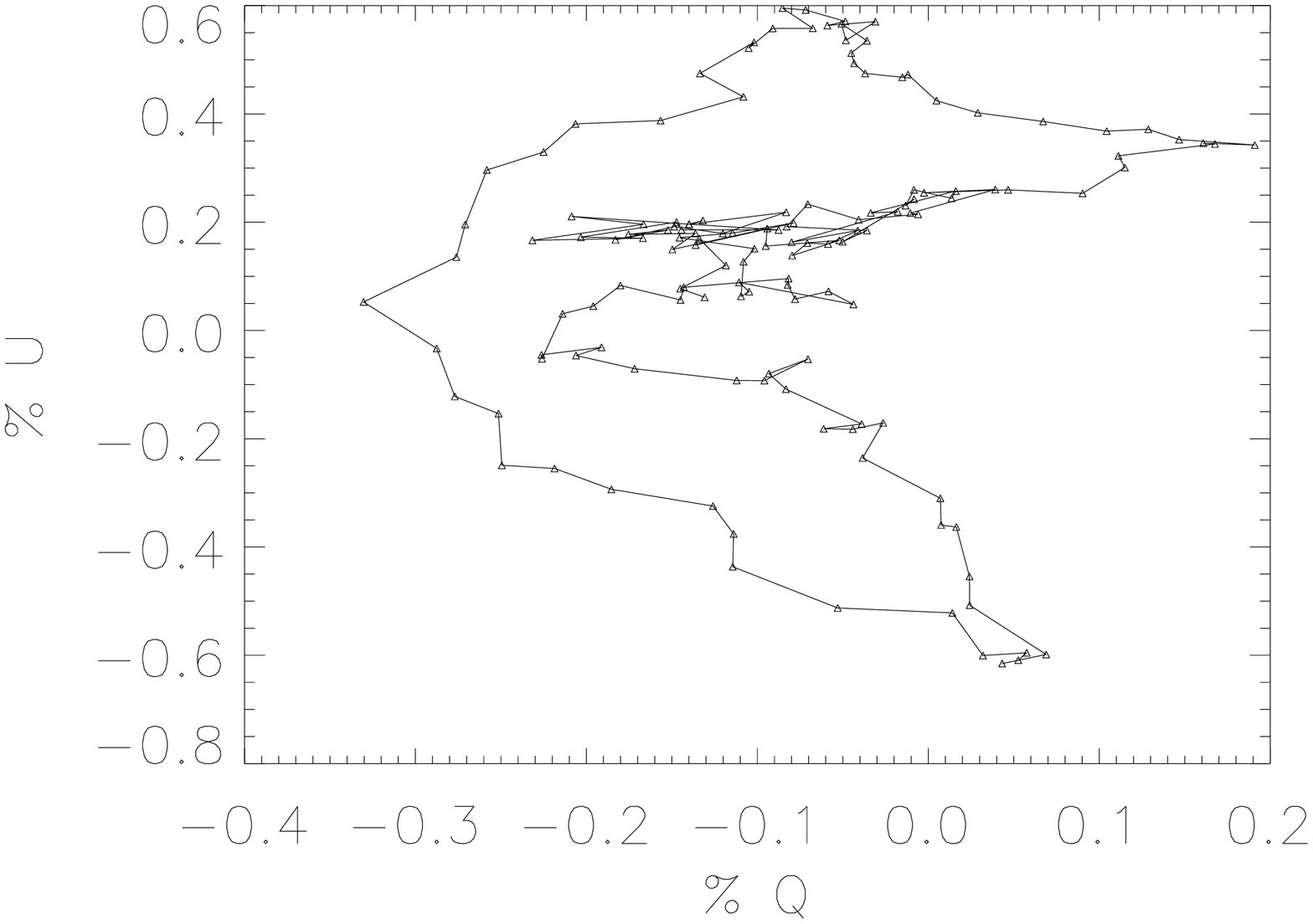} \\
\includegraphics[width=0.35\linewidth, angle=90]{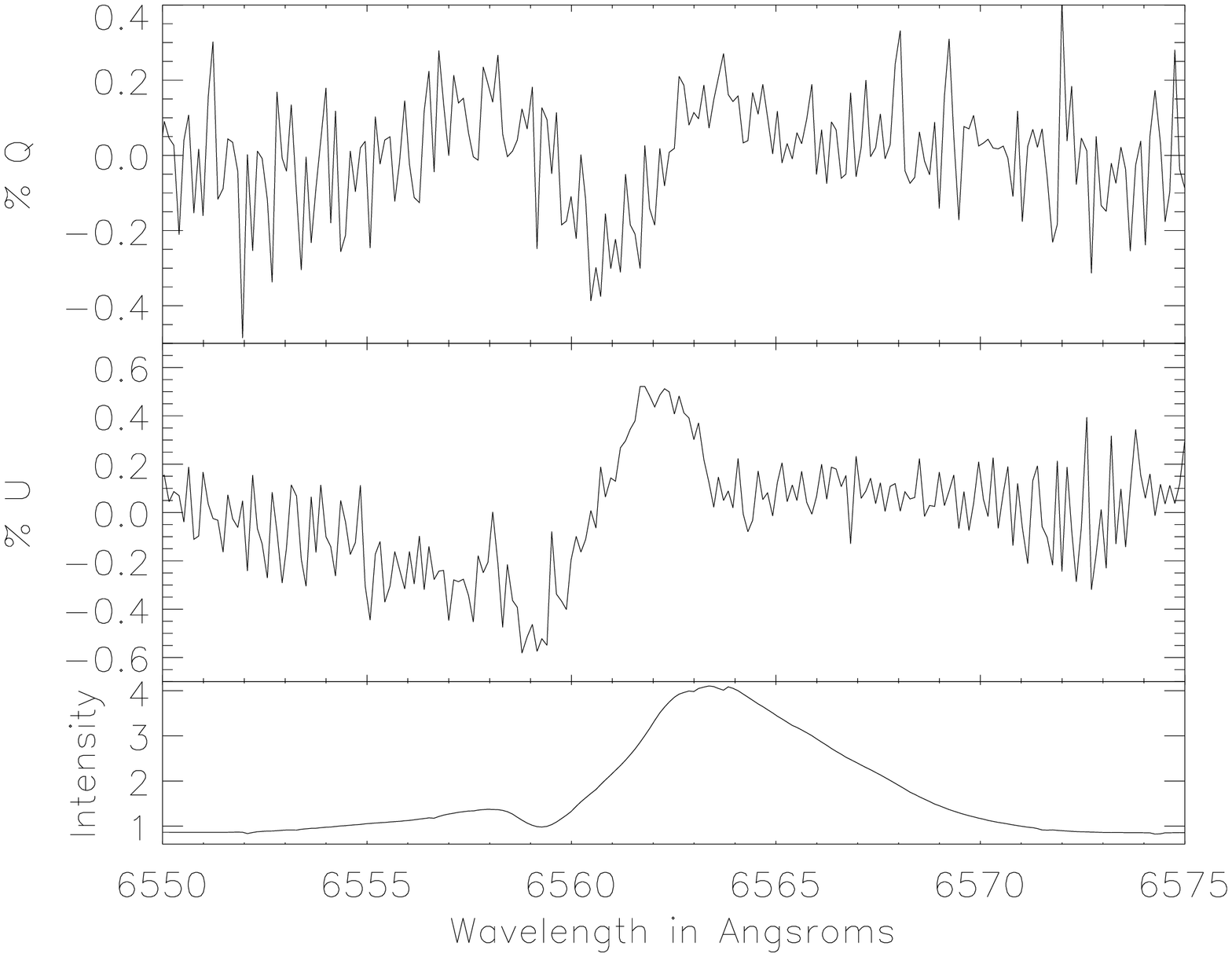}
\includegraphics[width=0.35\linewidth, angle=90]{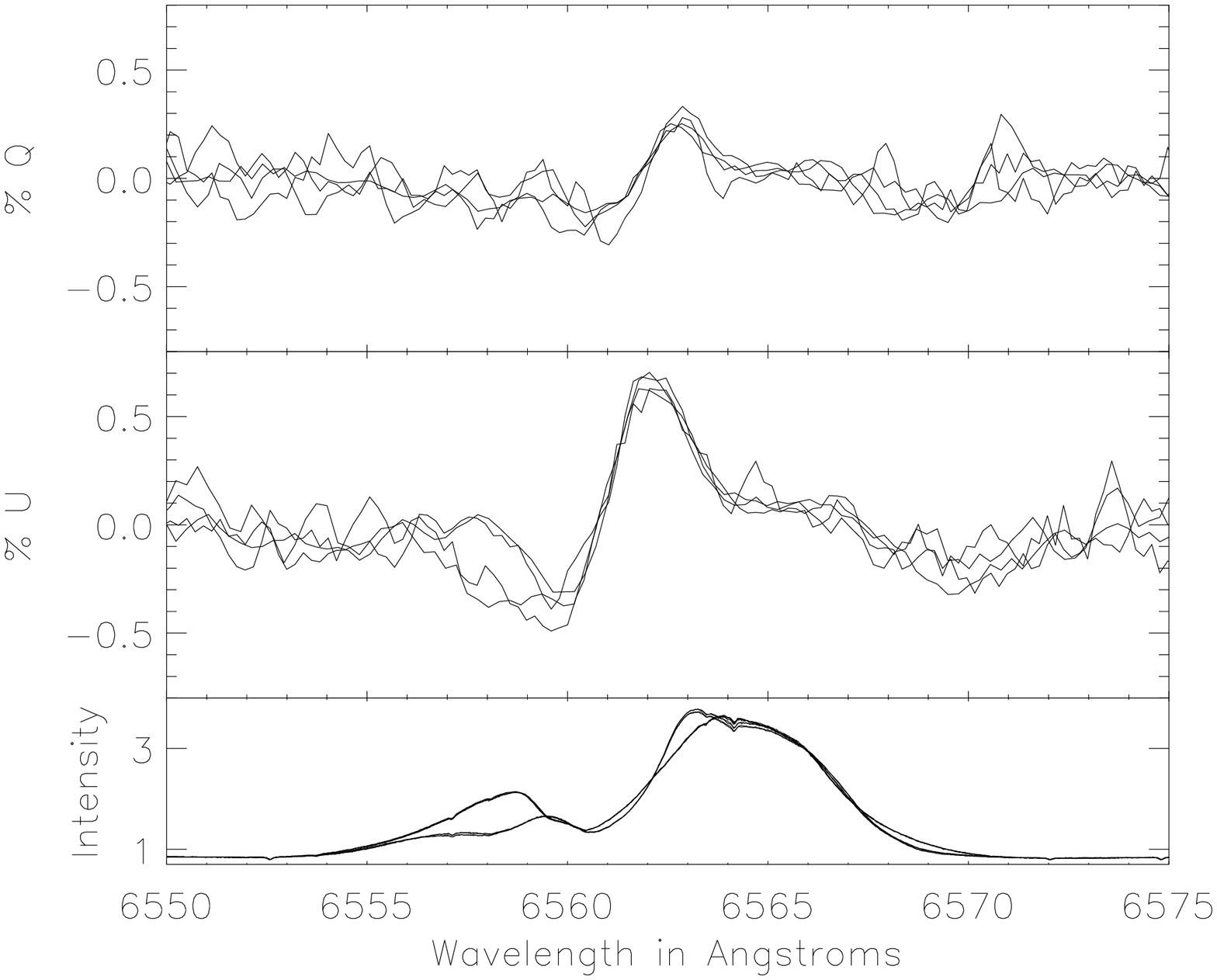}
\vspace{8mm}
\caption{The spectropolarimetry for our HD 163296 ESPaDOnS observations. From left to right {\bf a)} An example polarized spectrum for the HD 163296 H$_\alpha$ line. The spectra have been binned to 5-times continuum. The top panel shows Stokes q, the middle panel shows Stokes u and the bottom panel shows the associated normalized H$_\alpha$ line. There is clearly a detection in the blue-shifted absorption of a complex 0.2\% signature in q and a 0.7\% antisymmetric signature in u. {\bf b)} This shows q vs u from 6549.9{\AA} to 6565.8{\AA}.  The knot of points at (0.0,0.0) represents the continuum. {\bf c)} Shows archive ESPaDOnS data from August 13th 2006 again with a complex 0.2\% q and antisymmetric 0.5\% u. {\bf d)} shows the HiVIS observations on June 23 and 24th 2007.}
\vspace{8mm}
\label{fig:swp-hd163-esp}
\end{center}
\end{figure*}

\begin{figure*}
\begin{center}
\includegraphics[ width=0.35\linewidth, angle=90]{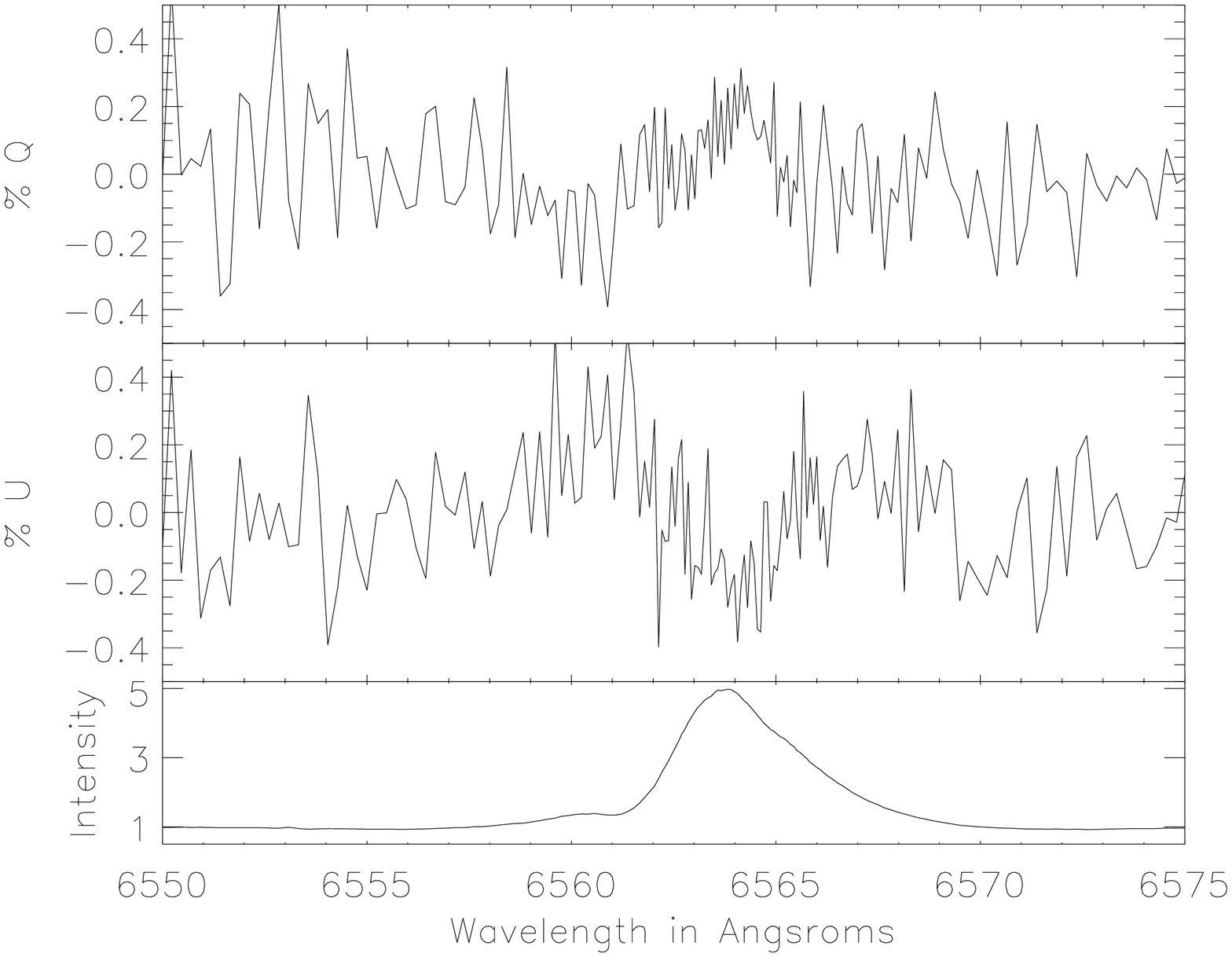}
\includegraphics[ width=0.35\linewidth, angle=90]{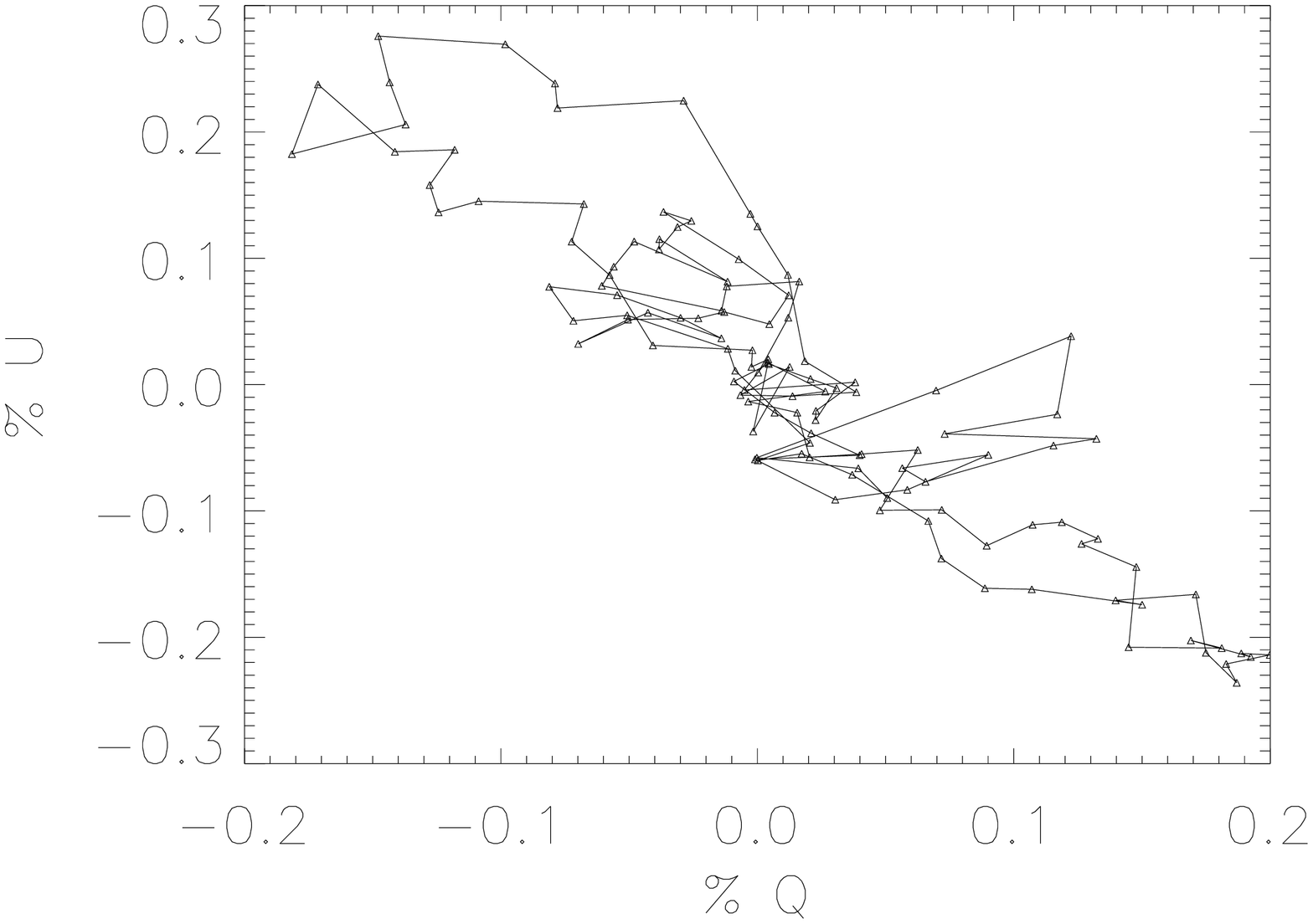} \\
\includegraphics[ width=0.35\linewidth, angle=90]{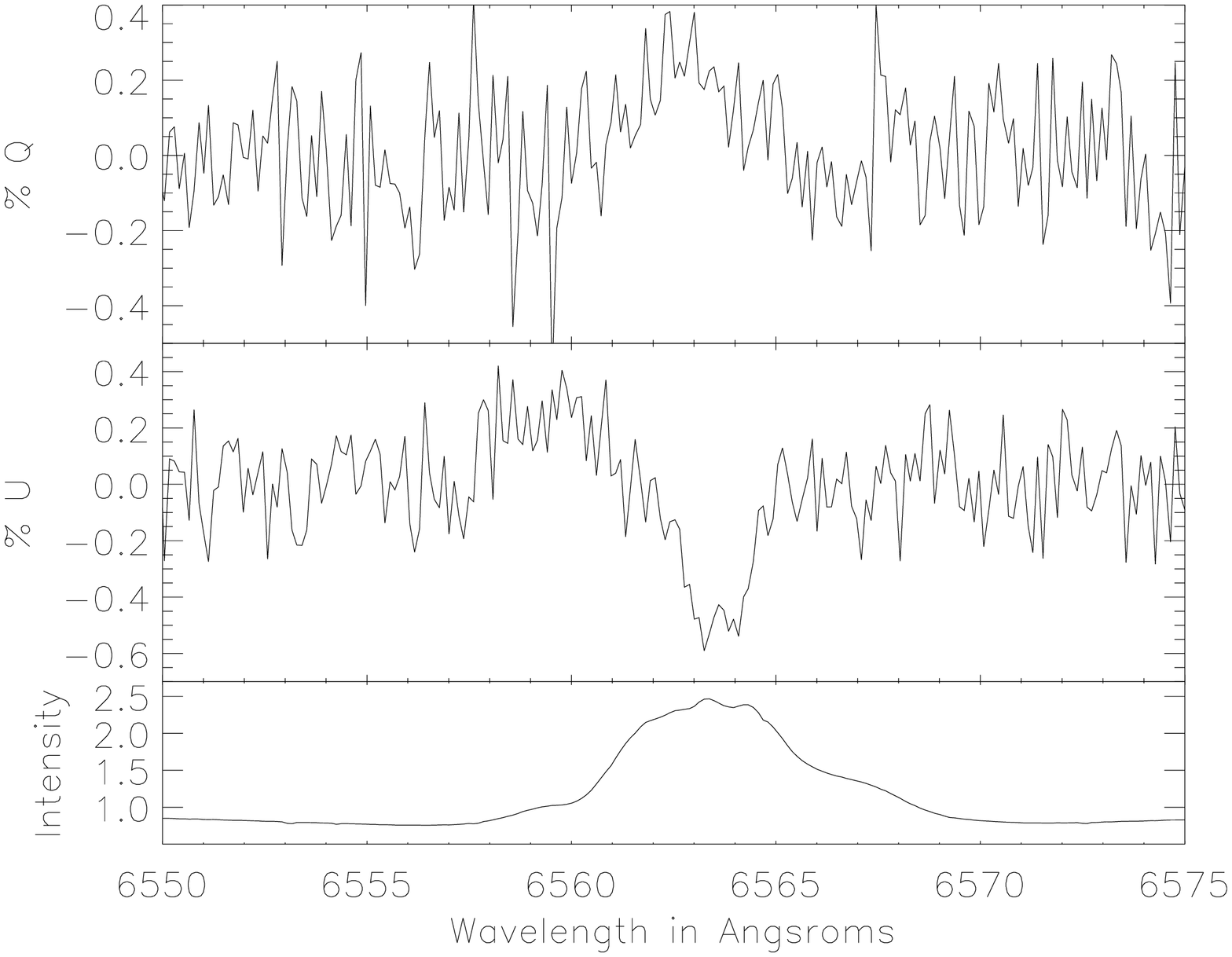}
\includegraphics[ width=0.35\linewidth, angle=90]{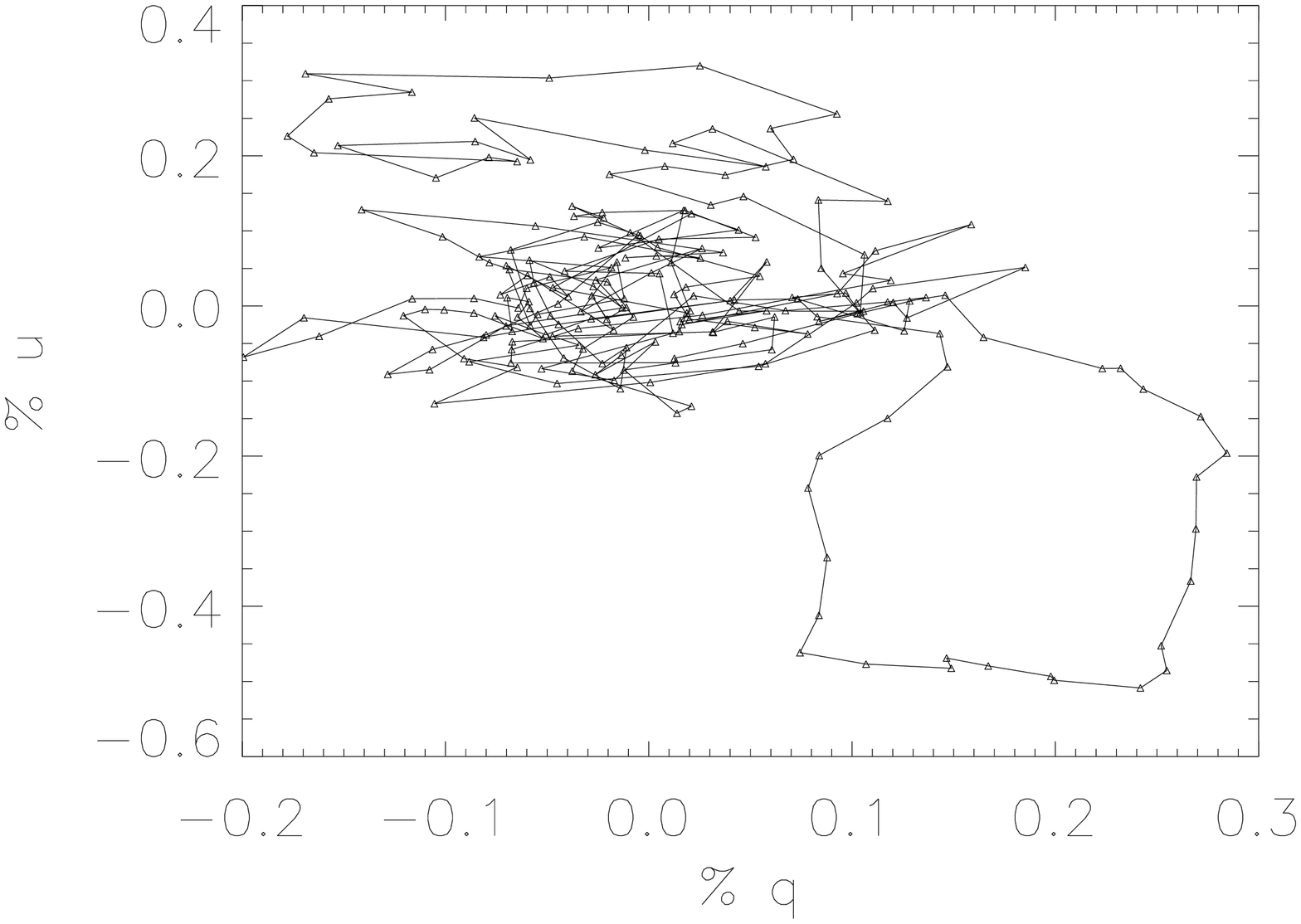}
\vspace{8mm}
\caption{The spectropolarimetry for our HD 179218 ESPaDOnS observations. From let to right: {\bf a)} An example polarized spectrum for the HD 179218 H$_\alpha$ line. The spectra have been binned to 5-times continuum. The top panel shows Stokes q, the middle panel shows Stokes u and the bottom panel shows the associated normalized H$_\alpha$ line. There is clearly a detection across the entire line of 0.2\% in q and u. {\bf b)} This shows q vs u from 6551.9{\AA} to 6568.9{\AA}.  The knot of points at (0.0,0.0) represents the continuum. {\bf c)} This shows ESPaDOnS spectropolarimetry taken on March 21st 2008. {\bf d)} This shows the corresponding qu-loop.}
\vspace{8mm}
\label{fig:swp-hd179}
\end{center}
\end{figure*}

	  H$_\alpha$ line spectropolarimetry from 1995 and 1996 was presented by Oudmaijer et al. (1999). The line profiles they observed were similar to the HiVIS observations and a line effect was detected in both observations. Their Jan. 1995 observtions showed a 0.4\% decrease in polarization across the red emission peak while the Dec. 1996 observations showed a depolarization of the same magnitude but with more complex form. They report continuum polarizations of 0.71\% and 0.65\% respectively. Bjorkman et al. 1998 found a continuum polarization of 0.7\% at 160$^\circ$ with an ``almost flat" wavelength dependence from 4000 to 9000{\AA}. From the flatness they concluded that electron scattering, not dust scattering is the polarizing mechanism. They found significant polarization in nearby stars of 0.1\% to 0.8\% and concluded that the nearby environment was complex and variable. Using a number of methods, including an estimate based on the 'depolarization' mechanism, they conclude that the inter-stellar polarization towards this star is 0.2\%. Vink et al. 2002 report a double-peaked line profile with complex signature with a 0.3\% amplitude on a continuum polarization of 0.65\%.    

	This star had a very strong change in polarization across the central absorptive component of the line. There was also a much smaller change in polarization seen occasionally across the one or both of the emission peaks. Figure \ref{fig:haebe-specpol2} shows 35 HiVIS measurements and there is a very clear change in the line center detected in nearly every data set. Figure \ref{fig:swp-mwc158-esp} shows one such example. There is a 1.5\% change in both q and u in the central absorptive component, even thought the absorption does not go below continuum. The qu-plot of the figure shows a very strong loop indicating that the change in q and u are not in phase (not at the same wavelengths) and that there is a significant asymmetric component to the Stokes u measurement. First q increases with a small decrease in u, then u increases strongly. There are ESPaDOnS archival observations from February 9th 2006 that show this effect very clearly. In this observation set, q is strongly asymmetric with a 0.5\% rise followed by a 0.8\% drop in the central absorption. Stokes u shows a strong decrease of 1.3\% that is centered on the red side of the central absorption. In both ESPaDOnS observations, there is also a small, broad decrease in both q and u of roughly 0.3\%, but only on the red side of the line. The polarization on the blue-shifted emission is identical to that of continuum. The qu-loops are dominated by only the central narrow wavelength range.

\subsection{HD 58647} 
 	
	Vink et al. 2002 report a 0.6\% increase in polarization in a very narrow range  (single resolution element) in the central absorption on a continuum of 0.1\%. This star is very similar to MWC 158 in that there is a large clear polarization in the central absorption trough, and a smaller change seen in the double-peaked emission. Figure \ref{fig:haebe-specpol2} shows 22 measurements where this can be seen. The magnitude of the signature is smaller, at best 1\% with 0.5\% being more typical. However, the intensity of the line is more than 4 times smaller than MWC 158. Figure \ref{fig:swp-hd} shows a typical spectrum - a 0.8\% change in q and a 0.4\% change in u. However, in this star the change in q and u are in phase producing a linear extension in the qu-plot of the figure.
	  
	Oudmaijer et al. 2001 presented multi-wavelength polarimetry as evidence for a disk-like structure in the ionized gas surrounding the object. The polarization was nearly flat with wavelength near 0.15\% but had U and I-band excesses of 0.05-0.1\%. Since dust alone would not produce both U and I polarization excesses, they conclude that the Serkowski-misfitting would not imply a dusty disk.  
		
	In the compilation plot of figure \ref{fig:haebe-specpol2}, the Stokes u plot shows a clear $\sim$ 0.2\% asymmetric signature in many of the observations.  This asymmetric u signature is also confined to the central absorption and the polarization at the red and blue emission peaks is identical to continuum within the noise.  An individual HiVIS example is shown in figure \ref{fig:swp-hd}.

\subsection{HD 163296 - MWC 275}
	  
	  There are several continuum polarization measurements of this star that show variability. Beskrovnaya et al. 1998 report R-band polarizations of roughly 0.3\% with a 7.5 day cyclic variation of 0.1\% in the July 17-30 1995 period. However, Oudmaijer et al. 2001 found a non-variable V-band polarization of only 0.02\% on 7 observations over the 1998-1999 period.
	  
	  This star had a very clear change in polarization in the blue side of the line profile, but the change extended all the way to the center of the emission. Figure \ref{fig:haebe-specpol1} shows 24 measurements all with similar form in q and u. A typical example is shown in figure \ref{fig:swp-hd163-esp}. The q spectrum shows a small increase in the blue shifted absorption as well as one centered on the emission. The u spectrum shows a very clear decrease then increase with the polarization returning to continuum by the emissive peak. The plot in qu-space is very complex. The two increases in q manifest themselves as the two right-most excursions while the u spectrum gives the vertical extent. The polarized flux also shows a very strong clear change. 
	  
	The asymmetry of the line shows the strong absorption on the blue-shifted side of the line profile. The polarization on the red side of the emission line is essentially the continuum value. The polarization on the blue-shifted side of the line profile shows a very strange morphology in both the qu-space and the polarized spectra. This star has ample evidence for stellar winds (out-flow) from other studies. In both HiVIS and ESPaDOnS archive observations, on the far blue component, q increases while u decreases. Moving towards the red, the Stokes u signature reverses sign while q returns to zero then increases without changing sign.

\subsection{HD 179218 - MWC 614}
	
	This star shows a small change in polarization extending from the blue-shifted absorption across the entire emission line. The change small, typically 0.3\% to 0.5\%. Figure \ref{fig:haebe-specpol1} shows 25 observations with the change clearest in the Stokes u spectra. An example spectrum is shown in figure \ref{fig:swp-hd179}. There is a 0.3\% increase in u with a 0.2\% decrease in q in the blue-shifted absorption. This then reverses to a 0.3\% decrease in u and a 0.2\% increase in q centered on the emissive peak. In the qu-plot of the figure this shows up as two linear extensions away from the continuum-knot at (0,0).
	
	Though this star does show evidence for significant line absorption, this spectropolarimetric signature is centered on the emissive component of the line and is as wide as the emission. The q and u signatures are in-phase, showing up as linear extensions in qu-space. In the blue-shifted absorption, the q and u effects are also phased and in the opposite direction as the changes across the emission peak.

\subsection{HD 150193 - MWC 863}

\begin{figure}
\begin{center}
\includegraphics[ width=0.75\linewidth, angle=90]{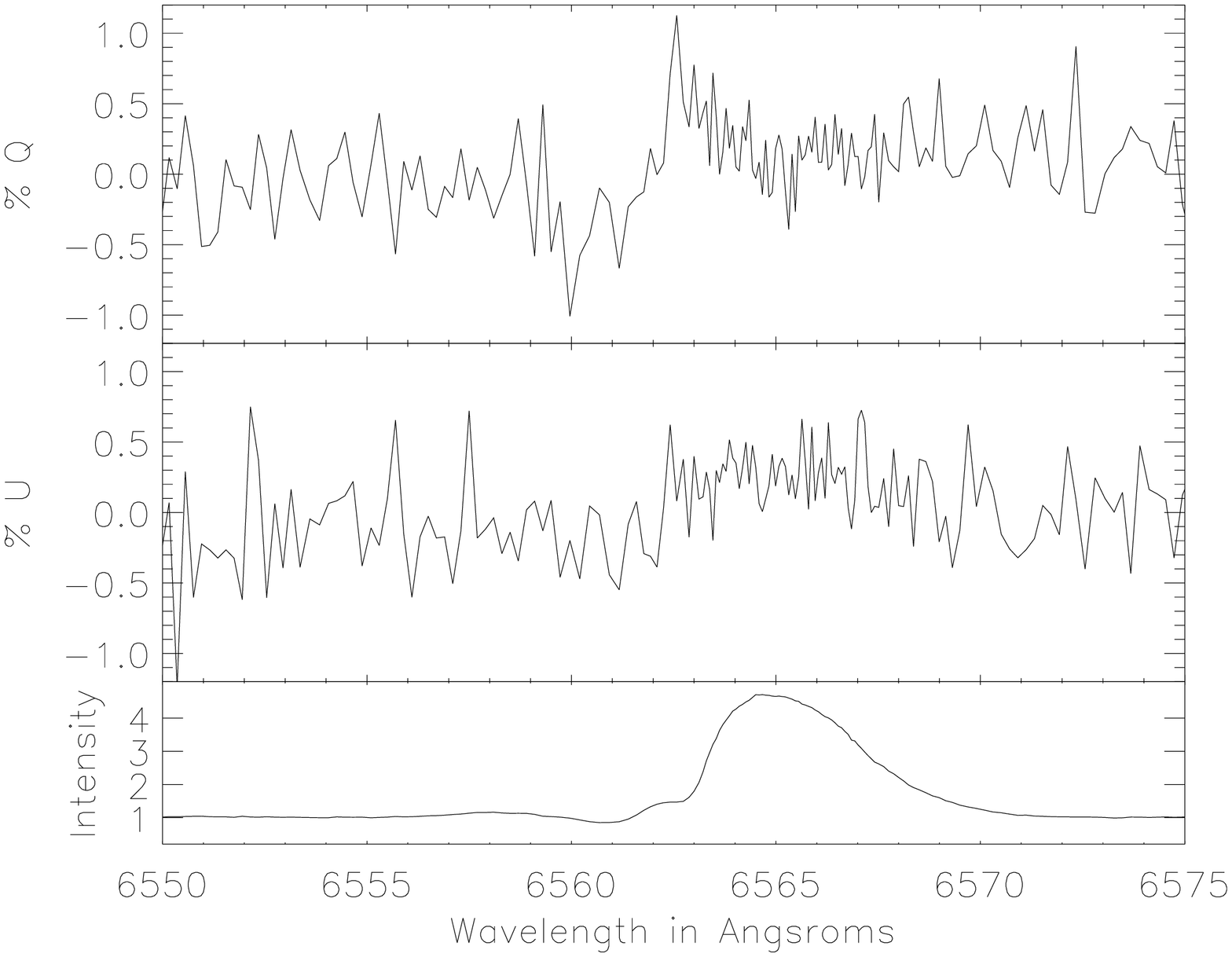} \\
\includegraphics[ width=0.75\linewidth, angle=90]{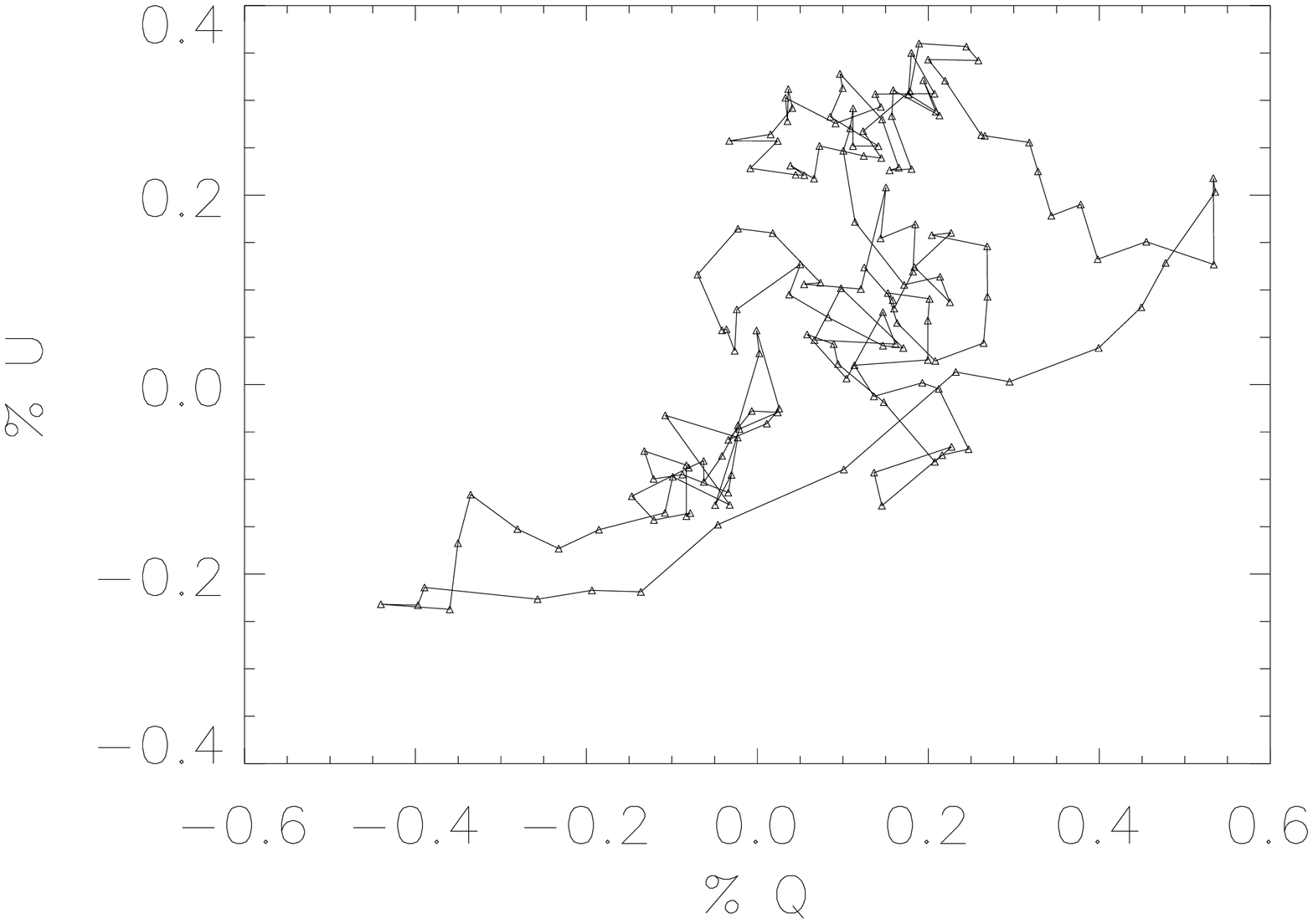} 
\caption{An example of HD 150193 spectropolarimetry. {\bf a)} shows an individual example polarized spectrum for the HD 150193 H$_\alpha$ line. The spectra have been binned to 5-times continuum. The top panel shows Stokes q, the middle panel shows Stokes u and the bottom panel shows the associated normalized H$_\alpha$ line. There is clearly a detection in the blue-shifted absorption of $\pm$0.5\% in q. {\bf b)} This shows q vs u from 6553.1{\AA} to 6578.3{\AA}. The knot of points at (0.0,0.0) represents the continuum.}
\label{fig:swp-hd150}
\end{center}
\end{figure}

	This star shows a small but significant detection extending from the blue shifted absorption to the emission profile in the 12 measurements of figure \ref{fig:haebe-specpol1}. This can be seen as a small deviation in Stokes u. This star is a bit noisier than many, but the signature is seen clearly in both q and u as the strong change on the blue side of the emission peak. An example of this signature is shown in figure \ref{fig:swp-hd150}. There is a clear change in q of 0.5\% first decreasing in the blue-shifted absorption and then increasing on the blue side of the emission peak. The u spectrum shows a much more broad change across the emission peak. Chavero et al. 2006 measured a R-band polarization of 5.0$\pm$0.5\% and show that the polarization deviates significantly from the traditional Serkowski law. This star had previously been identified as a polarized standard in Whittet et al. 1992 with an R-band polarization of 5.19$\pm$0.05\%.

	 There is a marginal broad effect seen in the HiVIS data of figure \ref{fig:swp-hd150} over the emission in Stokes u. The signature in Stokes q is $\sim$0.8\%, antisymmetric and blue-shifted. It is negative on the far blue side with a strong, rapid change to positive on the blue edge of the transition between absorption and emission.

\begin{figure*}
\begin{center}
\includegraphics[ width=0.35\linewidth, angle=90]{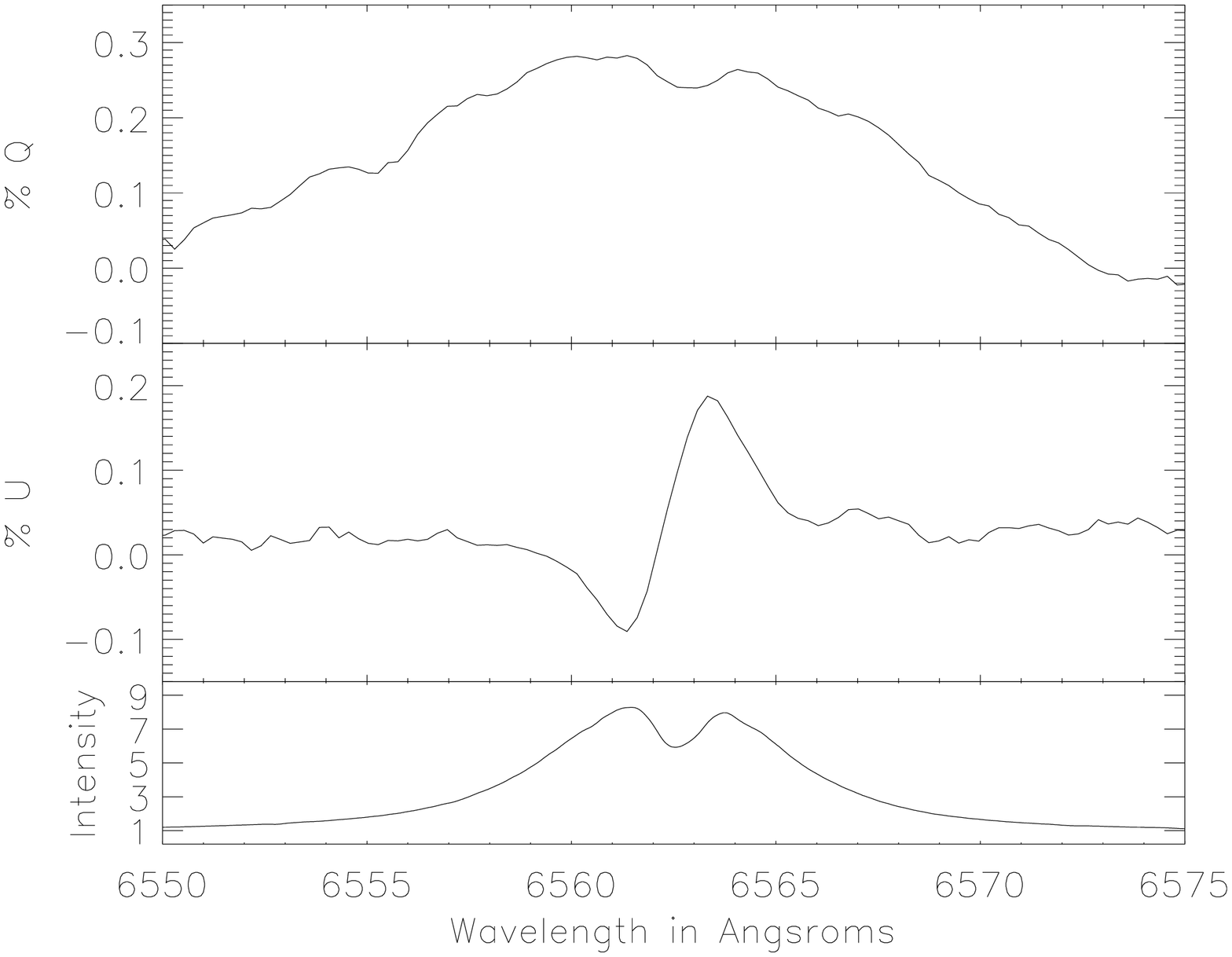}
\includegraphics[ width=0.35\linewidth, angle=90]{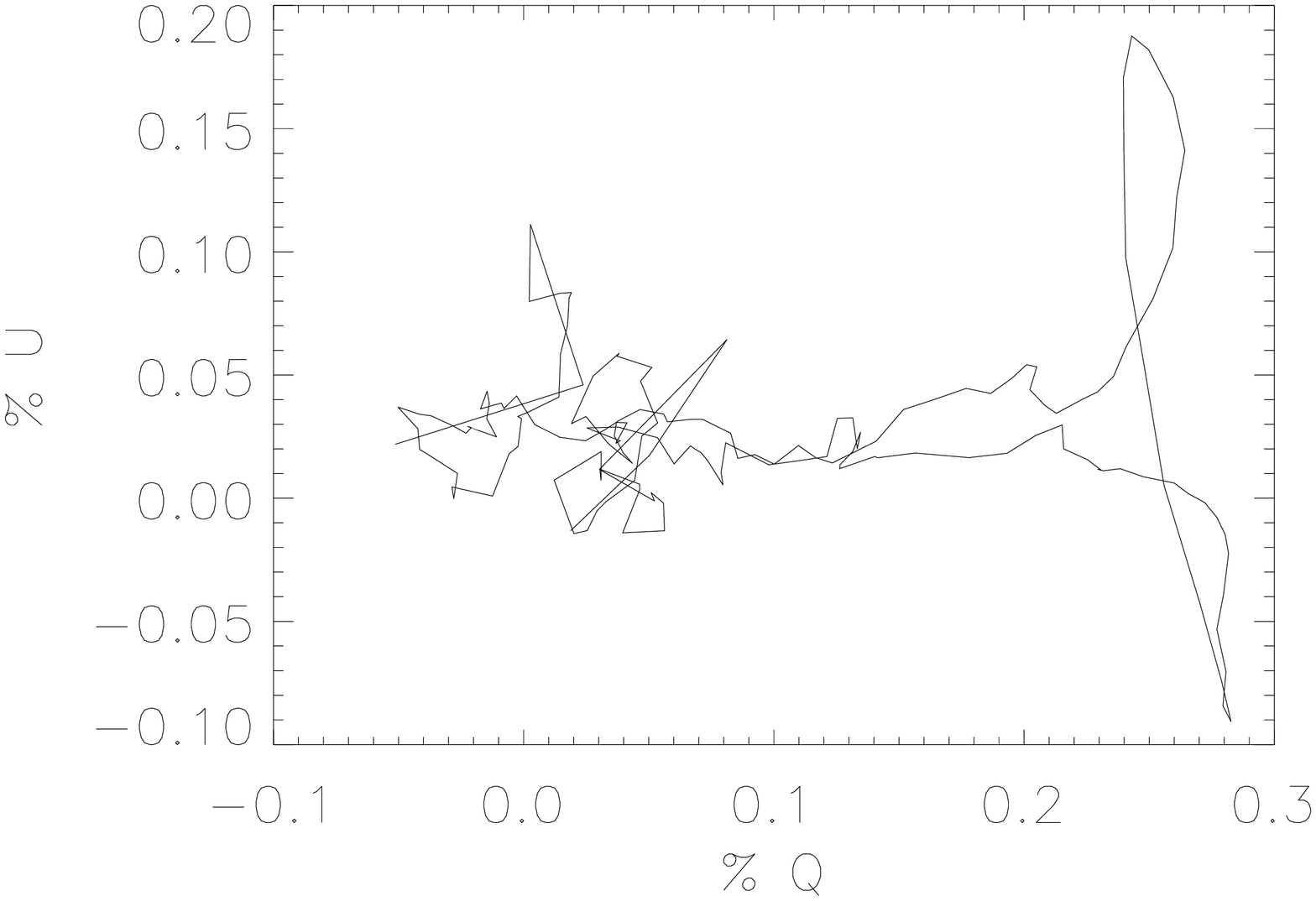} \\
\includegraphics[ width=0.35\linewidth, angle=90]{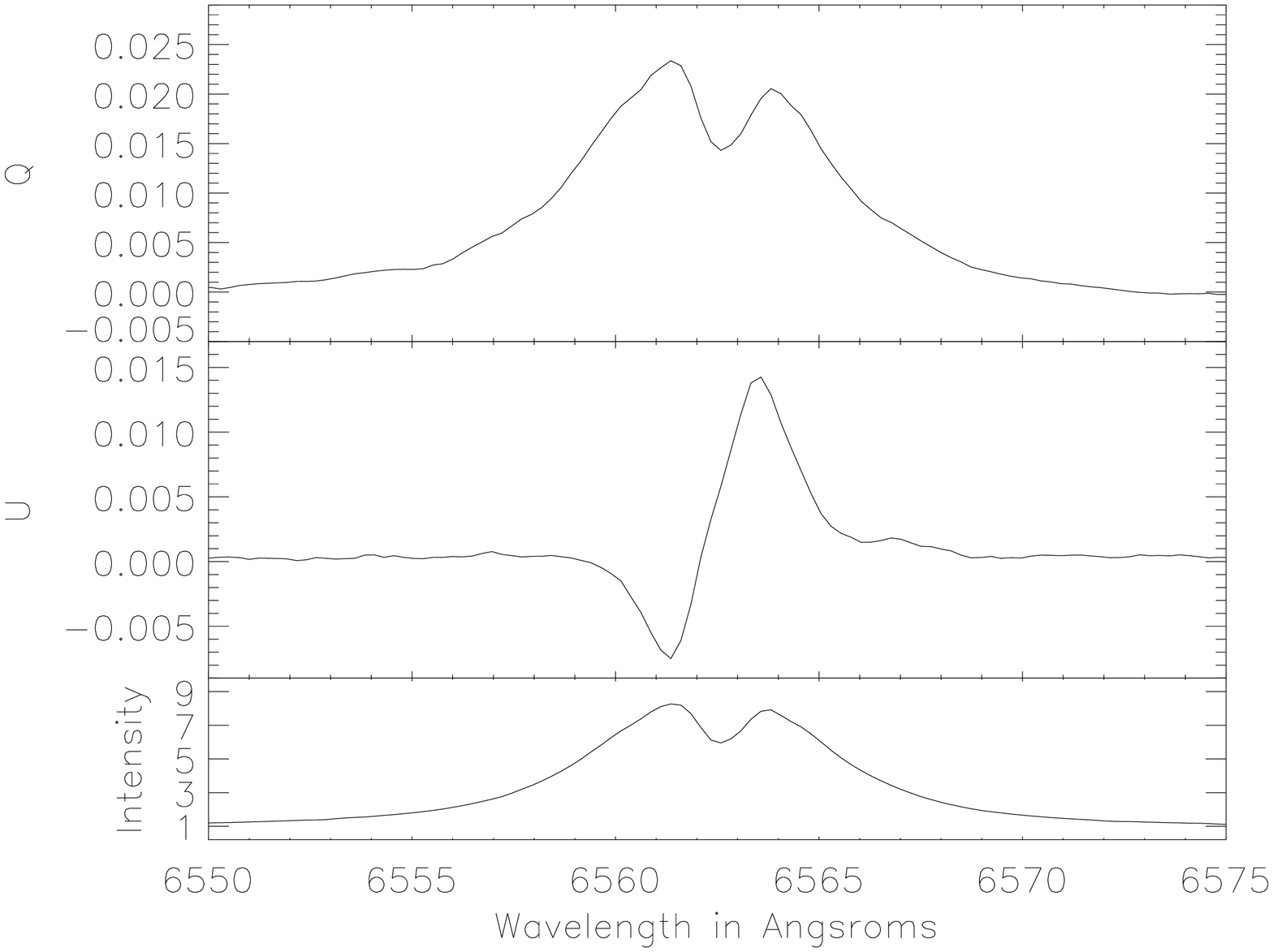}
\includegraphics[ width=0.35\linewidth, angle=90]{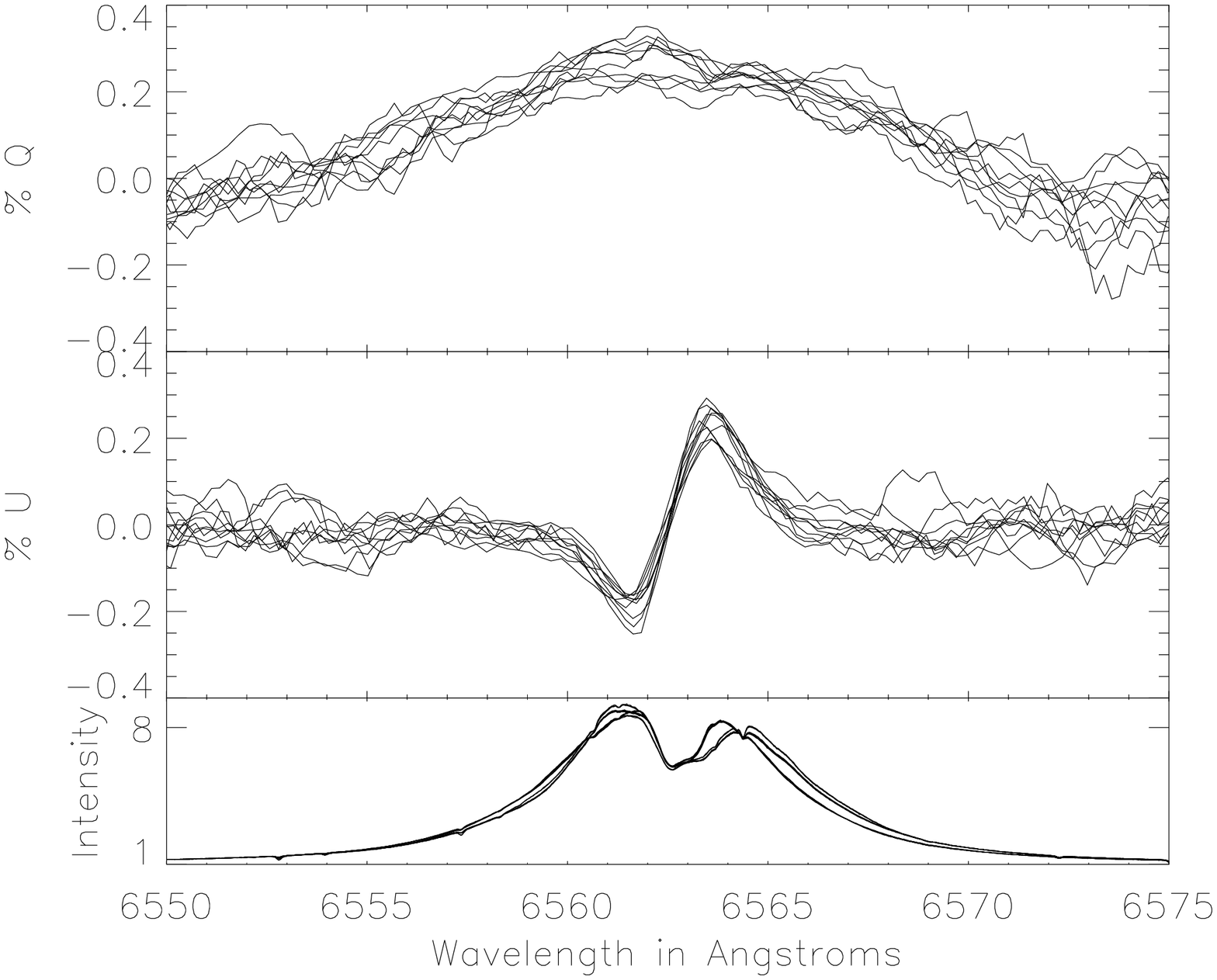}
\vspace{5mm}
\caption{This figure shows MWC 361 in detail. From left to right: {\bf a)} All polarized spectra averaged for the MWC 361 H$_\alpha$ line. The spectra have been individually rotated to a common frame and then averaged. The top panel shows Stokes q, the middle panel shows Stokes u and the bottom panel shows the associated averaged and normalized H$_\alpha$ line. {\bf b)} This shows q vs u from 6550{\AA} to 6575{\AA}. The knot of points at (0.0,0.0) represents the continuum. {\bf c)} The polarized flux q*I for the HiVIS observations. {\bf d)} ESPaDOnS observations for MWC 361 on August 1st and 3rd, 2006 and June 23rd and 24th 2007.}
\label{fig:swp-mwc361}
\end{center}
\end{figure*}

\begin{figure}
\begin{center}
\includegraphics[ width=0.79\linewidth, angle=90]{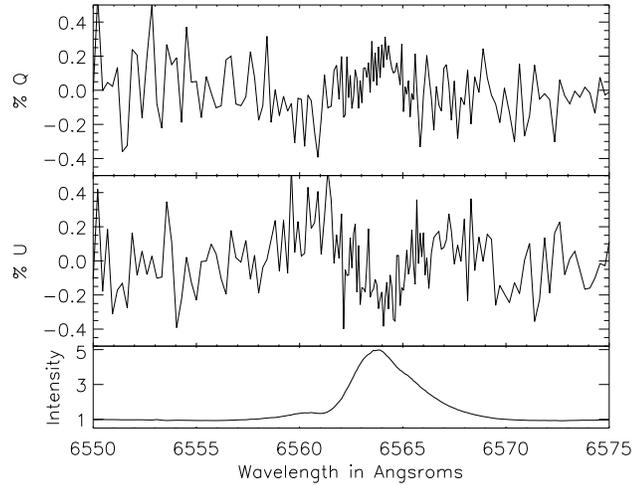}
\includegraphics[ width=0.79\linewidth, angle=90]{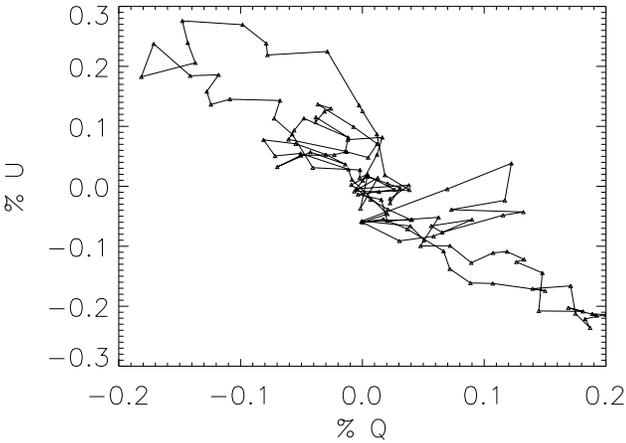} \\
\includegraphics[ width=0.79\linewidth, angle=90]{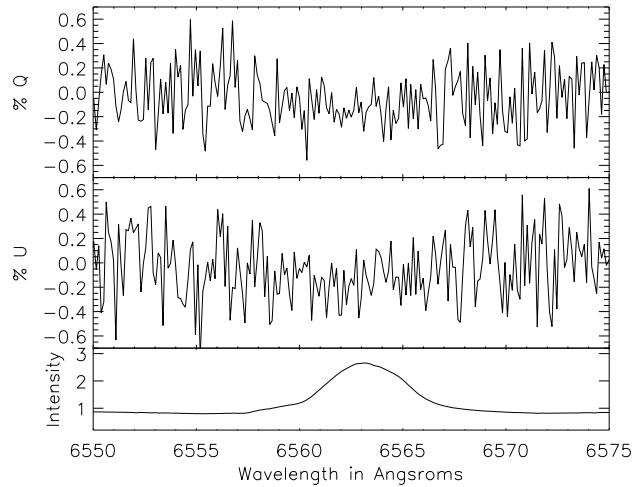}
\caption{This shows an example of MWC 758 spectropolarimetry. From left to right: {\bf a)} An example polarized spectrum for the MWC 758 H$_\alpha$ line. The spectra have been binned to 5-times continuum. The top panel shows Stokes q, the middle panel shows Stokes u and the bottom panel shows the associated normalized H$_\alpha$ line. There is clearly a detection in the blue-shifted absorption through line center of -1.0\% in u. {\bf b)} This shows q vs u from 6553.7{\AA} to 6569.7{\AA}.  The knot of points at (0.0,0.0) represents the continuum. {\bf c)} Archived ESPaDOnS observations for MWC 758 on February 9th, 2006.}
\label{fig:swp-mwc758}
\end{center}
\end{figure}

\begin{figure}
\begin{center}
\includegraphics[ width=0.8\linewidth, angle=90]{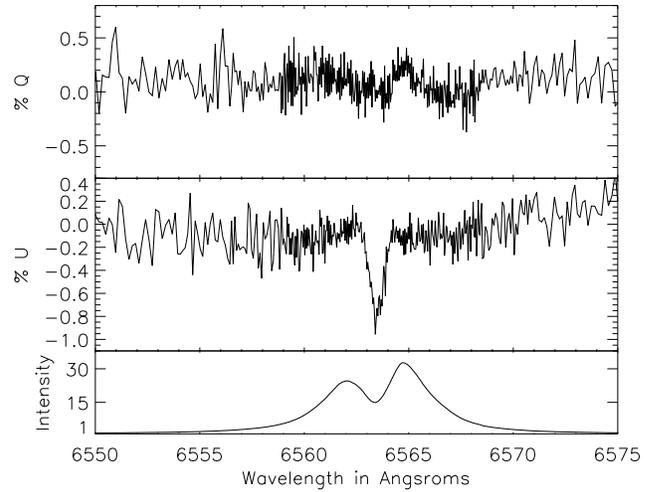} \\
\includegraphics[ width=0.8\linewidth, angle=90]{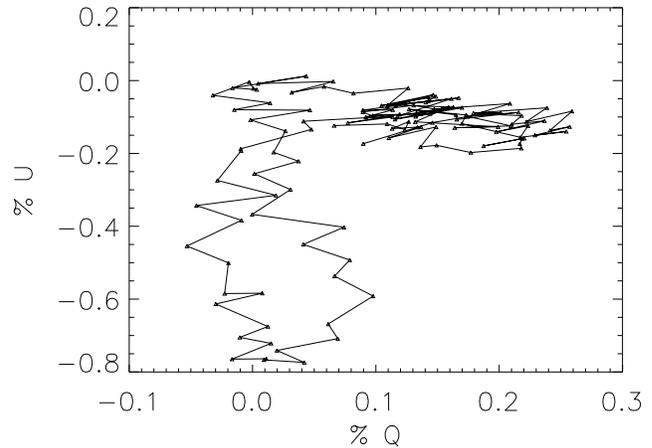} 
\caption{Spectropolarimetry for HD 45677. From left to right: {\bf a)} An example polarized spectrum for the HD 45677 H$_\alpha$ line. The spectra have been binned to 5-times continuum. The top panel shows Stokes q, the middle panel shows Stokes u and the bottom panel shows the associated normalized H$_\alpha$ line. There is clearly a detection of -0.2\% in q and -0.8\% in u. {\bf b)} This shows q vs u from 6560.0{\AA} to 6565.2{\AA}.  The knot of points at (0.0,0.0) represents the continuum.}
\label{fig:swp-hd456}
\end{center}
\end{figure}

\begin{figure*}
\begin{center}
\includegraphics[width=0.35\linewidth, angle=90]{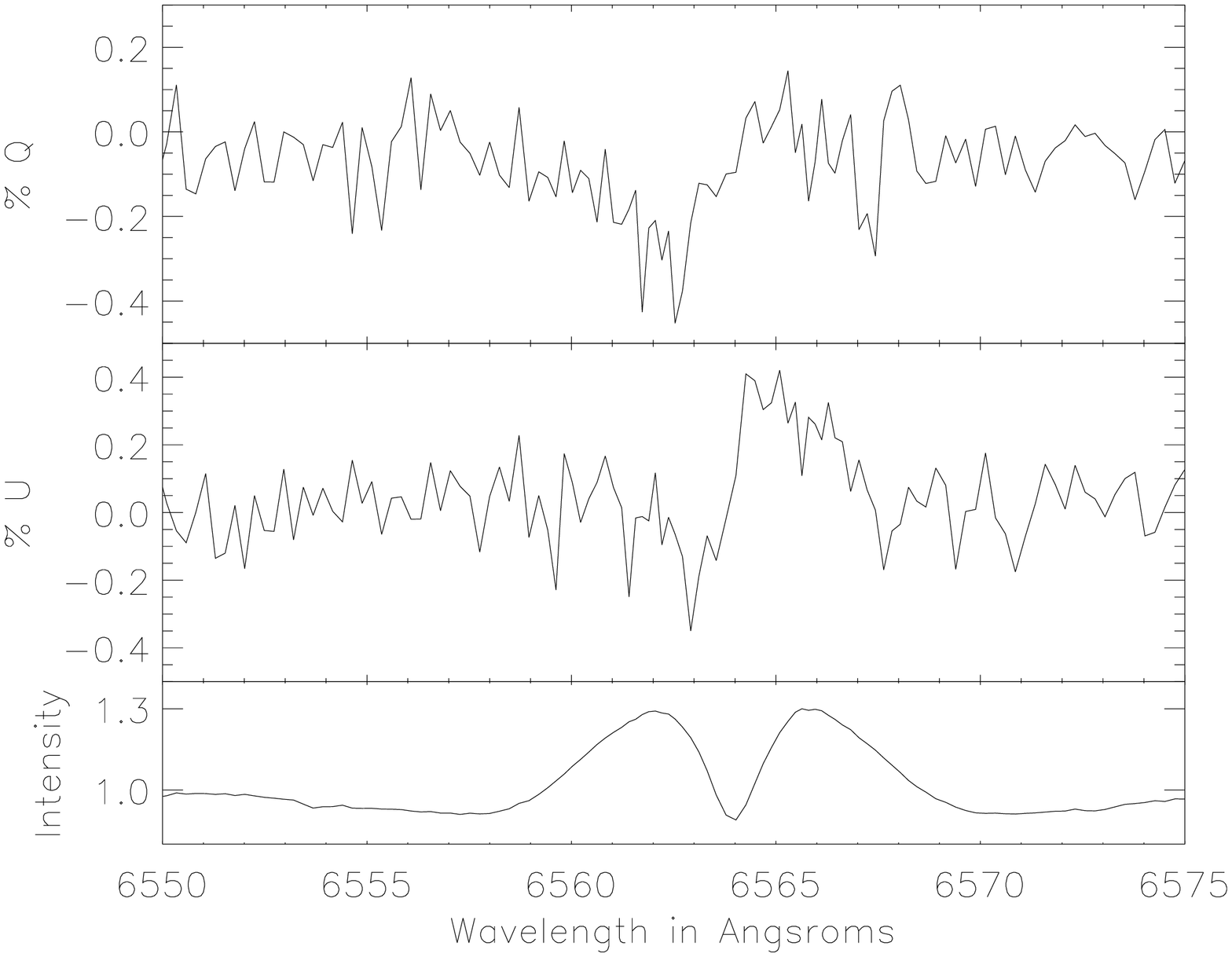}
\includegraphics[width=0.35\linewidth, angle=90]{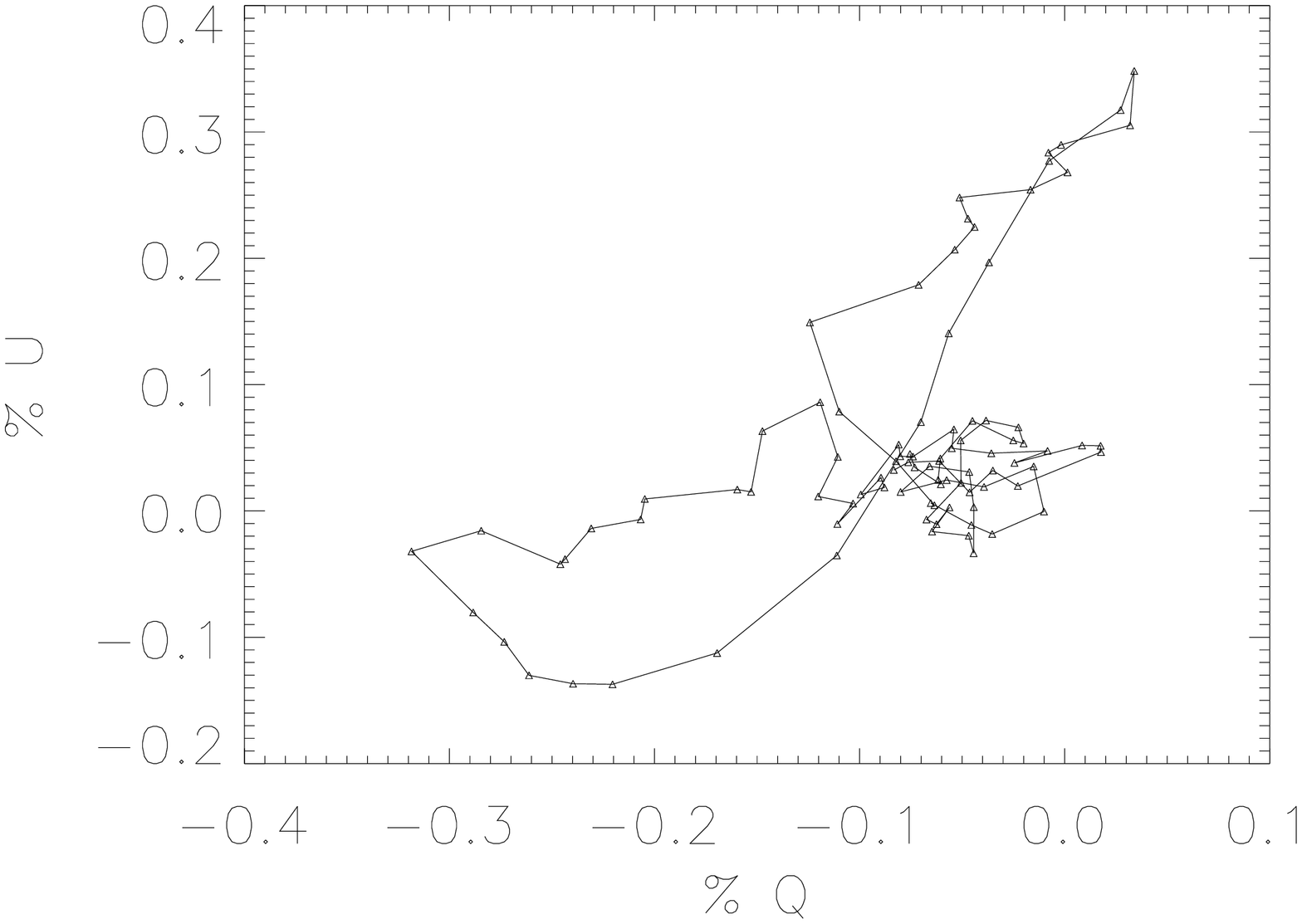} \\
\includegraphics[width=0.35\linewidth, angle=90]{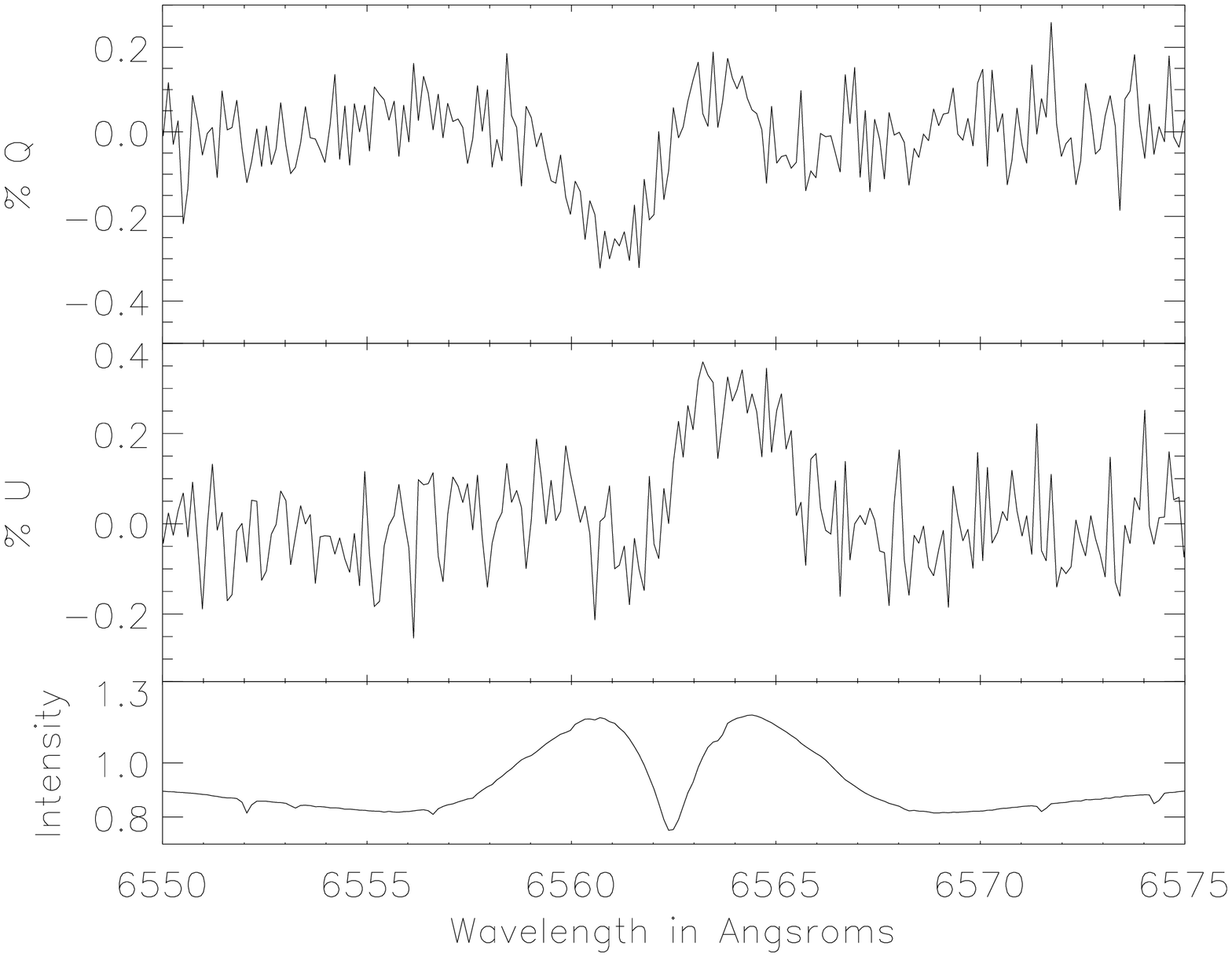}
\includegraphics[width=0.35\linewidth, angle=90]{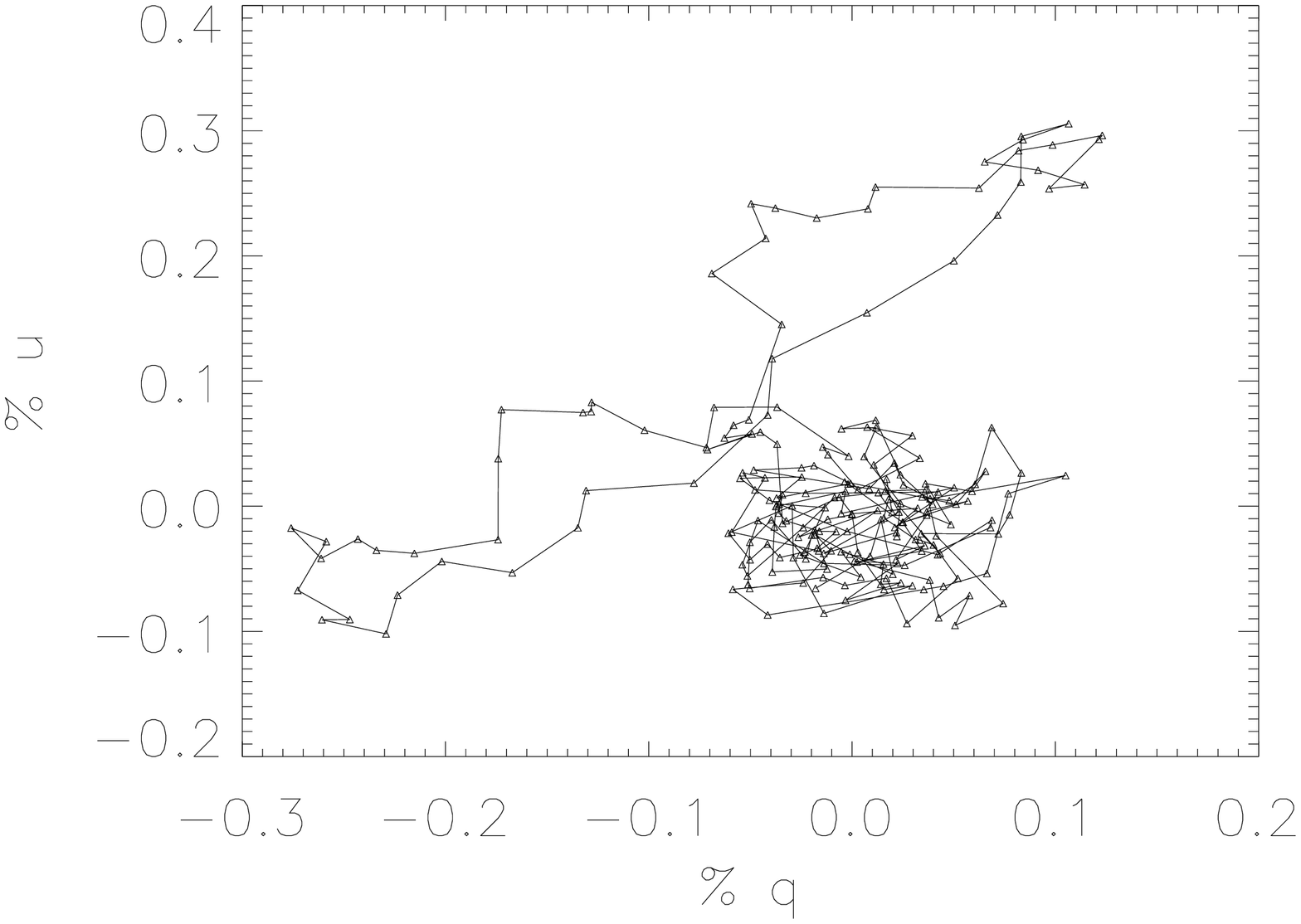} \\
\includegraphics[width=0.35\linewidth, angle=90]{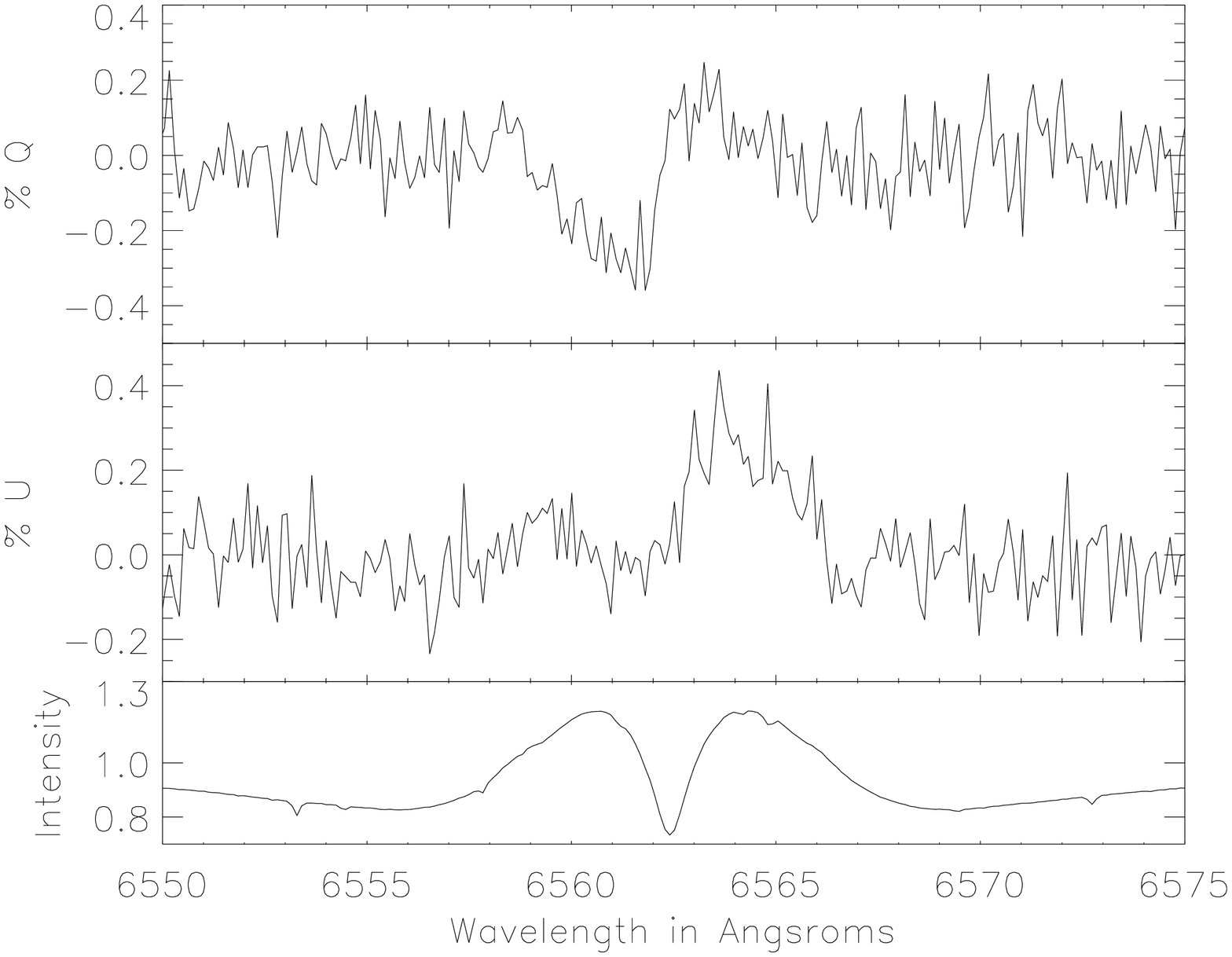}
\includegraphics[width=0.35\linewidth, angle=90]{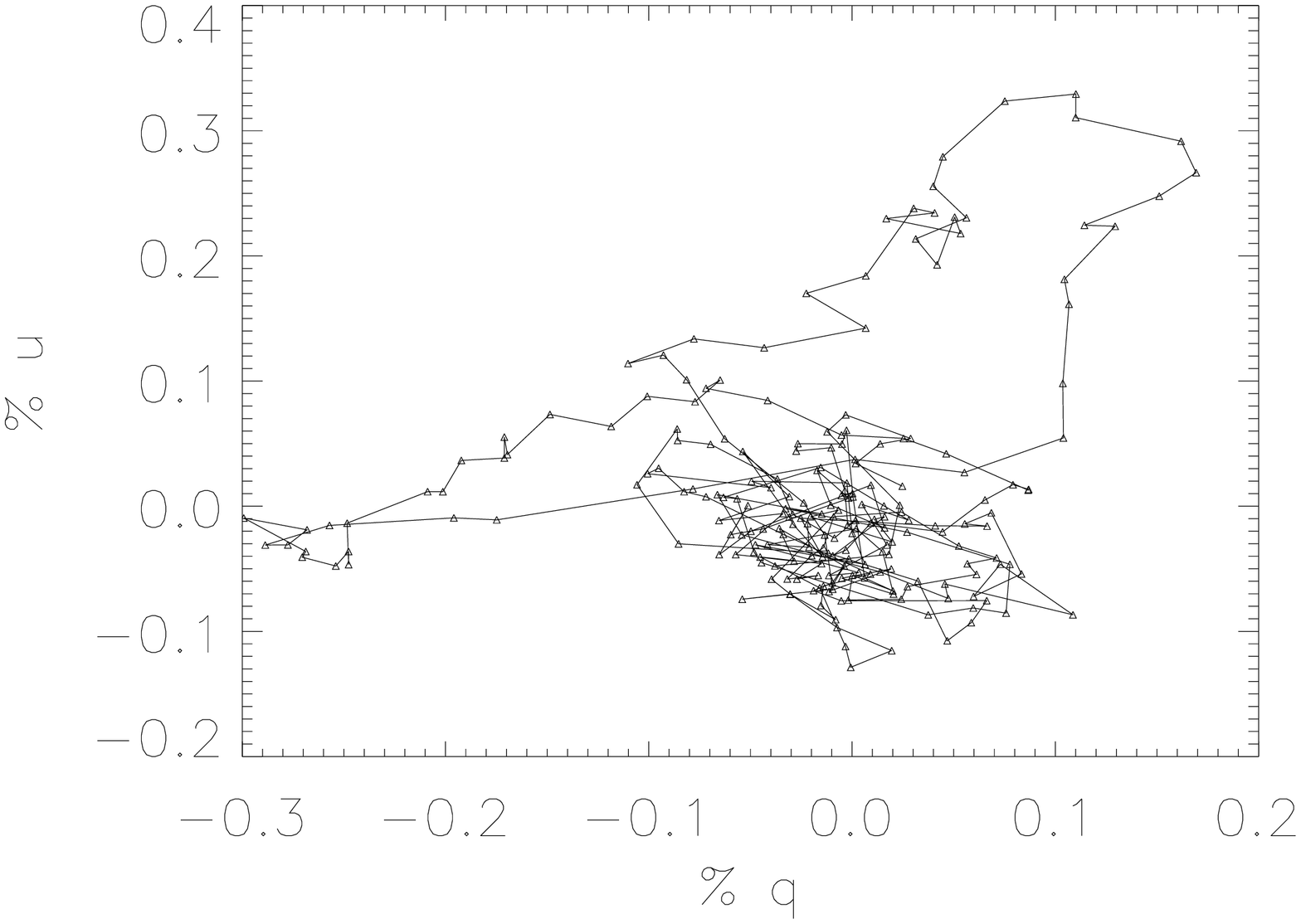}
\vspace{5mm}
\caption{Spectropolarimetry for 51 Oph Polarization. From left to right: {\bf a)} An example polarized spectrum for the 51 Oph H$_\alpha$ line. The spectra have been binned to 5-times continuum. {\bf a)} The top panel shows Stokes q, the middle panel shows Stokes u and the bottom panel shows the associated normalized H$_\alpha$ line. Both q and u show an antisymmetric signature of roughly 0.3\% {\bf b)} This shows q vs u from 6554.9{\AA} to 6574.3{\AA}. The knot of points at (0.0,0.0) represents the continuum. {\bf c)} The ESPaDOnS archive data for 51 Oph on August 13th, 2006. Both q and u again show antisymmetric signatures of 0.3\% {\bf d)} The corresponding qu plot, which looks fairly similar to that of HiVIS. {\bf e)} Shows ESPaDOnS data taken on March 20th 2008 and {\bf f)} is the corresponding qu-plot.}
\vspace{5mm}
\label{fig:swp-51oph}
\end{center}
\end{figure*}

\begin{figure}
\begin{center}
\includegraphics[width=0.75\linewidth, angle=90]{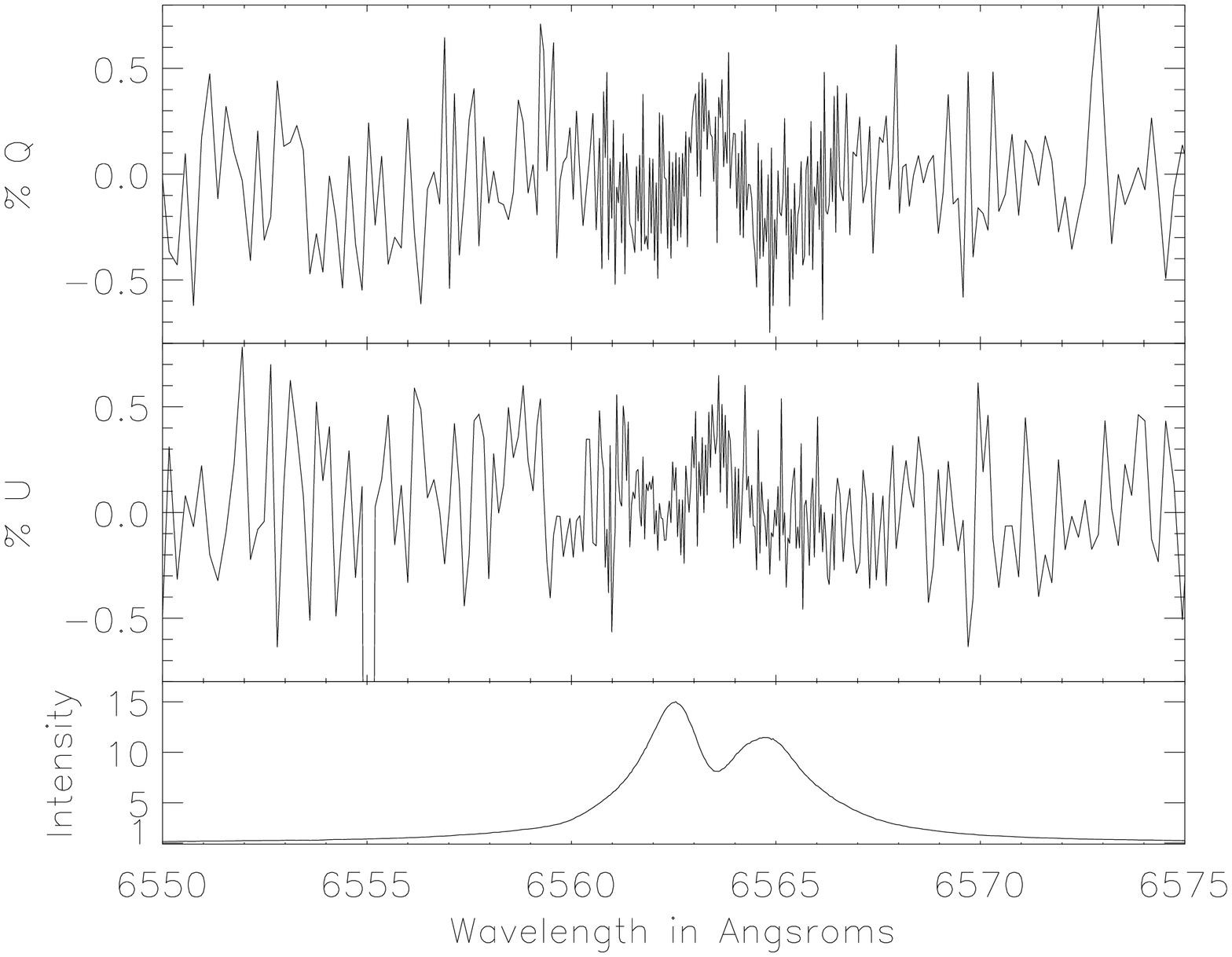} \\
\includegraphics[width=0.75\linewidth, angle=90]{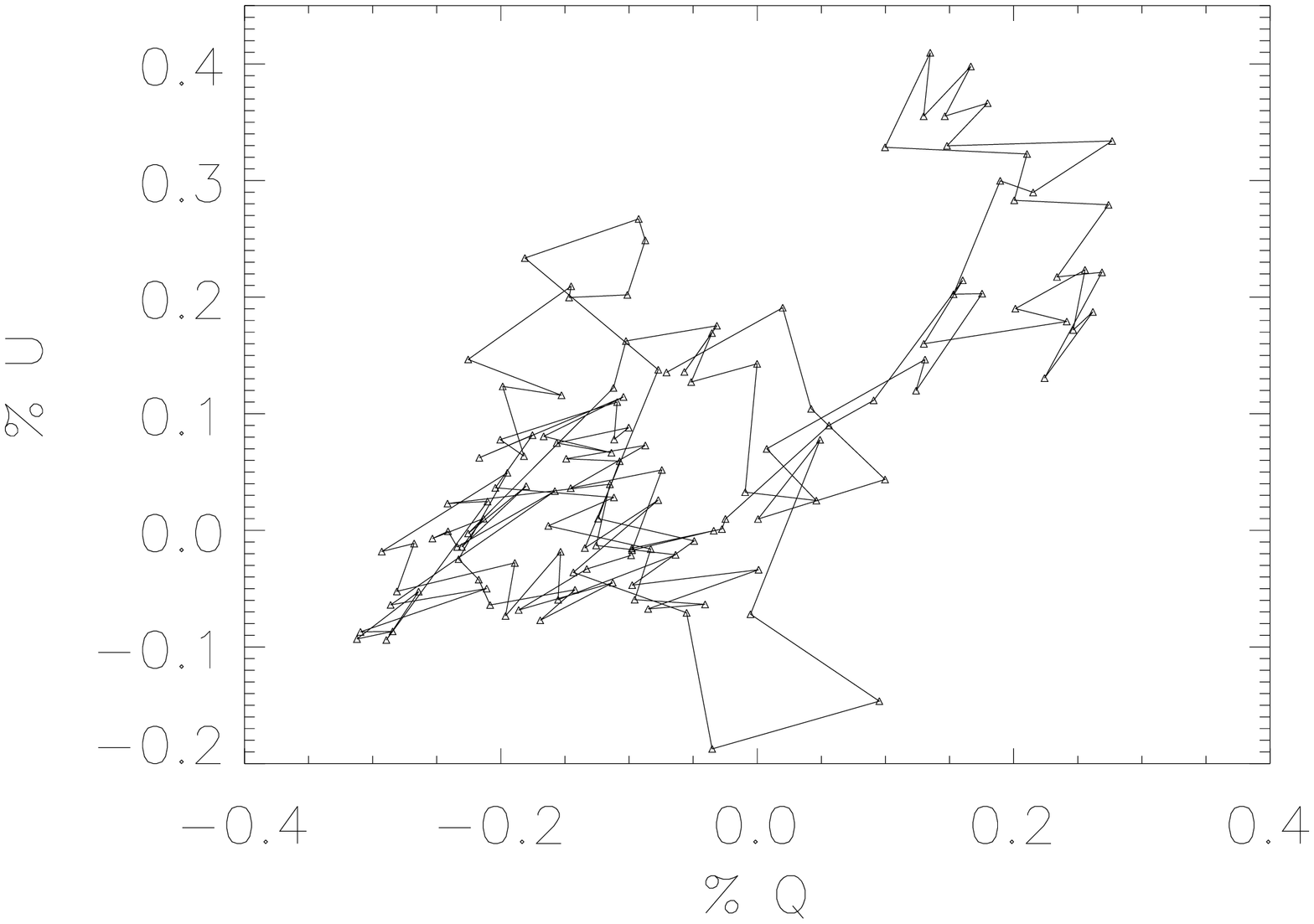}
\caption{Spectropolarimetry for MWC 147. From left to right: {\bf a)} An example polarized spectrum for the MWC 147 H$_\alpha$ line. The spectra have been binned to 5-times continuum. The top panel shows Stokes q, the middle panel shows Stokes u and the bottom panel shows the associated normalized H$_\alpha$ line. There is a fairly clear detection across the center of the absorption line. {\bf b)} This shows q vs u from 6561.4{\AA} to 6565.9{\AA}. The knot of points at (-0.2,0.0) represents the polarization on both emission peaks, which is significantly less than the average continuum. Since such a narrow wavelength range was chosen to avoid the noisy data outside the line core, the continuum knot is non-existant and is represented by the larger swarm of points generally around zero.}
\label{fig:swp-mwc147}
\end{center}
\end{figure}

\begin{figure*}
\begin{center}
\includegraphics[width=0.35\linewidth, angle=90]{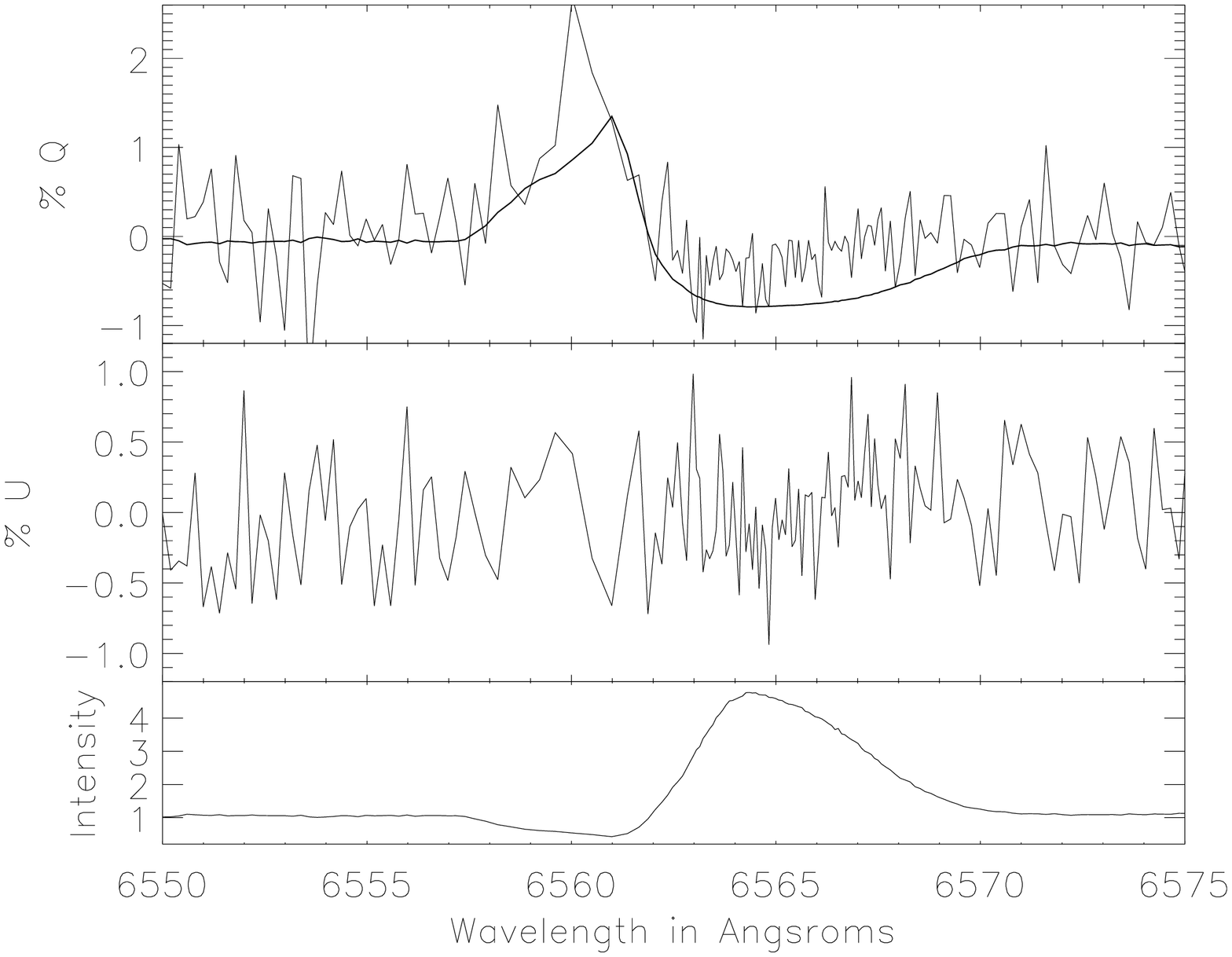} 
\includegraphics[width=0.35\linewidth, angle=90]{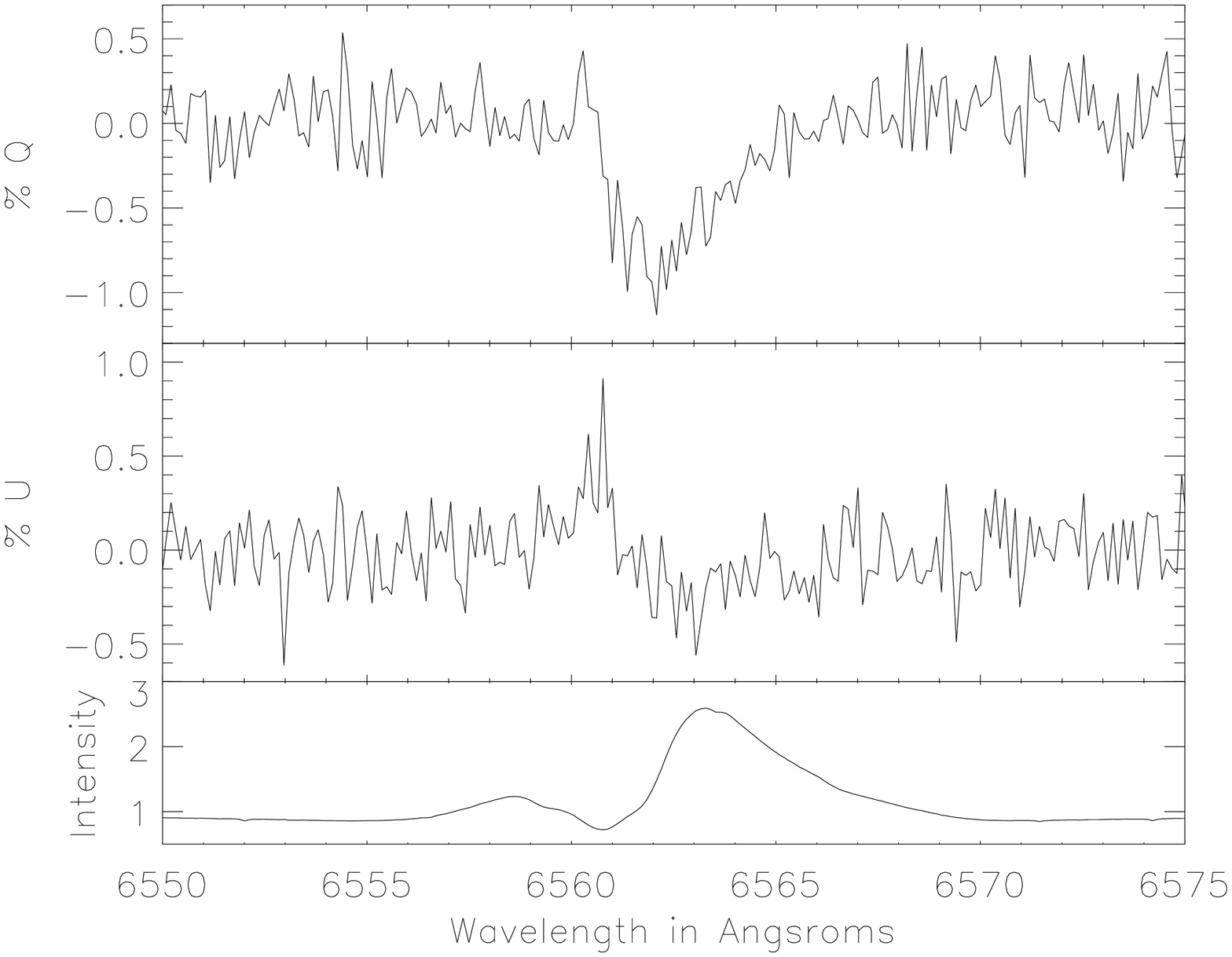} \\
\includegraphics[width=0.35\linewidth, angle=90]{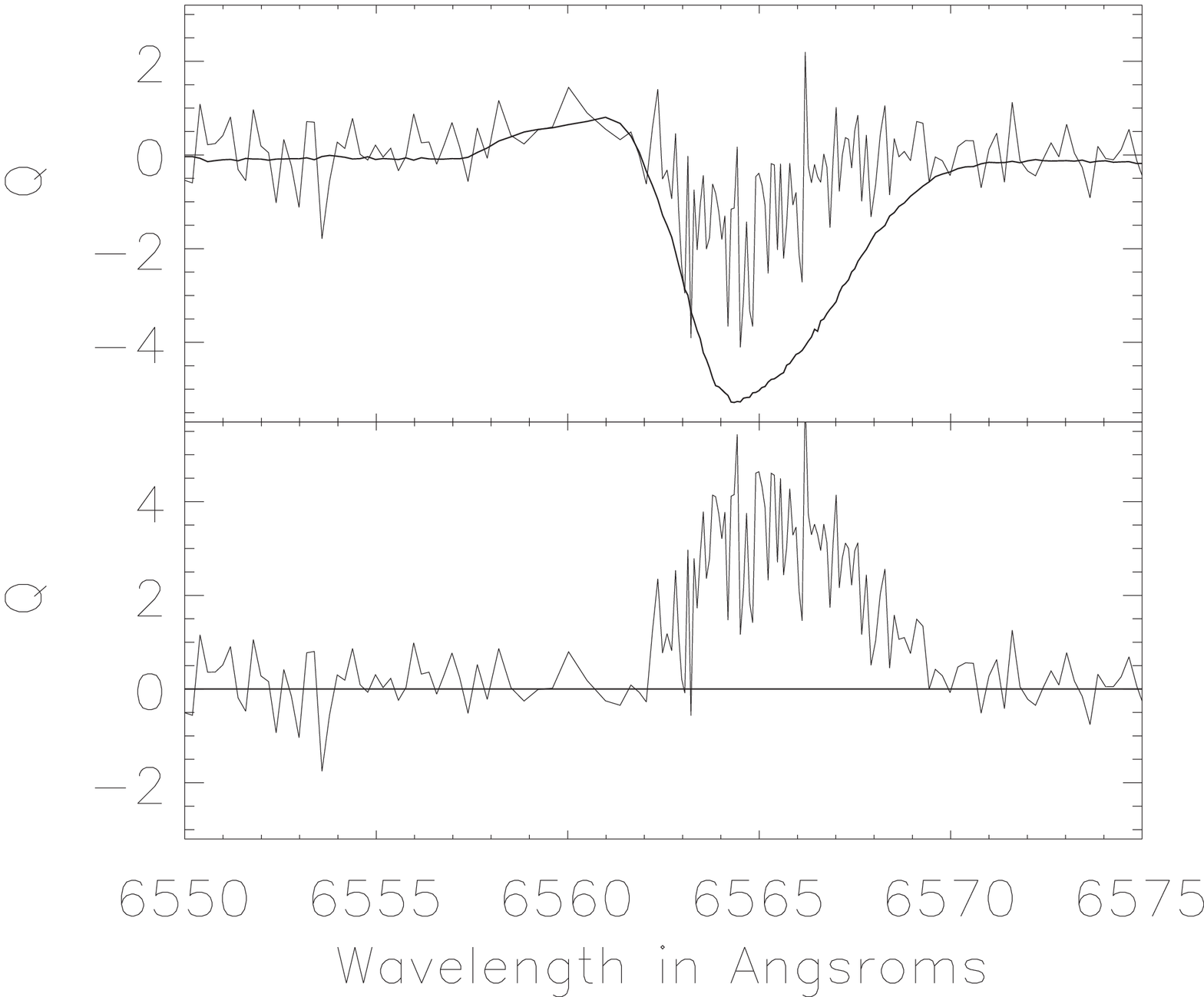} 
\includegraphics[width=0.35\linewidth, angle=90]{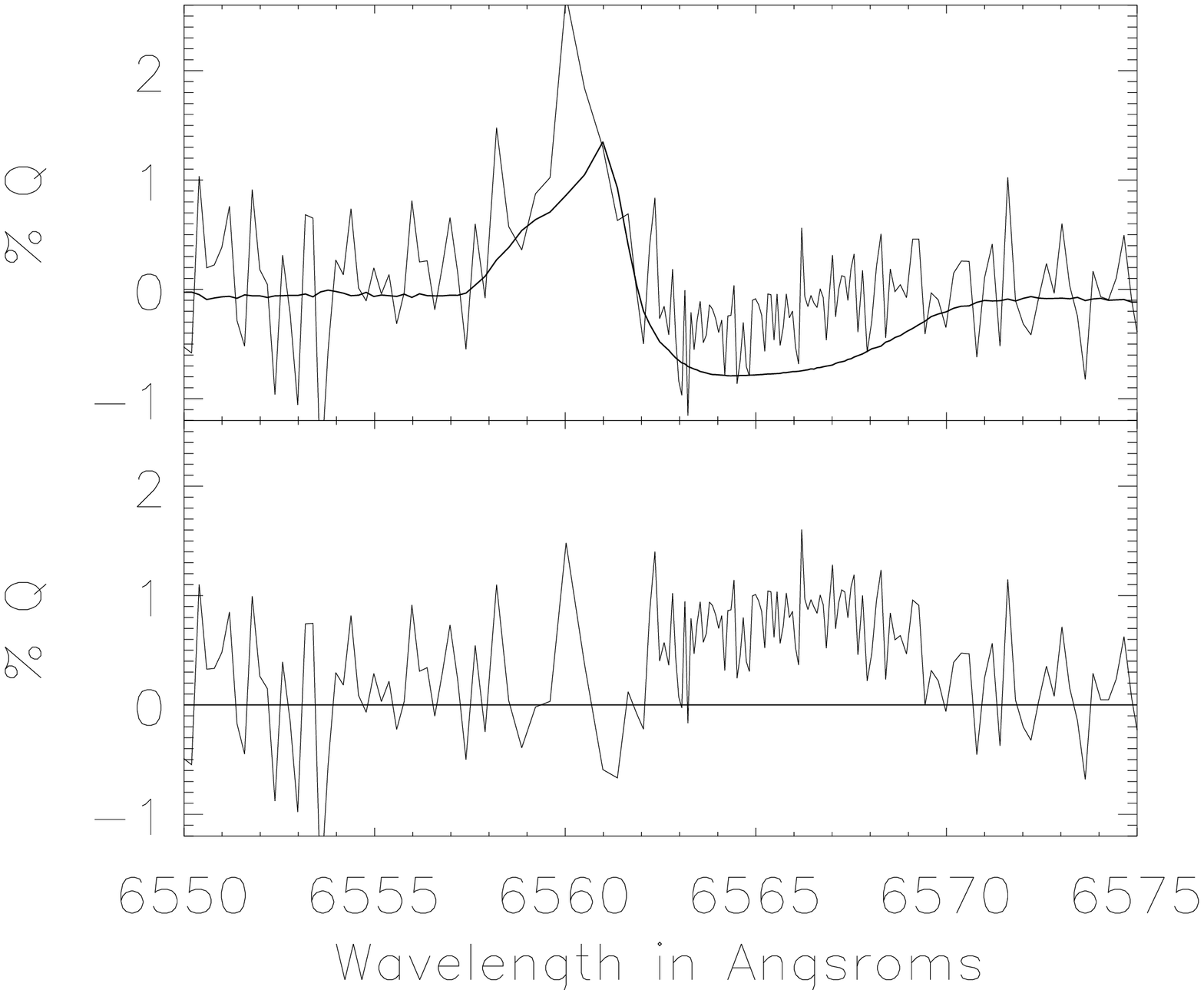}
\includegraphics[width=0.45\linewidth, angle=90]{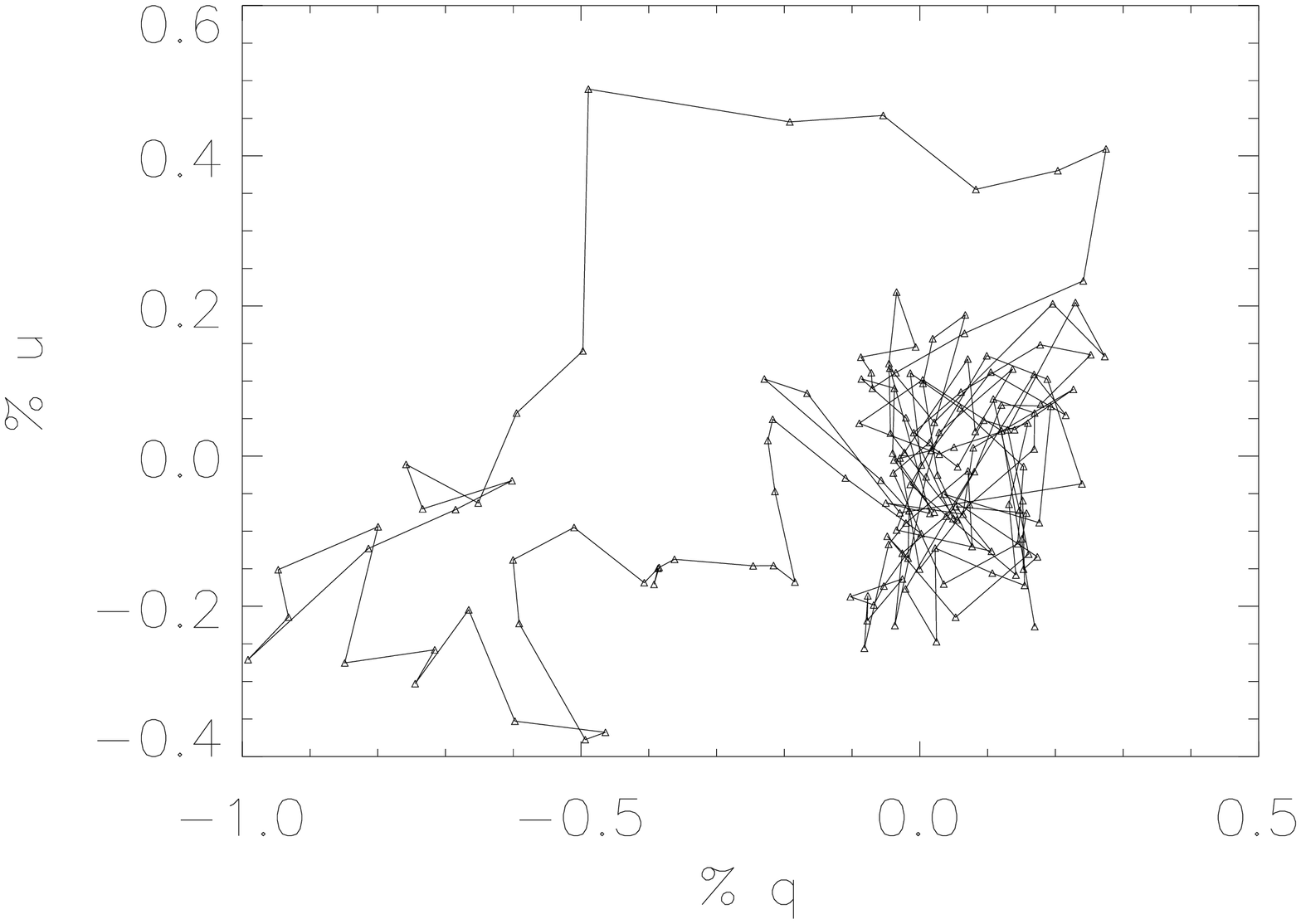}
\vspace{5mm}
\caption[HD 144432 Spectropolarimetry]{Polarized spectra for the HD 144432 H$_\alpha$ line from  {\bf a)} HiVIS on August 29th 2007 and  {\bf b)} ESPaDOnS on August 14th, 2006. The HiVIS polarized spectra have been binned to 5-times continuum. The top panel iin a) and b) shows Stokes q, the middle panel shows Stokes u and the bottom panel shows the associated normalized H$_\alpha$ line. There is clearly a detection in the blue-shifted absorption of -1.0\% in q and 0.5\% in u for the ESPaDOnS data and 2\% in q for the HiVIS data. {\bf c)} A 1/I subtraction of Stokes Q - the polarized flux (q*I) shown in the top box was adjusted to remove the 2\% polarization in the blue-shifted absorption. The dark line shows the fitted Stokes Q. After removal, the resulting Stokes Q is shown in the bottom box with a significant polarized flux on the side of the emission line. {\bf d)} Stokes q after the adjustment. The top box shows the original Stokes q with the removed 1/I signature. The bottom box shows the resulting Stokes q with the red-shifted residual. {\bf e)} The qu loop for the ESPaDOnS archive observations.}
\label{fig:swp-hd144-all}
\end{center}
\end{figure*}

\begin{figure*}
\begin{center}
\includegraphics[width=0.35\linewidth, angle=90]{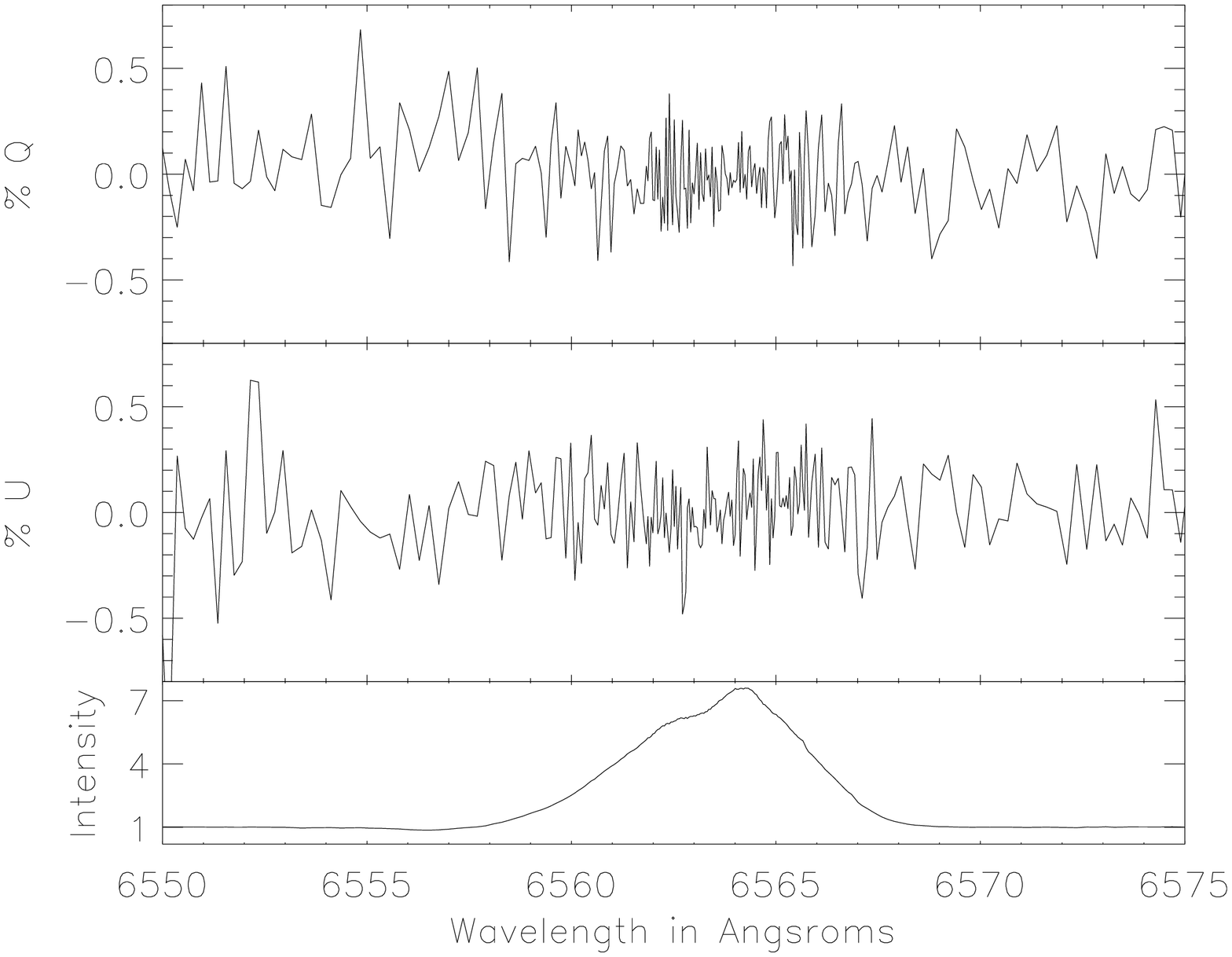}
\includegraphics[width=0.35\linewidth, angle=90]{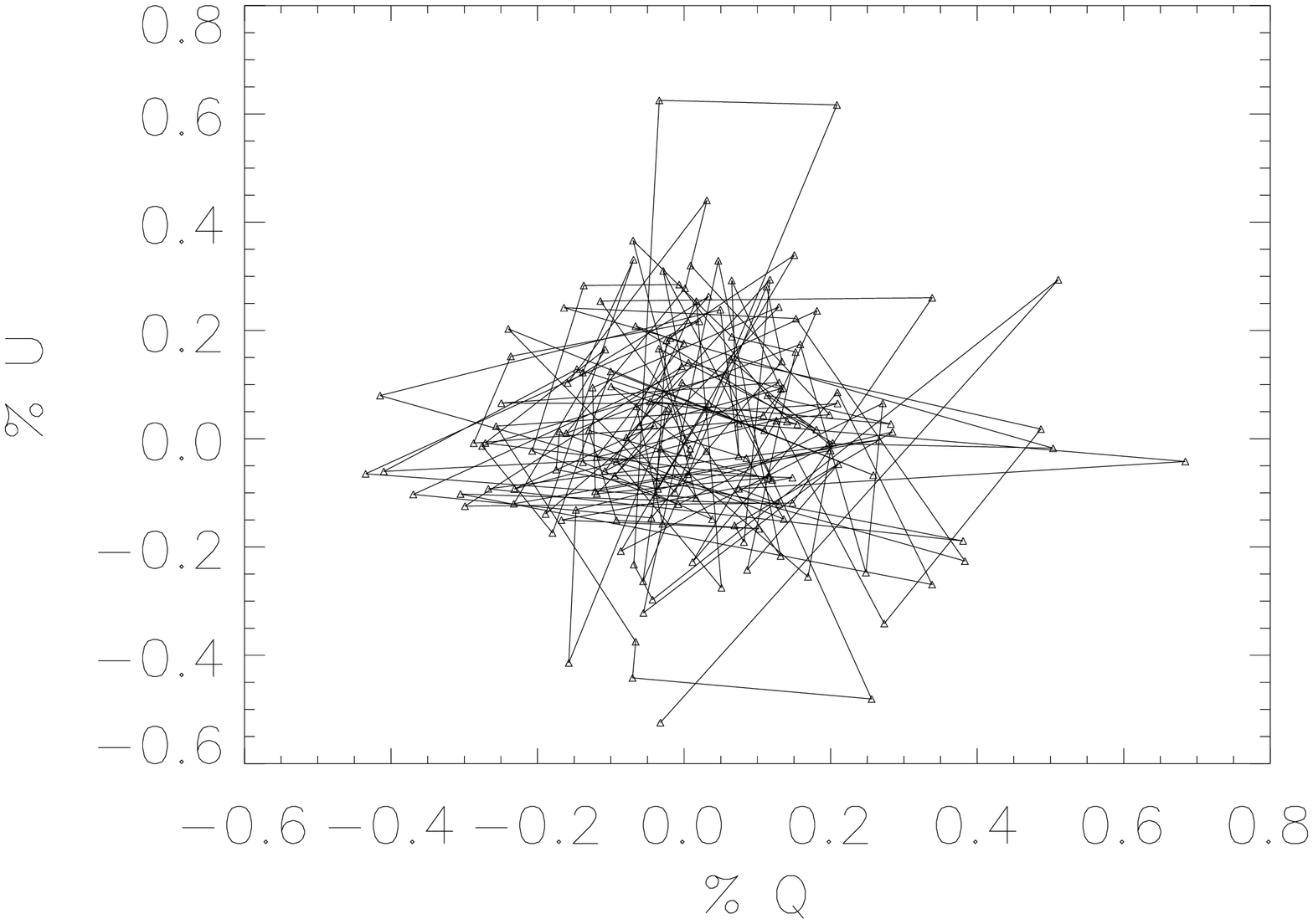} \\
\includegraphics[width=0.35\linewidth, angle=90]{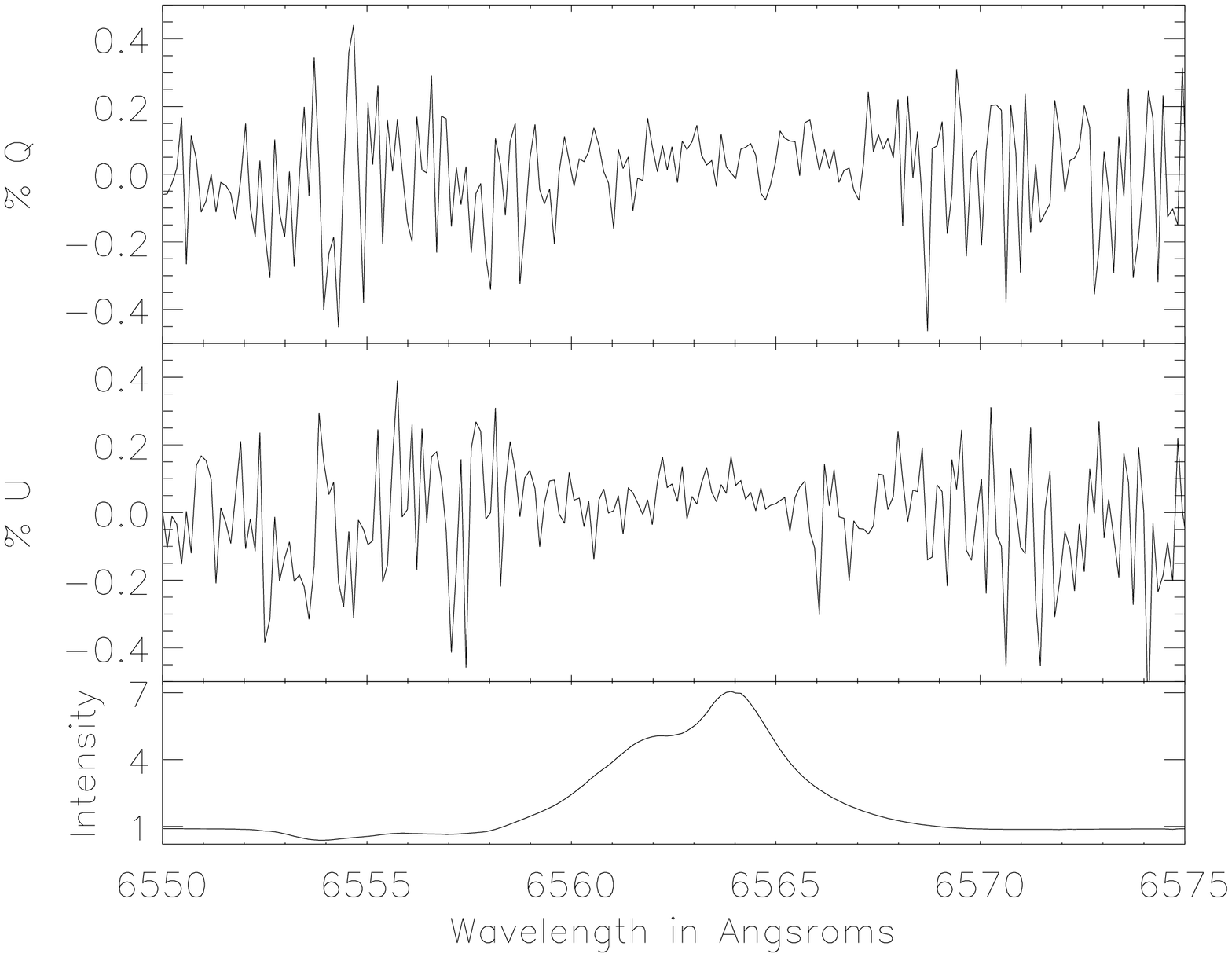}
\includegraphics[width=0.35\linewidth, angle=90]{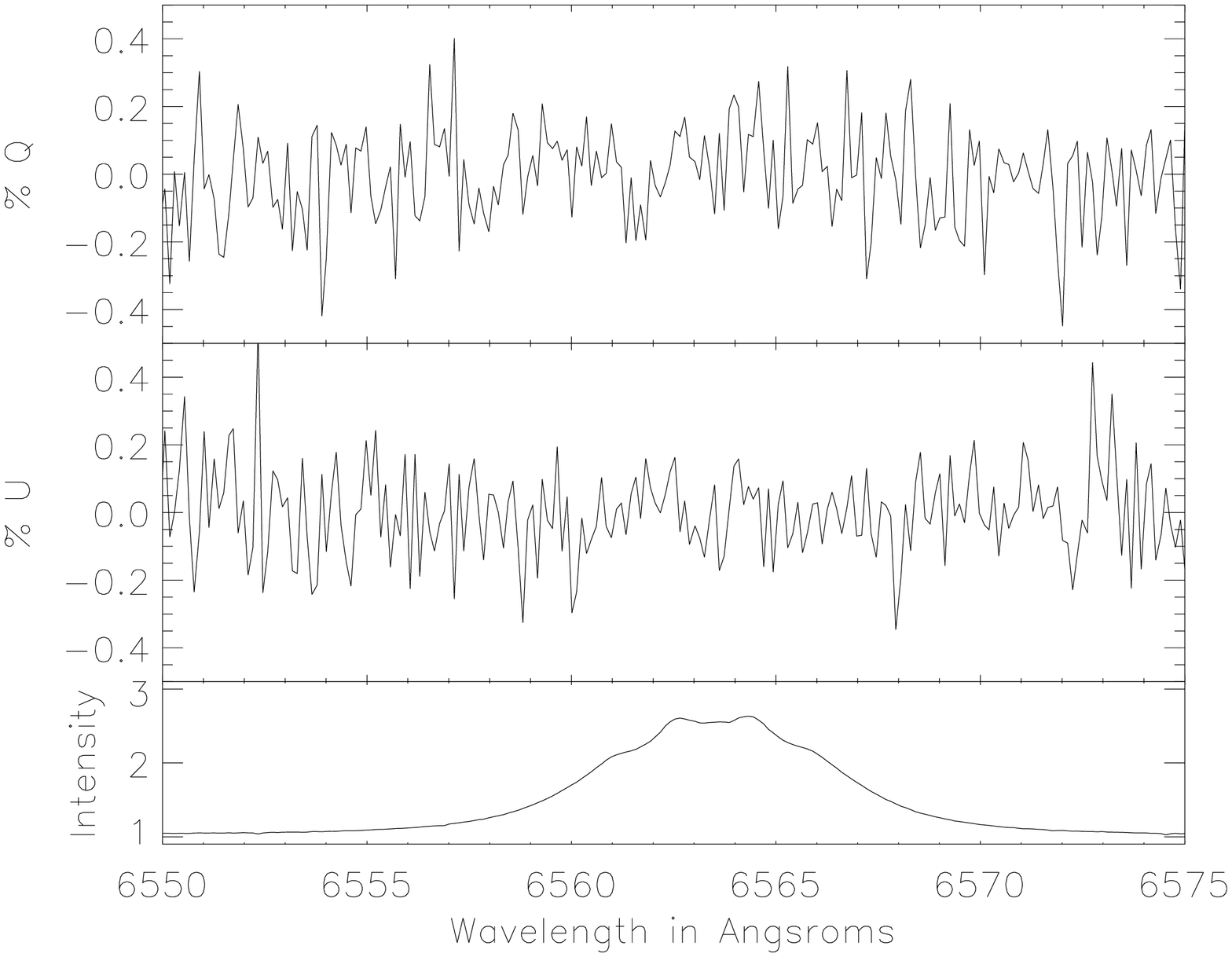} \\
\includegraphics[width=0.35\linewidth, angle=90]{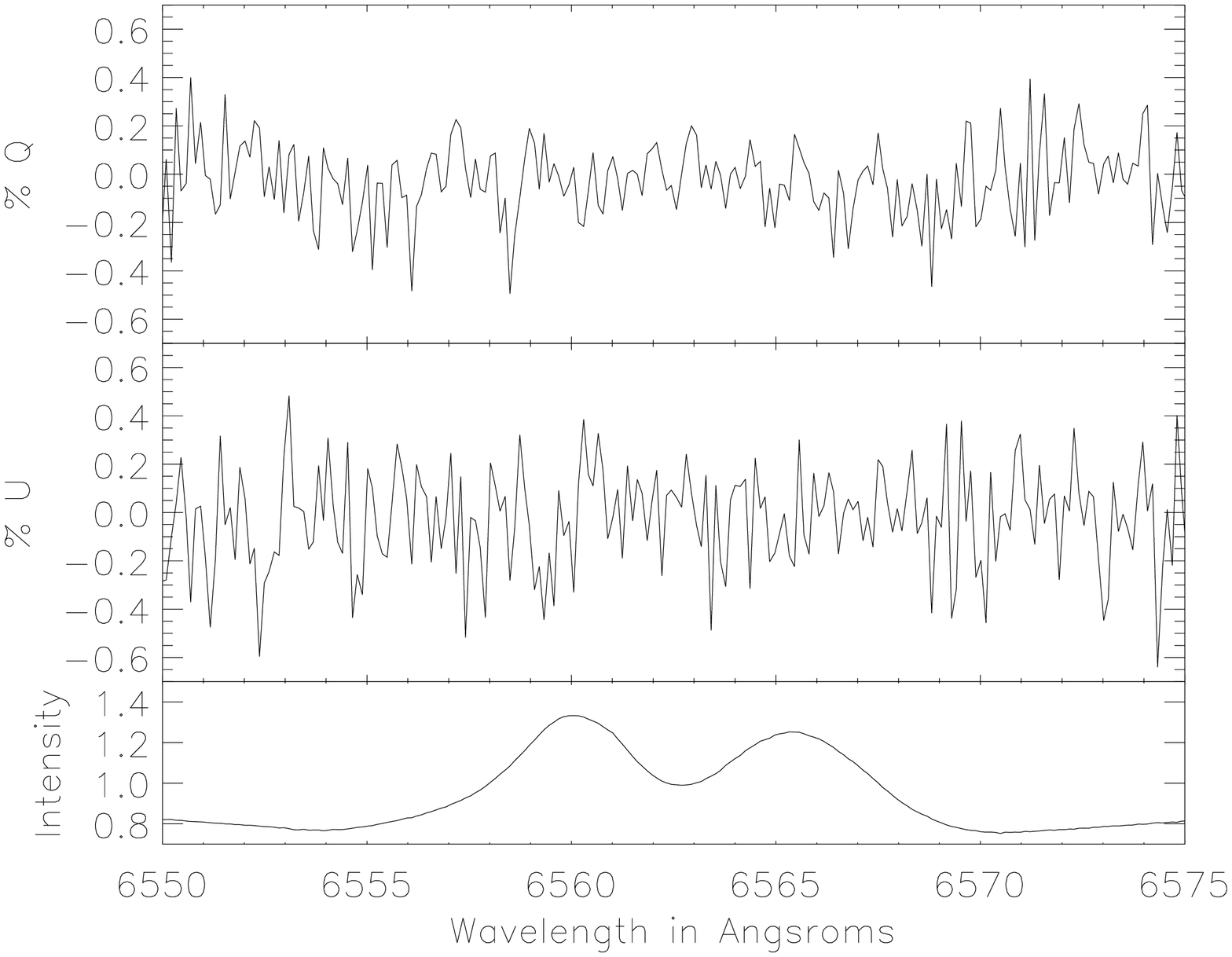}
\includegraphics[width=0.35\linewidth, angle=90]{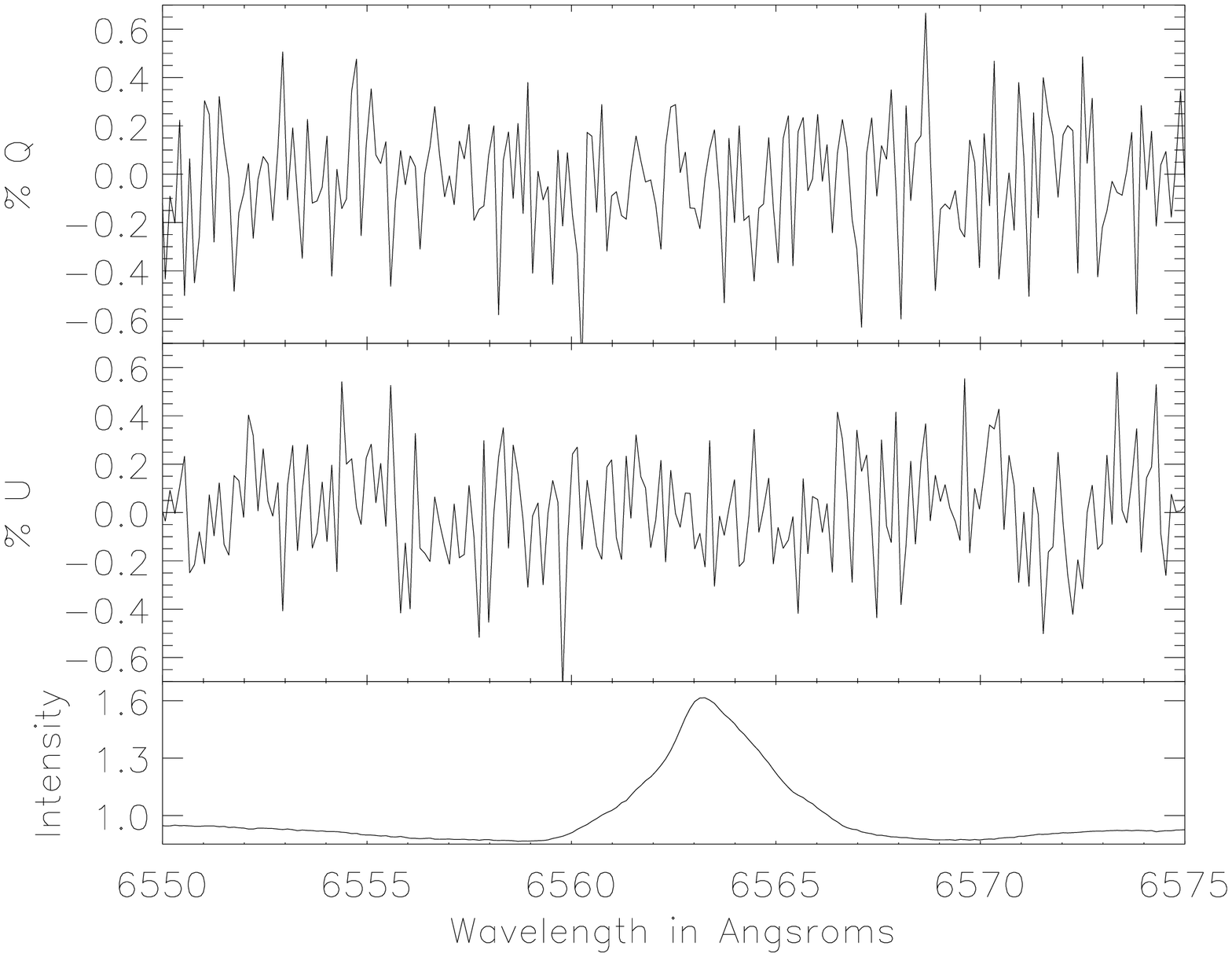}
\vspace{5mm}
\caption{Spectropolarimetric non-Detections. V1295 Aql HiVIS and Archive ESPaDOnS non-detections: V1295 Aql, GU Cma, HD 141569, HD 35929. From left to right: {\bf a)} An example polarized spectrum for the V1295 Aql H$_\alpha$ line. The spectra have been binned to 5-times continuum. The top panel shows Stokes q, the middle panel shows Stokes u and the bottom panel shows the associated normalized H$_\alpha$ line. {\bf b)} This shows q vs u from 6551.3{\AA} to 6565.6{\AA}. The following panels show the non-detections from the ESPaDOnS archive. Each panel shows Stokes q, Stokes u and the normalized intensity profile from top to bottom. {\bf a)} V1295 Aql from August 13th 2006. {\bf b)} GU Cma from February 9th 2006. {\bf c)} HD 141569 and {\bf d)} HD 35929 both from February 8th 2006.}
\vspace{5mm}
\label{fig:swp-nd-esp}
\end{center}
\end{figure*}

\subsection{HD 200775 - MWC 361}
	
	This star has been very stable in HiVIS polarization signature over the past few years. A broad spectropolarimetric signature was detected in all 32 measurements in figure \ref{fig:haebe-specpol2}. In Harrington \& Kuhn 2008, HiVIS and ESPaDOnS observations that showed the signature didn't change in over a year. The HiVIS observations did not show significant morphological changes from telescope polarization effects. When all HiVIS observations were rotated to a common frame, the morphology of the spectropolarimetric effects match the ESPaDOnS observations very well. Figure \ref{fig:haebe-specpol2} shows all the HiVIS data unrotated. In figure \ref{fig:swp-mwc361}, a least-squares rotation was applied to all measurements and then the average was taken. A very clear, very high signal-to-noise polarized spectrum shows a very broad 0.25\% change in q with a much more narrow asymmetric flip in u reaching -0.1\% and +0.15\%. The u measurements are slightly smaller in amplitude than that detected with ESPaDOnS, but the overall form matches perfectly. 
	
	The qu-plot of the figure illustrates the change very clearly - q increases, u goes low then high, and finally q returns to zero. The polarized flux shows that the Q flux is essentially a widened and flattened copy of the H$_\alpha$ line. The U spectrum has basically the same shape as the u spectrum but with an amplification.

	The emission line is quite strong while the polarization is relatively small, less-strongly peaked and significantly broader than the emission line. The asymmetric change seen in u complicates the picture. In this star, q does provide a strong linear extension, but while q is nearly constant in the emission peak, the u profile complicates this signature. Though this star does show evidence for absorption as the center of the line is notched, and in many observations asymmetrically so, the u polarization is much broader than the central notch.

\subsection{HD 36112 - MWC 758}

	This star did not have the highest signal-to-noise but it has a detectable signature. Figure \ref{fig:haebe-specpol1} shows all 11 measurements and there is a subtle effect that can be seen in the u spectrum. However, in an individual, higher signal-to-noise spectrum of figure \ref{fig:swp-mwc758} you can see a 1\% change in u. There is also a smaller effect across the q spectrum. The qu-plot shows the qu change clearly. The continuum knot is clear and there is a mostly linear extension out to (+0.6,-1.0). In the ESPaDOnS archival data from February 9th 2006, this star was a non-detection but the H$_\alpha$ line showed significantly less absorption than in almost all of the HiVIS observations. Beskrovnaya et al. 1999 report a continuum polarization of near 0.15\% but with very significant nightly variability of  0.4\%. This star did not show a spectropolarimetric effect in Vink et al. 2002.
		
	The polarization change occurs mainly on the blue-shifted transition from absorption to emission, and is not centered in the absorptive trough. However, the maximum decrease in u occurs not where there is maximum emission, but just at the edge of the blue-shifted absorption. In qu-space, the polarization changes are close to being in-phase but are not quite matched. There is significant width to the qu-loop arising from the small increase in q.

\subsection{HD 45677}

	In the line profiles presented in Oudmaijer \& Drew 1999, the H$_\alpha$ line in this star is single peaked and asymmetric, quite different than the double-peaked line profile observed. They report a continuum polarization of 0.33\% in January 1995 and 0.14\% in December 1996 with line/continuum ratios of 35 and 34 respectively. They also report a line-effect as an increase in the degree of polarization of roughly 0.4\% in both observations. Patel et al. 2006 present ISIS spectropolarimetry showing a line effect combined with literature polarimetry and photometry. The central portion of the polarization spectrum was discarded because of a systematic effect produced by sub-resolution focal shifts. The the polarized spectra show an increase in polarization of 0.4\% across both blue and red shifted emission peaks. However, after removing an estimated interstellar polarization based on the line-center polarization in Oudmaijer \& Drew 1999, they compute a polarization spectrum showing a broad depolarization effect of roughly 0.9\%. Grady et al. 1993 presented evidence of accreting gas in this system based on the UV spectra.

	This star has a large and clear signature in the central absorption of the emission line. It is detected at various amplitudes in all 7 measurements of figure \ref{fig:haebe-specpol2}. Figure \ref{fig:swp-hd456} shows a single example of this. The q spectrum shows a small amplitude change on the red emission peak and the u spectrum shows a nearly 1\% decrease in the central absorption. In the qu-plot the u decrease causes the 1\% vertical extension with a complete return to continuum before the small increase in q causes the horizontal extension. 
	
	The detection is quite strong and is mostly confined to the central absorptive notch in the emission line. As can be clearly seen in the qu-plot as well as the polarized flux, there is significant polarization across the entire line, but the functional form is quite strange. A strong change in q is seen around line center and a small, independent change in u is seen only on the redshifted emission peak.
	
\subsection{HD 158643 - 51 Oph}
	
	This star has a significant but small change across the entire line in all 3 measurements of figure \ref{fig:haebe-specpol2}. The amplitude is roughly 0.1\% in all the HiVIS measurements. Figure \ref{fig:swp-51oph} shows an individual example where the change can be seen as a decrease in both q and u on the blue-shifted emission and then an increase in both q and u on the red-shifted emission. The qu-plot illustrates this as nearly linear extensions in qu-space.  Chavero et al. 2006 report an R-band polarization of 0.35$\pm$0.05\% and show that a Serkowski law does not fit the stars polarization. Oudmaijer et al. 2001 found 0.47\% V-band polarization and a similar Serkowski mis-fit. They note that this star is sometimes listed as a Vega-type star. The polarization increases towards the blue, going from 0.3\% at 7000-8500{\AA} to 0.55\% at 3500{\AA} and 4500{\AA}. From this they conclude that circumstellar material (likely disks) cause the intrinsic polarization. They argue that the dust responsible for the polarization has small grains and is the same dust responsible for the IR emission.
	
	The red-shifted emission peak shows an increase in both q and u while the blue-shifted emission peak shows a decrease in both q and u. The transition between positive and negative changes occurs right at line center. While the changes are both in phase and show linear extensions in qu-space, they take two distinct excursions with opposite directions in wavelength.

\subsection{HD 259431 - MWC 147 - V700 Mon}

	The observations of Oudmaijer \& Drew 1999 show a marginal decrease in polarization of 0.2\% across the line with a continuum of 1.06\% and a line/continuum ratio of 11. In Mottram et al. 2007, the observations also showed a similar signature but with only six resolution elements across the bulk of the emission.  
	
	This star had a marginal detection in the center of the emission line in some but not all of the data sets. It's hard to see in the 8 measurements of figure \ref{fig:haebe-specpol2}, but in the single example of figure \ref{fig:swp-mwc147} the change can be seen as small amplitude changes in both q and u across the core of the emission line. The q spectrum shows a small decrease of roughly 0.2\% across both emission peaks with a return to continuum polarization just outside the emission peaks. The qu-plot shows a small wavelength range, 6561.4 to 6565.9{\AA}, where the clear change across the central-absorption is seen. The aparent continuum-knot is not centered at zero because the wavelength range selected only includes the line core and both the blue-shifted and red-shifted emission peaks cluster at polarizations of (-0.2,0.0). Adding a greater wavelength range would, with these noise statistics, make the small change from (0,0) across the emission and absorption invisible.

	The morphology of this detection is difficult to discuss because of its smallness and narrowness. In qu-space the extension in the central absorption is nearly linear but with a significant deviation away from linear at the very tip. The polarization in both blue and red emission peaks is only marginally different from zero. The change in q and u does seem to be phased, but there is width to the qu-loop. All that can be concluded at this point that there definitely is a small narrow signature.

\subsection{HD 144432}

	This star had only one good observation that is shown in figure \ref{fig:swp-hd144-all} from August 29th, 2007. There is a clear change in Stokes q of roughly 2\% in the absorptive component and 0.3\% in the blue side of the emission line. There is also archival ESPaDnS data from August 14th 2006 also shown in figure \ref{fig:swp-hd144-all}. In the year between observations, the absorption in H$_\alpha$ line increased and widened greatly to become a full P-Cygni profile. There is clearly a detection in the blue-shifted absorption of -1\% in q and 0.5\% in u for the ESPaDOnS data and 2\% in q for the HiVIS data. Initially, a 1/I error seemed like a likely candidate, but the calculations shown in figure \ref{fig:swp-hd144-all} show that a 1/I signature does not fit the polarization spectrum. To show this, a fit to the polarized flux was performed to remove the 2\% signature. The 2\% signature in absorption was removed completely but only after introducing a significant red-shifted residual. The archival ESPaDnS data also provided an independent measurement showing that the HiVIS results were robust since the magnitude and wavelengths of the polarization change were the same.
	
	In the ESPaDOnS archive data, there is a very significant decrease in q and a smaller decrease in u across the center of the emission line. The blue-shifted absorption does shows a change in the opposite sense of the change across the emission line. But, inspecting the peak decrease in q shows that this also occurs not in the emission peak but in the transition from emission to absorption. Also, there is some width to the qu-loop as seen in figure \ref{fig:swp-hd144-all}. The increases in q and u are not phased together, suggesting that the removal of unpolarized light by absorption is more complex than this simple assumption.

\subsection{HD 190073 - MWC 325 - V1295 Aql}

	This star had no detectable polarization signature in all 17 observations. The line profiles were consistently asymmetric with the blue-shifted absorption as ample evidence for outflowing material. However, even in heavily binned profiles with 0.2\% detection thresholds there was no significant polarization change. Figure \ref{fig:haebe-specpol1} shows all the data and examples of an individual spectrum are shown in figure \ref{fig:swp-nd-esp}. The qu-plot shows a simple continuum-knot centered at (0,0) with no significant asymmetry or deviation. It was also a non-detection in the archival ESPaDOnS data of Aug 13th 2006 shown in figure \ref{fig:swp-nd-esp} even though there is evidence for stronger absorption from the decreased line strength, deeper blue-shifted absorption, and a strongly blue-shifted absorption at 6554{\AA}.

\subsection{Non-Detections and Other Notes}

	GU CMa was a non-detection in the archival ESPaDOnS data of Feb 9th 2006 shown in figure \ref{fig:swp-nd-esp}. It was also a non-detection in Oudmaijer \& Drew 1999 with a continuum of 1.15\% and a line/continuum ratio of 3.  HD 141569 was a non-detection in the archival ESPaDOnS data of Feb 8th 2006. The emission from this star was always weak, being 1.3 in the archival data and 1.6 in the HiVIS observations. HD 35929 was a non-detection in the archival ESPaDOnS data of Feb 8th 2006. The star shows blue-shifted absorption though the emission from this star was weak. In the HiVIS observations, the line was somewhat stronger, 1.9 vs 1.6, with less clear absorption. It was also a non-detection in Vink et al. 2002. This star, as noted in section 2.14, has been studied in-depth spectroscopically and was, in 2004, determined to be an F2 III post-MS giant on the instability strip.

\subsection{Comments on Other Systems}
 	
	MWC 166 was a clear detection in Oudmaijer \& Drew 1999 of a 0.2\% decrease in polarization with a continuum of 1.15\% and a line/continuum ratio of 2.6. Mottram et al. 2007 show a decrease in polarization of 0.2\% across a continuum of 0.5\%. but with a singly-peaked asymmetric line profile. In two observations shown in figure \ref{fig:haebe-specpol2}, no effect at the 0.1\% magnitude is detected, though telescope polarization effects could be the cause as there are only a few observations at a few pointings. It should be noted that Oudmaijer et al. 2001 classify HD142666, HD 141569, and HD 163296 as HAeBe but also zero-age main-sequence. Il Cep was a non-detection in Mottram et al. 2007 as well as in this survey. XY Per showed a 0.5\% decrease in polarization in the central absorption from a continuum of 1.5\% in Vink et al. 2002 though HiVIS did not detect any signature. T Ori was reported to have a spectropolarimetric effect of 0.8\% on a continuum of 0.4\% in Vink et al. 2002. The polarization increase is detected in our survey at the wavelengths where there is spatially resolved, diffuse H$_\alpha$ emission (see Harrington \& Kuhn 2008 for more detail). HiVIS detected a much larger amplitude signature but at higher spectral resolution. Given the difference in relative flux contribution between the star and diffuse H$_\alpha$, these observations are compatible. However, since it cannot be certain what ontribution is intrinsic to the star, this star has not been included in the detections statistics. In Vink et al. 2002, KMS 27 showed a broad 0.2\% increase in polarization across the entire line on a continuum of 0.1\% with a P-Cygni type profile. This star is a non-detection in the survey, but with a lower signal-to-noise.

\subsection{Summary of Herbig Ae/Be Spectropolarimetry}

The HiVIS survey has been complemented by ESPaDOnS observations as well as archival ESPaDOnS data. Most of the windy sources and more than half the disky sources showed spectropolarimetric signatures. The signatures show very different morphologies even among stars of similar H$_\alpha$ line type.
	
	In the windy sources, especially AB Aurigae and MWC 480, polarization changes were only seen in the P-Cygni absorption trough for many observations. The emission peak and red wing had the same polarization as the continuum to with the detection threshold of typically $<$0.2\%. In other windy sources, such as HD 163296, MWC 120, MWC 758, HD 150193, and HD 144432 showed large polarization signatures, 1\% to 2\%, that were greatest in blue-shifted absorptive components even if those components did not go below continuum. HD 163296 and MWC 120 were also clearly variable in the absorptive polarization. There were several stars that were non-detections and one, HD 179218, that showed polarization across the entire line. There are difficulties classifying stars in this loose way and making rigorous statistical conclusions from such a small number of stars. However, more than half the windy stars showed significant polarization and general morphological considerations can be very useful to compare models. 
	
	In the disky systems, such as MWC 158, HD 58647 and HD 45677, similar spectropolarimetric effects are seen in absorption. The polarization in the absorptive component, near line center, show polarization signatures of 1\% to 2\% while the the emissive components have a polarization at or very near continuum. The qu-loops come almost entirely from the polarization effects in the central absorptive component. In other disky systems, like MWC 147 and 51 Oph, the polarization signature is much smaller but it spans most of the line width but with a complex linear structure in qu-space. 
	
	There are several main morphological considerations when trying to establish the presence of a certain spectropolarimetric effect. The depolarization effect should show a broad change in polarization whenever there is emission as this effect is by definition a dilution of polarized continuum by unpolarized line emission. In this model, absorption is said to preferentially remove unpolarized flux so the depolarization effect in absorption must act with an opposite sign of that in emission. The depolarization effect is always acting on the underlying continuum polarization, and is roughly inversely proportional to the dilution (the normalized line intensity). The morphology cannot include antisymmetric components in absorption.

	The disk scattering effect of also makes concrete morphological predictions (Wood et al. 1993, Vink 2005b). This effect redistributes scattered light in wavelength and produces a spectropolarimetric effect that is wider than the original emission line. A pure Keplarian disk produces symmetric spectropolarimetric profiles. The disk scattering theory produces polarization preferentially away from line center and clearly produces detectable polarization on the red side of the line profile. In the Wood et al. 1993 case of pure outward radial motion (a ``disk" that is expanding), there was polarization only on the red shifted side of the line profile, none on the blue-shifted side. The qu-loops corresponded to changes in position-angle that had the same width as the polarized spectra - wider than the emission line. There are also amplitude considerations. The disk-scattering effect does not have any natural amplitude and simply scales with the amount of scattered light. The similarity of the magnitude of detected spectropolarimetric effects would argue for similar scattering environments. It is also interesting to note that in Wood \& Brown 1994, the effect of thermal smearing is shown to be very significant for this disk-scattering effect in early-type stars. The scattering medium causes a broadening of the line effect because of the thermal motion of the scattering particles. This effect is not mentioned or modeled in Vink et al. 2005, suggesting that these models, if anything, underestimate the broadness and overestimate the amplitude of the disk-scattering effect, making narrow features more unlikely in this model.  
		
	 In most stars, there were several morphological arguments against these scattering theories. In all stars, there was polarization in absorption, even if there was no polarization at more than an order of magnitude less in any other part of the line. Also, the qu-loops correspond to a change in position angle that spans the entire line. This is simply not seen. ABAur is one such example. It is very difficult to match our observations with either the depolarization effect or the disk-scattering effects from scattering theory because of the complete lack of a spectropolarimetric effect outside the P-Cygni absorption. The emissive component of the line shows a polarization identical to the continuum within the noise.
	 
	 The McLean 1979 effect also has difficulty explaining a number of observations with strong absorptive and mild emissive effects. In the McLean model, the emission dilutes polarized continuum and absorption selectively removes unpolarized line flux. This model has absorption acting on the polarization in an opposite sense to emission. For example, in MWC 480, if the depolarization was acting, the unpolarized flux from the emission would simply create a linear extension in qu-space away from the intrinsic+interstellar polarization (the continuum value which we have set to zero) toward the interstellar polarization value. Since the qu-loop across the emission has a very significant width, and the effect is not centered on the emission, the depolarization effect has difficulty explaining these observations. In the MWC158 HiVIS data, and very clearly in the ESPaDOnS archive data, the spectropolarimetric effect is again inconsistent with the depolarization effect. Since the removal of unpolarized light in absorption will act opposite to the broad depolarization signature, the HiVIS u and archival q observations of figure \ref{fig:swp-mwc158-esp} cannot be simple depolarization. In addition, if somehow the depolarization effect was acting only on the red-shifted emission peak, the removal of unpolarized light in the central absorption would have to change the sign of the spectropolarimetric signature in absorption. This is clearly not the case in the archival u observations. These very narrow linear extensions only in the absorptive component, seen in many stars such as HD 58647, HD 45677 MWC 158 (HiVIS observations) or MWC 147, cannot be the simple depolarization effect as the emission is not causing any detectable change in the polarization properties of the line. 
	 
	 As far as the disk-scattering effect is concerned, since the broad polarization effect is present in both q and u for the red-shifted emission in some stars like MWC 158, a case may be advanced for a combined disk-scattering and depolarization effect. While a number of papers have suggested combining models, none have actually produced quantitative models.  Ignoring the absorptive component, these effects may be argued to represent the ``expanding" disk case of Wood et al. 1993. In that case there is a decrease in polarization at line center with an increase in only the red-shifted emission because the scattered light is entirely red-shifted by the out-flowing scattering particles. One can then claim that the absorptive effects complicate and mask the ``decrease" in central polarization and that the broad effect on the red-shifted side is from the out-flowing disk case. However, one would still expect that the preferential removal of unpolarized light at line-center would act in a sense opposite to the polarization expected for the disk model at that wavelength. However, in several cases antisymmetric profiles are observed in absorption. The antisymmetry cannot be produced by the preferential removal of light from a single occulting region.
	
	HD 163296  and HD 150193 show antisymmetric, blue-shifted components. This presents problems with both the depolarization and disk theories. For disk-scattering theory to work, the scattering particles in the disk must be inflowing (coherently as a disk) to produce a significant blue-shift in the scattered light. The depolarization effect also has difficulties because of the very complex morphology, both profiles having both increases and decreases in q or u, and the lack of any detectible effect on the red-shifted side of the line.

	In the context of depolarization theory, there were stars (MWC 361 and MWC 170) with a broad, centered signatures that are roughly consistent with this theory. If depolarization is the only effect acting, it produces linear extensions in qu-space. MWC 361 did show an additional antisymmetric Stokes u in the central absorption. More sophisticated modeling is necessary to show what the absorptive effects are doing.

	Variability of the absorptive component was consistently seen in the ESPaDOnS observations of HD 163296 and MWC 480. Also, 51 Oph is mildly variable across the entire profile but only at low level. Though HiVIS variability studies can only be done at identical telescope pointings on stars with a large amount of repeat observations, HiVIS data clearly shows variability of the absorptive component in AB Aurigae, MWC 120 and MWC 158. The fractional variability in the absorptive component is much larger than and is not necessarily correlated with the variation in the intensity profile. For instance, the magnitude of the variability is $\sim$0.2\% for HD 163296 over a small spectral range with a corresponding significant change in the intensity profile at the same wavelengths. On the opposite extreme is a variation of about 1\% polarization for MWC 480 in the P-Cygni absorption even though the intensity profile hardly changes structure.

\section{Be \& Emission-Line Star Comparison - Spectroscopy \& Spectropolarimetry}

	After observing so many stars that had very significant polarization effects that did not match the morphology of any current scattering theory, a shorter observing campaign of Be and other emission line stars was done for comparison and cross-checking. There is a much larger set of observations and theoretical work in the literature on Be stars. These stars are quite bright and have been the target of spectropolarimetric work since at least the 1970's. Since these stars are bright and have readily available data and theories, a set of comparative observations was done on these stars in 2007-2008. The extension of the program has provided a very useful comparison between star types. The spectropolarimetry is sigificantly different even between H$_\alpha$ lines that are quite similar.

	Classical Be stars are simply stars of B-type showing emission lines with a wide variety of  sub-types and polarizing mechanisms that must be considered (cf. Porter \& Rivinius 2003). Classical Be stars are rapidly rotating near-main-sequence stars with a gaseous circumstellar disk or envelope. There is no dust present in the disk so the assumed polarizing mechanism is electron scattering. However, the sub-type B[e] stars do have a strong dust signatures. Many of these stars have been resolved by interferometry and show flattened envelopes as well as intrinsic polarization perpendicular to the long-axis of the envelope (Quirrenbach et al. 1997). Even the stars themselves have been rotationally flattened as was observed for Achernar (Domiciano de Souza et al. 2003). 

	In the context of optically thin envelopes around hot stars, a few models to compute the continuum polarization were developed for asymmetric envelopes (Brown \& McLean 1977, McLean \& Brown 1978, Brown et al. 1978). These models were then applied to bright emission-line stars and the depolarization model was created (McLean \& Clark 1979, McLean 1979). These models have been extended and expanded over the years to include such things as finite-sized sources with limb darkening or winds (Cassinelli et al. 1987,  Brown et al. 1989, Hillier 1990, 1991, 1994, 1996). The models have been applied to specific stars to derive geometrical properties of circumstellar material (cf. Wood et al. 1997).

\begin{table}[!h,!t,!b]
\begin{center}
\begin{footnotesize}
\caption{Be/Emission-Line Stellar Properties \label{tab-obsbe}}
\begin{tabular}{lrccc}
\hline
\hline
{\bf Name}             & {\bf HD}  &{\bf MWC}&{\bf V}   & {\bf ST}                \\
\hline
\hline
$\gamma$ Cas    &   5394     &  9          & 2.4      &B0IVpe            \\         
25 Ori                    & 35439     &  110     & 4.9     & B1Vpe               \\
$\eta$ Tau            &  23630    & 74        & 2.9      &B7IIIe                    \\
11 Mon                 & 45725     & 143      & 4.6     & B3Ve                      \\
$\omega$ Ori       & 37490    & 117      & 4.6    & B3IIIe                   \\
Omi Pup                & 63462    & 186      & 4.5     & B1IV:nne             \\
$\kappa$ CMa    & 50013     & 155      & 3.5     & B1.5Ve             \\
$\alpha$ Col       & 37795     & 119      & 2.6      &B7IVe             \\
66 Oph                & 164284    & 278     & 4.8     & B2Ve               \\
31 Peg                & 212076    & 387      & 4.8     & B2IV-Ve               \\
11 Cam              & 32343       &  96      & 5.0       & B2.5Ve             \\
12 Vul                 & 187811    &  323     & 4.9       & B2.5Ve                 \\
HD 36408           & 36408      &             & 5.5     & B7IIIe                  \\
$\lambda$ Cyg  & 198183    & 352     & 4.6     & B5Ve            \\
$\zeta$ Tau        &  37202      & 115     & 3.0    & B2IV              \\
$\kappa$ Dra     & 109387    & 222     & 3.9     & B6IIIpe        \\
QR Vul                 & 192685    &            & 4.8     &  B3Ve              \\
\hline
$\psi$ Per             & 22192     &  69      & 4.3       & B5Ve           \\
10 CMa                 & 48917      &  152    & 5.2    & B2IIIe                     \\
Omi Cas               & 4180    &   8  & 4.5     & B5IIIe                     \\
18 Gem                & 45542    &  141     & 4.1     & B6IIIe                    \\
$\alpha$ Cam    &  30614     & 92    &  4.3     & O9.5Iae             \\
$\beta$ CMi        & 58715     & 178      &  2.9   & B8Ve             \\
C Per                  & 25940    &  81  & 4.0      & B3Ve              \\
$\kappa$ Cas   & 2905     &  7  & 4.2       & B1Iae         \\
$\phi$ And         & 6811     &  420   & 4.3      & B7Ve              \\
MWC 77              & 24479    & 77    & 4.9      & B9.5Ve                \\
$\xi$ Per            & 24912    &     & 4.0     & O7.5IIIe         \\
R Pup              &  68980      & 192   & 4.8  & B1.5IIIe   \\
Phecda            & 103287   &  583    & 2.4  & A0Ve    \\
\hline
3 Pup                  &62623    &  570     & 4.0     & A3Iabe           \\
\hline
$\epsilon$ Aur   & 31964  &              & 3.0  & A8Iab  \\
AC Her                & 170756&             & 7.6    & F4Ibpv \\
V856 Sco            & 144668 &            &   7.0  & A7IVe  \\
SS Lep                & 41511    & 519   & 5.0   & Asph     \\
U Mon                 &  59693    &          & 6.8    & K0Ibpv  \\
89 Her                 &163506   &          & 5.5     & F2Ibe   \\
\hline
\hline
\end{tabular}
\end{footnotesize}
\end{center}
\tablecomments{The Be and Emission-line stars. The columns list the name, HD catalog number, MWC catalog number, V magnitude and spectral type (ST) for each star. All information is from the Simbad Online Database.}
\end{table}

\begin{table}[!h,!t,!b]
\begin{center}
\begin{footnotesize}
\caption{Be and Emission-line Star H$_\alpha$ Results \label{be-res}}
\begin{tabular}{lcccc}
\hline        
\hline
{\bf Name}    &{\bf H$_\alpha$}      &  {\bf Effect?}      & {\bf Mag}            & ${\bf Type}$                  \\
\hline
\hline
$\zeta$ Tau           & 2.7         & Y                                             &  0.5\%                 &  Disk                            \\
MWC 143               &4.8          & Y                                             &  1.1\%                 &  Disk                            \\
25 Ori                     &2.4         & Y                                            &  0.5\%                  &  Disk                           \\
$\psi$ Per             &5.8          & Y                                             &  0.5\%                 &  Disk                            \\      
10 CMa                  &4.8         & Y                                             &  0.5\%                 &  Disk*                            \\
$\kappa$ CMa      &4.7         & Y                                             &  0.2\%                 &  Disk*                            \\
$\kappa$ Dra      & 4              & Y                                             &  0.2\%                 &  Disk*                            \\
Omi Cas                &6.4          & Y                                             &  0.2\%                 &  Disk*                            \\
Omi Pup                 &2.2          & Y                                             &  0.2\%                 &  Disk*                            \\
$\gamma$ Cas    &4.4          & Y                                             &  0.2\%                 &  Disk*                          \\
\hline
\hline
18 Gem                  &1.5         & Y                                           &   0.2\%                     & Disk                               \\
$\eta$ Tau             &2.2         & Y                                             &  0.15\%                 &  Disk                            \\
$\beta$ CMi           &7.4          & Y                                             &  0.07\%               &  Disk                            \\
$\alpha$ Col          &2.7         & Y                                             & 0.1\%                 &  Disk                             \\
R Pup                      & 7           &  Y                                            & 0.2\%                  &Disk*                            \\
\hline
\hline
31 Peg                   & 6.4         & N                                           &  $<$0.1\%                           &   Emis                            \\
11 Cam                   &7.5         & N                                            &   $<$0.1\%                          &   Emis                         \\
C Per                      &7.0          &  N                                         &     $<$0.1\%                        &  Emis                            \\
12 Vul                    & 1.3          & N                                            &   $<$0.1\%                         &   Disk                              \\
$\omega$ Ori        &2.2          &  N                                           &  $<$0.1\%                         &   Disk                            \\
66 Oph                   &2.1          & N                                           &    $<$0.1\%                           &   Disk                          \\
$\phi$ And             &1.9       & N                                            &     $<$0.1\%                          &  Disk**                           \\
MWC 77                &0.8          & N                                           &     $<$0.1\%                          & Disk**                       \\
HD 36408             &0.8          & N                                            &    $<$0.1\%                           & Disk**                        \\
QR Vul                  & 0.75         & N                                          &     $<$0.1\%                        &   Abs                             \\
$\lambda$ Cyg     & 0.7         & N                                           &     $<$0.1\%                         &   Abs                             \\
$\xi$ Per                 &0.85       & N                                            &     $<$0.1\%                          &  Abs                           \\
Phecda                  & 0.6          & N                                            &    $<$0.1\%                            & Abs                            \\
MWC 92                  &1.35       &  N                                           &     $<$0.1\%                         &  Other                            \\
$\kappa$ Cas       &1.5           &    N                                          &    $<$0.1\%                          &  Other                            \\ 
\hline
3 Pup                     & 3              & Y                                              &  2.5\%                         & Disk        \\
\hline
$\epsilon$ Aur   & 1.1             & Y                                             &  1.0\%                             &  \\
AC Her                & 2.0              & Y                                            & 0.8\%                             & \\
V856 Sco            & 2.1              & Y                                            & 1.5\%                            &  \\
SS Lep                & 2.5              & Y                                            & 0.8\%                            &  \\
U Mon                 &  2.9             & Y                                             & 0.5\%                            &  \\
89 Her                 &1.3               & Y                                            & 3.5\%                            &  \\
\hline
\hline
\end{tabular}
\end{footnotesize}
\end{center}
\tablecomments{The columns show the name, H$_\alpha$ line strengths, presence of a detectable spectropolarimetric line effect (Effect?), magnitude of the polarization effect (Mag) and H$_\alpha$ line morphology (Type). Where there were no detections, upper limits are typically 0.05\% to 0.1\% depending on observing conditions and binning. The top portion shows the 10 broad-signature spectropolarimetric detections. The middle portion shows the 5 smaller, more complex detections. The lower portion shows the non-detections.}
\end{table}

\subsection{Be Star H$_\alpha$ Line Profiles}

\begin{figure*}
\begin{center}
\includegraphics[width=0.23\linewidth, angle=90]{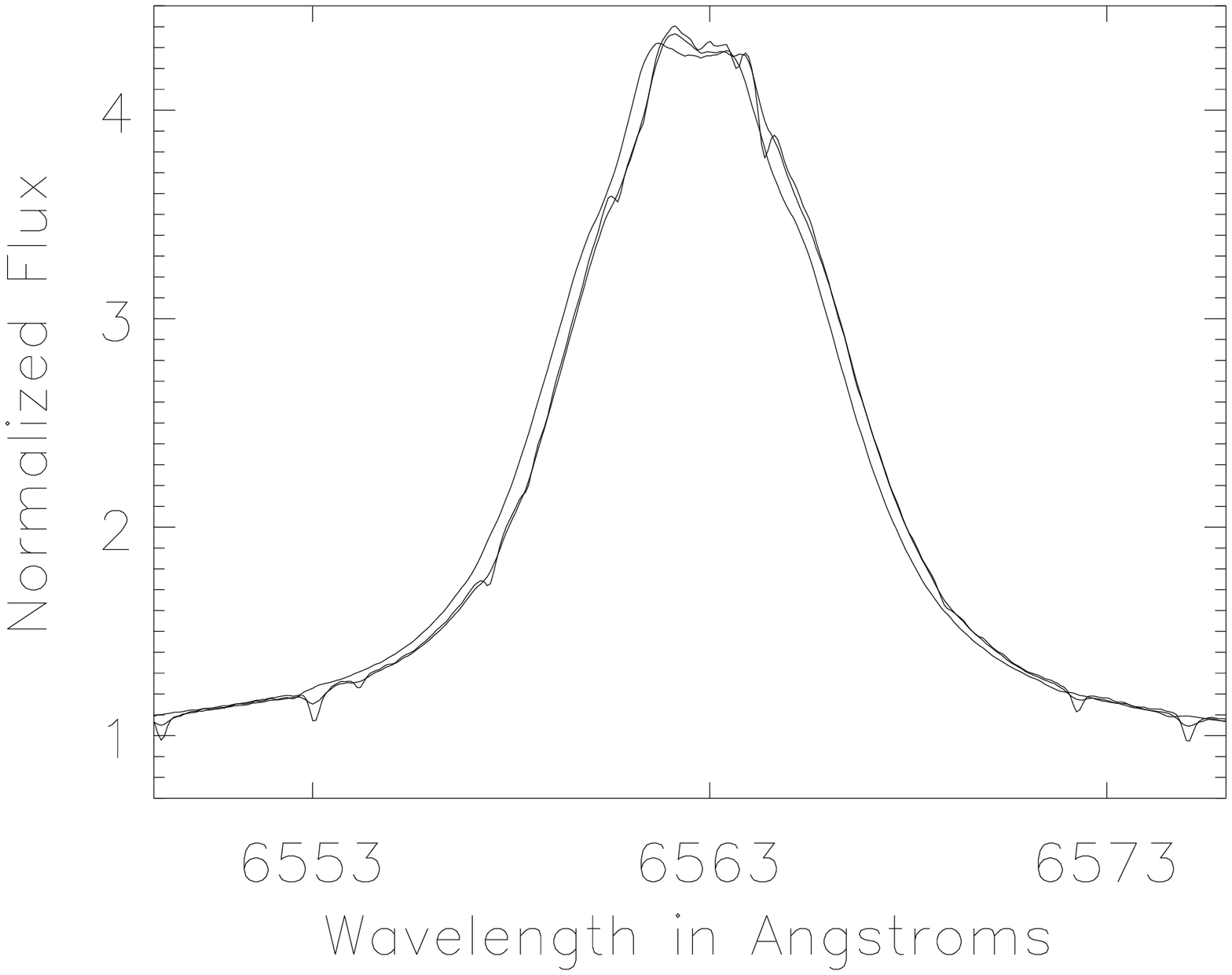}
\includegraphics[width=0.23\linewidth, angle=90]{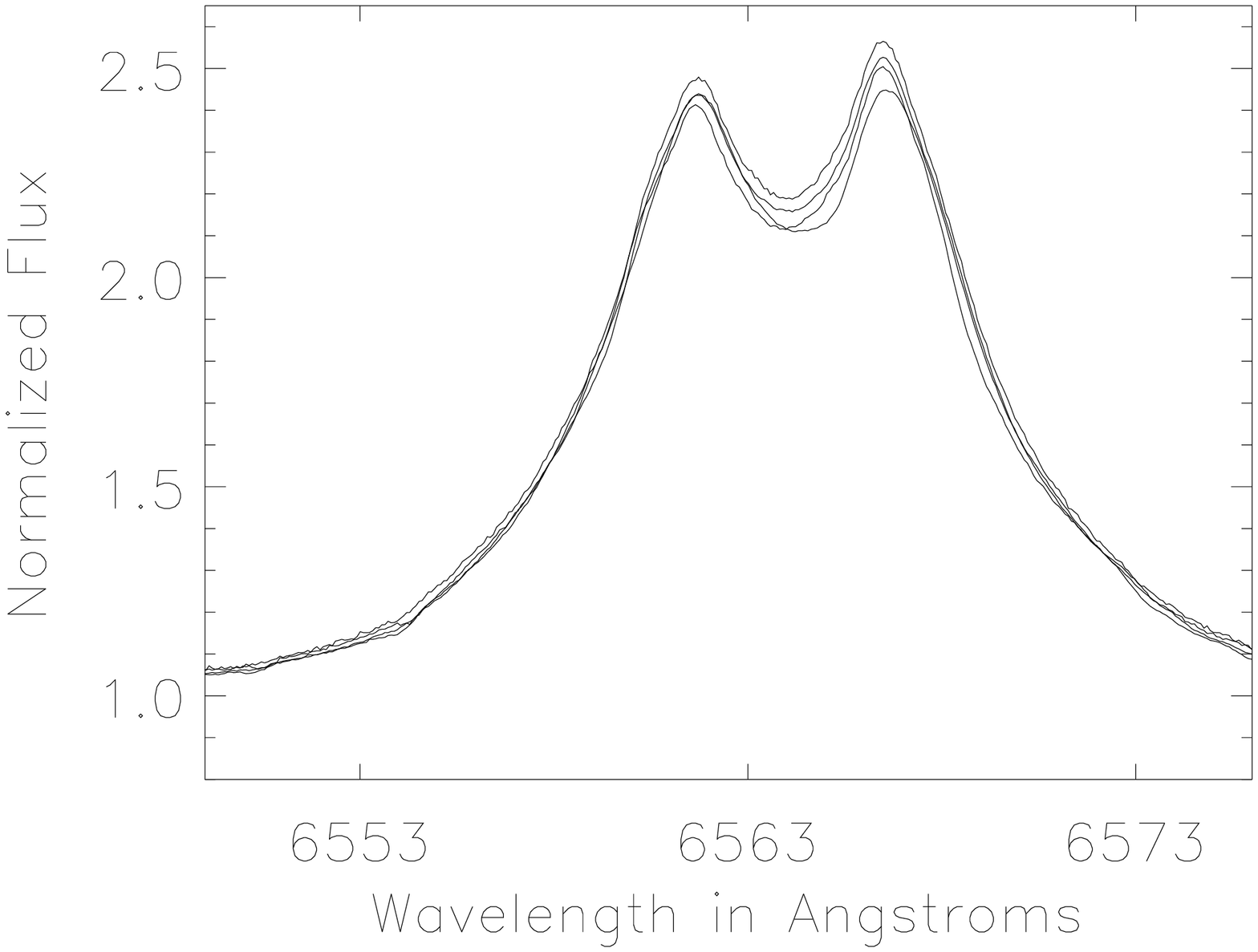}
\includegraphics[width=0.23\linewidth, angle=90]{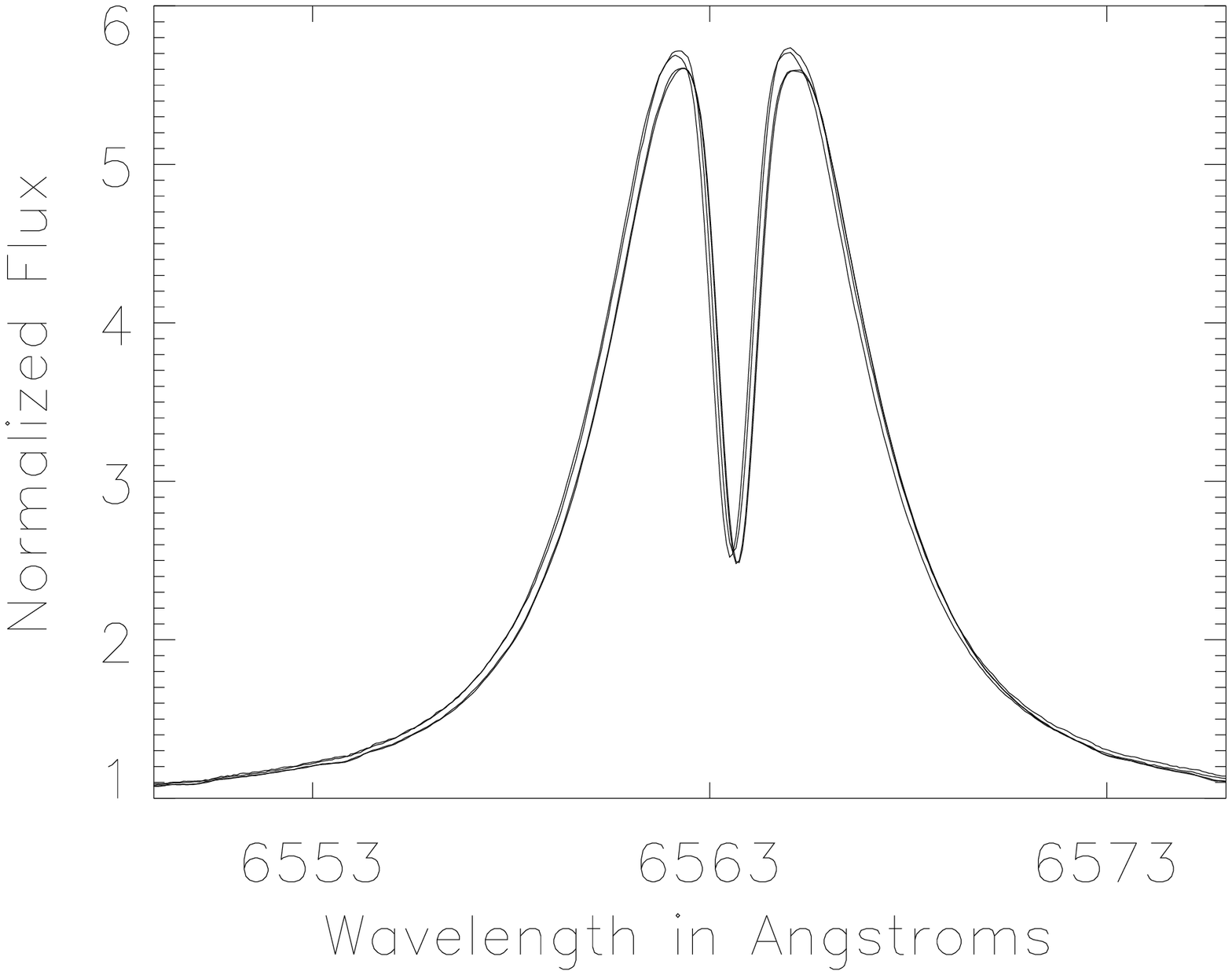} \\
\includegraphics[width=0.23\linewidth, angle=90]{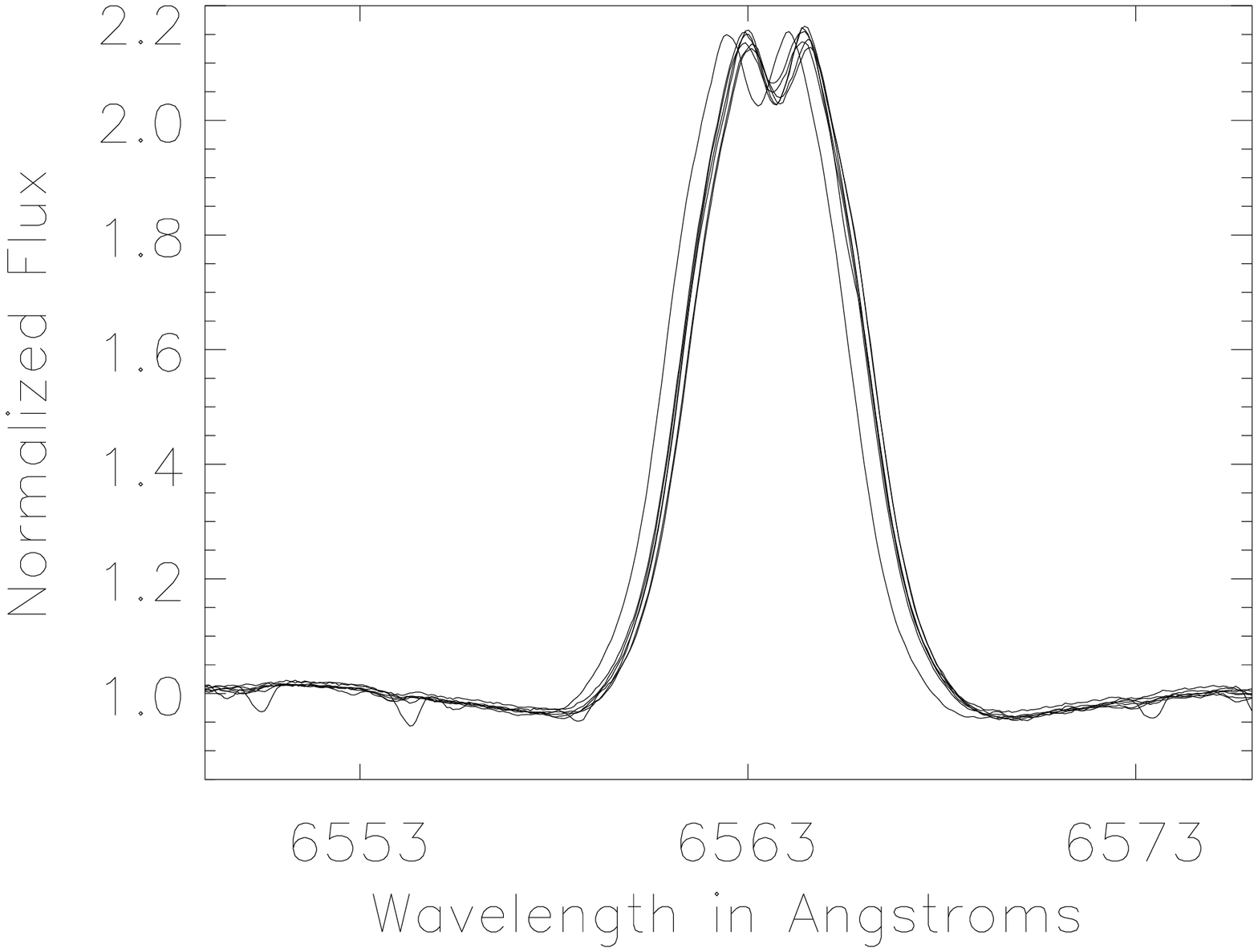}
\includegraphics[width=0.23\linewidth, angle=90]{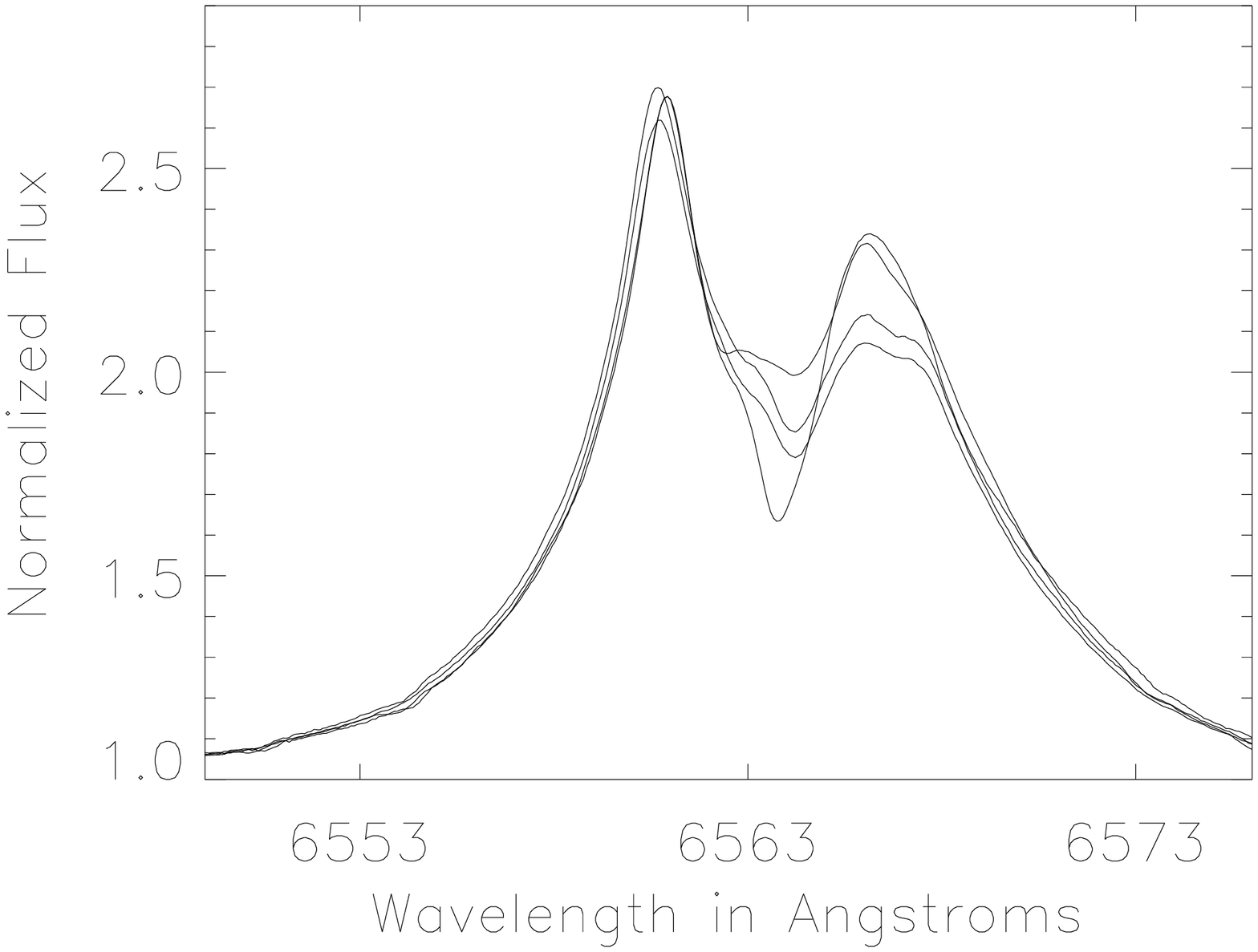}
\includegraphics[width=0.23\linewidth, angle=90]{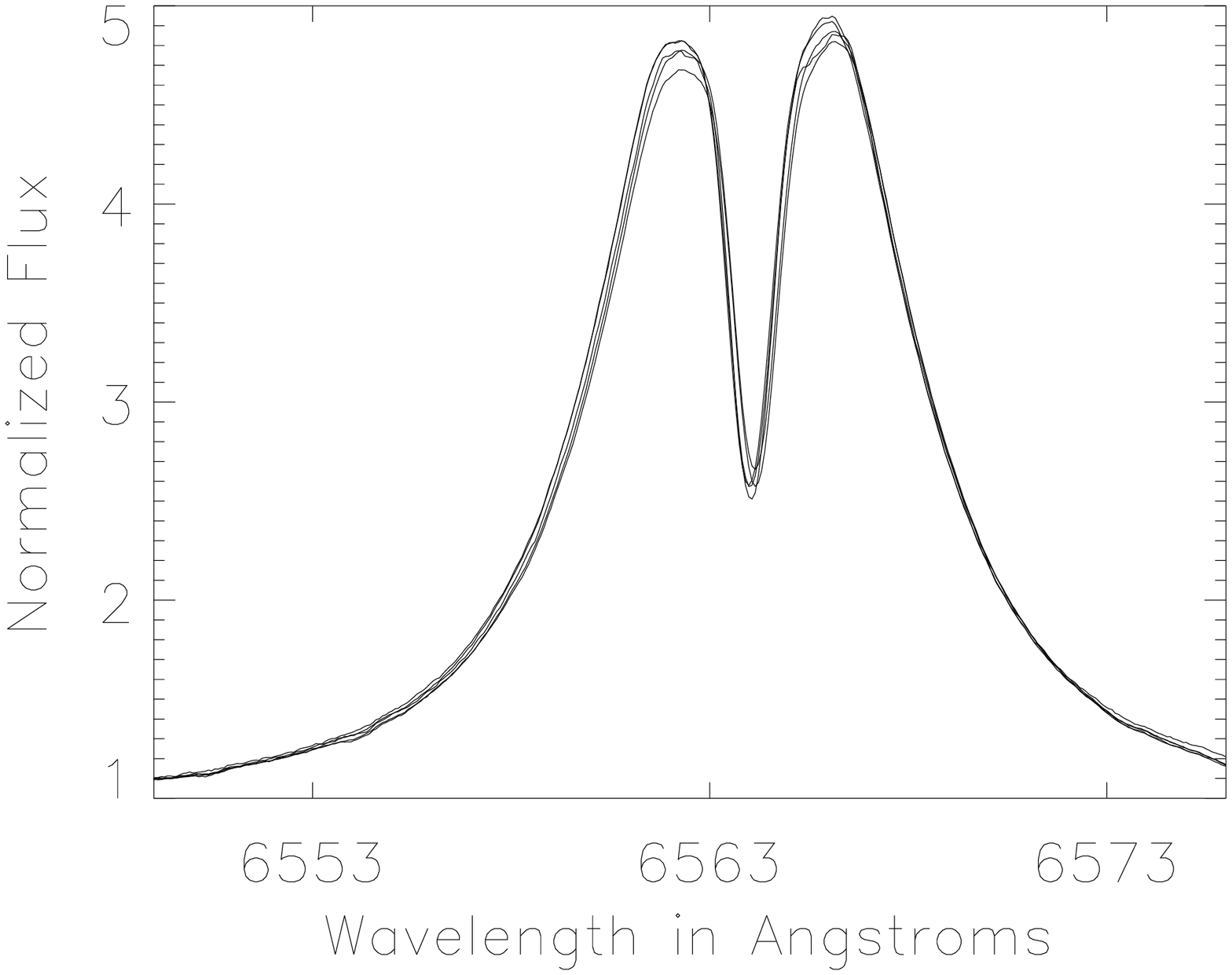} \\
\includegraphics[width=0.23\linewidth, angle=90]{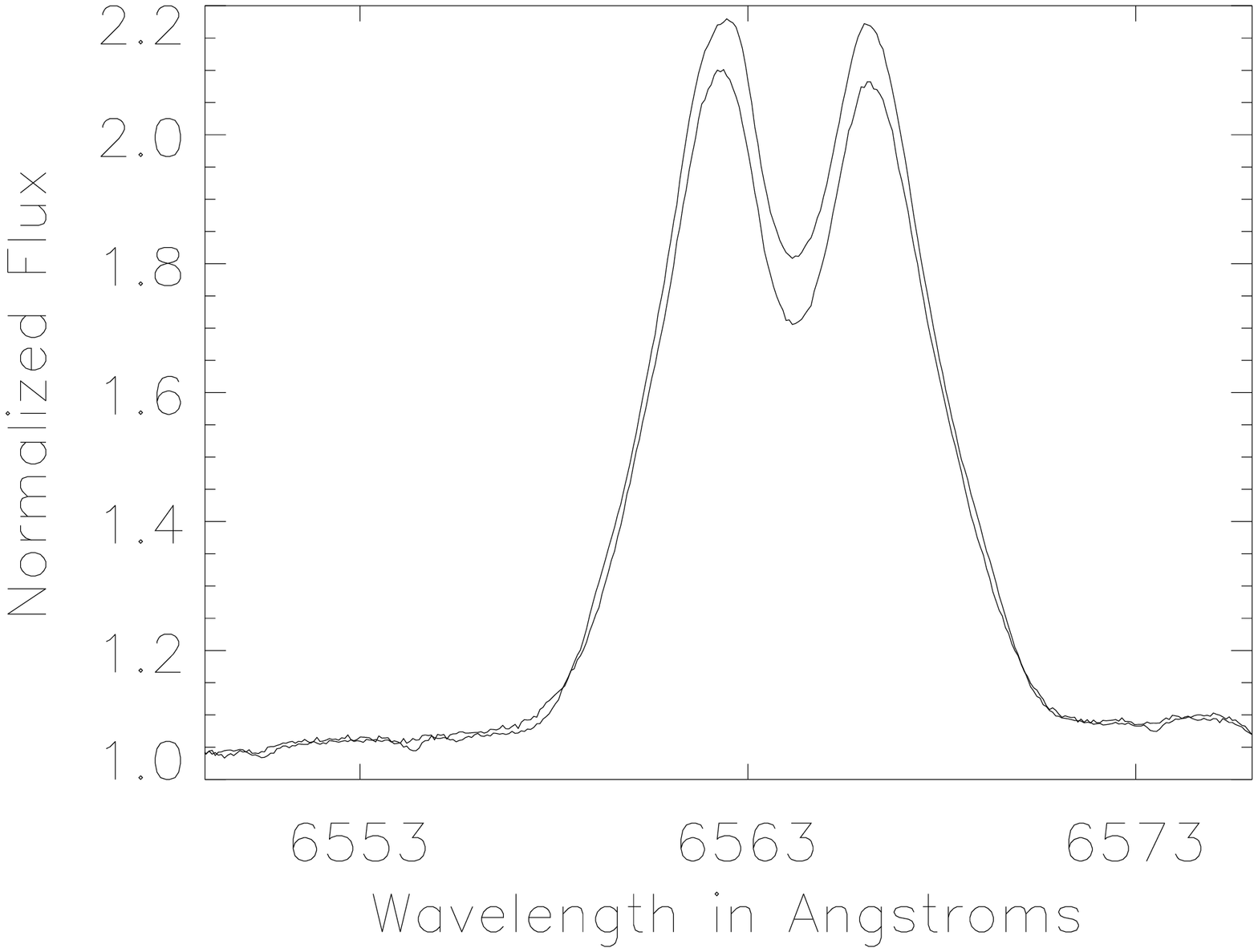}
\includegraphics[width=0.23\linewidth, angle=90]{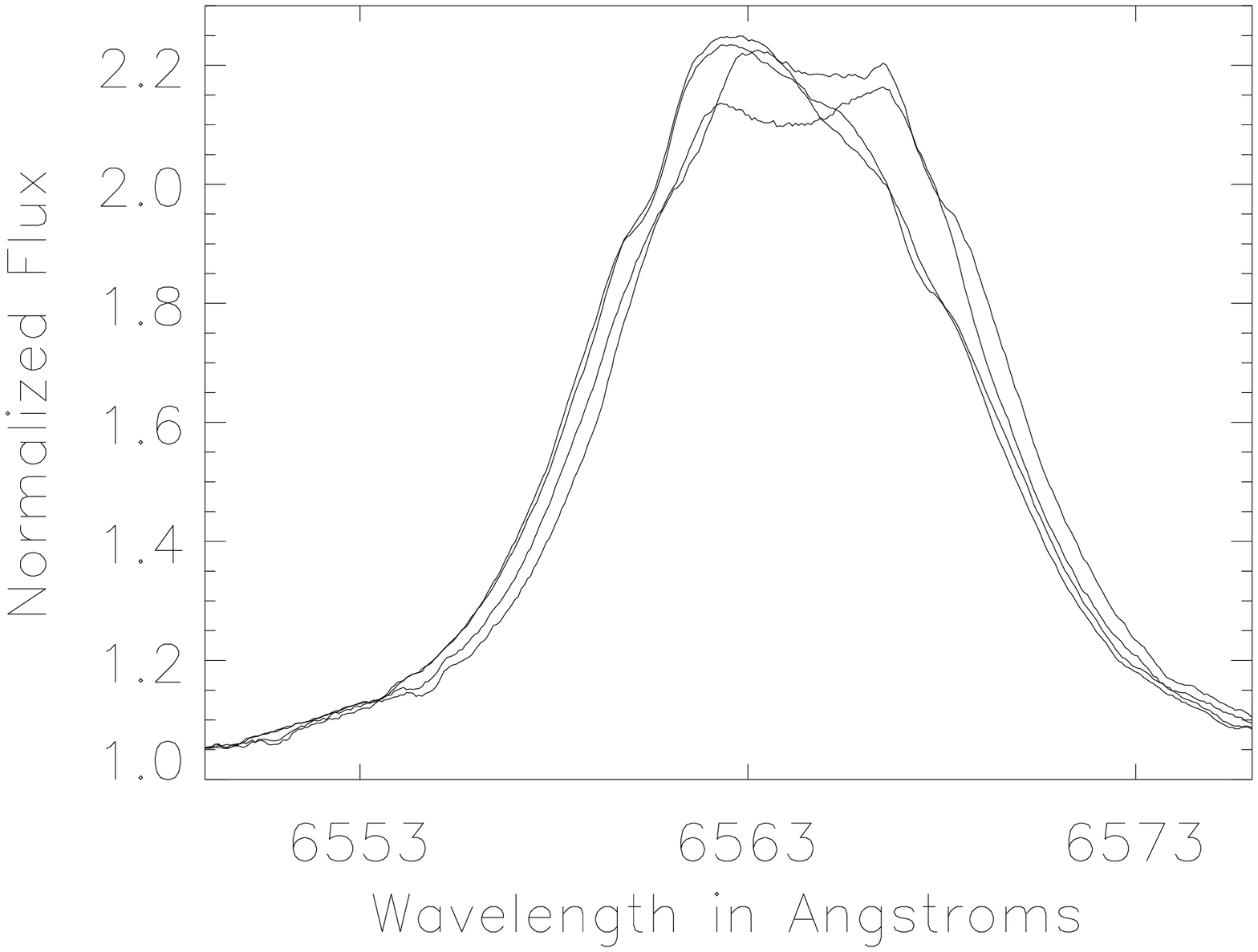}
\includegraphics[width=0.23\linewidth, angle=90]{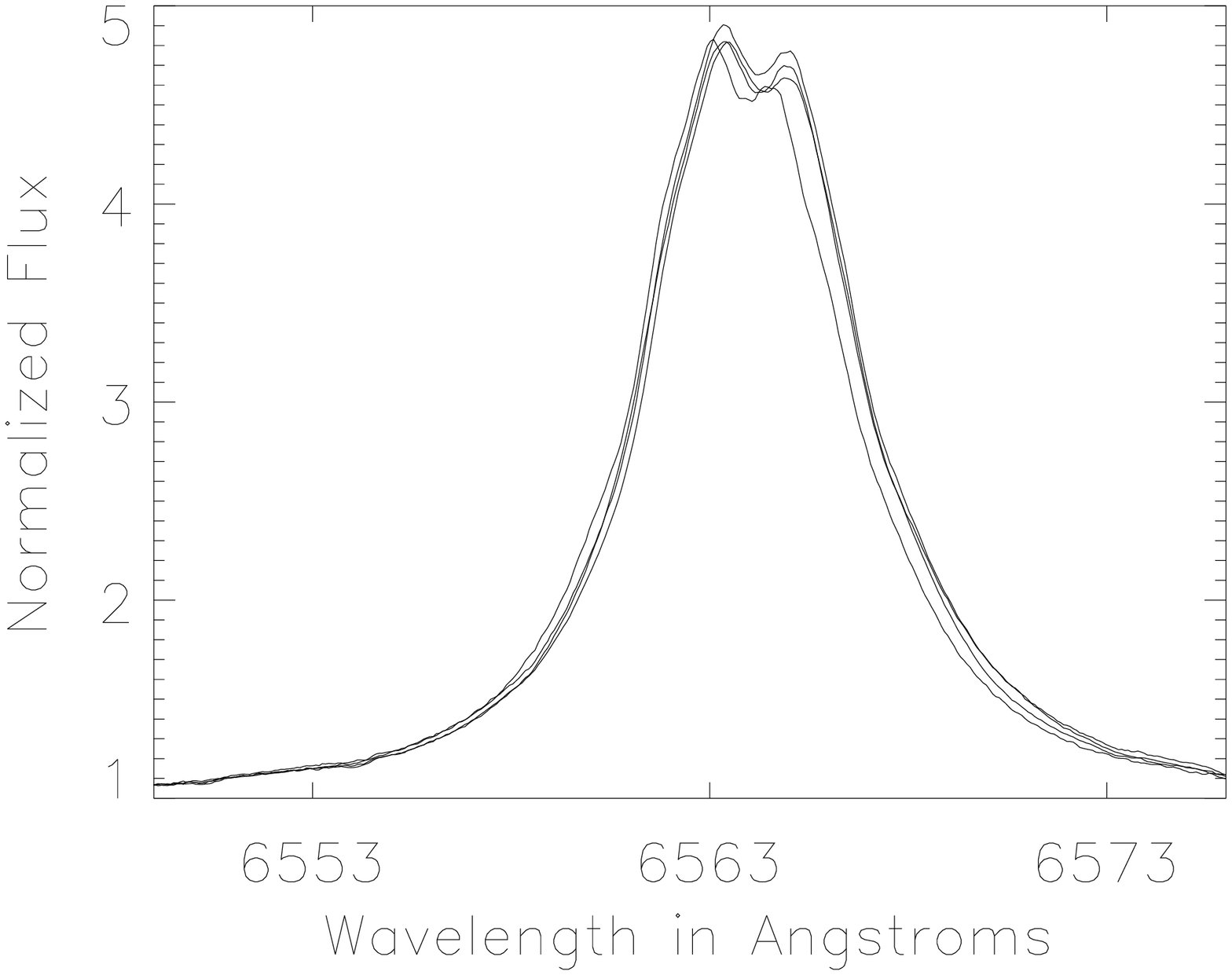} \\
\includegraphics[width=0.23\linewidth, angle=90]{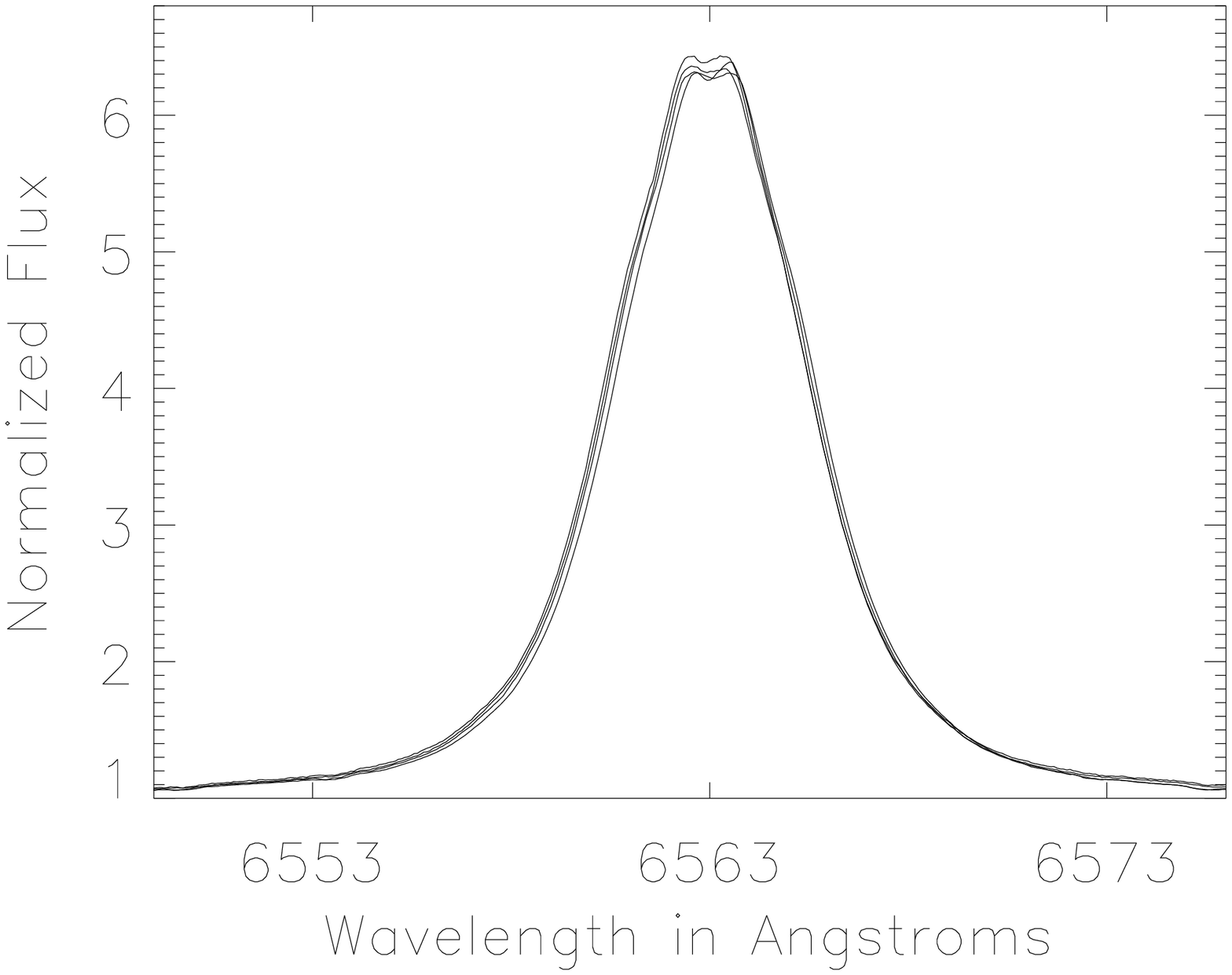}
\includegraphics[width=0.23\linewidth, angle=90]{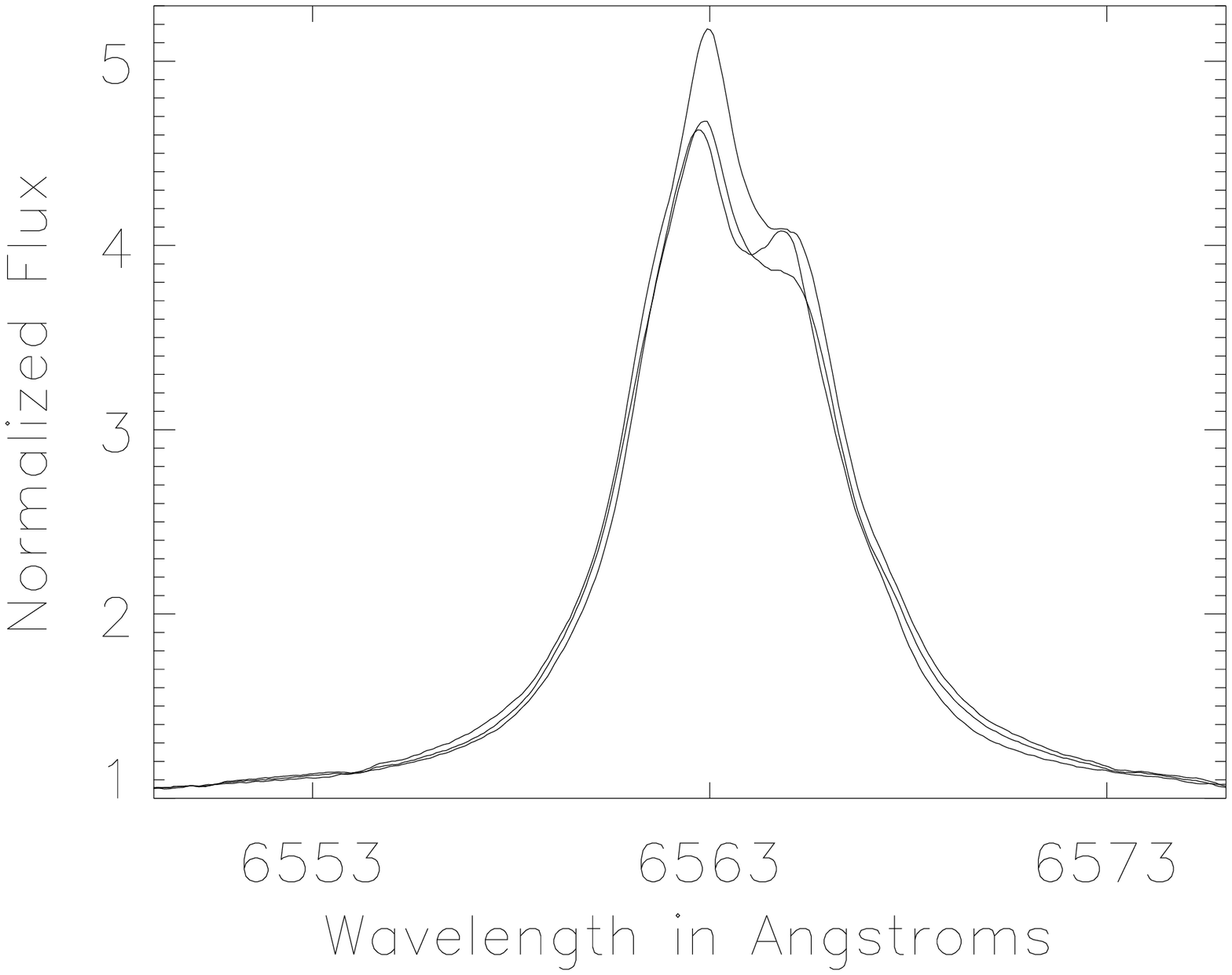}
\includegraphics[width=0.23\linewidth, angle=90]{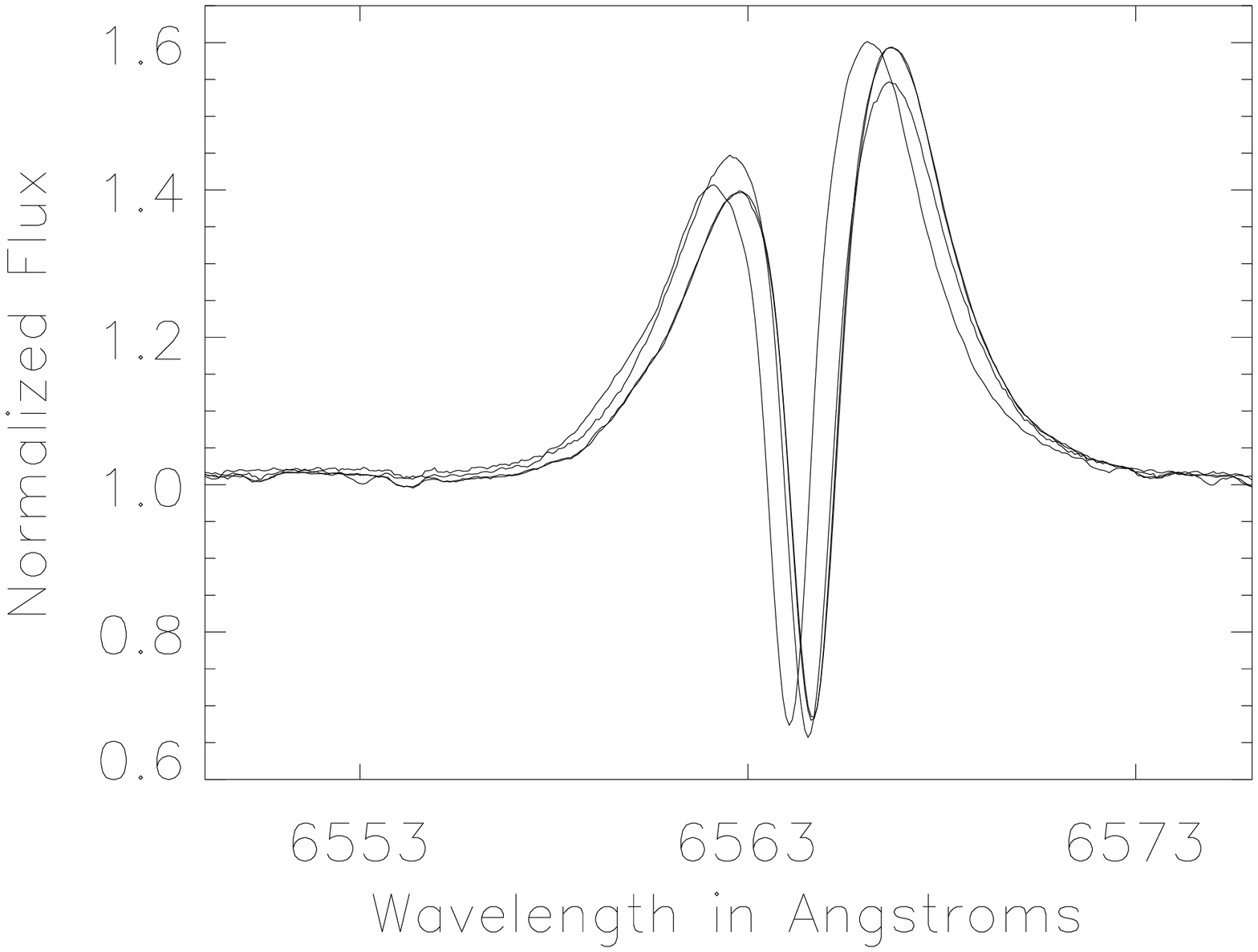} \\
\includegraphics[width=0.23\linewidth, angle=90]{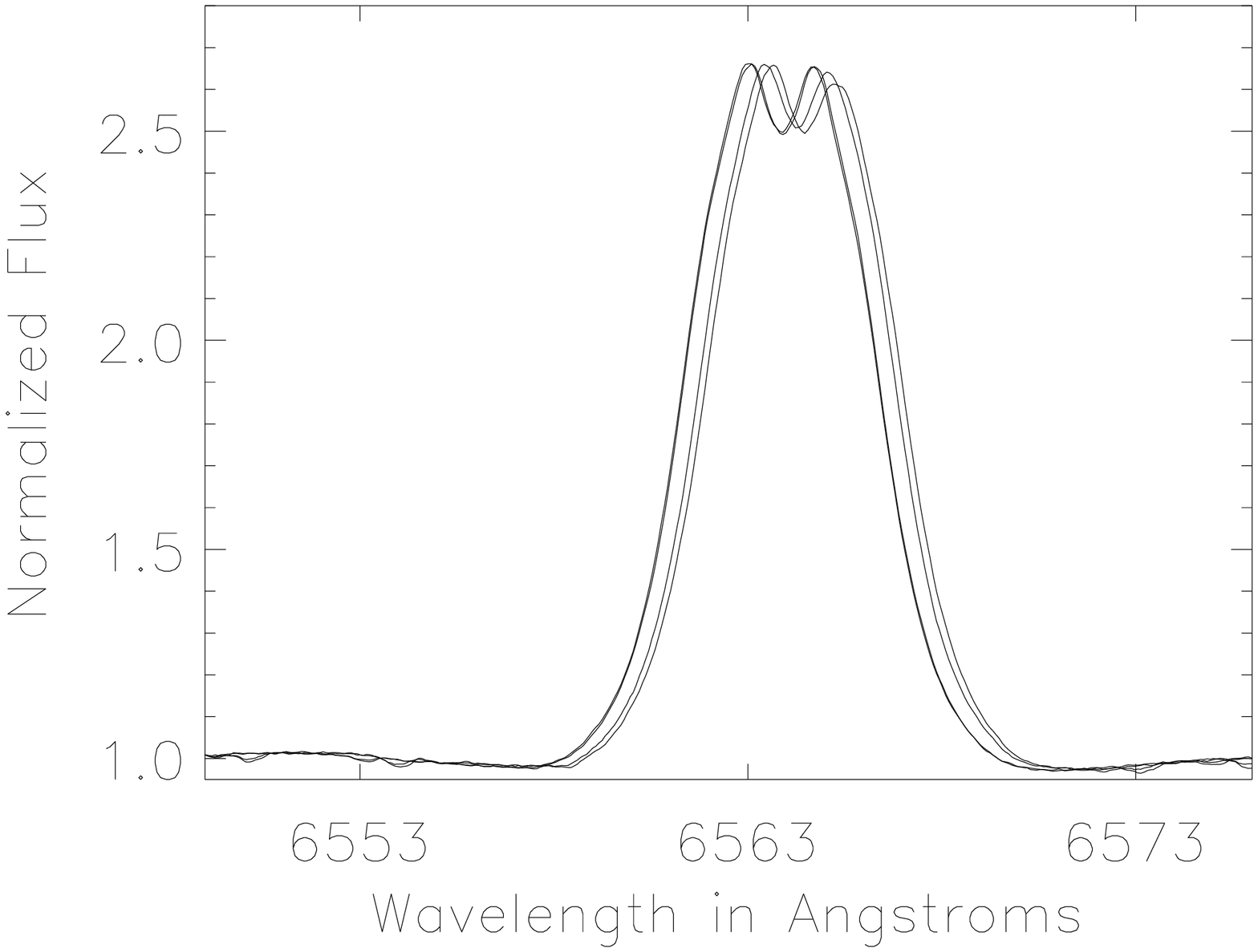}
\includegraphics[width=0.23\linewidth, angle=90]{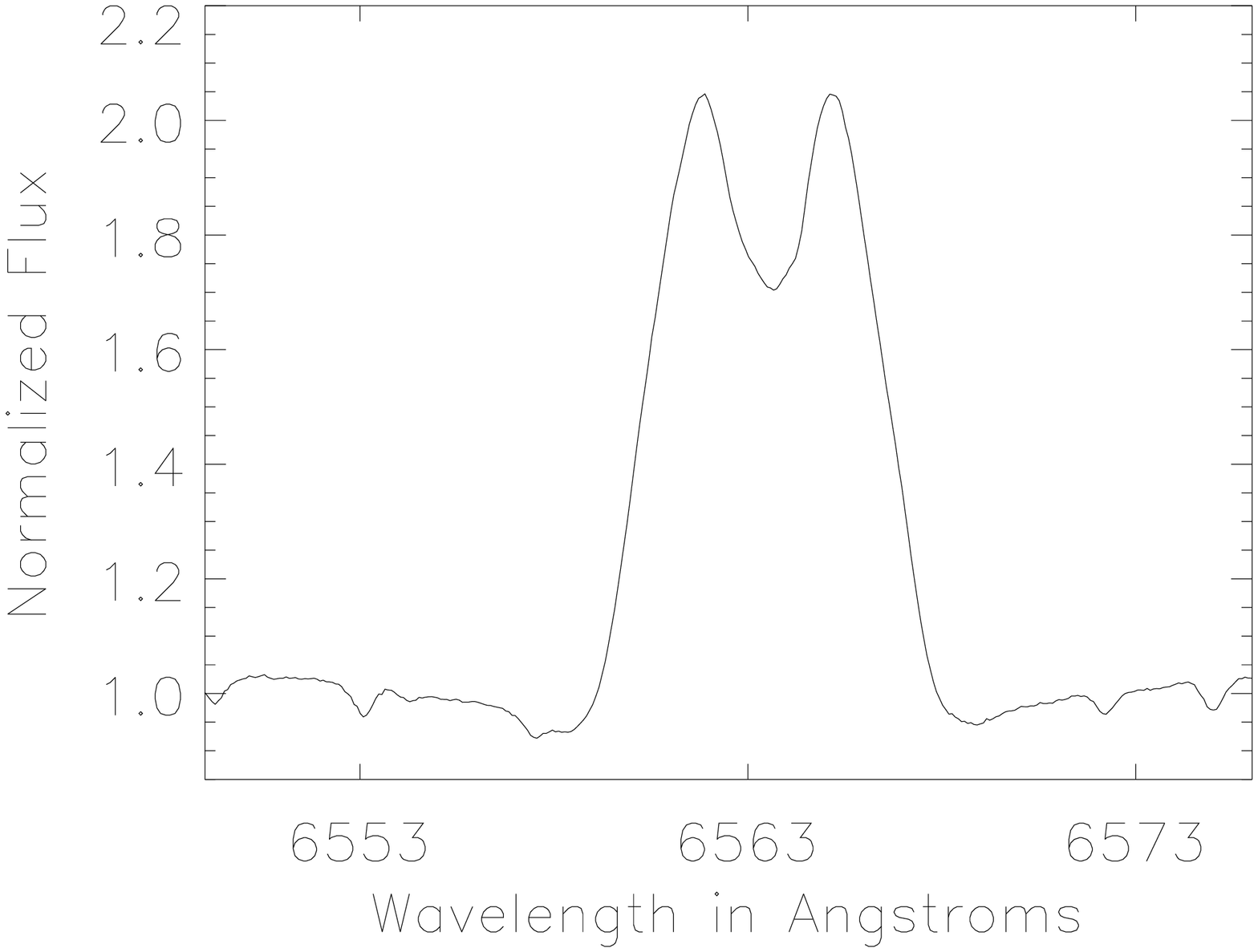}
\includegraphics[width=0.23\linewidth, angle=90]{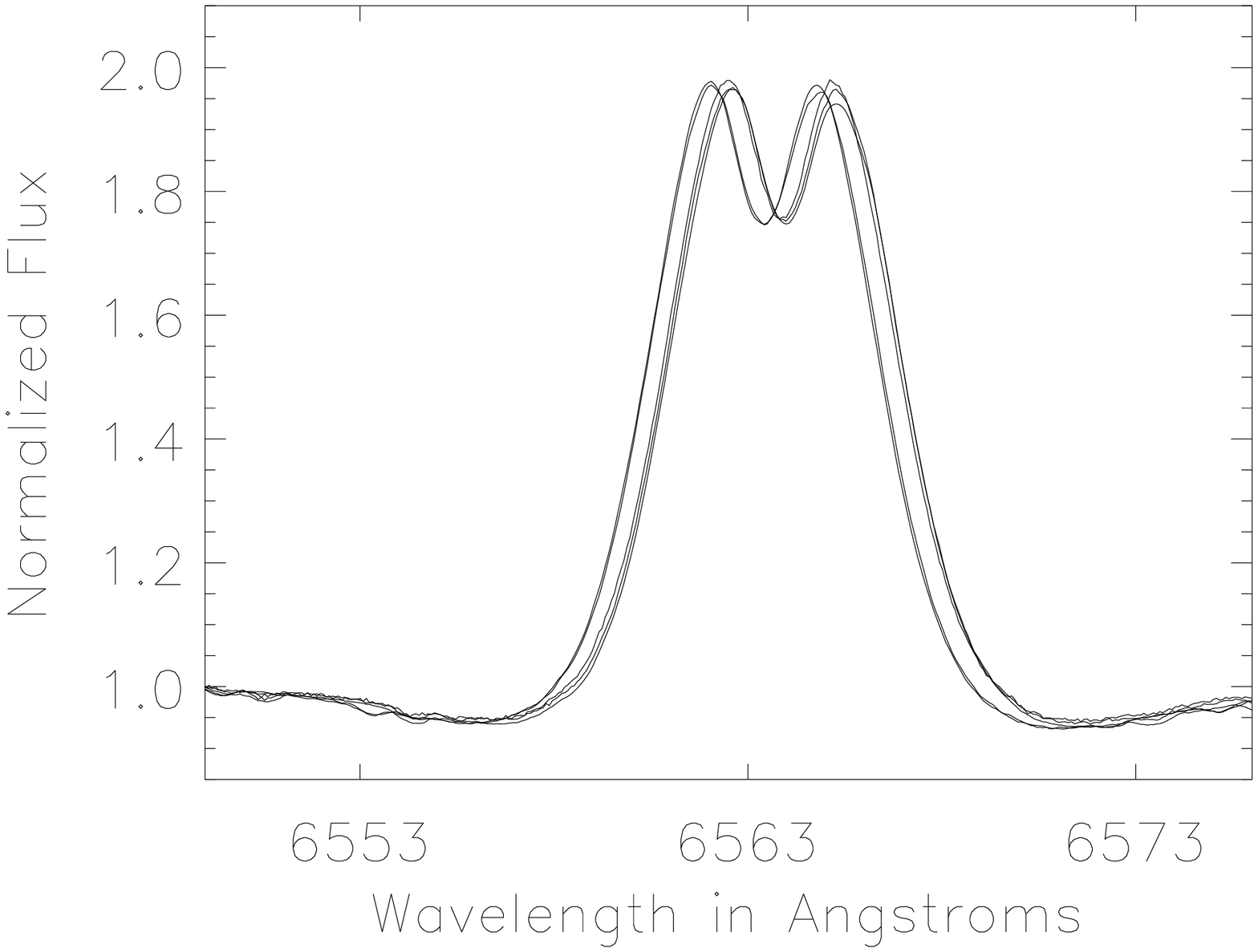}
\caption{Be Line Profiles I. The stars, from left to right, are: $\gamma$ Cas, 25 Ori, $\psi$ Per, $\eta$ Tau, $\zeta$ Tau, MWC 143, $\omega$ Ori, Omi Pup, 10 CMa, Omi Cas, $\kappa$ CMa, 18 Gem, $\alpha$ Col, 66 Oph and $\beta$ CMi.}
\label{fig:be-lprof1}
\end{center}
\end{figure*}

\begin{figure*}
\includegraphics[width=0.23\linewidth, angle=90]{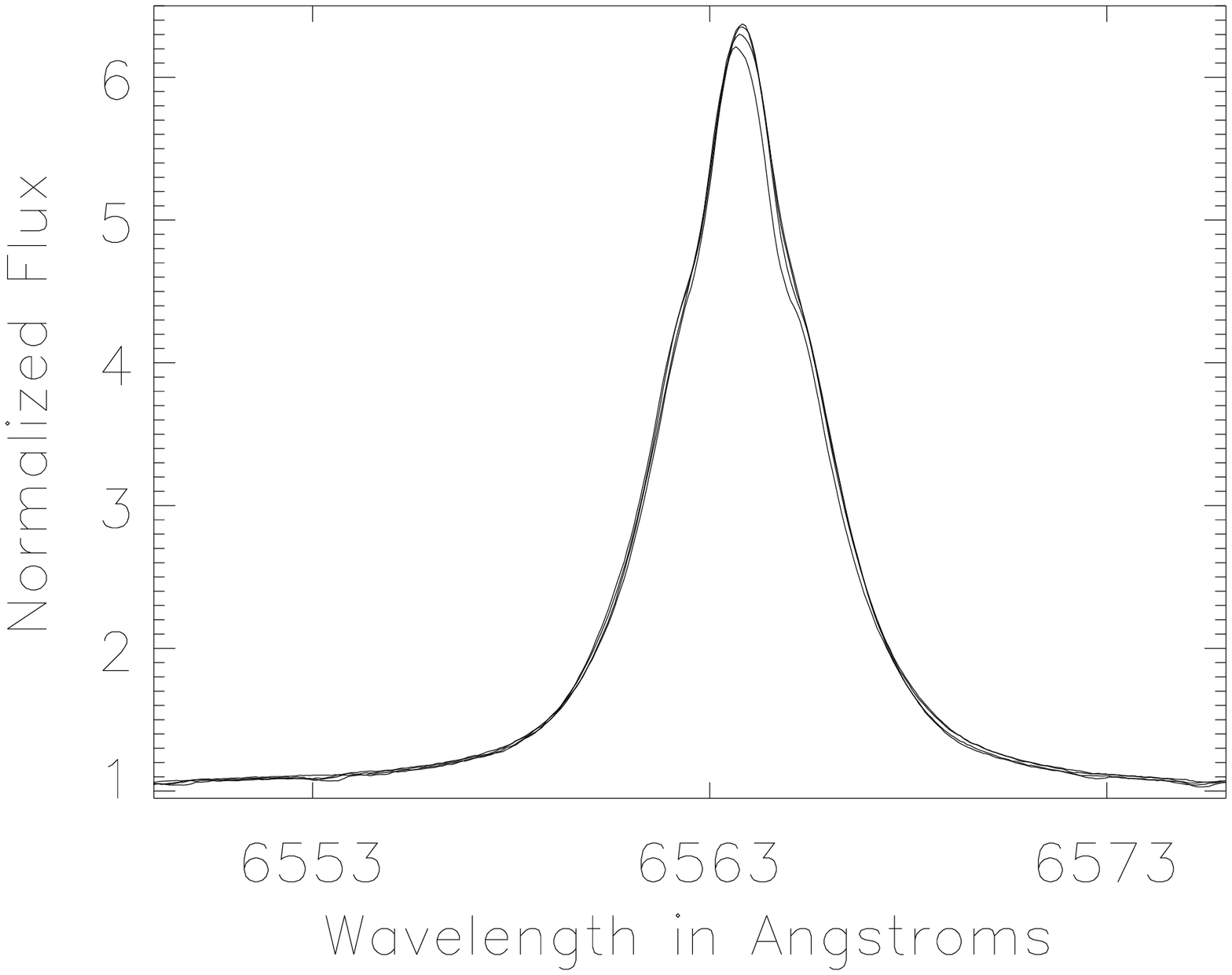}
\includegraphics[width=0.23\linewidth, angle=90]{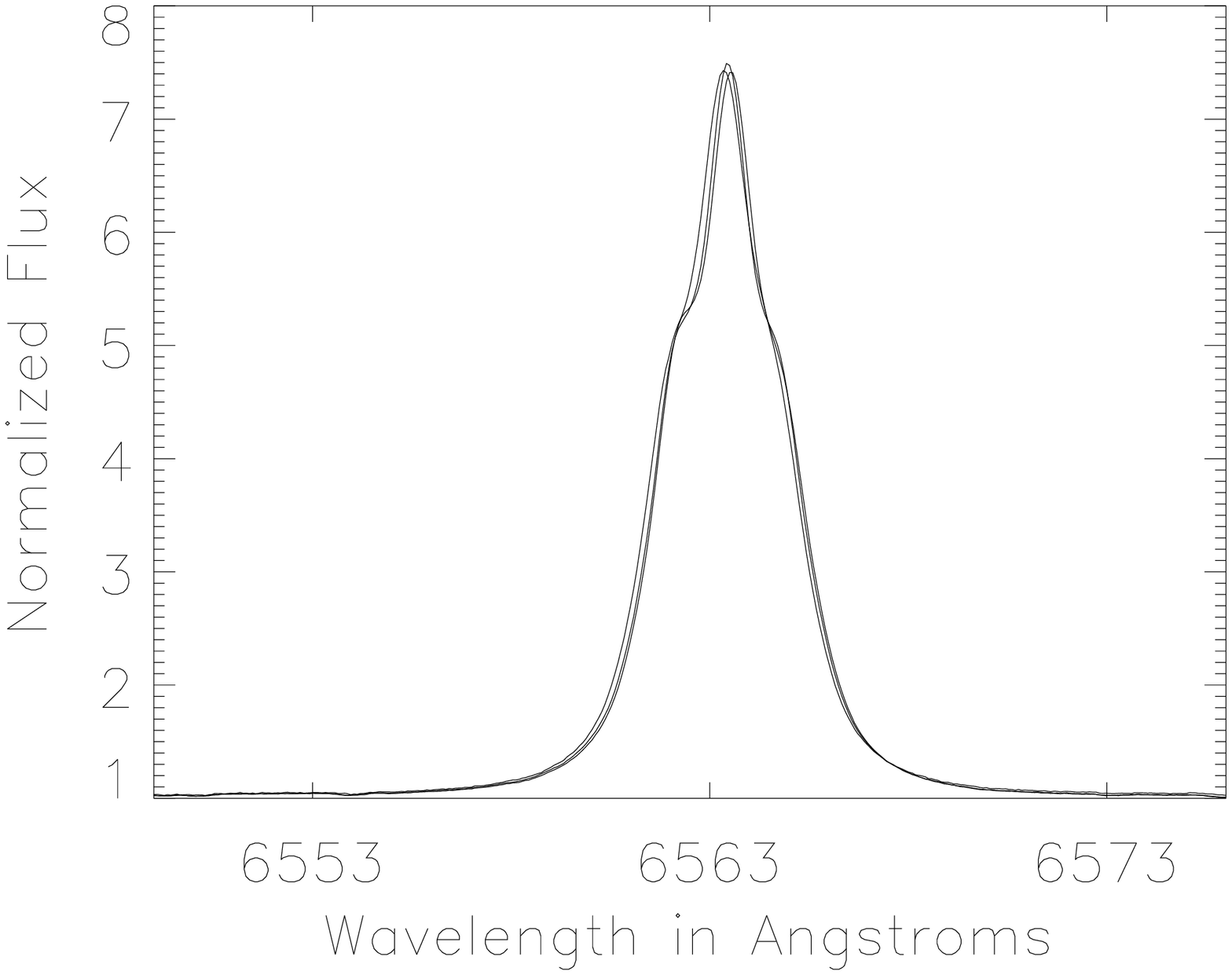}
\includegraphics[width=0.23\linewidth, angle=90]{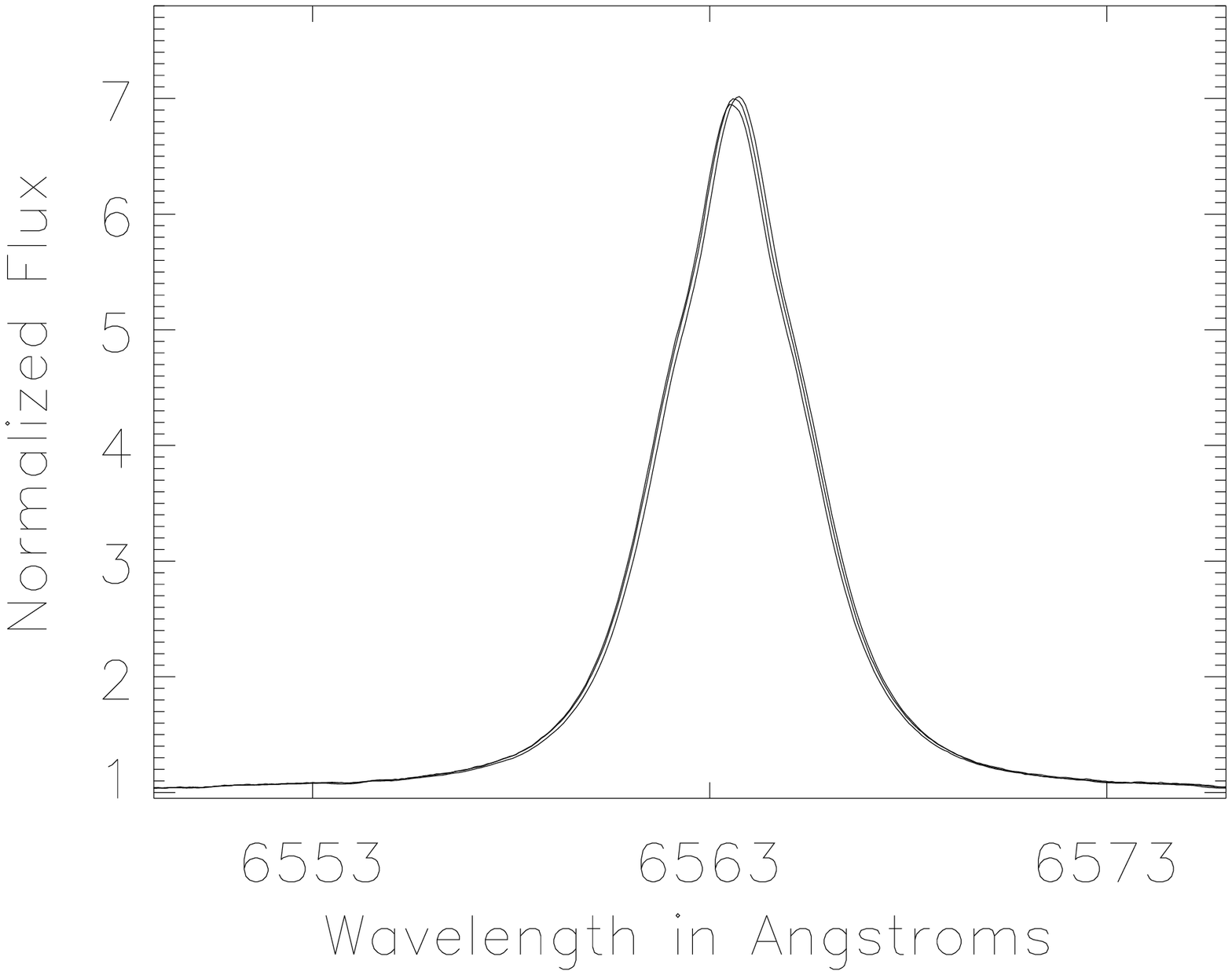} \\
\includegraphics[width=0.23\linewidth, angle=90]{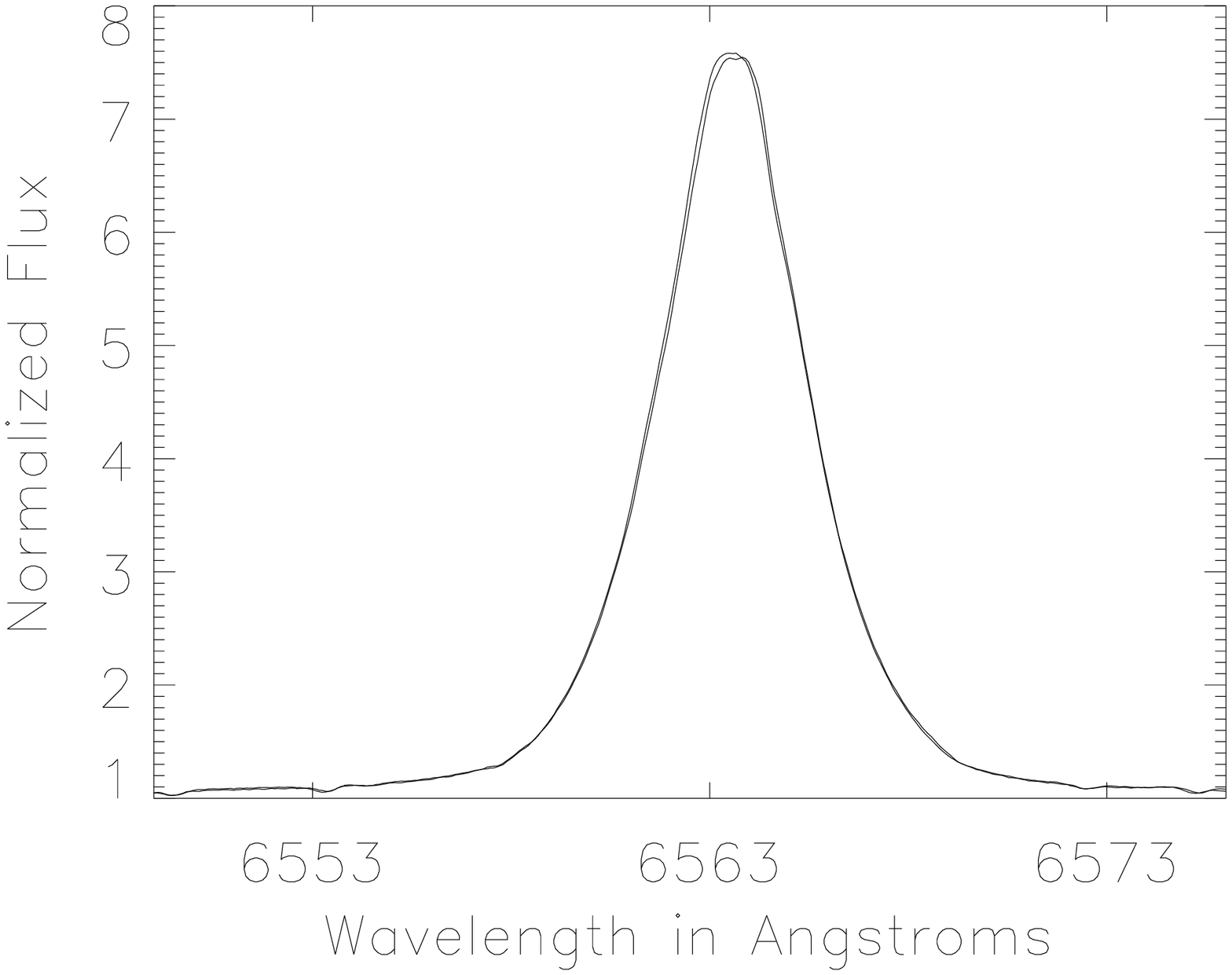}
\includegraphics[width=0.23\linewidth, angle=90]{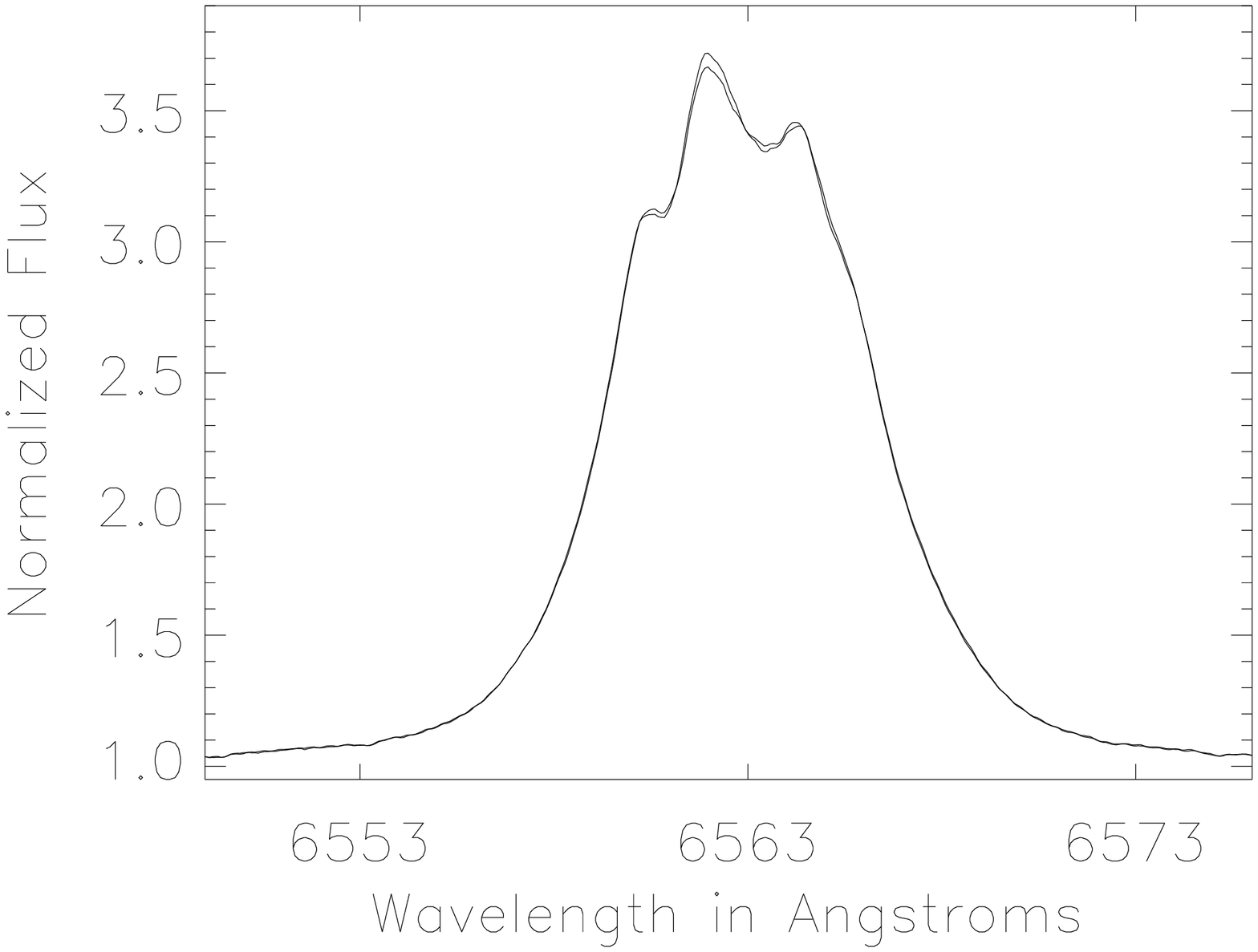}
\includegraphics[width=0.23\linewidth, angle=90]{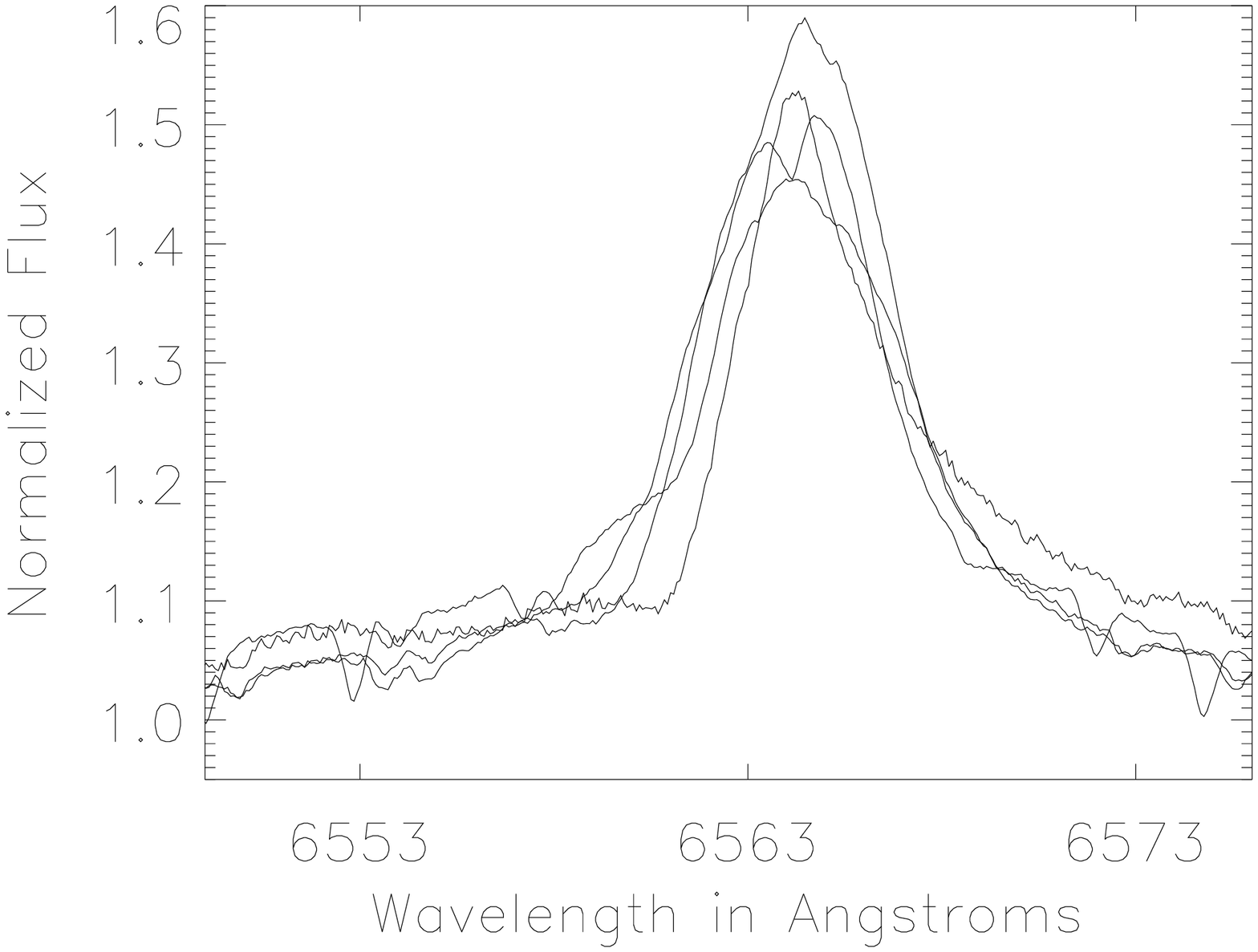} \\
\includegraphics[width=0.23\linewidth, angle=90]{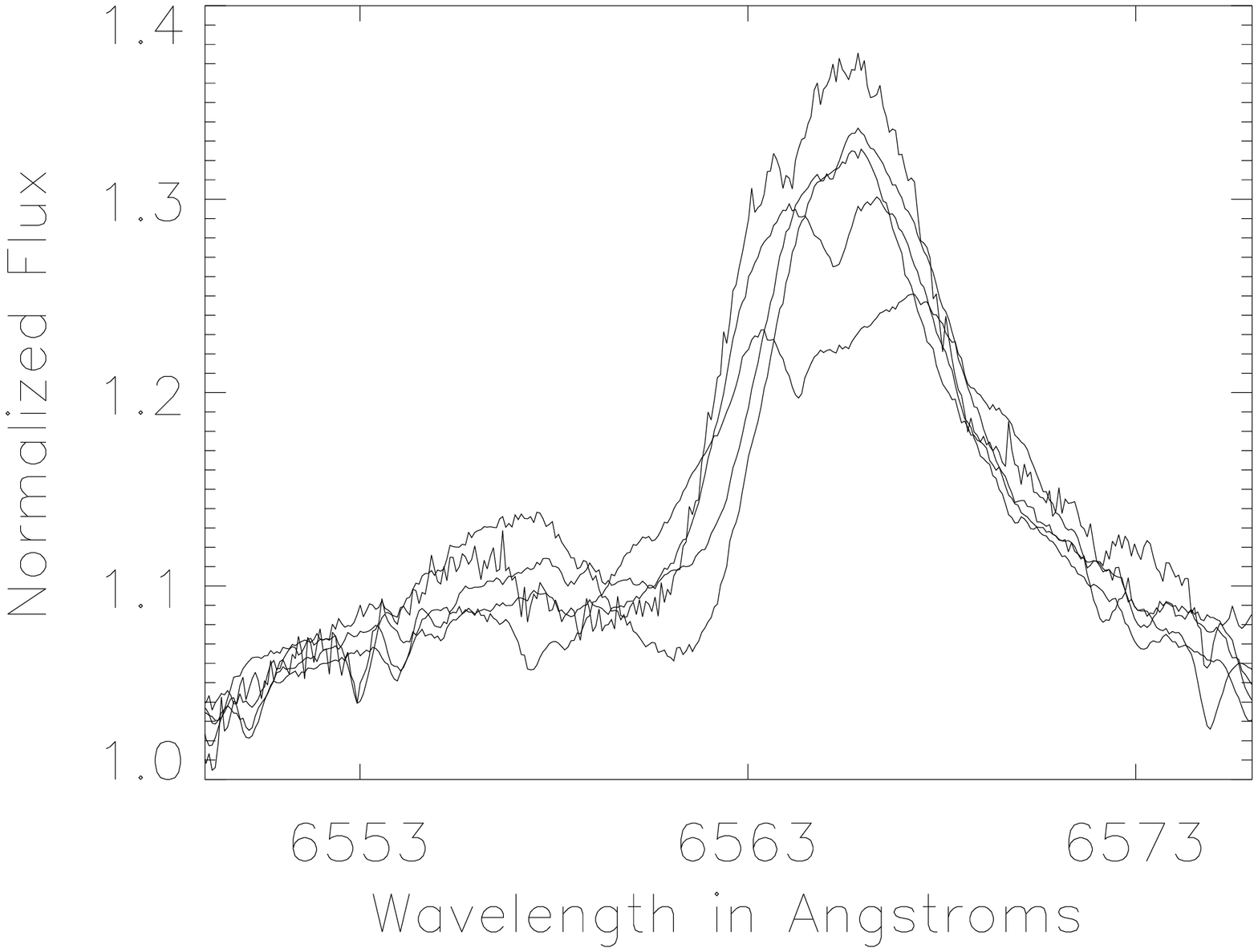}
\includegraphics[width=0.23\linewidth, angle=90]{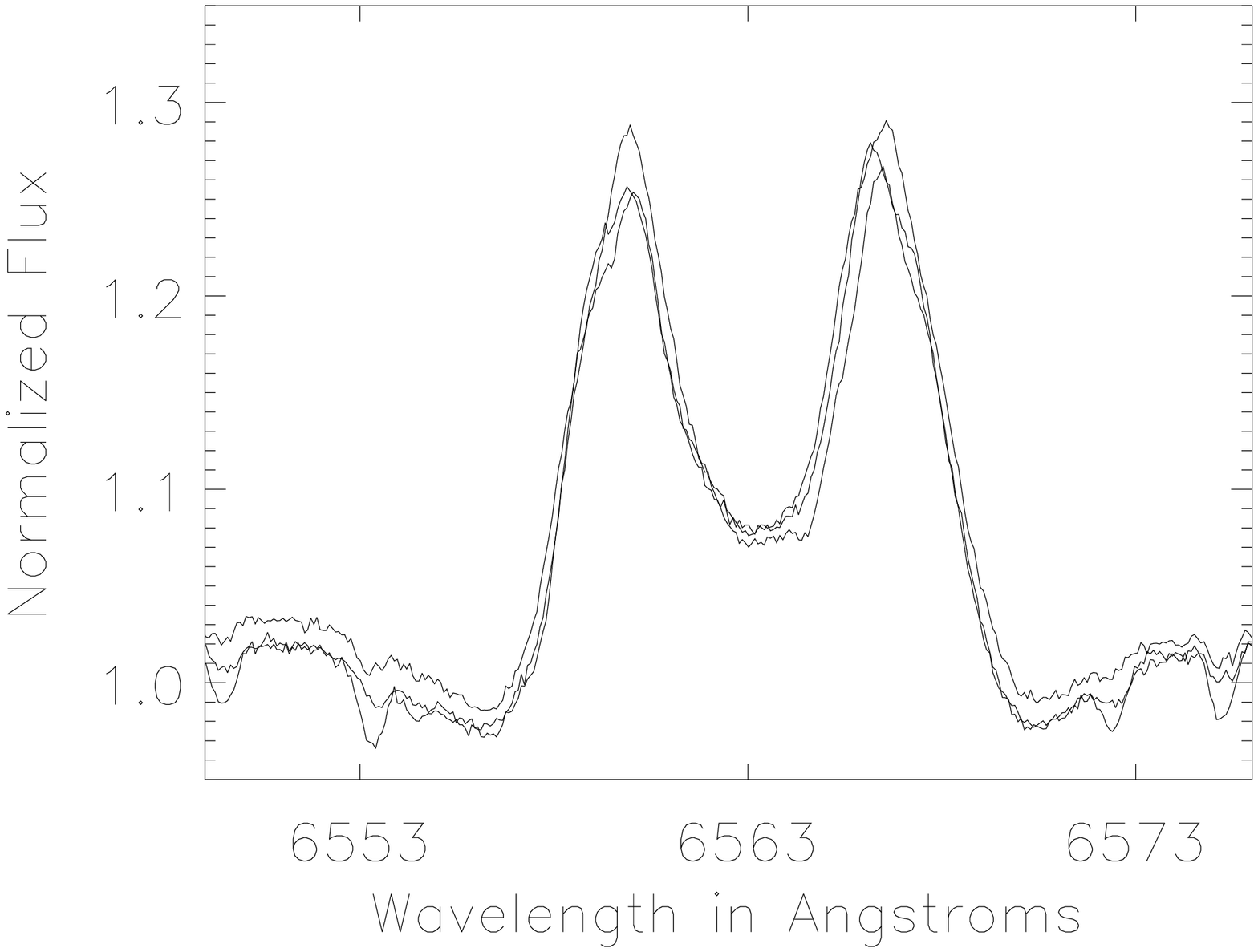}
\includegraphics[width=0.23\linewidth, angle=90]{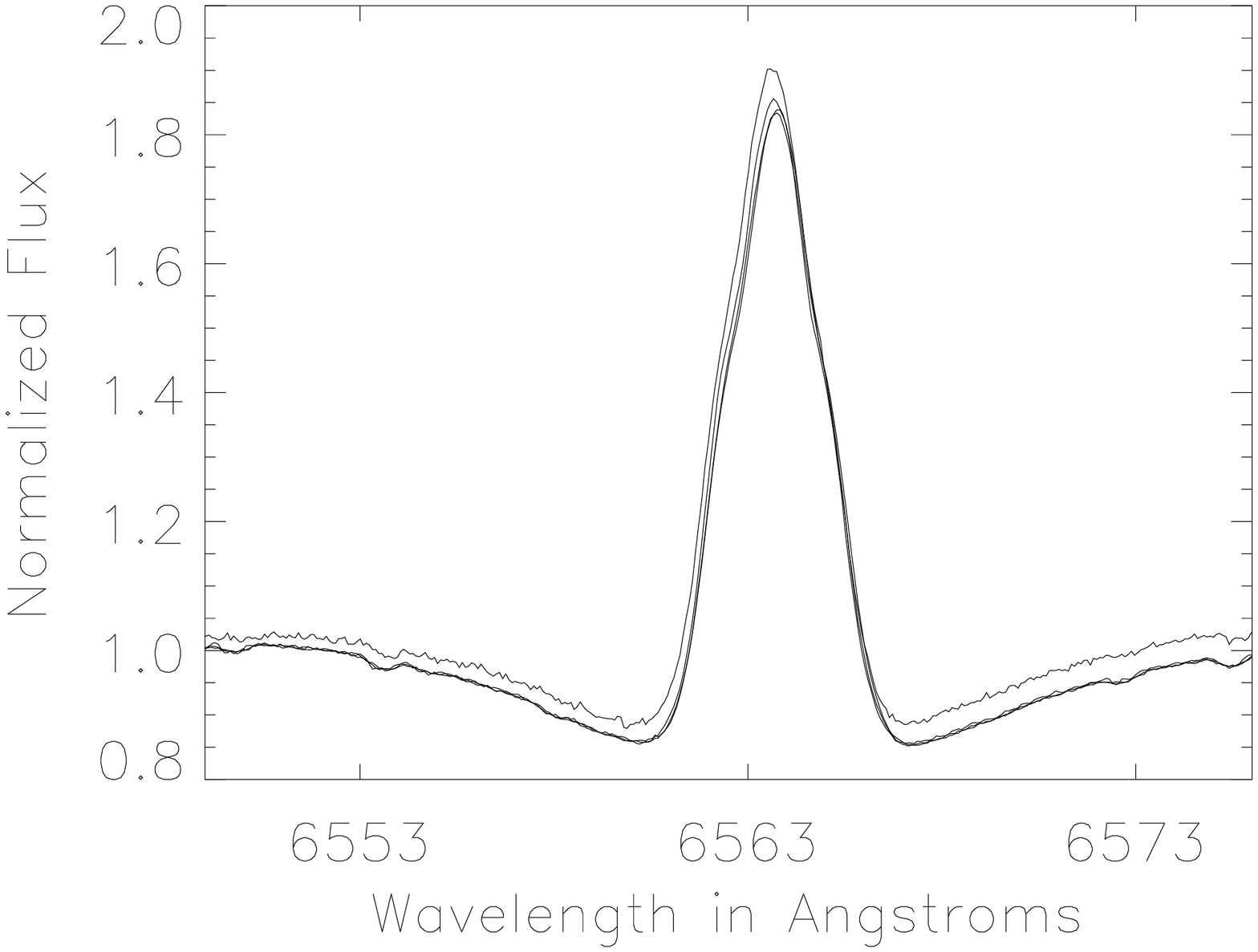} \\
\includegraphics[width=0.23\linewidth, angle=90]{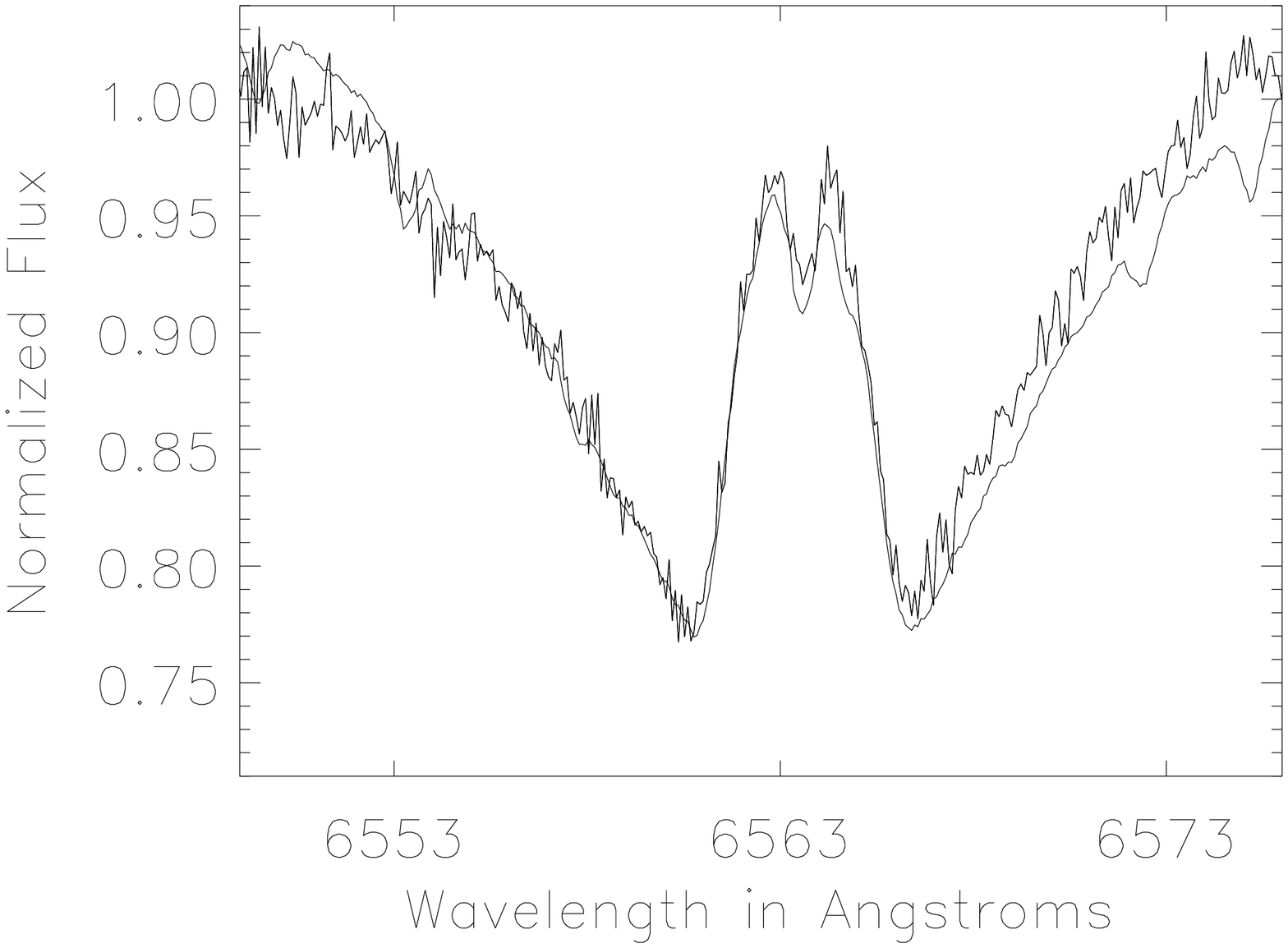}
\includegraphics[width=0.23\linewidth, angle=90]{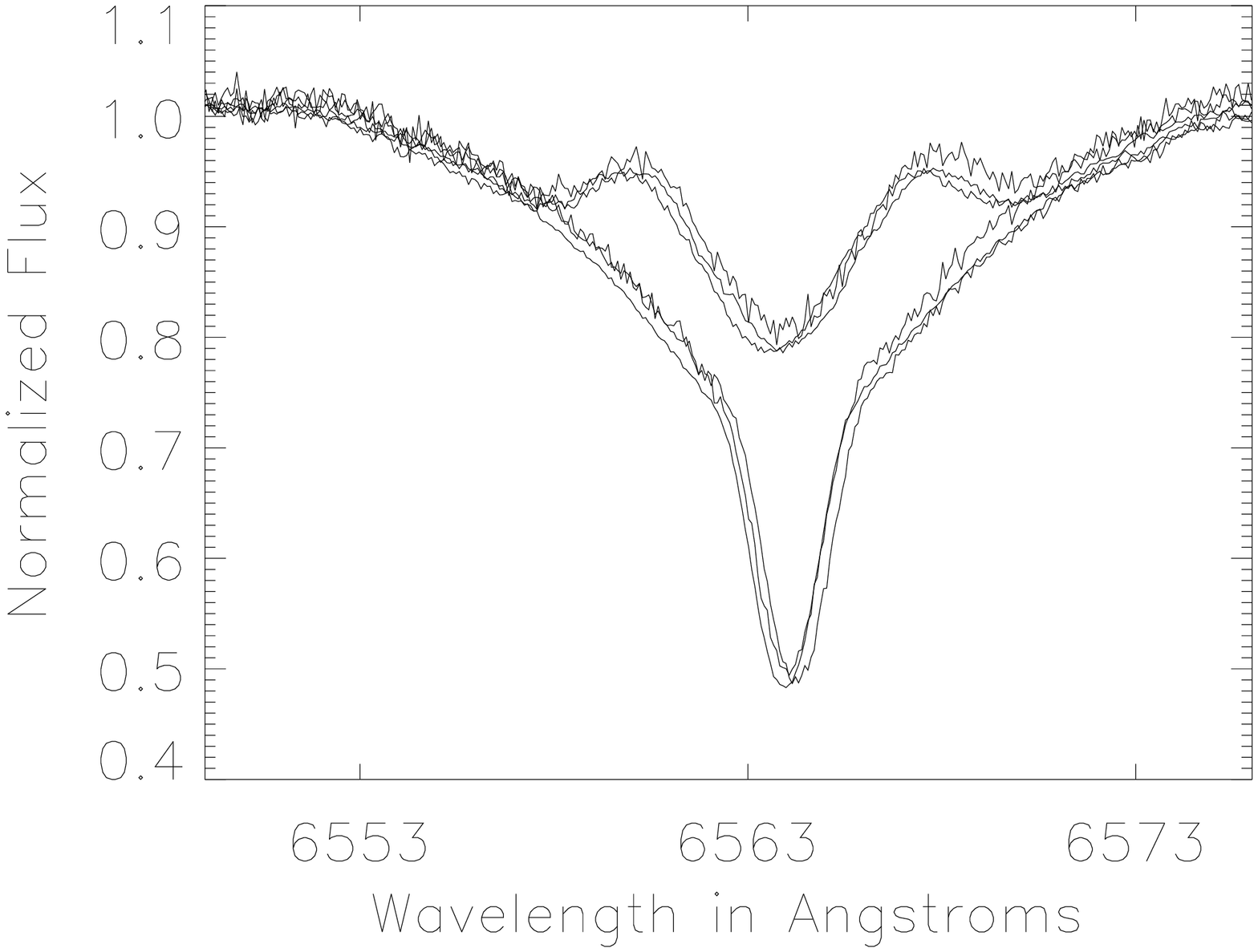}
\includegraphics[width=0.23\linewidth, angle=90]{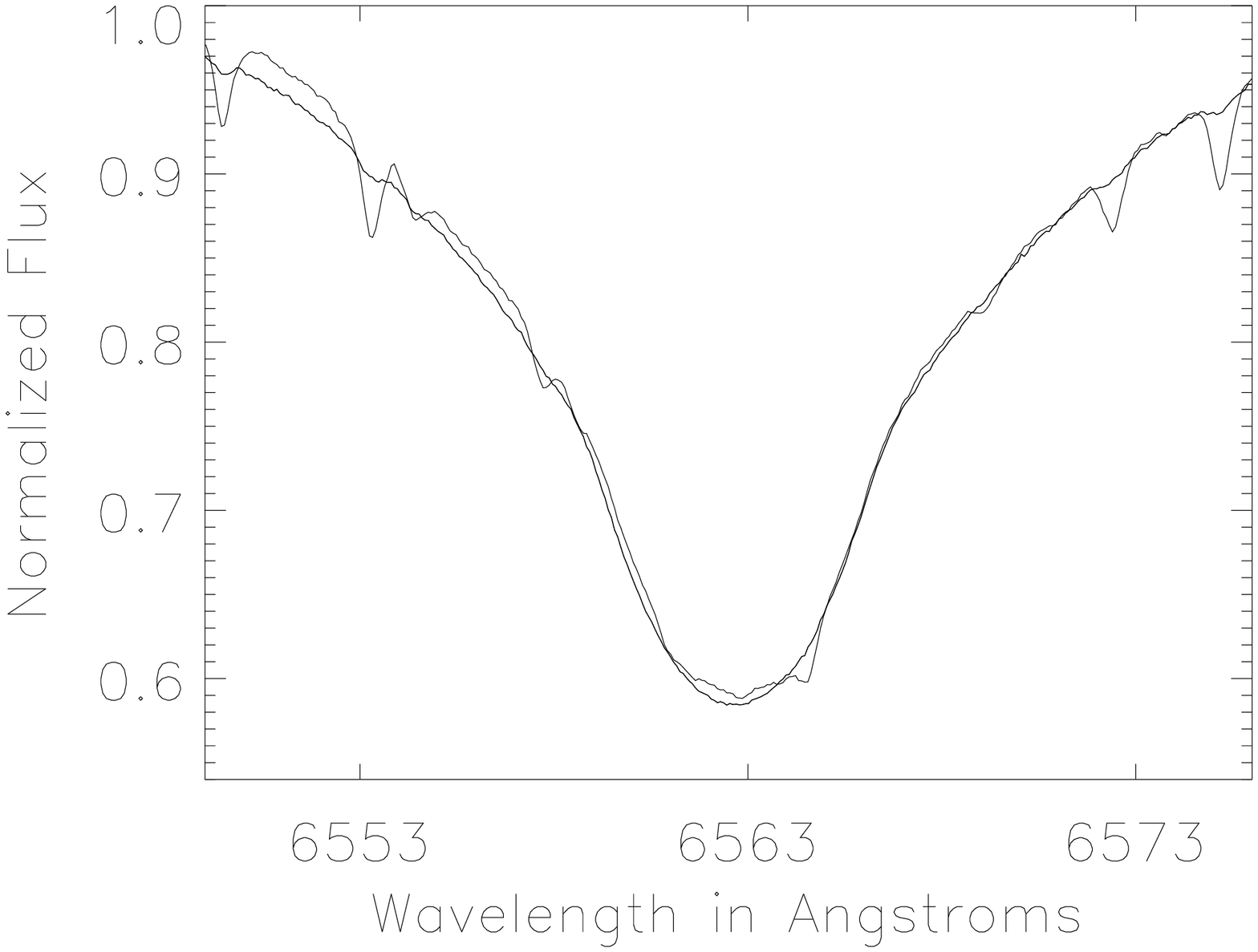} \\
\includegraphics[width=0.23\linewidth, angle=90]{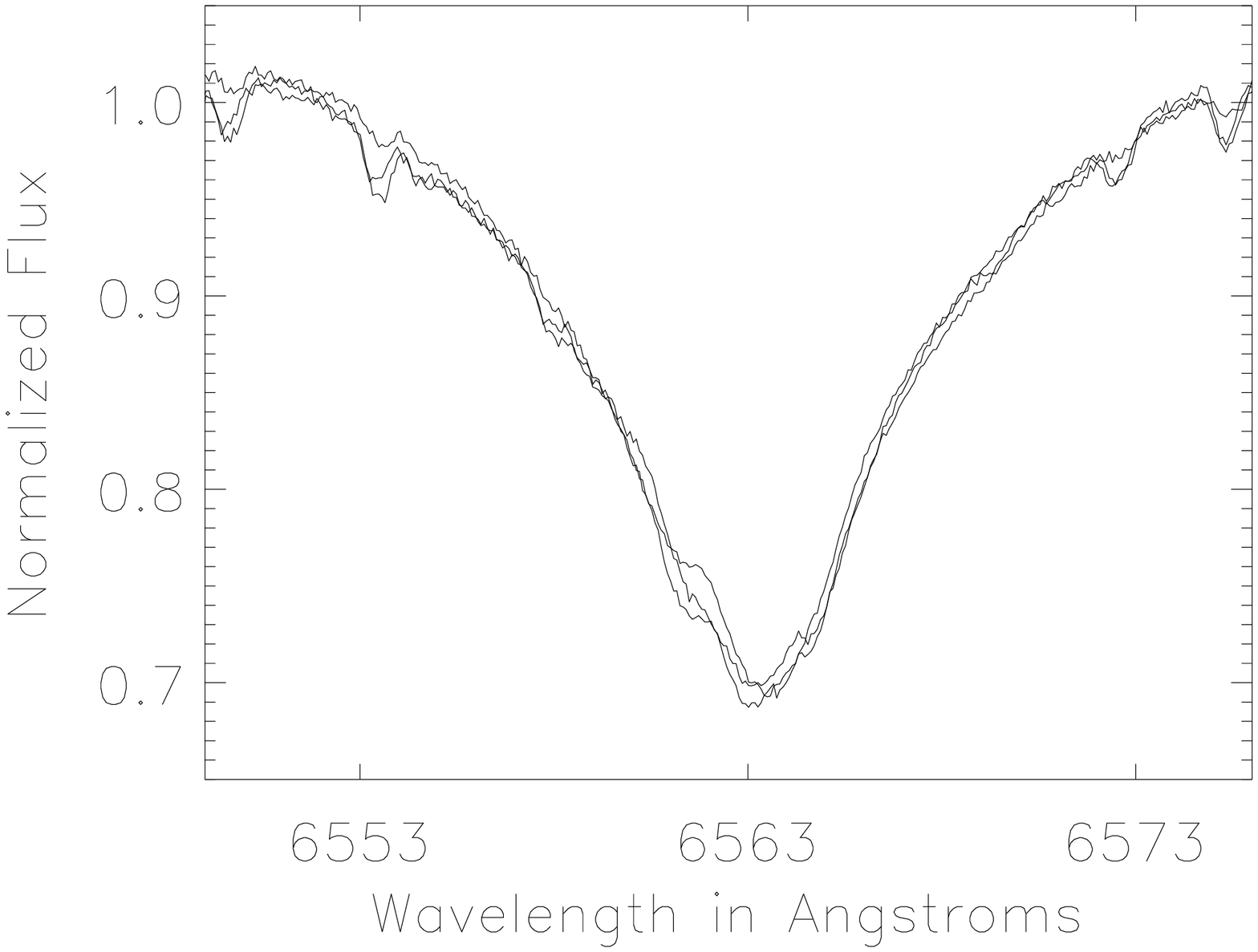}
\includegraphics[width=0.23\linewidth, angle=90]{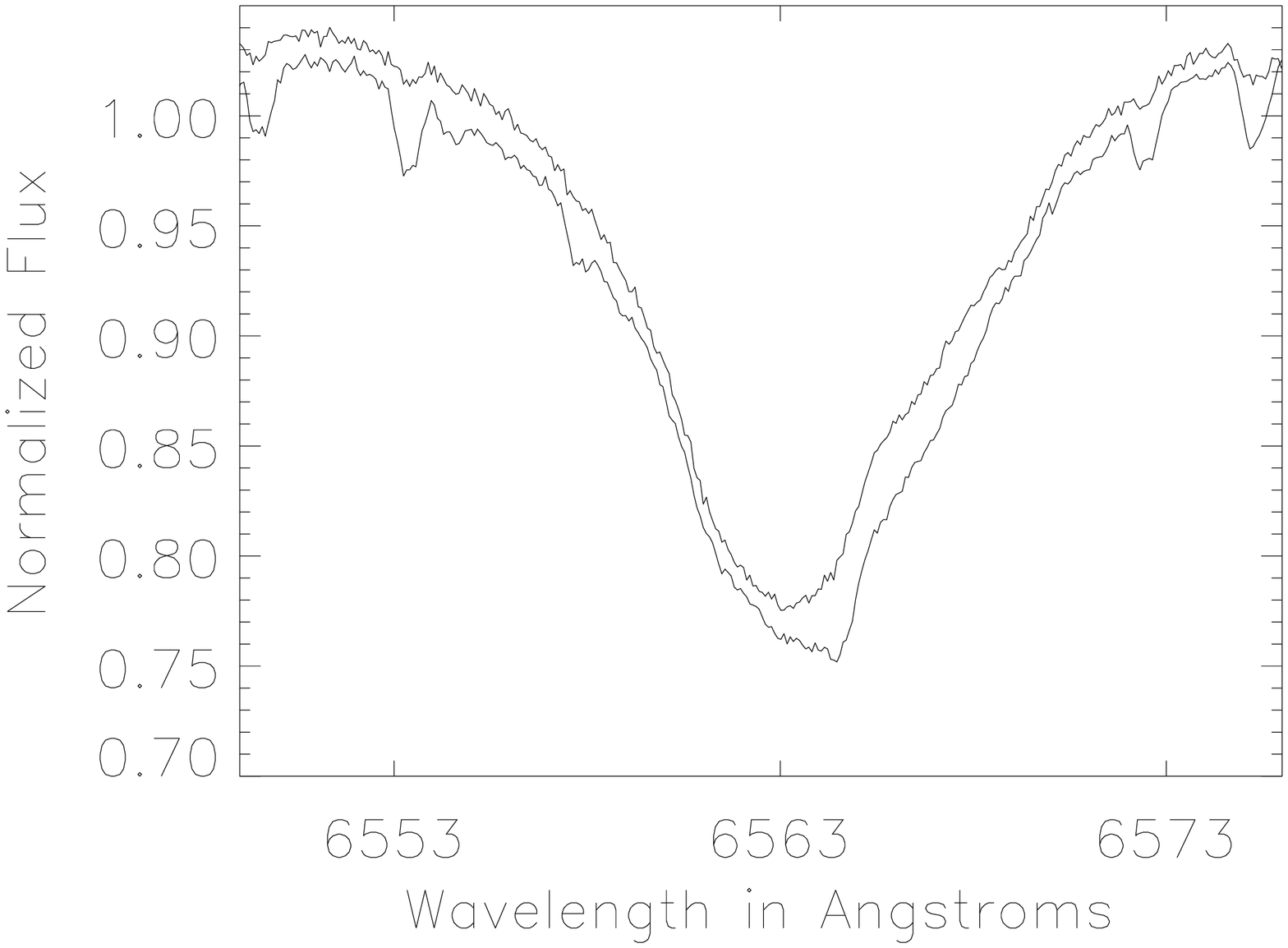}
\includegraphics[width=0.23\linewidth, angle=90]{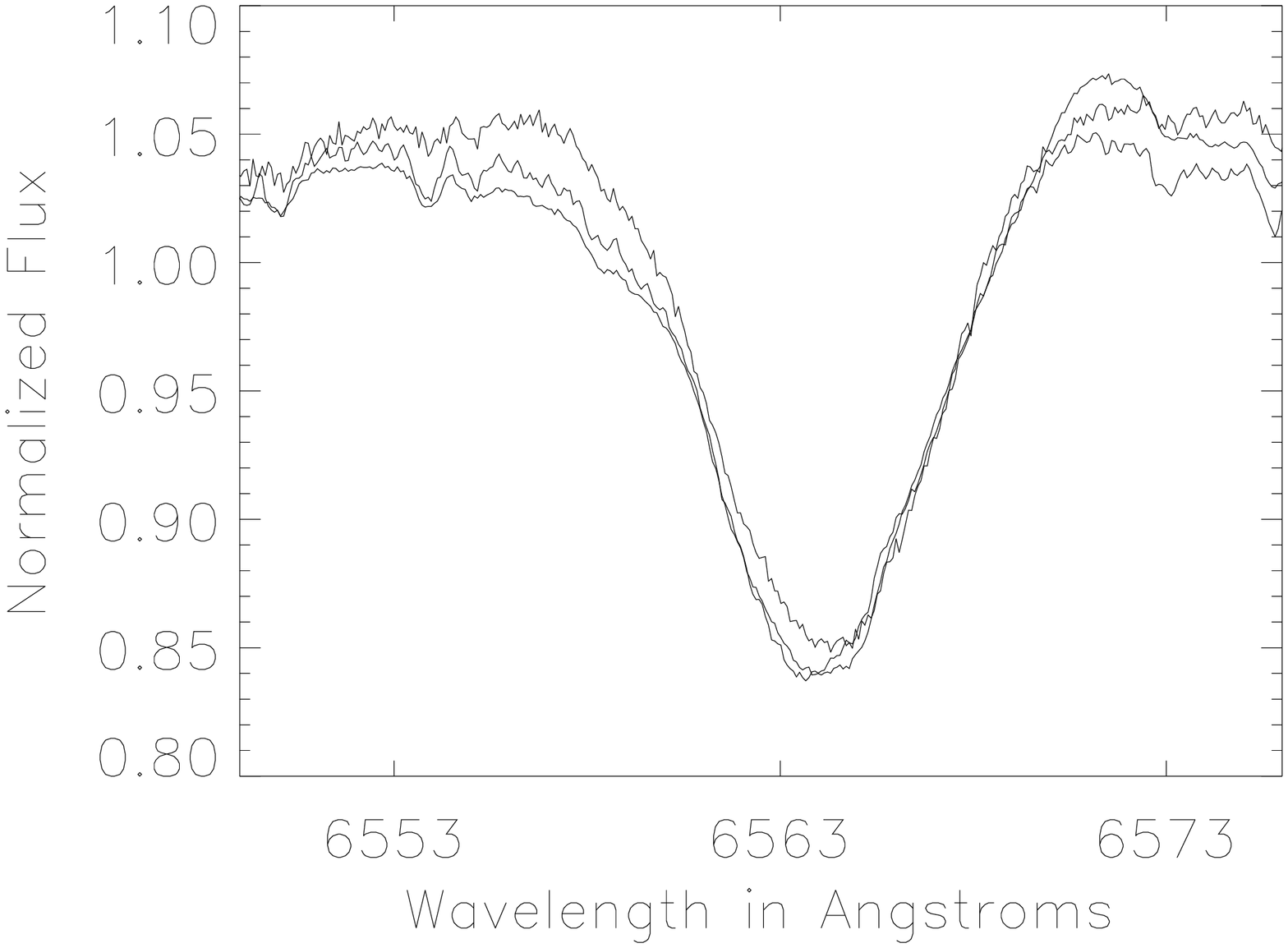}
\caption{Be Line Profiles II. The stars, from left to right, are: 31 Peg, 11 Cam, C Per, MWC 192, $\kappa$ Dra, $\kappa$ Cas, MWC 92, 12 Vul, $\phi$ And, MWC 77, both binary components of HD 36408, Phecda, $\lambda$ Cyg, QR Vul and $\xi$ Per.}
\label{fig:be-lprof2}
\end{figure*}

	Our observing campaign first focused on $\gamma$ Cas in November of 2006 but was then widened to include 30 targets. The H$_\alpha$ line profiles for these stars show a wide variety of profiles. Figures \ref{fig:be-lprof1} and \ref{fig:be-lprof2} show all the profiles compiled for each stars. Some show strong ``disky" signatures like $\psi$ Per and MWC 143. Others show a more simple symmetric emission-line such as 11 Cam or 31 Peg.

	Table \ref{tab-obsbe} lists the basic stellar properties. Essentially all the stars are B-type and are 5th magnitude or brighter. Although there is only a month or two baseline between observations for variability studies, most systems showed only small profile changes, if any. The most variable stars were $\zeta$ Tau, with its strong emission, $\kappa$ Cas and MWC 92 with their weaker emission lines. Many stars showed small variability of the emission, such as 25 Ori, $\omega$ Ori, $\kappa$ CMa, or 18 Gem, but the overall structure of the emission line did not change. This shows that during the observing window, the circumstellar region responsible for the emission did not change its structure significantly.  
	
	For the purposes of this study, which is concerned with the circumstellar material, the stars can be broken into subclasses based on their H$_\alpha$ line profiles. There are strong, symmetric emission lines such as 11 Cam in figure \ref{fig:be-lprof2} that have line:continuum ratios of 5-8 and don't show evidence for intervening absorption. There are the classical ``disky" systems such as MWC 143 that have strong emission with a strong, narrow central absorptive component. Another class will be distinguished in table \ref{tab-obsbe} called Disk* - these show less absorption than the classical ``disky" systems, but do show central, often variable absorption. The last type does not show significant emission, such as $\lambda$ Cyg or QR Vul.

\begin{figure*}
\begin{center}
\includegraphics[width=0.23\linewidth, angle=90]{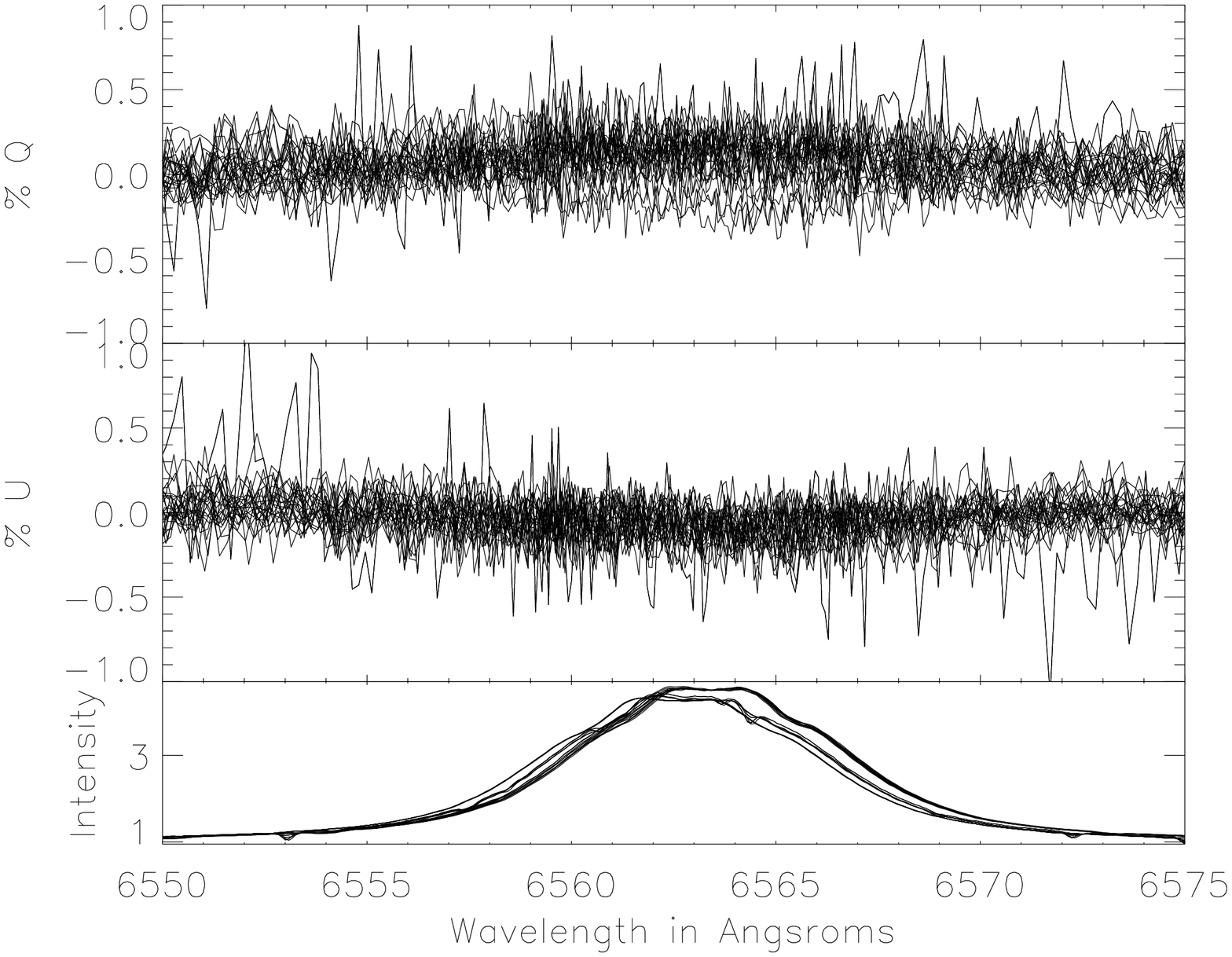}
\includegraphics[width=0.23\linewidth, angle=90]{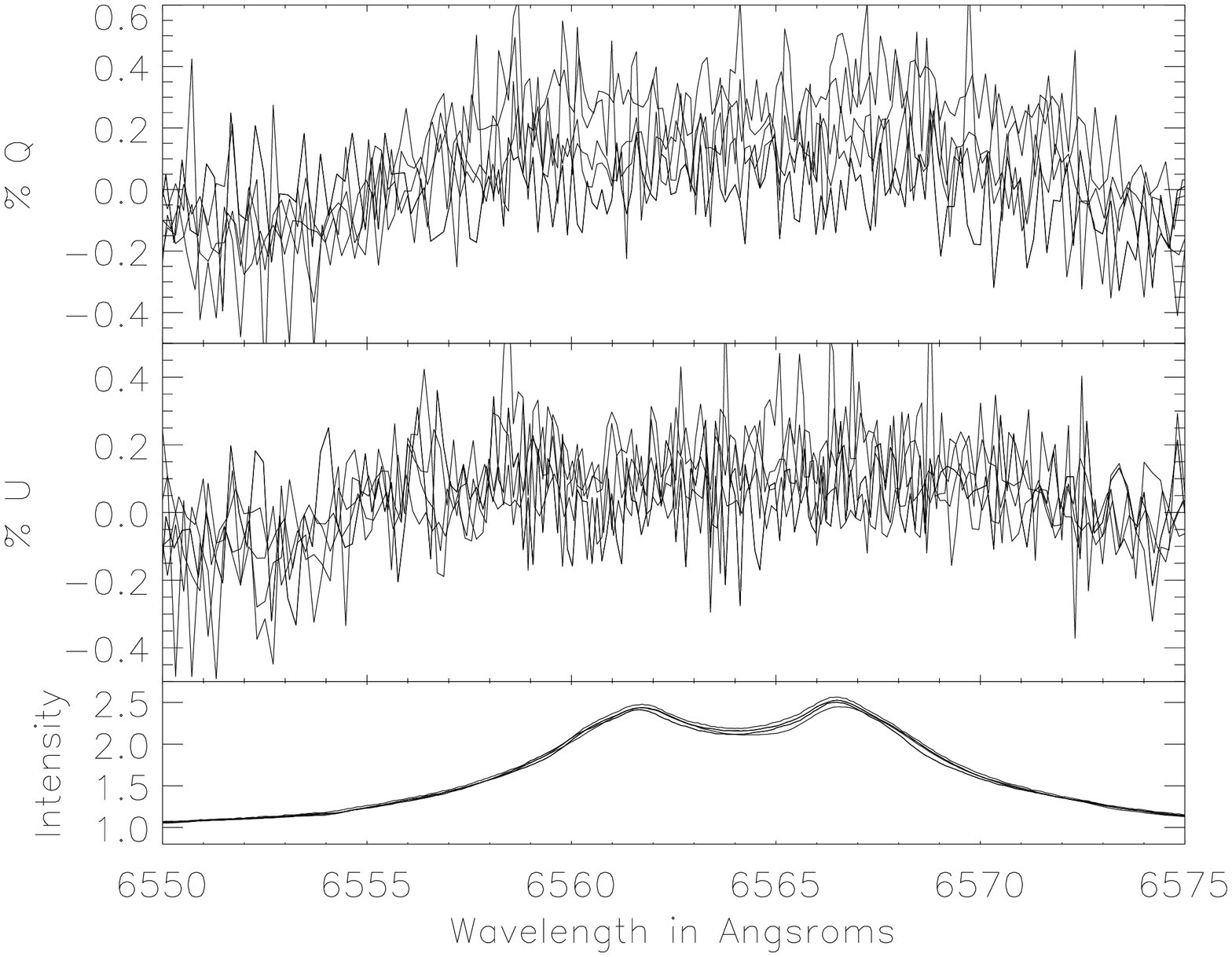}
\includegraphics[width=0.23\linewidth, angle=90]{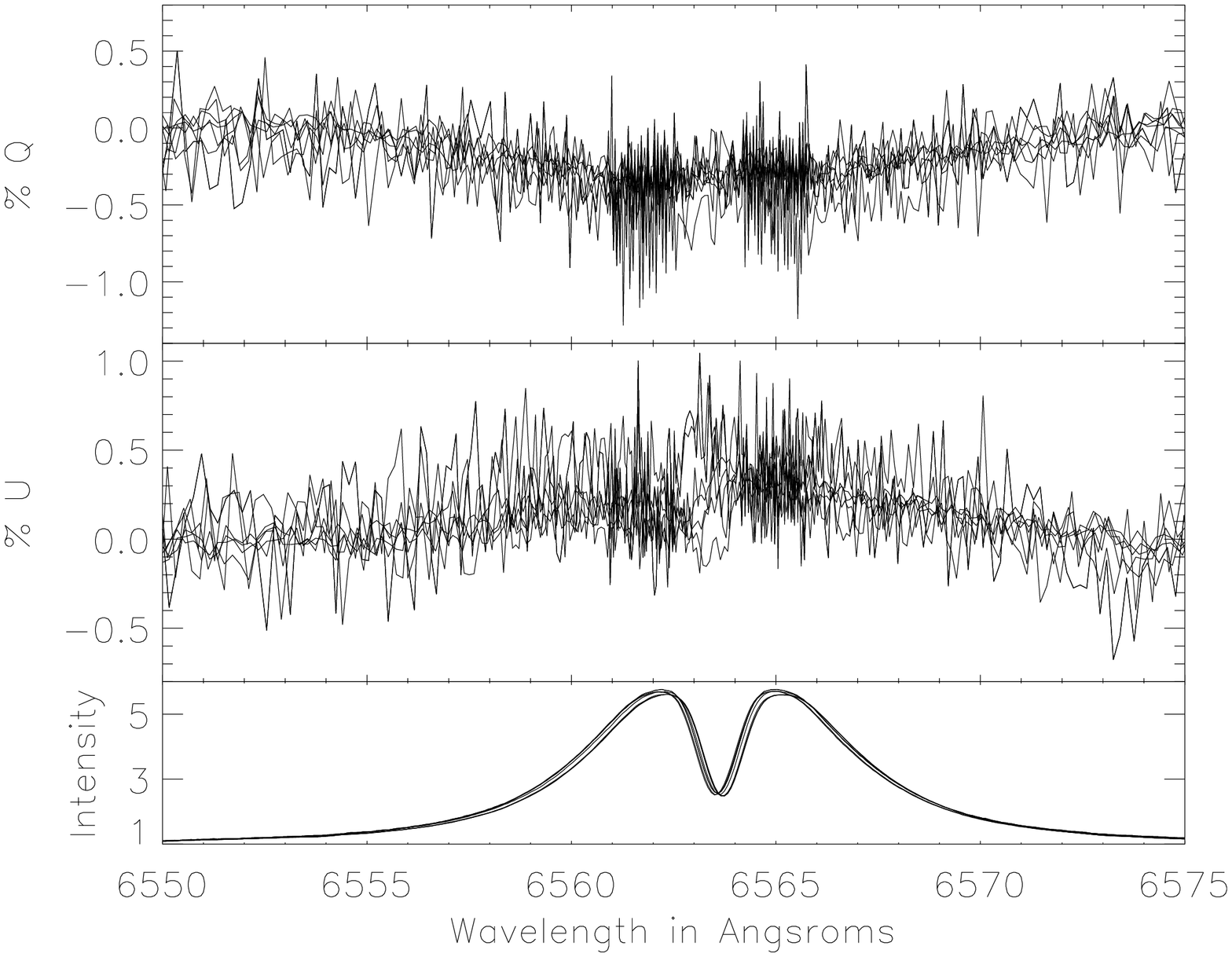} \\
\includegraphics[width=0.23\linewidth, angle=90]{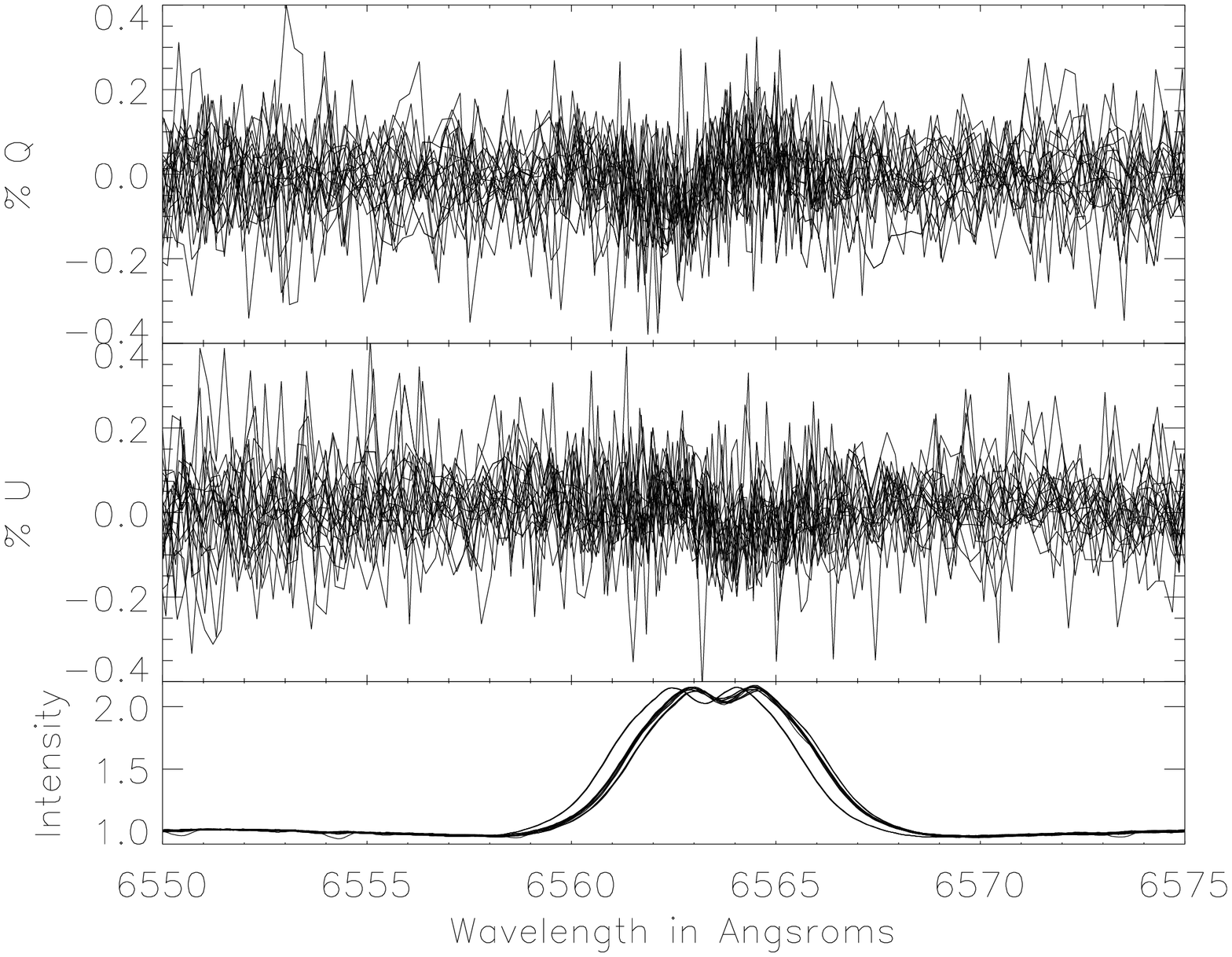}
\includegraphics[width=0.23\linewidth, angle=90]{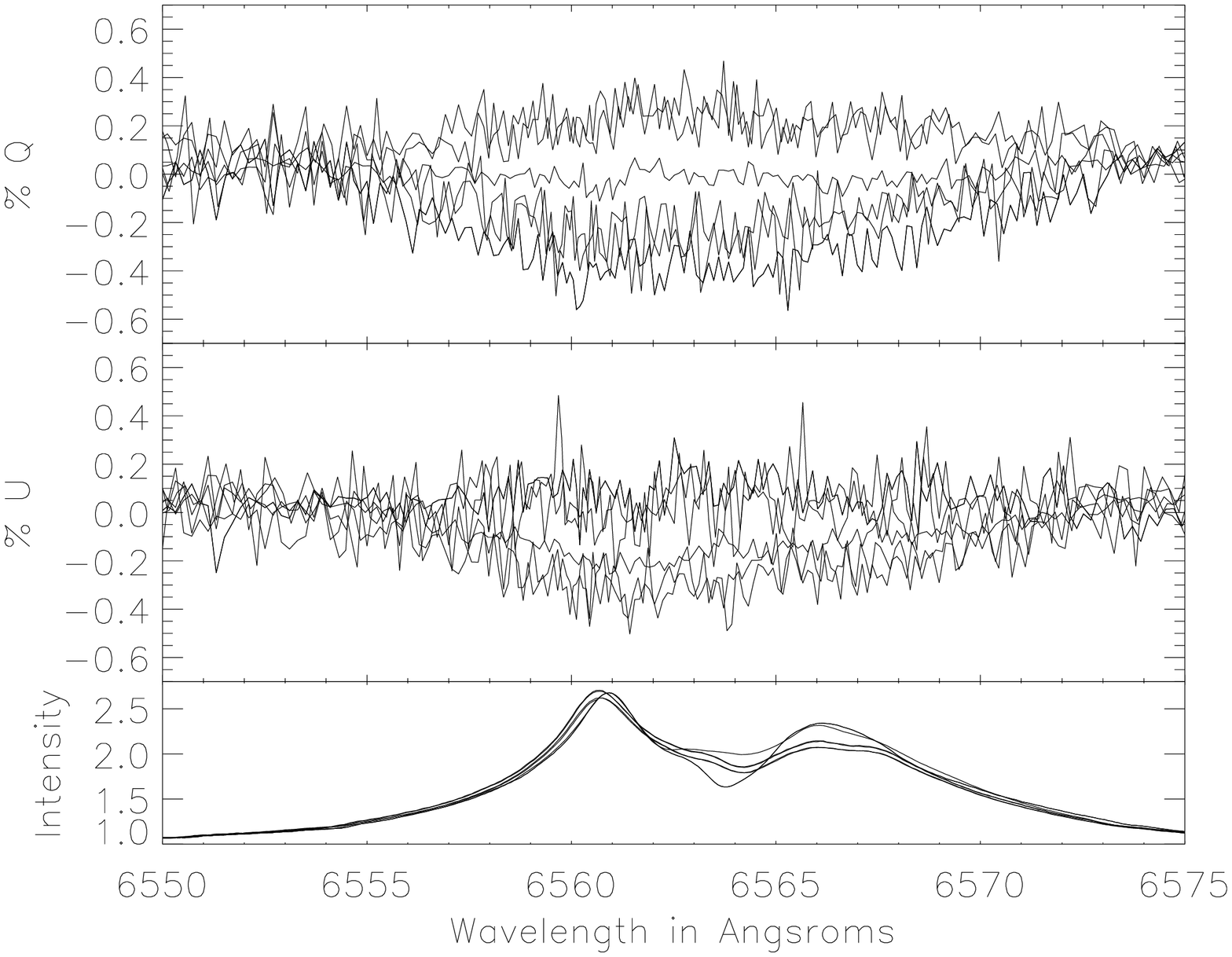}
\includegraphics[width=0.23\linewidth, angle=90]{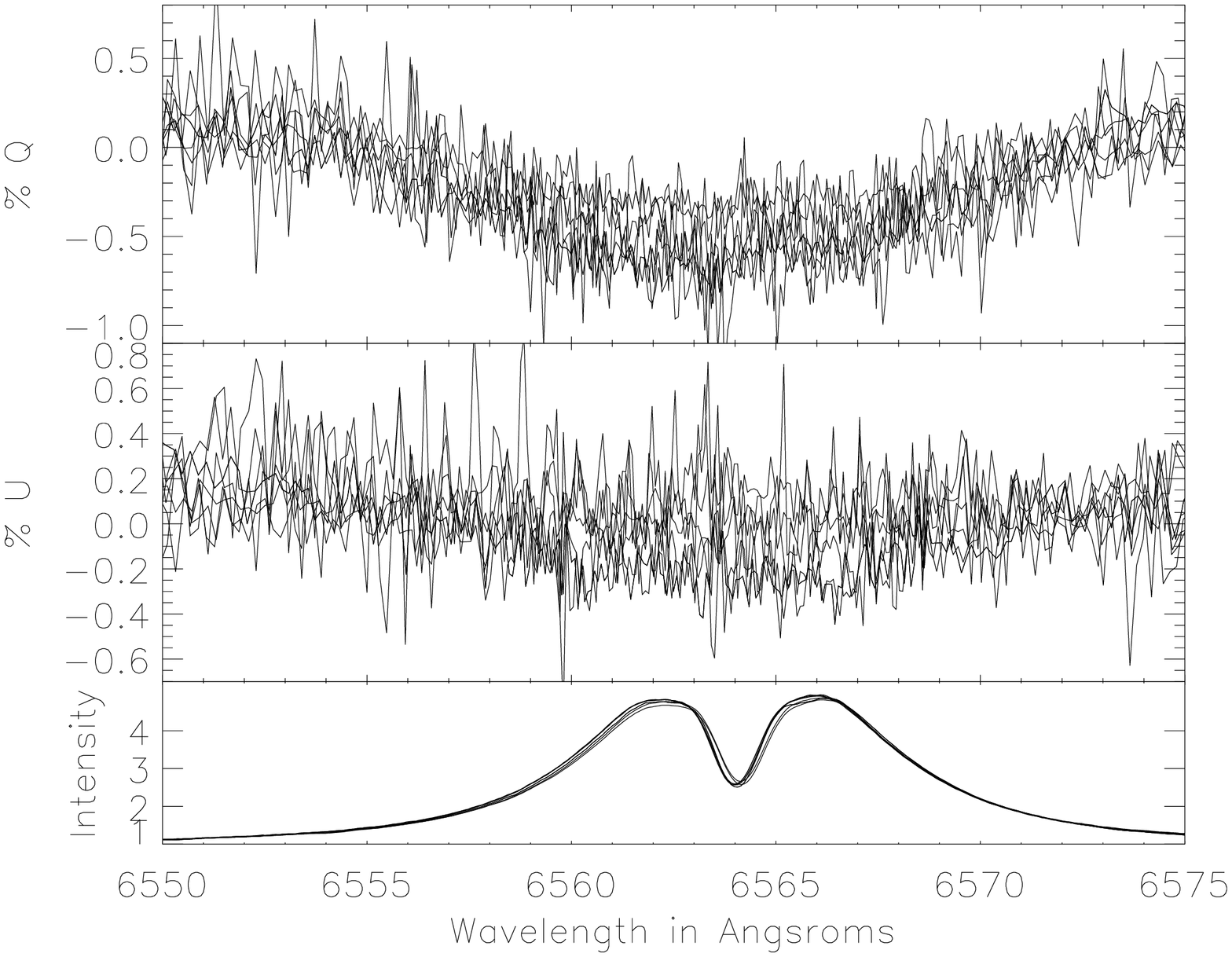} \\
\includegraphics[width=0.23\linewidth, angle=90]{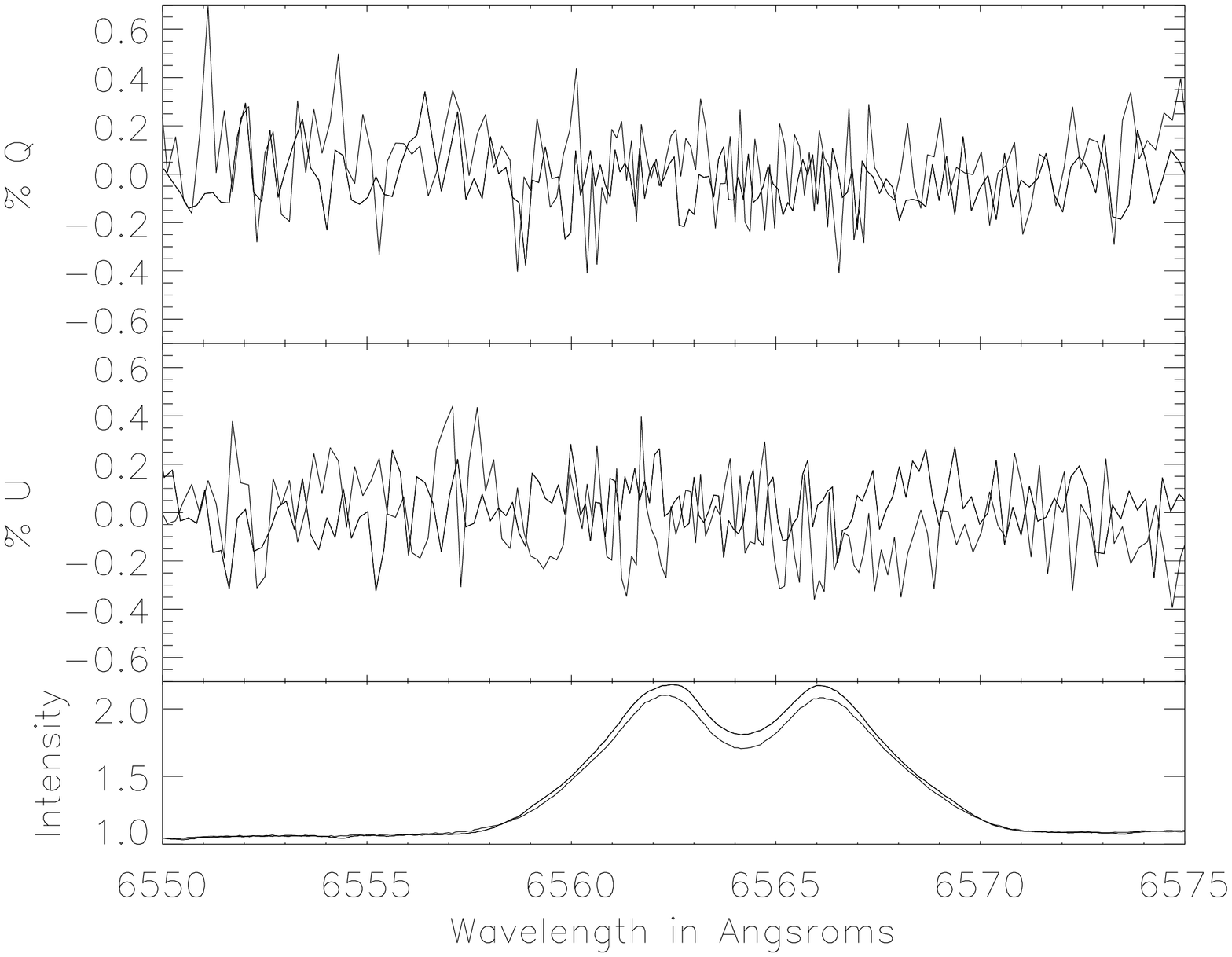}
\includegraphics[width=0.23\linewidth, angle=90]{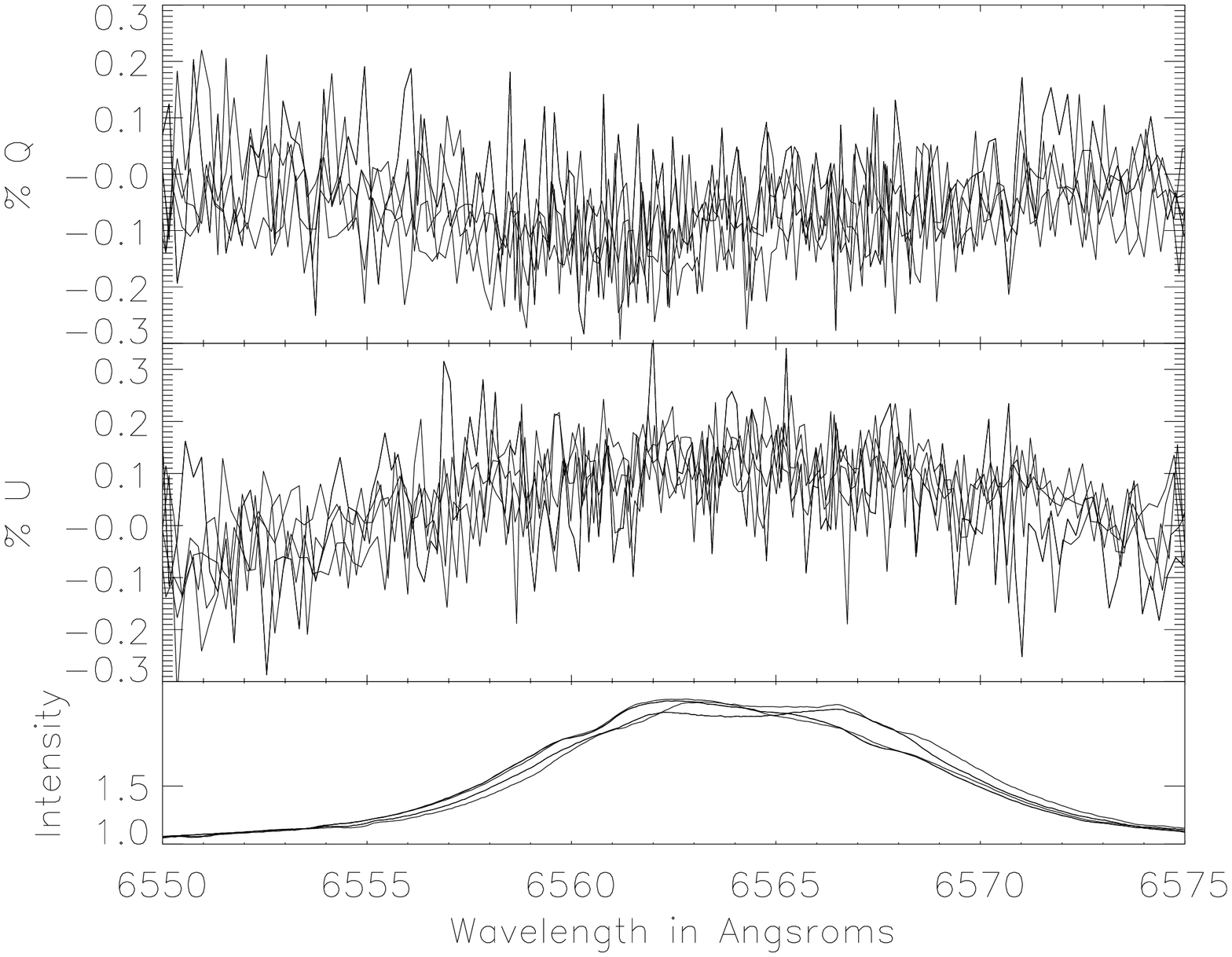}
\includegraphics[width=0.23\linewidth, angle=90]{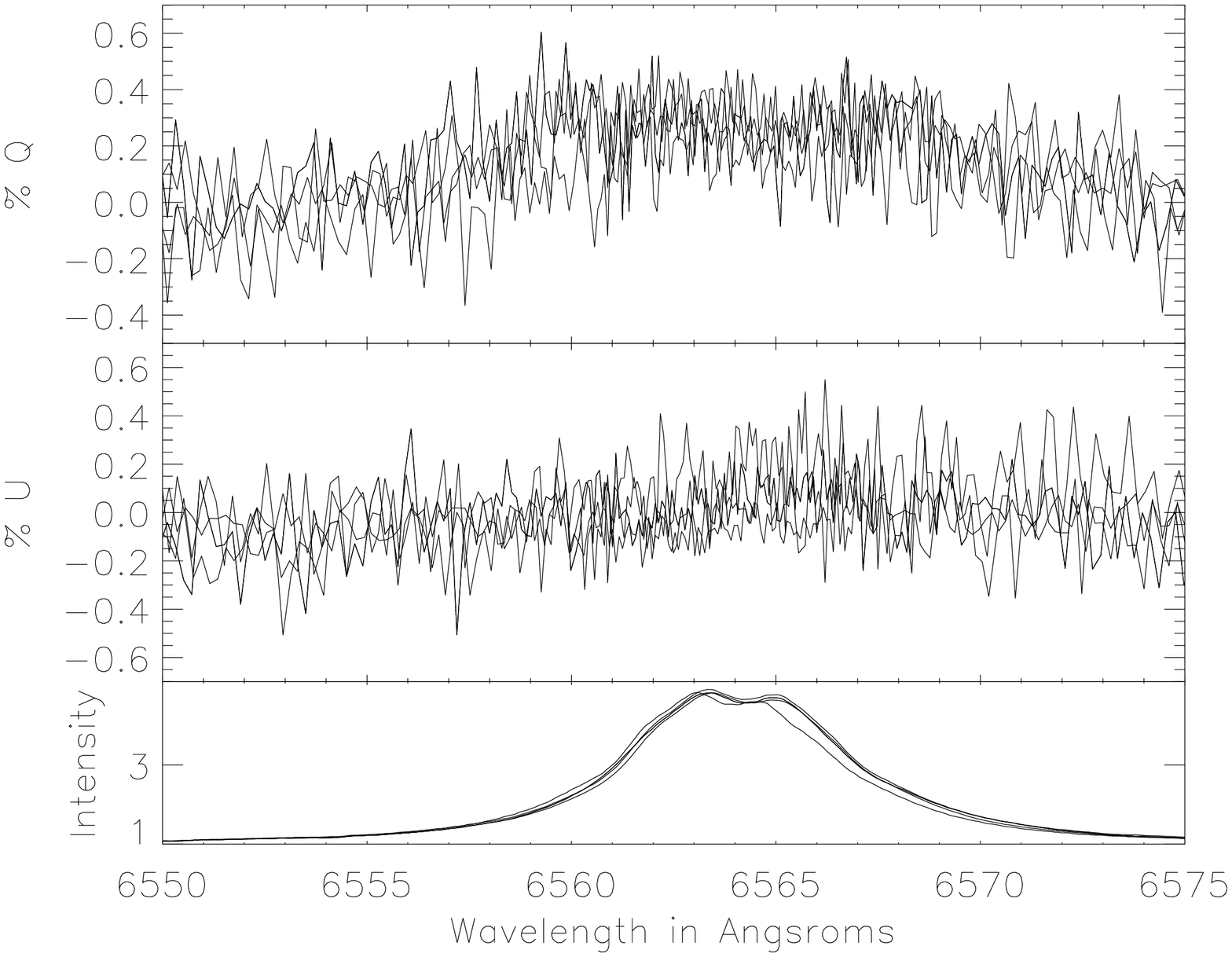} \\
\includegraphics[width=0.23\linewidth, angle=90]{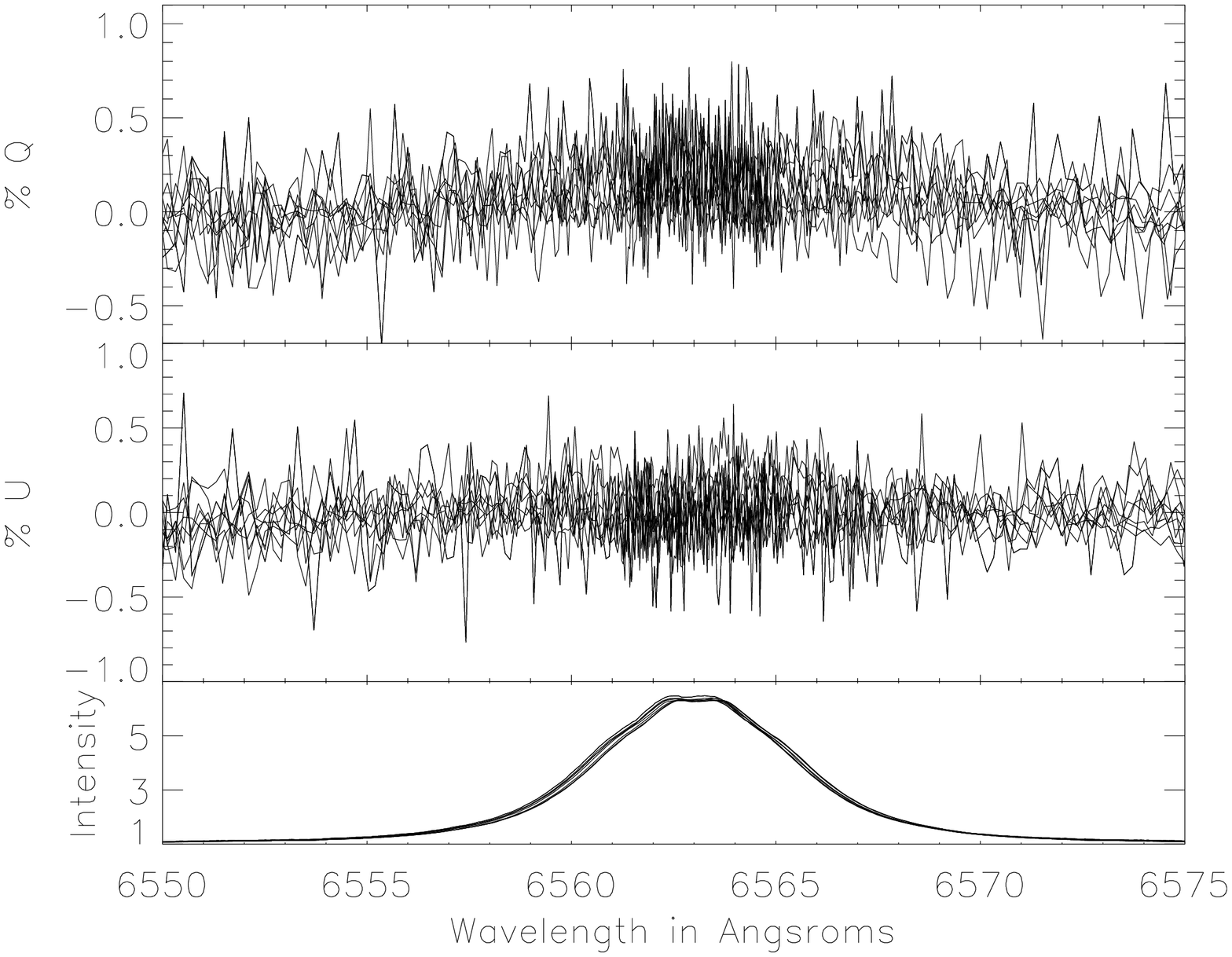}
\includegraphics[width=0.23\linewidth, angle=90]{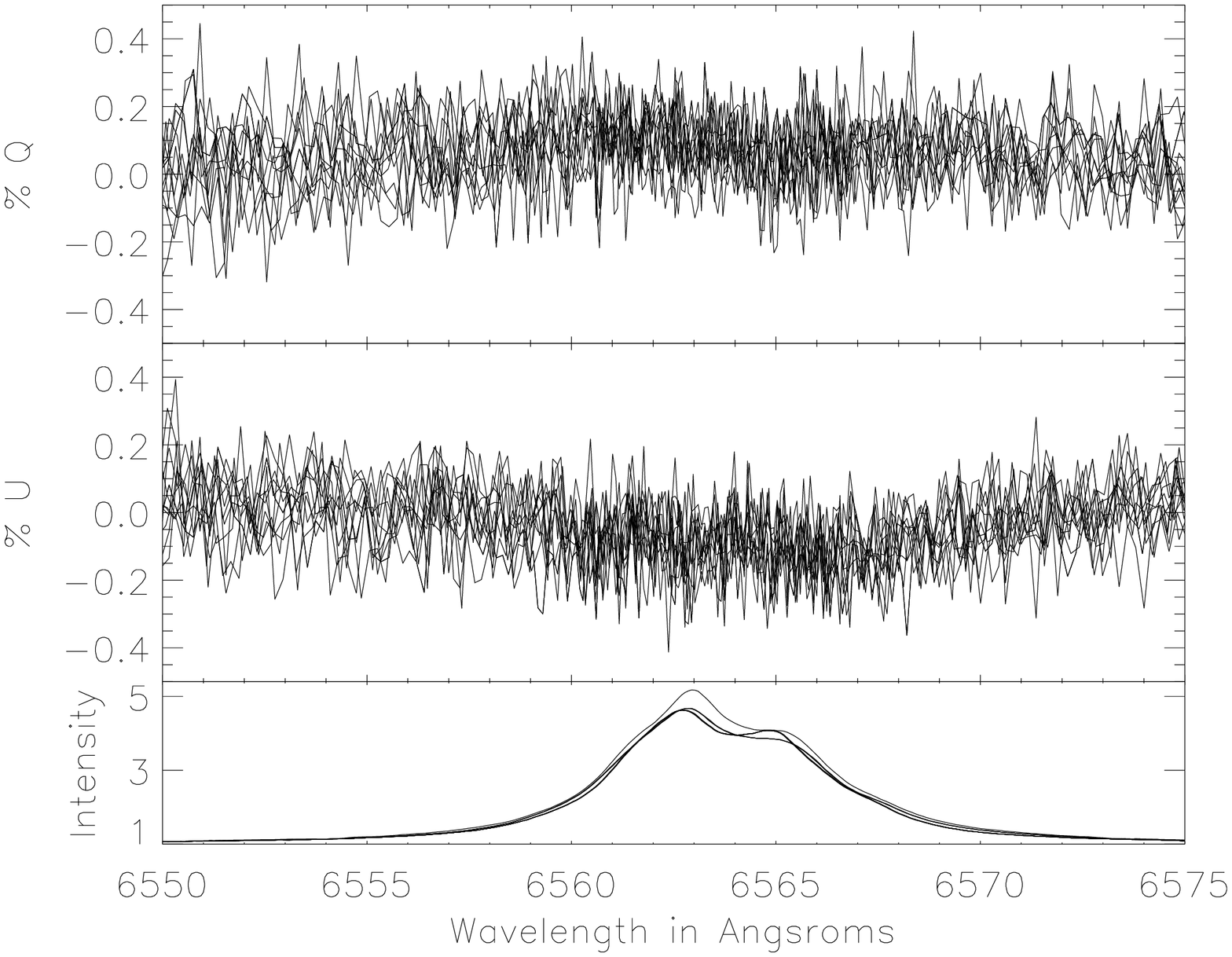}
\includegraphics[width=0.23\linewidth, angle=90]{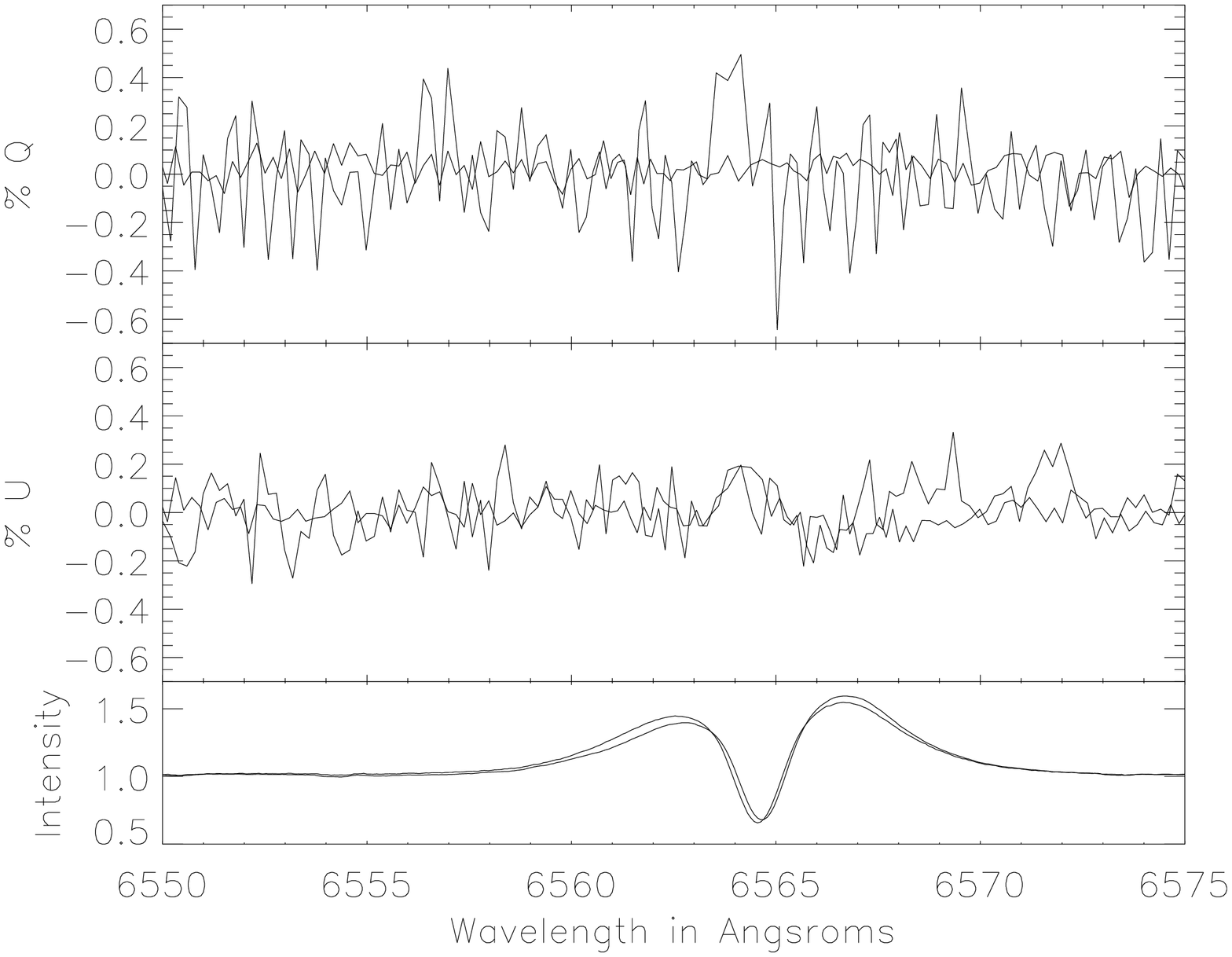} \\
\includegraphics[width=0.23\linewidth, angle=90]{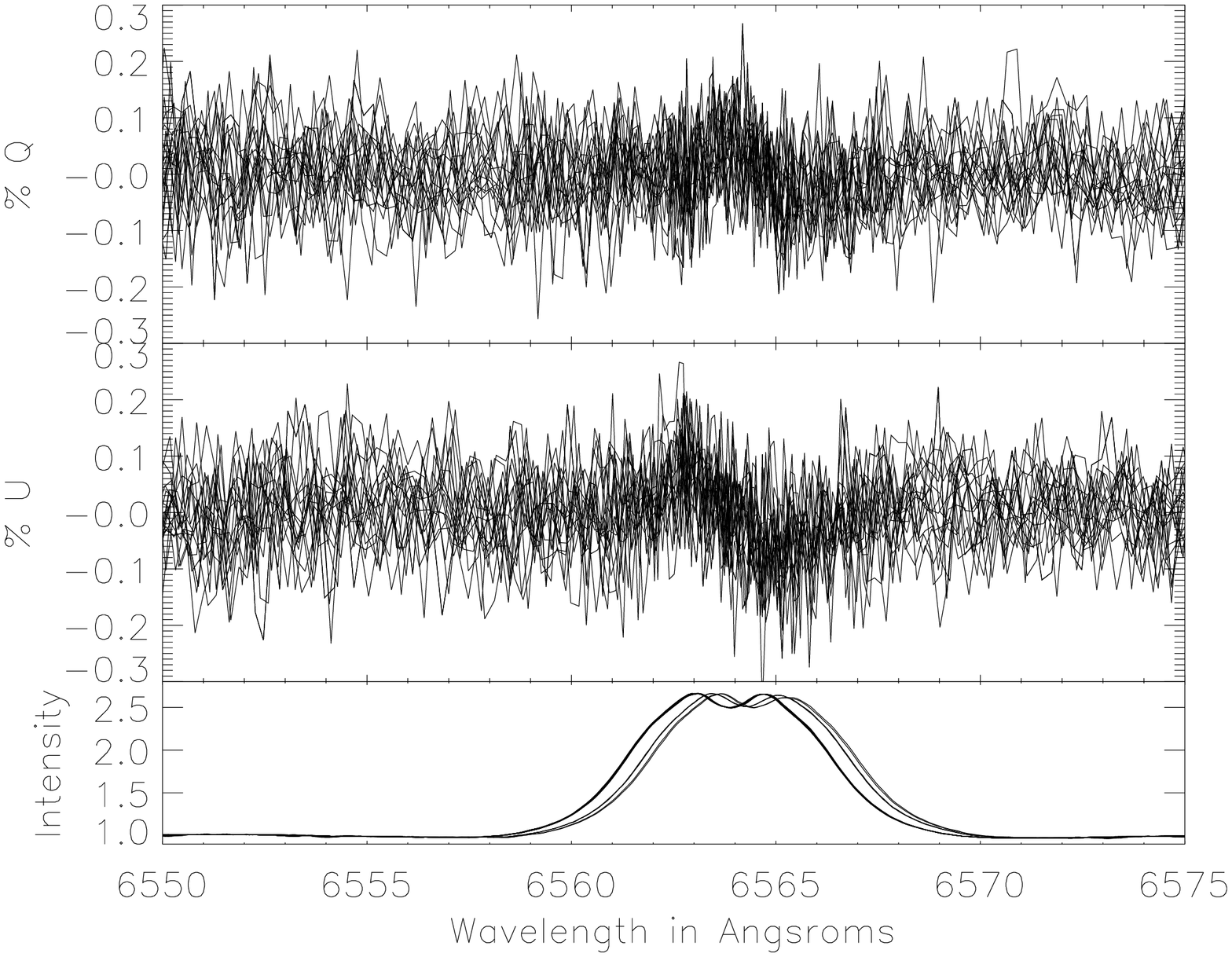}
\includegraphics[width=0.23\linewidth, angle=90]{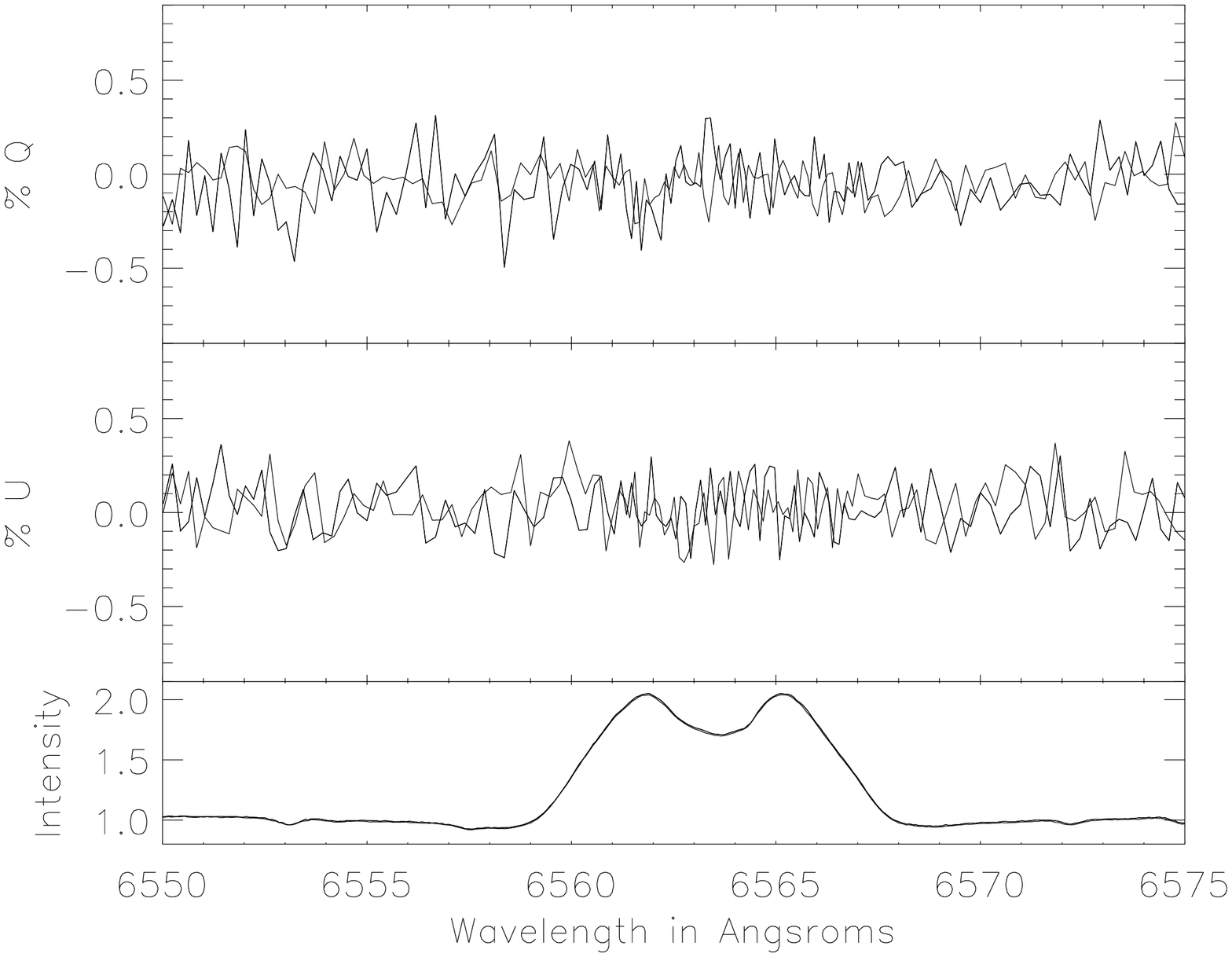}
\includegraphics[width=0.23\linewidth, angle=90]{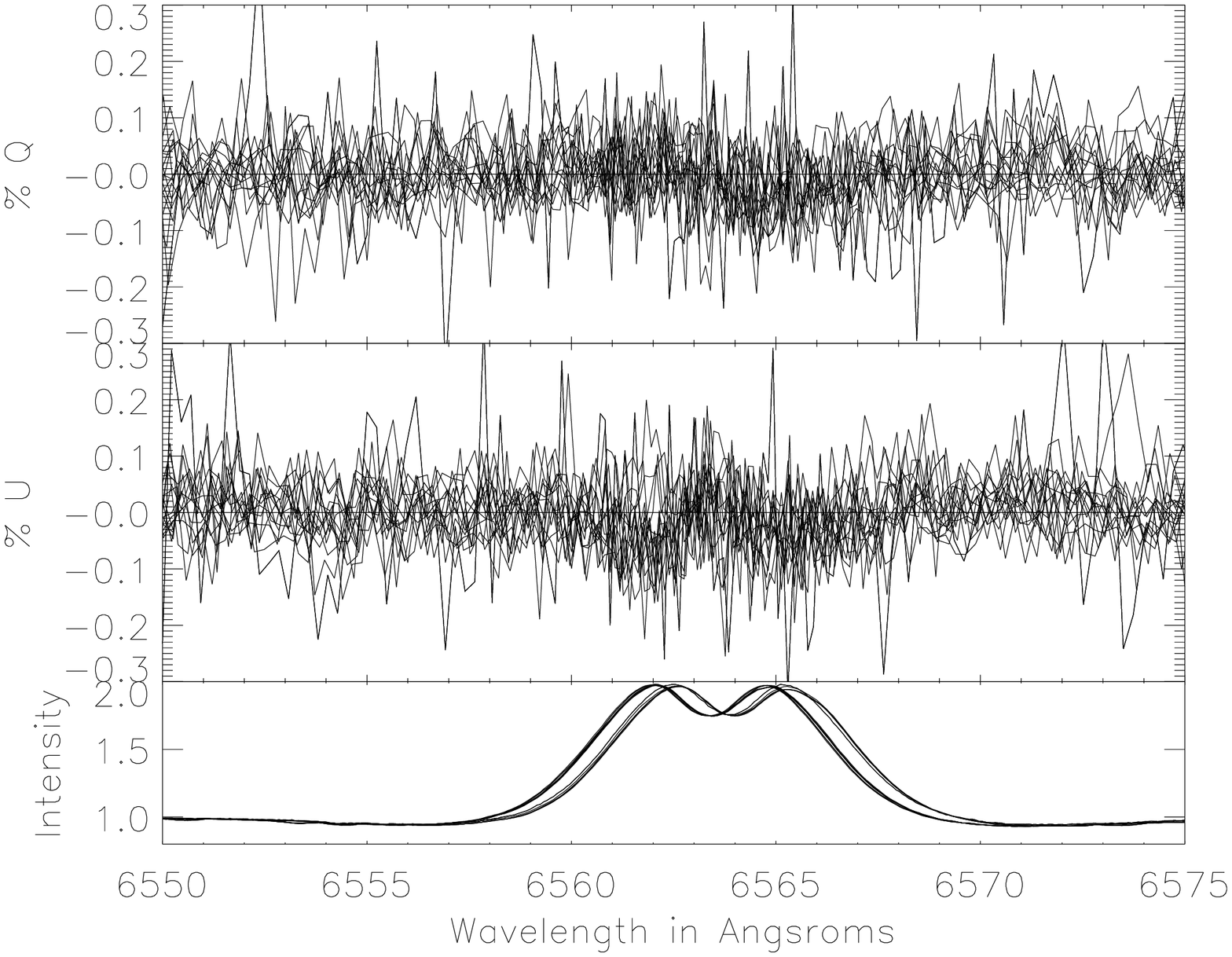}
\caption{Be Spectropolarimetric Profiles I. The stars, from left to right, are: $\gamma$ Cas, 25 Ori, $\psi$ Per, $\eta$ Tau, $\zeta$ Tau, MWC 143, $\omega$ Ori, Omi Pup, 10 CMa, Omi Cas, $\kappa$ CMa, 18 Gem, $\alpha$ Col, 66 Oph and $\beta$ CMi.}
\label{fig:be-specpol1}
\end{center}
\end{figure*}

\begin{figure*}
\begin{center}
\includegraphics[width=0.23\linewidth, angle=90]{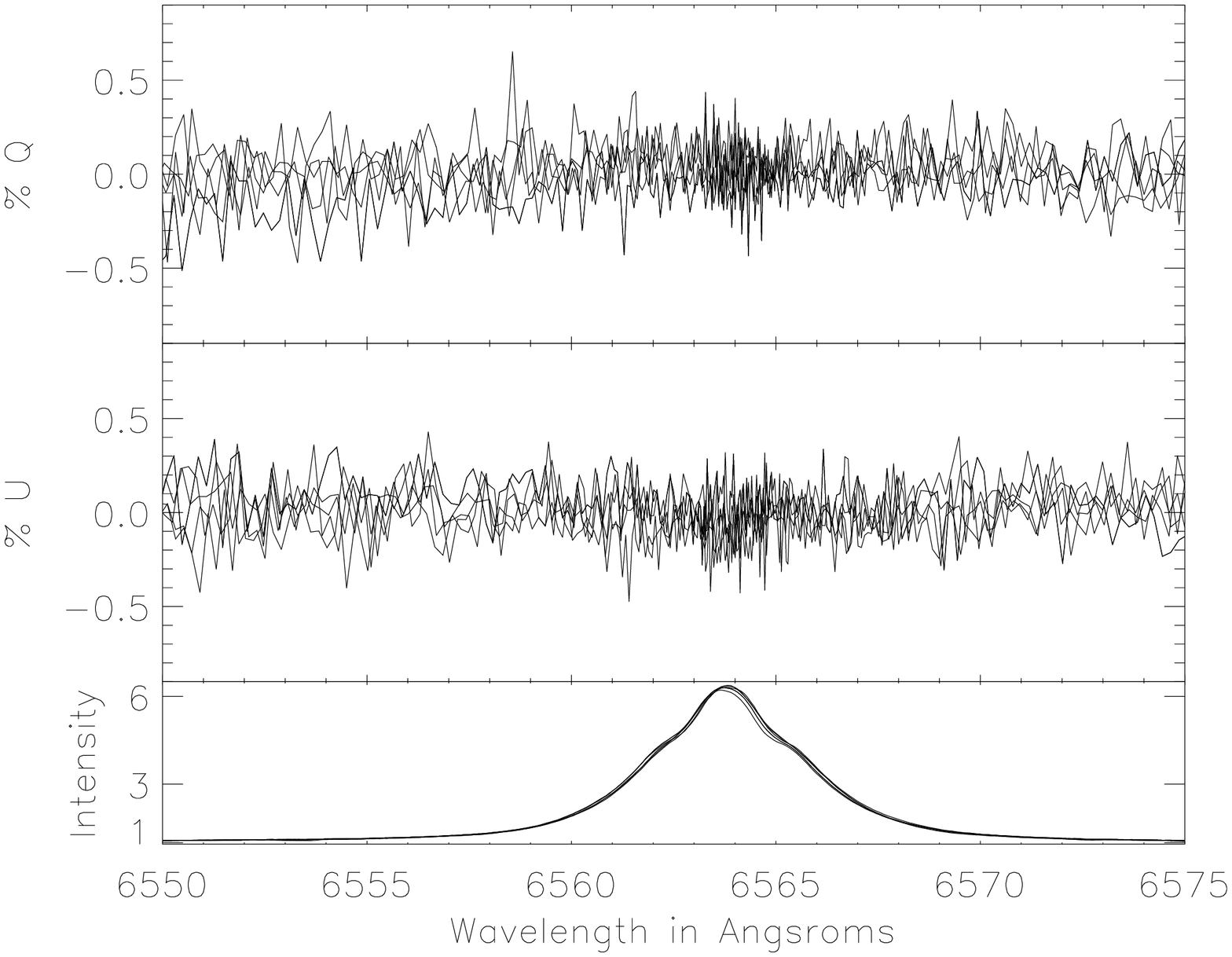}
\includegraphics[width=0.23\linewidth, angle=90]{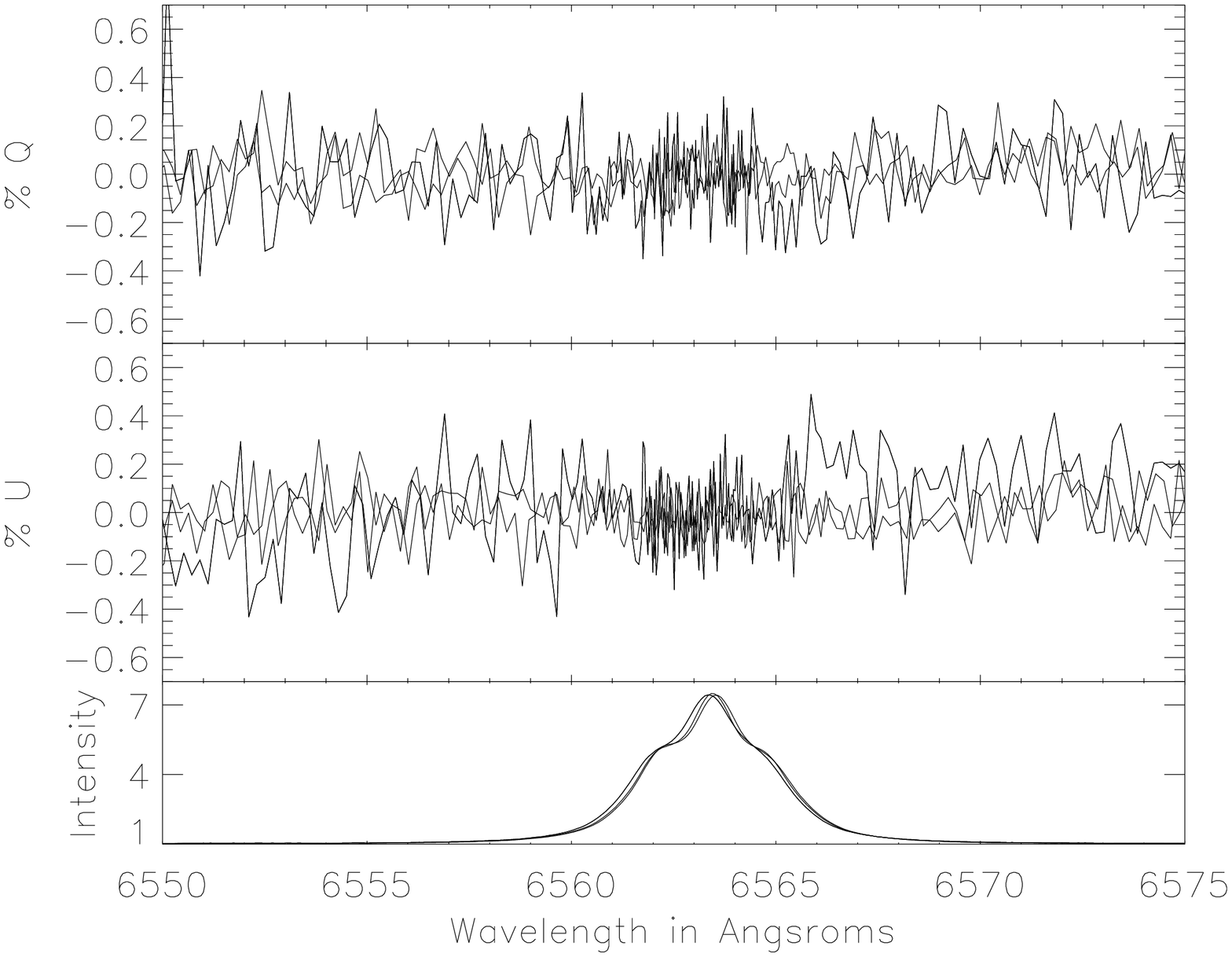}
\includegraphics[width=0.23\linewidth, angle=90]{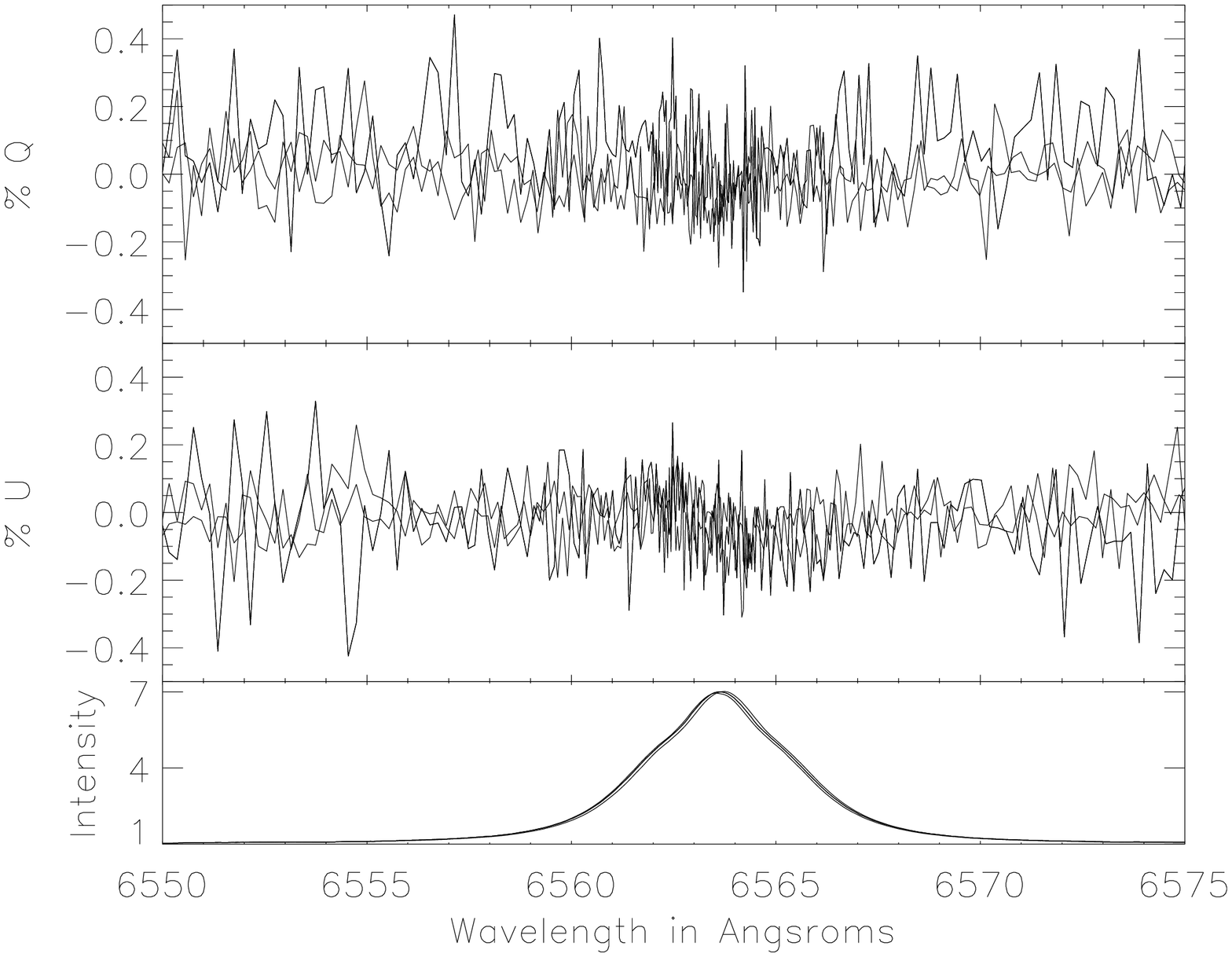} \\
\includegraphics[width=0.23\linewidth, angle=90]{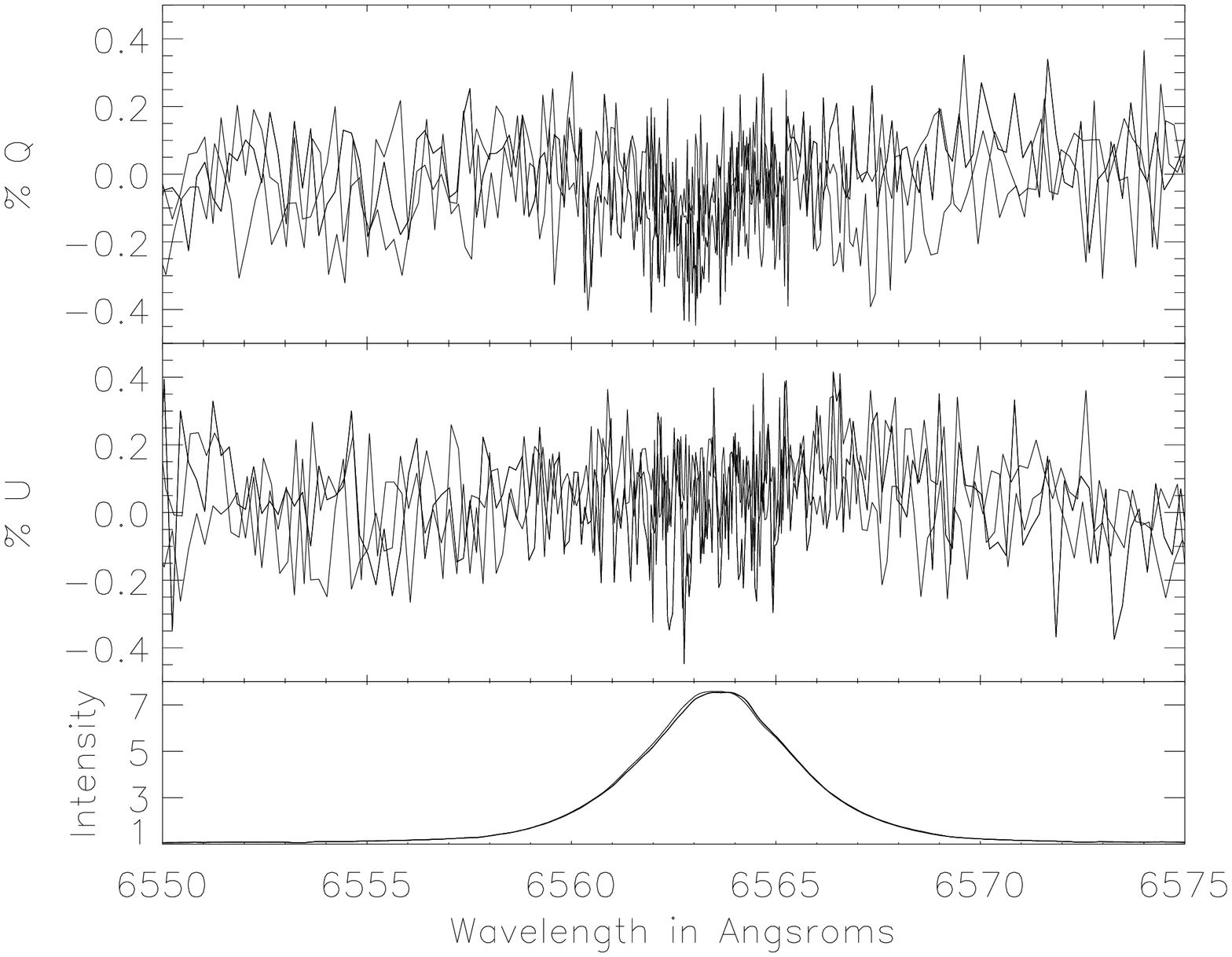}
\includegraphics[width=0.23\linewidth, angle=90]{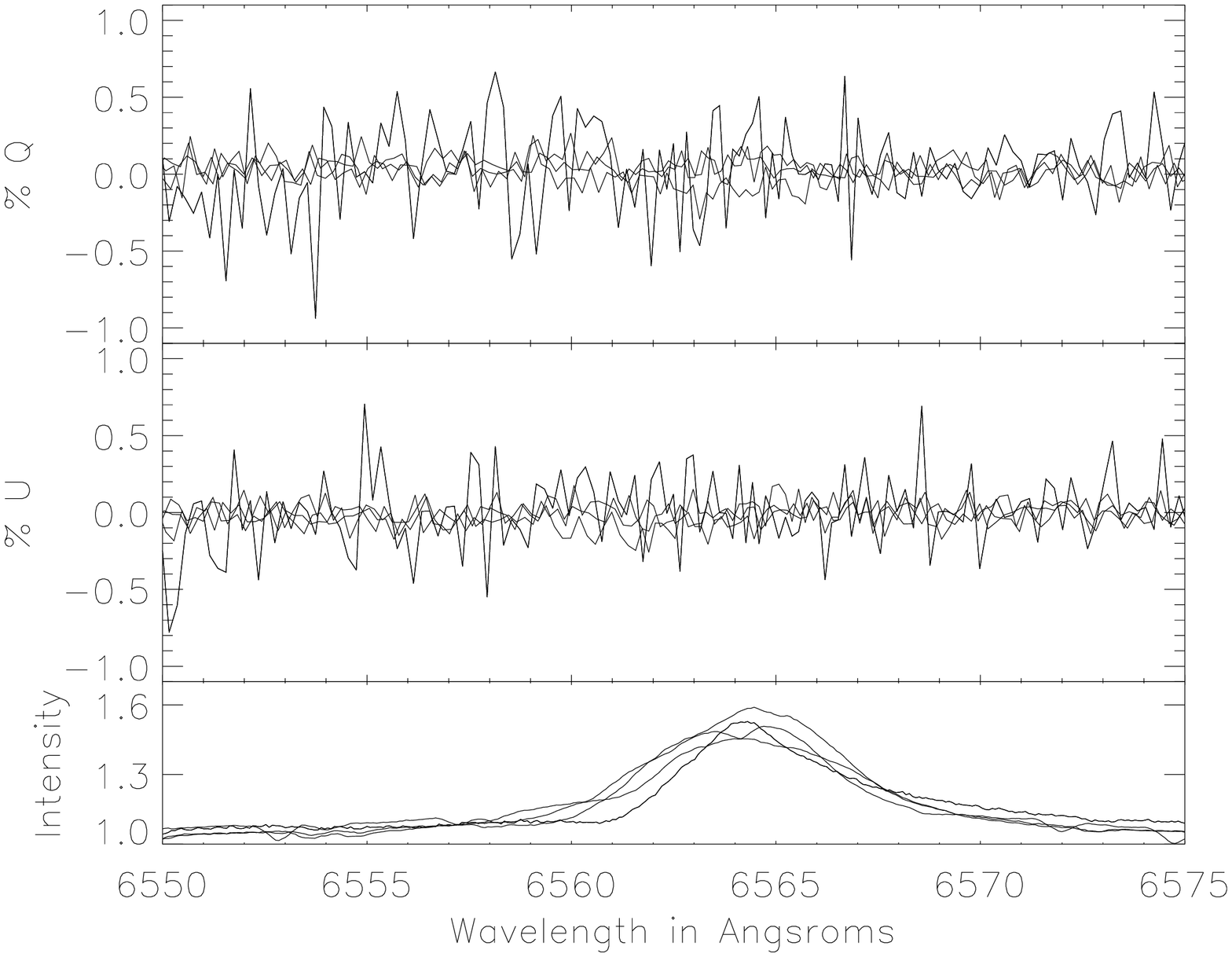}
\includegraphics[width=0.23\linewidth, angle=90]{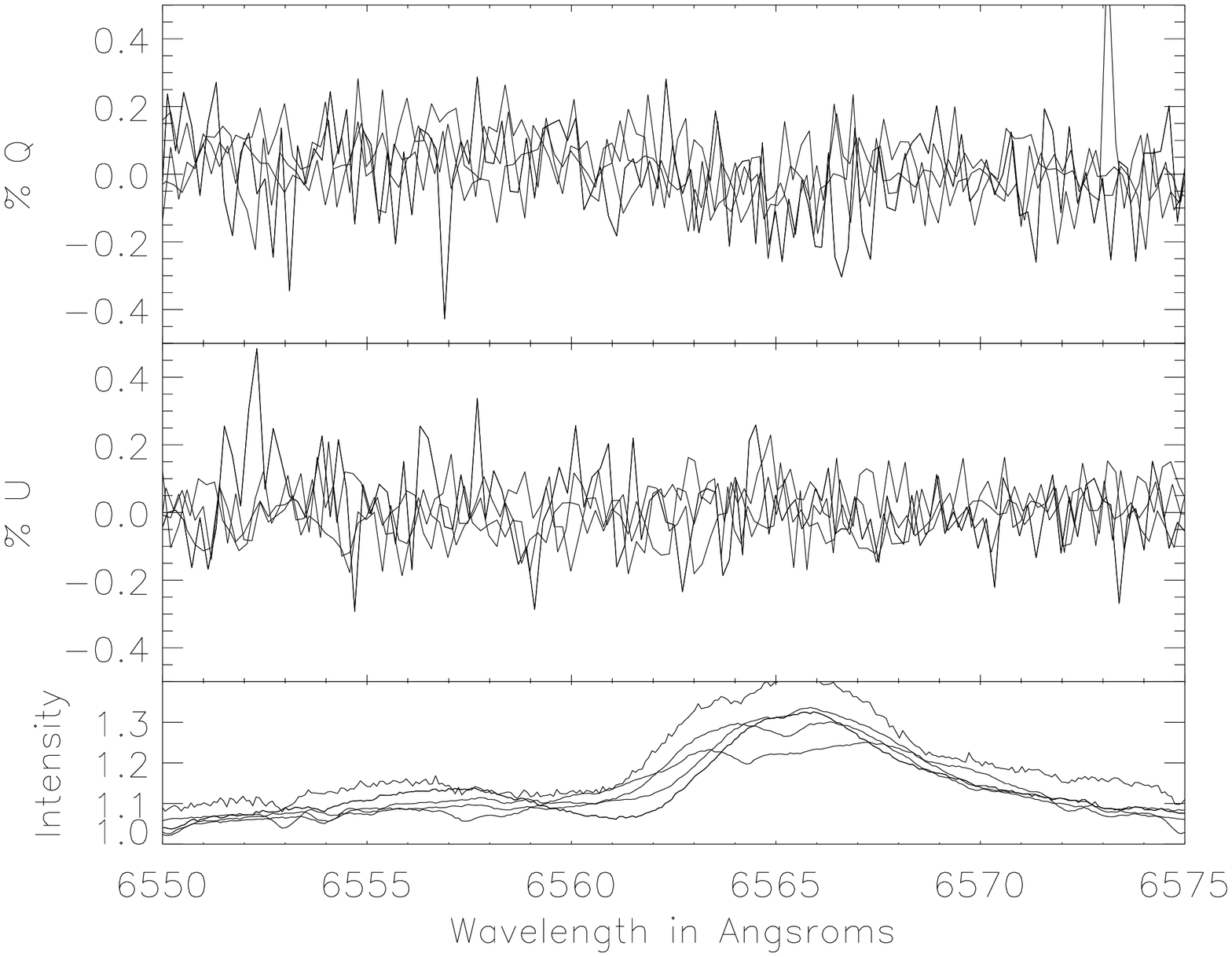} \\
\includegraphics[width=0.23\linewidth, angle=90]{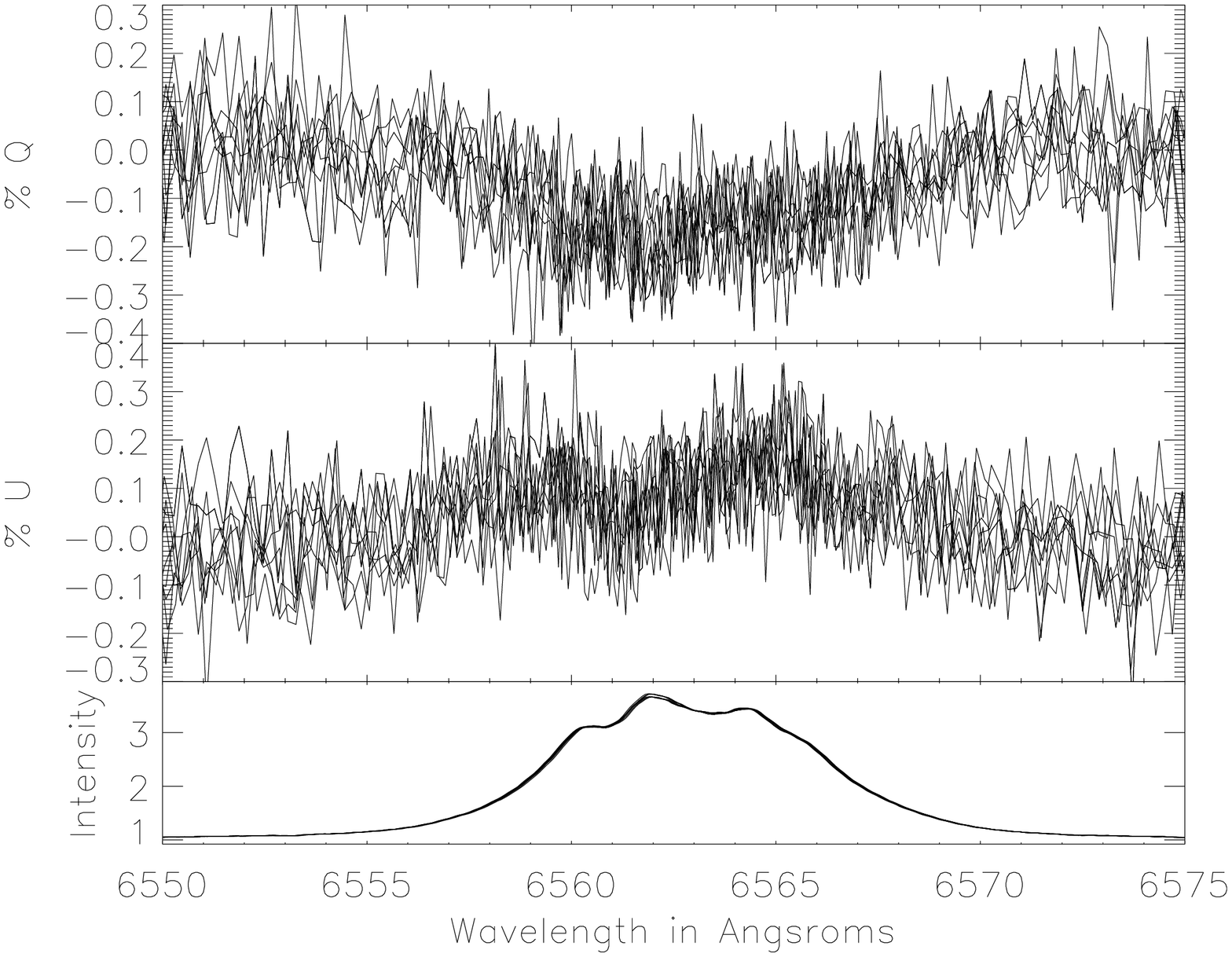}
\includegraphics[width=0.23\linewidth, angle=90]{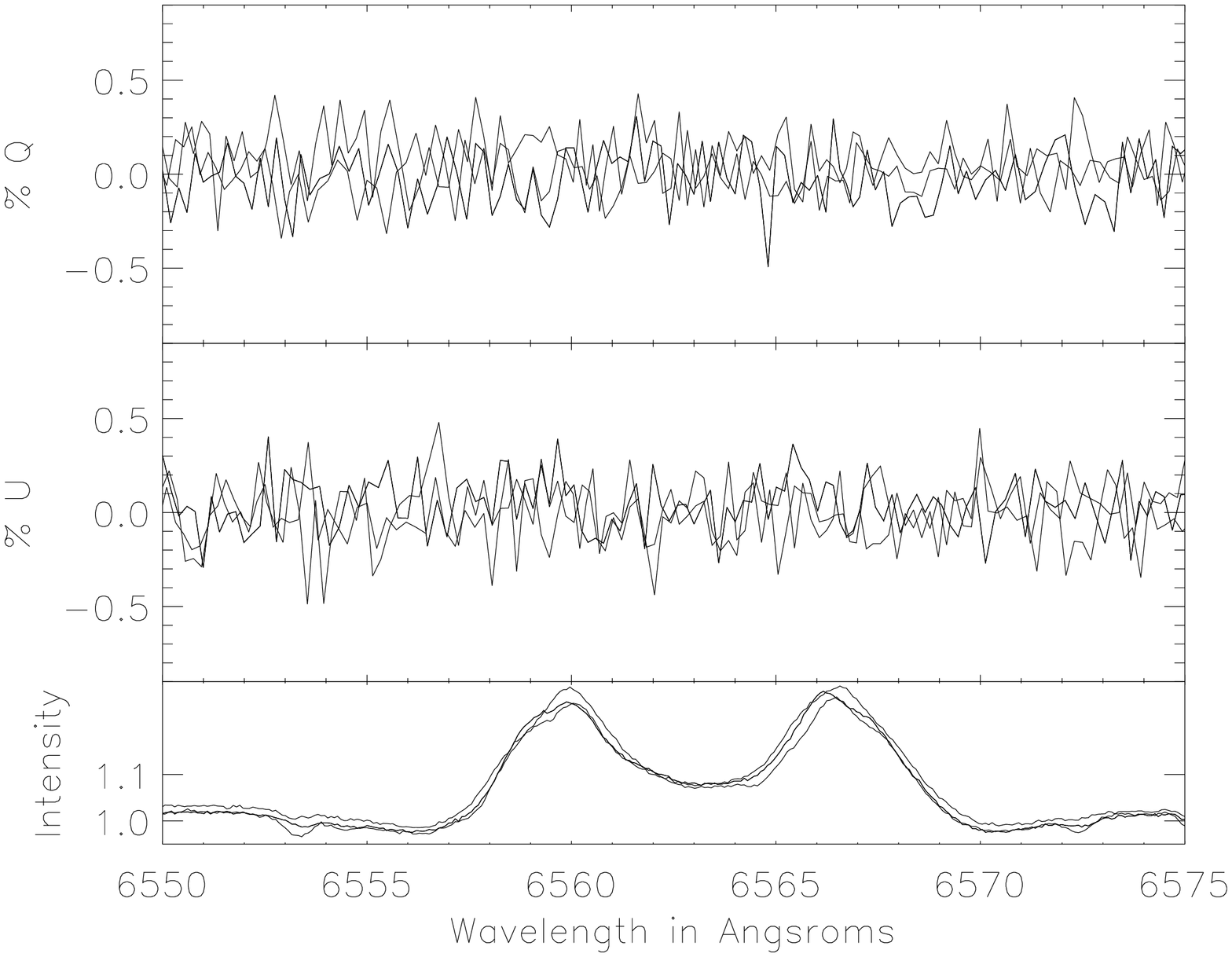}
\includegraphics[width=0.23\linewidth, angle=90]{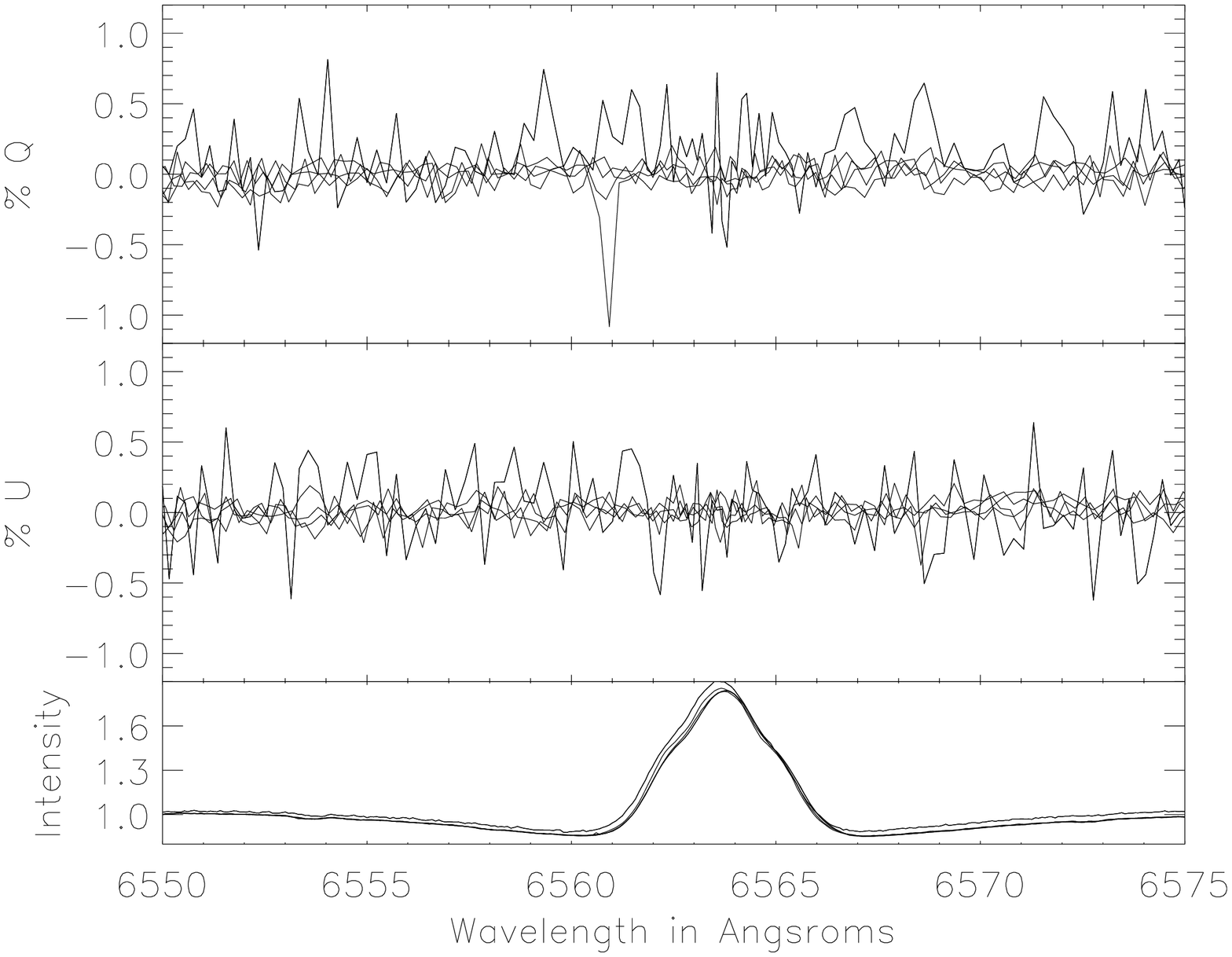} \\
\includegraphics[width=0.23\linewidth, angle=90]{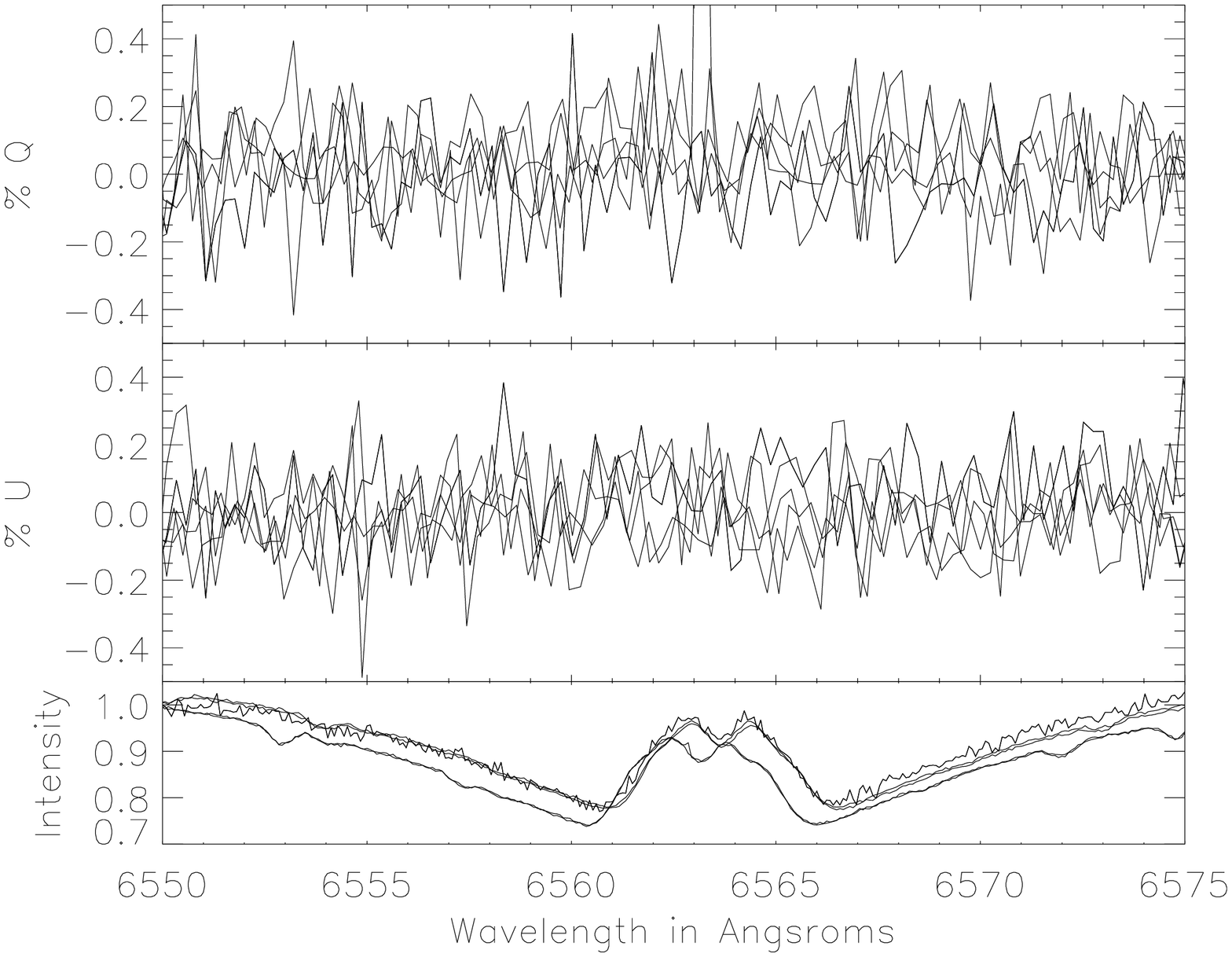}
\includegraphics[width=0.23\linewidth, angle=90]{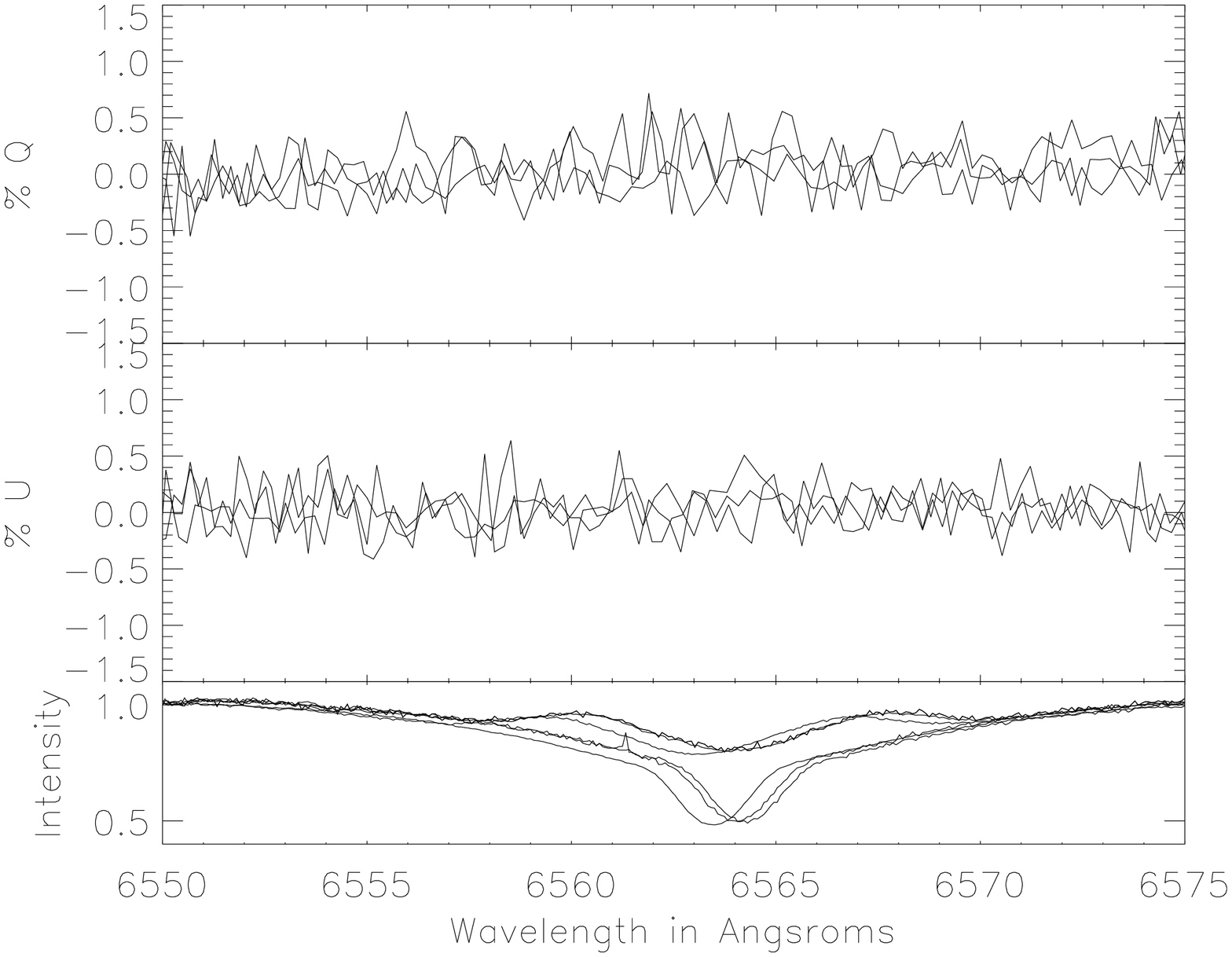}
\includegraphics[width=0.23\linewidth, angle=90]{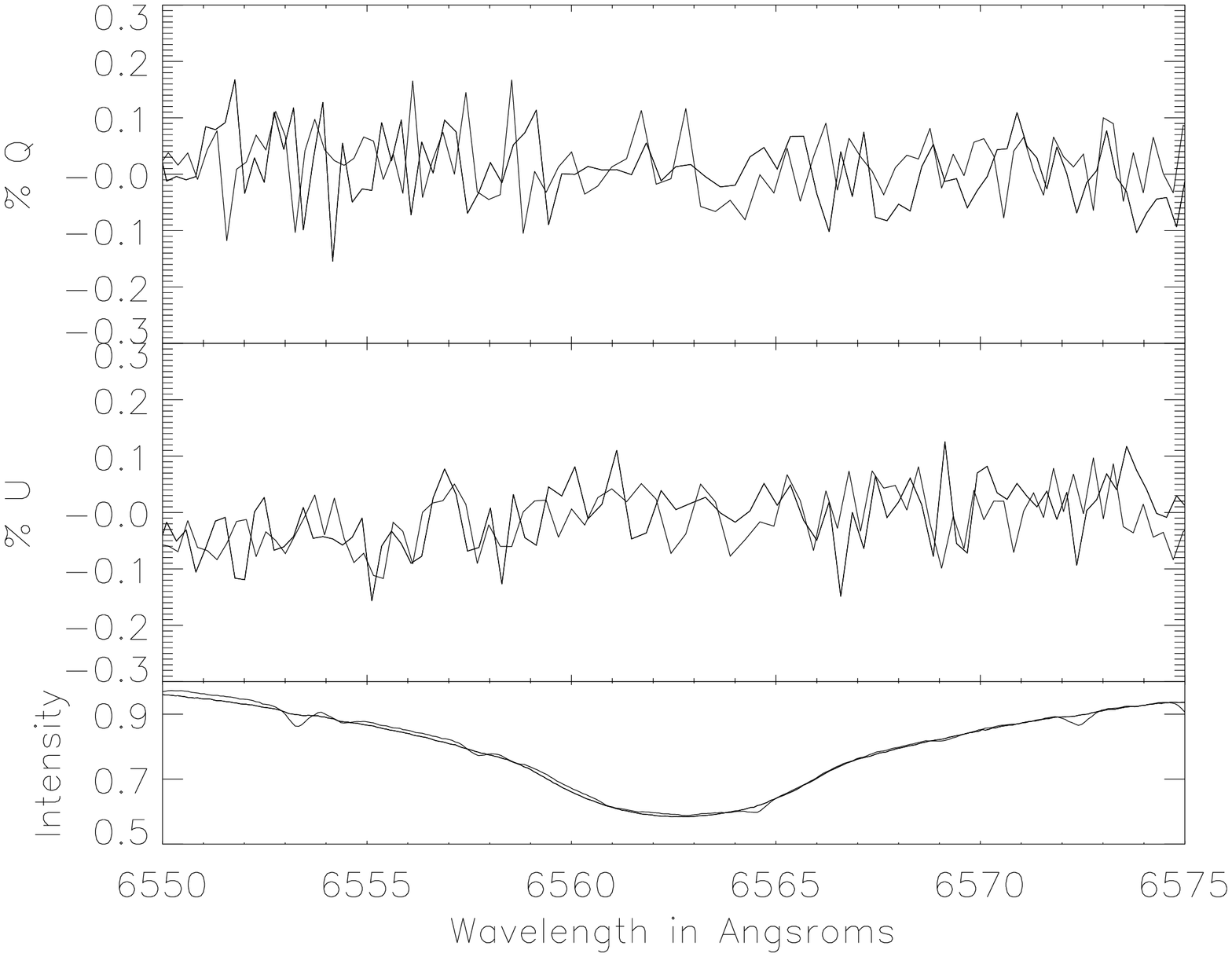} \\
\includegraphics[width=0.23\linewidth, angle=90]{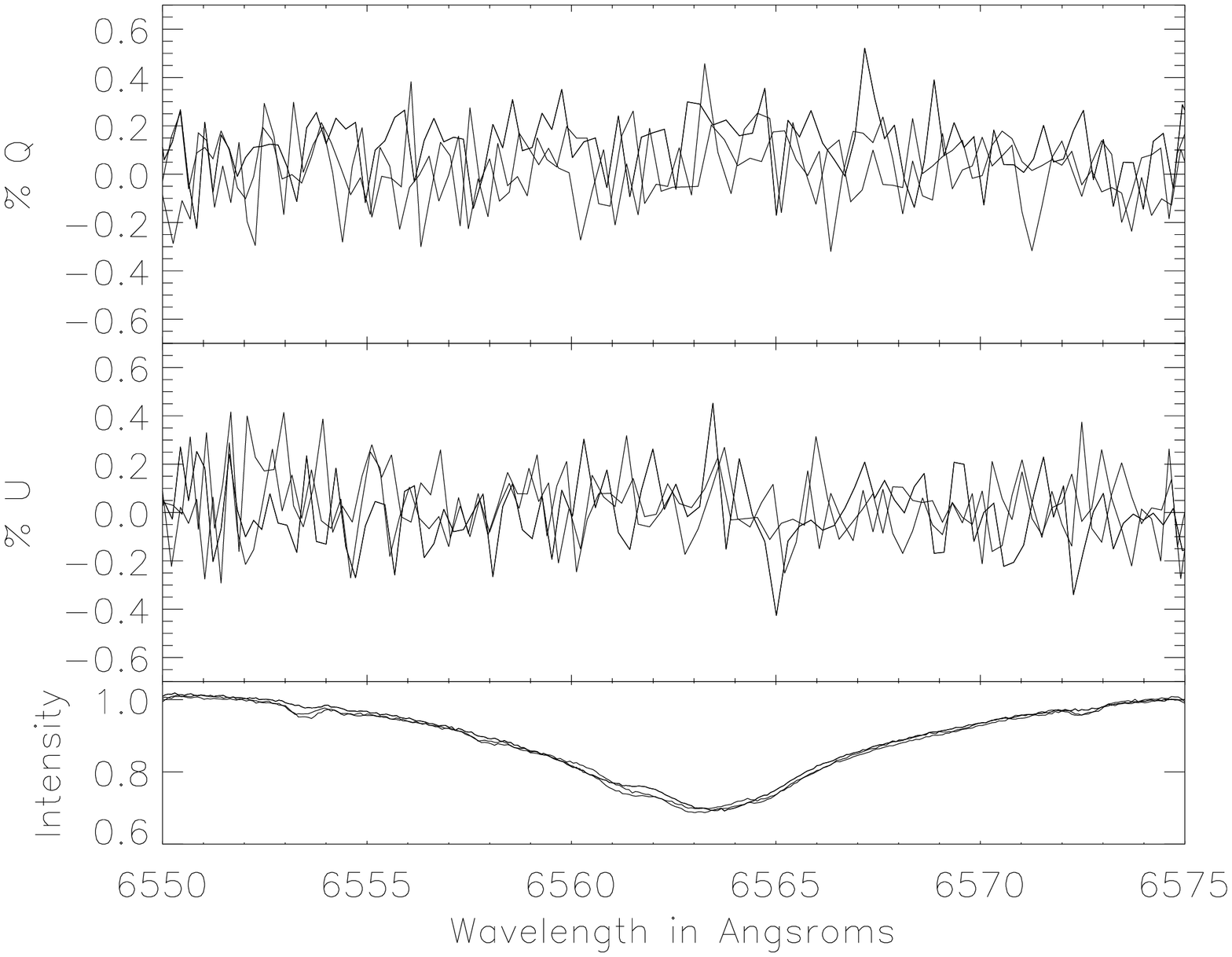}
\includegraphics[width=0.23\linewidth, angle=90]{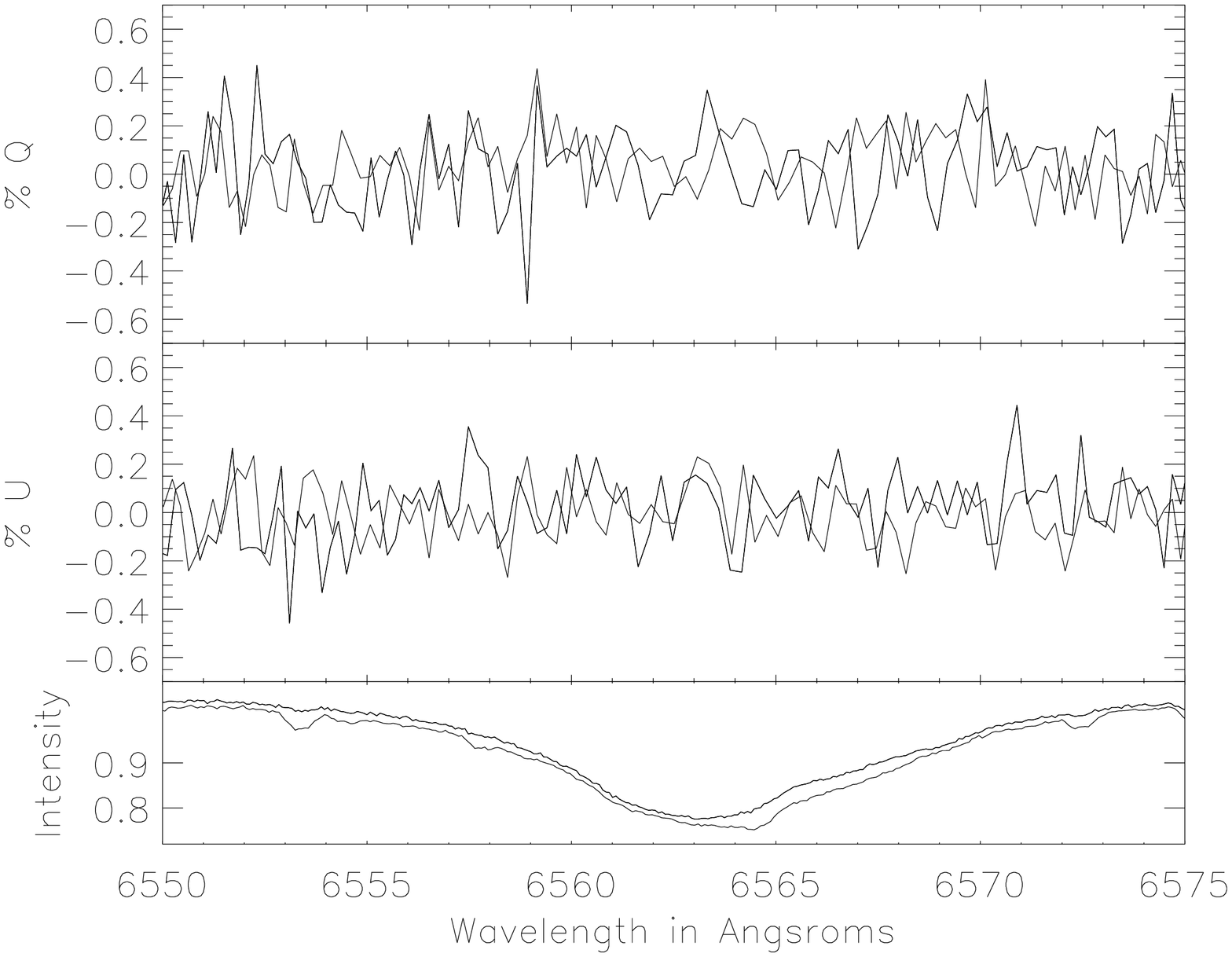}
\includegraphics[width=0.23\linewidth, angle=90]{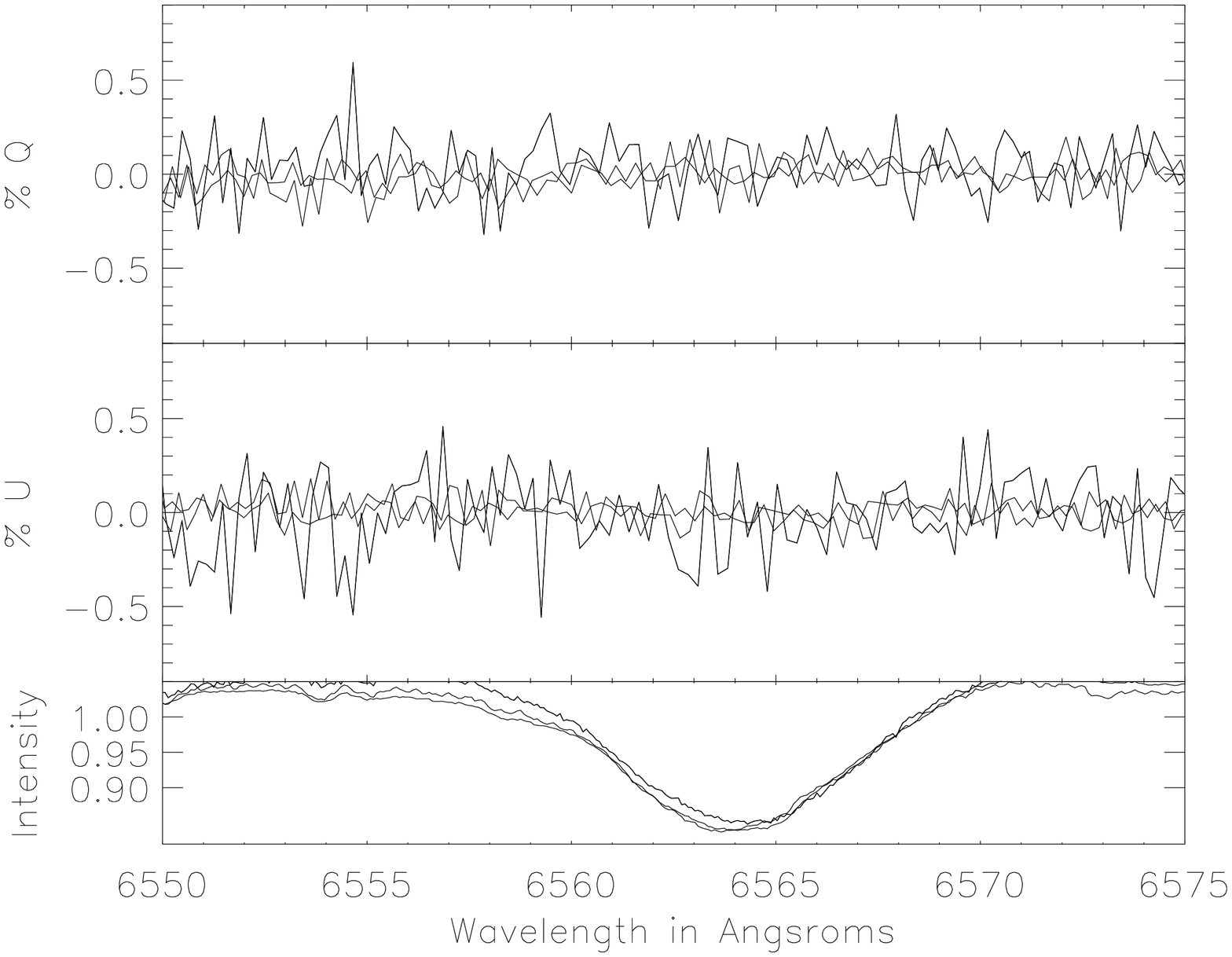}
\caption{Be Spectropolarimetric Profiles II. The stars, from left to right, are: 31 Peg, 11 Cam, C Per, MWC 192, $\kappa$ Dra, $\kappa$ Cas, MWC 92, 12 Vul, $\phi$ And, MWC 77, both binary components of HD 36408, Phecda, $\lambda$ Cyg, QR Vul and $\xi$ Per.}
\label{fig:be-specpol2}
\end{center}
\end{figure*}

\subsection{Be and Emission-Line Star Spectropolarimetry}

	In contrast to the detected HAe/Be morphologies, many Be and Emission-line star signatures showed broad, smooth polarization changes spanning the entire width of the line with the polarization change centered on the line. Although none of these stars show P-Cygni profiles and direct comparison with many HAe/Be detections is not possible, there are several  ``disky" systems provide a robust comparison.
	
	The magnitude of the polarization change in these stars was also smaller on average, with 0.5\% or less being typical. Some stars showed signatures over 1\%, but the polarization signature was much broader than the line and extended out into the line wings. Even though many of the H$_\alpha$ lines in systems with broad spectropolarimetric detections had strong absorption features, there were fewer stars that showed strong deviations from the broad polarization morphology. In 10/30 stars, very clear detections of the broad polarization change can be seen. Of these 10, four had significant spectropolarimetric effects in and around absorptive features.

	The compiled spectropolarimetry is shown in figures \ref{fig:be-specpol1} and \ref{fig:be-specpol2}. The detections are outlined in table \ref{be-res}. There are 30 total targets. Of these, ten stars that show the broad signature, five stars show narrow low-amplitude polarization effects, and the other 15 stars are non-detections at the 0.05\% to 0.1\% level. This detection rate can be broken down into subclasses. The broad signatures range in magnitude from a barely detectable 0.1\% to 1.1\%. All of the stars with detected signatures show evidence of intervening absorption in their line profiles, either by the flattened or notched line centers, or by having a full-blown ``disky" line. The stars with broad spectropolarimetric effects do show morphological deviations from the typical depolarization profile, but the polarization is quite different from the polarization-in-absorption of the Herbig Ae/Be systems. In four of the 10 broad-signature stars, more complex effects in and around absorptive components of the line can be seen. These effects can be large in amplitude but would not have been observed at low spectral resolution. These are observed in addition to the broad ``depolarization'' effect, unlike the Herbig Ae/Be stars. The five narrow detections do not fit the depolarization description and will be discussed in more depth later.

\subsection{$\gamma$ Cas - A Clear Example}

	$\gamma$ Cas is a very well-studied target that has an extended envelope and a strong H$_\alpha$ emission line. Quirrenbach et al. 1993 resolved the H$_\alpha$ emission region using interferrometry. An elliptical Gaussian model fits the observations with an axial ratio of 0.74 and an angular FWHM of 3.2mas. This flattened envelope is exactly the kind imagined in the original studies on continuum polarization from circumstellar envelopes.

\begin{figure}
\begin{center}
\includegraphics[width=0.75\linewidth, angle=90]{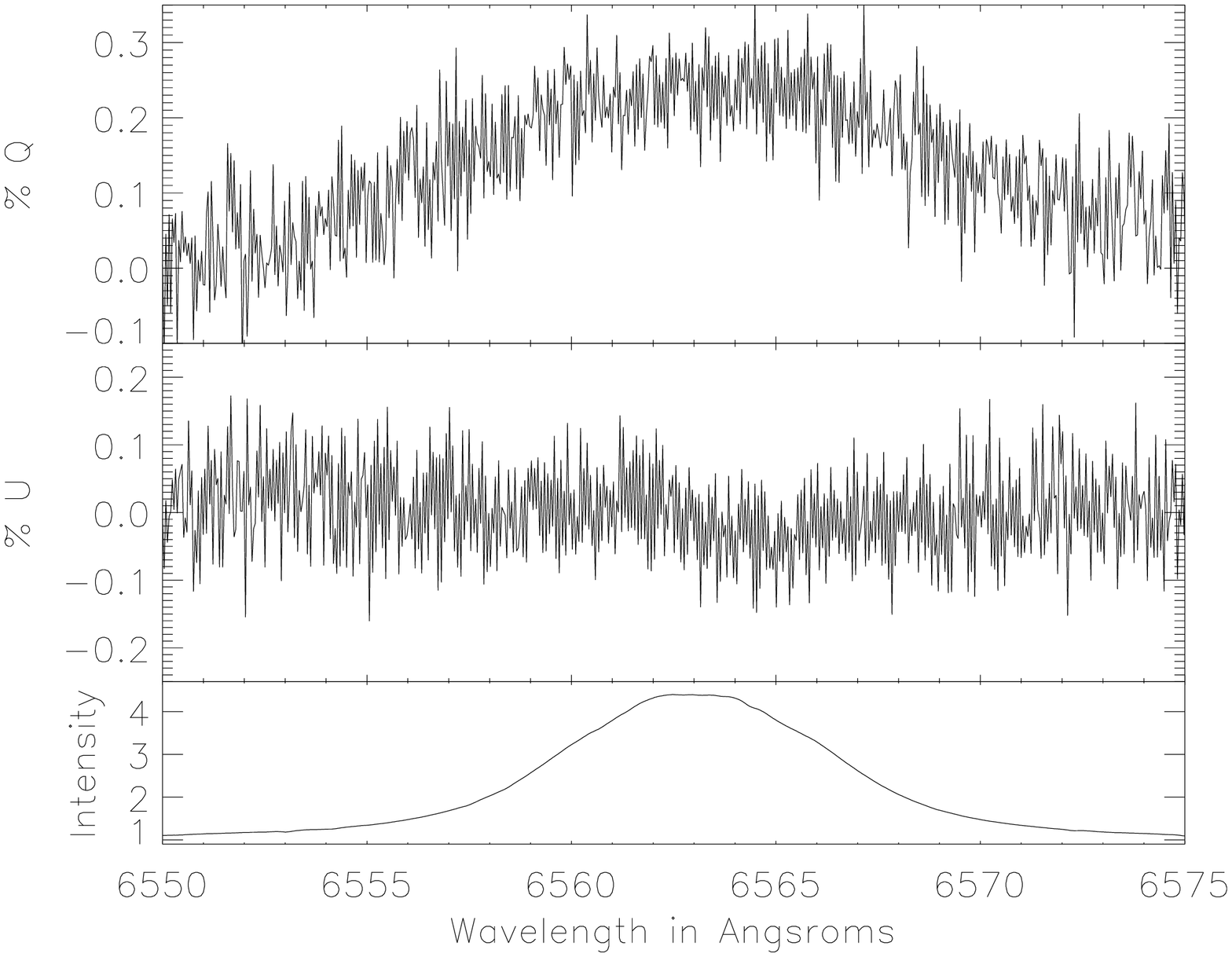} \\
\includegraphics[width=0.75\linewidth, angle=90]{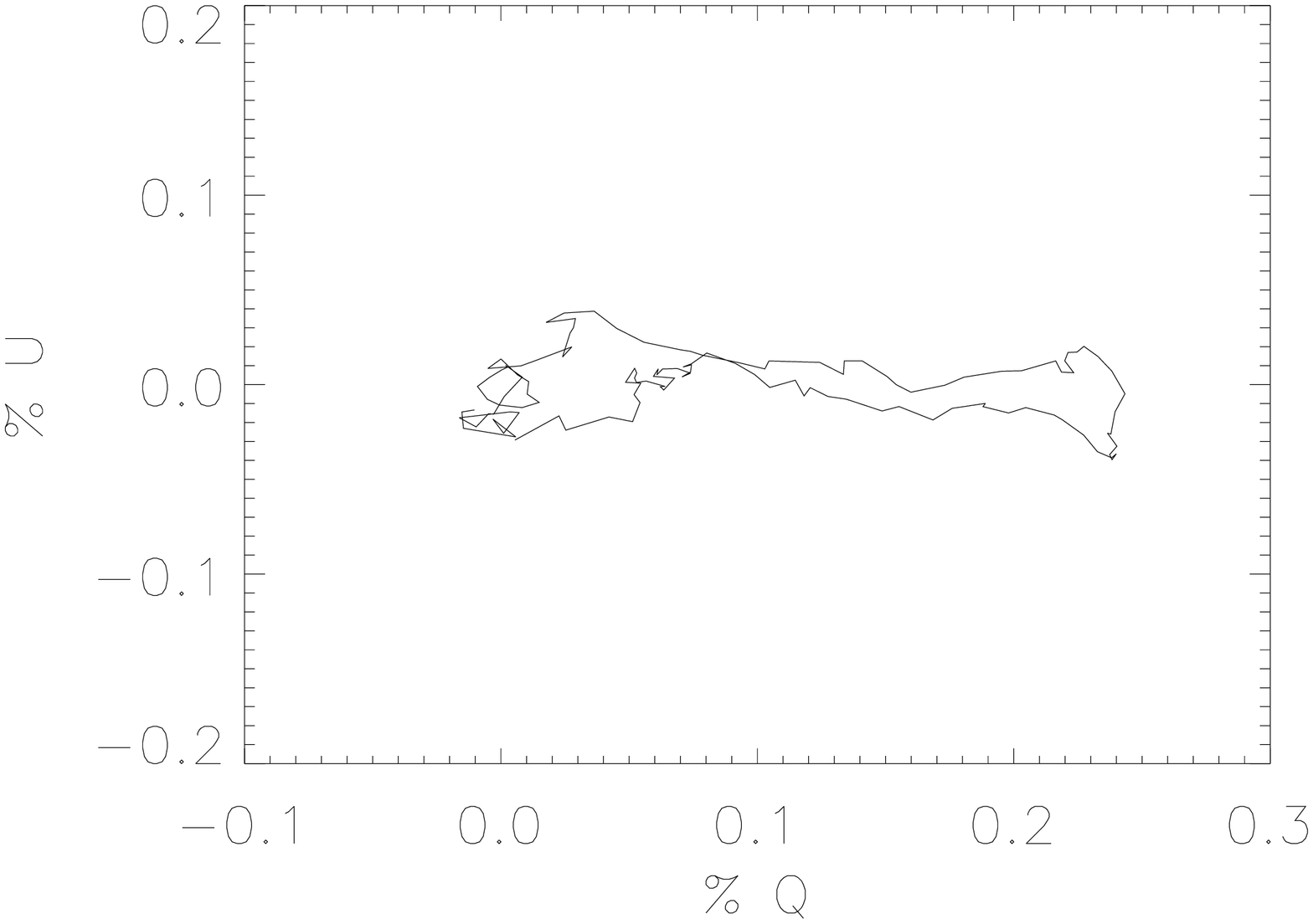}
\caption{$\gamma$ Cas as a broad-effect example. {\bf a)} The average $\gamma$ Cas Spectropolarimetry at full resolution. All observations have been rotated to maximize Stokes q in line-center then averaged {\bf b)} the corresponding qu plot - a linear extension. The +q orientation is an artifact of the arbitrary alignment.}  
\label{fig:bebroadex}
\end{center}
\end{figure}

	$\gamma$ Cas was monitored quite heavily over a year and we can use this target as a clear example for the entire class. This star did not show very significant H$_\alpha$ line-profile or spectropolarimetric variations throughout the monitoring period. All the spectropolarimetric observations were taken without binning and aligned in qu-space to maximize Stokes q in the central 50 pixels. These polarized spectra all had similar morphology but slightly varying magnitudes because of the varying telescope polarization properties at the many different telescope pointings used. The resulting 22 spectra were then averaged to make a single polarized spectra with only 0.07\% rms error at full spectral resolution (no bin-by-flux was applied to the observations). Figure \ref{fig:bebroadex} shows the spectropolarimetry. The polarization signature is almost entirely in the +q spectrum. Though the spectra have been arbitrarily rotated to maximize +q, this figure shows that the signature is almost entirely linear in qu-space. In the qu-loop plot, this polarization change becomes a simple horizontal excursion. The polarization change is roughly 0.3\% and is morphologically broader than the line-profile itself. There is a very small width to the extension which comes from a very slight variation in the u spectrum on the red side of the emission line, as seen in the middle panel of the spectropolarimetry in figure \ref{fig:bebroadex}. A decrease in polarization of the same 0.2\% magnitude had been reported in commissioning runs with the William-Wehlau spectropolarimeter (Eversberg et al. 1998), though at lower spectral resolution.
	
	Since the signature is broad and is a single linear excursion in qu-space, this star fits well with the traditional ``depolarization"  effect from scattering theory. The change in polarization is smaller in magnitude than the known intrinsic continuum polarization. The initial polarization studies all found a similar depolarization in continuum polarizations of: 1.0\% McLean \& Brown 1978, 0.8\% McLean 1979, 0.5\% in Poeckert \& Marlborough 1977.

\begin{figure*}
\begin{center}
\includegraphics[width=0.23\linewidth, angle=90]{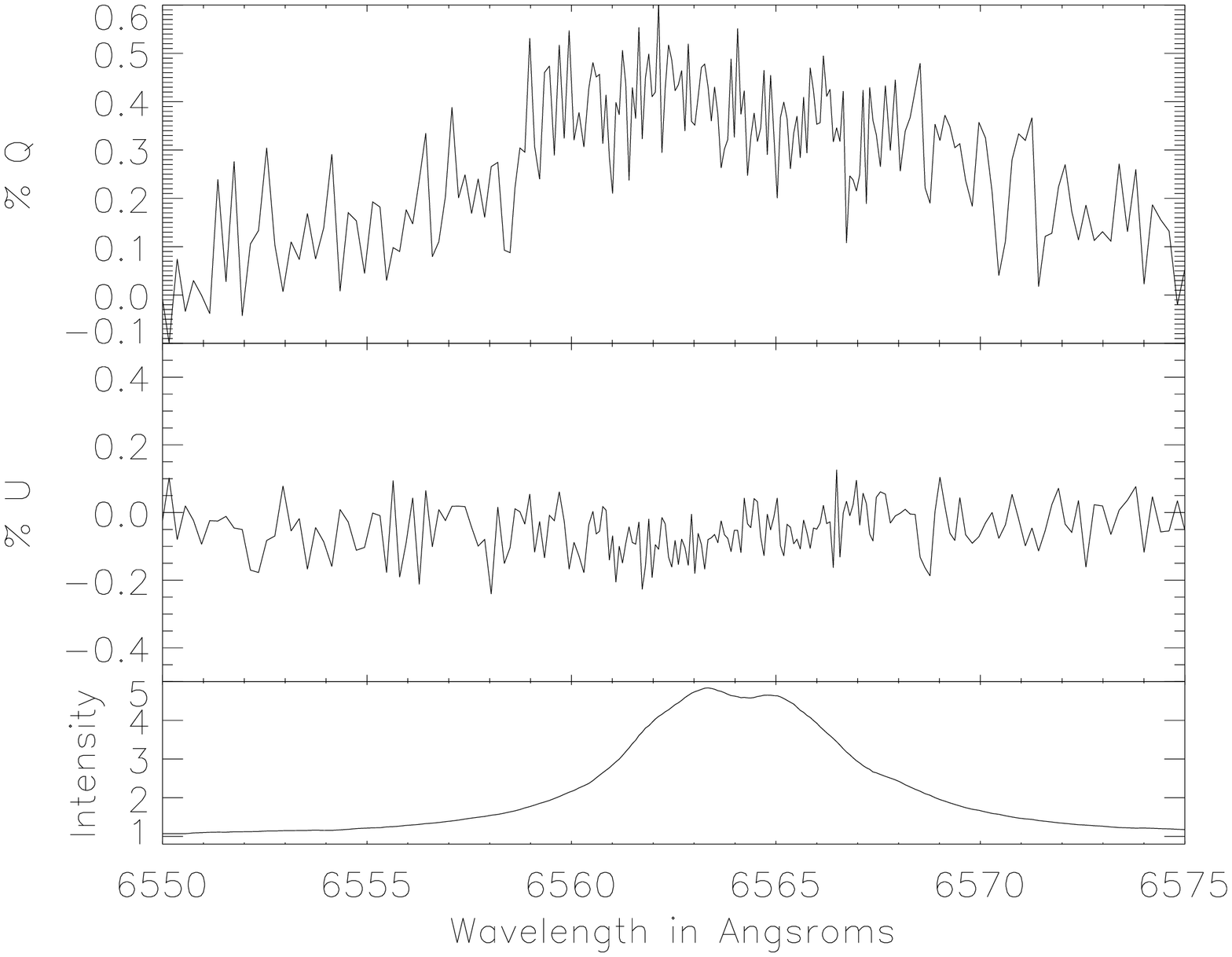}
\includegraphics[width=0.23\linewidth, angle=90]{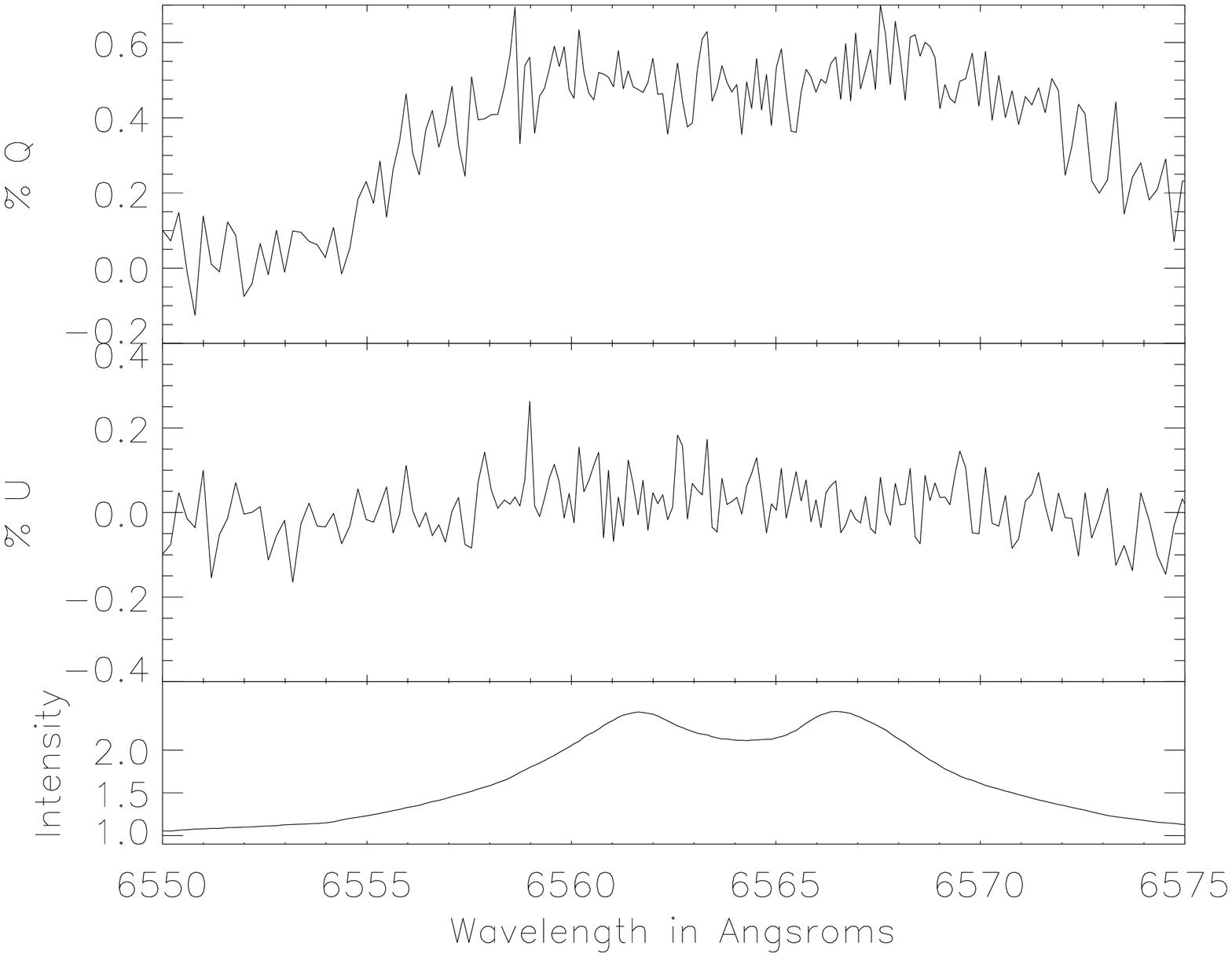}
\includegraphics[width=0.23\linewidth, angle=90]{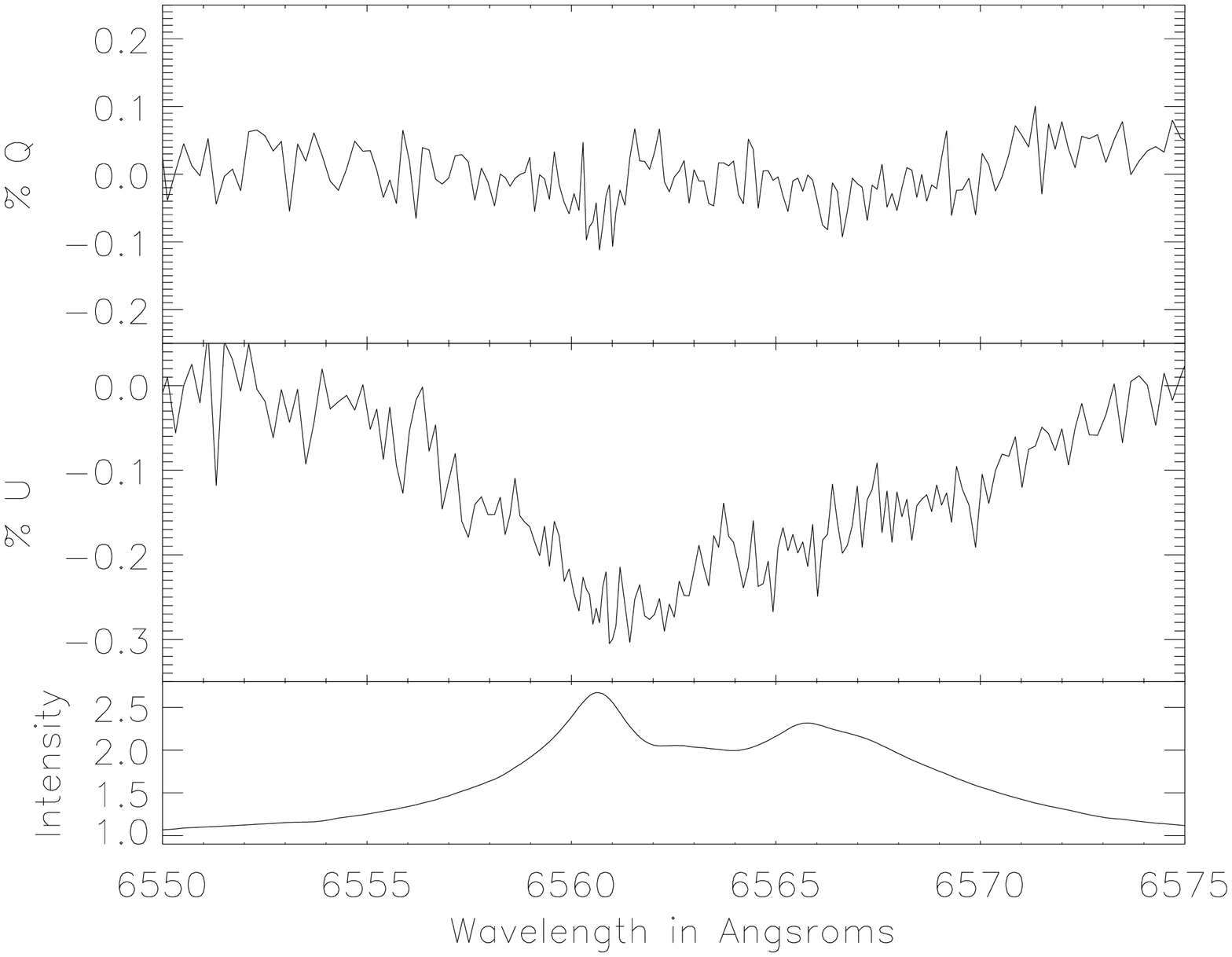} \\
\includegraphics[width=0.23\linewidth, angle=90]{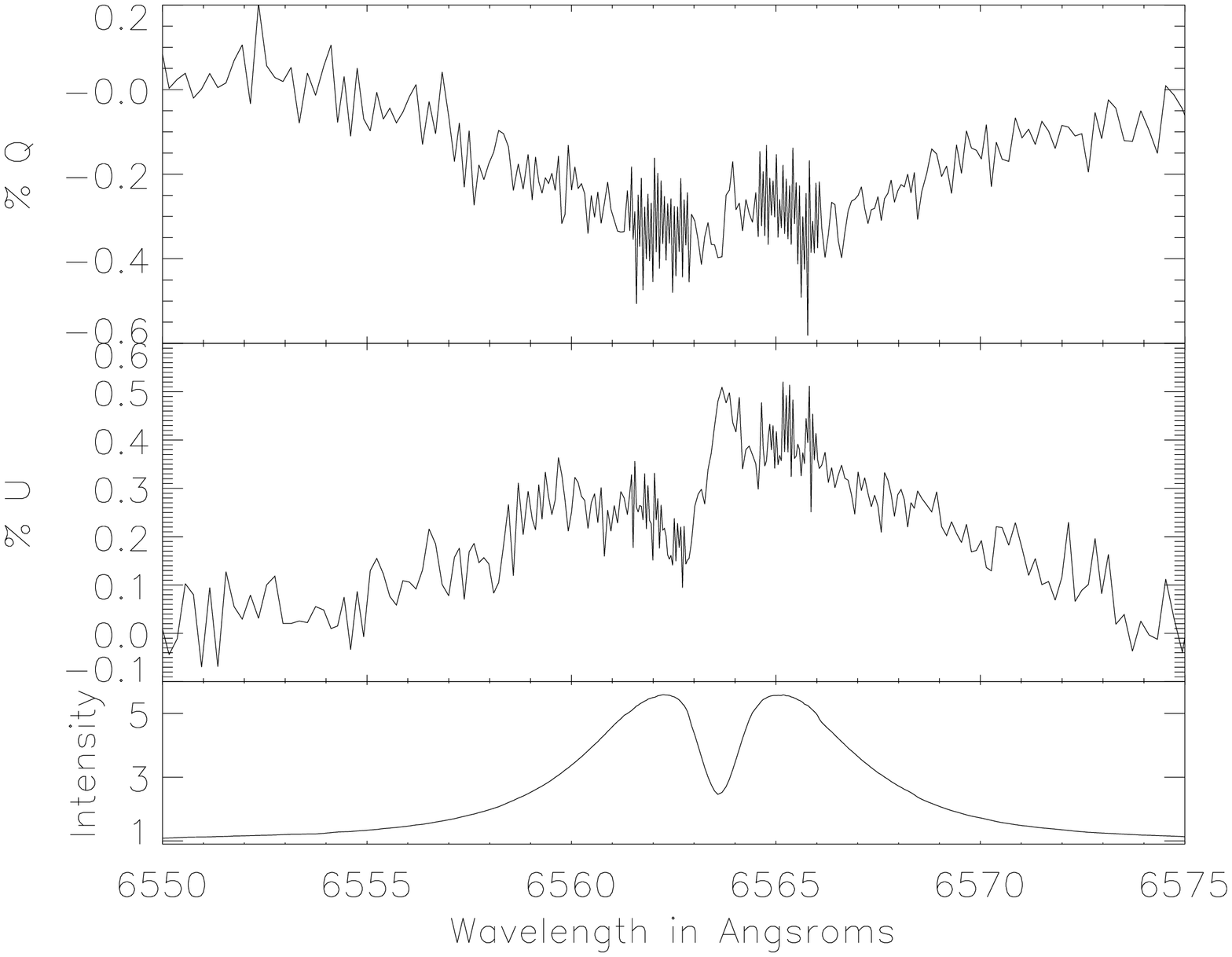}
\includegraphics[width=0.23\linewidth, angle=90]{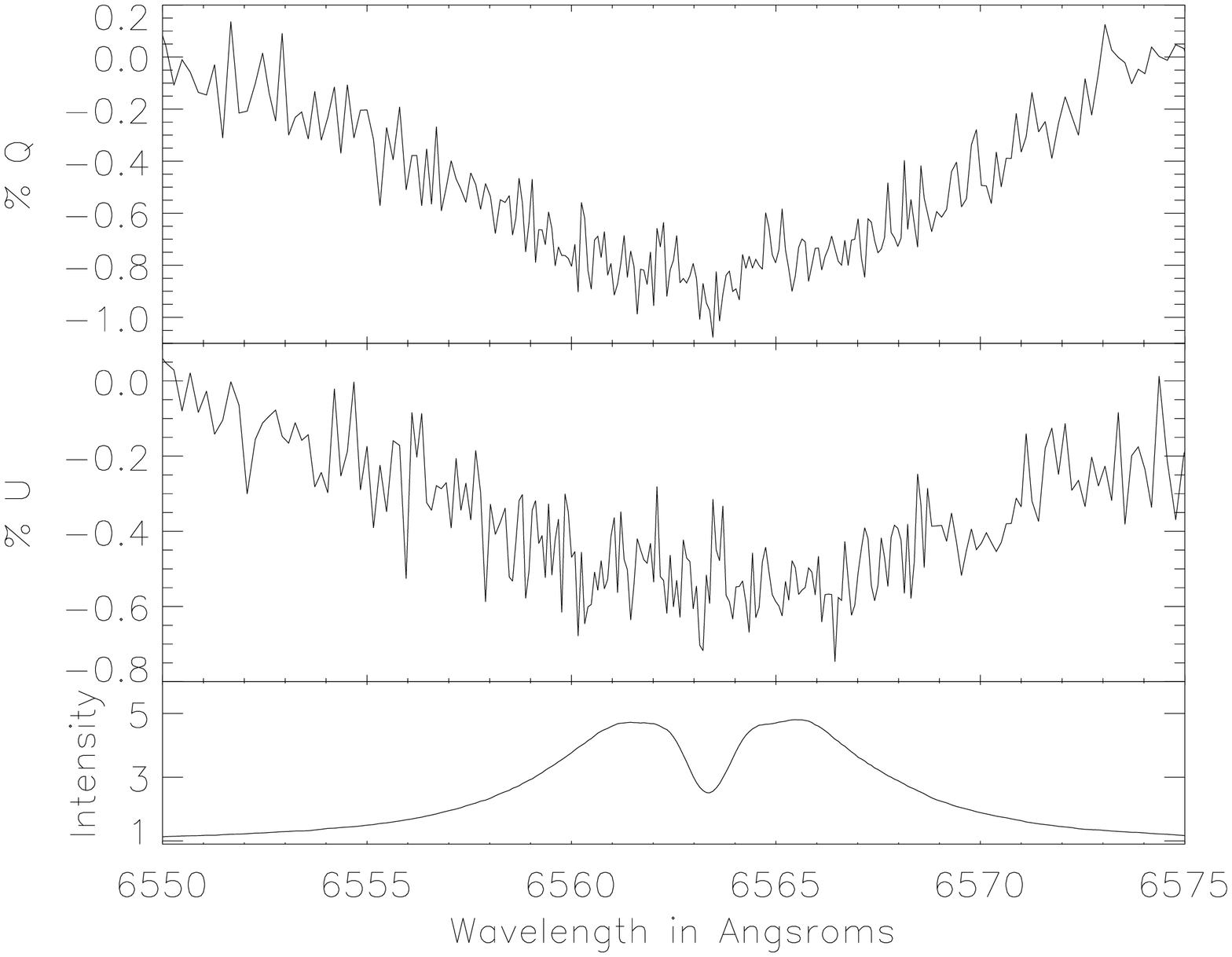}
\includegraphics[width=0.23\linewidth, angle=90]{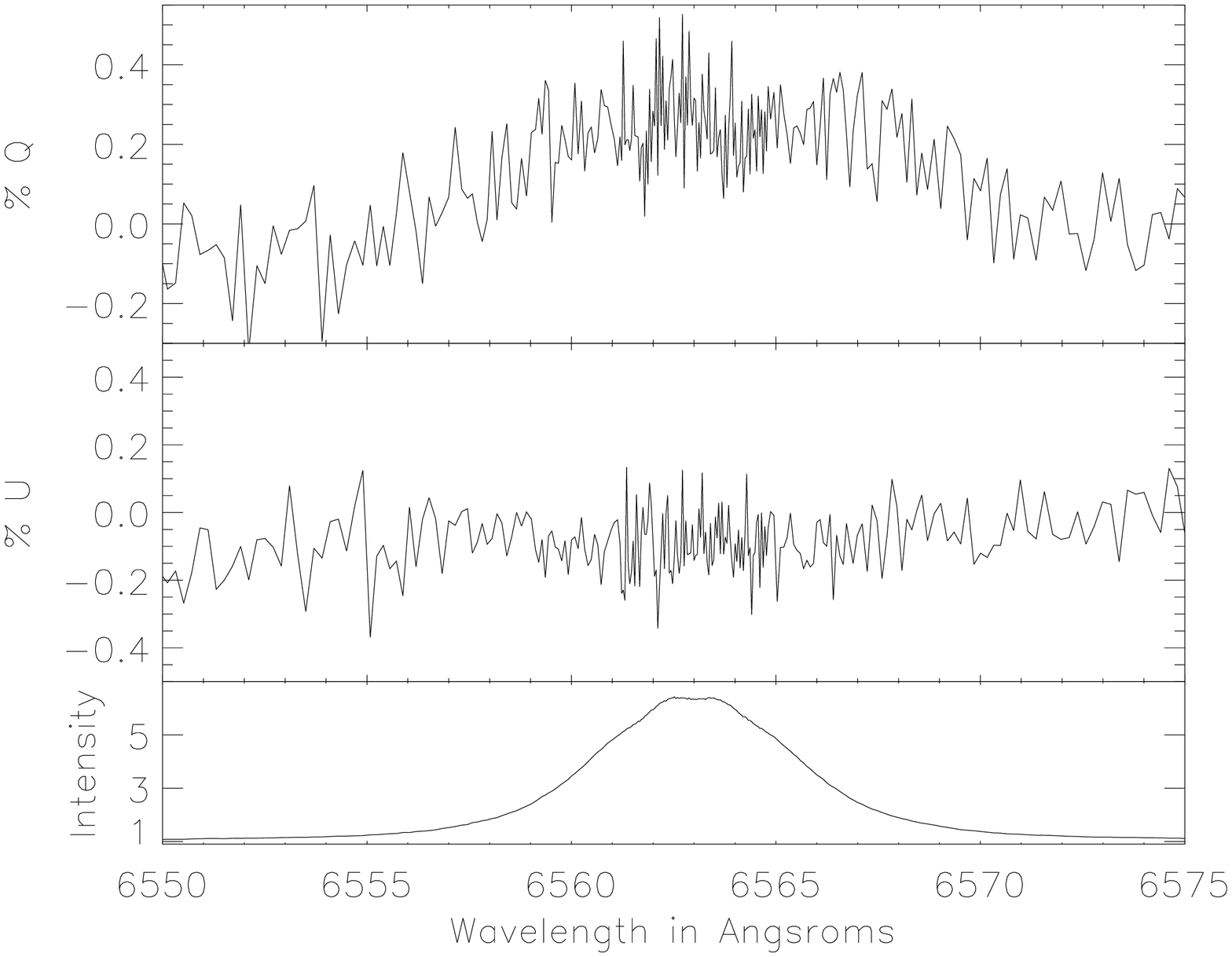} \\
\includegraphics[width=0.23\linewidth, angle=90]{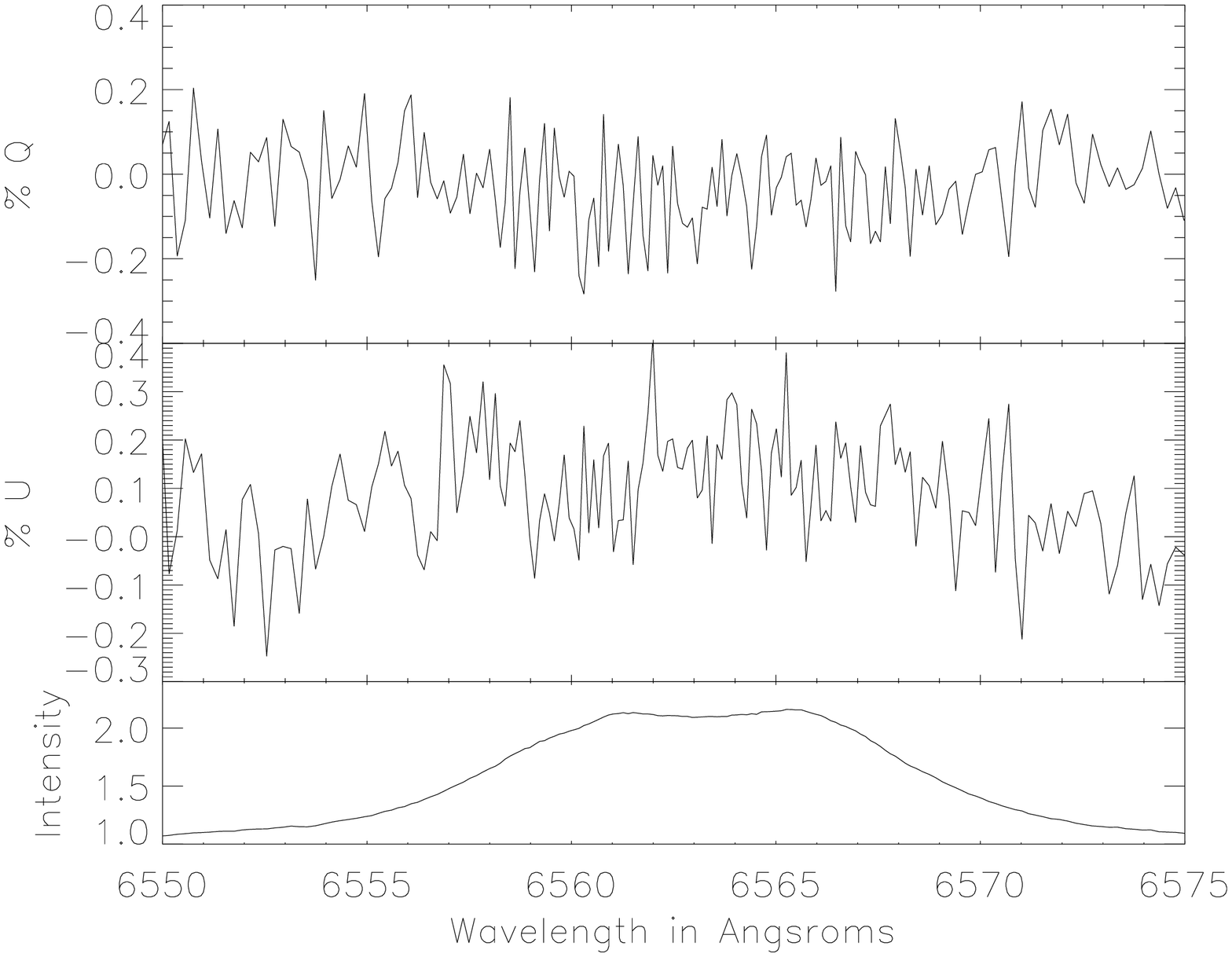}
\includegraphics[width=0.23\linewidth, angle=90]{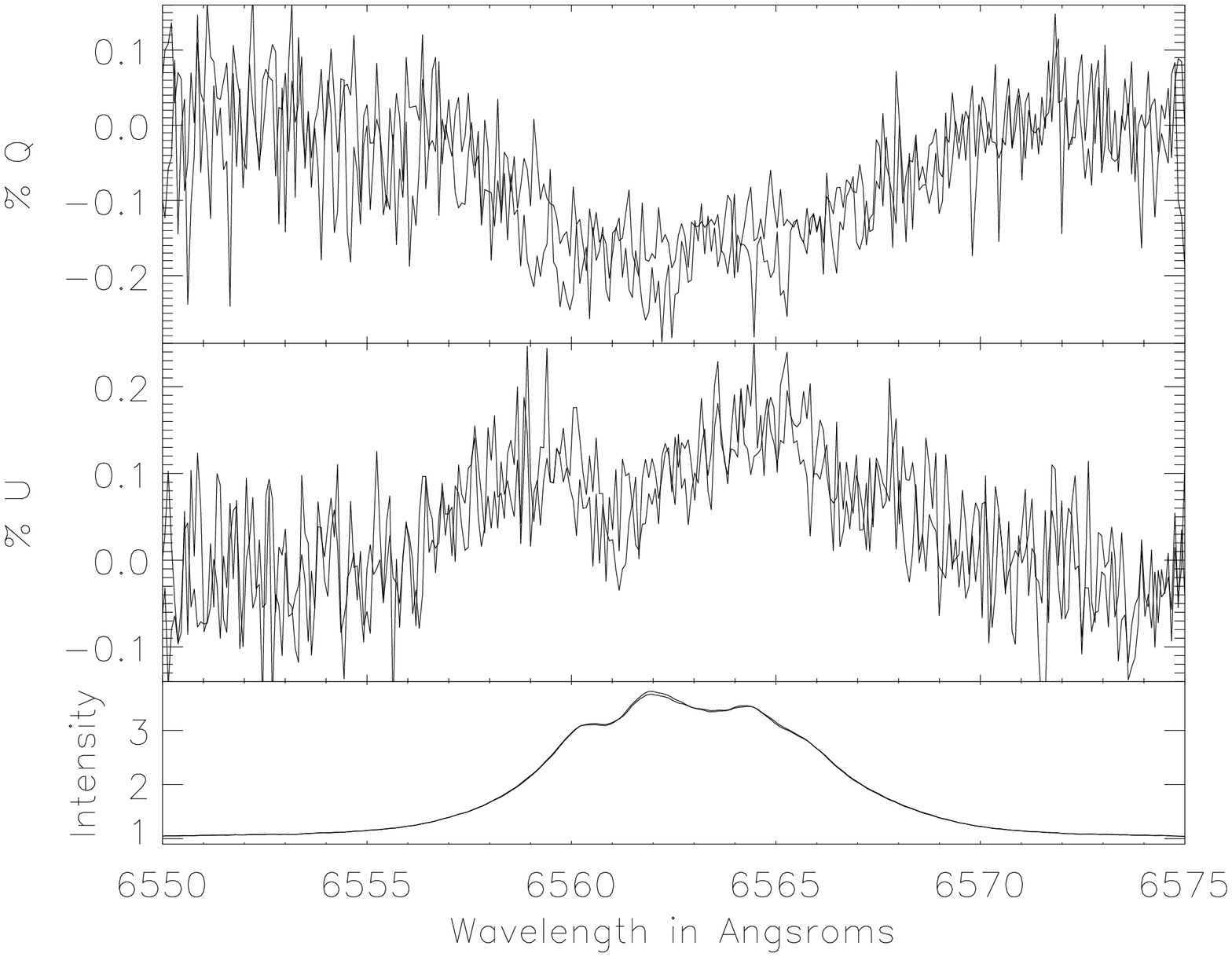}
\includegraphics[width=0.23\linewidth, angle=90]{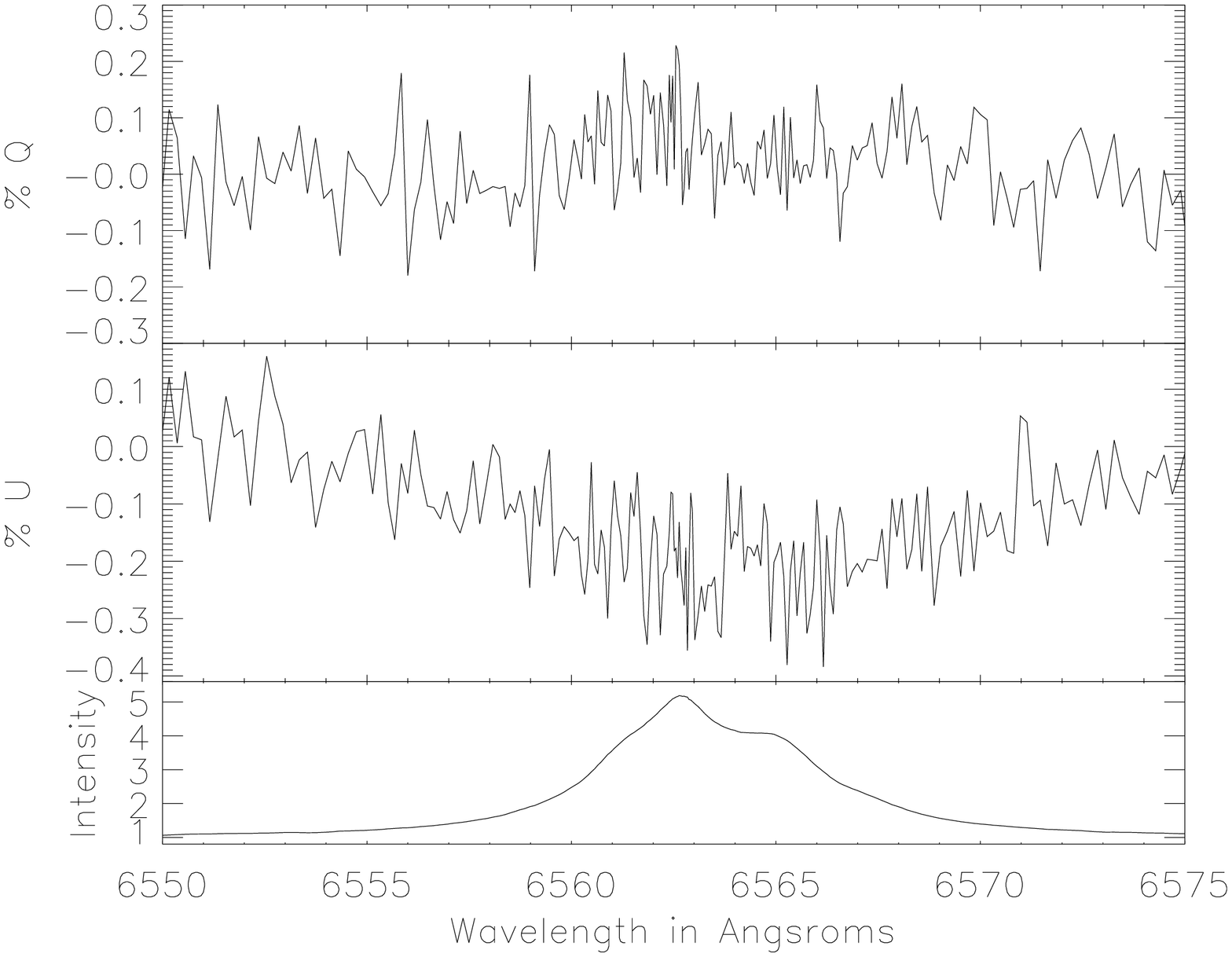}
\caption{Individual spectropolarimetric examples of HiVIS detections for 9 of the 10 stars with broad spectropolarimetric signatures. The 10th, $\gamma$ Cas, is shown in the previous figure. The individual detections are from left to right: {\bf a)}10 CMa  {\bf b)} 25 Ori.  {\bf c)} $\zeta$ Tau {\bf d)} $\psi$ Per {\bf e)} MWC 143  {\bf f)} Omi Cas.  {\bf g)} Omi Pup {\bf h)} $\kappa$ Dra {\bf i)} $\kappa$ CMa. The detections are all more broad than the H$_\alpha$ line itself. $\psi$ Per and $\kappa$ Dra show fairly strong morphological changes across the line and there is significant evidence for absorption. Omi Pup and 25 Ori are severely "flat-topped". $\zeta$ Tau is highly asymmetric with the strongest polarization change on the blue side of the emission line. $\kappa$ Dra shows two independent observations with significant morphological deviations across the line.}  
\label{fig:bebroad}
\end{center}
\end{figure*}

\begin{figure*}
\begin{center}
\includegraphics[width=0.35\linewidth, angle=90]{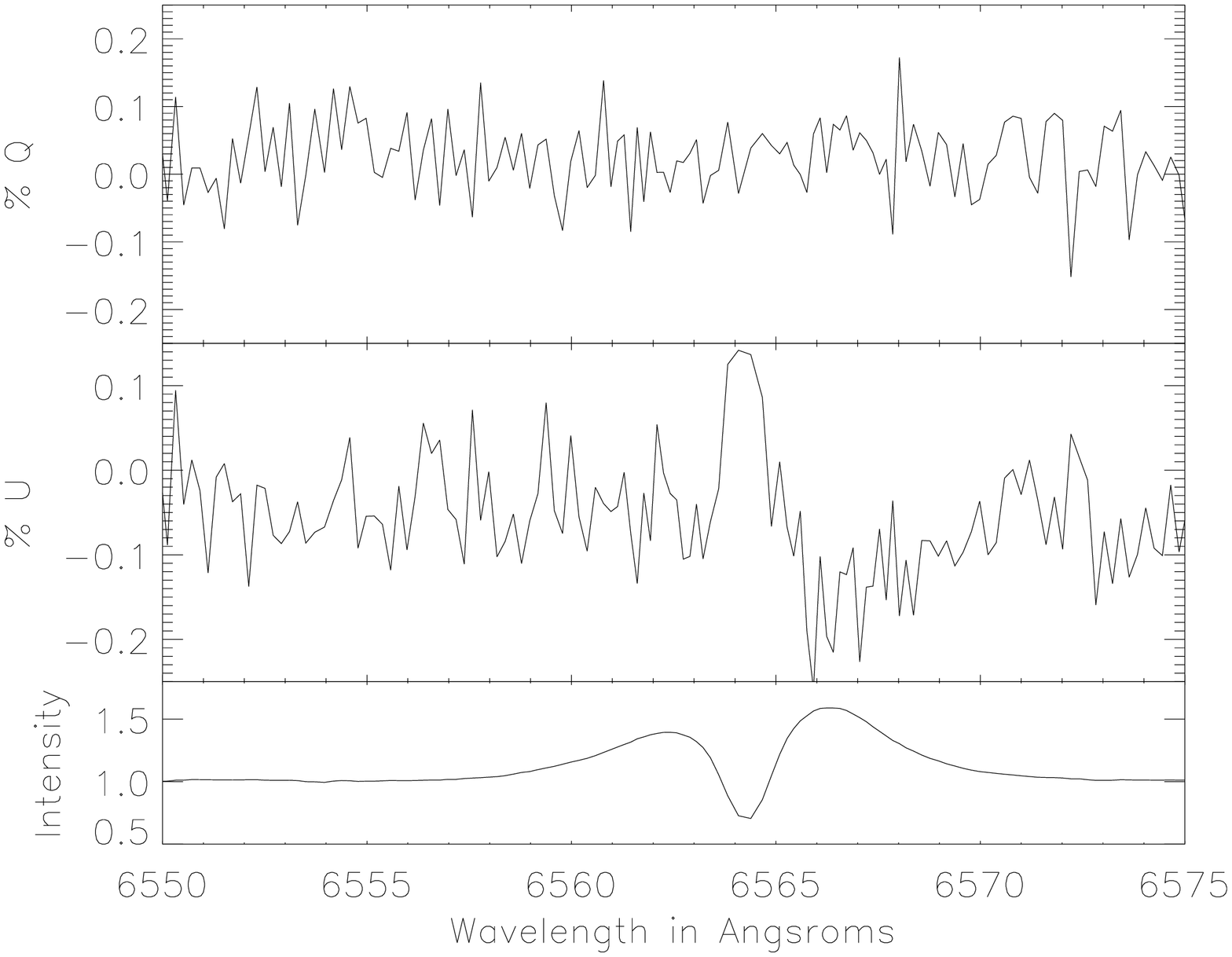}
\includegraphics[width=0.35\linewidth, angle=90]{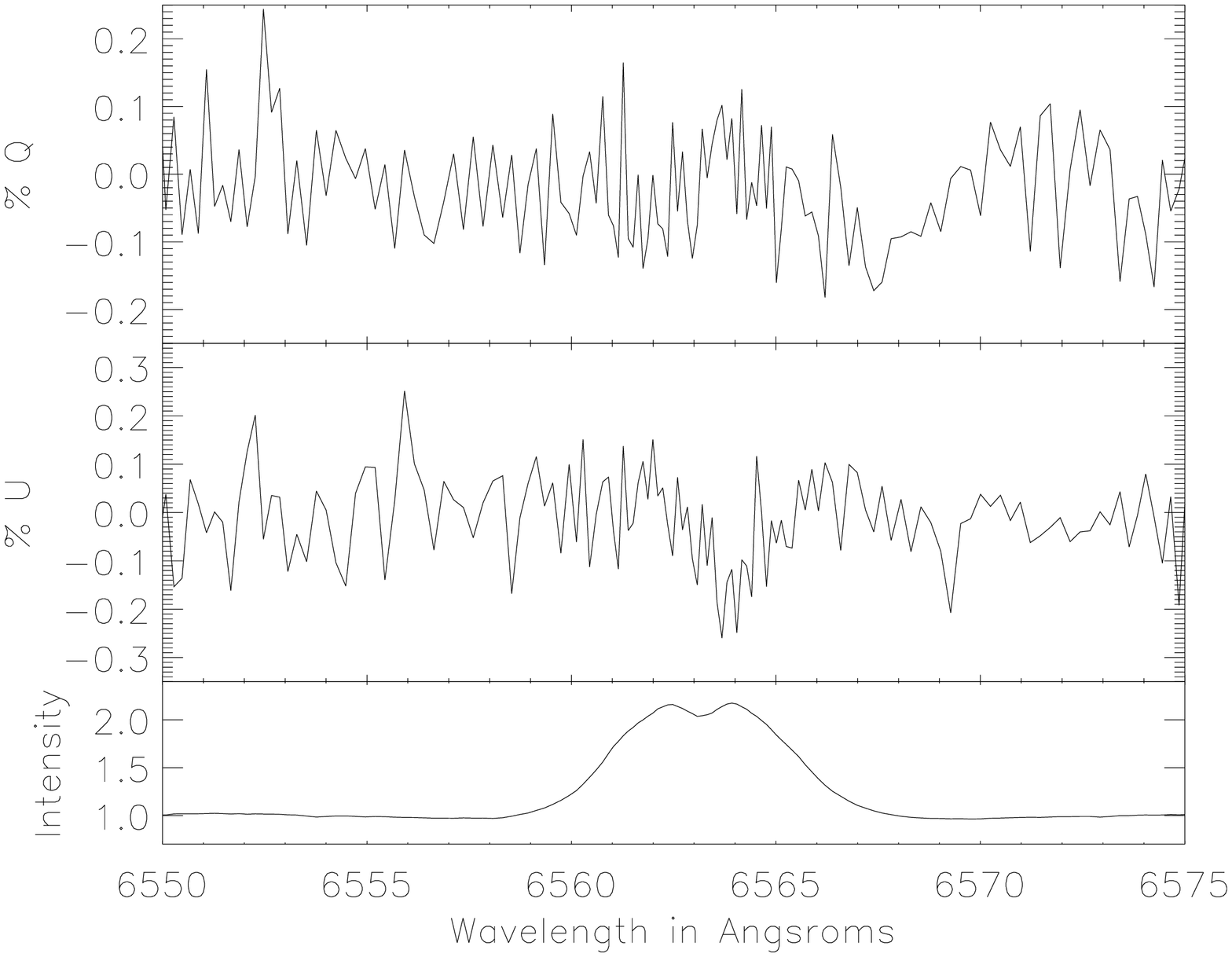} \\
\includegraphics[width=0.35\linewidth, angle=90]{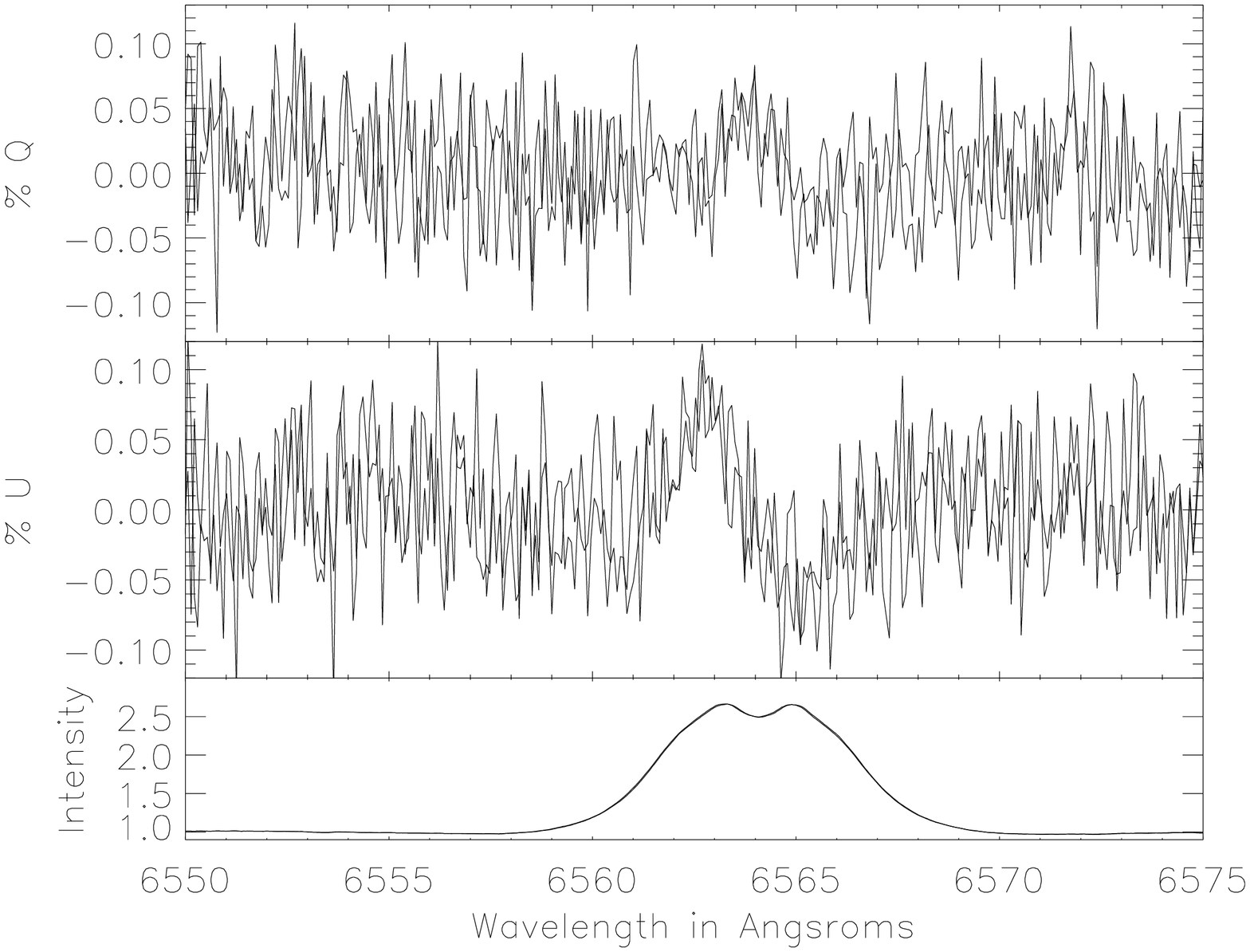}
\includegraphics[width=0.35\linewidth, angle=90]{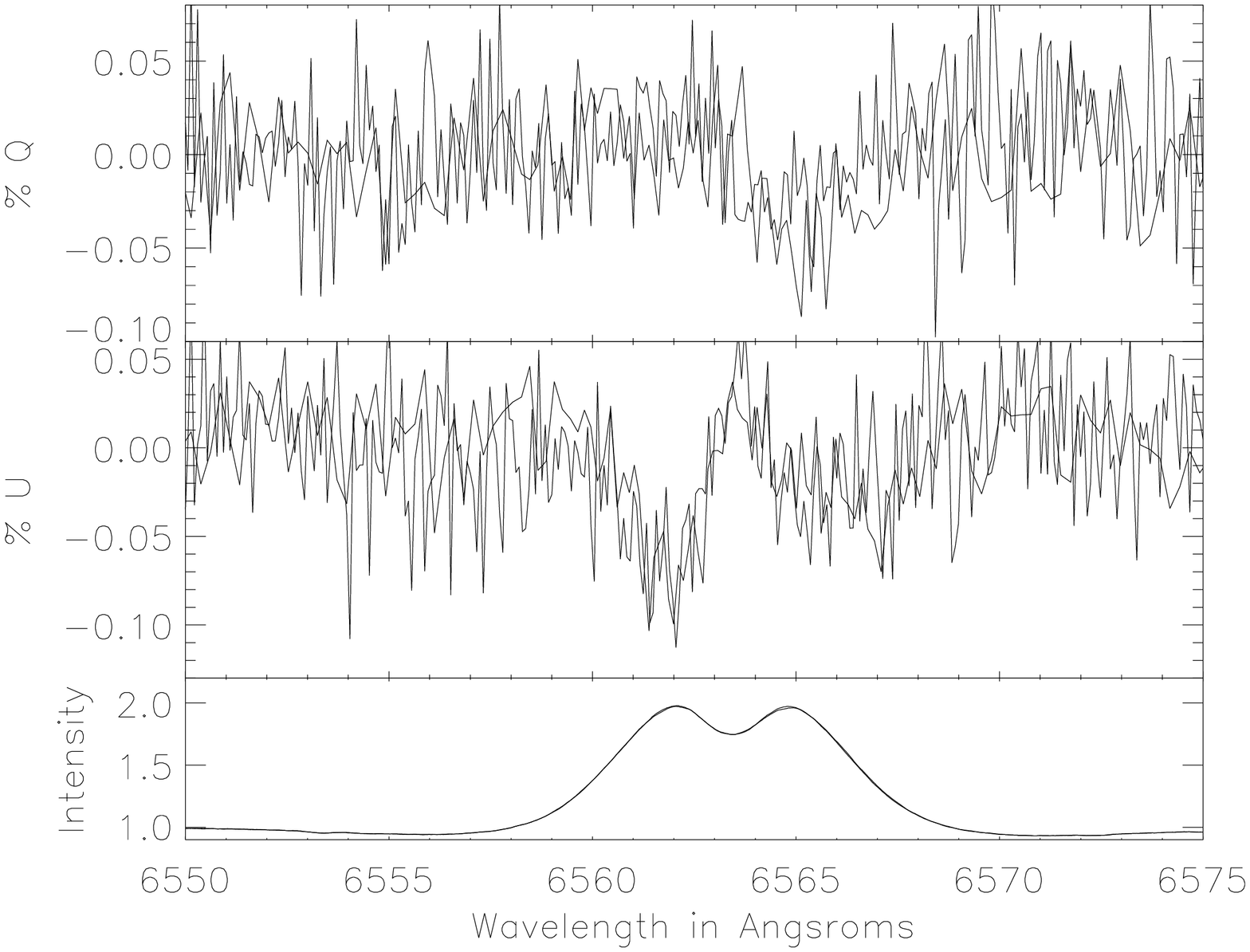} \\
\includegraphics[width=0.45\linewidth, angle=90]{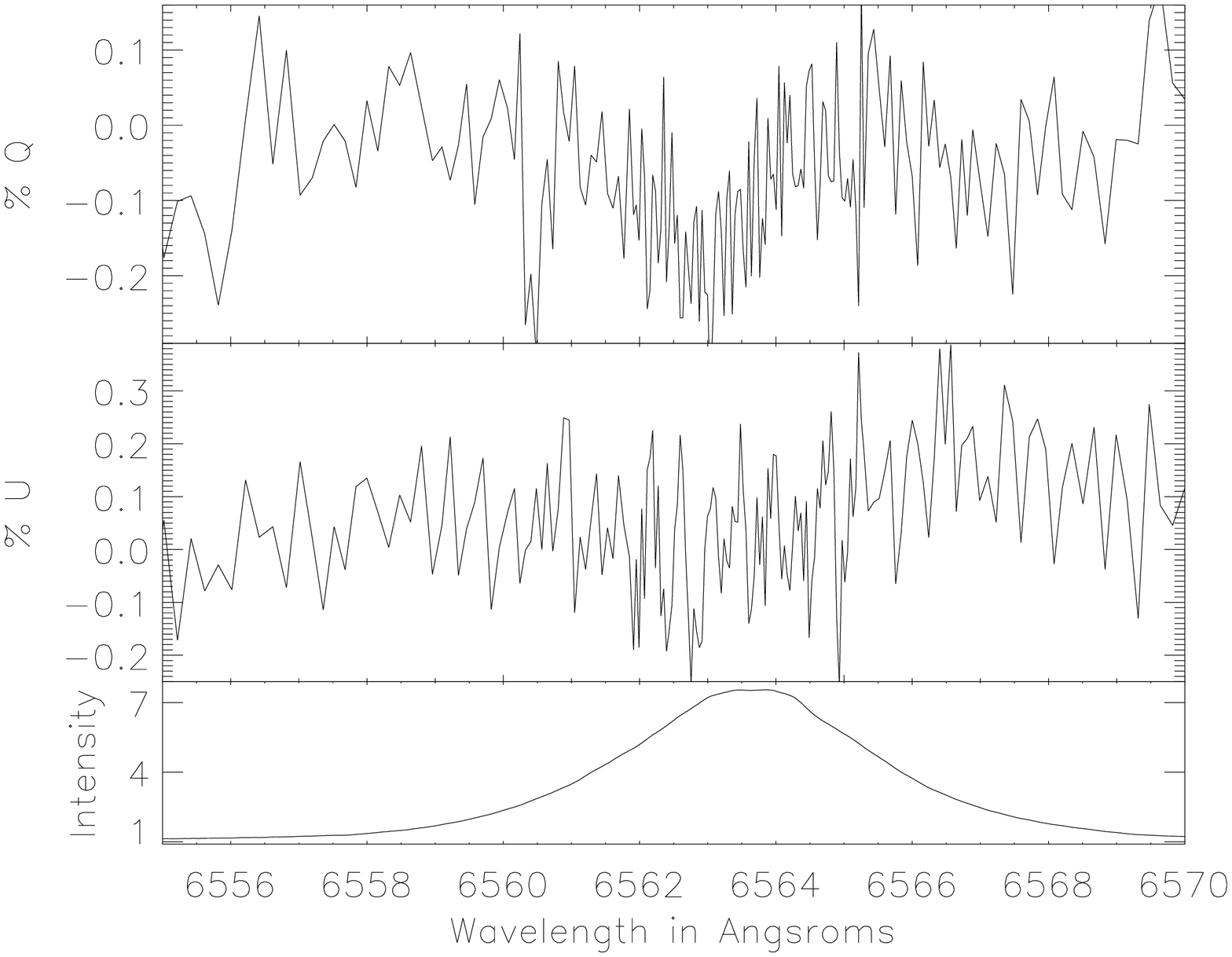}
\caption{More complex spectropolarimetric detections. The five smaller, more complex detections, from left to right, are: {\bf a)}18 Gem  {\bf b)} $\eta$ Tau {\bf c)} $\alpha$ Col {\bf d)} $\beta$ CMi {\bf e)} R Pup. The signature in 18 Gem is reminiscent of several Herbig Ae/Be stars. Both $\alpha$ Col and $\beta$ CMi show two detections of similar signal-to-noise taken on two separate nights at the same pointing. The detections match beautifully despite the differing conditions.}   
\label{fig:otherbe}
\end{center}
\end{figure*}

\subsection{Clear Detections}

	The other nine targets with broad detections show very similar ``depolarization" morphology. All of them show broad changes that extend to the wings of the line as seen in figure \ref{fig:bebroad}. However, a more detailed examination is necessary. The example of $\gamma$ Cas illustrates the shape of these broad signatures, but the overall magnitude and morphology varies between the targets. For example, MWC 143 has a very large 1.1\% deviation (0.9\% q, 0.6\% u) but Omi Pup is barely detectable at 0.1\%. However, the main point that must be emphasized is that there is a very clear polarimetric morphological difference between these stars and Herbig Ae/Be stars. These stars tend to show a polarization variation over the broad line profile, if there is any polarization signature, while the often heavily obscured Herbig Ae/Be systems show obvious  ``polarization-in-absorption." There is a very broad signature apparent in 10 of 30 stars with many showing some additional morphological features. Another 5 show more complex signatures worth describing in detail. This is probably not surprising since these systems have presumably weaker nearby extinction as they show only subtle absorptive polarization variations.   
	There are some morphological differences that are worth mentioning specifically. In 25 Ori, the change in Stokes q is ``flat-topped" across the line center. The star $\zeta$ Tau shows a noticeable asymmetry, being larger in the blue side of the line. There is a small, narrow asymmetric spectropolarimetric signature in the absorptive part of the $\psi$ Per ``disky" profile. On the blue side of the central absorption there is a relative decrease with respect to the broad signature and a relative increase on the red side of the absorption. These morphological effects show that there is more to be learned about circumstellar environment. Qualitatively we might expect that a combination of scattering and absorptive linear polarization might account for these results. 

	Other spectropolarimetric studies of some of these stars have been done as well. $\omega$ Ori was a non-detection in Oudmaijer \& Drew 1999 with a continuum polarization of 0.30\% and a non-detection in Vink et al. 2002 with a continuum of 0.27\%. It should be noted that these effects extend to the other hydrogen lines. For instance, Oudmaijer et al. 2005 report a broad 0.4\% change across Pa$_\beta$ at 1.28$\mu$m in $\zeta$ Tau. This is only moderately less than the signature observed in H$_\alpha$.

	There are interesting interferrometric constraints on some of these stars as well. $\zeta$ Tau is a very well-studied target. Quirrenbach et al. 1994 resolved the H$_\alpha$ emission region using interferrometry and created a maximum-entropy map of the circumstellar region. An elliptical Gaussian model fits the observations with an axial ratio of 0.30 and an angular FWHM of 3.55mas. The map shows a highly elongated structure that was interpreted as a near edge-on disk. In a later interferrometric and spectropolarimetric survey, Quirrenbach et al. 1997 report resolving H$_\alpha$ emission regions in $\gamma$ Cas, $\phi$ Per, $\eta$ Tau, $\zeta$ Tau, and $\beta$ CMi with angular sizes of 1.5 to 3.5mas. These measurements show that the H$_\alpha$ emission is quite extended and is much larger than the stellar-radii scale. All of the stars were point-sources in the nearby continuum interferrometric measurements. In light of these observations, the depolarization effect from extended and less-scattered H$_\alpha$ emission is extremely likely. However, the simple models used to date are not enough to fit the morphologies observed here completely.

	Other stars show smaller, more complex signatures. There were five stars with detected spectropolarimetric signatures of very small magnitude: 18 Gem, $\alpha$ Col, $\beta$ CMi, R Pup, and $\eta$ Tau. There were a number of 18 Gem and $\eta$ Tau observations that were non-detections, as seen in figure \ref{fig:otherbe}, but the signal-to-noise was not as good. The 18 Gem signature is a small antisymmetric change in Stokes u across the center of the line. The signature for $\eta$ Tau is a small drop in Stokes u across line-center, as seen in figure \ref{fig:otherbe}. In $\alpha$ Col and $\beta$ CMI, the detection was antisymmetric, spanned the line and was stable over two nights observations at the same telescope pointing. Both nights of data are shown in figure \ref{fig:otherbe}. R Pup showed a small drop in q across the emission blue side of the simple emission line. 
	
	All of these detections do not fit in to the ``broad" morphology of the typical Be depolarization signature. The detections in $\alpha$ Col and $\beta$ CMi are extremely small, having an amplitude of $<$0.1\% with antisymmetric components. However, they do not fit in to the scheme of the disk-scattering theory either. The signatures are not double-peaked and symmetric and are not significantly wider than the H$_\alpha$ line. All that can be conclude is that more detailed modeling and investigation is necessary.

\subsection{Detections in other systems}

	The ESPaDOnS archive has many stars with linear spectropolarimetric detections. In the context of this paper, comparing and contrasting spectropolarimetric signatures from different stellar systems, it is very useful to show some of the morphologies for different types of stars. The stars discussed here range from A to K spectral types. Most of the systems we present are consistent with polarization being affected by absorptive components of the line profiles.  
	
\begin{figure*}
\begin{center}
\includegraphics[width=0.32\linewidth, angle=90]{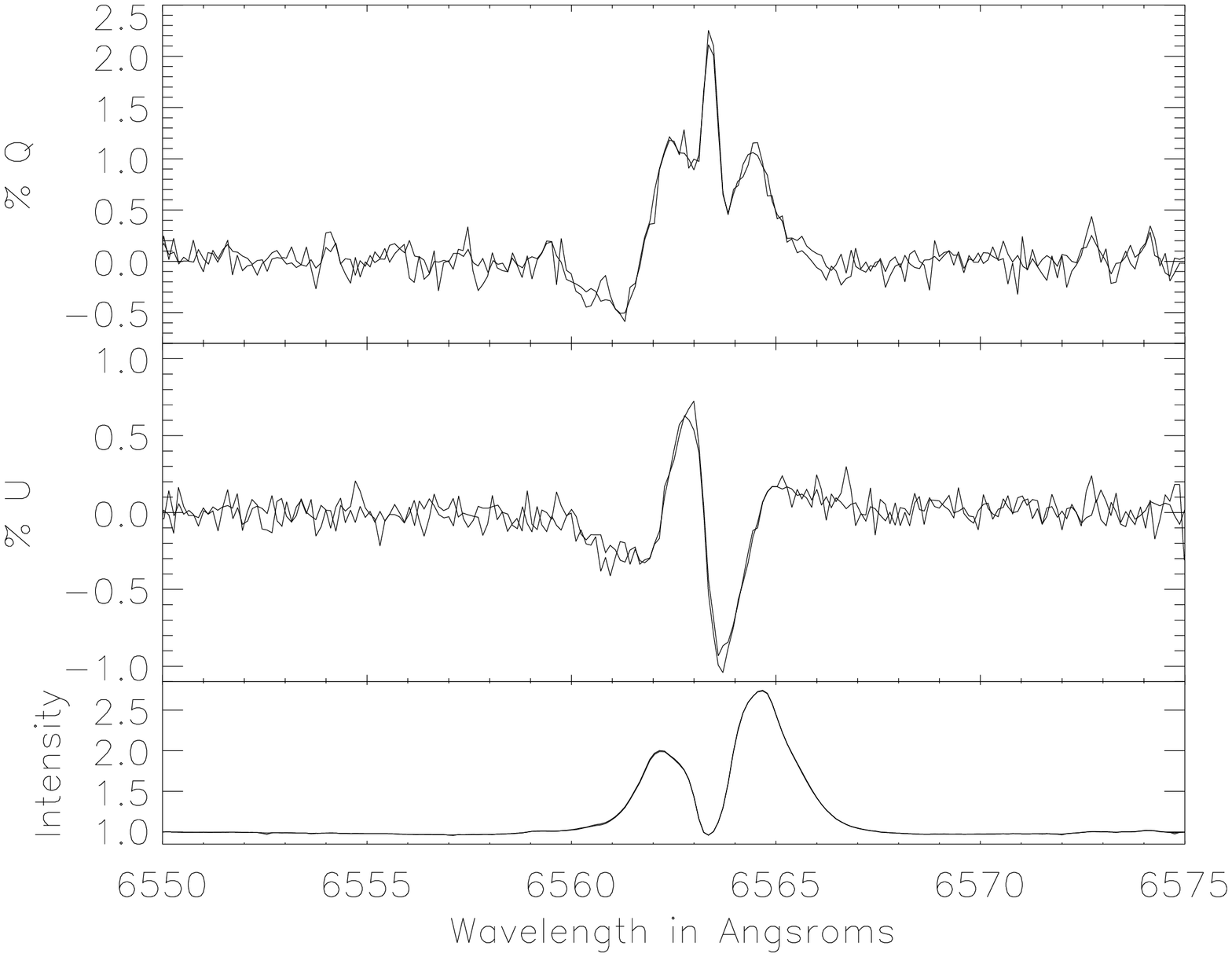}
\includegraphics[width=0.32\linewidth, angle=90]{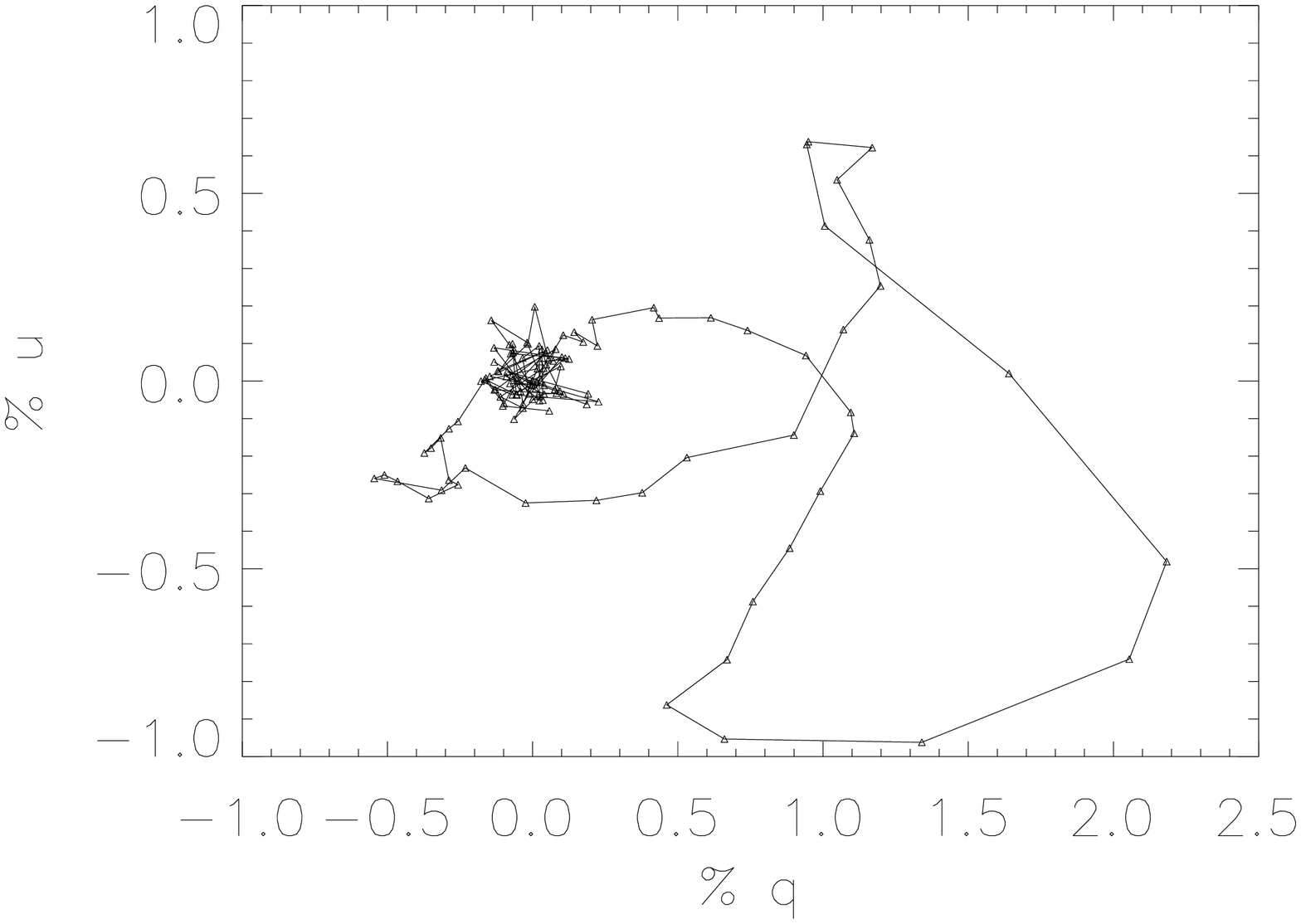} \\
\includegraphics[width=0.32\linewidth, angle=90]{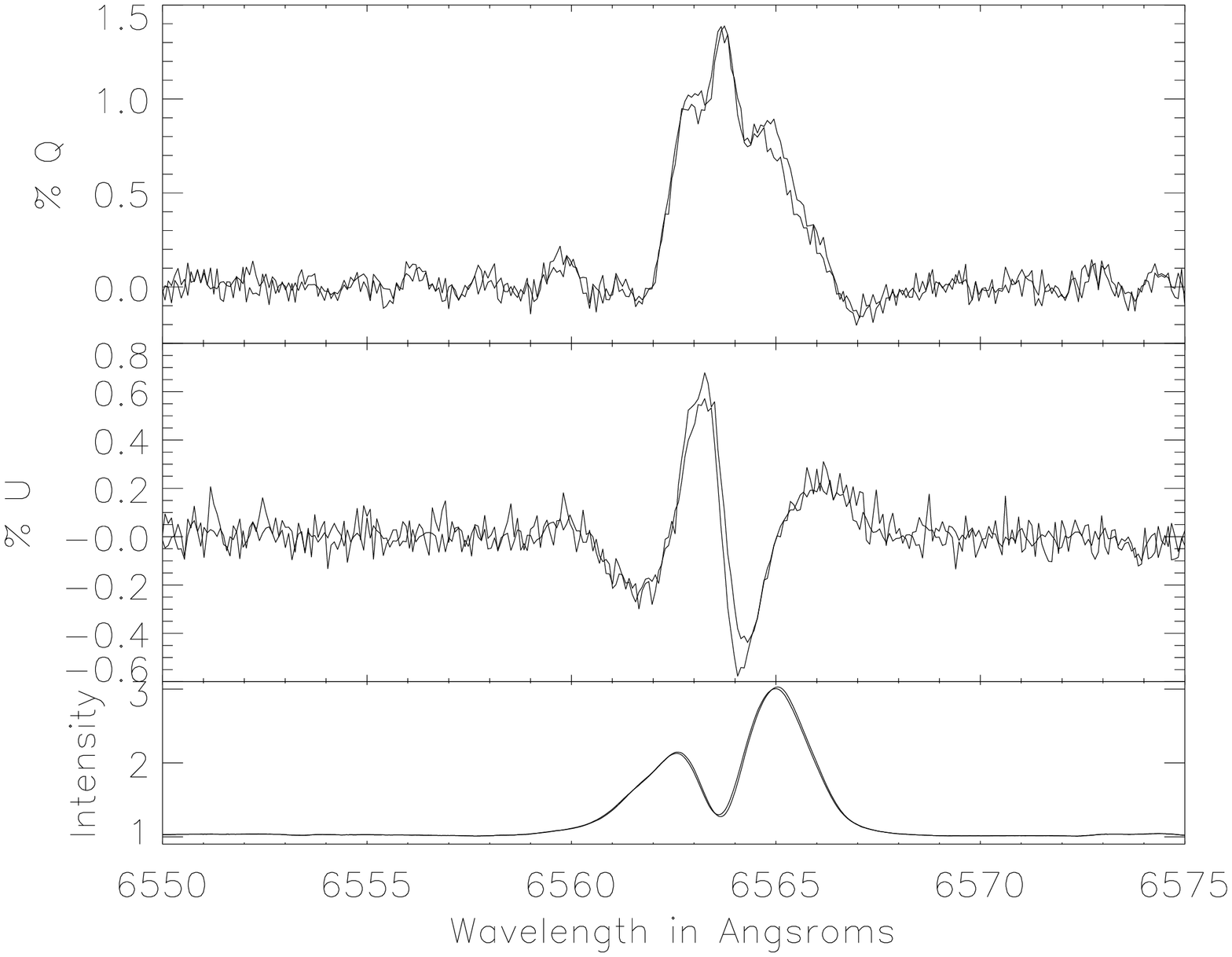}
\includegraphics[width=0.32\linewidth, angle=90]{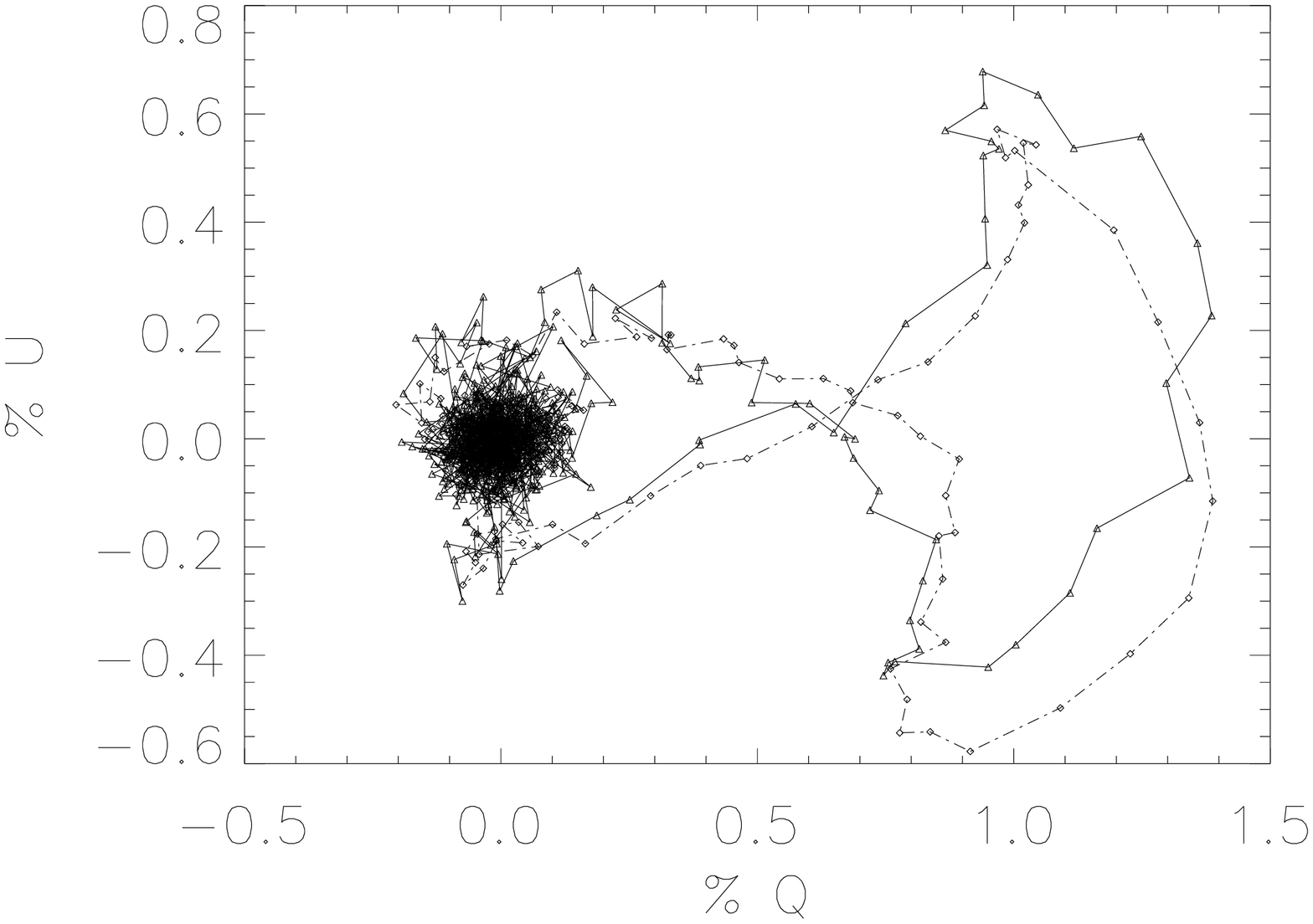}
\caption{The archive ESPaDOnS \& HiVIS spectropolarimetry for 3 Pup. From left to right: {\bf a)} The ESPaDOnS archive data for 3 Pup on February 7th \& 8th, 2006 and {\bf b)} the corresponding QU plot for the higher S/N observation. Both observations match nearly perfectly and the qu-loops show the complicated structure of this signature. {\bf c)} Shows the HiVIS observations from late 2007 and {\bf d)} the two corresponding qu-loops. The observations were taken with only a small difference in pointing, but this difference is enough to rotate the qu-loops by roughly 10$^\circ$. Both have a nearly identical double-looped figure-8 form, but the loops themselves are slightly rotated with respect to each other.}
\label{fig:swp-3pup-esp}
\end{center}
\end{figure*}

	3 Pup (MWC 570) is an emission line star of spectral type A3Iab (Simbad) with a ``disky" H$_\alpha$ line. This star provides yet another comparison between the different star types as it has an A spectral type because it has a distinctly different spectropolarimetric profilie. There is good quality archival ESPaDOnS data from February 7th and 8th 2006 showing a very large complex H$_\alpha$ signature. The polarization change, shown in figure \ref{fig:swp-3pup-esp} is centered on the ``disky" emission line but spans the entire width of the line. The polarization in the two emissive peaks is more than 0.5\% increasing to over 2\% in the central absorption. The qu-plot shows a very complex structure. There are multiple components of narrow-wavelength range in Stokes q with a strongly antisymmetric Stokes u. In qu-space this translates into multiple loops. In the corresponding HiVIS observations, shown in figure \ref{fig:swp-3pup-esp}, there is a similar overall morphology and strength, but with a slightly different exact shape. There has been no rotation applied to the HiVIS observations, but the alignment is close. The qu-loops show the differences much more clearly, regardless any rotation that could be applied to the data. The overall amplitude of the HiVIS observations is lower, and there is much more symmetry in the HiVIS loops. There double-loop structure is preserved, with a pattern somewhat resembling a figure-8 shape. In the ESPaDOnS archive observations, the small third loop near (1.5,0.5) is quite clear whereas a simple cluster of points is only seen in the HiVIS loops.  However, the loss of the loop could be entirely caused by the bin-by-flux routine that is automatically applied to the HiVIS observations, while it was not applied to the ESPaDOnS data. The third loop at a q of 1.5\% is at the wavelengths of the central absorption, where HiVIS binning is roughly 5 pixels (since the threshold was 5x continuum). 

	In the context of the scattering theories, this star is quite difficult to explain in any framework currently available. The polarization is very strong compared to most other observations. The form is also very complex, having strong deviations and asymmetries. The polarization is also strong across the entire width of the line: absorption, emission and wings. The overall intensity of the H$_\alpha$ line is quite low, being only 3 times continuum, and the central absorption is not extremely strong either, returning only to near continuum. This example provides yet another good example for the need for new, more detailed theories.
		
\begin{figure*}
\begin{center}
\includegraphics[width=0.35\linewidth, angle=90]{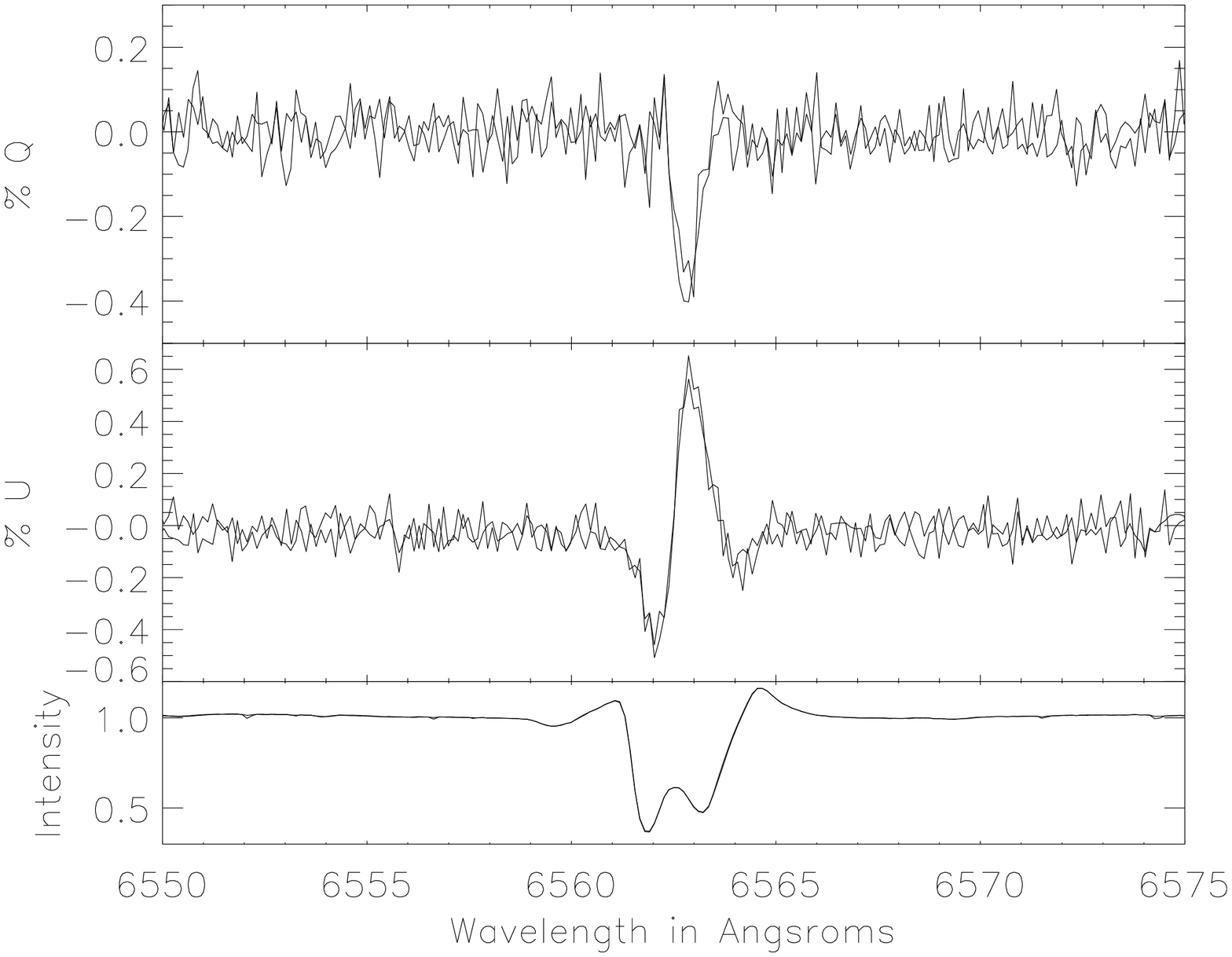}
\includegraphics[width=0.35\linewidth, angle=90]{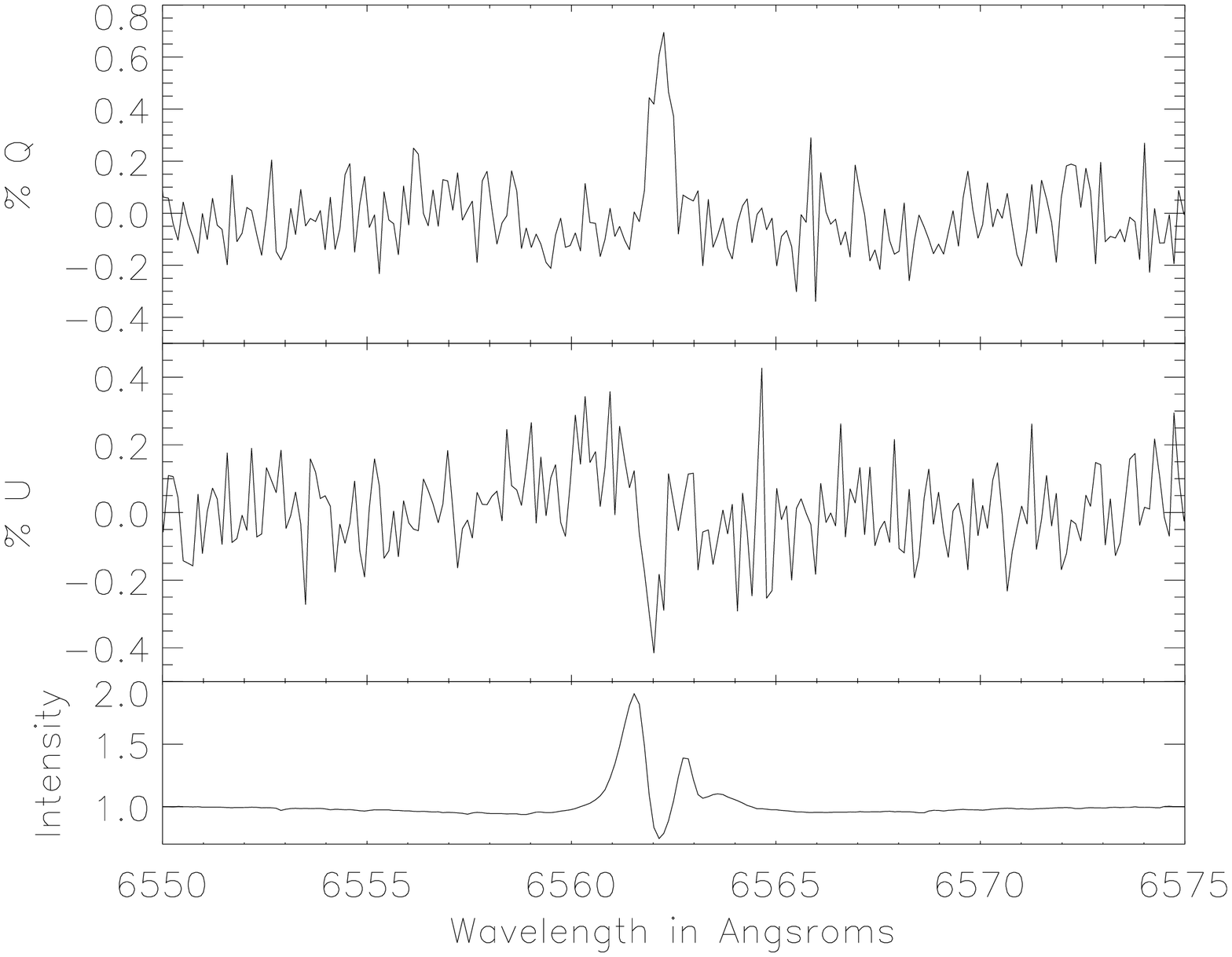} \\
\includegraphics[width=0.35\linewidth, angle=90]{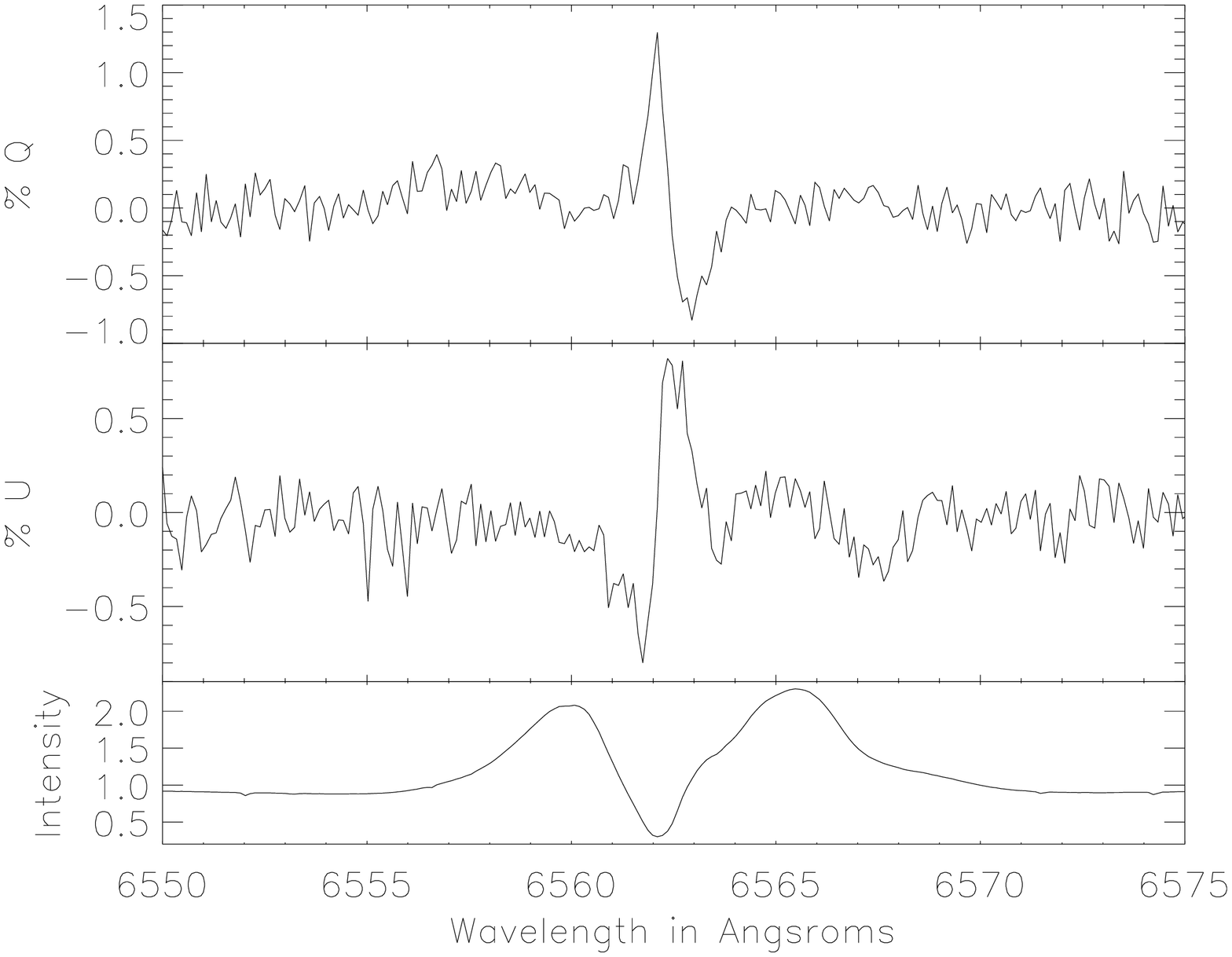}
\includegraphics[width=0.35\linewidth, angle=90]{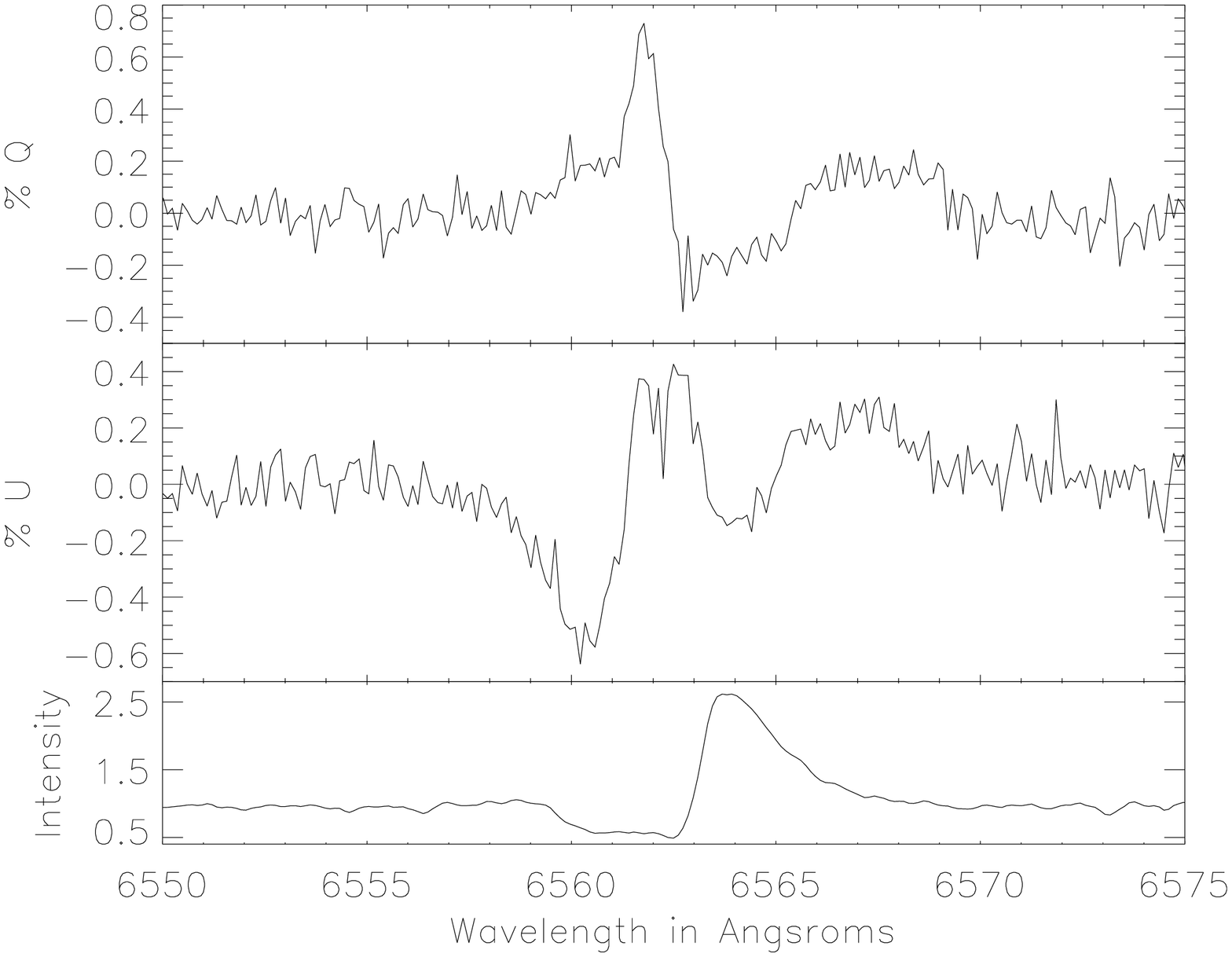} \\
\includegraphics[width=0.35\linewidth, angle=90]{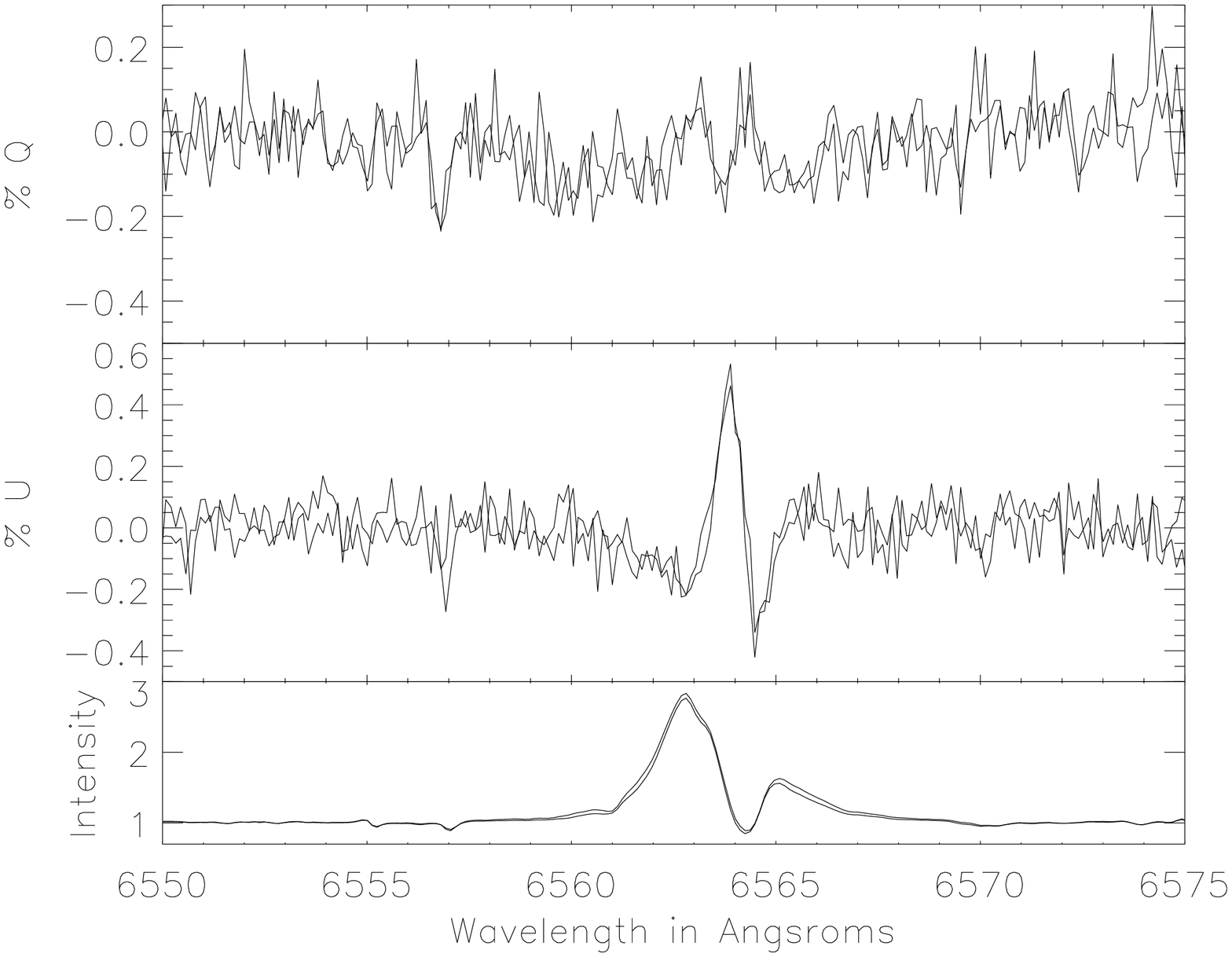}
\includegraphics[width=0.35\linewidth, angle=90]{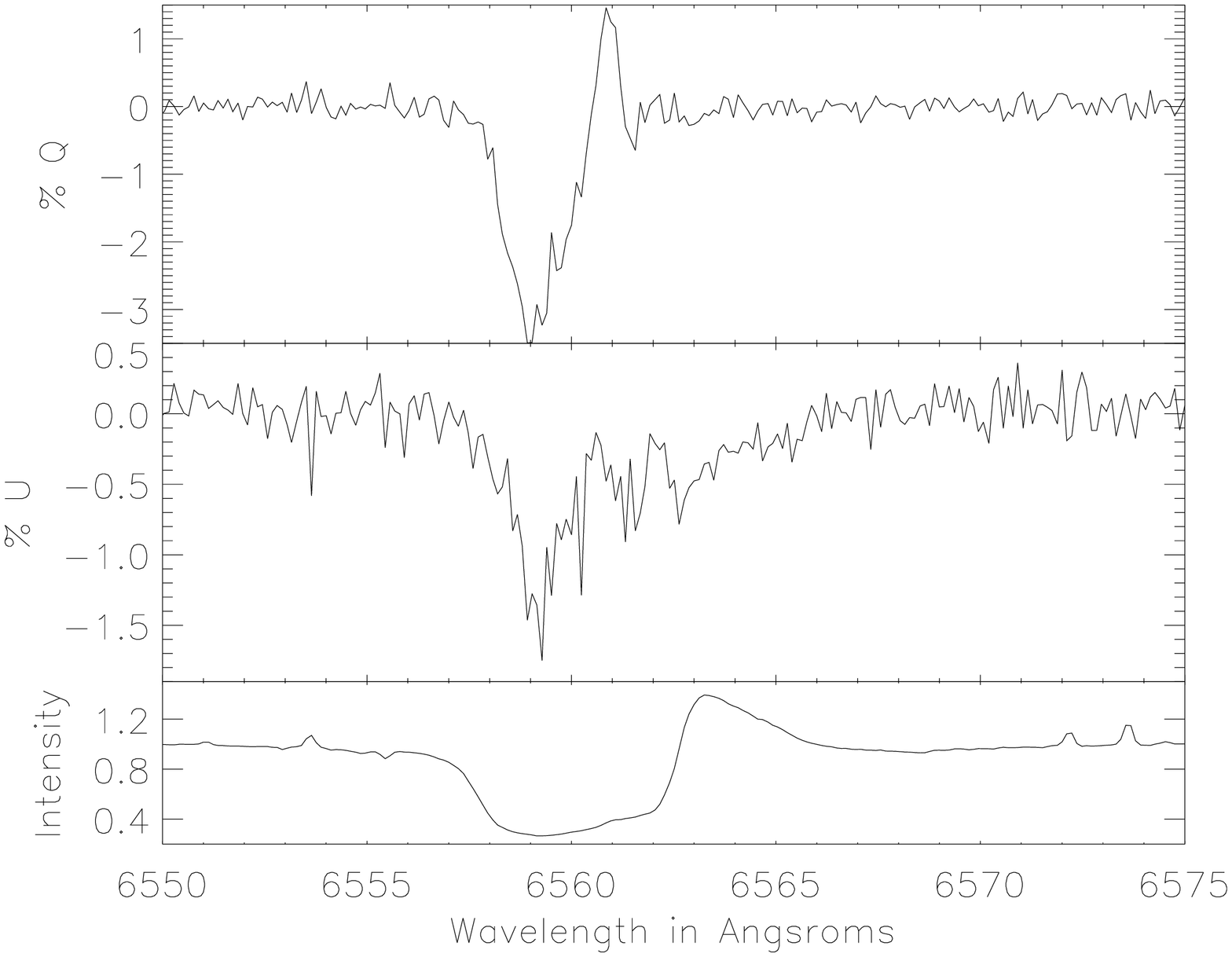}
\caption{Archive ESPaDOnS spectropolarimetry for other systems. From left to right: {\bf a)} This shows $\epsilon$ Aurigae on February 7th and 8th 2006 and {\bf b)} Shows AC Her on February 7th 2006. {\bf c)} Shows HD144668 on August 14th 2006. {\bf d)} Shows SS Lep on February 9th 2006. {\bf e)} Shows U Mon on February 7th and 8th 2006. {\bf f)} Shows 89 Her on February 7th 2006.}
\label{fig:swp-astar-esp}
\end{center}
\end{figure*}

	The archival spectropolarimetry for several additional targets is worth investigating for evidence of polarization-in-absorption. Figure \ref{fig:swp-astar-esp} shows six systems with clear spectropolarimetric detections. 

	Epsilon Aurigae (HR1605, HD31964) is an Algol-type eclipsing binary of spectral type A8Iab (Simbad). The H$_\alpha$ line for this star has a complex shape with evidence for emission and overlying absorption. There is good quality archival ESPaDOnS data from February 7th and 8th 2006 also showing a strong and complex H$_\alpha$ signature. The polarization change is mostly symmetric about line-center and spans the entire width of the absorptive component of the line. The polarization has an amplitude of roughly 1\% almost exactly at line-center. 
		
	AC Her (HD 170756) is an RV-Tau type variable star with a spectral type F4Ibpv (Simbad). There is archival data from February 7th 2006 shown in figure \ref{fig:swp-astar-esp}. The H$_\alpha$ line has a peak intensity of roughly twice continuum and is double peaked with absorptive effects clearly overlying emission. The spectropolarimetric effect is roughly 0.8\% in magnitude (+0.7\% q, -0.3\% u) and is concentrated almost entirely in the central absorptive component. This is very similar to some HAeBe stars such as HD 58647 that show a simple morphology in absorption.
	
	V856 Sco (HD 144668, HR 5999) is a Delta-Scut type variable star of spectral type A7IVe (Simbad). There is archival data from August 14th 2006 shown in figure \ref{fig:swp-astar-esp}. The H$_\alpha$ line is fairly wide with an amplitude of roughly two times the continuum. There is a clear absorptive component that creates an asymmetric, double peaked profile with absorptive effects clearly overlying emission. The spectropolarimetric effect is quite large in amplitude, being over 1\%. This star is reminiscent of the Herbig star MWC 158 with the strong anti-symmetric signatures being almost entirely confined to the central absorptive component. 

	SS Lep (17 Lep, HD 41511, MWC 519, HR 2148) is listed as an emission-line star of spectral type Apsh (Simbad). There is archival data from February 9th 2006 shown in figure \ref{fig:swp-astar-esp}. The H$_\alpha$ line has a P-Cygni type profile with an amplitude of 2.5 times the continuum. There is a strong, wide blue-shifted absorptive component. The spectropolarimetric effect is large in amplitude, being 0.5\%. The effect is definitely strongest in absorption, but does span the entire width of the line and extends quite far to the red side of the line. 
	
	U Mon (HD 59693) is an RV-Tau type variable star with a K0Ibpv spectral type (Simbad). There is archival data from February 7th and 8th 2006 shown in figure \ref{fig:swp-astar-esp}. The H$_\alpha$ line is fairly wide with an amplitude of roughly three times the continuum. There is a clear, mildly red-shifted absorptive component that creates an asymmetric, double peaked profile. The spectropolarimetric effect is clearly complex, being roughly 0.5\% in Stokes u at peak amplitude. This star is also has a relatively narrow effect around the absorptive component.

	89 Her (HD 163506, HR 6685) is a post-AGB star of spectral type F2Ibe (Simbad). There is archival data from February 7th 2006 shown in figure \ref{fig:swp-astar-esp}. The H$_\alpha$ line has a very wide, strong absorptive component and a small red-shifted emissive component. The spectropolarimetric effect is complex, being antisymmetric in q and monotonic in u with a very large amplitude of over 3\%(-3\% q, -1.5\% u). This polarization effect is clearly strongest in the absorptive component, though the Stokes u spectrum does show a gradual decrease toward zero extending to the red-shifted component of the line.

	All of these detections in various stellar types show that there are a wealth of spectropolarimetric morphologies and that the polarization in the absorptive components must be understood in order to make meaningful constraints on the circumstellar environment.

\subsection{Be/Emission-Line Star Comparison Summary}

	The Be and emission line spectropolarimetry shows remarkably different spectropolarimetric profiles when compared to the Herbig Ae/Be stars even with the same shape and magnitude H$_\alpha$ line profile. The Be signatures are smaller in magnitude and typically span the entire line (10/30). The morphology of the Be spectropolarimetry does not seem to be only tied to the strong absorptive components of the line directly because the polarization effect is present across the entire line. There is some morphological change tied to the presence or strength of the absorption in four of the 10 broad spectropolarimetric profiles, but this has a smaller amplitude than detected in the Herbig Ae/Be systems. These small amplitude effects would not be observed at lower spectral resolution and deserve more detailed analysis. Systems with both strong and weak absorption in both strong and weak H$_\alpha$ lines showed the broad spectropolarimetric effects. However, Be stars with either symmetric unobscured emission lines or absorption lines did not show any detectable polarization changes. Herbig Ae/Be stars in contrast show a polarization that is much stronger and is directly tied to the absorptive component of the line. In windy systems with P-Cygni type line profiles the polarization can be over 2\% in the blue-shifted absorption trough. In many cases, especially AB Aurigae and MWC 480, the polarization at the emissive peak is identical to the continuum polarization within the 0.1\% noise. In both Be and HAe/Be stars, there are a wealth of spectropolarimetric morphologies even for the same line type. In other stellar types, such as post-AGB and RV-Tau type stars, there is very significant absorptive polarization with a complex morphology.
	
\begin{figure*}
\begin{center}
\includegraphics[width=0.35\linewidth, angle=90]{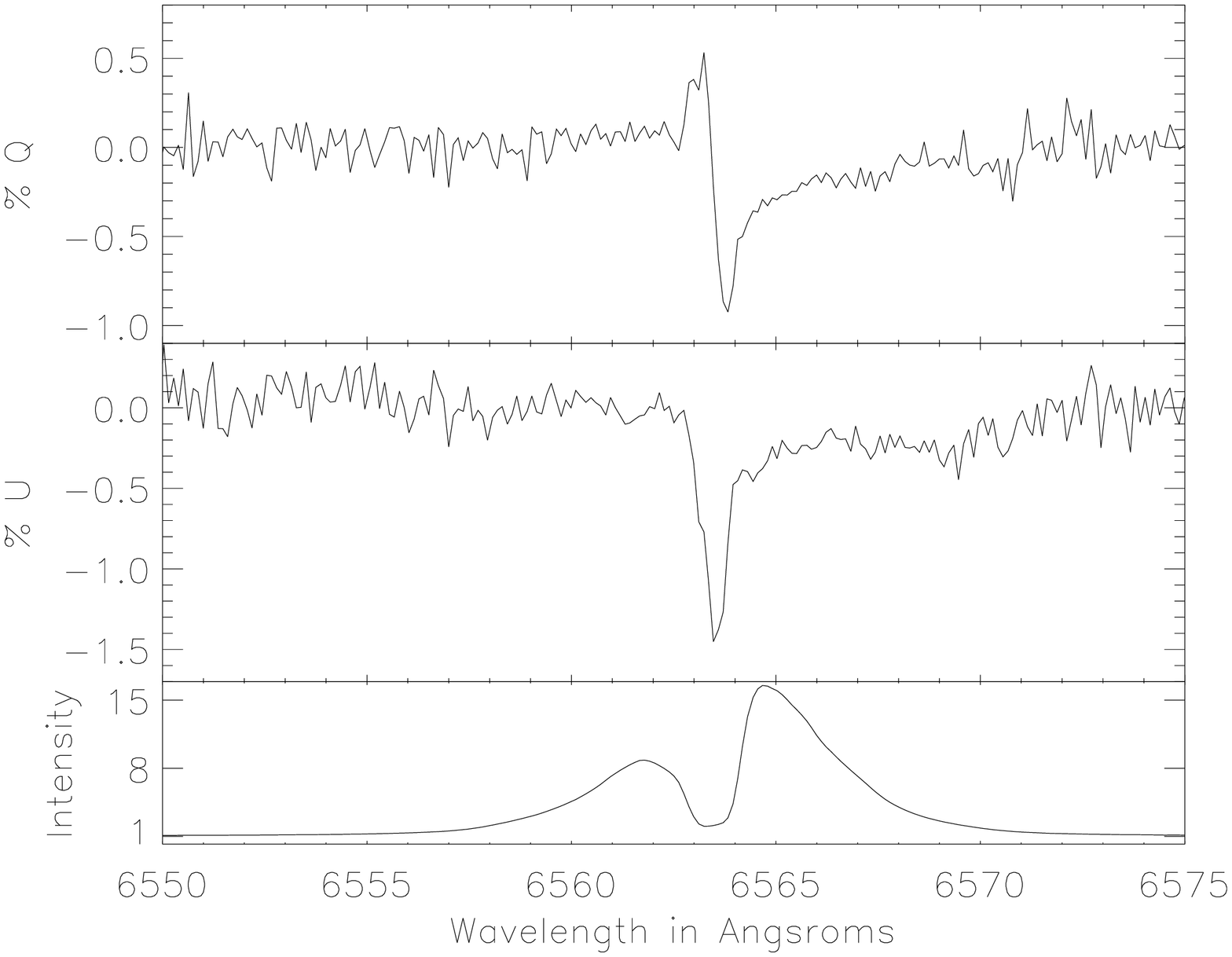}
\includegraphics[width=0.35\linewidth, angle=90]{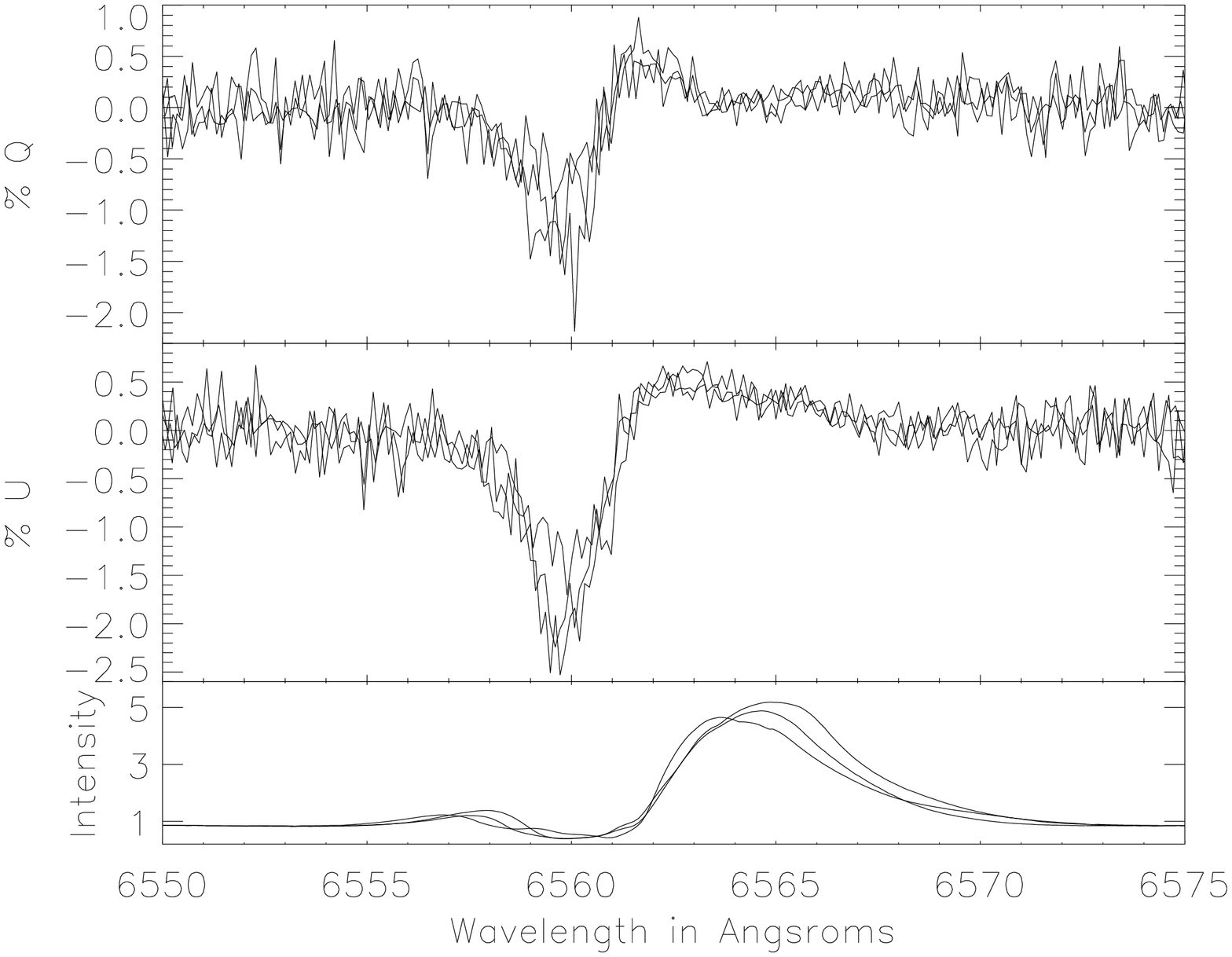} \\
\includegraphics[width=0.35\linewidth, angle=90]{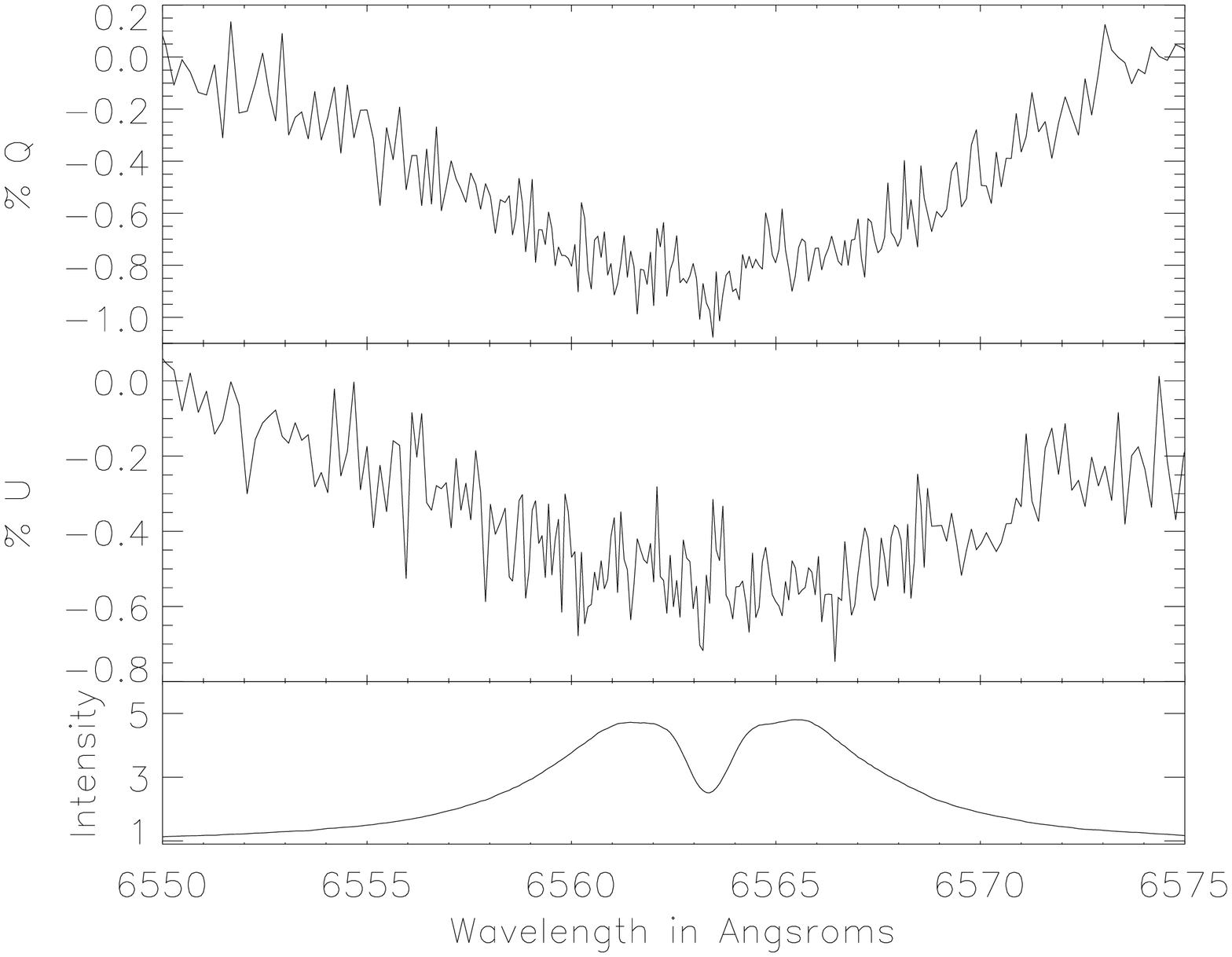}
\includegraphics[width=0.35\linewidth, angle=90]{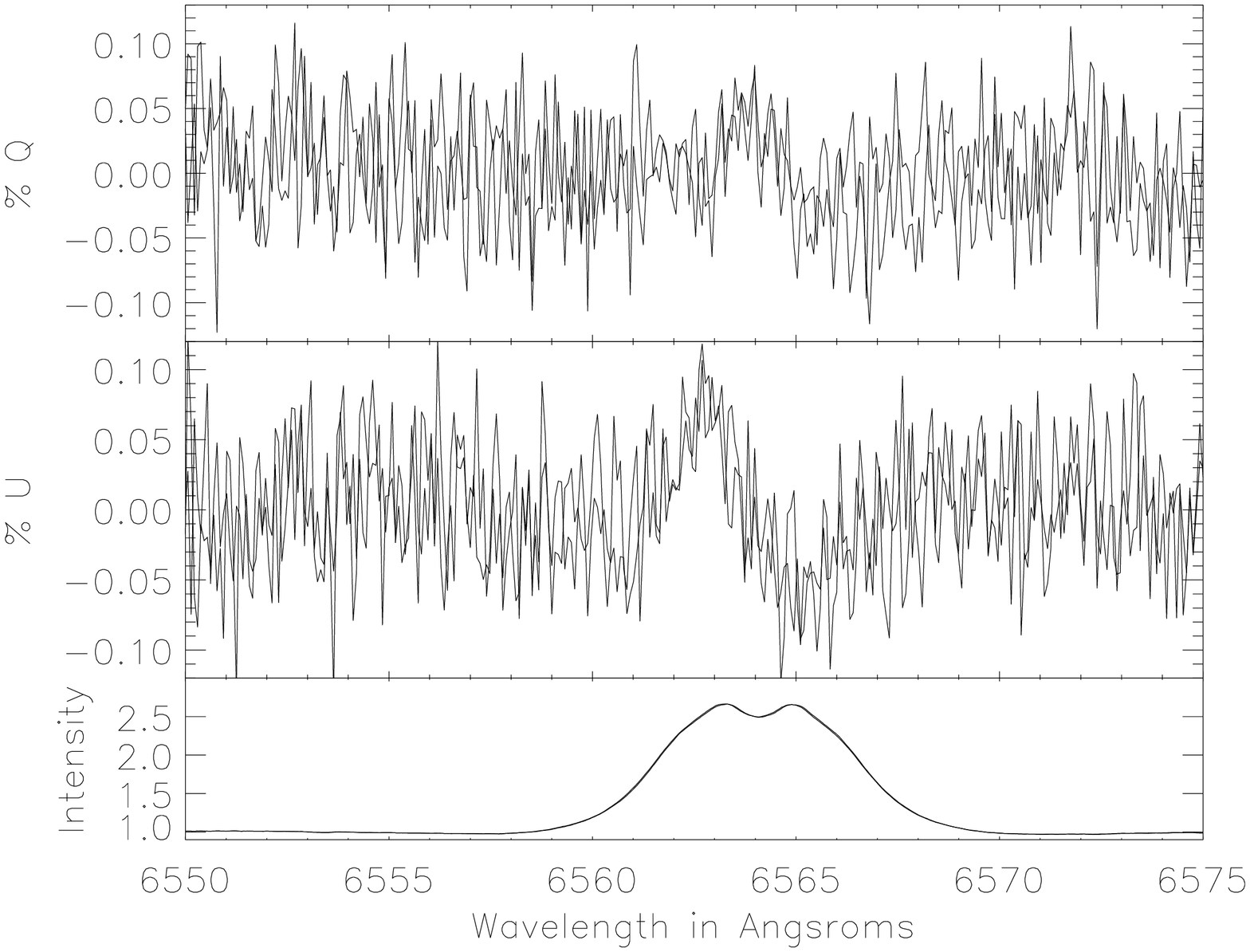} \\
\includegraphics[width=0.35\linewidth, angle=90]{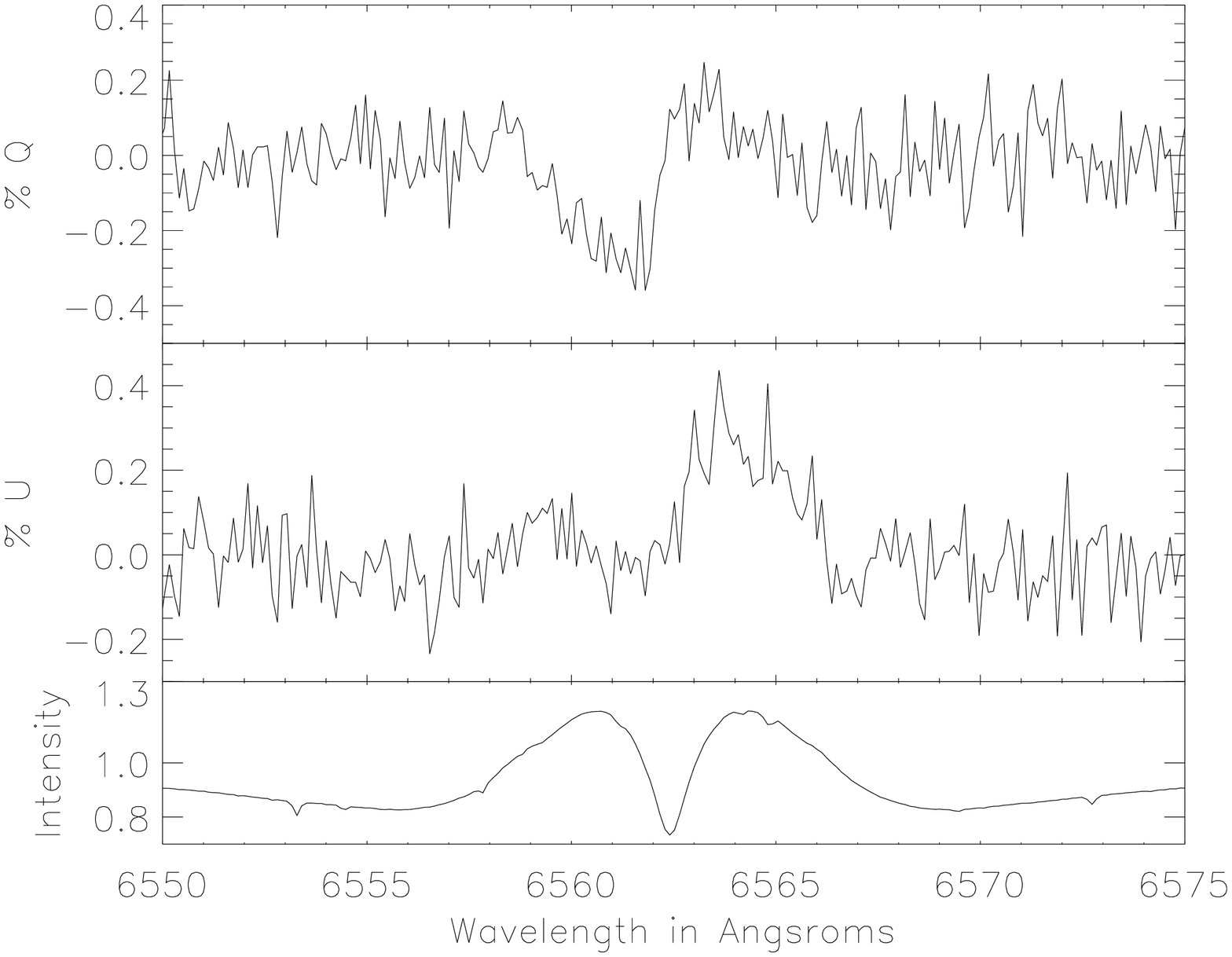}
\includegraphics[width=0.35\linewidth, angle=90]{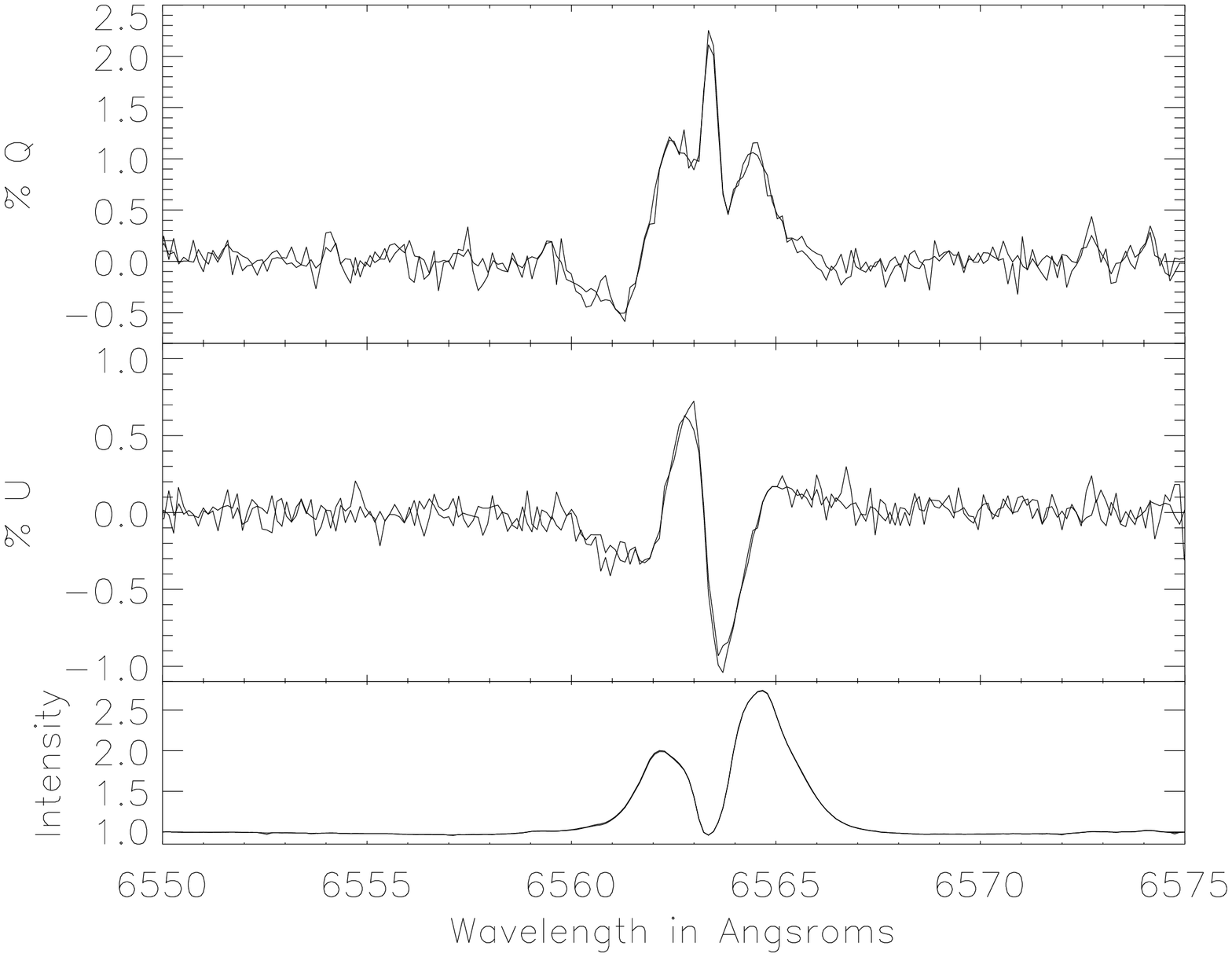}
\caption{A side-by-side comparison of the spectropolarimetric morphology of individual stars. From left to right: {\bf a)} The ESPaDOnS archive data for MWC 158 on February 9th, 2006. This is a ``disky" HAeBe star with a strong spectropolarimetric effect in the central absorption. {\bf b)} The ESPaDOnS archive data for MWC 480 on February 7th, 8th, and August 13th 2006 This is a ``windy" HAeBe star with a strong spectropolarimetric effect in the blue-shifted absorption. {\bf c)} The HiVIS data for MWC 143 - a Be type star showing a broad spectrpolarimetric signature that shows no effect across the central absorption. {\bf d)} The HiVIS data for $\alpha$ Col - a Be star with a smaller, more complex spectropolarimetric signature. {\bf e)} The ESPaDOnS archive data for 51 Oph on March 20th 2008. This is a HAe star with a more complex spectropolarimetric effect that is not only in the central absorption. {\bf f)} The ESPaDOnS archive data for 3 Pup on February 7th \& 8th, 2006. This is an A-type emission-line star with a very strong, complex spectropolarimetric signature. The morphology is completely different between these stars. The polarization-in-absorption of the HAe/Be stars is seen in a) and b). There are many Be/emission-line stars with broad spectropolarimetric signatures like c) but a significant number with smaller more complex detections similar to d). Within each class there are exceptions to these general trends, as in e) 51 Oph. These morphologies are also worth exploring in other types of stars, as was shown by 3 Pup in f).}
\label{fig:swp-sidebyside}
\end{center}
\end{figure*}

	A direct comparison between six different systems is shown in figure \ref{fig:swp-sidebyside}. Though all stars have H$_\alpha$ lines with absorption features, very different spectropolarimetric morphologies are seen. MWC 158 and MWC 480 illustrate the typical Herbig Ae/Be effects. The polarization changes are very strong in the absorptive components of the line and the qu-loop morphologies are dominated by the absorptive regions. MWC 143 is a Be star that shows a broad clear spectropolarimetric signature that is wider than the emission line itself and shows no significant deviation from this broad spectropolarimetric trend in the central absorption. This is a ``disky'' system that shows no effect in absorption, a clear change from the typical Herbig Ae/Be systems. The Be star $\alpha$ Col illustrates the 5/30 stars that show complex, more narrow effects that have not been seen in Be stars at lower spectral resolution or lower precision. 51 Oph is classified as a Herbig Ae/Be star but shows antisymmetric, lower-amplitude components similar to Be stars. There is not a pronounced effect in the absorptive component of the line. This is unlike most of the Herbig Be stars such as HD 58647 which show a strong spectropolarimetric signature only in the very center of the central absorption.	The star 3 Pup is an A-type star not in either Be or HAeBe classes and shows a very strong, complex signature that is stable during the epochs of our observations. The polarization signature is as broad as the emission line and has many complex components. 
	
	Comparison of these three stars highlights the range of effects seen in this survey. The detections show that, although the broad spectropolarimetric effect seems common (10/30) in Be and other B-type emission-line stars, the effect is certainly not common in the Herbig Ae/Be stars. One common thread in those Herbig Ae/Be stars was the consistent presence of spectropolarimetric signatures in and around absorptive components of the H$_\alpha$ line. In many of the stars, there were very large polarization changes in absorptive components while the wavelengths of strongest emission showed no polarization deviations from continuum. The Be and emission-line stars do have their own morphological complexities that are worth investigating, and 5 of 30 detections that do not follow the broad morphological description and 4 of 10 ``broad'' signatures show additional absorptive effects. These more complex detections, in some cases, look very similar to some Herbig Ae/Be detections. Now that a very clear morphological difference has been established between Herbig Ae/Be stars and other emission-line stars, a new theory will be explored that may better fit the ``polarization-in-absorption" morphology of the Herbig Ae/Be stars.

\section{Discussion}

	This paper presents a large amount of HiVIS observations taken on over 100 nights on 29 Herbig Ae/Be stars, 30 Be and emission-line targets as well as several alternatively classified objects to compile a linear spectropolarimetric survey far larger in scope than any survey performed to date.  Archival data as well as our own observations with the ESPaDOnS spectropolarimeter were used to supplement the HiVIS survey and to prove the robustness of the combined dataset. The ESPaDOnS observations represent an order-of-magnitude gain in resolution from any survey done to date. The observing campaign from 2004 to 2008 was outlined in section 2. The spectroscopic observations of the Herbig Ae/Be stars showed, as expected, a wealth of line morphologies. The H$_\alpha$ line of these stars is typically strong and quite variable. The emission lines can be a few to thirty times continuum with many different absorptive components on top of the emission. Spectroscopic measurements for these stars over many time-scales were presented and discussed. A few short-term variability studies were reported illustrating the dynamic circumstellar environment in these stars. 
	
	Current theories of spectropolarimetric line profiles, based on disk or envelope scattering effects were discussed in section 3. The current disk-scattering theory predicts symmetric, broad, double-peaked spectropolarimetric signatures in the orbiting thin-disk case. The corresponding qu-loops show position-angle changes that are equally broad in wavelength. In the case of outward radial motion, the disk-scattering theory predicts a shift of the spectropolarimetric profiles toward the red side of the spectral line. The depolarization theory is essentially a dilution of continuum polarization by unpolarized emission. This effect produces broad signatures that are roughly inversely proportional to the amount of emission. 

	In section 4, the spectropolarimetric survey of the Herbig Ae/Be stars was presented. The survey showed that roughly two thirds of the HAeBe stars showed polarization effects in the absorptive components of the H$_\alpha$ line. Both the windy-type and disky-type stars showed this polarization-in-absorption effect in the blue-shifted or line-center absorptive components respectively. The magnitude of the effect was typically 0.1\% to 2\% with several stars showing complex morphologies and complex geometries in qu-space. Only one star, HD 179218, showed a spectropolarimetric effect across the entire H$_\alpha$ line. These observations are difficult to explain using current scattering theories. The depolarization effect causes a broad, monotonic change in polarization that is roughly inversely proportional to emission and is a linear extension in qu-space. The disk-scattering effect shows broad, symmetric morphology for orbiting material and a shift toward red wavelengths for out-flowing material. Both of these effects do not fit the observed morphology of polarization-in-absorption on the blue-shifted side of the windy-stars or the polarization-in-central-absorption-only for the disky-stars. The amplitude of the observed effects is also a consideration. Scattering theory has no natural amplitude, but most of the detections fall in the 0.3\% to 1.5\% range. 
	
	A  survey of 30 Be and emission-line stars was performed and presented in section 5. This was also complemented by studies of 6 other-type stars such as post-AGB and RV-Tau types, as well as an in-depth look at 3-Pup. These Be and emission-line stars had a very different type of spectropolarimetric morphology. In 10 of 30 stars there was a broad, monotonic spectropolarimetric signature present across the entire H$_\alpha$ line. This type of signature is typical of the depolarization effect. However, in these 10 broad detections, four showed more complex morphologies in addition to the simple depolarization tied to absorptive components of the emission line. In another 5 of 30 stars there were antisymmetric or other more complex spectropolarimetric signatures that are inconsistent with depolarization theory. This smaller survey of Be stars is in good agreement with the large collection of previous work done to date on these stars at lower resolution. The broad spectropolarimetric morphology is reproduced, but the higher resolution, higher signal-to-noise and larger sample of this survey shows that there are still some additional uncertainties about the exact form of the spectropolarimetric effects in some of these stars. One third of the detections (5/15) do not fit the simple depolarization framework. Four of 10 broad signatures show significant deviations of the line from the simple depolarization morphology in absorptive components that would only have been seen at high resolution with high signal-to-noise.
	
	The inability of scattering theories to fit the ``polarization-in-absorption" seen in the Herbig Ae/Be stars is good evidence for intrinsic absorptive polarization effects. This survey has already inspired optical pumping calculations (Kuhn et al. 2007) that demonstrate such a mechanism.

	Detailed models based on any paradigm assuming any realistic circumstellar environments do not yet predict spectropolarimetric line profiles. The best example of this is the ``polarization-in-absorption" seen primarily in the Herbig Ae/Be stars or the more complicated profiles observed in Be and other emission-line stars. Unusual linear spectropolarimetric profiles were seen in both HAeBe and Be classes as well as in other A-type stars such as 3 Pup and $\epsilon$ Aur. An intrinsic absorptive polarization mechanism seems essential  for explaining many of these observations as the Herbig Ae/Be detections are clearly inconsistent with disk-scattering theories. However, as this large survey shows, there is a broad range of linear spectropolarimetric ``signatures''  in these stellar systems waiting to be deciphered.

\acknowledgements

	This program was partially supported by the NSF AST-0123390 grant, the University of Hawaii and the AirForce Research Labs (AFRL). Some of this research used the facilities of the Canadian Astronomy Data Centre operated by the National Research Council of Canada with the support of the Canadian Space Agency. This archive was very helpful by providing the released ESPaDOnS data.

	This program also made use of observations obtained at the Canada-France-Hawaii Telescope (CFHT) which is operated by the National Research Council of Canada, the Institut National des Sciences de l'Univers of the Centre National de la Recherche Scientifique of France, and the University of Hawaii. These CFHT observations were reduced with the dedicated software package Libre-Esprit made available by J. -F. Donati. The Simbad data base operated by CDS, Strasbourg, France was very useful for compiling stellar properties.


\begin{thebibliography}{}
\bibitem{ash99}    Ashok N. M., et. al., 1999, IAU Circ. 7130
\bibitem{bai06}      Baines D. et al., 2006, MNRAS, 367, 737
\bibitem{bes94}    Beskrovnaya N.G. et al., 1994, A\&A, 287, 564
\bibitem{bes95}    Beskrovnaya N.G. et al., 1995, A\&A, 298, 585
\bibitem{bes98}    Beskrovnaya N.G. et al., 1998, A\&AS, 127, 243
\bibitem{bes99}   Beskrovnaya N.G. et al., 1999, A\&A, 343, 163
\bibitem{bes04}     Beskrovnaya N.G. \& Pogodin M.A., 2004, A\&A, 414, 955
\bibitem{bjo98}       Bjorkman K.S. et al., 1998, ApJ, 509, 904
\bibitem{boh93}    Bohm T. \& Catala C., 1993, A\&AS, 101, 629
\bibitem{boh95}    Bohm T. \& Catala C., 1995, A\&A, 301, 155
\bibitem{boh96}    Bohm T. et al., 1996, A\&A Sup., 120, 431
\bibitem{bou97}    Bouret J.-C., Catala C., \& Simon T., 1997, A\&A, 328, 606
\bibitem{bro77}     Brown J.C. \& McLean I.S., 1977, A\&A, 57, 141
\bibitem{bro78}     Brown J.C. et al., 1978, A\&A, 68, 415
\bibitem{bro89}     Brown J.C. et al., 1989, A\&A, 68, 415
\bibitem{car07}     Carmona A. et al., 2007, ApJ, 344, 341
\bibitem{cas87}      Cassinelli J.P. et al., 1987, ApJ, 317, 290
\bibitem{cat87}       Catala C. \& Kunasz P.B., 1987, A\&A, 174, 158
\bibitem{cat89}     Catala C. et al. 1989, A\&A, 221, 273
\bibitem{cat99}       Catala C. et al., 1999, A\&A, 345, 884
\bibitem{cat07}       Catala C. et al., 2007, A\&A, 462, 293
\bibitem{che06}    Chavero C. et al., 2006, A\&A, 452, 921
\bibitem{dev00}    Devine D. et al., 2000, ApJ, 542, L115
\bibitem{dom03}   Domiciano de Souza A. et al., 2003, A\&A, 407, L47
\bibitem{don99}   Donati J.F. et al., 1999, A\&AS, 134, 149
\bibitem{eve98}      Eversberg T. et al., 1998, PASP, 110, 1356
\bibitem{fer95}       Fern\'{a}ndez M. et al., 1995, A\&AS, 114, 439
\bibitem{fu07}       Fu H.-H et al., 1997, AJ, 114, 1623
\bibitem{fuk04}      Fukagawa M. et al., 2004, ApJ, 605, L53
\bibitem{gar06}     Garcia Lopez R. et al., 2006, A\&A, 459, 837
\bibitem{gra93}     Grady C.A. et al., 1993, ApJ, 415, L39
\bibitem{gra00}     Grady C.A. et al., 2000, ApJ, 544, 895
\bibitem{gra05}     Grady C.A. et al., 2005, ApJ, 630, 958
\bibitem{hap72}    Happer, W. 1972, Rev. Mod. Phys. 44, 169
\bibitem{har00}     Harries T.J., 2000, MNRAS, 315, 722
\bibitem{har06}      Harrington D.M. et al., 2006, PASP, 118, 845
\bibitem{har07_01}    Harrington D.M. et al., 2007, Icarus, 187, 177
\bibitem{har07_02}     Harrington D.M. \& Kuhn J.R. 2007, ApJL, 667, L89
\bibitem{har08}     Harrington D.M. \& Kuhn J.R., 2008, PASP, 120, 89
\bibitem{hil93}     Hillenbrand L.A. et al., 1992, ApJ, 397, 613
\bibitem{hil90}       Hillier D.J., 1990, A\&A, 231, 116
\bibitem{hil91}       Hillier D.J., 1991, A\&A, 247, 455
\bibitem{hil94}       Hillier D.J., 1994, A\&A, 289, 492
\bibitem{hil96}       Hillier D.J., 1996, A\&A, 308, 521
\bibitem{hu6a}        Hubrig S. et al., 2006, A\&A, 446, 1089
\bibitem{hu6b}       Hubrig S. et al., 2006b, MNRAS, 371, 1953
\bibitem{ign04}      Ignace R. et al., 2004, ApJ, 609, 1018
\bibitem{ism05}    Ismailov N.Z. \& Aliyeva A.A., 2005, IBVS, 5634
\bibitem{jas98}      Jaschek C. \& Andrillat T., 1998, A\&AS, 128, 475
\bibitem{koz03}    Kozlova O.V. et al., 2003, Astrofizika, 46, 331
\bibitem{koz04}    Kozlova O.V., 2004, Astrofizika, 47, 287
\bibitem{koz06}   Kozlova O.V., 2006, Astrofizika, 49, 81
\bibitem{kuh07}    Kuhn J.R. et al., 2007, ApJL, 668, L63
\bibitem{liu07}        Liu W.M. et al., 2007, ApJ, 658, 1164
\bibitem{man97b}  Mannings V. et al., 1997, Nature, 388, 555
\bibitem{man06}   Manoj P. et al., 2006, ApJ, 653, 657 
\bibitem{mar00}    Marconi M. et al., 2000 A\&A, 355, L35
\bibitem{mcl78}     McLean I.S. \& Brown J.C., 1978, A\&A, 69, 291
\bibitem{mcl79_01}     McLean I.S. \& Clarke D., 1979, MNRAS, 186, 245
\bibitem{mcl79_02}    McLean I.S., 1979, MNRAS, 186, 265
\bibitem{mee06}   Meech K.J. et al., 2005, Science, 310, 265
\bibitem{mir98}      Miroshnichenko A.S. et al., 1998, PASP, 110, 883
\bibitem{mir99_01}    Miroshnichenko A.S. et al., 1999, MNRAS, 302, 612
\bibitem{mir99_02}    Miroshnichenko A.S. et al., 2004, A\&A, 427, 937
\bibitem{mot07}    Mottram J.C. et al., 2007, MNRAS, 377, 1363
\bibitem{oud99}    Oudmaijer R.D., \& Drew J.E., 1999, MNRAS, 305, 166
\bibitem{oud01}   Oudmaijer R.D., et al., 2001, A\&A, 379, 564
\bibitem{oud05}   Oudmaijer R.D. et al., 2005, MNRAS, 364, 725
\bibitem{pat06}    Patel M. et al., 2006, MNRAS, 373, 1641
\bibitem{pir97}        Pirzkal N. et al., 1997, ApJ, 481, 392
\bibitem{poe76}     Poeckert R. \&  Marlborough L.M., 1976, ApJ, 206, 182
\bibitem{poe77}     Poeckert R. \&  Marlborough L.M., 1977, ApJ, 218, 220
\bibitem{poe78}     Poeckert R. \&  Marlborough L.M., 1978, ApJ, 220, 940
\bibitem{poe79}     Poeckert R. \&  Marlborough L.M., 1979, ApJ, 233, 259
\bibitem{pog94}     Pogodin M.A., 1994, A\&A, 282, 141
\bibitem{pog97}     Pogodin M.A., 1997, A\&A, 317, 185
\bibitem{pog00}     Pogodin M.A. et al., 2000, A\&A, 359, 299
\bibitem{pog04}     Pogodin M.A. et al., 2004, A\&A, 417, 715
\bibitem{pog05}     Pogodin M.A. et al., 2005, A\&A, 438, 239
\bibitem{pon00}    Pontefract M. et al., 2000, MNRAS, 319, L19
\bibitem{por03}    Porter J.M. \& Rivinius R., 2003, PASP, 115, 1153
\bibitem{qui93}      Quirrenbach A. et al., 1993, ApJ, 416, L25
\bibitem{qui94}      Quirrenbach A. et al., 1994, A\&A, 283, L13
\bibitem{qui97}      Quirrenbach A. et al., 1997, ApJ, 479, 477
\bibitem{rod01}    Rodgers B.M., 2001, Univ. Washington PhD Thesis
\bibitem{sit81}      Sitko M.L., 1981, ApJ, 247, 1024
\bibitem{ski93}      Skinner S.L. et al., 1993, ApJS 87, 217
\bibitem{the94}      Th\'e P.S. et al., 1994, A\&AS, 104, 315
\bibitem{van98}    van den Ancker M.E. et al., 1998, A\&A, 330, 145
\bibitem{vin02}      Vink J.S. et al., 2002, MNRAS, 337, 356
\bibitem{vin05a}   Vink, J.S., 2005a, A\&A, 430, 213
\bibitem{vin05b}   Vink J.S. et al., 2005b, MNRAS, 359, 1049
\bibitem{wad07}   Wade G.A. et al., 2007, MNRAS, 376, 1145
\bibitem{wat98}    Waters L.B.F.M. \& Waelkens C., 1998, Ann. Rev. A\&A, 36, 233
\bibitem{whi92}    Whittet D.C.B. et al., 1992, ApJ, 386, 562
\bibitem{wo93}     Wood K. et al., 1993, A\&A, 271, 492
\bibitem{wo94}     Wood K. \& Brown J.C., 1994, A\&A, 291, 202
\bibitem{wo97}     Wood K. et al., 1997, ApJ, 477, 926
\end{thebibliography}
\end{document}